\documentclass[english]{article}
\usepackage[T1]{fontenc}
\usepackage[latin9]{inputenc}
\usepackage[letterpaper]{geometry}
\usepackage{color}
\usepackage{array}
\usepackage{verbatim}
\usepackage{float}
\usepackage{amsthm}
\usepackage{xcolor}
\usepackage{lipsum}
\usepackage{xfrac} 
\usepackage{graphicx,amsmath,amssymb}
\usepackage{epstopdf}
\usepackage{lineno}

\makeatletter

\newcommand{\lyxmathsym}[1]{\ifmmode\begingroup\def\b@ld{bold}
  \text{\ifx\math@version\b@ld\bfseries\fi#1}\endgroup\else#1\fi}

\providecommand{\tabularnewline}{\\}
\theoremstyle{plain}

\makeatother

\usepackage{babel}

\begin{document}

\title{A correlation between the dark content of elliptical galaxies and their ellipticity.}

\author{A. Deur\\
~\\
University of Virginia, Charlottesville, VA 22904. USA} 

\date{}

\maketitle
\begin{abstract}
Observations indicate that the baryonic matter of galaxies is
surrounded by vast dark matter halos, which nature remains unknown. 
This document details the analysis of the results published in Ref.~\cite{Deur:2013baa} 
reporting an empirical correlation between the ellipticity of elliptical galaxies and 
their dark matter content. Large and homogeneous samples
of elliptical galaxies for which their dark matter content is inferred were selected
using different methods. Possible methodological biases in the 
dark mass extraction are alleviated by the multiple methods employed.
Effects from galaxy peculiarities are minimized by a homogeneity
requirement and further suppressed statistically. After forming
homogeneous samples (rejection of galaxies with signs
of interaction or dependence on their environment, of peculiar elliptical
galaxies and of S0-type galaxies) a clear correlation emerges. 
Such a correlation is either spurious --in which case it signals an ubiquitous 
systematic bias in elliptical galaxy observations or their analysis--
or genuine --in which case it implies in particular that at equal luminosity,
flattened medium-size elliptical galaxies are on average five times
heavier than rounder ones, and that the non-baryonic
matter content of medium-size round galaxies is small. It would also
provides a new testing ground for models of dark matter and galaxy formation. 
\end{abstract}
\tableofcontents

\section{Scope of this document}

This document is an archival article that details the analysis
performed in Ref.~\cite{Deur:2013baa} to investigate
the relation between the amount of dark matter in elliptical galaxies and their shape. 
A significant correlation was found and reported in~\cite{Deur:2013baa}. 
Here, we provide the details of the analysis method, the data used and
the systematic studies conducted to understand the nature of the correlation
and the influences of specific factors such as the environment of the
galaxies. We also provide a similar analysis as that performed in~\cite{Deur:2013baa},
but for the baryonic mass rather than the dark mass. No similar correlation is found. 

\section{Introduction}
Dark matter is an essential ingredient of cosmology. It provides together 
with dark energy for a consistent 
description of many large-scale features of the universe~\cite{Komatsu}. 
However, at the galactic and semi-galactic scales, open questions remain~\cite{Auger 2, Napolitano 09}.
There, galaxies are depicted as  constituted of a vast dark matter halo surrounding a smaller baryonic component. 
One method to advance our understanding
at the galactic scale is to look for relationships between the dark matter content
of a galaxy and its observed luminous matter. Useful empirical relations have been found, e.g. that of
Tully-Fisher~\cite{Tully-Fisher} for spiral galaxies which
links the galactic rotation speed to the galaxy absolute luminosity. For
elliptical galaxies, a prominent empirical correlation is the 
``Fundamental Plane"~\cite{Djorgovski, Dressler}
which combines the Kormendy~\cite{Kormendy} and the Faber-Jackson~\cite{Faber-Jackson} 
relations and links the galactic effective radius $R_{eff}$,
the surface brightness and the dispersion
of stellar velocities $\sigma$.  Intriguingly, none of these correlation directly  involve
dark matter. Equally  puzzling is the observation that some elliptical galaxies harbor
little dark matter~\cite{Romanowsky}.
We discuss here the correlation between the defining 
characteristic of elliptical galaxies --the ellipticity, on which e.g. their
Hubble sequence  is entirely founded-- and their relative
dark matter content, expressed as the total galactic mass (baryonic+dark)
normalized to luminosity ($\sfrac{M}{L}$ ratio). 

The conventional Hubble classification
groups galaxies into four broad morphological categories 
based on visual appearance: spiral, elliptical (with smooth featureless distribution), 
lenticular (with a bright central bulge and external disk
but no spiral structure) and irregular (without well-defined structure).
In this study, we focus on elliptical galaxies. Although those
generally have tri-axial ellipsoidal structures and ellipticities depending
on radii, to first approximation the geometric structure of elliptical galaxies can generally
be simply modeled as oblate ellipsoids with constant ellipticities
($\varepsilon$). Those span a wide range,
from $\varepsilon=0$ (round galaxies) to $\varepsilon=0.7$ (highly flattened ellipsoids).
This essentially continuous variation of $\varepsilon$ 
allows us to investigate how the general elliptical galaxy characteristics evolve with $\varepsilon$.
A caveat, however, is that only the ellipticity projected on our observation plane 
(hereafter $\varepsilon_{apparent}$, the apparent ellipticity) is observed. The true
ellipticity (henceforth $\varepsilon_{true}$) can be inferred only by detailed modeling of the galaxy. Furthermore
$\sfrac{M}{L}$ for elliptical galaxies are difficult to measure comparatively to disk galaxies~\cite{Romanowsky}.
These difficulties can be circumvented statistically: in a large and homogeneous samples
of elliptical galaxies one can minimize the effects of galaxy peculiarities by a homogeneity
requirement. The large number of galaxies further suppresses these effects statistically. 
Performing this analysis with $\sfrac{M}{L}$ extracted using different
methods allows to minimize possible systematic bias associated with a particular method. 
Finally, the projection problem can also be addressed statistically: it is straightforward to model its 
overall effect on the studied correlation and correcting for it. 

The actual shape of the dark halo remains an open question. A correlation between
the amount of dark matter and the ellipticity of galaxies offers
an opportunity to experimentally address this question as well as
the fundamental questions of the interaction of baryonic and dark
matter, and the present tensions between the standard model of galaxy
formation and the data.

Throughout this document, the $\sfrac{M}{L}$ are expressed in solar unit $\sfrac{M_{\odot}}{L_{\odot}}$
and to characterize the shape of a galaxy (assumed to be oblate) we
use its axis ratio $\sfrac{R_{min}}{R_{max}}=1-\varepsilon$ where
$R_{min}$ is the minor radius and $R_{max}$ the major radius.
Finally, $DM$ will stand for Distance Modulus, not dark matter.

\section{Choice of data set and method}

We used 42 separate publications that provide $\sfrac{M}{L}$ or related ratios
for at least several elliptical galaxies. We also considered 24 other articles,
but without using their results for reasons given in Appendix~\ref{sec:Other-data}.
Overall, after selecting appropriate galaxies, we obtained 255 different
galaxies for a total of 685 data points. Tables listing the galaxies
used are given in appendices~\ref{sec:Summary-table NGC} to~\ref{sec:Summary-table-for other lenses}.
The last appendix presents our most detailed analysis, including a
study of the interrelation  between the various variables characterizing
the galaxies to check whether the $\sfrac{M}{L}$ correlation with $\sfrac{R_{min}}{R_{max}}$ that was eventually identified 
could be a consequence of other correlations. We grouped the publications
according to the methods used to extract $\sfrac{M}{L}$ (or related ratios
such as $\sfrac{M_{total}}{M_*}$ with $M_*$ the part of the galactic mass from stars, 
or the dark matter fraction $DMf=1-\sfrac{M_*}{M_{total}}$).
In the following sections we describe the methods, their advantages
and caveats. In the subsections, we summarize our analysis for each
data set and its specificities. We classified the results into reliability
groups defined as:
\begin{itemize}
\item Group 1: reliable results. 
\item Group 2: somewhat reliable. 
\item Group 3 somewhat less reliable. 
\item Group 4: less reliable. 
\end{itemize}
Authors whose work belong to groups 4 or 3 should not be offended: 
this grouping is pertinent solely in the context of the work relevance
to our study, and not in any other sense: all the results are published
in peer-reviewed journals (except for~\cite{Capaccioli}) and
thus should be reliable. This reliability criterion was used in Section~\ref{sec:Global-results}
to weight the results when combining them. This procedure is coarse
and might introduce some subjectivity but in practice, its effects turned out 
to be numerically small because lower reliability groups typically have lower 
statistics. After combining the results in Section~\ref{sec:Global-results},
correcting them (Section~\ref{sec:Projection-correction}) if necessary
for the fact that the correlation was studied vs the projected
axis ratio (apparent axis ratio $\sfrac{R_{min}}{R_{max}}|_{apparent}$) rather
than the intrinsic one (true axis ratio $\sfrac{R_{min}}{R_{max}}|_{true}$),
we present and discuss the global results in Section~\ref{sec:Global-results}.

\subsection{General selection criteria\label{sub:General-selection-criteria}}

For each publication, we chose of subset of the analyzed elliptical galaxies as homogeneous
as possible. This was done for two reasons: \\
1) it reduces the noise coming from effects peculiar to given galaxies 
which would obscure a possible correlation with  point-to-point uncorrelated variations;\\
2) it avoids biasing the studied correlation
with systematic effects from a class of galaxies, e.g. giant elliptical
galaxies (point-to-point correlated effect). \\
Our selection criteria are more strictly applied to sets of local galaxies and have to be
relaxed for sets of distant galaxies because these are not
characterized as accurately. (Typically, one can use only distant galaxies for strong
lensing analyses or those using the Fundamental Plane time-evolution,
see e.g. Ref.~\cite{Kelson}). 

The local and distant selection criteria
are listed bellow, with galaxy characteristics obtained from either the NASA/IPAC Extragalactic Database (NED)
\cite{NED} or from the publication that provided the $\sfrac{M}{L}$.
It is important to take notice that the selection criteria have been decided 
before carrying out the analysis. Thus, the analysis is ``blind'' with minimal subjective bias.

\subsubsection{Local galaxies \label{sub: selection criteria Local-galaxies}}

When possible, only medium size elliptical galaxies were selected.
Those tend to be ``disky'' and to have almost  isotropic random
velocities.  We also required the galaxies to be undisturbed because
galaxy interactions often compromise the applicability of the formulae used
to extract the dark matter content (e.g. strong lensing equations, virial theorem,
hydrostatic equilibrium equations). We reject galaxies
that are, according to the NASA/IPAC Extragalactic Database (NED)
\cite{NED} or the article in which $\sfrac{M}{L}$ is calculated: 
\begin{itemize}
\item Lenticular galaxies (S0-type);
\item Active Galactic Nucleus (AGN) galaxies because an AGN may signal a recent disturbance of
the galaxy. Furthermore, AGNs emit at all
wavelengths and can consequently bias our study by lowering the $\sfrac{M}{L}$
ratios;
\item LINERS galaxies (because they may be due to AGN), Seyfert (Sy) and BL
Lacertae objects (BLLAC) active galaxies, for the same reasons as
AGN;
\item Peculiar galaxies, or any galaxy listed in the Arp catalogue~\cite{Arp} of peculiar
galaxies;
\item Giant elliptical galaxies (D), supergiant elliptical galaxies (cD), 
Brightest Cluster Galaxies (BrClG), EXG~\cite{EXG CXG VCXG and XE note}
and XE galaxies. These galaxies belong to different classes of elliptical
galaxies, tend to be triaxial and moreover, the ``boxy'' giant galaxies 
are characterized by anisotropic random velocities. 
In addition, the determination of their amount of dark matter 
could be skewed by contribution from the cluster or group the
galaxy belong to;
Further reasons to reject large elliptical galaxies are provided in~\cite{Cappellari 2006};
\item Compact elliptical galaxies (cE);
\item E? galaxies in the NED because the lack of definite morphology assignment may reflect
poor measurements and may contaminate our galaxy set with non-elliptical galaxies;
\item Transition-type (E+) and galaxies formerly described as elliptical but now identified as spiral galaxies;
\item HII emission galaxies, because it may signal a recent disturbance. In any case, the presence of HII regions 
is unusual for elliptical galaxies, making those to fulfill  the ``peculiar galaxy'' rejection criterion.
 HII emission might also bias the $\sfrac{M}{L}$ determination because newly born blue stars 
increase significantly the luminosity $L$;
\item VCXG galaxies~\cite{EXG CXG VCXG and XE note} since they
show signs of disturbed hydrostatic equilibrium.
\item NELG (narrow emission line galaxy) galaxies since it may signal a recently disturbed galaxy.
\end{itemize}
We keep LERG (low excitation radio galaxies) and WLRG (weak emission-line
radio-galaxies) since we saw no obvious reason to exclude them.

\subsubsection{Distant galaxies \label{cut distant galaxies}}

Distant galaxies are usually not well enough characterized to apply
the above criteria. Nevertheless, they are very useful to include in our study since 
typically only distant galaxies are available to apply the 
strong lensing method e.g.~\cite{Auger 1} or~\cite{Jiang-Kochanek}
or that of the Fundamental Plane time-evolution, e.g.~\cite{Kelson}
or~\cite{Rettura}. The rejection criteria for distant galaxies are:
\begin{itemize}
\item Massive galaxies, with typically $M\gtrsim5\times10^{11}\mbox{M}_{\odot}$.
This criterion should minimize the amount of contamination of our sample by 
cD, D or BrClG galaxies. The choice for mass
selection is based on the work of Ferreras {\it et al.}~\cite{Ferreras2}:
the authors noted that galaxies with $M_{tot}>10^{12}M_{\odot}$ behave
differently that those with $M_{tot}\ll10^{12}M_{\odot}$;
\item Galaxies with relatively low velocity dispersions, $\sigma \leq225$~km.s$^{-1}$,
if S0 and elliptical galaxies are not separated but considered altogether in the publication, 
or if the classification may not be reliable enough. This should 
suppress possible S0 contamination: S0 tend to have $\sigma\leq225$
km.s$^{-1}$. We verified that, within a sample of well identified
local elliptical galaxies, rejecting the genuine elliptical galaxies with $\sigma\leq225$
km.s$^{-1}$ does introduce a bias of the $\sfrac{M}{L}$ vs $\sfrac{R_{min}}{R_{max}}$ correlation,
see Section~\ref{sub:Systematic-studies Prugniel-Simien}. Hence,
this criterion is adequate to reject S0 without biasing our study.
\end{itemize}
In addition, if some of the characteristics listed in Section~\ref{sub: selection criteria Local-galaxies}
are available (such as AGN or known interaction with another galaxy),
then these galaxies are also rejected.

\subsection{Uncertainties \label{Uncertainties}}

Since all the data sets will be eventually combined in a single overall 
determination of $\sfrac{M}{L}$ vs. $\sfrac{R_{min}}{R_{max}}$, the uncertainties for 
$\sfrac{M}{L}$ and $\sfrac{R_{min}}{R_{max}}$ must be estimated consistently lest galaxy samples 
with optimistic determinations of their uncertainties will be 
given an unwarranted preponderance. In addition, some 
publications do not provide any uncertainties and these then need to be assessed
without the detailed knowledge of the analyses carried in the publication. 
To obtain a consistent determination of the uncertainties, we employ
the {\it{unbiased estimate}}, i.e, we re-scaled the uncertainties% 
\footnote{If the uncertainties are not provided in the publication, we assume $\Delta \sfrac{M}{L}$
to be proportional to $\sfrac{M}{L}$, assume no uncertainty on $\sfrac{R_{min}}{R_{max}}$, 
and then apply the {\it{unbiased estimate}} method.} 
 so that, when a fit of $\sfrac{M}{L}$ vs. $\sfrac{R_{min}}{R_{max}}$ is
performed, its $\sfrac{\chi^{2}}{ndf}$ is set to unity. This supposes
a gaussian dispersion of the data of a given galaxy set. For
the fit, we choose it to be linear for simplicity and to minimize the number of fit parameters. 
If the true dependence of $\sfrac{M}{L}$ vs. $\sfrac{R_{min}}{R_{max}}$ is not linear, then
the $\sfrac{\chi^{2}}{ndf}$ would increase which in turn would increase
the final uncertainties when $\sfrac{\chi^{2}}{ndf}$ is set to unity. Hence,
the linearity assumption is accounted for in the final uncertainty.
%When the dark matter content was provided without uncertainty, we
%assigned one proportional to the dark matter content.

\noindent A caveat of forcing $\sfrac{\chi^{2}}{ndf}$ to unity when fitting
$\sfrac{M}{L}$ vs {\it apparent} axis ratio comes from the fact that a large
part of the data scatter is not due to the data gaussian dispersion.
Rather, it is due to the random projection of the real 3D shape of
the galaxy onto our 2D observation plan, see Section~\ref{sec:Projection-correction}
and in particular Fig.~\ref{Flo:R/R projection effect1-1} or~\ref{Flo:R/R projection effect2-1}.
Still, we adopte the procedure of forcing $\sfrac{\chi^{2}}{ndf}=1$ because
1) it is the simplest method to estimate uncertainties when data
are provided without uncertainties; 2) we can apply the same procedure
to all data sets for consistency; and 3) it was a conservative procedure,
as discussed above. 

Finally, conscious that some assumptions underlie the fit  procedure and the unbiased 
estimate method, we independently assessed the degree of  correlation using 
the Pearson  correlation coefficient.

\subsection{Systematic studies}
In sections~\ref{sec:Data-sets-using virial theo} to~\ref{sec:Data-sets-using strong lensing}, 
we summarize the analyses made on each galaxy set for
obtain $\sfrac{M}{L}$ vs $\sfrac{R_{min}}{R_{max}}$. When possible,
the systematic effect associated with a particular galaxy characteristic was also 
studied. This was done when the set contained a large enough number of galaxies and  
the galaxy characteristic was provided. We list below these auxiliary studies.
\begin{itemize}
\item Effect of galactic metallicity  using the Lauer data~\cite{Lauer} based on the virial theorem. (no effect) ;
\item Influence of the correlation between $\sfrac{M}{L_B}$ and central luminosity density, using the Lauer data~\cite{Lauer} based on the virial theorem. (Large effect);
\item The effect of luminosity was also studied using the Auger {\it et al.} data~\cite{Auger 1} based on lensing. This also check the correlation with galaxy boxy/disky shape: more luminous galaxies tend to be boxy and less ones tend to be disky. No correlation between the $DMf$ and the luminosity/boxy-disky galaxy character was observed. 
\item Effect rejecting bona-fide galaxies with low velocity dispersion, using the Prugniel \& Simien data~\cite{Prugniel}  based on the virial theorem. (No effect);
\item Effects of LINERS, using the Prugniel \& Simien data~\cite{Prugniel}  based on the virial theorem. (No effect);
\item Effect of ellipticity projection, using Bertola {\it et al.}~\cite{Bertola93} and Pizzella {\it et al.}~\cite{Pizeella}, both based on analyzing embedded gas disks, and  the Barnabe {\it et al.}~\cite{Barnabe} data based on lensing. (Large effect, except for the Barnabe {\it et al.}~\cite{Barnabe} data).
\item The effect of environment was studied using the Auger {\it et al.}~\cite{Auger 1}, Barnabe {\it et al.}~\cite{Barnabe} and Cardone {\it et al.}(2009, 2011)~\cite{Cardone09, Cardone11} data sets based on lensing. Galaxies residing in clusters tend to show a smaller $DMf$ vs axis ratio correlation (smaller slope) with more jitter (larger relative uncertainty). The lensing data set from Cardone {\it et al.}(2009)~\cite{Cardone09} shows, however, the opposite trend.  

\item The effect of choosing a particular initial mass function (IMF) to interpret the data can be checked with 
the Auger {\it et al.}~\cite{Auger 1} and Barnabe {\it et al.}~\cite{Barnabe} data sets based on lensing, 
and the Deason {\it et al.}~\cite{Deason} data employing PNe and GC.
See also the Cappellari {\it et al.}~\cite{Cappellari 2006} data based on using light profile measurements and stellar orbit modeling.
\end{itemize}

\section{Data sets using virial theorem \label{sec:Data-sets-using virial theo}}

Galactic mass-to-light ratio can be extracted using the virial theorem. Its
simple form (scalar virial theorem) gives $\sfrac{M}{L}=\sfrac{c \sigma^{2}}{Ir}$
where $\sigma$ is the stellar velocity dispersion, $c$ a proportionality coefficient, 
$r$ a given radius (e.g. the effective radius $R_{eff}$ or the core radius) %$r_{c}$)
and $I$ is the surface brightness. The value of $c$ is in principle the same
for galaxies of an homogeneous sample. The formula assumes
spherical symmetry and a virialized system. The ellipticity of the
galaxy can be accounted for using the tensor virial theorem, see~\cite{BMS}
and references within. 
The data discussed in this section were obtained using ellipticity
corrected virial formulae. When the published data did not account
for ellipticity, we used the Bacon {\it et al.}~\cite{BMS}
ellipticity corrections formula (simplified formula using central dispersion
only and $\sfrac{R_{min}}{R_{max}}|_{true}=\sfrac{R_{min}}{R_{max}}|_{apparent}$). This
correction, derived analytically, is important. It was independently
verified by the results of van der Marel~\cite{van der Marel 1991},
see Section~\ref{sub:van-der-Marel91}. In addition, without this
correction, the averaged $\sfrac{M}{L}$ would be too small compared to an
expected value around 8 $\sfrac{M_{\odot}}{L_{\odot}}$. We remark that the
modification to the (spherical) Newton Shell Theorem to include ellipticity
yields a similar figure, see Section~\ref{sec:Data-sets-using X-ray}.

Caveats for this method of extracting $\sfrac{M}{L}$ are:
\begin{itemize}
\item Anisotropy effects are generally not accounted for;
\item Only $\sfrac{M}{L}$ for the galaxy inner part are obtained. This one tends
to be rounder and dominated by baryonic matter, thereby biasing and
diluting the effect we study;
\item A constant $\sfrac{M}{L}$ is assumed. Although it is now established that
$\sfrac{M}{L}$ varies significantly with the galaxy radius, see e.g.~\cite{Kronawitter, Thomas, van der Marel 1991, Capaccioli, Magorrian01,
Nagino, Napolitano, Bertola93}, a constant
$\sfrac{M}{L}$ is a good assumption for $r<R_{eff}$. 
\end{itemize}
Another potential caveat of this method is that unphysical correlations
between $\sfrac{M}{L}$ and $\sfrac{R_{min}}{R_{max}}$ can be induced by anisotropic
star motions. However, this should not be an issue because, as assessed
in Ref.~\cite{BMS}, the effect is small. Furthermore anisotropies
have little effects on the slopes of Figs. 1a and 1b of~\cite{BMS}
whilst affecting primarily the absolute values. Because we are investigating
a dependence of $\sfrac{M}{L}$ on $\sfrac{R_{min}}{R_{max}}$, the absolute scale of $\sfrac{M}{L}$
is of secondary importance. Therefore, possible effects of the
anisotropies are not critical. Furthermore, the absolute values of
$\sfrac{M}{L}$ will be normalized  to the expected $\sfrac{M}{L}$=8 $\sfrac{M_{\odot}}{L_{\odot}}$
when all the data sets are combined, see Section~\ref{sec:Global-results}.
In addition, bright galaxies --which tend to display these anisotropies--
are discarded from our samples by our selection criteria. Our assessment 
that anisotropies should be of small concern was {\it a posteriori} justified
by checking that the results from~\cite{BMS} --which used a large
set of galaxies and the virial method-- displayed no significant $\sfrac{M}{L}$
vs brightness correlation, see the correlation study in appendix
\ref{sec:Detailed-analysis-of Bacon et al}.

\subsection{Bacon {\it et al.} (1985)\label{sub:Bacon-et-al.}}

Bacon {\it et al.}~\cite{BMS} extracted $\sfrac{M}{L_B}$ ($\sfrac{M}{L}$
in the B-band) for 197 early-type galaxies. We analyzed these data
thoroughly  using additional selection criteria%
\footnote{We required  in addition to the selection criteria described in~\ref{sub: selection criteria Local-galaxies}
that the galaxy characteristics listed in the 1985 Ref.~\cite{BMS}
are compatible with the up-to-date (as of 2008, when the analysis was performed) characteristics provided
in the NED~\cite{NED}.} 
and studying possible correlations between various galaxy characteristics
that could have biased our study. The full analysis is presented in
appendix~\ref{sec:Detailed-analysis-of Bacon et al}. Depending on
data availability, ellipticity corrections formulae of various accuracies
were used.

\begin{figure}
\centering
\includegraphics[scale=0.5]{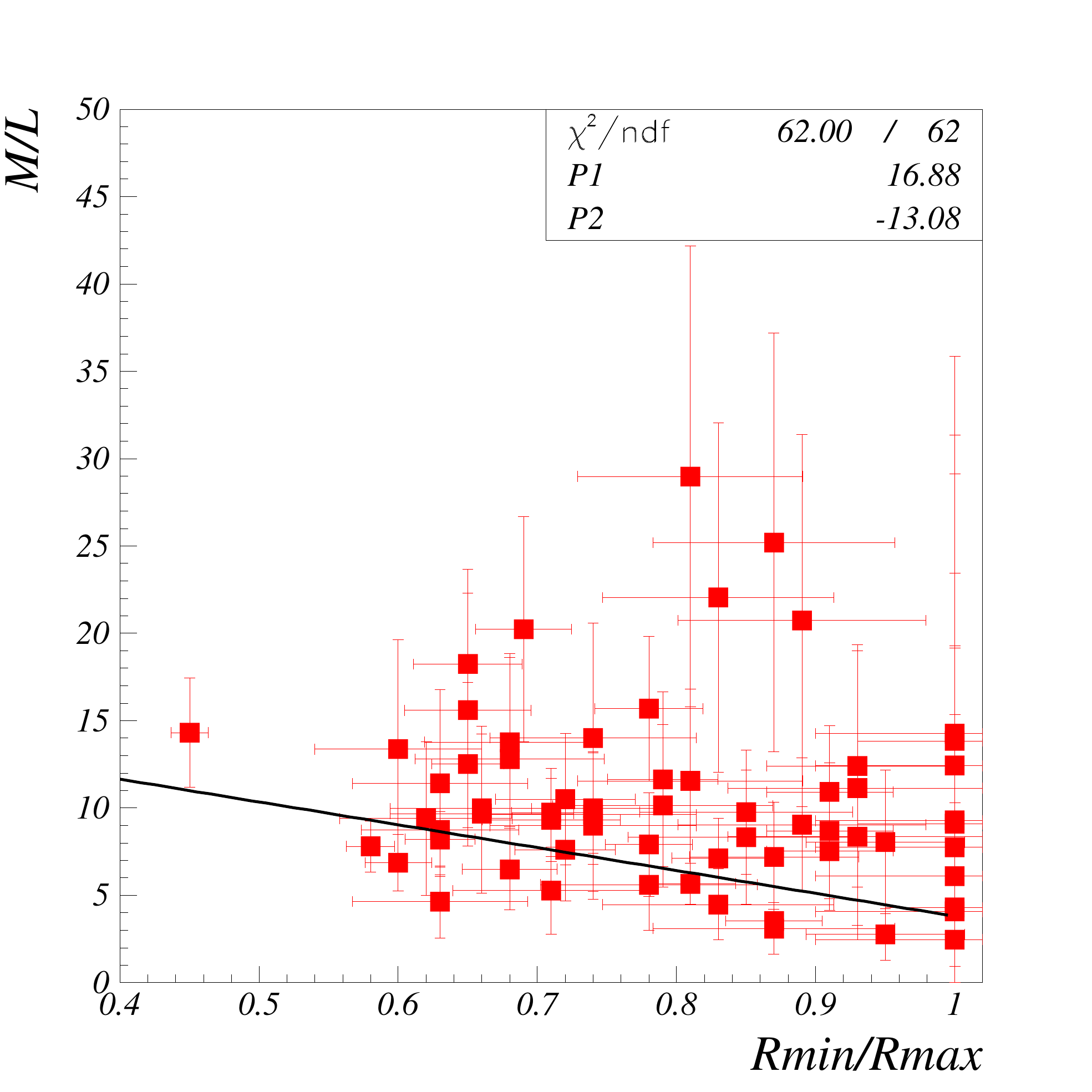}
\vspace{-0.5cm} \caption{\label{fig: bacon 1}
$\sfrac{M}{L_B}$ vs apparent axis ratios from Bacon {\it et al.}~\cite{BMS}
for Sample 1 (ellipticity corrections from the $\sigma^{2}$ isotropic
method). In this figure and the 40 others that follow, the straight
line shows the best linear fit to the data. The $\sfrac{\chi^{2}}{ndf}$
is given in the top right box together with the $\sfrac{M}{L}$-intercept  at $\sfrac{R_{min}}{R_{max}}=0$
($P1$) and the slope ($P2$).}
\end{figure}

\begin{figure}
\centering
\includegraphics[scale=0.5]{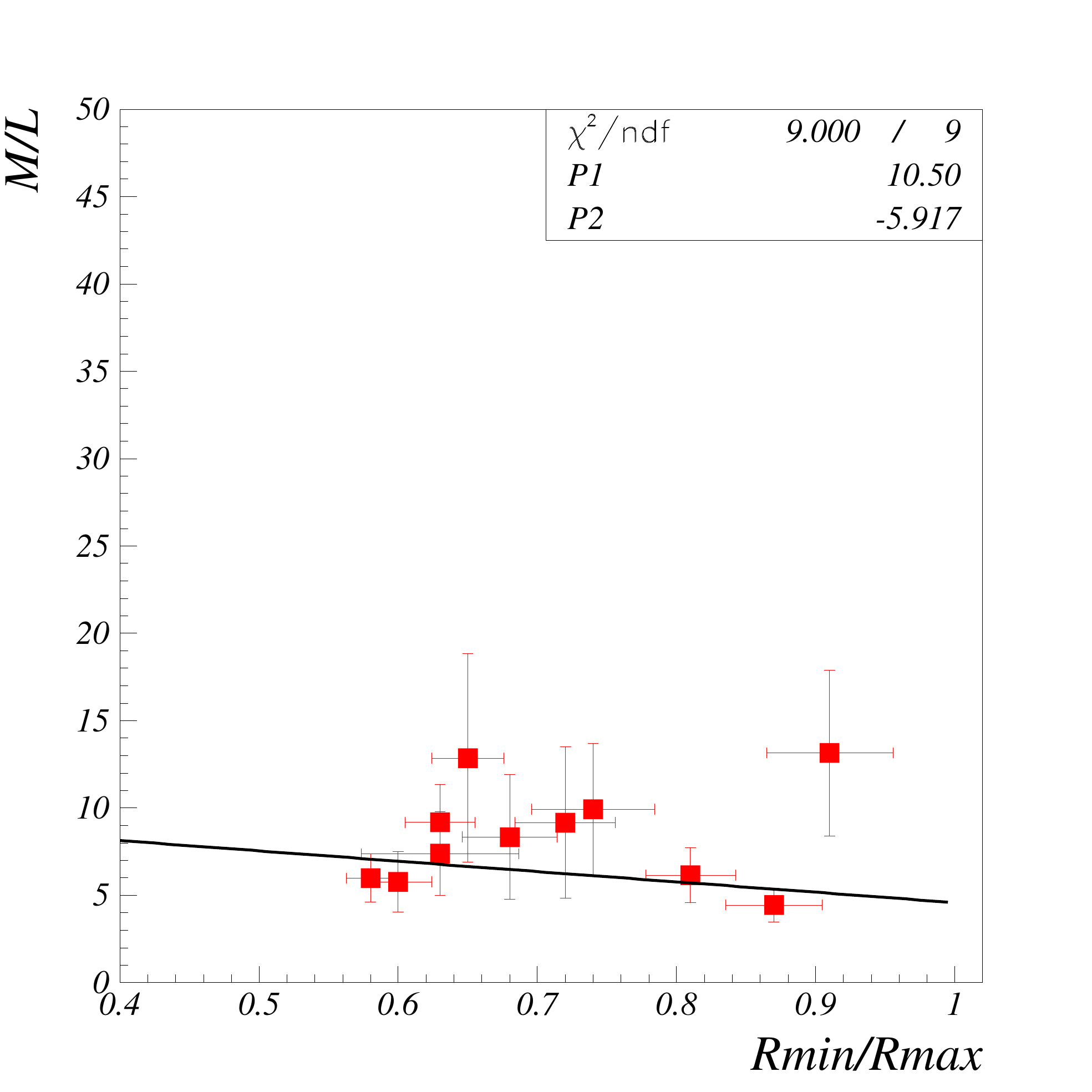}
\vspace{-0.5cm} \caption{\label{fig: bacon 2}
$\sfrac{M}{L_B}$ vs apparent axis ratios from Bacon {\it et al.}~\cite{BMS}
for Sample 2 (ellipticity corrections from the $\mu^{2}$ isotropic
method).}
\end{figure}

\begin{figure}
\centering
\includegraphics[scale=0.5]{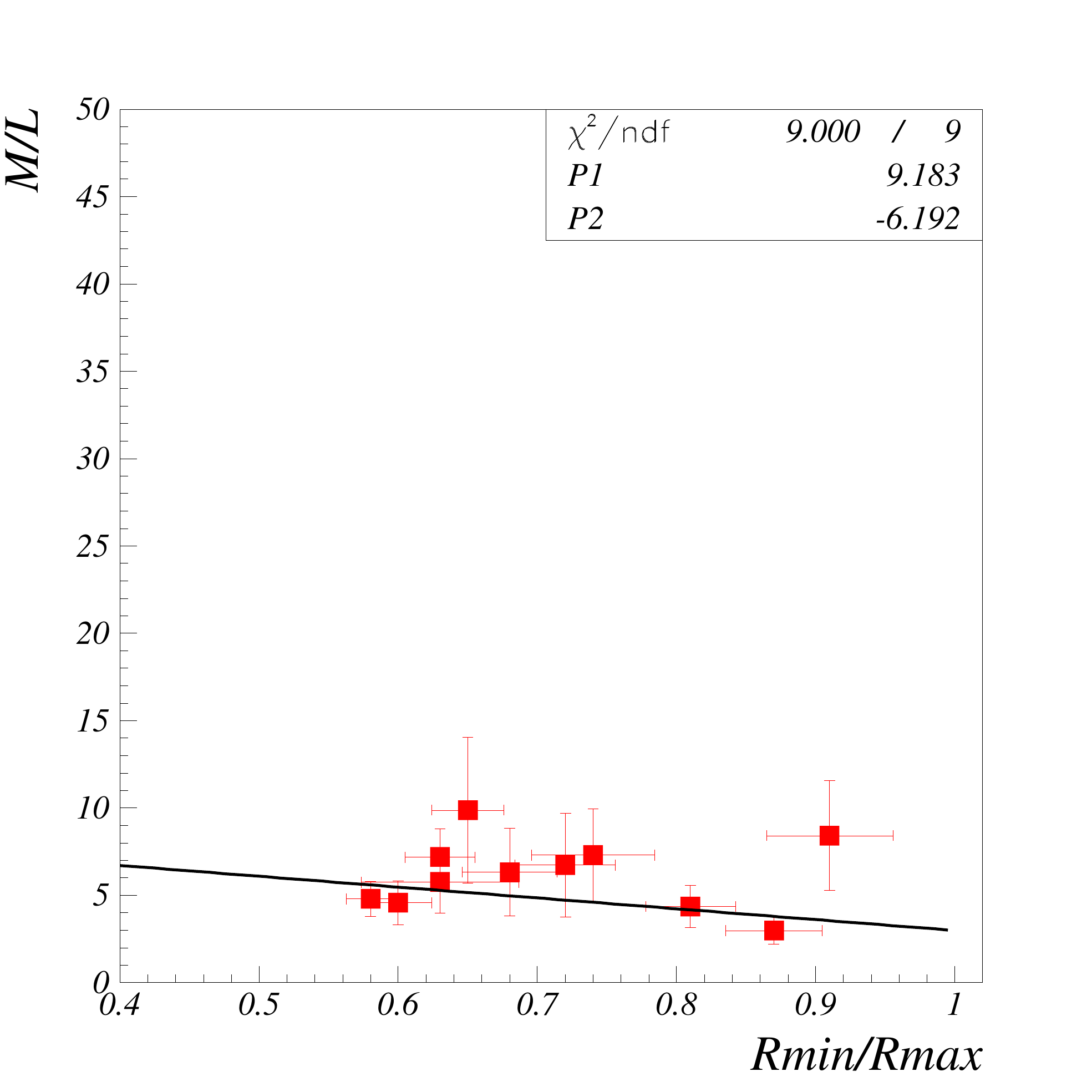}
\vspace{-0.5cm} \caption{\label{fig: bacon 3}
$\sfrac{M}{L_B}$ vs apparent axis ratios from Bacon {\it et al.}~\cite{BMS}
for Sample 2 (ellipticity corrections from the $\mu^{2}$ anisotropic
method).}
\end{figure}

After applying the selection criteria, we were left with of 64 elliptical
galaxies in our main sample (Sample 1). Better ellipticity corrections
are available for 11 galaxies (Sample 2). The results of fits to these
data using the Bacon {\it et al.} values for $\sfrac{R_{min}}{R_{max}}$ and
distance moduli are shown in Figs.~\ref{fig: bacon 1},~\ref{fig: bacon 2}
and~\ref{fig: bacon 3} and are:

$\sfrac{M}{L_{B}}=(-13.08\pm2.97)\sfrac{R_{min}}{R_{max}}|_{apparent}+(16.88\pm2.32)$
for Sample 1 (ellipticity corrections from $\sigma^{2}$ isotropic
method). 

$\sfrac{M}{L_{B}}=(-5.91\pm4.67)\sfrac{R_{min}}{R_{max}}|_{apparent}+(10.50\pm3.48)$
for Sample 2 (ellipticity corrections from $\mu^{2}$ isotropic method, where $\mu^{2}$ is the quadratic
sum of the galaxy velocity and its central velocity dispersion $\sigma$, see~\cite{BMS}). 

$\sfrac{M}{L_{B}}=(-6.19\pm3.59)\sfrac{R_{min}}{R_{max}}|_{apparent}+(9.18\pm2.64)$
for Sample 2 (ellipticity corrections from $\mu^{2}$ anisotropic method). 

The uncertainties on $\sfrac{R_{min}}{R_{max}}$ were taken as the difference
between the Bacon {\it et al.} and NED values. The $\sfrac{M}{L_B}$ uncertainties
are from Bacon {\it et al.,} rescaled so that $\sfrac{\chi^{2}}{ndf}=1$ ({\it{unbiased estimate}}).

\noindent There are several advantages to this data set: \\
$\bullet$ The virial formula used to obtain the published $\sfrac{M}{L}$ is 
corrected for galactic ellipticity;\\ 
$\bullet$ The large data set (197 galaxies) allowed
strict criteria to be used to remove unsuitable or suspicious galaxies;\\ 
$\bullet$ Three methods were used to obtain $\sfrac{M}{L}$; \\ 
$\bullet$ Triaxial galaxies were excluded
from the original data set, as they are less suited for the virial
method. 

The analysis is dated (1985) but we have used only data consistent
with the more recent NED numbers, so we do not consider this to be
an important caveat. We assigned the results to group 2 reliability
for Samples 1 and 2 with ellipticity corrections from the $\mu^{2}$
isotropic method, and reliability group 1 for Sample 2 with ellipticity
correction from the $\mu^{2}$ anisotropic method.

\subsection{Bender {\it et al.} (1989)}

\begin{figure}
\centering
\includegraphics[scale=0.5]{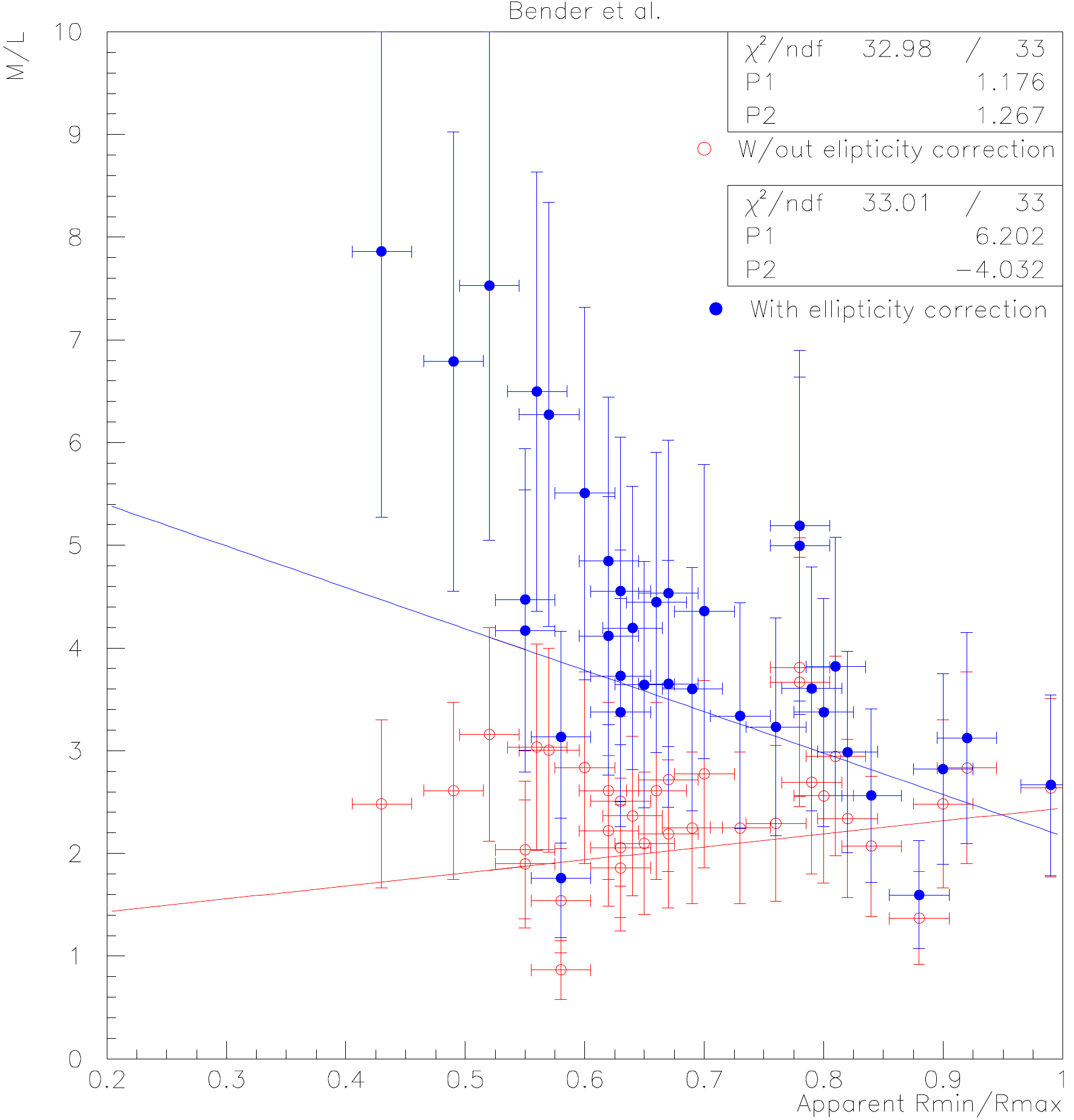}
\vspace{-0.4cm} \caption{\label{fig: bender}$\sfrac{M}{L_B}$ vs apparent axis ratios from Bender {\it et al.}~\cite{Bender}
before applying ellipticity corrections (empty red circles) and after
(blue filled circles).
}
\end{figure}

Bender {\it et al.}~\cite{Bender} analysis assumed spherical
symmetry. There was 109 galaxies in the original sample. After selection,
we retained 35 galaxies. $\sfrac{M}{L_B}$ vs apparent axis ratio is shown
in Fig.~\ref{fig: bender}. The fit result for the 35 selected galaxies
(using NED values for $\sfrac{R_{min}}{R_{max}}$ with a $\pm0.025$ uncertainty)
is:

$\sfrac{M}{L_{B}}=(-4.03\pm1.44)\sfrac{R_{min}}{R_{max}}|_{apparent}+(6.20\pm1.07)$. 

The $\sfrac{M}{L_B}$ uncertainty was assigned to be proportional to $\sfrac{M}{L_B}$
with the overall scale factor adjusted so that $\sfrac{\chi^{2}}{ndf}=1$.

Specific caveats of this analysis are: the analysis is dated; There
were frequent large discrepancies between the values used in the original
analysis and the NED values (e.g. redshift, ellipticity or luminosity values); No uncertainty was provided. 

We remark that the $\sfrac{M}{L_B}$ values without ellipticity corrections
appear to be too low: in average, $\left\langle \sfrac{M}{L_B}\right\rangle =2.05\pm0.08$.
Using the more modern value for the Hubble constant $H_{0}=70$ km/s/Mpc
rather than 50 km/s/Mpc as used by the authors, we obtained $\left\langle \sfrac{M}{L_B}\right\rangle =1.46\pm0.06$.
This is well below the expected (minimal) value of $\left\langle \sfrac{M}{L_B}\right\rangle \sim4$
$\sfrac{M_{\odot}}{L_{\odot}}$ for an elliptical galaxy without dark matter
contribution, and $\left\langle \sfrac{M}{L_B}\right\rangle \sim8$ $\sfrac{M_{\odot}}{L_{\odot}}$
with dark matter contribution. This proves  the necessity of the ellipticity
corrections. We assigned the results to group 3 reliability.

\subsection{Kelson {\it et al.} (2000)}

\begin{figure}
\centering
\includegraphics[scale=0.5]{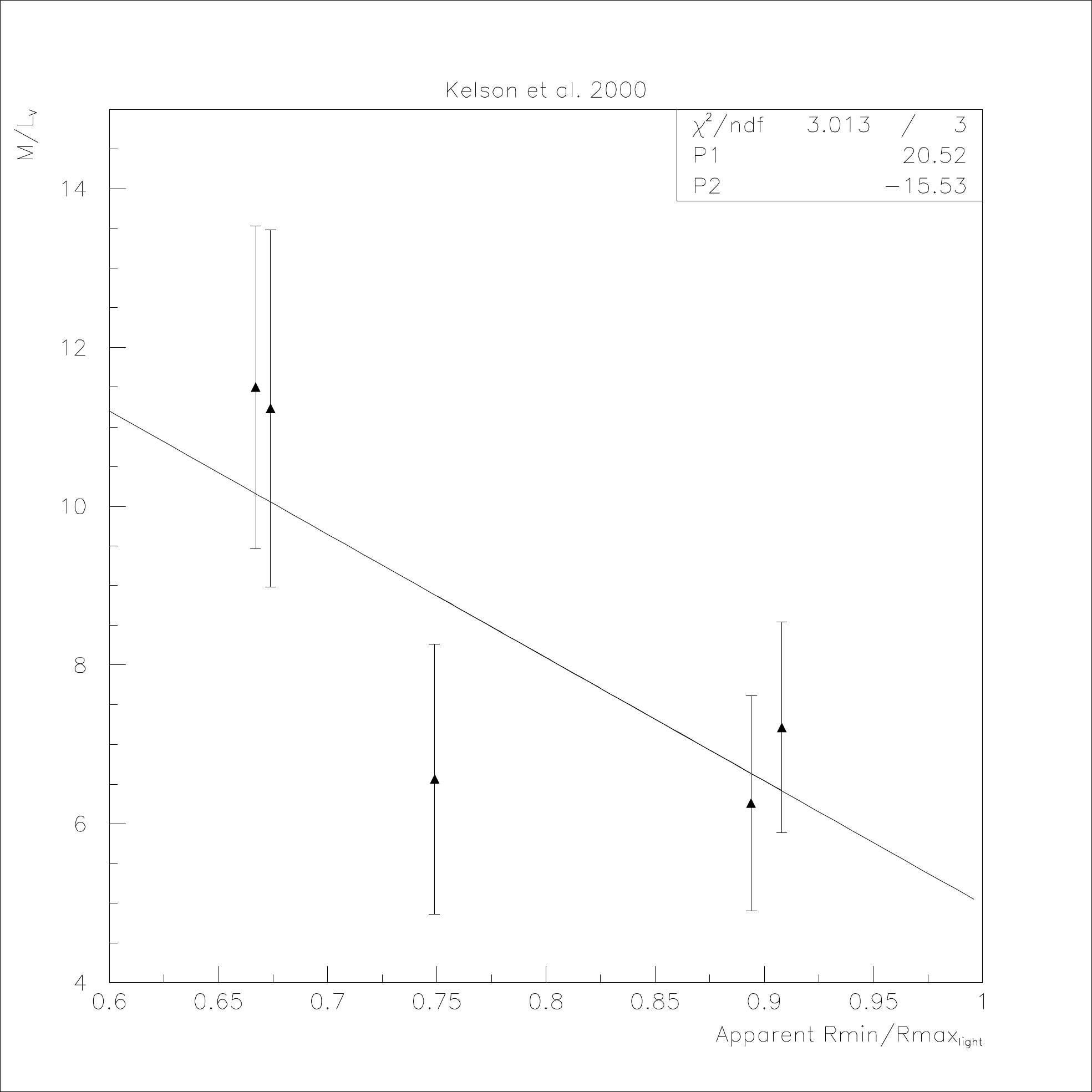}
\vspace{-0.4cm} \caption{\label{fig: Kelson}$\sfrac{M}{L_B}$ vs apparent axis ratio from Kelson {\it et al.}~\cite{Kelson}
(after applying ellipticity corrections).
}
\end{figure}

Kelson {\it et al.}~\cite{Kelson} determined the internal kinematics,
length scale and surface brightness of 53 galaxies from cluster CL1358+62,
11 of them elliptical galaxies. The data were used to form $\sfrac{M}{L_B}$.
Applying an absolute magnitude selection $M_{B}\leq-19.5$ to minimize
the contamination from boxy galaxies would have left only one available
galaxy. Relaxing the selection to $M_{B}\leq-20$ left 5 galaxies
to study. We applied the ellipticity corrections from Bacon {\it et
al.}~\cite{BMS}. The $\sfrac{M}{L_B}$ vs apparent axis ratio is shown
in Fig.~\ref{fig: Kelson}. Uncertainties were slightly scaled to
force $\sfrac{\chi^{2}}{ndf}=1$. %
 The best linear fit is:

$\sfrac{M}{L_B}=(-15.53\pm7.21)\sfrac{R_{min}}{R_{max}}|_{apparent}+(20.52\pm5.95)$.

The specific caveats of this analysis are that the galaxies are not
as well characterized (morphology, spectra) as local ones, and they
belong to a dense cluster. Thus, they may not be in virial equilibrium,
and we cannot apply the usual strict criteria that select galaxies likely to be in virial
equilibrium. Consequently, we assigned these data
to group 4 reliability.

{\footnotesize 
\noindent For completeness, the fit before ellipticity
ellipticity corrections yields $\sfrac{M}{L}=(-2.29\pm4.79)\sfrac{R_{min}}{R_{max}}|_{apparent}+(7.78\pm3.79)$.}
{\footnotesize \par}

\subsection{Lauer  (1985)}

\begin{figure}
\centering
\includegraphics[scale=0.5]{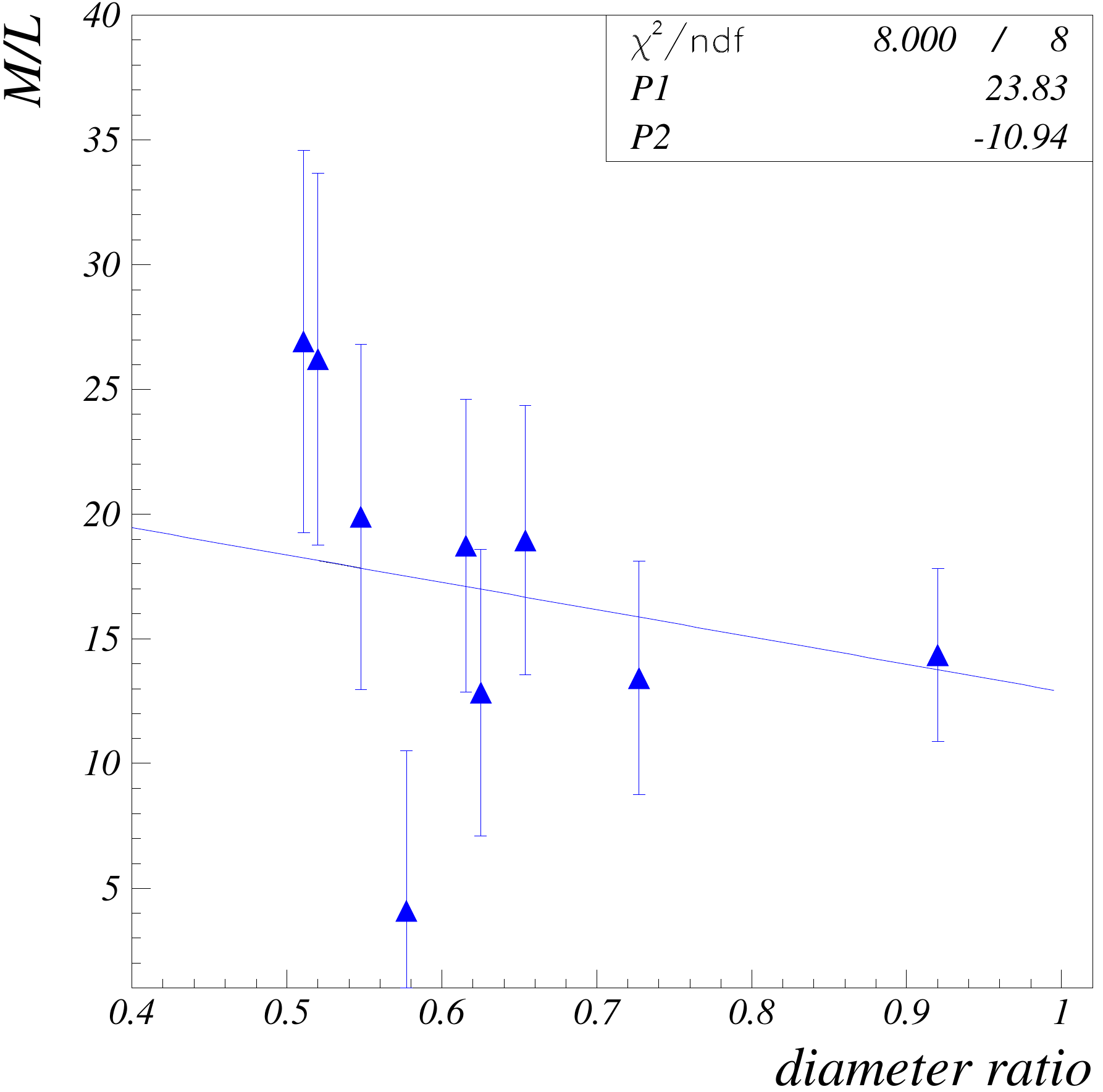}
\vspace{-0.5cm} \caption{\label{fig:Lauer results class1-2-3}$\sfrac{M}{L_B}$ ratio vs diameter ratio $\sfrac{R_{min}}{R_{max}}$, from Lauer
\cite{Lauer}, for the 10 galaxies of our sample. The original
Lauer data is corrected for ellipticity and uncertainties are assigned
so that $\sfrac{\chi^{2}}{ndf}=1$. 
}
\end{figure}

The Lauer data set~\cite{Lauer} contains 42 galaxies. $\sfrac{M}{L_B}$
was extracted from the stellar velocity dispersion in the galaxy core,
assuming a spherical star distribution. Thus, the $\sfrac{M}{L}$ were determined
for radii $r\lesssim0.03R_{eff}$. The $\sfrac{M}{L_B}$ ratios were given
without uncertainty estimates. The uncertainty we assigned did not
follow our standard procedure: we assigned a constant uncertainty
of $\Delta(\sfrac{M}{L})=3.0$ rather than $\Delta(\sfrac{M}{L})\propto \sfrac{M}{L}$
because of the low $\sfrac{M}{L}$ outlier at $\sfrac{R_{min}}{R_{max}}=0.58$, see
Fig.~\ref{fig:Lauer results class1-2-3}. With the standard procedure,
the outlier would have had a very small uncertainty and would have
driven the fit. The value $\Delta(\sfrac{M}{L})=3.0$ is chosen so that
$\sfrac{\chi^{2}}{ndf}=1$. Lauer classified galaxies in 3 classes, depending
on the quality of the core resolution. Class 1 is for well resolved
cores, class 2 for partially resolved cores and class 3 for unresolved
cores. In addition to $\sfrac{M}{L_B}$ for the core, Lauer provided a secondary
analysis that determined more globally $\langle \sfrac{M}{L_B}\rangle$ using effective
radius rather than the core radius, which extends his analysis beyond
the galaxy core. However, because his main analysis concerns galaxy
cores, we will consider only $\sfrac{M}{L_B}$ rather than $\langle \sfrac{M}{L_B}\rangle$.
We applied our usual selection criteria with, as in Section~\ref{sub:Bacon-et-al.},
the added requirement  that the galaxy characteristics listed in Lauer's
1985 publication are compatible with the one in
NED~\cite{NED}, with the exception of the axis ratio because it
was not provided in~\cite{Lauer}. We used the NED values for
the axis ratios. The sample was reduced to 10 galaxies after selection.
If we had restricted the sample to galaxies with resolved (class 1)
or partially resolved (class 2) cores, only two galaxies would have
remained (one in class 1: NGC 720 and one class 2: NGC 7619). Hence
we did not apply this additional requirement and used galaxies from
the 3 classes indiscriminately.

The $\sfrac{M}{L_B}$ dependence with axis ratio is shown in Fig~\ref{fig:Lauer results class1-2-3}.
The fit results gives:

$\sfrac{M}{L_{B}}=(-10.94\pm10.18)\sfrac{R_{min}}{R_{max}}|_{apparent}+(23.83\pm7.80)$. 

The data set has several specific caveats: it is dated and no ellipticity
correction was originally done. The $\sfrac{M}{L_B}$ ratios were computed
for the galaxy cores and given without uncertainty estimates. The
uncertainty was determined following a particular procedure rather
than our standard one due to an outlier having a $\sfrac{M}{L}$ value close to 0. Because
of the above caveats, we assigned the results to group 4 reliability.

{\footnotesize For information we note that:}{\footnotesize \par}
\begin{itemize}
\item {\footnotesize Before applying ellipticity corrections the fit was:
$\sfrac{M}{L_{B}}=(+10.2\pm6.7)\sfrac{R_{min}}{R_{max}}|_{apparent}+(2.8\pm4.5)$. }{\footnotesize \par}
\item {\footnotesize Results using the more recent NED distances~\cite{NED}
are: $\sfrac{M}{L_{B}}=(-5.35\pm10.16)\sfrac{R_{min}}{R_{max}}|_{apparent}+(21.17\pm7.78)$.}{\footnotesize \par}
\item {\footnotesize Results for the global mass to light ratio $\langle \sfrac{M}{L_B}\rangle$
are: $\langle \sfrac{M}{L_B}\rangle=(+14.70\pm14.73)\sfrac{R_{min}}{R_{max}}|_{apparent}+(7.71\pm11.29)$.
Using the more recent NED distances: $\langle \sfrac{M}{L_B}\rangle=(+24.56\pm17.67)\sfrac{R_{min}}{R_{max}}|_{apparent}+(2.51\pm13.53)$.
The opposite correlation sign appears to be spurious and due to
correlations between the luminosity density $\rho_{e}$ and $\sfrac{R_{min}}{R_{max}}$ on
the one hand and $\rho_{e}$ and $\langle \sfrac{M}{L_B}\rangle$ on the other. 
This is discussed next.}{\footnotesize \par}
\end{itemize}

\paragraph{Correlations}

\subparagraph{Correlation with metallicity }

Lauer noticed a correlation between galaxy metallicity $\mbox{M}_{\mbox{g}_{2}}$
and $\sfrac{M}{L}$. Indeed we found $\sfrac{M}{L}=(61.0\pm24.9)\mbox{M}_{\mbox{g}_{2}}-(6.4\pm7.6)$
and $\langle \sfrac{M}{L}\rangle=(55.2\pm24.3)\mbox{M}_{\mbox{g}_{2}}-(4.6\pm7.4)$. However,
we found no clear correlation between $\mbox{M}_{\mbox{g}_{2}}$ and
$\sfrac{R_{min}}{R_{max}}$. Consequently, the relation between $\mbox{M}_{\mbox{g}_{2}}$
and $\sfrac{M}{L}$ should not induce the correlation oberved between $\sfrac{M}{L}$
and $\sfrac{R_{min}}{R_{max}}$.

\subparagraph{Correlation with central luminosity density}

Lauer also signaled a correlation between the luminosity density $\rho$
and $\sfrac{M}{L}$. We found $\sfrac{M}{L}=(-6.5\pm1.6)\rho_{0}+(22.2\pm2.6)$
and $\langle \sfrac{M}{L}\rangle=(7.3\pm1.8)\rho_{e}-(3.0\pm4.4)$. Notice the opposite
correlations between central and effective densities. There
is no clear correlation between $\rho_{0}$ and $\sfrac{R_{min}}{R_{max}}$,
whilst a correlation exists between $\rho_{e}$ and $\sfrac{R_{min}}{R_{max}}$:
$\sfrac{R_{min}}{R_{max}}=(0.14\pm0.04)\rho_{e}+(0.40\pm0.10)$. Thus,
the correlation between $\langle \sfrac{M}{L}\rangle$ and $\sfrac{R_{min}}{R_{max}}$ may
be biased by, or even due to, the ($\rho_{e},\sfrac{R_{min}}{R_{max}}$) and
($\langle \sfrac{M}{L}\rangle,\rho_{e}$) correlations. This can explain the opposite (positive)
sign of the ($\sfrac{R_{min}}{R_{max}}$, $\langle \sfrac{M}{L}\rangle$) correlation compared
to the (negative sign) ($\sfrac{R_{min}}{R_{max}}$, $\sfrac{M}{L}$) correlation:
subtracting the $\rho_{e}$ correlation from $\langle \sfrac{M}{L_B}\rangle=(+14.70\pm14.73)\sfrac{R_{min}}{R_{max}}|_{apparent}+\mbox{cst}$,
we get $\langle \sfrac{M}{L}\rangle=(-37.4\pm44.9)\sfrac{R_{min}}{R_{max}}|_{apparent}+\mbox{cst}$, consistent
with the negative slope in $\sfrac{M}{L_{B}}=(-10.94\pm10.18)\sfrac{R_{min}}{R_{max}}|_{apparent}+(23.83\pm7.80)$
and the expectation that $\sfrac{M}{L}$ below $R_{eff}$ should not vary strongly.
We will not use the $\langle \sfrac{M}{L_B}\rangle$ results in the global analysis of
Section~\ref{sec:Global-results}.

\subsection{Leier (2009) }

Leier~\cite{Leier09} analyzed distant galaxies using both the
strong lensing and the virial methods. These results are discussed
in Section~\ref{sub:Leier-data-(2009)}. The result based on the virial
method were assigned to group 2 reliability.

\subsection{Prugniel \& Simien (1996)\label{sub:Prugniel-and-Simien}}

\subsubsection{Main study}

Prugniel \& Simien~\cite{Prugniel} compiled kinematics and
photometric data for 371 early-type galaxies. They remarked that this
set might be biased toward flat galaxies since those are preferably
chosen for major axis kinematic measurements. This should be of no
consequence in our context. We selected adequate elliptical galaxies
following the usual morphology and emission criteria. We did not use
the galaxies from the PCG catalog, although some seem to be good
elliptical galaxies, since they all have low velocity dispersions.
The set obtained contains 102 galaxies.

\begin{figure}
\centering
\includegraphics[scale=0.5]{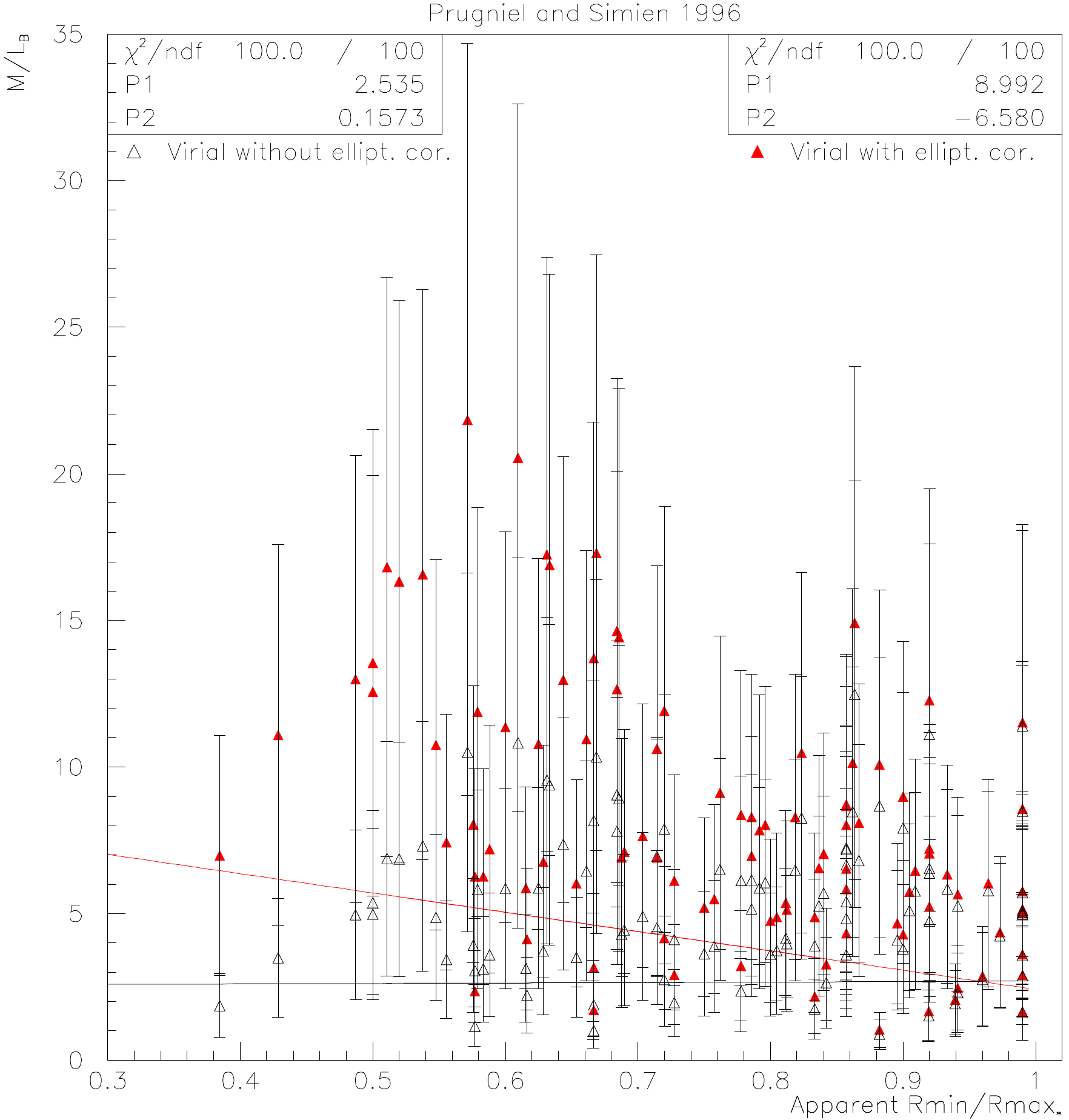}
\vspace{-0.4cm} \caption{\label{fig: prugniel}$\sfrac{M}{L_B}$ vs apparent axis ratios from Prugniel \& Simien~\cite{Prugniel}
before applying ellipticity corrections (empty triangles) and after
(red filled triangles).
}
\end{figure}

We formed $ $$\sfrac{M}{L_B}\propto\sigma_{0}/L_{B}^{\sfrac{1}{2}}I_{e}^{\sfrac{1}{2}}$ from
the virial theorem and corrected it for ellipticity using the virial
tensor correction formula from Bacon {\it et al.}~\cite{BMS}.
The galaxy type and characteristics, including axis ratio, were obtained
from NED. The $\sfrac{M}{L}$ uncertainties were assumed to be proportional
to $\sfrac{M}{L}$ and set so that $\sfrac{\chi^{2}}{ndf}=1$. $\sfrac{M}{L_B}$ vs apparent
axis ratio is shown in Fig.~\ref{fig: prugniel}. %
 The best linear fit is:

$\sfrac{M}{L_B}=(-6.58\pm1.98)\sfrac{R_{min}}{R_{max}}|_{apparent}+(8.99\pm1.68)$.

The advantage of this analysis is that the data are numerous and relatively
recent compared to the other extractions of  $\sfrac{M}{L}$ that used the virial theorem. The
only specific caveat of this analysis is that no uncertainties are
provided. We assigned these results to group 2 reliability. 

{\footnotesize For information, the result before applying the ellipticity
corrections is: $\sfrac{M}{L_B}=(+0.16\pm1.19)\sfrac{R_{min}}{R_{max}}|_{apparent}+(2.54\pm0.93)$.}{\footnotesize \par}

\subsubsection{Systematic studies\label{sub:Systematic-studies Prugniel-Simien}}

\paragraph*{Effect of a velocity dispersion selection}

The large sample of well identified elliptical galaxies provided the
opportunity to check the effect of removing from the sample good elliptical
galaxies with low velocity dispersion $(\sigma<225$~km.s$^{-1}$).
Such selection is applied in the analyses of distant galaxies that
have less reliable type identification, in order to minimize possible
S0 contamination. However, this selection may at the same time also remove genuine elliptical
galaxies. Thus, it is important to check if a low velocity dispersion selection produces a bias. 
Once the $\sigma>225$~km.s$^{-1}$ criterion was applied, only 44 galaxies out of 112 remained.
The mean value of $\sfrac{M}{L}$ increased significantly because $\sfrac{M}{L}\propto\sigma$.
However, normalizing the fit result to the same average $\langle \sfrac{M}{L}\rangle$ value
as in the main study yielded a correlation similar to that of the main study:

$\sfrac{M}{L_B}=(-7.10\pm1.05)\sfrac{R_{min}}{R_{max}}|_{apparent}+(9.57\pm0.89)$,

\noindent  This indicates that no bias arises from the the $\sigma>225$
km.s$^{-1}$ criterion. This test is statistically significant because
more than 50\% of the 112 galaxies has been removed. (We performed
the same study on data without ellipticity corrections and reached
the same conclusion).

\paragraph*{Effects of LINERs}

We added 17 LINER elliptical galaxies to the main sample. The best
fit from this larger sample including LINERs agrees well with the main
result given in previous section. However, this is not statistically
significant because the sample size increased only by 13\% (from
112 to 127 galaxies). We also checked the results of the linear fit
using only LINERs and found it compatible with the main data (again
with large statistical uncertainty). However, we notice that: 
\begin{itemize}
\item The LINER distribution may be biased toward apparently rounder galaxies:
94$\pm24$\% of LINER have $\sfrac{R_{min}}{R_{max}}|_{apparent}>0.65$, for only
76$\pm8$\% of our non-LINER sample. If true, this could bias
the magnitude of the $\sfrac{M}{L}$ and $\sfrac{R_{min}}{R_{max}}$ correlation in
the case of the LINER if that correlation is not linear (e.g. a weaker correlation at 
high $\sfrac{R_{min}}{R_{max}}$ values would lower more the average $\sfrac{M}{L}$ vs 
$\sfrac{R_{min}}{R_{max}}$ slope in the case of LINERs).
\item The shape of the $\sfrac{M}{L}$ vs $\sfrac{R_{min}}{R_{max}}$ distribution is similar
for LINER and non-LINER. The are both flat before ellipticity corrections
and thus similar after applying identical ellipticity corrections. 
\end{itemize}

\subsection{Rettura{\it et al.} (2006)}

\begin{figure}
\centering
\includegraphics[scale=0.5]{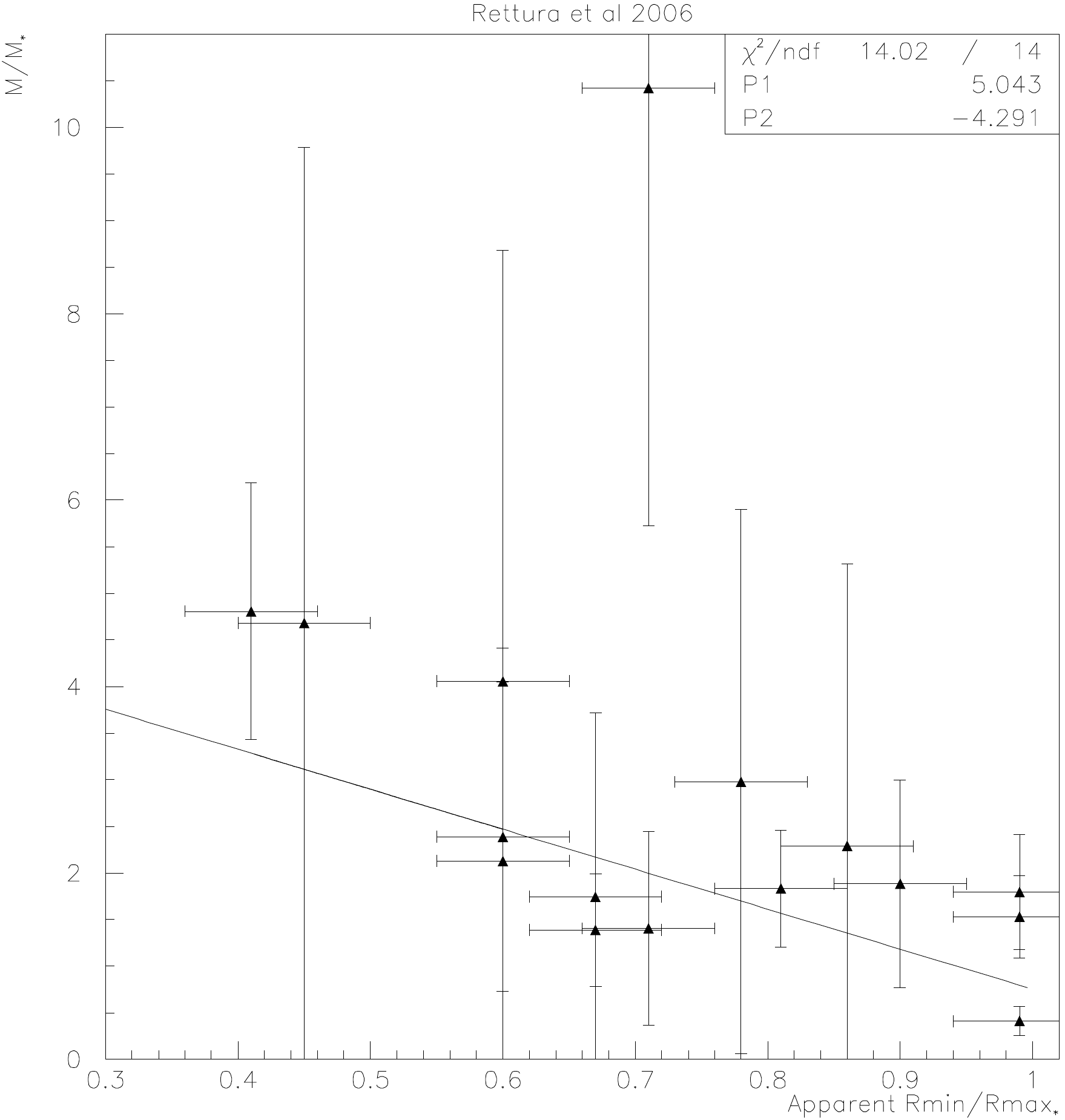}
\vspace{-0.4cm} \caption{\label{fig: rettura}$\sfrac{M_{dyn}}{M_*}$ vs apparent axis ratio from Rettura {\it et al.}
\cite{Rettura} after applying ellipticity corrections. The uncertainties
on $\sfrac{M_{dyn}}{M_*}$ have been scaled so that $\sfrac{\chi^{2}}{ndf}\simeq1$.
}
\end{figure}

Rettura {\it et al.}~\cite{Rettura} studied the relation between
galaxy stellar masses $M_*$ and dynamical masses $M_{dyn}$ for
a sample of 37 elliptical, 5 lenticular and 6 bulge-dominated
spiral galaxies. $M_*$ was obtained by matching the spectral energy distribution
to composite stellar population model templates obtained from stellar
population models in which a Kroupa Initial Mass Function -IMF-~\cite{Kroupa}
was assumed. Several models were used but the results are not strongly
model-dependent. $M_*$ includes masses from stars and stars remnants
(white dwarves, Black Holes,...) masses. $M_{dyn}$ was computed assuming
that galaxies are spheroidal non rotation-supported systems in virial
equilibrium: $M_{dyn}\propto\sigma_{0}^{2}R_{eff}$. From the 37 elliptical galaxies, 
we formed a sample excluding known AGN, two interacting giant galaxies belonging
to the CL1252 cluster, O-II emission galaxies, E+A galaxies, and large
X-ray emission galaxies. Furthermore, because the morphology of the
galaxies is not as well known as for local galaxies, we excluded galaxies
with velocity dispersion $\sigma_{0}<200$~km.s$^{-1}$ to minimize possible S0
contamination (this selection could {\it{a priori}} bias our investigation
because of the relation $M_{dyn}\propto\sigma_{0}^{2}$. However,
our study in Section~\ref{sub:Prugniel-and-Simien} indicated no such
bias). Our final sample contains 16 galaxies. The axis ratios are
obtained from~\cite{van der wel},~\cite{di Serego} or
\cite{Holden}. We assumed a $\pm0.05$ uncertainty on them. We
applied the ellipticity corrections from Bacon {\it et al.}~\cite{BMS}.
The $\sfrac{M_{dyn}}{M_*}$ vs apparent axis ratio is shown in Fig.~\ref{fig: rettura}.
 The best linear fit is:

$M_{dyn}/M_*=(-4.29\pm1.36)\sfrac{R_{min}}{R_{max}}|_{apparent}+(5.04\pm1.25)$.

Advantages of this analysis are that the stellar component was calculated
differently from the other analyses and the data are relatively recent.
A caveat is the additional model dependency
(stellar population model, IMF choice) of the method. In addition, the galaxies
are not as well characterized (morphology, spectra) as local ones.
We assigned these data to group 3 reliability.

{\footnotesize For completeness, the fit before ellipticity corrections
yields $M_{dyn}/M_*=(-1.36\pm0.67)\sfrac{R_{min}}{R_{max}}|_{apparent}+(2.03\pm0.59)$).}{\footnotesize \par}

\subsection{van der Wel{\it et al.} (2006)}

\begin{figure}
\centering
\includegraphics[scale=0.5]{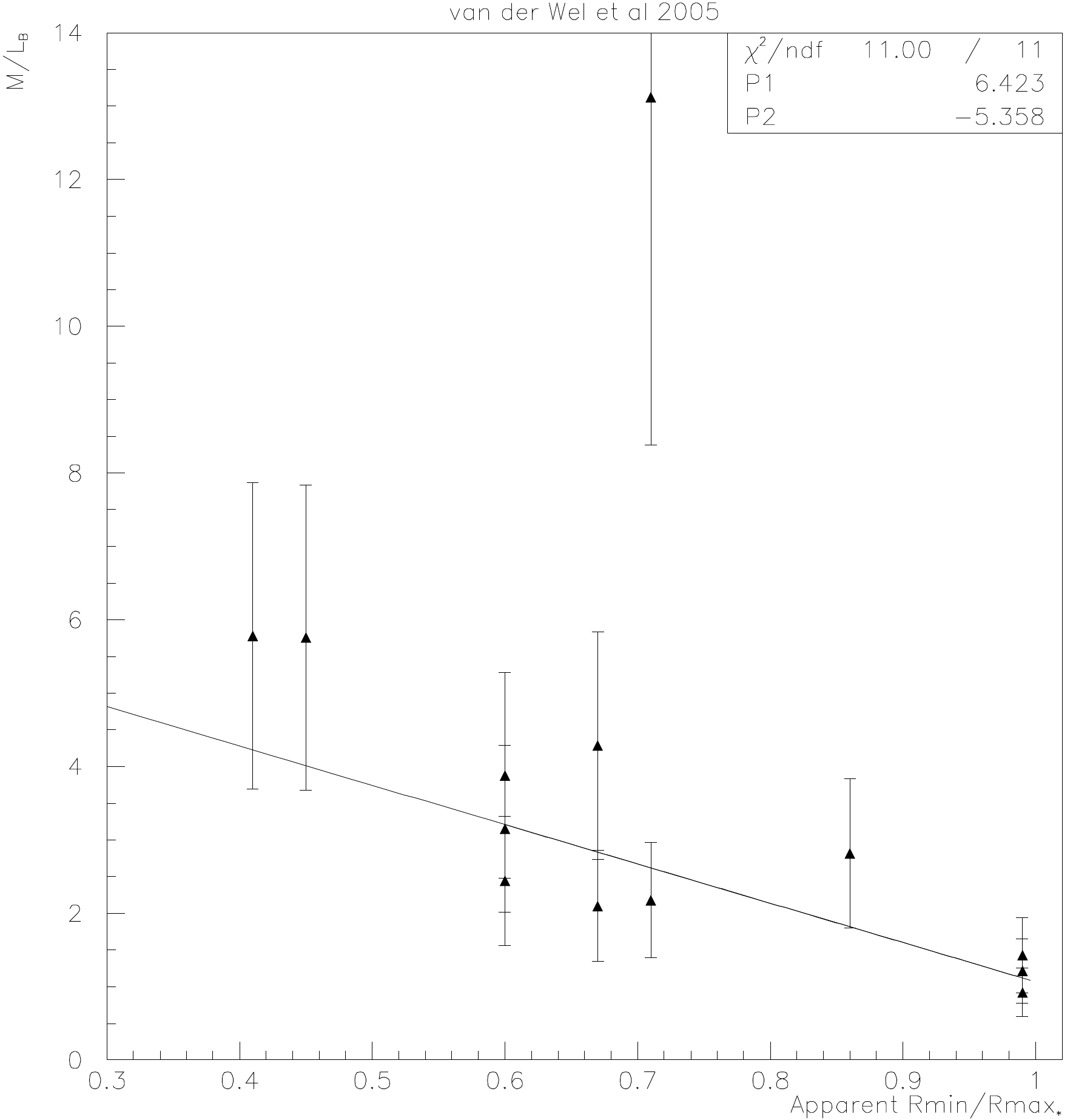}
\vspace{-0.4cm} \caption{\label{fig: vanderwel}$M_{dyn}/L_{B}$ vs apparent axis ratio from van der Wel {\it et
al.}~\cite{van der wel} after applying ellipticity corrections. The {\it{unbiased estimate}} method 
(forcing $\sfrac{\chi^{2}}{ndf}=1$ by rescaling  uncertainties has been applied.
}
\end{figure}

In Ref.~\cite{van der wel}, van der Wel {\it et al.} used a similar
galaxy set as Rettura {\it et al.}~\cite{Rettura}. The $\sfrac{M}{L_B}$
vs apparent axis ratio is shown in Fig.~\ref{fig: vanderwel}. We
apply the ellipticity corrections from Bacon {\it et al.}~\cite{BMS}.
There are 13 galaxies in our sample. %
 The best linear fit is:

$\sfrac{M}{L_B}=(-5.36\pm1.24)\sfrac{R_{min}}{R_{max}}|_{apparent}+(6.42\pm1.12)$.

As for Rettura {\it et al.}~\cite{Rettura}, we assign these
data to group 3 reliability.

{\footnotesize For completeness, the fit before ellipticity correction
yields $\sfrac{M}{L_B}=(-1.12\pm0.75)\sfrac{R_{min}}{R_{max}}|_{apparent}+(2.28\pm0.57)$)}{\footnotesize \par}

\section{Data using light profile measurements and stellar orbit modeling}

This approach relies on modeling the stellar dynamics of galaxies
using photometry data at different radii (light profile). Potentials
for stellar and dark components are assumed and the light of the stars
moving in the total potential is simulated. Parameters of the potentials
are adjusted until the models match the data. We applied on the data
sets the usual homogeneity selection criteria. However, when models
did not rely on virial or isothermal equilibrium, we have relaxed
the standard selection and include LINERS, AGN and Seyfert galaxies.
An advantage of this method is that the true ellipticity of the galaxy
can be inferred and accounted for in the determinations of $\sfrac{M}{L}$. 
A caveat is that the results are necessary model-dependent.

\subsection{Cappellari {\it et al.} SAURON IV data set (2006)\label{sec:Cappellari06}}

\begin{figure}
\centering
\includegraphics[scale=0.5]{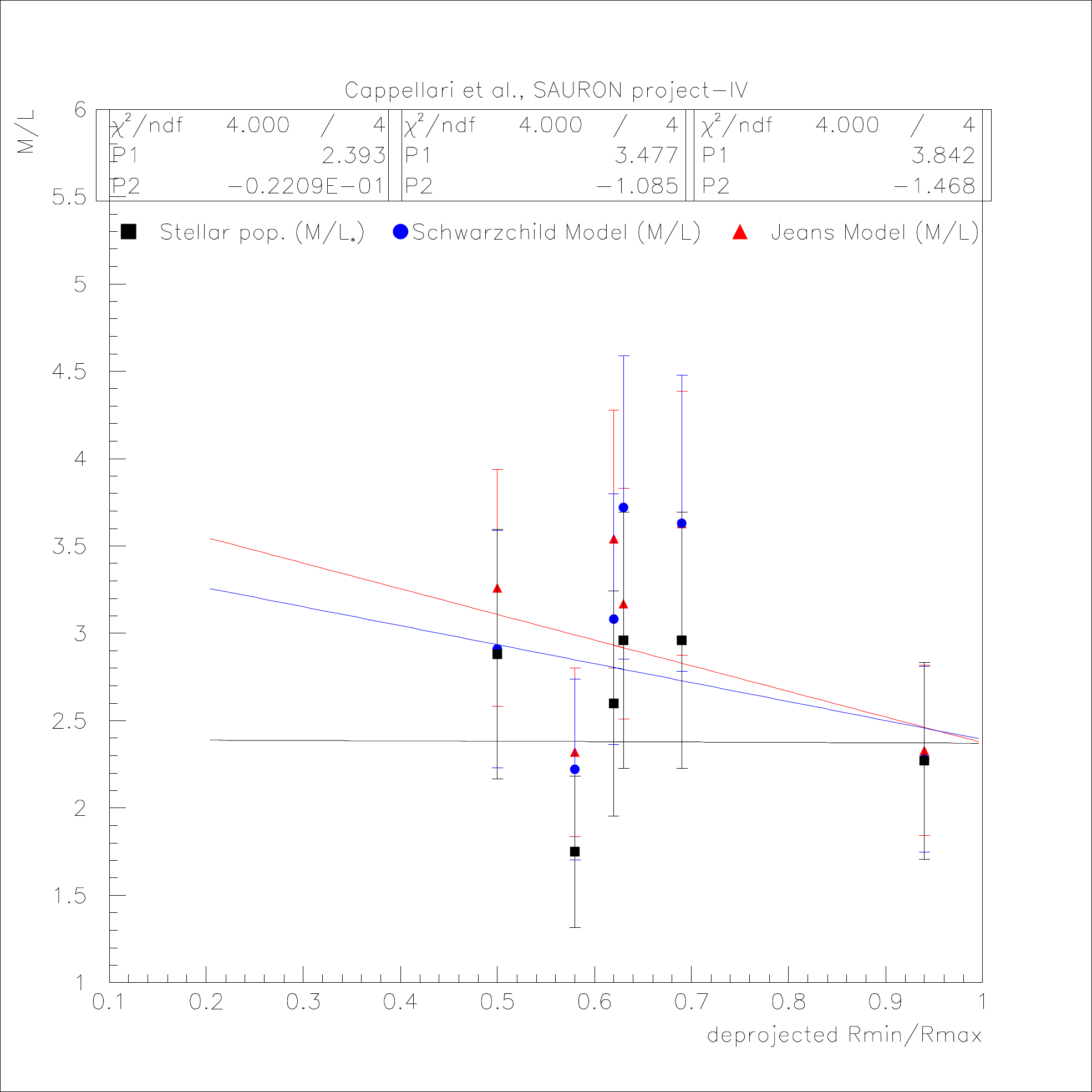}
\vspace{-0.4cm} \caption{\label{fig:sauron-4}$\sfrac{M}{L}$ vs intrinsic axis ratio $\sfrac{R_{min}}{R_{max}}|_{true}$
from M. Cappellari {\it et al}.~\cite{Cappellari 2006}. The
red triangles, blue circles and black squares are for $\left.\sfrac{M}{L}\right|_{jeans}$,
$\left.\sfrac{M}{L}\right|_{Schw}$ and $\sfrac{M}{L}_*$ respectively. 
}
\end{figure}

Cappellari {\it et al.}~\cite{Cappellari 2006} used stellar
photometry data to provide $\sfrac{M}{L}$ at $r=R_{eff}$ for 25 nearby bright
elliptical galaxies. Careful extraction of $\sfrac{M}{L}$ is argued, leading
the authors to estimate a 6\% error on $\sfrac{M}{L}$ and 10\% on stellar
mass to light ratios, $\sfrac{M_*}{L}$. The true ellipticity was input in the calculations.
It was obtained by modeling the photometry data. The authors chose
to work with the velocity dispersion at $R_{eff}$, $\sigma_{e}$,
rather than the central velocity dispersion $\sigma_{0}$ because
$\sigma_{e}$ is less sensitive to the aperture used for the observation device, and to details of
orbitals. Out of an initial sample of 48 galaxies the authors selected
25 galaxies, requesting them to have accurate distance determination,
a Hubble Space Telescope WFPC2 photometry and no strong evidence of
a bar. After our selection we obtained a set of 6 galaxies. The authors
provided two estimates of $\sfrac{M}{L}$. One is based on a 2-integral Jeans
model (numerically accurate but less general) and one on a 3-integral
Schwarzschild model~\cite{Schwarzschild} (numerically noisier
but more general). The authors cautioned  that large elliptical galaxies
(more massive and slow rotators, redder and more metal rich) tend
to be tri-axial, which may alter the $\sfrac{M}{L}$ determination. The fits
to the $\sfrac{M}{L}$ vs ellipticity are shown in Fig.~\ref{fig:sauron-4}.
The best fits, slightly  corrected for the stellar $\sfrac{M_*}{L}$ dependence
with the intrinsic axis ratio $\sfrac{R_{min}}{R_{max}}|_{true}$ are:

$\left.\sfrac{M}{L}\right|_{Jeans}=(-1.47\pm1.56)\sfrac{R_{min}}{R_{max}}|_{true}+(3.84\pm1.10)$
and

$\left.\sfrac{M}{L}\right|_{Schw}=(-1.09\pm1.68)\sfrac{R_{min}}{R_{max}}|_{true}+(3.48\pm1.17)$. 

\noindent The fit result for the stellar population is $\sfrac{M}{L}_*=(-0.02\pm1.72)\sfrac{R_{min}}{R_{max}}|_{true}+(2.39\pm1.16)$. 

The analysis is based on a robust $\sfrac{M}{L}$ extraction, with true ellipticity
and projection effect accounted for. We assigned the results to group
1 reliability.

\subsection{Cappellari {\it et al.} ATLAS project (2013)\label{sub:Cappellari-et-al. 2013}}

\begin{figure}
\centering
\includegraphics[scale=0.47]{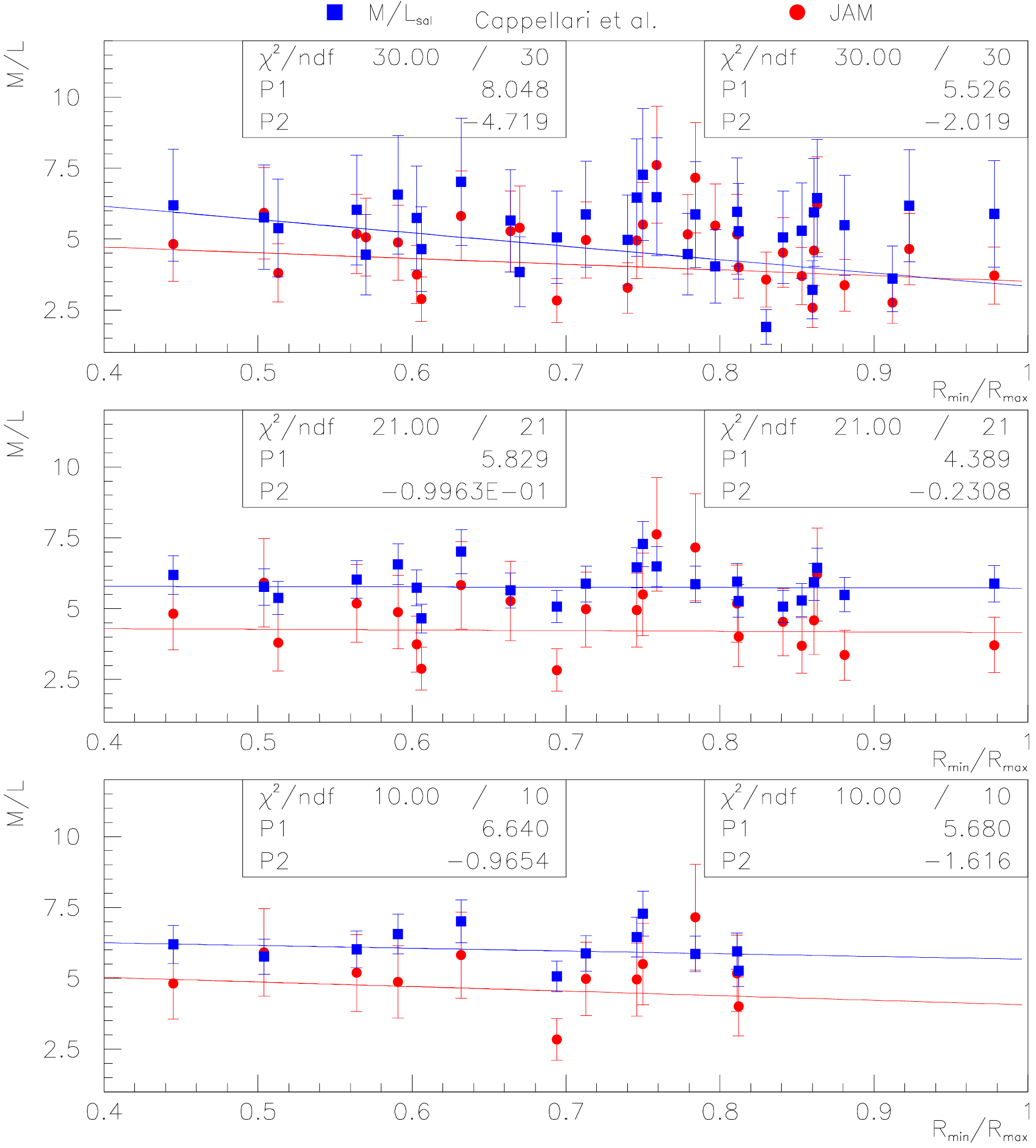}
\vspace{-0.4cm} \caption{\label{fig:cappellari13}$\sfrac{M}{L}$ vs intrinsic axis ratio
from M. Cappellari {\it et al}.~\cite{Cappellari 2013a, Cappellari 2013b}.  
The top panel shows the full sample, the middle panel a subsample with galaxy models  of quality 1 to 3,
and the bottom panel a subsample with model quality 2 or 3. 
The blue squares represent data obtained with a Salpeter IMF for modeling the light distribution,
and the red circles were obtained with the JAM model. The best fit parameters are provided in the 
left boxes (Salpeter IMF) or right boxes (JAM model). 
}
\end{figure}

The data used here come from two publications from Cappellari 
{\it et al.}~\cite{Cappellari 2013a, Cappellari 2013b}.
The authors analyzed a set of 32 elliptical galaxies using either the Salpeter IMF or the JAM model for 
modeling the light distribution. The authors also provide a quality factor for their extracted $\sfrac{M}{L}$
and ellipticities, classifying 
each galaxy model from 0 (lower quality) to 3 (better quality). 
Thus, in addition to the full galaxy sample, we also considered two subsamples, the largest containing 23
galaxies with models of quality 1 to 3, 
and a smaller sample of 12 galaxies with models of quality 2 or 3.
(The quality 3 model provides a sample of only 4 galaxies, a too small sample to be of use).
The fits to the $\sfrac{M}{L}$ vs  axis ratio are shown for the full sample and 
subsamples in Fig.~\ref{fig:cappellari13}.
The best fits for these samples are:

Quality 0-3, Salpeter IMF: $\left.\sfrac{M}{L}\right|=(-4.72\pm 2.19)\sfrac{R_{min}}{R_{max}}|_{true}+(8.04\pm1.69)$

Quality 0-3, JAM model:~ $\left.\sfrac{M}{L}\right|=(-2.02\pm 1.48)\sfrac{R_{min}}{R_{max}}|_{true}+(5.53\pm1.13)$

---

Quality 1-3, Salpeter IMF: $\left.\sfrac{M}{L}\right|=(-0.10\pm 0.97)\sfrac{R_{min}}{R_{max}}|_{true}+(5.83\pm0.30)$

Quality 1-3, JAM model:~ $\left.\sfrac{M}{L}\right|=(-0.23\pm 1.72)\sfrac{R_{min}}{R_{max}}|_{true}+(4.39\pm0.54)$

---

Quality 2-3, Salpeter IMF: $\left.\sfrac{M}{L}\right|=(-0.97\pm 1.07)\sfrac{R_{min}}{R_{max}}|_{true}+(6.64\pm1.07)$

Quality 2-3, JAM model:~ $\left.\sfrac{M}{L}\right|=(-1.62\pm 3.23)\sfrac{R_{min}}{R_{max}}|_{true}+(5.68\pm2.22)$

~

The $\sfrac{M}{L}$ used in the analysis are robust and the true ellipticities are assessed consistently. 
Thus, we assigned the results to group 1 reliability.

As the four (sub)samples are not independent, we can use only one of them 
when combining all the data reported in this article to obtain an overall correlation. 
We used the results from the full sample (32 galaxies). 

\subsection{Kronawitter {\it et al.} (2000)\label{sec:Kronawitter}}

\begin{figure}
\centering
\includegraphics[scale=0.5]{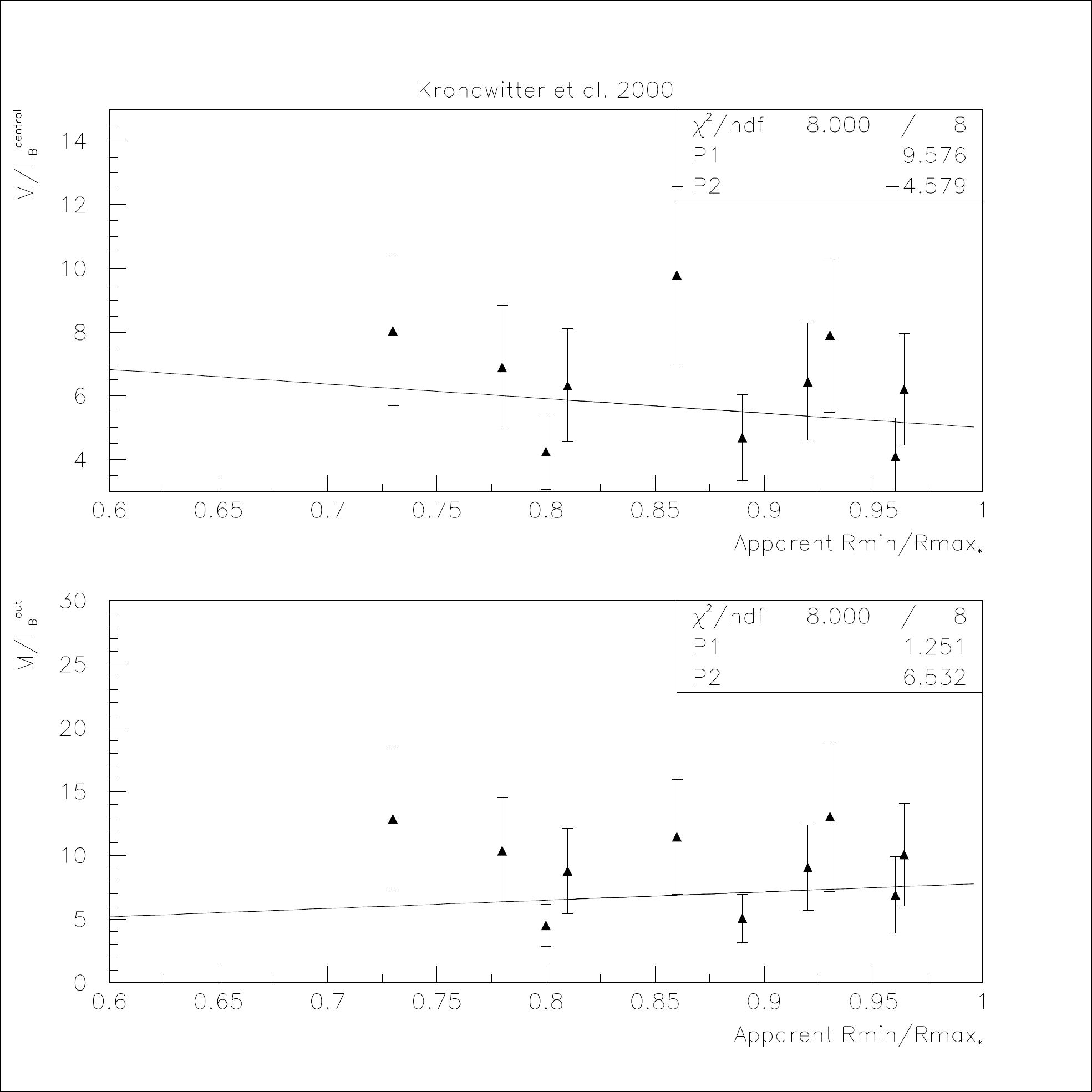}
\vspace{-0.4cm} \caption{\label{fig:kronawitter}$\sfrac{M}{L_B}$ vs apparent axis ratio from Kronawitter {\it et al.}~\cite{Kronawitter}.
The top (bottom) panel is for the inner (outer) $\sfrac{M}{L_B}$. 
}
\end{figure}

Kronawitter {\it et al.}~\cite{Kronawitter} used non-parametric
spherical models to map the gravitational fields of 21 round elliptical
galaxies. From the maps, $\sfrac{M}{L_B}$ in the inner (central $\sfrac{M}{L}$)
and outer (outermost kinematic data point) parts of the galaxies were
calculated. The technique used here did not rely on isothermal or
virial equilibrium. We thus rejected only the cD, S0, peculiar, cE,
SA0 and Arp galaxies present in the initial sample, keeping in particular
LINERS, AGN and Seyfert galaxies. We remark that after standard selection,
only three elliptical galaxies would have remained. This relaxed selection
provided a sample of 10 galaxies. To estimate the model uncertainty,
we took the difference between low and high $\sfrac{M}{L}$ values provided
in~\cite{Kronawitter}. In addition, we added an uncertainty
proportional to $\sfrac{M}{L_B}$ to account for possible experimental uncertainties
that would be common to the low and high estimates. The uncertainties
were adjusted so that $\sfrac{\chi^{2}}{ndf}\simeq1$ for the best fits. The authors
estimated the effects of non-sphericity on their data and we corrected
their $\sfrac{M}{L_B}$ based on their numbers. Fig.~\ref{fig:kronawitter} shows $\sfrac{M}{L}_{B}$
vs apparent axis ratio. The best
fits to the data are:

$\sfrac{M}{L_B}|_{inner}=(-4.58\pm7.04)\sfrac{R_{min}}{R_{max}}|_{apparent}+(9.58\pm6.17)$.

$\sfrac{M}{L_B}|_{outer}=(+6.53\pm14.48)\sfrac{R_{min}}{R_{max}}|_{apparent}+(1.25\pm12.48)$. 

The $\sfrac{M}{L_B}|_{outer}$ ratio was originally determined at the outermost
kinematic data point with $r_{max}$ varying between 0.45$R_{eff}$
and 2.95$R_{eff}$, a domain where the dependence of $\sfrac{M}{L}$ with radius
becomes important. To obtain the $\sfrac{M}{L}$ at the same radius
value ($r=1.25R_{eff}$, the average value for $r_{max}$), we interpolated
or extrapolated the authors results. 

The caveat of this data set is that the galaxies are chosen preferentially round. In
the case of $\sfrac{M}{L_B}|_{outer}$, the smaller uncertainties on two
galaxies (NGC3640 and NGC3379) drives the fit results. An advantage
of the analysis is the correction for (apparent) ellipticity. We assigned
the results to group 1 reliability.

\subsection{Magorrian {\it et al.} (1998)\label{sub:Magorrian98}}

\begin{figure}
\centering
\includegraphics[scale=0.5]{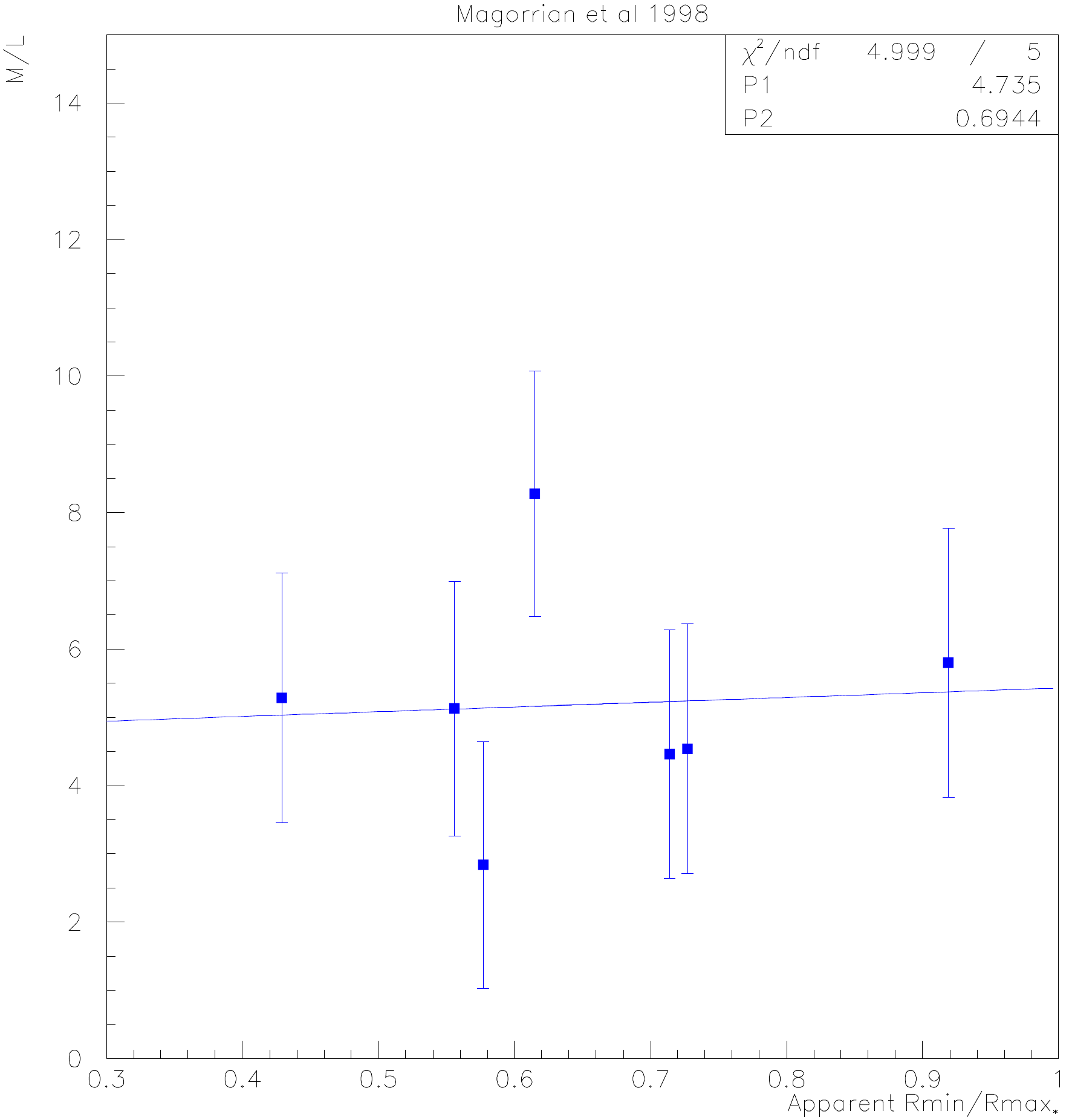}
\vspace{-0.4cm} \caption{\label{fig: magorrian97}$\sfrac{M}{L}$ vs apparent axis ratio from Magorrian {\it et al.}~\cite{Magorrian98}. 
}
\end{figure}

Magorrian {\it et al.}~\cite{Magorrian98} modeled 32 galaxies
to obtain $\sfrac{M}{L}$ and central black hole mass $M_{BH}$ (the main goal
of~\cite{Magorrian98}). The models assumed that galaxies are
axisymmetric with an arbitrary inclination (not provided in the article),
have a constant $\sfrac{M}{L}$, and have a central black hole. The models did not
include anisotropies. The black hole, stellar and dark matter gravity
fields were obtained respectively from a fit to Hubble Space Telescope
(HST) photometry data, a given $M_{BH}$ and assuming a given constant
$\sfrac{M}{L}$. Jeans equations were used to obtain velocity and velocity
dispersion profiles that were compared with ground based data. The
ground based and HST data do no extend further than $R_{eff}$ and
are typically a few tenths of $R_{eff}$. After our standard selection,
7 galaxies remained. We used axis ratios from NED. The $\sfrac{M}{L}$ uncertainty
was symmetrize  and rescaled so that $\sfrac{\chi^{2}}{ndf}=1$. $\sfrac{M}{L}$ vs
apparent axis ratio is shown in Fig.~\ref{fig: magorrian97}. %
 The best fit is:

$\sfrac{M}{L}=(+0.69\pm4.95)\sfrac{R_{min}}{R_{max}}|_{apparent}+(3.26\pm4.73)$.

A caveat is that the models do not include anisotropies. $\sfrac{M}{L}$ is
assumed to be constant, which is now known to be incorrect in general
but is probably valid in the radial range addressed in this study.
Advantages of the analysis are the corrections for ellipticity and
inclination. We assigned the results to group 1 reliability.

\subsection{Thomas {\it et al.} (2007, 2011) and Wegner {\it et al.} (2012)\label{sub:Thomas-et-al.}}

Thomas {\it et al.}~\cite{Thomas} studied 19 galaxies from
the Coma cluster. Dark and luminous masses were obtained within $R_{eff}$
using a dynamical model employing Schwarzschild orbit superposition
techniques~\cite{Schwarzschild}. Galaxies were assumed to be
axisymmetric and in dynamical equilibrium. The flattening of the galaxies
was included in the model, although only 3 possible values for the
galaxy inclinations were considered: 90$^\circ$ (assumes $\sfrac{R_{min}}{R_{max}}|_{true}=\sfrac{R_{min}}{R_{max}}|_{apparent}$),
minimum inclination (assumes the galaxy is a E7) and intermediate
inclination (assumes galaxy is a E5). A Log or a Navarro-Frenk-White
(NFW)~\cite{NFW} profile was used for dark matter distribution and it was
assumed that the baryonic mass profile is the same as the light emission 
profile (mass follows light assumption). Using these mass profiles,
the potential was calculated and all possible orbits were determined.
The light profiles obtained were compared to the data and the best
fit determined the best model. The authors cautioned that, although this
procedure should determine the most likely of the 3 possible inclinations,
the selection procedure may be biased toward the 90$^\circ$ case. The
uncertainty on $\sfrac{M}{L}$ was calculated by accounting for all possible
inclinations. We rescaled it so that our fit $\sfrac{\chi^{2}}{ndf}$ is 1.
The authors computed the lens characteristics of their galaxies, although
there is no known lensing observation from them, in order to compare
their results to the SLACS strong lensing results, see sections~\ref{sub:Auger-et-al.}
and~\ref{sec:Barnabe-et-al.}. From this they provided, in the $\mbox{R}_{\mbox{c}}$-band,
a dark matter fraction ($DMf$) within the Einstein radius $R_{Ein}$. 

\begin{figure}
\centering
\includegraphics[scale=0.9]{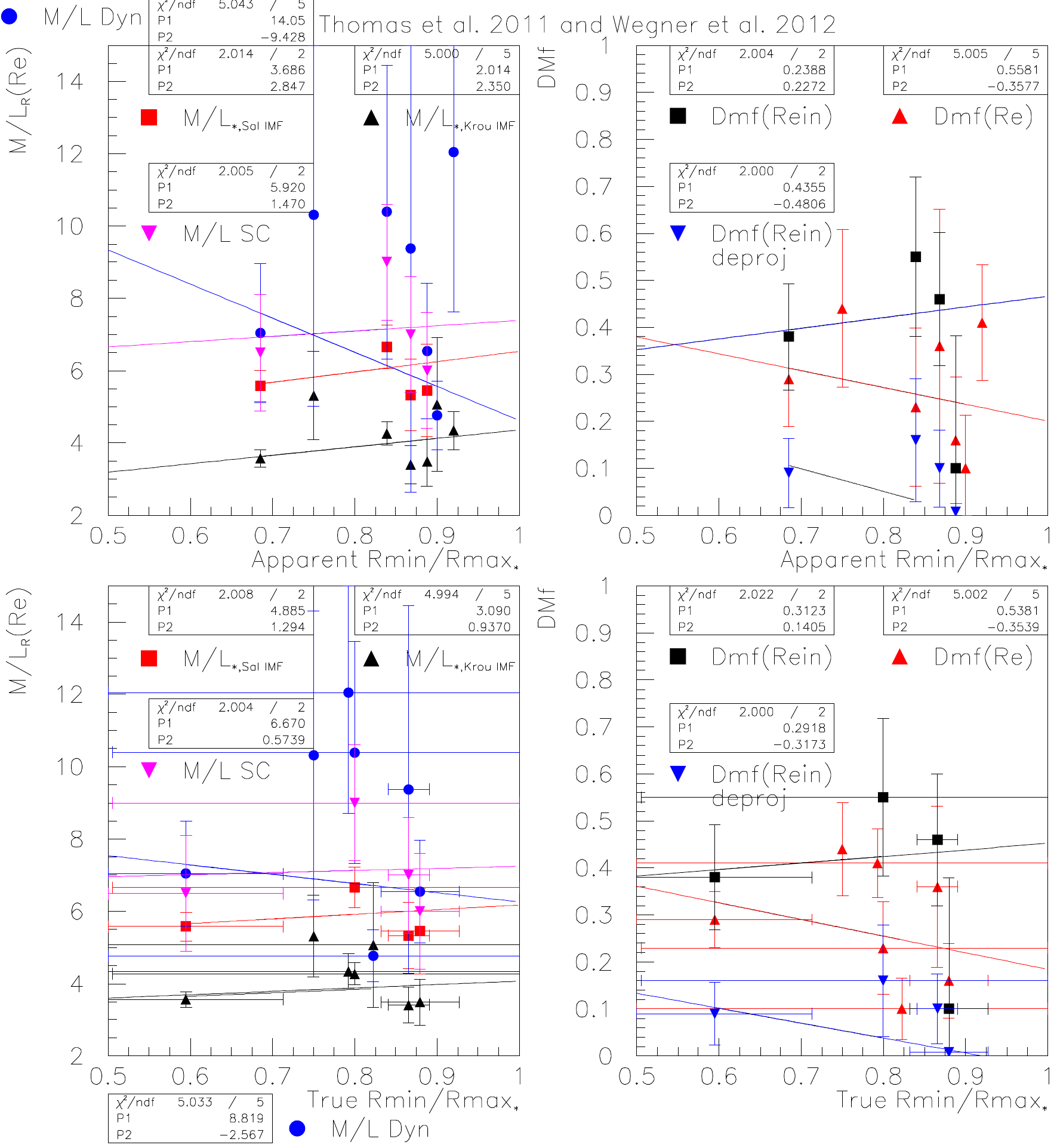}
\vspace{-0.4cm} \caption{\label{fig: thomas}Mass over light ratios and $DMf$ vs apparent and true axis ratios
from Thomas {\it et al.}~\cite{Thomas} and Wegner {\it et
al.}~\cite{Wegner}. The top left (resp. right) panel shows various
$\sfrac{M}{L}$ (resp.  $DMf$) vs apparent axis ratio. Bottom row: same as top but
for true axis ratio. The quantities of interest are $\sfrac{M}{L_{dyn}}$ (blue
filled circles) and $DMf(Re)$ (red filled triangles). Other quantities
are shown for information.}
\end{figure}

The same group that reported in~\cite{Thomas} used the same
methods and tools to analyze 8 early-type galaxies from the Abell
262 cluster (Wegner {\it et al.}~\cite{Wegner} ). We can thus
add these data to the ones of~\cite{Thomas}. Differences between
the two works include: 
\begin{itemize}
\item The Abell 262 cluster is less densely populated than the Coma cluster, so these
galaxies are less liable to interaction with environment;
\item The values of the Hubble constant is 69 in~\cite{Thomas} and
70 in~\cite{Wegner}. We have corrected the data from~\cite{Wegner}
to account for this difference.
\item The galaxies discussed in~\cite{Thomas} tend to be flat whilst
the ones in~\cite{Wegner} are selected to be round.
\end{itemize}
Four galaxies from~\cite{Thomas} and three galaxies from~\cite{Wegner}
passed our selection criteria. The mass over light ratios and $DMf$
within $R_{eff}$ are shown in Fig.~\ref{fig: thomas}. $\sfrac{M}{L_{dyn}}$
is obtained from the relation $\sfrac{M}{L_{dyn}}=\frac{\sfrac{M_*}{L_{dyn}}}{(1-DMf_{dyn})}$. 
The best fits are:

$\sfrac{M}{L_{R_{eff},dyn}}=(-9.43\pm9.74)\sfrac{R_{min}}{R_{max}}|_{apparent}+(14.05\pm8.43)$, 

$DMf_{R_{eff},dyn}=(-0.36\pm0.54)\sfrac{R_{min}}{R_{max}}|_{apparent}+(0.62\pm0.25)$ 

\noindent for the apparent axis ratio and: 

$\sfrac{M}{L_{R_{eff},dyn}}=(-2.57\pm6.80)\sfrac{R_{min}}{R_{max}}|_{true}+(8.82\pm5.30)$,

$DMf_{R_{eff},dyn}=(-0.35\pm0.36)\sfrac{R_{min}}{R_{max}}|_{true}+(0.54\pm0.27)$\\
for the true axis ratio.

The fit parameters obtained with apparent and true axis ratios are compatible, as expected since the uncertainties
cover all 3 assumed inclinations. 

The advantage of this analysis is that the effect of ellipticity was
fully accounted for, although necessarily in a model dependent way
and with only 3 possible values for the galaxy inclination. The caveat
of this study is that the galaxies belong to a cluster. They thus
may be subject to their environment. The results were assigned to
group 1 reliability.

{\footnotesize For information, we show other quantities in Fig.~\ref{fig: thomas}
: the }{\it \footnotesize stellar}{\footnotesize  $\sfrac{M_*}{L}$ computed
using a Kroupa or Salpeter IMF (available only from~\cite{Thomas});
$\sfrac{M}{L_{sc}}$: this quantity assumes that the total mass follows light,
i.e, $\sfrac{M}{L}$ is independent of the galaxy radius. This assumption has
been ruled out for $r>R_{eff}$ by many recent studies, including
the one of Thomas }{\it \footnotesize et al}{\footnotesize .; Projected
and deprojected $DMf(R_{Ein})$: these quantities are available only
from~\cite{Thomas}. They depend on the choice of linear relation
between $R_{Ein}$ and $\sigma_{eff}$, which is rather arbitrary.
The procedure is adequate in the context of~\cite{Thomas} but
somewhat arbitrary for our purpose. }{\footnotesize \par}

{\footnotesize The best fits for these quantities are:}{\footnotesize \par}

{\footnotesize $\sfrac{M_*}{L_{Krou}}=(+2.35\pm1.75)\sfrac{R_{min}}{R_{max}}|_{apparent}+(2.01\pm1.37)$, }{\footnotesize \par}

{\footnotesize $\sfrac{M_*}{L_{Sal}}=(+2.87\pm3.82)\sfrac{R_{min}}{R_{max}}|_{apparent}+(3.69\pm2.92)$, }{\footnotesize \par}

{\footnotesize $\sfrac{M}{L_{sc}}=(+1.47\pm9.88)\sfrac{R_{min}}{R_{max}}|_{apparent}+(5.92\pm8.14)$,}{\footnotesize \par}

{\footnotesize $DMf_{proj}(R_{ein)}=(+0.22\pm0.85)\sfrac{R_{min}}{R_{max}}|_{apparent}+(0.24\pm0.67)$, }{\footnotesize \par}

{\footnotesize $DMf _{deproj}(R_{ein)}=(-0.48\pm0.12)\sfrac{R_{min}}{R_{max}}|_{apparent}+(0.44\pm0.10)$ }{\footnotesize \par}

\noindent{\footnotesize  for the apparent axis ratios and }{\footnotesize \par}

{\footnotesize $\sfrac{M_*}{L_{Krou}}=(+0.94\pm1.48)\sfrac{R_{min}}{R_{max}}|_{true}+(3.09\pm1.05)$, }{\footnotesize \par}

{\footnotesize $\sfrac{M_*}{L_{Sal}}=(+1.30\pm2.62)\sfrac{R_{min}}{R_{max}}|_{true}+(4.89\pm1.85)$,}{\footnotesize \par}

{\footnotesize $\sfrac{M}{L_{sc}}=(+0.57\pm7.02)\sfrac{R_{min}}{R_{max}}|_{true}+(6.67\pm5.57)$.
(We will use this result for our global analysis with a reliability
factor 2)}{\footnotesize \par}

{\footnotesize $DMf_{proj}(R_{Ein)}=(+0.14\pm0.60)\sfrac{R_{min}}{R_{max}}|_{true}+(0.31\pm0.44)$,}{\footnotesize \par}

{\footnotesize $DMf _{deproj}(R_{Ein)}=(-0.32\pm0.27)\sfrac{R_{min}}{R_{max}}|_{true}+(0.29\pm0.24)$ }{\footnotesize \par}

\noindent{\footnotesize  for the true axis ratios.}{\footnotesize \par}

\subsection{van der Marel (1991)\label{sub:van-der-Marel91}}

\begin{figure}
\centering
\includegraphics[scale=0.5]{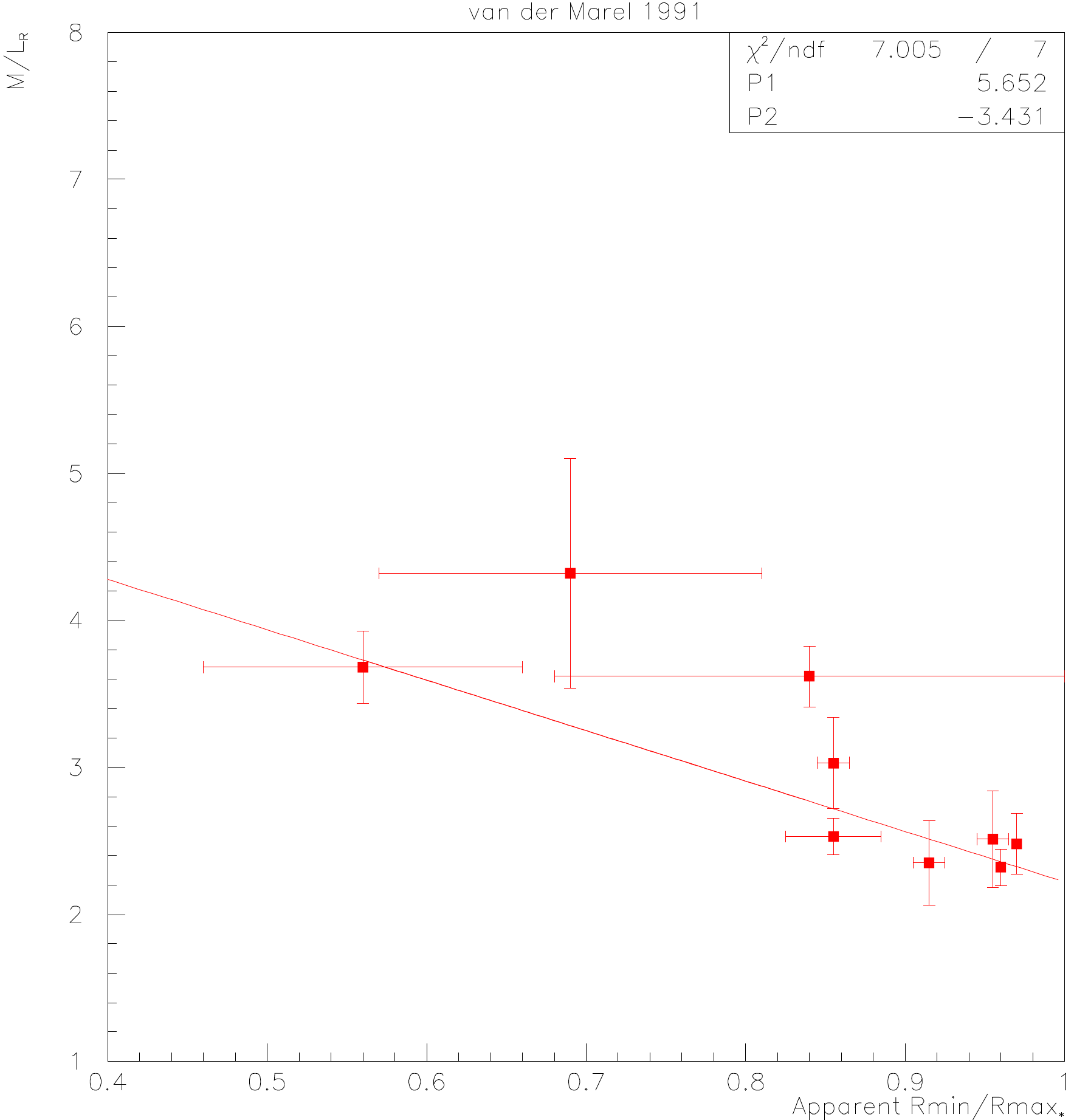}
\vspace{-0.4cm} \caption{\label{fig: van der marel 1991}$\sfrac{M}{L_R}$ vs axis ratio from van der Marel~\cite{van der Marel 1991}.
}
\end{figure}

In~\cite{van der Marel 1991} van der Marel constructed axisymmetric
dynamical models for 37 local early-type galaxies selected for their
high quality photometry. The models include anisotropy and rotation
but kept $\sfrac{M}{L}$ constant with radius. $\sfrac{M}{L}$ was obtained by
fitting the major and minor axis kinematics, which is more accurate
than core fitting using the virial theorem. However, the author checked
his work by also computing $\sfrac{M}{L}$ using the virial tensor formula
of Bacon {\it et al.}~\cite{BMS} which includes ellipticity
but assumes $\sfrac{M}{L}$ to be constant with radius. The two results, $\sfrac{M}{L}|_{\mbox{\scriptsize{Virial tensor}}}$
and $\sfrac{M}{L}|_{\mbox{\scriptsize{axisy. dyn. model}}}$ correlate well but the virial
results are in average $\sfrac{2}{3}$ smaller. The good correlation confirms Bacon{\it 
et al.}~\cite{BMS} formula, which we used throughout Section~\ref{sec:Data-sets-using virial theo} 
to correct results that did not account for the galaxy ellipticity. 
To form $\sfrac{M}{L}$, the mass density was obtained by deprojecting the luminosity profile
and assuming $\sfrac{M}{L}=\mbox{constant}$. From that density, the gravitational
potential was calculated by solving the Poisson equation. Finally, the Jeans equations,
which assume hydrostatic equilibrium, were solved. $\sfrac{M}{L}$ was calculated
for two inclination angles: $i=90^\circ$ and $i=60^\circ$ but the difference
is small:  $\sfrac{M}{L}$ is larger by about 10\% for axis ratios of about
0.7 and, as expected, disappears for axis ratios near 1. The $\sfrac{M}{L}$ used here are
the average between the $i=90^\circ$ and $i=60^\circ$ results. 
The author computed the axis ratios
using his models. We used the average of his result with the NED axis
ratio, the difference being taken as uncertainty. The $\sfrac{M}{L_R}$ ratios vs apparent
axis ratios are shown in Fig.~\ref{fig: van der marel 1991} for the 9 galaxies
that passed selection. In addition
to the standard selection criteria%providing a sample as homogeneous as possible
%and complying with the hydrostatic equilibrium requirement, 
, we further rejected S0 candidates (NGC636, 3557, 4494 and IC0179) and S (NGC1370)
galaxies, although they are listed as elliptical galaxies in NED.
 The best fit is:
 
$\sfrac{M}{L_R}=(-3.43\pm0.92)\sfrac{R_{min}}{R_{max}}|_{apparent}+(5.65\pm0.84$).

\noindent We rescaled the $\sfrac{M}{L}$ uncertainties so that $\sfrac{\chi^{2}}{ndf}=1$. 
The advantage of the analysis is that care has been taken in accounting
and studying the effects of apparent ellipticities, inclination angles,
anisotropies and rotations. The caveats are that the analysis is dated,
but this is balanced by the author's galaxy selection based on the
high quality of their photometry. The models assumed hydrostatic equilibrium,
which we enforced by strict selection that removes about two thirds
of the original galaxy sample. Finally a constant $\sfrac{M}{L}$ is assumed. This
is known to be incorrect at large radii, as the author and many other
studies found, but this is reasonable at radius up to $R_{eff}$.
We assigned the results to group 1 reliability. 

\subsection{van der Marel and van Dokkum (2007a)}

\begin{figure}
\centering
\includegraphics[scale=0.5]{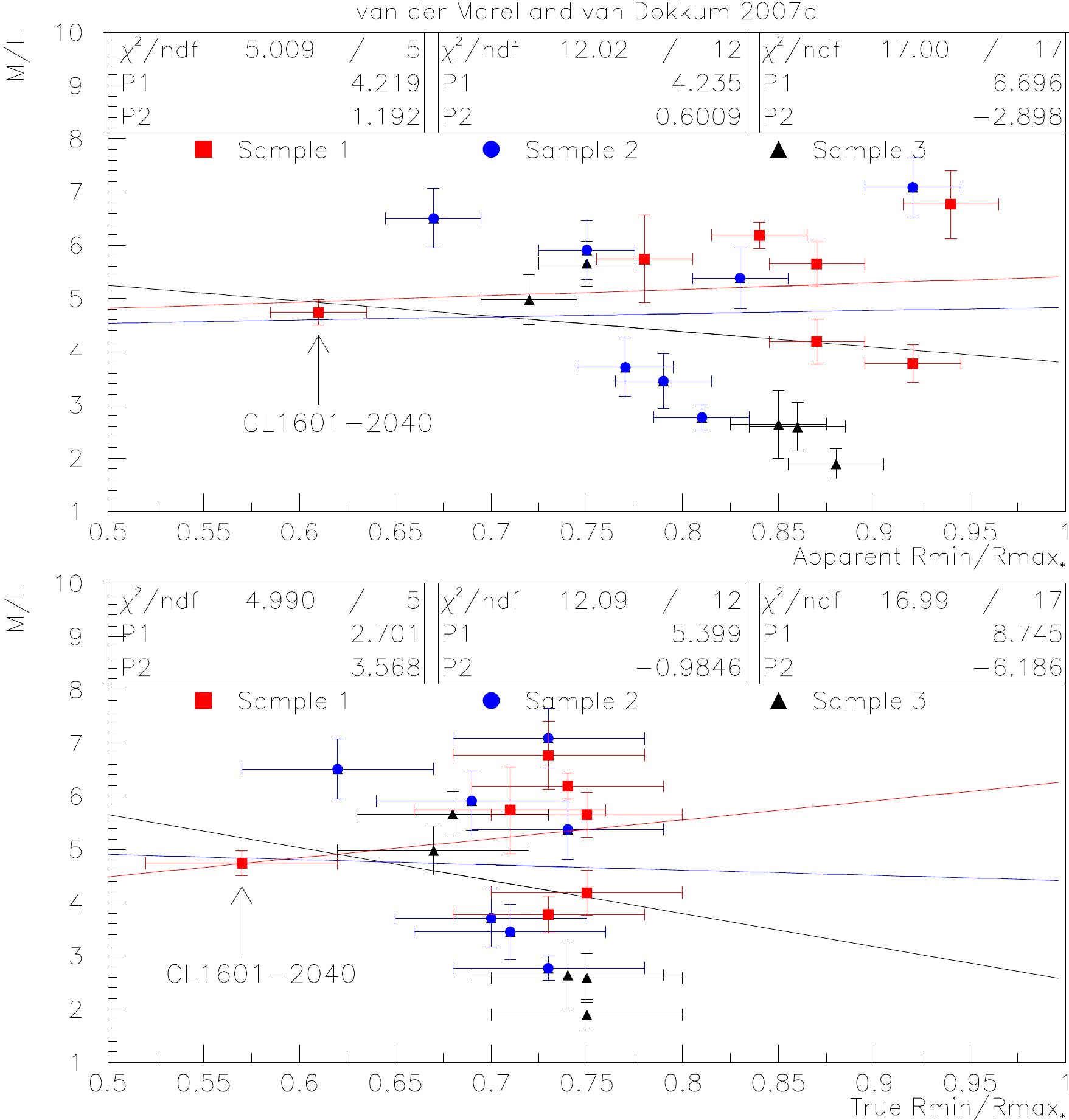}
\vspace{-0.4cm} \caption{\label{fig: van der marel 2007a}$\sfrac{M}{L}$ vs apparent (top) and  ``true'' (bottom) axis ratios from
van der Marel and van Dokkum~\cite{van der Marel 2007a}.
}
\end{figure}

In Ref.~\cite{van der Marel 2007a}, van der Marel and van Dokkum studied
25 early-type galaxies using stellar rotation and velocity dispersion
data. The galaxies belong to 3 clusters at redshift $z\simeq0.5$
(except for one field galaxy. But we will ignore it as it is classified
as a E/S0). The data were interpreted with oblate axisymmetric 2-integral
dynamical models based on Jeans' equations of hydrostatic equilibrium.
$\sfrac{M}{L_B}$ was assumed to be constant with radius. This assumption has 
been ruled out by many recent studies 
(e.g.~\cite{Kronawitter, Thomas, van der Marel 1991, Capaccioli, Magorrian01, Nagino, Napolitano, Bertola93}), 
although it appears reasonable in the radius range considered here ($r<R_{eff}$).
Using the inferred probability distribution of true axis ratio for
elliptical galaxies, the authors computed for each galaxy a most-likely
true axis ratio. This one was calculated by taking the average over
all possible values of the true axis ratio for a galaxy ($0.3\leq \sfrac{R_{min}}{R_{max}}|_{true}\leq \sfrac{R_{min}}{R_{max}}|_{apparent}$)
weighted by the true axis ratio probability distribution. The resulting
 ``true'' axis ratio may not be the actual one but is a reasonable
guess that follows the expected gaussian distribution of true axis
ratios. The galaxies in~\cite{van der Marel 2007a} were identified
only by visual inspection. To address the possibility of S0 contaminating the galaxy set,
we estimated the probability to be a S0 using the expected characteristics
of S0 ($\sigma$ tends to be lower, high rotational support and strong
ellipticity dependence with radius. We did not consider the additional
fact that ellipticities of S0 are large to not bias our study). Based
upon this criteria, of the 25 galaxies 7 were unlikely to be S0 (none
of those are listed as S0 in~\cite{van der Marel 2007a}), 12
were possibly S0 (4 listed as S0 or E/S0 in~\cite{van der Marel 2007a})
and 6 were likely to be S0 (1 of those is a known S0/Sb). Assuming
that about half of the 12 possible S0 are indeed S0, the 25 early-type
sample contained about 13 E and 12 S0, agreeing with the known repartition
of the local E and S0 galaxies. Removing the 5 galaxies listed in
\cite{van der Marel 2007a} as visually identified as S0, E/S0
or S0/Sb, we obtained Sample 3 of 19 galaxies. Removing from it 5
galaxies that are likely to be S0 based on our criteria, we obtained
Sample 2 (15 galaxies). Finally considering only the galaxies unlikely
to be S0, we obtained Sample 1 (7 galaxies). The $\sfrac{M}{L_B}$ ratios
vs apparent and true axis ratios are shown in Fig.~\ref{fig: van der marel 2007a}.
We can see that the S0 rejection criteria removed the low $\sfrac{M}{L}$ points,
possibly biasing our study. To address this, we did not use this criteria
for the determination of the slopes but only to estimate the uncertainty
due to possible S0 contamination. %
 The best fits are:

$\sfrac{M}{L_B}=(+1.19\pm3.31)\sfrac{R_{min}}{R_{max}}|_{apparent}+(4.22\pm2.65)$
for Sample 1,

$\sfrac{M}{L_B}=(+3.57\pm5.12)\sfrac{R_{min}}{R_{max}}|_{"true"}+(2.70\pm3.55)$
for Sample 1,

$\sfrac{M}{L_B}=(+0.60\pm3.83)\sfrac{R_{min}}{R_{max}}|_{apparent}+(4.24\pm3.06)$
for Sample 2,

$\sfrac{M}{L_B}=(-0.98\pm6.04)\sfrac{R_{min}}{R_{max}}|_{"true"}+(5.40\pm4.24)$
Sample 2,

$\sfrac{M}{L_B}=(-2.90\pm3.84)\sfrac{R_{min}}{R_{max}}|_{apparent}+(6.70\pm3.10)$
for Sample 3,

$\sfrac{M}{L_B}=(-6.19\pm5.99)\sfrac{R_{min}}{R_{max}}|_{"true"}+(8.75\pm4.23)$
for Sample 3. 

\noindent All in all using the samples 1 and 2 to estimate the S0
contamination to the main sample (sample 3), we obtain:

$\sfrac{M}{L_B}=(-2.90\pm5.60)\sfrac{R_{min}}{R_{max}}|_{apparent}+(6.70\pm3.97)$,

$\sfrac{M}{L_B}=(-6.19\pm11.45)\sfrac{R_{min}}{R_{max}}|_{"true"}+(8.75\pm7.38)$. 

\noindent The uncertainties from~\cite{van der Marel 2007a} were rescaled
so that $\sfrac{\chi^{2}}{ndf}=1$. Advantages of this study are that ellipticity
was mostly accounted for in the calculations, and a true ellipticity
was tentatively given. Caveats of this analysis are the lack of good galaxy
identification (S0 contamination) and of features revealing
possible disturbance, whereas the $\sfrac{M}{L}$ extraction
assumes hydrostatic equilibrium. Thus, we could
not implement our standard selection of undisturbed galaxies. An uncertainty
for the S0 contamination was assessed using tight selection. Other
caveats are the redshift $z\simeq0.5$ may cause systematic
differences between the sample and local galaxies, and that $\sfrac{M}{L}$
was assumed to be constant. Finally, the positive fit slopes
are driven by a single galaxy of large ellipticity and small uncertainty:
CL1601-2040. As can be seen in Fig.~\ref{fig: van der marel 2007a},
all slopes would be negative without CL1601-2040. However, we
had no reason to exclude it from the study. The {\it{unbias estimate}}
method partly accounted  for this outlier. Since the S0 and outlier issues are reflected
in the final uncertainty, the results are still assigned to group
2 reliability rather than a lower reliability group.

\subsection{van der Marel and van Dokkum (2007b)}

\begin{figure}
\centering
\includegraphics[scale=0.5]{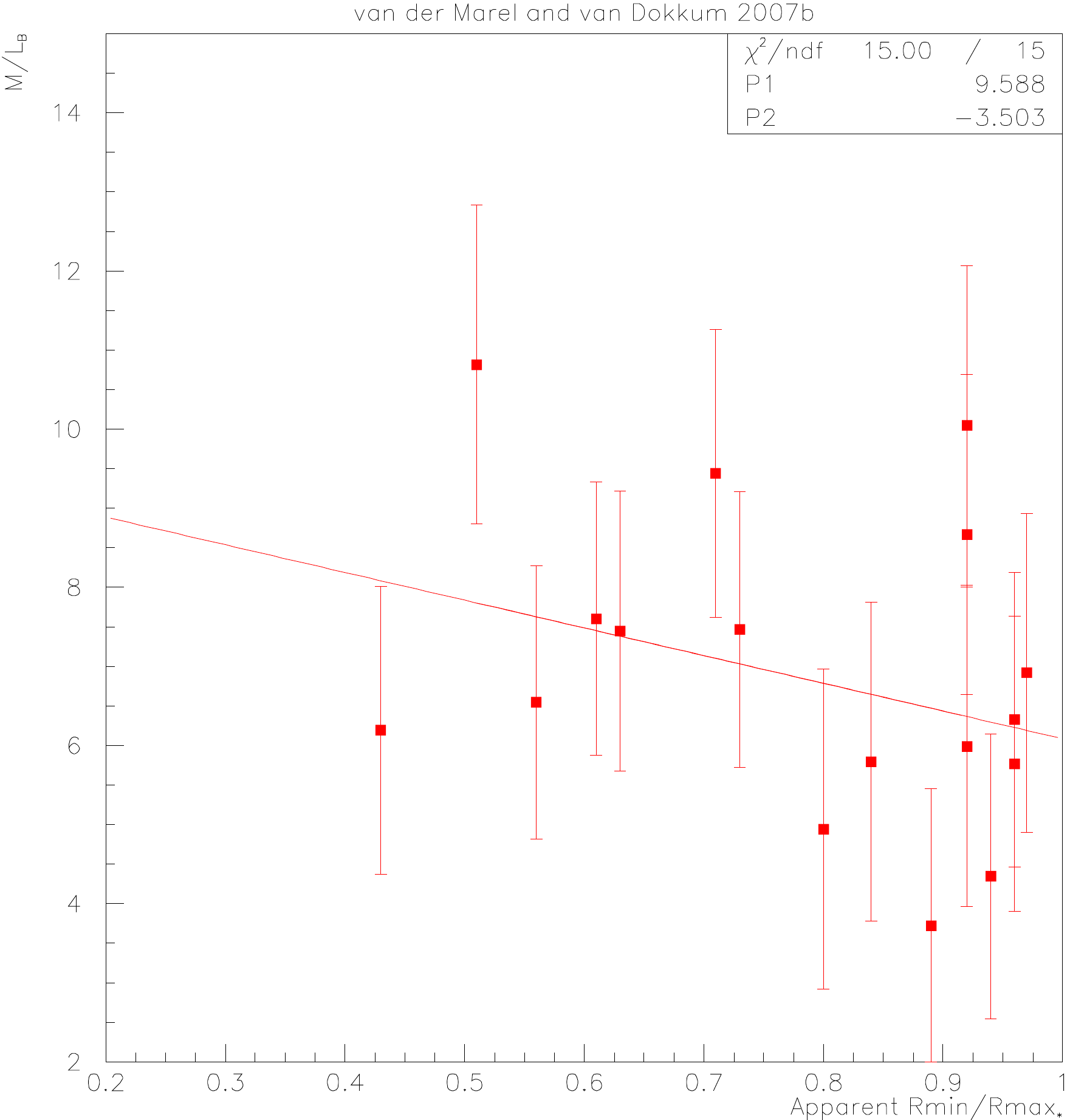}
\vspace{-0.4cm} \caption{\label{fig: van der marel 2007b}$\sfrac{M}{L}$ vs apparent axis ratio 
from van der Marel and van Dokkum~\cite{van der Marel 2007b}.
}
\end{figure}

In order to compare the $z\simeq0.5$ results of the previous section
with results from local galaxies, van der Marel and van Dokkum compiled
from the literature a homogenized $\sfrac{M}{L_B}$ for 60 early-type galaxies
\cite{van der Marel 2007b}. The $\sfrac{M}{L_B}$ were corrected using
a consistent distance determination and, if needed to be, re-expressed
in the B-band. The redshift of the galaxies forming the set  is $z\simeq0.005$. All the literature
sources compiled in~\cite{van der Marel 2007b} were also used
in the our present study: van der Marel: Section~\ref{sub:van-der-Marel91},
Magorrian {\it et al.}: Section~\ref{sub:Magorrian98}, Gebhardt
{\it et al.}: Section~\ref{sec:Other-data}, Kronawitter {\it et
al.}: Section~\ref{sec:Kronawitter} and Cappellari {\it et al.}:
Section~\ref{sec:Cappellari06}. However, it is still interesting
to consider the compilation of~\cite{van der Marel 2007b} because
results have been homogenized. Applying our standard selection to
the initial 60 galaxies, we obtained a sample of 17 galaxies. The
sources contribute with similar statistics (van der Marel: 8 galaxies,
Magorrian {\it et al.}: 5 galaxies, Kronawitter {\it et al.}: 4
galaxies, Gebhardt {\it et al.}: 6 galaxies, Cappellari {\it et
al.}: 6 galaxies). Hence, our final sample is a representative
average rather than being strongly bias toward a particular work.
The $\sfrac{M}{L_B}$ ratio vs apparent axis ratio is shown in Fig.~\ref{fig: van der marel 2007b}.
 The best fit gives:

$\sfrac{M}{L_B}=(-3.50\pm2.62)\sfrac{R_{min}}{R_{max}}|_{apparent}+(9.59\pm2.08)$.

The advantages and caveats of these data are already discussed in
the sections~\ref{sub:van-der-Marel91} (group 1 reliability),~\ref{sub:Magorrian98}
(group 1 reliability),~\ref{sec:Other-data} (data unused, due to
low statistics and correlations),~\ref{sec:Kronawitter} (group 1
reliability) and~\ref{sec:Cappellari06} (group 1 reliability). Most
of these data were already used for our analysis, although they have
been homogenized and averaged when a particular galaxy was available from several
of the literature sources. The results are assigned to group 1 reliability
but partly weighted out when we considere the global results incorporating
all $\sfrac{M}{L}$ vs axis ratio, see Section~\ref{sec:Global-results}.

\section{Data sets using Planetary Nebulae and Globular Clusters}

Planetary Nebulae (PNe) and Globular Clusters (GC) provide well resolved
tracers extending to large radii. They offer the possibility to examine 
individual stars (PNe) or compact  stelar assemblies (GC) to robustly assess
the gravitational potential over a large radial range. Caveats of
this method in our context is that the ellipticity determined from
photometry might not match the ellipticity at large radii, and that
the tracers at large radii may be influenced by the cluster or group
hosting the galaxy. A general concern with GC is that
they may be decoupled from the evolution their host  galaxy: they
are good tracers of the overall gravity field but may have a different
rotational structure and whilst PNe tend to follow the light distribution,
GC tend to be more extended. However, this is unimportant in the context
of our study. Furthermore, since GC tend to belong to heavier galaxies
which typically do not fulfill the selection criteria, the role of GC in our study is minimal.

All the literature sources used below assumed spherical symmetry.
To account for ellipticity, we noted that the ellipticity corrections
to the virial method (Section~\ref{sec:Data-sets-using virial theo})
and the X-ray method (Section~\ref{sec:Data-sets-using X-ray}) are
close, as can be seen by comparing our Fig.~\ref{fig: newton shell correction}
to the figures 1a and 1b of Bacon {\it et al.}~\cite{BMS}.
We assumed the same correction to the $\sfrac{M}{L}$ extracted from PNe and
GC data.

\subsection{Capaccioli, Napolitano and Arnaboldi (1992)\label{sub:Capaccioli}}

\begin{figure}
\centering
\includegraphics[scale=0.5]{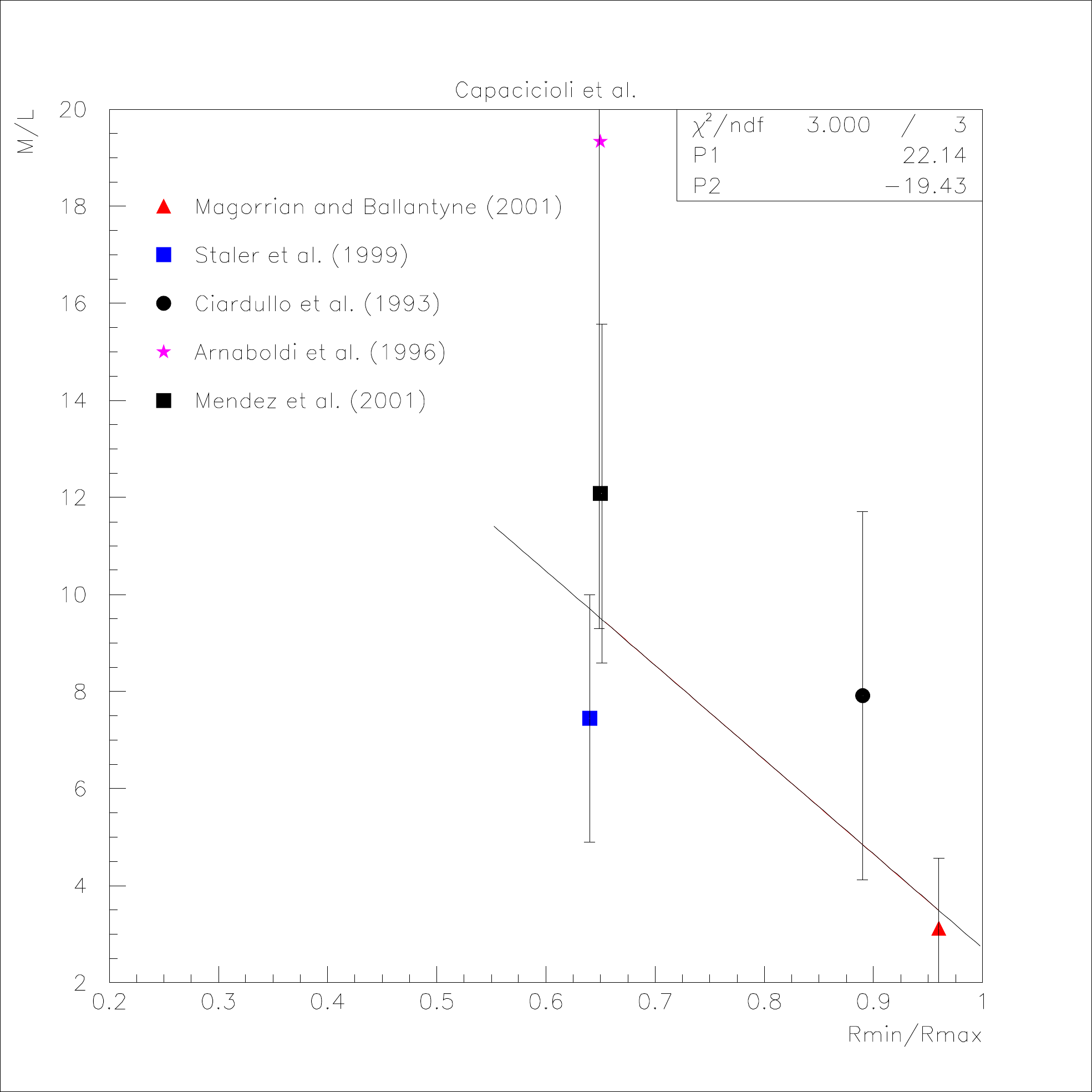}
\vspace{-0.4cm} \caption{\label{fig:Capacioli}$\sfrac{M}{L_B}(R_{eff})$ ratio vs apparent axis ratio $\sfrac{R_{min}}{R_{max}}$
for the compilation of Capaccioli, Napolitano and Arnaboldi~\cite{Capaccioli}.
The various symbols refer to the original authors of the $\sfrac{M}{L}$ extraction,
see references in~\cite{Capaccioli}.
}
\end{figure}

Capaccioli, Napolitano and Arnaboldi~\cite{Capaccioli} provided
a compilation from the literature on $\sfrac{M}{L_B}$. The mass to light
ratio for 14 galaxies was obtained from PNe tracers (no GC)
at two different radii (inner and outer $\sfrac{M}{L_B}$), so that the $\sfrac{M}{L_B}$
gradient could be calculated. 
The radii at which the $\sfrac{M}{L_B}$ were given differ from galaxy to
galaxy so we interpolated linearly between the inner and outer $\sfrac{M}{L_B}$
to obtain a value at $R_{eff}$ for all the galaxies. 
Following the authors, we used uncertainties
of 10\% and 30\% for the inner and outer $\sfrac{M}{L_B}$ respectively,
and propagated them to $R_{eff}$. None of the galaxies in~\cite{Capaccioli}
passed our selection criteria. Relaxing the selection to include all
galaxies except the disrupted ones, galaxies showing possible interactions
and clear non-elliptical galaxies, we obtained a set of 5 galaxies:
NGC 1379 (maybe a spiral), 1700 (EXG), 3379 (LINER, VCXG), 4406 (SO(3)/E3)
and 4697 (LLAGN). The fit to the $\sfrac{M}{L_B}$ vs axis ratio is shown
in Fig.~\ref{fig:Capacioli}. The best fit is:

$\sfrac{M}{L_{B}}(R_{eff})=(-19.43\pm7.85)\sfrac{R_{min}}{R_{max}}|_{apparent}+(22.14\pm6.82)$. 

Caveats are that the data were obtained from difference sources
(which may result in author-to-author bias) and include galaxies
of unsure type, cD, LINERS and LLAGN. We assigned the results to group
3 reliability.

{\footnotesize For information, the best fit before ellipticity corrections
is:} 

{\footnotesize $\sfrac{M}{L_{B}}(R_{eff})=(-9.62\pm5.87)\sfrac{R_{min}}{R_{max}}|apparent+(12.59\pm5.10)$. }{\footnotesize \par}

\subsection{Deason {\it et al.} (2012)\label{sub:Deason}}

\begin{figure}
\centering
\includegraphics[scale=0.5]{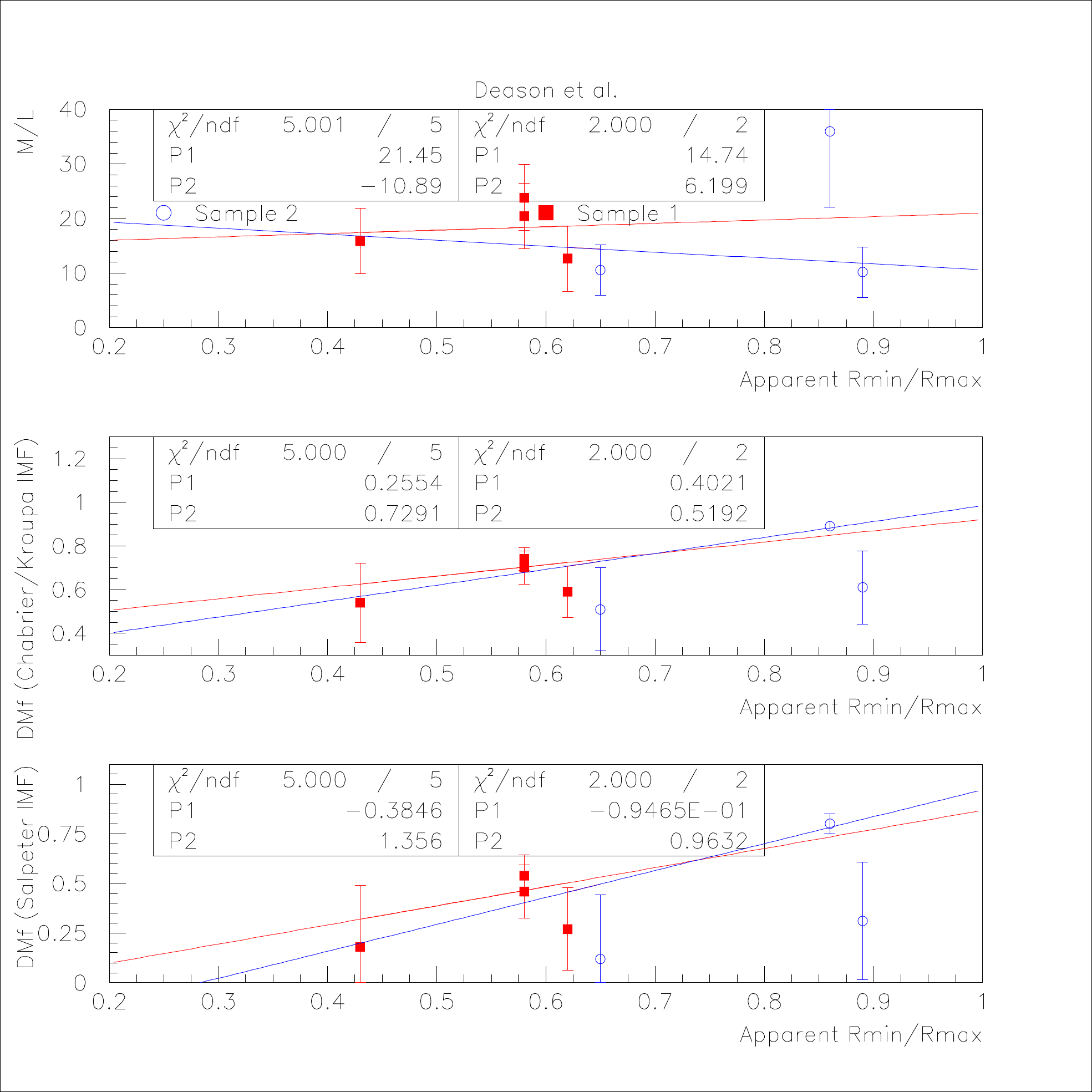}
\vspace{-0.4cm} \caption{\label{fig:Deason}$\sfrac{M}{L}$ and  $DMf$ vs diameter ratio $\sfrac{R_{min}}{R_{max}}$
for the Deason {\it et al.} analysis~\cite{Deason}. The red
and blue lines represent the best fits to the data of Sample 1 and
2 respectively. The $DMf$ values on the middle and bottom panels
were obtained with different IMF models.
}
\end{figure}

Deason {\it et al.}~\cite{Deason} employed PNe
and GC as kinematic tracers to provide $\sfrac{M}{L}$
and  $DMf$ integrated up to a radius of 5$R_{eff}$. The
$DMf$ was extracted model dependently using either
a Chabrier/Kroupa~\cite{Chabrier}/\cite{Kroupa} or a Salpeter
\cite{Salpeter} Initial Mass Function (IMF). The authors studied
15 galaxies using a spherically symmetric model. The model allowed
to form line of sight velocity distributions that were matched to
the measured ones using a maximum likelihood method. It was assumed
that the tracers' orbit are dominated by random motion rather than
systematic motion. Furthermore, any possible galaxy rotation was ignored and simple
power-law profiles for tracer density and potential were assumed.
Velocity anisotropies  were accounted for. The potentials and anisotropy
parameters were determined by the best fit to the measurements of the tracer line of sight
velocity distribution. From this, the mass was extracted.
The usual galaxy selection yielded a first sample (Sample 1) of 4
good elliptical galaxies (NGC 821, 1344, 3377 and 4564). For this
analysis, we can relax the selection to include LINERS, Sy, VCXG and
LLAGN galaxies. We obtained then Sample 2 with 3 additional galaxies
NGC 3379, 4374 and 4697. $DMf$ vs $\sfrac{R_{min}}{R_{max}}$ is shown on
Fig.~\ref{fig:Deason}. The fits yield:

$\sfrac{M}{L}=(+6.20\pm41.49)\sfrac{R_{min}}{R_{max}}|_{apparent}+(14.74\pm23.12)$
for sample 1 and

$\sfrac{M}{L}=(-10.89\pm17.71)\sfrac{R_{min}}{R_{max}}|_{apparent}+(21.45\pm12.73)$
for sample 2.

We assigned the $\sfrac{M}{L}$ result to group 1 reliability and used Sample 2 when combining all
the publication together into one global result in Section~\ref{sec:Global-results}.

{\footnotesize For information, the best fits before ellipticity corrections
are: \\ for sample 1:}{\footnotesize \par}

{\footnotesize $\sfrac{M}{L}=(+18.86\pm24.59)\sfrac{R_{min}}{R_{max}}|_{apparent}+(-0.17\pm13.70)$,}{\footnotesize \par}

{\footnotesize $DMf=(+0.52\pm1.12)\sfrac{R_{min}}{R_{max}}|_{apparent}+(0.40\pm0.65(\mbox{fit)}\pm0.49$
for the Chabrier/Kroupa IMF and}{\footnotesize \par}

{\footnotesize $DMf=(+0.96\pm1.91)\sfrac{R_{min}}{R_{max}}|_{apparent}+(-0.08\pm1.10$
for the Salpeter IMF.}{\footnotesize \par}

{\footnotesize \noindent For Sample 2:}{\footnotesize \par}

{\footnotesize $\sfrac{M}{L}=(+5.88\pm13.72)\sfrac{R_{min}}{R_{max}}|_{apparent}+(5.41\pm9.86)$,}{\footnotesize \par}

{\footnotesize $DMf=(+0.73\pm0.30)\sfrac{R_{min}}{R_{max}}|_{apparent}+(0.26\pm\pm0.25$
for the Chabrier/Kroupa IMF and}{\footnotesize \par}

{\footnotesize $DMf=(+1.37\pm0.59)\sfrac{R_{min}}{R_{max}}|_{apparent}+(-0.38\pm0.50$
for the Salpeter IMF.}{\footnotesize \par}

\subsection{Magorrian and Ballantyne (2001)\label{sec:Magorrian-and-Ballantyne}}

Magorrian and Ballantyne~\cite{Magorrian01} used PNe data to
compute $\sfrac{M}{L}(r)$ for 18 round early-type galaxies. The following
method was employed: the velocity dispersion $\sigma(r)$
was calculated assuming a $\sfrac{M}{L}(r)$ form and  using Jeans equations and the luminosity density. This
one was obtained from the measured surface brightness deprojected
assuming a spherically symmetric galaxy. The obtained $\sigma(r)$
was fit to measured stellar or PNe velocity dispersions. The best
fit determined the free parameters of the assumed $\sfrac{M}{L}(r)$ form. 

\begin{figure}
\centering
\includegraphics[scale=0.5]{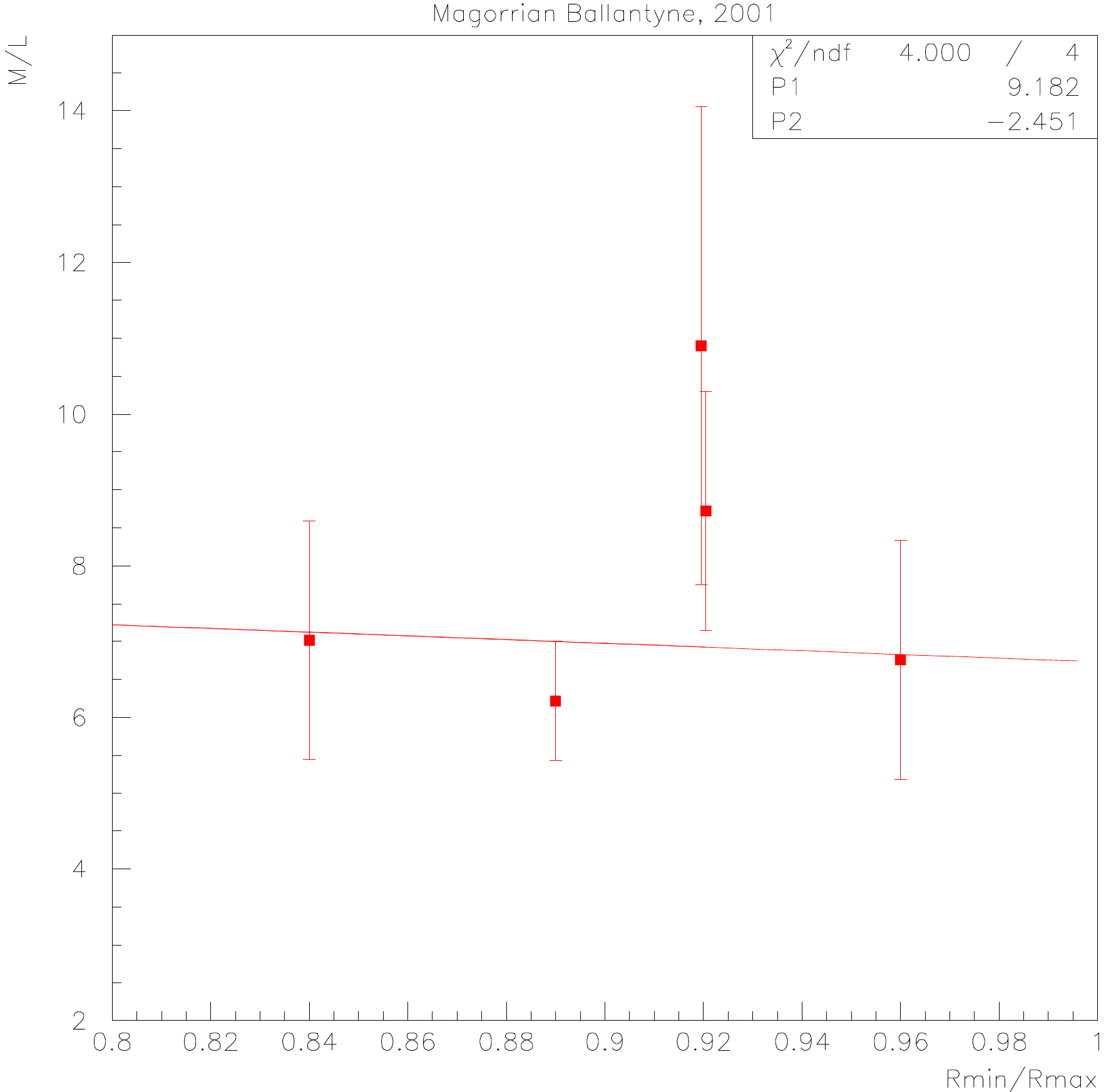}
\vspace{-0.4cm} \caption{\label{fig: Magorrian}$\sfrac{M}{L}$ vs apparent axis ratio from the data analyzed  by Magorrian
and Ballantyne~\cite{Magorrian01}. 
}
\end{figure}

After we applied our standard selection, 6 elliptical galaxies remained. We used
values of $\sfrac{M}{L}(r)$ integrated up to $R_{eff}$. The $\sfrac{M}{L}$ vs stellar
axis ratios are shown in Fig.~\ref{fig: Magorrian}. %
 The best fit yields:

$\sfrac{M}{L}=(-2.45\pm8.60)\sfrac{R_{min}}{R_{max}}|_{apparent}+(9.18\pm7.57)$. 

Advantages of this analysis are that the authors provided a detailed
study of the effect of anisotropy and spherical asymmetry assumptions.
They concluded that anisotropies have a small effect on $\sfrac{M}{L}$ whilst
the spherical symmetry is a critical assumption with large effects
on $\sfrac{M}{L}$. The difficulty is that the effects of anisotropies or flattening
are not distinguishable on the observed velocity dispersion data whilst,
as just said, the magnitude of their influence on the $\sfrac{M}{L}$ value
is different. According to~\cite{Magorrian01} their spherically
symmetric model underestimates $\sfrac{M}{L}$. We assigned the result to reliability
group 1.

{\footnotesize For information, the best fit before ellipticity corrections
is:} 

{\footnotesize $\sfrac{M}{L}=(+4.38\pm7.99)\sfrac{R_{min}}{R_{max}}|_{apparent}+(2.34\pm7.03)$. }{\footnotesize \par}

\subsection{Romanowsky {\it et al.} (2003)}

\begin{figure}
\centering
\includegraphics[scale=0.5]{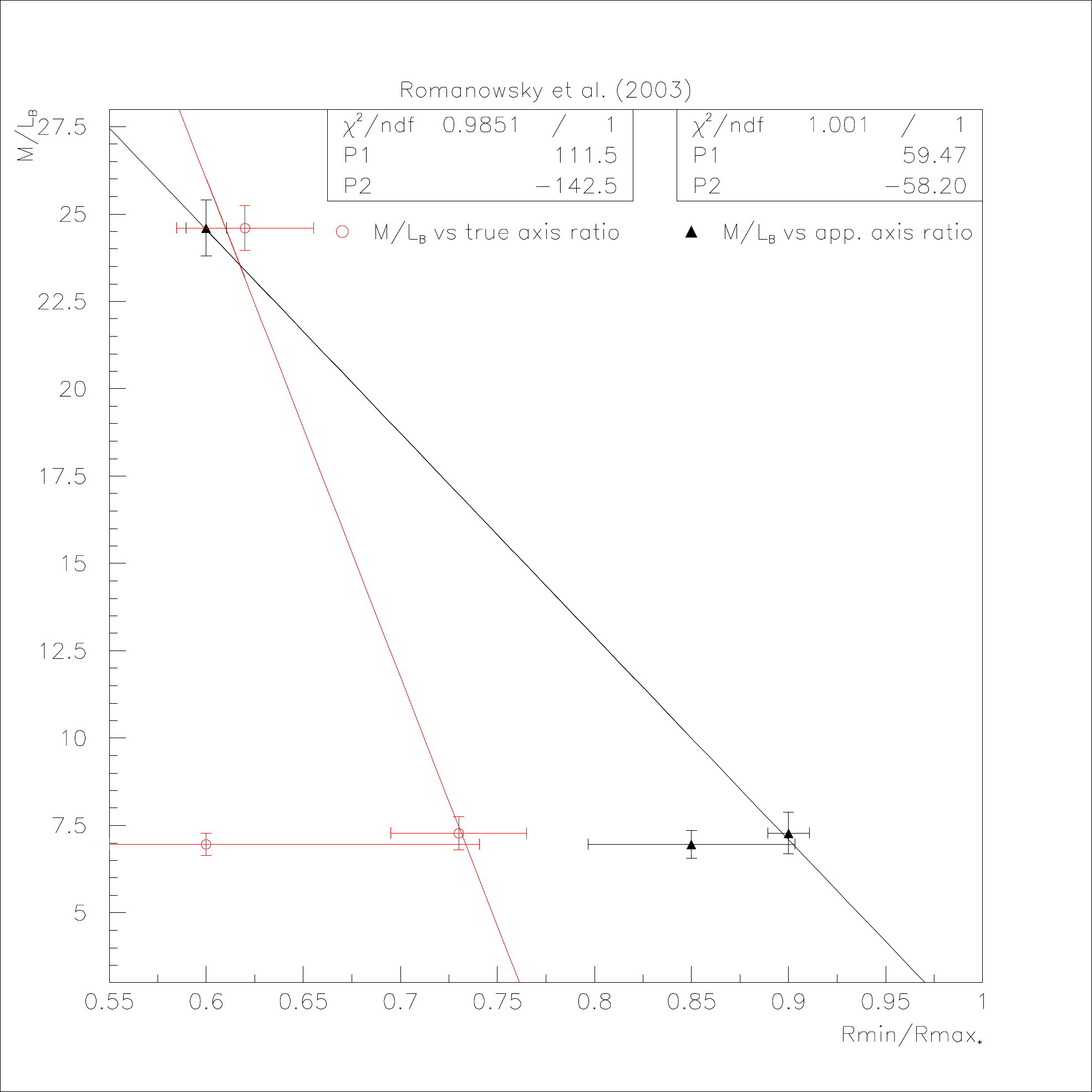}
\vspace{-0.4cm} \caption{\label{fig: romanowsky}$\sfrac{M}{L}$ from Romanowsky {\it et al.}~\cite{Romanowsky} vs apparent
axis ratio (full black triangles) and inferred true axis ratio (open
red circles). 
}
\end{figure}

Romanowsky {\it et al.}~\cite{Romanowsky} used kinematics of
PNe to extract the rotation curves of three elliptical
galaxies, each of them possessing  about a hundred of observable  PNe. The possible
effect of anisotropies was studied with a spherical Jeans model in
which an anisotropy parameter spans its full possible range. From
the model, $\sfrac{M}{L_B}(r=5R_{eff})$ were obtained. Furthermore, a different
method was used with NGC3379 to investigate the effect of velocity
anisotropy variation with radius. The true axis ratios of the galaxies
investigated in~\cite{Romanowsky} had been inferred in~\cite{Cappellari 2006}
,~\cite{Foster} and~\cite{Statler}. The $\sfrac{M}{L_B}$ vs
axis ratios are shown in Fig.~\ref{fig: romanowsky}. %
 The best fits are:

$\sfrac{M}{L_B}(r=5R_{eff})=(-58.20\pm4.40)\sfrac{R_{min}}{R_{max}}|_{apparent}+(59.47\pm3.48)$,

$\sfrac{M}{L_B}(r=5R_{eff})=(-142.47\pm63.47)\sfrac{R_{min}}{R_{max}}|_{true}+(111.50\pm42.86)$, \\
where we rescaled the uncertainties on both $\sfrac{M}{L}$ and $\sfrac{R_{min}}{R_{max}}$
so that the fit $ $$\sfrac{\chi^{2}}{ndf}=1$. An advantage of this work
is the detailed study of the effect of anisotropies using two different
techniques. We assigned the results to group 1 reliability.

{\footnotesize For information, the best fits before ellipticity corrections
are:} 

{\footnotesize $\sfrac{M}{L_B}(r=5R_{eff})=(-28.99\pm4.42)\sfrac{R_{min}}{R_{max}}|_{apparent}+(32.30\pm3.56)$,}{\footnotesize \par}

{\footnotesize $\sfrac{M}{L_B}(r=5R_{eff})=(-68.78\pm33.08)\sfrac{R_{min}}{R_{max}}|_{true}+(56.83\pm22.38)$.}{\footnotesize \par}

\section{Data sets using X-ray \label{sec:Data-sets-using X-ray}}

The extraction of a galaxy mass $M(r)$ from X-ray emitting interstellar
medium (ISM) data assumes that 1) the hot gas is in hydrostatic equilibrium
under the galaxy gravitational potential: $\sfrac{dP}{dr}=-GM(r)\rho(r)/r^{2}$,
and 2) that the gas follows the perfect gas law: $Pm=\rho kT(r)$.
Here, $P$ is the ISM gas pressure, $\rho$ its density, $T$ its
temperature, $m$ the gas molecule average mass ($m\simeq582$~MeV) and 
$k$ is the Boltzmann constant.
$M(r)$ can be extracted by combining the two previous equations:
$M(r)=-\frac{kTr}{Gm}\frac{dln(\rho)+dln(T)}{dln(r)}$. 

In the context of our study, this method has caveats. X-ray emission occurs mostly in
giant galaxies, usually the one dominating its group or
cluster (cD galaxies). Hot gas is highly collisional %i.e, it departs from the perfect gas law
and can be heated by inner processes (electromagnetic emissions) or outer influences
(galaxy interacting with its environment) and hence may be off equilibrium.
In fact, results may correlate with the galaxy environment,
as found in Das {\it et al.}~\cite{Das}. 

\begin{figure}
\centering
\hspace{-1.08cm}
 \includegraphics[scale=0.45]{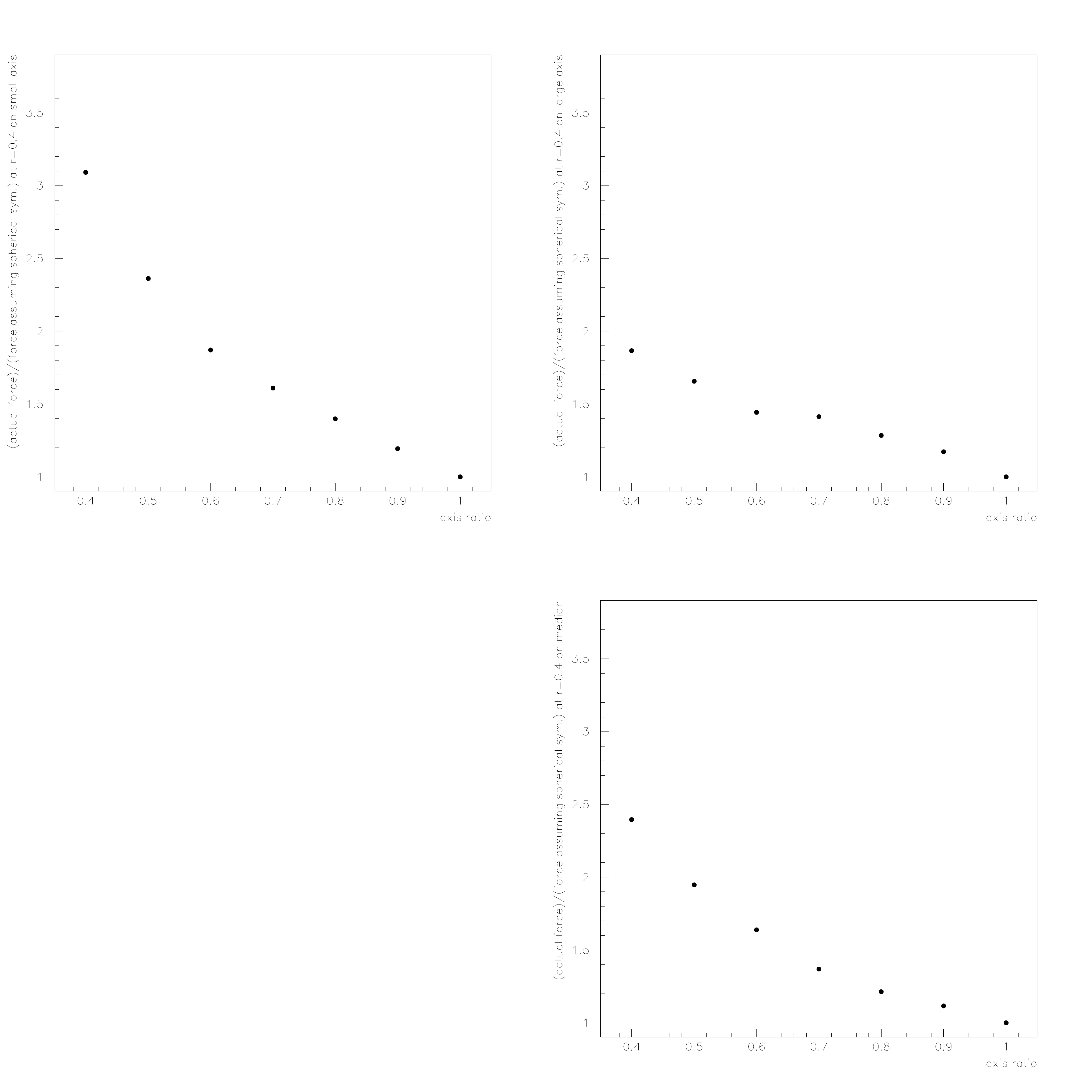}
\vspace{-0.4cm} \caption{\label{fig: newton shell correction}Gravitational force in function of the axis ratio for an oblate spheroid
of constant density, normalized to the force one would (erroneously)
obtain assuming Newton's shell theorem. Top left plot: the force is
computed at point on a small axis at a radius r=0.4$R_{max}$. Top
right plot: same as left but the force is computed at a r=0.4$R_{max}$
point on a large axis. Bottom plot: same as top plots but for the
force on a point on the median between a small and a large axis (r=0.4$R_{max}$). 
}
\end{figure}

All publications used here assumed spherical symmetry. We can partly
account for ellipticity effects by replacing Newton's shell theorem
used to derive $\sfrac{dP}{dr}=-GM(r)\rho(r)/r^{2}$ by a relation
accounting for the ellipticity. In Fig.~\ref{fig: newton shell correction},
we show the gravitational force in function of the axis ratio of an
oblate spheroid of constant density, normalized to the force one would
obtain if it were computed from Newton's shell theorem. We set the
mass of the spheroid to be independent of the axis ratio. Thus the
density increases with ellipticity. This is relevant for a galaxy
of known mass (luminosity) and unknown density%
\footnote{To obtain similar results at constant density (relevant to galaxies
of known density but unknown luminosity), one should take the results
of Fig.~\ref{fig: newton shell correction} and multiply them by the
axis ratio $\sfrac{R_{min}}{R_{max}}$, because an oblate spheroid volume is
(4/3)$\pi r_{min}^{2}r_{max}$. However, this case is irrelevant to
us. %
}. Although this simple correction is only accounting for part
of the ellipticity effects, we have used the median case in Fig.~\ref{fig: newton shell correction}
to correct the X-ray data. 

\subsection{Fukazawa {\it et al.} (2006)\label{sec:Fukazawa}}

\begin{figure}
\centering
\includegraphics[scale=0.5]{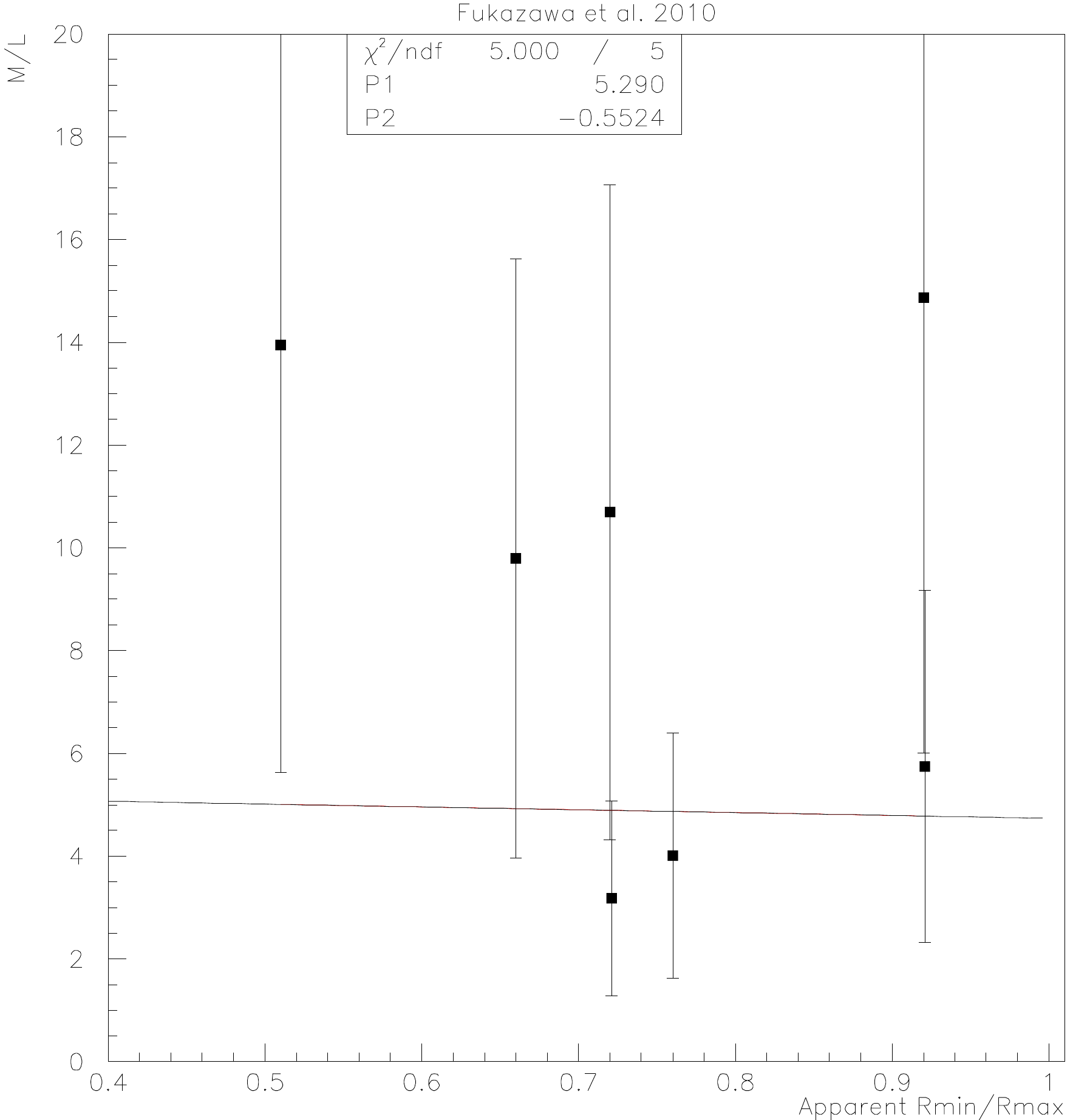}
\vspace{-0.4cm} \caption{\label{fig:fukazawa}$\sfrac{M}{L}$ vs apparent axis ratio for Fukazawa {\it et al.}~\cite{Fukazawa}. 
}
\end{figure}

X-ray data from Chandra X-ray Observatory satellite  were used by Fukazawa {\it et al. }~\cite{Fukazawa}
to extract the temperature profiles of 53 elliptical galaxies. Each
luminosity profile was obtained assuming a de Vaucouleur law~\cite{de Vaucouleur}.
Values for $\sfrac{M}{L}$ were provided at $R_{eff}$. After standard selection,
our final sample contains 7 adequate elliptical galaxies. $\sfrac{M}{L}$
vs axis ratio (from NED) is shown in Fig.~\ref{fig:fukazawa}. 
No uncertainties were provided in~\cite{Fukazawa}, so we assumed them to be proportional to their
corresponding $\sfrac{M}{L}$, and adjusted them so that the best fit has $\sfrac{\chi^{2}}{ndf}=1$
The best
fit to the data yields:

$\sfrac{M}{L_B}(R_{eff})=(-0.55\pm15.44)\sfrac{R_{min}}{R_{max}}|_{apparent}+(5.29\pm11.74)$.

The caveats of these results are as follow. Our correction for ellipticity
is basic (see Section~\ref{sec:Data-sets-using X-ray}). There is no correction for the asymmetry of the gas distribution
(due e.g. to AGN jets), although this particular caveat is alleviated
by our selection requirement. No uncertainties were provided. Because of the simplicity of
the ellipticity corrections, the possible gas asymmetry effects and
the absence of provided uncertainties, we assigned the results to
group 2 reliability. 

{\footnotesize For information, the fit before ellipticity
correction is:}{\footnotesize \par}

{\footnotesize $\sfrac{M}{L_B}(R_{eff})=(+0.38\pm11.45)\sfrac{R_{min}}{R_{max}}|_{apparent}+(3.54\pm8.52)$.}{\footnotesize \par}

\subsection{Nagino and Matsushita (2009)\label{sec:Nagino-and-Matsushita}}

Nagino and Matsushita~\cite{Nagino} used X-ray data from the XMM-Newton
and Chandra satellites  to extract $\sfrac{M}{L}$ integrated up to 3 different radii (0.5$R_{eff}$,
3$R_{eff}$ and 6$R_{eff}$) for 22 early-type galaxies. To obtain
$M(r)$, the temperature and density profiles were deprojected assuming the 
ISM to be spherically  symmetric. Likewise, the luminosity was deprojected to
obtain the stellar mass profile assuming spherical symmetry and a
de Vaucouleur profile~\cite{de Vaucouleur}. After standard selection,
4 galaxies remain. $\sfrac{M}{L}$ was extracted both in the B-band and
K-band. We assumed the uncertainties to be proportional to the corresponding $\sfrac{M}{L}$
values, with $\sfrac{\chi^{2}}{ndf}=1$. However, only
two points are available for 6$R_{eff}$ so $\sfrac{\chi^{2}}{ndf}$ is undefined.
In that case, we used the average of the uncertainties from 0.5$R_{eff}$
and 3$R_{eff}$. The axis ratios are from NED. %
\begin{figure}
\centering
\includegraphics[scale=0.5]{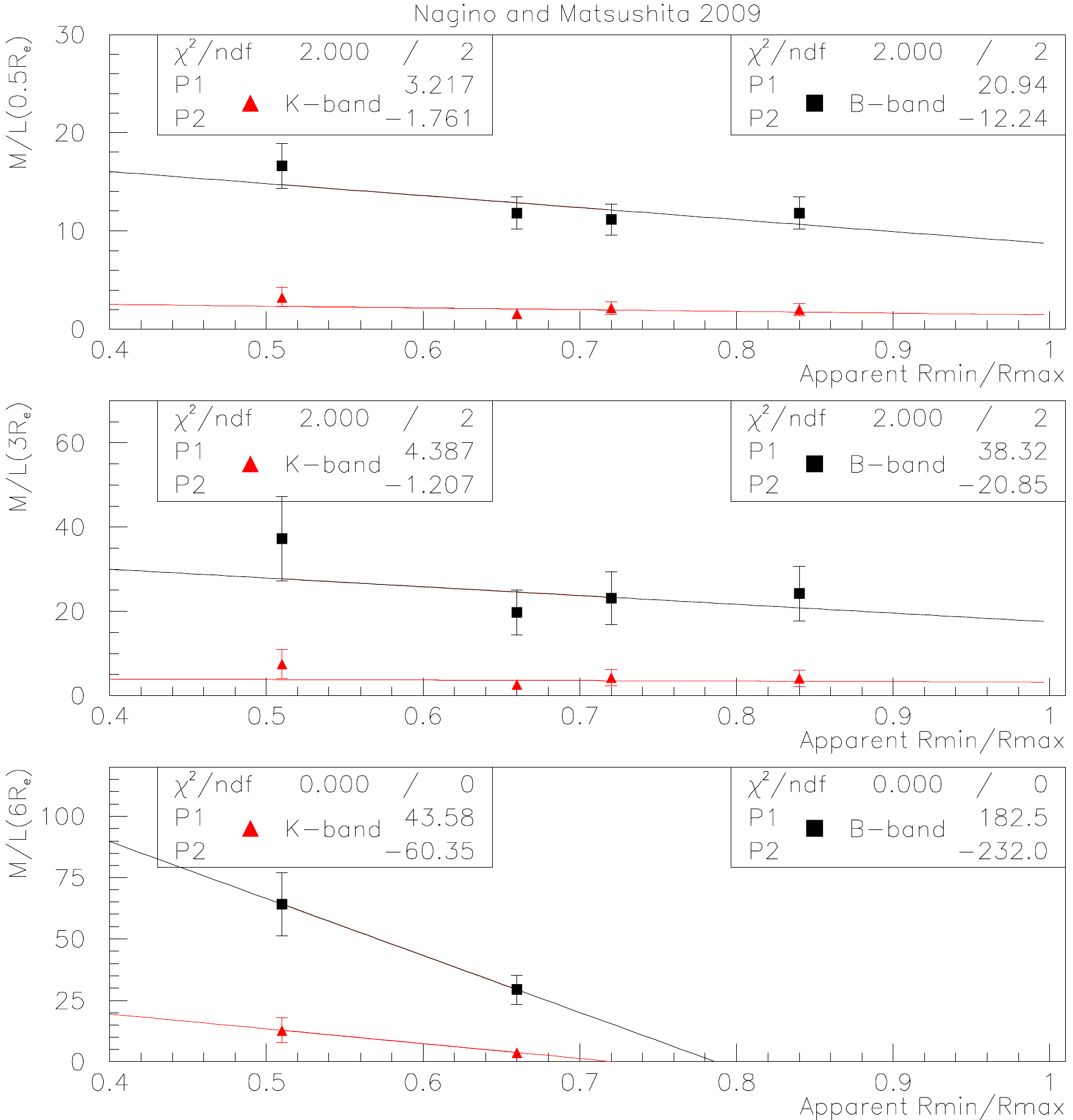}
\vspace{-0.4cm} \caption{\label{fig: nagino}$\sfrac{M}{L}$ vs apparent axis ratio for Nagino and Matsushita~\cite{Nagino}.
The top panel is for $\sfrac{M}{L}$ integrated up to $0.5R_{eff}$. The black
(red) symbols and fits are for $\sfrac{M}{L}$ determined in the B-band (K-band).
The middle and bottom panels are for $\sfrac{M}{L}$ integrated up to $3R_{eff}$
and 6$R_{eff}$ respectively. There are only 2 data points for $\frac{M}{L}(\leq6R_{eff})$
(no data for NGC4349).
}
\end{figure}
 The best fits (shown in Fig.~\ref{fig: nagino}) are:

$\sfrac{M}{L_B}(<0.5R_{eff})=(-12.24\pm8.41)\sfrac{R_{min}}{R_{max}}|_{apparent}+(20.94\pm6.01)$,

$\sfrac{M}{L_K}(<0.5R_{eff})=(-1.76\pm3.26)\sfrac{R_{min}}{R_{max}}|_{apparent}+(3.21\pm2.32)$,

$\sfrac{M}{L_B}(<3R_{eff})=(-20.85\pm33.35)\sfrac{R_{min}}{R_{max}}|_{apparent}+(38.32\pm23.74)$,

$\sfrac{M}{L_K}(<3R_{eff})=(-1.21\pm10.36)\sfrac{R_{min}}{R_{max}}|_{apparent}+(4.39\pm7.31)$.

\noindent The lines going through the 2 data points at 6$R_{eff}$
are:

$\sfrac{M}{L_B}(<6R_{eff})=(-232.0\pm94.0)\sfrac{R_{min}}{R_{max}}|_{apparent}+(182.5\pm59.9)$, 

$\sfrac{M}{L_K}(<6R_{eff})=(-60.35\pm34.68)\sfrac{R_{min}}{R_{max}}|_{apparent}+(43.58\pm22.52)$. \\
As for the results from Section~\ref{sec:Fukazawa}, we assigned these
results to reliability group 2.\\
{\footnotesize For information, the fit results before ellipticity
correction are:}{\footnotesize \par}

{\footnotesize $\sfrac{M}{L_B}(<0.5R_{eff})=(3.43\pm3.79)\sfrac{R_{min}}{R_{max}}|_{apparent}+(6.38\pm2.58)$,}{\footnotesize \par}

{\footnotesize $\sfrac{M}{L_K}(<0.5R_{eff})=(0.36\pm1.94)\sfrac{R_{min}}{R_{max}}|_{apparent}+(1.18\pm1.33)$,}{\footnotesize \par}

{\footnotesize $\sfrac{M}{L_B}(<3R_{eff})=(5.02\pm19.22)\sfrac{R_{min}}{R_{max}}|_{apparent}+(13.55\pm13.14)$,}{\footnotesize \par}

{\footnotesize $\sfrac{M}{L_K}(<3R_{eff})=(1.05\pm6.91)\sfrac{R_{min}}{R_{max}}|_{apparent}+(1.833\pm4.73)$.}{\footnotesize \par}

\noindent {\footnotesize The lines going through the 2 data points
at 6$R_{eff}$ are:}{\footnotesize \par}

{\footnotesize $\sfrac{M}{L_B}(<6R_{eff})=(-88.00\pm44.54)\sfrac{R_{min}}{R_{max}}|_{apparent}+(78.48\pm27.76)$, }{\footnotesize \par}

{\footnotesize $\sfrac{M}{L_K}(<6R_{eff})=(-27.33\pm17.25)\sfrac{R_{min}}{R_{max}}|_{apparent}+(20.64\pm11.08)$. }{\footnotesize \par}

\section{Data sets using warm or cold gas disk dynamics}

Cold hydrogen gas (HI) can be found in some elliptical galaxies. It
extends to large distances ($r>10R_{eff}$). $\sfrac{M}{L}$ is determined
using ionized gas disks embedded in the galaxies as kinematical tracers.
Ionized warm gas can also be used but warm gas disks are less extended
and thus  can be determined only in the inner regions. The $\sfrac{M}{L}$
are obtained the same way as for spiral galaxies, i.e, using a model
with visible and dark component potentials  to match the disk rotation
curve. Thus virial or hydrostatic equilibrium is not assumed and our
usual selection criteria can be relaxed. A caveat for our study with
this method of $\sfrac{M}{L}$ determination is that extended HI rings are located
mostly in S0 galaxies. In most cases elliptical galaxies with HI disks
are gas rich, and thus are unusual  among elliptical galaxies.

\subsection{Bertola {\it et al.} (1991 and 1993)\label{sub:Bertola-et-al.}}

\begin{figure}
\centering
\includegraphics[scale=0.5]{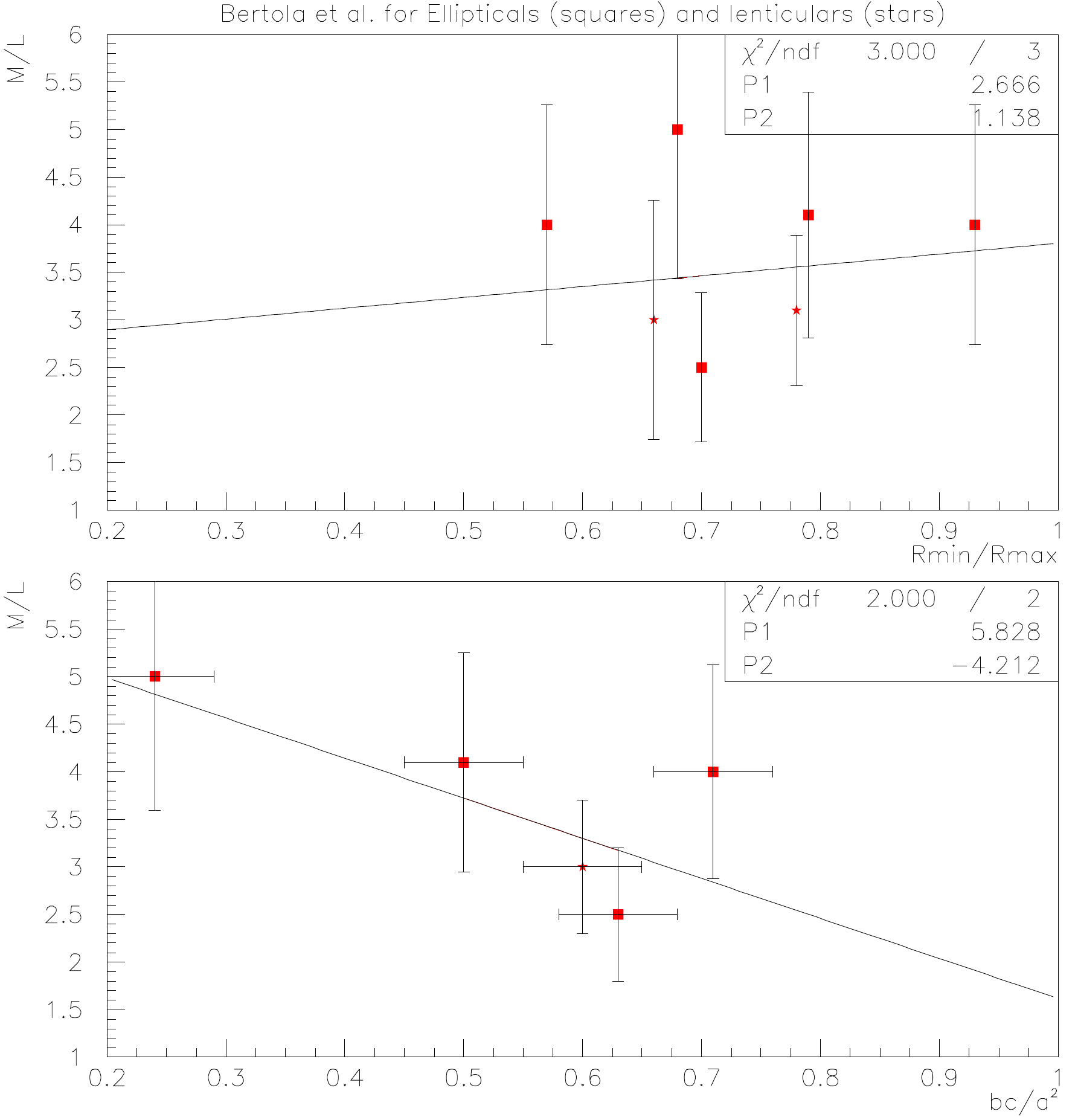}
\vspace{-0.4cm} \caption{\label{fig:bertola}$\sfrac{M}{L}$ vs apparent (top plot) and intrinsic axis ratios,
defined as $bc/a^{2}$, with $a,b$ and $c$ the radii of the elliptical
galaxy triaxial shape model. Data are from, Bertola {\it et al.}
\cite{Bertola93},~\cite{Bertola91}. The square symbols
represent  elliptical galaxies. The $S_0$ galaxies
are indicated  (star symbols), but not included in the fits. 
}
\end{figure}

Bertola {\it et al.}~\cite{Bertola93} provided $\sfrac{M}{L_B}$
for inner regions ($0.3R_{eff}<r<0.9R_{eff}$) of 5 elliptical
and 2 $S_0$ galaxies using ionized gas disks. They also determined
values for the outer regions of 4 elliptical galaxies using HI disks
($3.3R_{eff}<r<10R_{eff}$). They had already used the same technique
to obtain the inner $\frac{M}{L_B}(r=1.3R_{eff}$) for NGC5077, see~\cite{Bertola91},
which we added to the sample. All galaxies are LINERS and/or Sy type.
Their true ellipticities were obtained by fitting data with a
tri-axial model for four of the elliptical galaxies and one of the
lenticular galaxies. 
The best fit for the 4 elliptical galaxies is (see Fig.~\ref{fig:bertola}):

$\sfrac{M}{L_{B}}=(-4.21\pm3.55)\sfrac{bc}{a^{2}}+(5.83\pm2.09)$ for
the 4 elliptical galaxies,\\
where $a,b$ and $c$ are the radii of the triaxial shape model for the elliptical galaxy.
Uncertainties were not given. We assumed that $\Delta(\sfrac{M}{L})\propto \sfrac{M}{L}$,
with the proportionality constant such as $\sfrac{\chi^{2}}{ndf}=1$ in the
fit. We assigned a $\pm0.05$ uncertainty on the intrinsic  ``axis
ratio'' $bc/a^{2}$. We did not include the $S_0$ galaxy in
the fit reported above. Including it changes little
the results, except that the correlation is more marked. Using the
5 elliptical galaxies for which $\sfrac{M}{L}$ are provided and determining
the correlation with the {\it apparent} axis ratio $\sfrac{R_{min}}{R_{max}}$,
we found:\\
 $\sfrac{M}{L_{B}}=(+1.14\pm4.64)\sfrac{R_{min}}{R_{max}}|_{apparent}+(2.67\pm3.42)$.\\
This illustrates the importance of the galaxy projection effect.

The data analysis accounted for the true shapes of the galaxies, so
that there was no need of projection correction. The analysis used
special galaxies (presence of a gas disks, Sy2, LINERS) and no uncertainty
was given. On the other hand, the method is robust (similar to $\sfrac{M}{L}$
determination for spiral galaxies). We assigned the result to group 1 reliability.

\subsection{Pizzella {\it et al.} (1997)}

\begin{figure}
\centering
\includegraphics[scale=0.5]{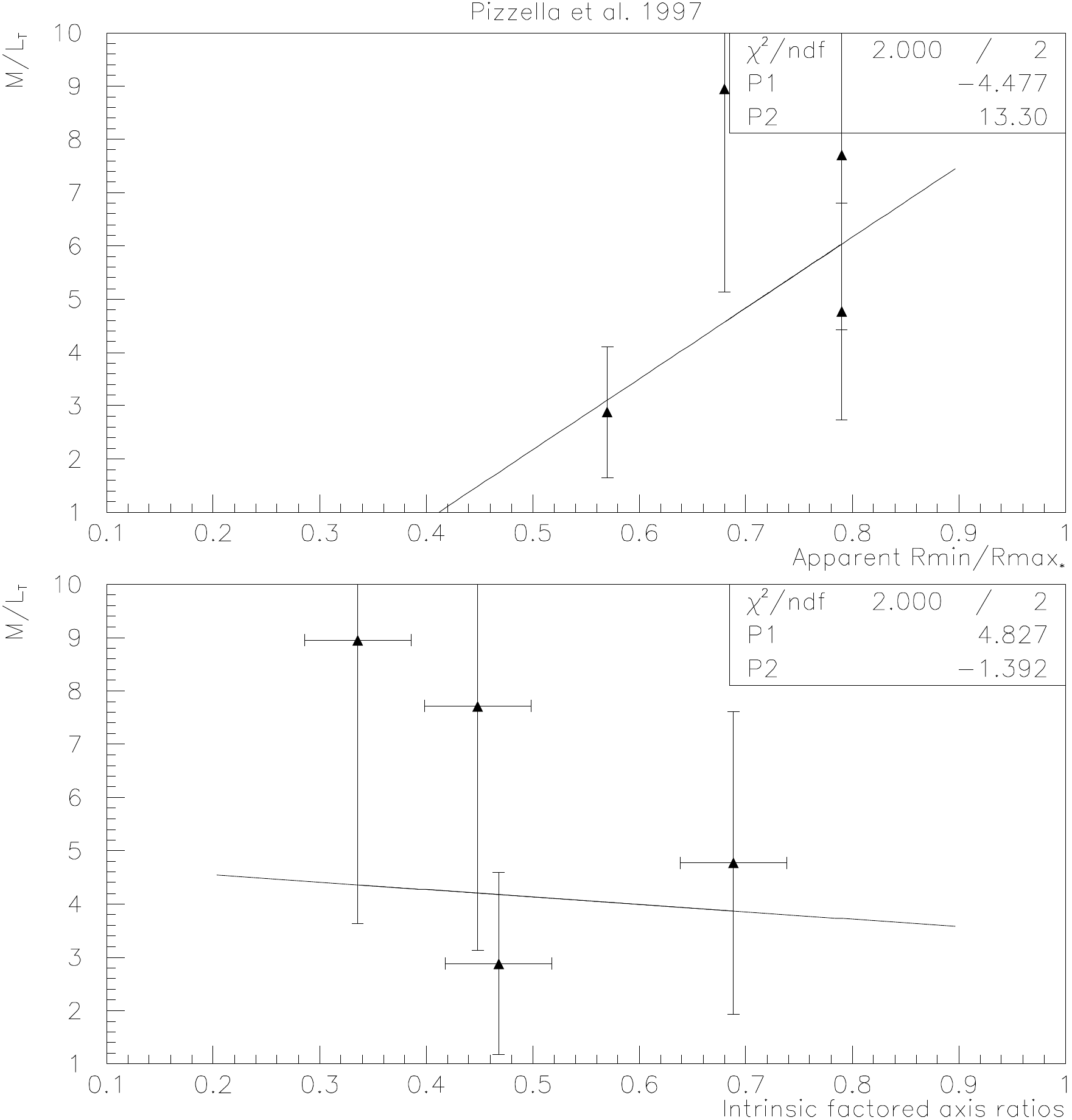}
\vspace{-0.4cm} \caption{\label{fig: pizzella}$\sfrac{M}{L_T}$ from Pizzella {\it et al.}~\cite{Pizeella} vs apparent
(top panel) and intrinsic (bottom panel) axis ratios.
}
\end{figure}

Pizzella {\it et al.}~\cite{Pizeella} used the velocity field
measurements on ionized warm gas disks embedded in 4 elliptical galaxies
to obtain their  gravitational potentials. For each galaxy, an intrinsic
triaxial galactic shape was obtained, which was argued by the authors
to not be strongly model-dependent. The galactic mass integrated
up to the maximum available radius was obtained from the velocity
field. The modeling worked well for 3 galaxies but not as well for
NGC7097 (which has a counter-rotating core). All galaxies are LINERS
and two (NGC2974 and 5077) are Sy galaxies. We kept NGC7097 because
its results are published nonetheless and its intrinsic ellipticity
($1-q_{0}p_{0}$, where $q_{0}$ and $p_{0}$ are the intrinsic axis
ratios of the triaxial galaxy) is close to the one given by Bertola
{\it et al.}~\cite{Bertola93}. $\sfrac{M}{L_T}$ was obtained from
the potential and the light density profile. $\sfrac{M}{L_T}$ vs apparent
and intrinsic axis ratios are shown in Fig.~\ref{fig: pizzella}. 

\noindent The best fits are:

$\sfrac{M}{L_T}(R_{eff}/2)=(+13.30\pm9.58)\sfrac{R_{min}}{R_{max}}|_{apparent}+(-4.48\pm6.27)$,

$\sfrac{M}{L_T}(R_{eff}/2)=(-1.39\pm13.13)q_{o}p_{o}+(4.83\pm6.80)$.\\
The uncertainties on $\sfrac{M}{L_T}$ were scaled so that $\sfrac{\chi^{2}}{ndf}=1$,
and a $\pm0.05$ uncertainty was assumed for the intrinsic axis ratio. 
The bias for using apparent rather than intrinsic axis ratios  is clear
(the effect of using the true shape rather than the projected one
to get $\sfrac{M}{L}$ is already accounted for). We assigned the
results to reliability group~1.

\section{Data sets using strong lensing\label{sec:Data-sets-using strong lensing}}

Strong lensing has become a standard technique to extract $\sfrac{M}{L}$. Its
advantage is that the determination of $M(R_{Ein})$, the total mass
within the Einstein radius $R_{Ein}$, is in principle model independent.
Obtaining $\sfrac{M}{L}$ at other radii necessitates assuming a mass profile.
Since strong lensing galaxies are distant, a light profile must
be assumed too. Nevertheless, the method is the least model
dependent way to obtain a galactic total mass. Virial or hydrostatic
equilibria are not assumed, so our tight selection can be relaxed,
keeping in mind that evidence of starburst or H-II emission can bias
$\sfrac{M}{L}$ by affecting the luminosity. Selection enforcing a homogeneous
galaxy sample (i.e, discarding cD, S0, cE galaxies...) is still necessary.
 Although equilibrium is not required, internal or external shears
due to interaction with nearby neighbors can bias the determination
of the mass. They are usually corrected for by the authors, although
such correction is not unambiguous. Lensing galaxies are often found
in clusters or groups. Our systematic study of the effects of environment
shows that the galaxies in groups or clusters display more dispersion 
around their $\sfrac{M}{L}$ vs $\sfrac{R_{min}}{R_{max}}$ correlation, see Sections
\ref{sub:Auger-et-al.},~\ref{sec:Barnabe-et-al.} and~\ref{sub:Cardone-et-al.2011}
(we found the opposite in Section~\ref{sub:Cardone-et-al.2009}
but it may be a statistical fluctuation). This is expected if unaccounted
galaxy-galaxy interactions cause a data jitter. In addition, the correlation slope
seems larger for isolated galaxies.

A general caveat of this technique is that the probability of observing
lensing from a galaxy is low so the known ones are distant. Consequently,
their detailed characteristics are not known. In addition, distant
galaxies may be at different evolution stages compared to the local
ones. Furthermore, lensing equations bias samples toward heavier elliptical
galaxies on the one hand, and higher ellipticity on the other hand.
whilst the galaxies are at large redshifts $z$, we chose to not correct
for a possible $\sfrac{M}{L}$-$z$ correlation because it is unclear whether
it comes from structure evolution (as investigated e.g. in~\cite{Keeton}),
or from structure and sample bias (lensing sample biased toward flatter
galaxies, with flatness correlating with $\sfrac{M}{L}$).

\subsection{Auger {\it et al.} (2010) \label{sub:Auger-et-al.}}

Auger {\it et al.}~\cite{Auger 1} modeled the light and mass
of 73 early-type galaxies based on photometric and strong lensing
data from the Sloan Lens ACS (SLACS) survey~\cite{Bolton (SLACS)}.
The galaxies are located at $0.06<z<0.51$ redshifts. The data are interpreted with
a de Vaucouleur profile~\cite{de Vaucouleur} for photometry
and a SIE (Singular Isothermal Ellipsoid) for the mass profile. The interpretation 
includes the effects of (apparent) ellipticity. From this, $DMf$ was
estimated at $R_{eff}/2$. S0 galaxies were identified so we did not
apply the standard $\sigma\geq225$~km.s$^{-1}$ criterion for this
analysis (and all other analyses using SLACS data).

\subsubsection{Analysis on full sample of Elliptical galaxies}

\begin{figure}
\centering
\includegraphics[scale=0.5]{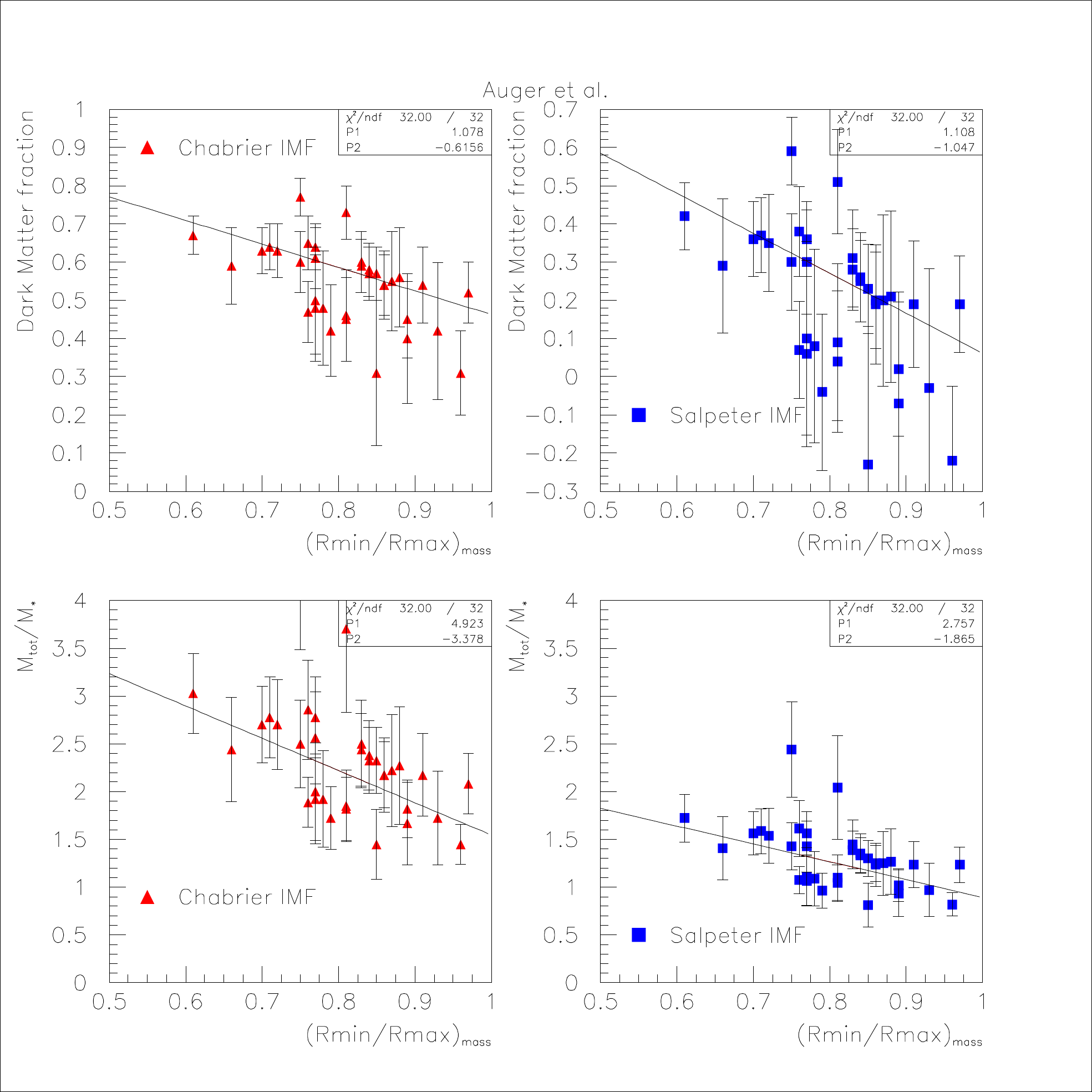}
\vspace{-0.4cm} \caption{\label{fig:Auger}$DMf$ (top plots) and $\sfrac{M_{tot}}{M_*}$ (bottom plots)
vs mass axis ratio $\sfrac{R_{min}}{R_{max}}$ for the Auger {\it et
al.} data set~\cite{Auger 1}. The left plots are for mass SIE
models using a Chabrier IMF and the right ones are for a Salpeter IMF.
The lines are the best linear fits to the data.
}
\end{figure}

We kept 34 elliptical galaxies, rejecting S0 galaxies (we relied on
the identification from~\cite{Auger 2}) and minimizing the contamination
of giant galaxies by requesting $M<10^{12}M_{\odot}$ (the masses
were taken from $ $\cite{Grillo09}). Mass axis ratios
$\sfrac{R_{min}}{R_{max}}|_{mass}$,\footnote{
$\sfrac{R_{min}}{R_{max}}|_{mass}$ is the axis ratio of the modeled mass system.}
 apparent axis ratios $\sfrac{R_{min}}{R_{max}}|_*$
and luminosities are from~\cite{Bolton (SLACS)}. If none were provided for
a given galaxy, it was rejected. In addition, we rejected  the galaxies
signaled by the authors as significant fit outliers in the size-$\sigma$-$M_*$
correlation space. The  $DMf$ vs axis is shown
in Fig.~\ref{fig:Auger}. The best fits give, for the results derived
with a SIE model using a Chabrier IMF~\cite{Chabrier}:

$DMf=(-0.62\pm0.17)\sfrac{R_{min}}{R_{max}}|_{mass}+(1.08\pm0.13)$

\noindent and, for results derived with a Salpeter IMF~\cite{Salpeter}:

$DMf=(-1.05\pm0.28)\sfrac{R_{min}}{R_{max}}|_{mass}+(1.11\pm0.22)$.

\noindent For the mass ratio, we have:

$\sfrac{M_{tot}}{M_*}=(-3.38\pm0.79)\sfrac{R_{min}}{R_{max}}|_{mass}+(4.92\pm0.66)$
(Chabrier),

$\sfrac{M_{tot}}{M_*}=(-1.86\pm0.46)\sfrac{R_{min}}{R_{max}}|_{mass}+(2.76\pm0.38)$
(Salpeter).

\noindent Using apparent axis ratio rather than mass, we obtain for
the Chabrier IMF:

$DMf=(-0.12\pm0.17)\sfrac{R_{min}}{R_{max}}|_{apparent}+(0.69\pm0.12)$

\noindent and, for the Salpeter IMF:

$DMf=(-0.22\pm0.27)\sfrac{R_{min}}{R_{max}}|_{apparent}+(0.46\pm0.21)$,

$\sfrac{M_{tot}}{M_*}=(-1.47\pm0.70)\sfrac{R_{min}}{R_{max}}|_{apparent}+(3.30\pm0.57)$
Chabrier),

$\sfrac{M_{tot}}{M_*}=(-0.79\pm0.40)\sfrac{R_{min}}{R_{max}}|_{apparent}+(1.84\pm0.32)$
(Salpeter).

\noindent The published uncertainties were rescaled slightly so that
$\sfrac{\chi^{2}}{ndf}=1$. We assigned the results using a Chabrier IMF to
reliability group 1.

\subsubsection{Luminosity and environment study}

Subsamples of galaxies allowed us to study the effect of luminosity
(correlating with the boxy/disky galactic shape) and environment on
these results:
\begin{enumerate}
\item Subsample 1 contains elliptical galaxies not found in clusters according
to Treu {\it et al.}~\cite{Treu09} (17 elliptical galaxies);
\item Subsample 2 contains elliptical galaxies found in clusters (17 elliptical
galaxies);
\item Subsample 3 contains elliptical galaxies with magnitude $M_{B}\gtrsim-19.5$.
This selects less massive disky galaxies (25 elliptical galaxies);
\item Subsample 4 contains elliptical galaxies with magnitude $M_{B}\lesssim-19.5$
and no generic selection $M>10^{12}M_{\odot}$ on the total mass.
This selects more massive boxy galaxies (17 elliptical galaxies);
\item Subsample 5 contains elliptical galaxies with magnitude $M_{B}\gtrsim-19.5$
and not found in clusters (13 elliptical galaxies).
\end{enumerate}

\paragraph{Subsample 1 results}

The best fit to the data obtained using a Chabrier IMF and mass axis
ratio is:

$DMf=(-0.81\pm0.20)\sfrac{R_{min}}{R_{max}}|_{mass}+(1.22\pm0.16)$ 

\noindent and, for the Salpeter IMF and mass axis ratio:

$DMf=(-1.43\pm0.34)\sfrac{R_{min}}{R_{max}}|_{mass}+(1.40\pm0.27)$.

\noindent Using apparent axis ratio, we obtain for the Chabrier IMF:

$DMf=(-0.68\pm0.26)\sfrac{R_{min}}{R_{max}}|_{apparent}+(1.10\pm0.20)$ 

\noindent and, for the Salpeter IMF:

$DMf=(-1.13\pm0.45)\sfrac{R_{min}}{R_{max}}|_{apparent}+(1.13\pm0.35)$.

\paragraph{Subsample 2 results}

The best fit to the data obtained using a Chabrier IMF and mass axis
ratio is:

$DMf=(-0.45\pm0.25)\sfrac{R_{min}}{R_{max}}|_{mass}+(0.97\pm0.20)$ 

\noindent and, for the Salpeter IMF and mass axis ratio:

$DMf=(-0.79\pm0.43)\sfrac{R_{min}}{R_{max}}|_{mass}+(0.79\pm0.34)$.

\noindent Using apparent axis ratio, we obtain for the Chabrier IMF:

$DMf=(+0.02\pm0.20)\sfrac{R_{min}}{R_{max}}|_{apparent}+(0.60\pm0.15)$ 

\noindent and, for the Salpeter IMF:

$DMf=(+0.02\pm0.34)\sfrac{R_{min}}{R_{max}}|_{apparent}+(0.30\pm0.26)$.

\paragraph{Subsample 3 results}

The best fit to the data obtained with a Chabrier IMF and mass axis
ratio is:

$DMf=(-0.41\pm0.23)\sfrac{R_{min}}{R_{max}}|_{mass}+(0.89\pm0.19)$ 

\noindent and, for the Salpeter IMF and mass axis ratio:

$DMf=(-0.66\pm0.38)\sfrac{R_{min}}{R_{max}}|_{mass}+(0.76\pm0.31)$.

\noindent Using apparent axis ratio, we obtain for the Chabrier IMF:

$DMf=(+0.08\pm0.14)\sfrac{R_{min}}{R_{max}}|_{apparent}+(0.50\pm0.12)$ 

\noindent and, for the Salpeter IMF:

$DMf=(+0.10\pm0.24)\sfrac{R_{min}}{R_{max}}|_{apparent}+(0.16\pm0.19)$.

\paragraph{Subsample 4 results}

The best fit to the data obtained using a Chabrier IMF and mass axis
ratio is:

$DMf=(-0.41\pm0.19)\sfrac{R_{min}}{R_{max}}|_{mass}+(0.94\pm0.14)$ 

\noindent and, for the Salpeter IMF and mass axis ratio:

$DMf=(-0.73\pm0.34)\sfrac{R_{min}}{R_{max}}|_{mass}+(0.89\pm0.25)$.

\noindent Using apparent axis ratio, we obtain for the Chabrier IMF:

$DMf=(-0.22\pm0.28)\sfrac{R_{min}}{R_{max}}|_{apparent}+(0.80\pm0.21)$ 

\noindent and, for the Salpeter IMF:

$DMf=(-0.35\pm0.48)\sfrac{R_{min}}{R_{max}}|_{apparent}+(0.62\pm0.36)$.

\paragraph{Subsample 5 results}

The best fit to the data derived with Chabrier IMF and mass axis ratio
is:

$DMf=(-0.62\pm0.25)\sfrac{R_{min}}{R_{max}}_{mass}+(1.06\pm0.20)$ 

\noindent and, for results derived with Salpeter IMF and mass axis
ratio:

$DMf=(-1.13\pm0.43)\sfrac{R_{min}}{R_{max}}_{mass}+(1.14\pm0.35)$.

\noindent Using apparent axis ratio, we obtain for the Chabrier IMF:

$DMf=(-0.15\pm0.36)\sfrac{R_{min}}{R_{max}}|_{apparent}+(0.68\pm0.29)$ 

\noindent and, for the Salpeter IMF:

$DMf=(-0.23\pm0.63)\sfrac{R_{min}}{R_{max}}|_{apparent}+(0.41\pm0.50)$.

\subsubsection{Conclusion}

The subsample results are generally compatible with the full sample
ones. There is no clear difference between samples of luminous/boxy
and of fainter/disky galaxies. Selecting field galaxies increased
the $DMf$ slope whilst the ones in groups/cluster appeared to have
a smaller slope. The correlation for field galaxies is also clearer
(smaller relative uncertainty for the slope)

The caveats of the analysis are that apparent axis ratios $\sfrac{R_{min}}{R_{max}}$
were used. However, from the results of Barnabe {\it et al. }~\cite{Barnabe},
considering intrinsic axis ratios rather than apparent ones has a
small effect: the change in slope is compatible with 0: $(-18\pm59)\%$
for the Chabrier results and $(-33\pm70)\%$ for the Salpeter IMF.
The results and the mass axis ratios are model dependent. The IMF
dependency can be estimated by comparing the Salpeter and Chabrier
IMF results.

\subsection{Barnabe {\it et al.} (2011) \label{sec:Barnabe-et-al.} }

\begin{figure}
\centering
\includegraphics[scale=0.5]{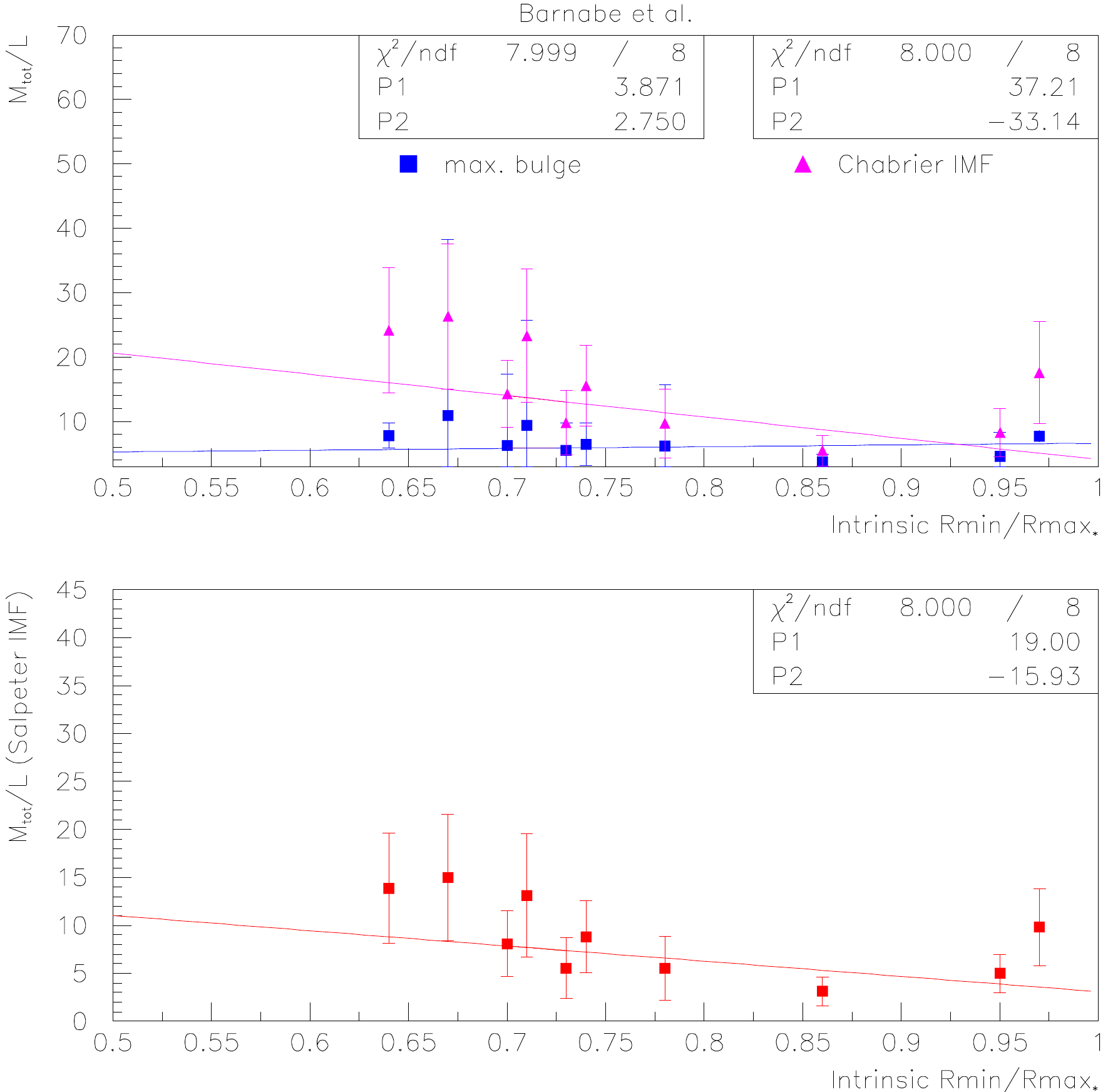}
\vspace{-0.4cm} \caption{\label{fig:Baranabe}$\sfrac{M}{L}$ vs true luminous axis ratio $\sfrac{R_{min}}{R_{max}}|_{true}$
from Barnabe {\it et al.}~\cite{Barnabe}. The top (bottom)
plot is for the stellar population synthesis model using a Chabrier
(Salpeter) IMF. The lines give the best fits to the data.
}
\end{figure}

Barnabe {\it et al.}~\cite{Barnabe} modeled the dark matter
and stellar contents within about 0.5$R_{eff}$ for 16 E and S0 galaxies
from the SLACS survey. In addition, they used spectroscopic observations
from other sources than SLACS to further constrain their models. The
stellar content was obtained with a Stellar Population Synthesis model
using a Chabrier or a Salpeter IMF. Independently, they obtained
a lower limit on $DMf$ by using a  ``maximum bulge
hypothesis''. The models provide intrinsic mass axis ratios and
luminous axis ratios. We selected 10 elliptical galaxies, the rest
being S0 galaxies or E for which $M>10^{12}M_{\odot}$ (masses from
$ $\cite{Grillo09}). Among the 10 elliptical galaxies, 2
are fast rotators and 8 are slow rotators, 4 belong to clusters and
6 are more isolated. We formed $\sfrac{M}{L}$ using $DMf$ and $\sfrac{M_*}{L}$.
The best fits yield for the $\sfrac{M}{L}$ correlation with intrinsic luminous
axis ratio (see Fig.~\ref{fig:Baranabe}):

$\sfrac{M}{L}=(-33.1\pm16.7)\sfrac{R_{min}}{R_{max}}|_{true(light)}+(37.2\pm13.9)$
for the data computed with a Chabrier IMF,

$\sfrac{M}{L}=(-15.9\pm9.8)\sfrac{R_{min}}{R_{max}}|_{true(light)}+(19.0\pm8.2)$
for the data computed with a Salpeter IMF. 

By fitting samples containing only isolated galaxies or galaxies belonging
to clusters we check the effects of the environment on these results.
The best fits yield for the galaxies not found in clusters:

$\sfrac{M}{L}=(-59.2\pm16.4)\sfrac{R_{min}}{R_{max}}|_{true(light)}+(56.24\pm13.2)$
for the data obtained with a Chabrier IMF,

$\sfrac{M}{L}=(-33.6\pm9.4)\sfrac{R_{min}}{R_{max}}|_{true(light)}+(32.0\pm7.5)$
with a Salpeter IMF,

\noindent and for the galaxies found in clusters:

$\sfrac{M}{L}=(-47.9\pm22.9)\sfrac{R_{min}}{R_{max}}|_{true(light)}+(55.8\pm20.8)$
for the data computed with a Chabrier IMF,

$\sfrac{M}{L}=(-26.3\pm13.4)\sfrac{R_{min}}{R_{max}}|_{true(light)}+(31.1\pm12.3)$
with a Salpeter IMF. 

\noindent The isolated galaxies have a clearer, and possibly
stronger, correlation as seen already in Section~\ref{sub:Auger-et-al.}.

Finally, we checked the effect of the viewing angle projection on our
analysis by repeating the analysis done on sample 1 but using the
apparent light axis ratios rather than the intrinsic  one:

$\sfrac{M}{L}=(-39.0\pm11.9)\sfrac{R_{min}}{R_{max}}|_{apparent(light)}+(44.8\pm10.7)$
for the data computed with a Chabrier IMF,

$\sfrac{M}{L}=(-21.1\pm7.18)\sfrac{R_{min}}{R_{max}}|_{apparent(light)}+(24.7\pm6.5)$
with a Salpeter IMF. 

\noindent The projection effect is small and the fit values of the
intrinsic and apparent cases are largely compatible. When the intrinsic
axis ratio is used rather than the apparent one, the $\sfrac{M}{L}$ vs axis
ratio slope decreases by about $(18\pm59)\%$ for the Chabrier result
and by about $(33\pm70)\%$ for the Salpeter result. (The uncertainties
determination assumes that the Chabrier and Salpeter data are uncorrelated,
which is not true. Hence, the 59\% and 70\% uncertainties are overestimated.)

Advantages of this analysis are that it was done in term of real axis
ratio and it used additional spectroscopic data. We assigned the results
to group 1 reliability.

\subsection{Cardone {\it et al.} (2009)\label{sub:Cardone-et-al.2009}}

Cardone {\it et al.}~\cite{Cardone09} examined 21 early-type
galaxies from the SLACS survey. Among them, 18 are elliptical galaxies
(identification from~\cite{Auger 2}). The strong-lensing/photometry
method was used to extract the $DMf$ up to the effective and Einstein
radii, with a flexible $\sfrac{M}{L}$ ansatz that interpolated between several
types of halo models. The preferred halo was selected by $\chi^{2}$
minimization. This reduces the model dependence of the $DMf$ determination
compared to other strong-lensing/photometry results. Spherical symmetry
was assumed, the ellipticity of the galaxies being accounted for only
in the deprojection procedure to obtain luminosity profiles. The stellar
$\sfrac{M_*}{L}$ was obtained from a Stellar Population Model using a Chabrier
IMF. We formed $\sfrac{M}{L}$ from $\sfrac{M_*}{L}$ and $DMf$. The apparent and mass
axis ratios were taken from~\cite{Bolton (SLACS)}. We compared
the luminosities reported by the authors with those in~\cite{Bolton (SLACS)}
and rejected galaxies disagreeing by more than $\sim20\%$, and galaxies
for which luminosities were not given in~\cite{Bolton (SLACS)}.
We also rejected 3 giant galaxies ($M>10^{12}M_{\odot}$, masses
from~\cite{Grillo09}). All in all, we kept 13 galaxies (Main
Sample). We split the Main Sample in two subsamples containing field
galaxies (subsample 1, with 9 elliptical galaxies) and galaxies found in
clusters (subsample 2, with 4 elliptical galaxies)~\cite{Treu09}.
The fits to $DMf$ or $\sfrac{M}{L}$ vs $\sfrac{R_{min}}{R_{max}}$, shown in Fig~\ref{fig:cardone21},
yield, for the quantities obtained at $R_{eff}$ (uncertainties are
from the 95\% CL values of~\cite{Cardone09} scaled so that
$\sfrac{\chi^{2}}{ndf}=1$): 
\begin{figure}
\centering
\includegraphics[scale=0.5]{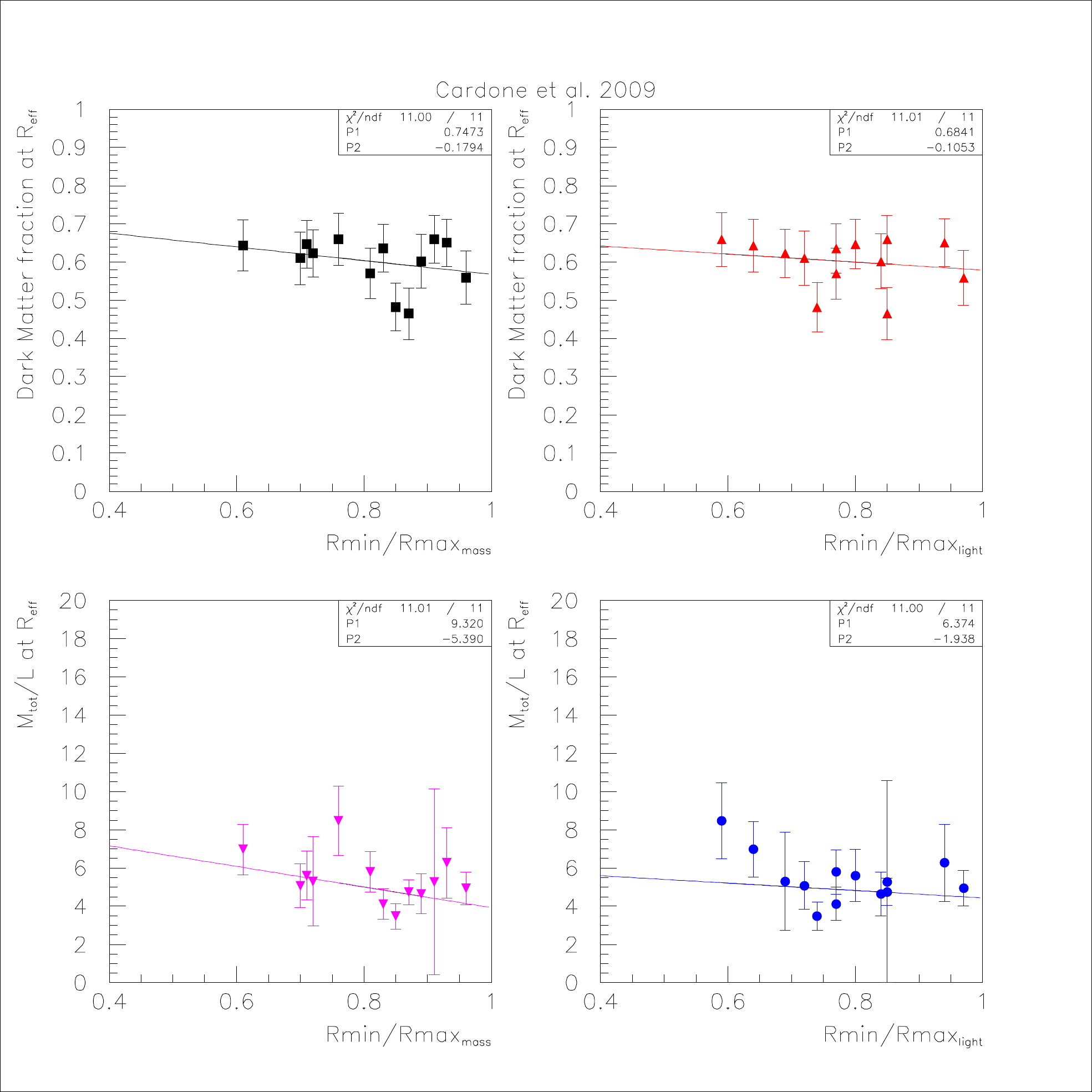}
\vspace{-0.4cm} \caption{\label{fig:cardone21}Dark matter content vs axis ratio for Cardone {\it et al.}~\cite{Cardone09}.
Top plots: $DMf$ vs mass axis ratio (left) and apparent
$\sfrac{R_{min}}{R_{max}}$ (right): bottom plots, $\sfrac{M}{L}$ vs
mass $\sfrac{R_{min}}{R_{max}}$ (left) and apparent $\sfrac{R_{min}}{R_{max}}$
(right).
}
\end{figure}
$DMf=(-0.18\pm0.18)\sfrac{R_{min}}{R_{max}}|_{mass}+(0.75\pm0.15)$,

$DMf=(-0.11\pm0.18)\sfrac{R_{min}}{R_{max}}|_{apparent}+(0.68\pm0.14)$,

$\sfrac{M}{L}=(-5.39\pm3.40)\sfrac{R_{min}}{R_{max}}|_{mass}+(9.32\pm2.85)$,

$\sfrac{M}{L}=(-1.94\pm3.52)\sfrac{R_{min}}{R_{max}}|_{apparent}+(6.37\pm2.82)$.

\noindent For the isolated galaxies (subsample 1), the best fits
are:

$DMf=(-0.08\pm0.30)\sfrac{R_{min}}{R_{max}}|_{mass}+(0.66\pm0.25)$,

$DMf=(0.20\pm0.31)\sfrac{R_{min}}{R_{max}}|_{apparent}+(0.42\pm0.26)$,

$\sfrac{M}{L}=(-1.68\pm3.97)\sfrac{R_{min}}{R_{max}}|_{mass}+(6.13\pm3.40)$,

$\sfrac{M}{L}=(+3.91\pm3.21)\sfrac{R_{min}}{R_{max}}|_{apparent}+(1.48\pm2.65)$.

\noindent For the galaxies found in clusters, the best fits are:

$DMf=(0.00\pm0.14)\sfrac{R_{min}}{R_{max}}|_{mass}+(0.64\pm0.10)$,

$DMf=(-0.13\pm0.10)\sfrac{R_{min}}{R_{max}}|_{apparent}+(0.73\pm0.07)$,

$\sfrac{M}{L}=(-13.25\pm8.83)\sfrac{R_{min}}{R_{max}}|_{mass}+(15.47\pm6.82)$,

$\sfrac{M}{L}=(-22.93\pm6.10)\sfrac{R_{min}}{R_{max}}|_{apparent}+(21.73\pm4.39)$.

The trend is that, contrarily to the results in sections~\ref{sub:Auger-et-al.},
\ref{sec:Barnabe-et-al.} and~\ref{sub:Cardone-et-al.2011}, isolated
galaxies have a weaker correlation than clustered ones. The low statistics
may be the cause of this disagreement with the other studies.

Results obtained at the Einstein radius $R_{Ein}$ yield similar slopes
($R_{Ein}$ and $R_{eff}$ results are correlated). However, they
should be less model dependent\footnote{
The $DMf$ and $\sfrac{M}{L}$ have still some model dependence because of the
assumed luminosity and star mass profiles.}
than the results at $R_{eff}$.  In average for our sample,
$R_{Ein}=0.6R_{eff}$. The best fits are:

$DMf=(-0.11\pm0.15)\sfrac{R_{min}}{R_{max}}|_{mass}+(0.67\pm0.12)$,

$DMf=(-0.07\pm0.14)\sfrac{R_{min}}{R_{max}}|_{apparent}+(0.64\pm0.11)$,

$\sfrac{M}{L}=(-4.57\pm2.79)\sfrac{R_{min}}{R_{max}}|_{mass}+(8.39\pm2.32)$,

$\sfrac{M}{L}=(-2.50\pm3.00)\sfrac{R_{min}}{R_{max}}|_{apparent}+(6.59\pm2.40)$.

\noindent For the isolated galaxies:

$DMf=(-0.05\pm0.24)\sfrac{R_{min}}{R_{max}}|_{mass}+(0.62\pm0.21)$,

$DMf=(0.15\pm0.27)\sfrac{R_{min}}{R_{max}}|_{apparent}+(0.45\pm0.23)$,

$\sfrac{M}{L}=(-1.32\pm2.95)\sfrac{R_{min}}{R_{max}}|_{mass}+(5.65\pm2.51)$,

$\sfrac{M}{L}=(+2.54\pm2.83)\sfrac{R_{min}}{R_{max}}|_{apparent}+(2.42\pm2.35)$.

\noindent For the galaxies found in clusters, the best fits are:

$DMf=(0.00\pm0.08)\sfrac{R_{min}}{R_{max}}|_{mass}+(0.62\pm0.06)$,

$DMf=(-0.10\pm0.66)\sfrac{R_{min}}{R_{max}}|_{apparent}+(0.68\pm0.45)$,

$\sfrac{M}{L}=(-12.08\pm7.70)\sfrac{R_{min}}{R_{max}}|_{mass}+(14.12\pm5.98)$,

$\sfrac{M}{L}=(-20.74\pm1.34)\sfrac{R_{min}}{R_{max}}|_{apparent}+(19.74\pm0.97)$.

An advantage of this analysis is the lesser model dependence claimed
by the authors. A caveat is that the model assumes spherical symmetry.
We assigned the results to group 1 reliability.

\subsection{Cardone {\it et al.} (2011) \label{sub:Cardone-et-al.2011}}

Cardone {\it et al.}~\cite{Cardone11} studied 59 early-type
galaxies from the SLACS survey using a Secondary Infall Model which
describes the collapse and virialization of a spherical halo. This
theoretically sound 
model was for the first time employed to describe
light and dark matter profiles. This allowed us, by comparing to other
dark matter extractions, to investigate their model dependence. 
The study provided the total masses within $\sim R_{eff}/2$.
The $DMf$ was extracted using the stellar $\sfrac{M_*}{L}$ results from Auger
{\it et al}. (with Salpeter IMF)~\cite{Auger 1}. We formed
the $\sfrac{M}{L}$ using the luminosities given in~\cite{Cardone11}.
Comparing them with the luminosities from~\cite{Bolton (SLACS)},
we found a 10\% shift between the two results. We retained galaxies
for which, apart for this 10\% shift, luminosities from~\cite{Cardone11}
and~\cite{Bolton (SLACS)} agree within 15\%. We also rejected
3 E/So galaxies, galaxies for which axis ratios were not available
from~\cite{Bolton (SLACS)} and giant galaxies ($M>10^{12}M_{\odot}$, 
masses from~\cite{Grillo09}). All in all, we obtained a sample
of 36 galaxies. Using either mass or apparent axis ratios,
the best fits are (see Fig~\ref{fig:cardone11}): 
\begin{figure}
\centering
\includegraphics[scale=0.5]{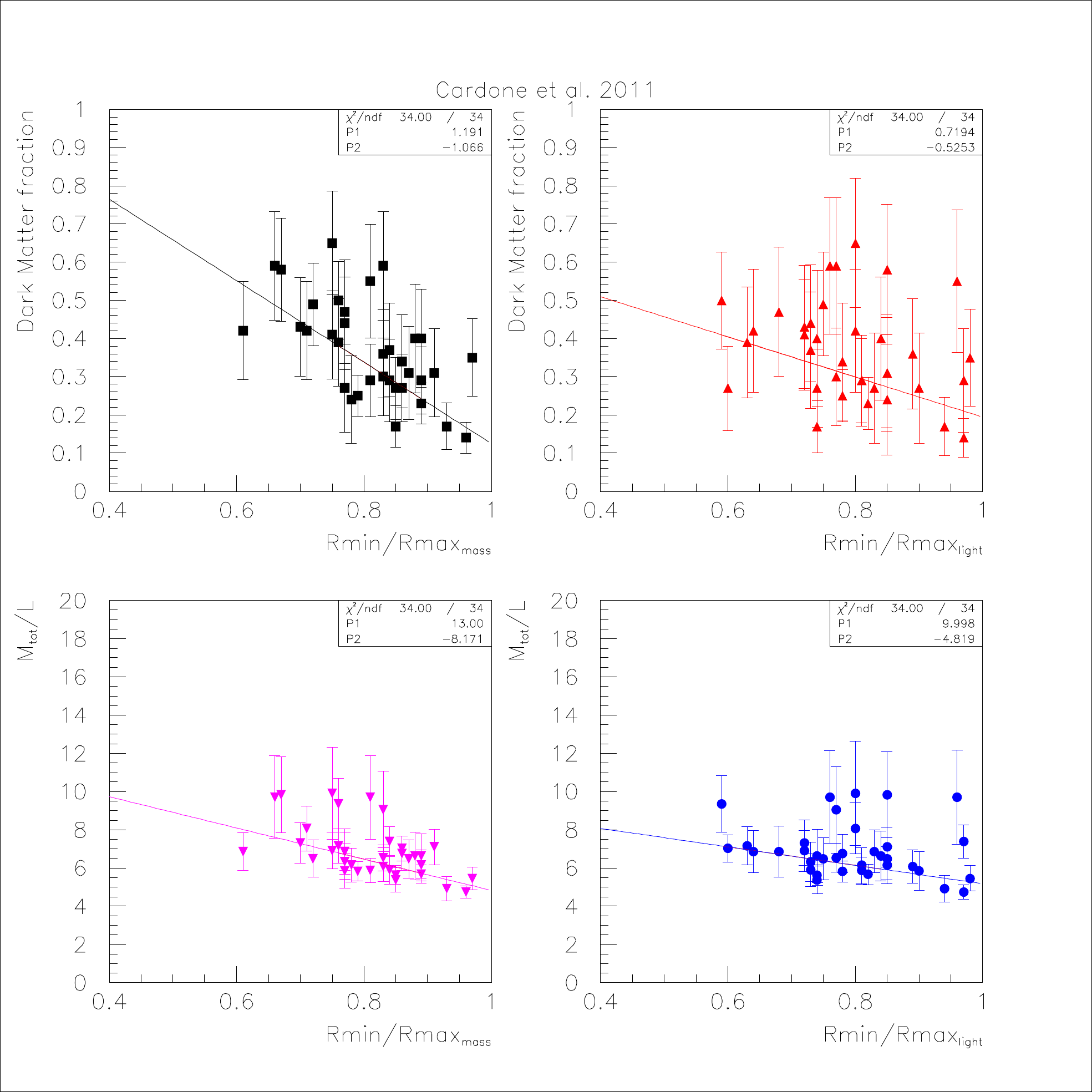}
\vspace{-0.4cm} \caption{\label{fig:cardone11}Dark matter content vs axis ratio for Cardone {\it et al.} (2011)
\cite{Cardone11}. Top plots: $DMf$ vs mass $\sfrac{R_{min}}{R_{max}}$
(left) and apparent $\sfrac{R_{min}}{R_{max}}$ (right). Bottom plots: same as top plots but for 
$\sfrac{M}{L}$.
}
\end{figure}

$DMf=(-1.07\pm0.19)\sfrac{R_{min}}{R_{max}}|_{mass}+(1.19\pm0.16)$,

$DMf=(-0.52\pm0.18)\sfrac{R_{min}}{R_{max}}|_{apparent}+(0.72\pm0.15)$,

$\sfrac{M}{L}=(-8.17\pm1.60)\sfrac{R_{min}}{R_{max}}|_{mass}+(13.00\pm1.37)$,

$\sfrac{M}{L}=(-4.82\pm1.32)\sfrac{R_{min}}{R_{max}}|_{apparent}+(10.00\pm1.09)$.

We checked the influence of environment by analyzing 
2 subsamples in which galaxies that are isolated (20 galaxies) or
belong to clusters (16 galaxies), according to~\cite{Treu09}.
For the isolated galaxies, the best fits are:

$DMf=(-1.12\pm0.28)\sfrac{R_{min}}{R_{max}}|_{mass}+(1.12\pm0.28)$,

$DMf=(-0.46\pm0.33)\sfrac{R_{min}}{R_{max}}|_{apparent}+(0.65\pm0.27)$,

$\sfrac{M}{L}=(-6.47\pm2.51)\sfrac{R_{min}}{R_{max}}|_{mass}+(11.61\pm2.15)$,

$\sfrac{M}{L}=(-1.46\pm2.56)\sfrac{R_{min}}{R_{max}}|_{apparent}+(7.24\pm2.05)$.

\noindent For the galaxies found in clusters, the best fits are:

$DMf=(-0.43\pm0.27)\sfrac{R_{min}}{R_{max}}|_{mass}+(0.67\pm0.22)$,

$DMf=(-0.23\pm0.25)\sfrac{R_{min}}{R_{max}}|_{apparent}+(0.50\pm0.19)$,

$\sfrac{M}{L}=(-2.97\pm2.02)\sfrac{R_{min}}{R_{max}}|_{mass}+(8.74\pm1.64)$,

$\sfrac{M}{L}=(-4.40\pm1.52)\sfrac{R_{min}}{R_{max}}|_{apparent}+(9.73\pm1.18)$.

\noindent Isolated galaxies display a stronger and clearer correlation
than clustered ones, as already seen in sections~\ref{sub:Auger-et-al.}
and~\ref{sec:Barnabe-et-al.}. 

This analysis is useful because it employed an original and physically
founded model. Its caveats are its assumed spherical
symmetry and usage  of the old Salpeter IMF. We assigned the results
to group 2 reliability.

\subsection{Faure {\it et al.} (2011)\label{sub:Faure}}

Faure {\it et al.}~\cite{Faure 2011} examined 20 strong-lensing
candidates from the COSMOS survey~\cite{COSMOS} and provided
$DMf$ within effective and Einstein Radii for 12 of them. The lens
candidates were tentatively identified as elliptical and $ $$S0$
galaxies. The galaxies are distant (redshift $0.34\leq z\leq1.13$).
Standard strong-lensing and photometry methods to extract $DMf$ were used,
with a Synthesis Model to infer the stellar mass and a SIE model for
the total mass. The authors assumed that mass follows light: a Sersic
profile~\cite{Sersic} was used for both the mass and light distributions.
Consequently, mass axis ratio and apparent axis ratios are equal.
Effect of the environment was accounted for in the model. We formed
$\sfrac{M_{tot}}{M_*}$ from the authors'  $M_{tot}$ and $M_*$. Out of
the 12 galaxies, we rejected two for which, according to~\cite{Faure 2011},
the model does not fit well the data (0047-5023 and 0050+4901) and
one (5914-1219) for which the mass suggests that it is a giant. Furthermore,
we rejected a galaxy (5921+0638) for which the ellipticity from~\cite{Faure 2011}
disagrees with~\cite{Faure 2008} because of less reliable modeling.
In addition, its low mass and low velocity dispersion suggest that it
is a S0. Finally, we rejected a galaxy (0038+4133) with an unphysical
$DMf<0$ and for which the low mass and velocity dispersion suggests
again that it is a S0. The final sample contains 7 galaxies. The fits
to $DMf$ or $\sfrac{M}{M_*}$ vs $\sfrac{R_{min}}{R_{max}}$ yield,\footnote{ 
We recall that it is assumed that $\sfrac{R_{min}}{R_{max}}|_{mass}=\sfrac{R_{min}}{R_{max}}|_{apparent}$.}
see Fig.~\ref{fig:faure}:
\begin{figure}
\centering
\includegraphics[scale=0.5]{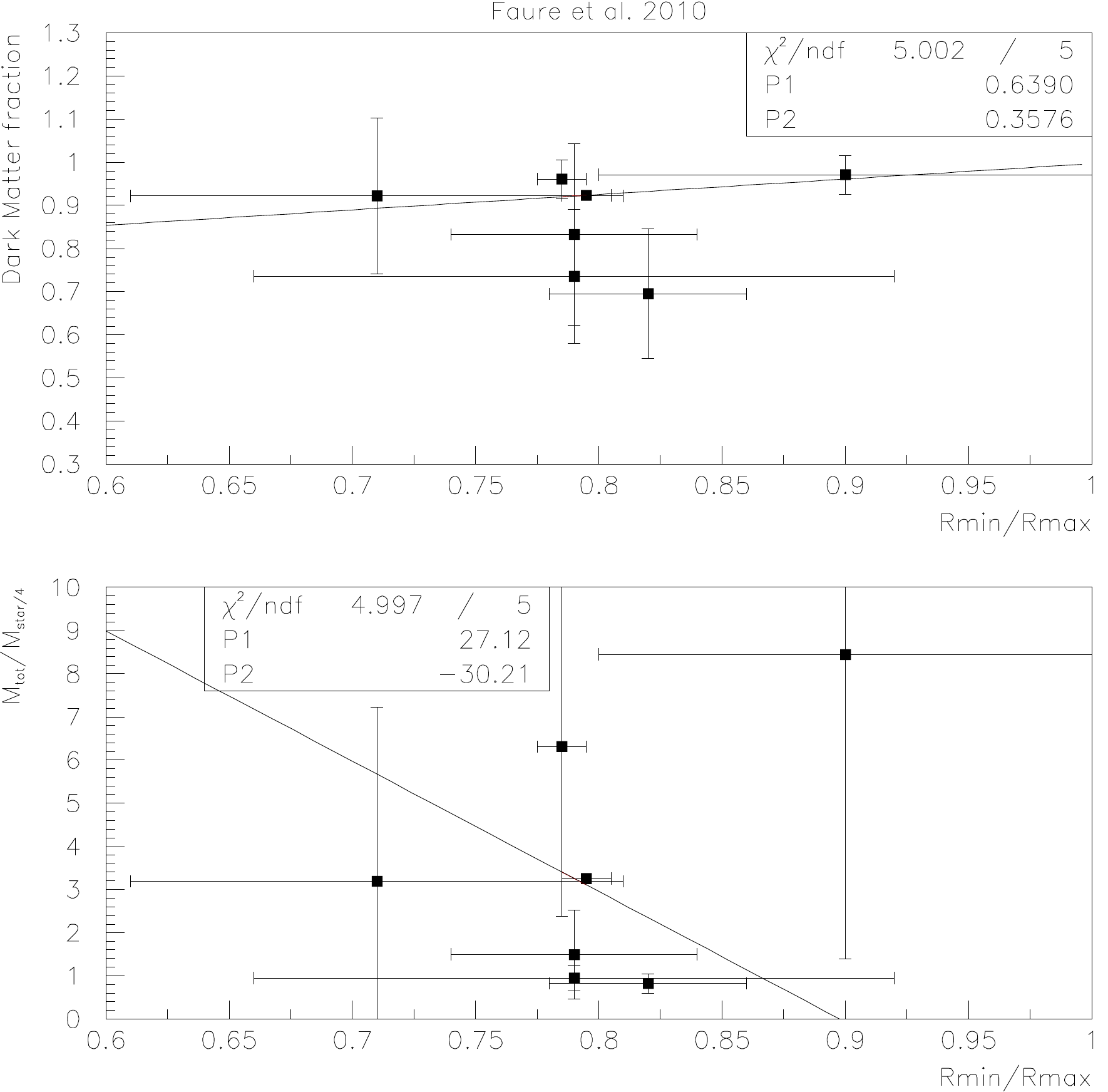}
\vspace{-0.4cm} \caption{\label{fig:faure}Results based on Faure {\it et al.}~\cite{Faure 2011}.
Top plot: $DMf$ vs $\sfrac{R_{min}}{R_{max}}$. Bottom
plot: $\sfrac{M_{tot}}{M_*}$ vs $\sfrac{R_{min}}{R_{max}}$. 
}
\end{figure}

$DMf=(+0.36\pm0.10)\sfrac{R_{min}}{R_{max}}+(0.63\pm0.08)$,

$\sfrac{M_{tot}}{M_*}=(-30.21\pm6.13)\sfrac{R_{min}}{R_{max}}+(27.12\pm4.88)$.

To compare with other works providing $\sfrac{M}{L_B}$ we can use $\sfrac{1}{4}\sfrac{M_{tot}}{M_*}\simeq \sfrac{M}{L_B}$
because for elliptical galaxies, $\sfrac{M_*}{L}_{B}\simeq4$ typically. The
specific caveats of this analysis are the  ``mass follows light'' assumption
(i.e, $\sfrac{M}{L}$ constant with $r$), which is believed to be invalid
for $r>R_{eff}$, and that the model assumed spherical symmetry. We
assigned the results to group 1 reliability.

\subsection{Ferreras, Saha and Williams (2005)}

\begin{figure}
\centering
\includegraphics[scale=0.5]{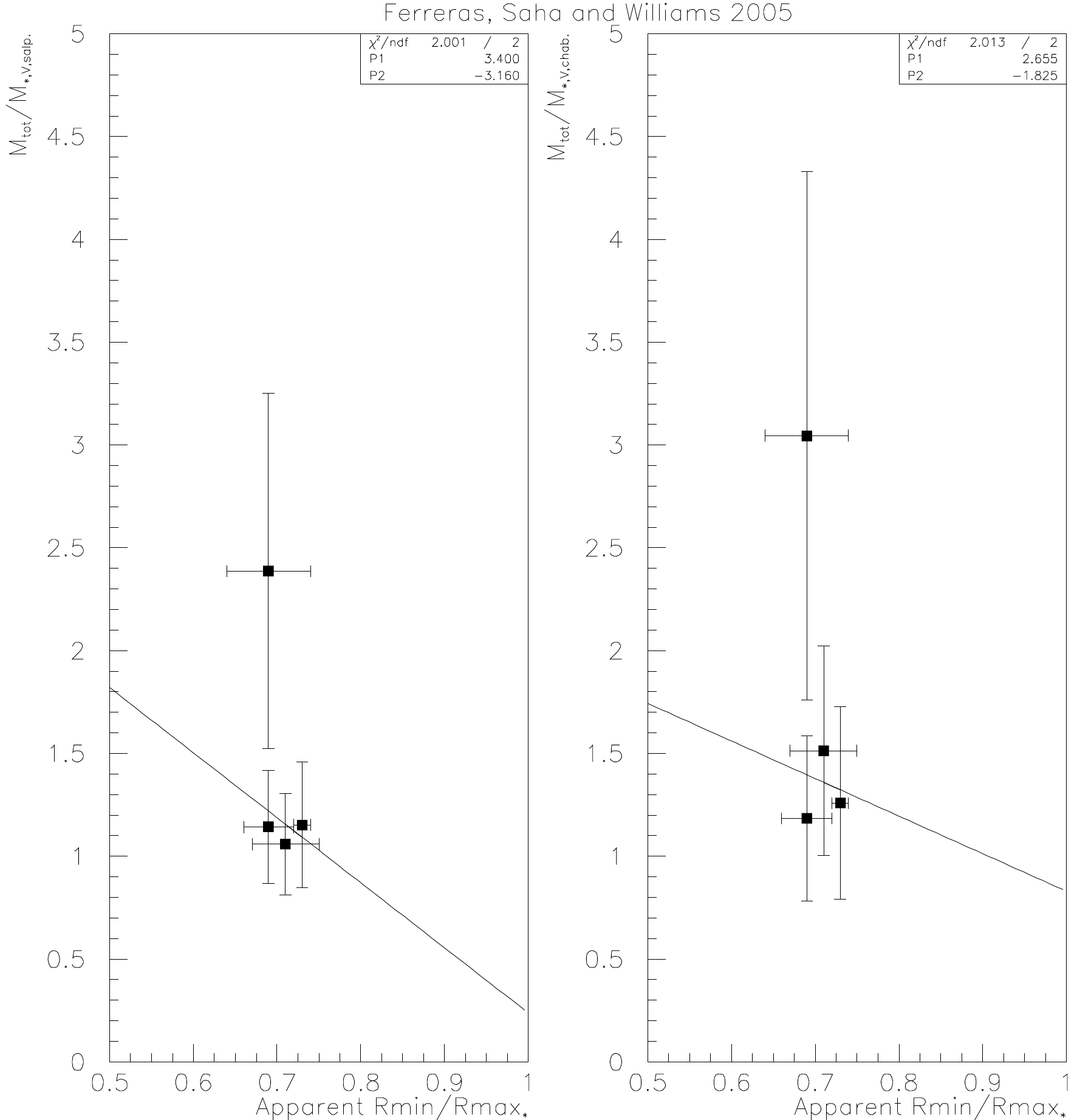}
\vspace{-0.4cm} \caption{\label{fig:ferreras2}$\sfrac{M_{tot}}{M_*}$ vs apparent and mass axis ratios from Ferreras {\it et
al.}~\cite{Ferreras2}. The right panel is with $M _*$ determined
with a Chabrier IMF and the left panel with a Salpeter IMF.
}
\end{figure}

Ferreras, Saha and Williams~\cite{Ferreras2} compared the Einstein 
masses $M_{Ein}$ of 18 strong-lensing early-type galaxies to their
visible masses $M_*$ obtained from photometry. The stellar model
used had its parameters varied over a large volume of the parameter
space, which should make the $M_*$ estimates robust~\cite{Ferreras2}.
$M_*$ is given in the V-band and depends on the choice of IMF.
The authors determined $M_*$ with a Chabrier IMF but also provided it
with a Salpeter IMF to assess the uncertainty attached to the choice
of IMF. The masses given are contained within $r_{lens}$, which is
a few $R_{eff}$: for our 4-galaxies sample 1.6$R_{eff}\leq r_{lens}\leq3.3R_{eff}$,
with in average $r_{lens}=1.8R_{eff}$. To select our sample, we
rejected a galaxy known to be a spiral or S0 galaxy  (B2237). To further suppress
possible S0 contamination, we selected galaxies with $\sigma\geq220$
km.s$^{-1}$ (the $\sigma$ are from~\cite{van de Ven}, we slightly
relaxed the usual $\sigma\geq225$~km.s$^{-1}$ otherwise only 2 galaxies
would remain in the final sample). This rejected J0951, B0952, B1009,
J1017, J1411, B1422 and B2149. We excluded galaxies with $M_{tot}>10^{12}M_{\odot}$
(B1104 and J1417). We rejected B0818 since it may be interacting
with its environment and had no axis ratio available from literature.
We rejected B1030 and B1608 because of interaction with environment.
Finally B1115 was excluded as a peculiar galaxy. There are only 4 galaxies
remaining after this selection (B0047, B0142, J0414 and B1520). We
plot $\sfrac{M_{tot}}{M_*}$ vs axis ratio in Fig.~\ref{fig:ferreras2}.
The uncertainty is assigned so that $\sfrac{\chi^{2}}{ndf}$=1. The best fits
to the data yield:

$\sfrac{M_{tot}}{M_{*,V}}=(-1.83\pm5.78)\sfrac{R_{min}}{R_{max}}|_{apparent}+(2.67\pm4.09)$
for the Chabrier IMF, 

$\sfrac{M_{tot}}{M_{*,V}}=(-3.16\pm3.69)\sfrac{R_{min}}{R_{max}}|_{apparent}+(2.61\pm3.40)$
for the Salpeter IMF. \\
The results were assigned to group 1 reliability. 

\subsection{Ferreras, Saha and Burles  (2008)\label{sub:Ferreras,Saha, Burles}}

Ferreras, Saha and Burles~\cite{Ferreras} fit photometric and
strong lensing data from the SLACS survey~\cite{Bolton (SLACS)}
to extract $\sfrac{M_{tot}}{M_*}$ for 9 E and S0 galaxies. The fit functions
are based on models (de Vaucouleur profile for photometry and a SIE
model for the mass profile). We kept 4 galaxies, rejecting 2 S0 galaxies
(identifications from~\cite{Auger 2}), one galaxy for which
$M_{tot}>10^{12}M_{\odot}$ and 2 galaxies for which the integrated
$\sfrac{M_{tot}}{M_*}$ was determined at significantly larger radius (1.66$R_{eff}$
and 2.01$R_{eff}$). The other $\sfrac{M_{tot}}{M_*}$ were determined at
$(\sim1\pm0.2)R_{eff}$. Ellipticities from~\cite{Ferreras}
differ from the mass and apparent ellipticities from the SLACS survey
article~\cite{Bolton (SLACS)}. We assign an uncertainty on $\sfrac{R_{min}}{R_{max}}$
equals to the difference between the~\cite{Bolton (SLACS)} and~\cite{Ferreras}
values. The redshift correction to $\sfrac{M_{tot}}{M_*}$ is small: there
is at most a redshift difference of $\Delta z=0.17$ among the galaxies
of the sample. From Keeton {\it et al. }~\cite{Keeton}, this
implies a modification of $\sfrac{M_{tot}}{M_*}$ of 5\%, therefore we neglected
it. The $\sfrac{M_{tot}}{M_*}$ dependence with apparent and mass $\sfrac{R_{min}}{R_{max}}$
are shown in Fig.~\ref{fig:ferreras1}. The best fit results are:
\begin{figure}
\centering
\includegraphics[scale=0.5]{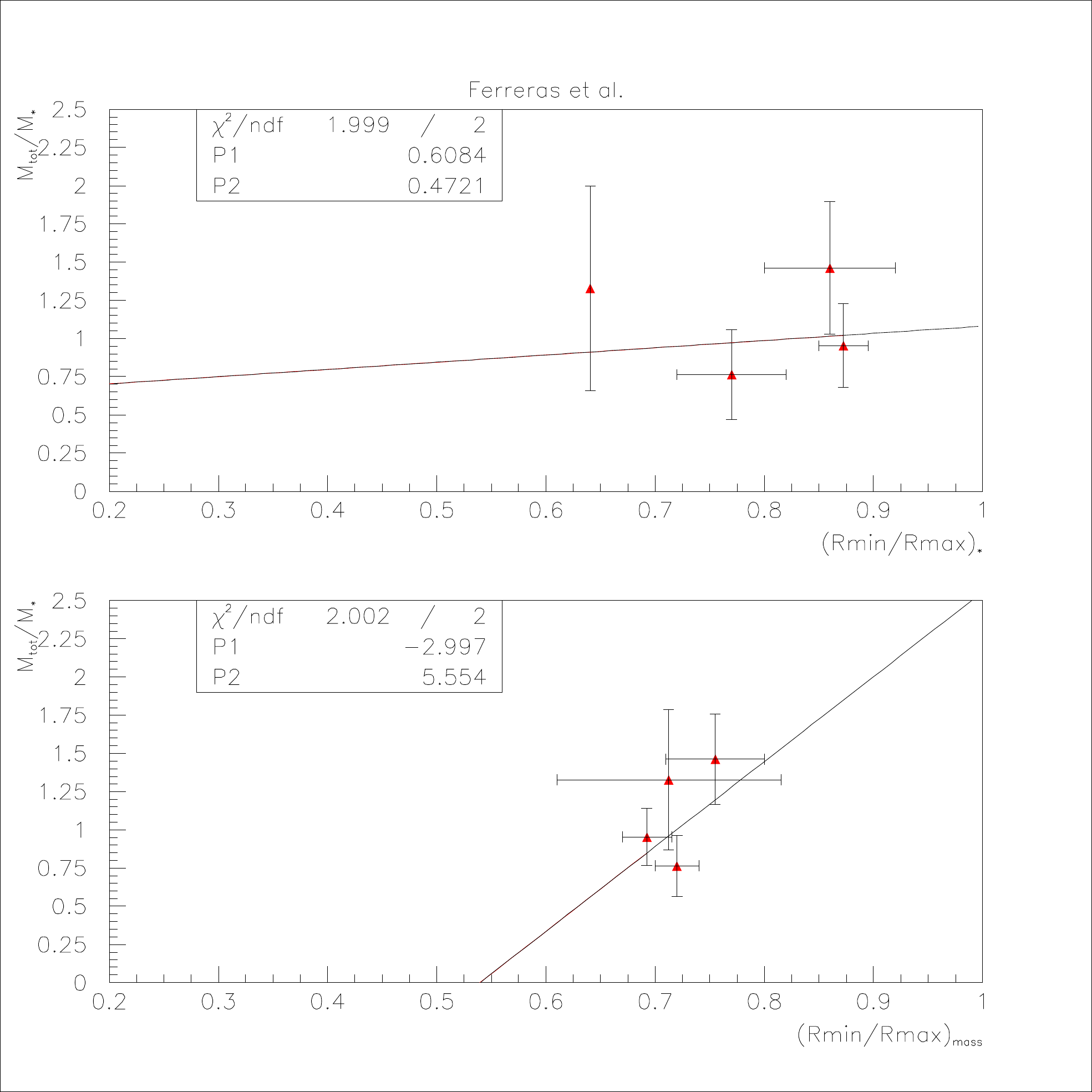}
\vspace{-0.4cm} \caption{\label{fig:ferreras1}$\sfrac{M_{tot}}{M_*}$ vs apparent (top panel) and mass (bottom panel) 
axis ratios for Ferreras, Saha and Burles~\cite{Ferreras}. 
}
\end{figure}

$\sfrac{M_{tot}}{M_*}=(+0.47\pm2.63)\sfrac{R_{min}}{R_{max}}|_{apparent}+(0.61\pm2.16)$
and

$\sfrac{M_{tot}}{M_*}=(+5.55\pm3.22)\sfrac{R_{min}}{R_{max}}|_{mass,}+(-3.00\pm2.30)$.\\
We assigned the results to group 1 reliability.

\subsection{Grillo {\it et al.} (2009) \label{sub:Grillo}}

\begin{figure}
\centering
\includegraphics[scale=0.5]{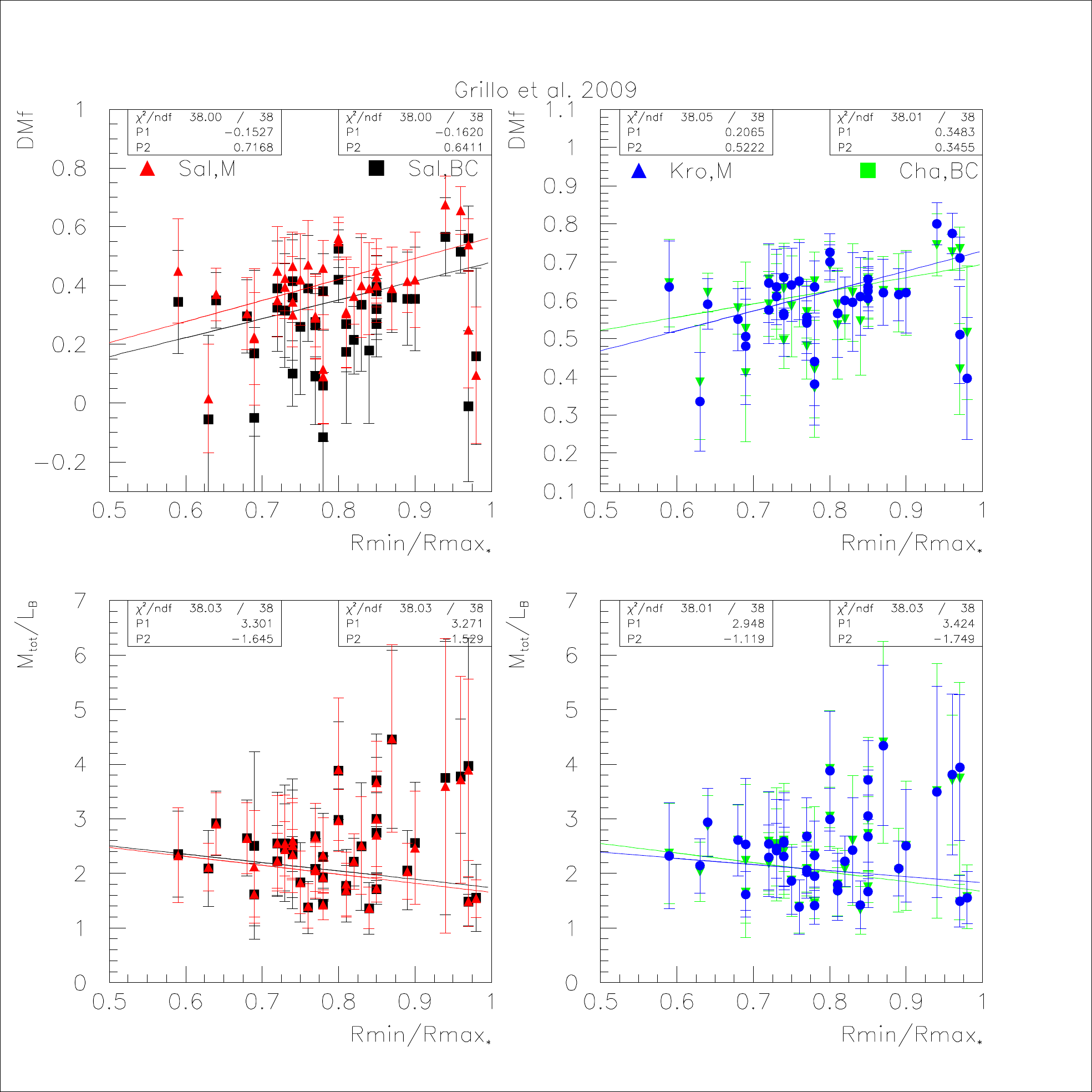}
\vspace{-0.4cm} \caption{\label{fig:grillo}$DMf$ (top plots) and $\sfrac{M}{L_B}$ (bottom plots) vs
apparent axis ratio from Grillo {\it et al.}~\cite{Grillo09}.
The red triangles are for luminous models using a Salpeter IMF and
a Maraston color metallicity template (SMT). The black squares are
for a Salpeter IMF and a Bruzual \& Charlot SMT. The blue circles
are for a Kroupa IMF and a Maraston SMT. The inverted green triangles
are for a Chabrier IMF and a Bruzual \& Charlot SMT.
}
\end{figure}

Grillo {\it et al. }~\cite{Grillo09} studied the stellar and
dark matter content of 57 early-type galaxies from the SLACS survey.
A Stellar Composite Model was used to obtain the stellar mass $M_*$
using two different sets of metallicity template and three different
IMF (Salpeter~\cite{Salpeter}, Kroupa~\cite{Kroupa},
Chabrier~\cite{Chabrier}). The best choice was selected
by the best fit to the galactic spectral energy density measurements.
The total magnitude for each galaxy was obtained using a spherically
symmetric de Vaucouleur profile. The total mass within $R_{Ein}$
was calculated  {\it{via}}  a SIE model, although the authors noted that
with their choice of normalization, the mass should not depend on
the ellipticity. After our standard SLACS selection, we retained 40
galaxies. The values of axis ratios used are from~\cite{Bolton (SLACS)}.
All uncertainties were symmetrized and scaled so that for our fits, $\sfrac{\chi^{2}}{ndf}=1$. 
The fits, shown in Fig.~\ref{fig:grillo}, yield: 

$M_{tot}(<R_{Ein})/L_{B}=(-1.65\pm0.97)\sfrac{R_{min}}{R_{max}}|_{apparent}+(3.30\pm0.78)$
for the luminous model using a Salpeter IMF and a Maraston solar metallicity
template (SMT)~\cite{Maraston}. The corresponding $DMf$ is:

$DMf(<R_{Ein})=(+0.72\pm0.20)\sfrac{R_{min}}{R_{max}}|_{apparent}+(-0.15\pm0.17)$.

$M_{tot}(<R_{Ein})/L_{B}=(-1.52\pm1.18)\sfrac{R_{min}}{R_{max}}|_{apparent}+(3.27\pm0.94)$
for a Salpeter IMF and a Bruzual \& Charlot SMT~\cite{Bruzual}.
The corresponding $DMf$ is:

$DMf(<R_{Ein})=(+0.64\pm0.23)\sfrac{R_{min}}{R_{max}}|_{apparent}+(-0.16\pm0.19)$.

$M_{tot}(<R_{Ein})/L_{B}=(-1.12\pm1.10)\sfrac{R_{min}}{R_{max}}|_{apparent}+(2.95\pm0.87)$
for a Kroupa IMF and a Maraston SMT~\cite{Maraston}. The corresponding
$DMf$ is:

$DMf(<R_{Ein})=(+0.52\pm0.14)\sfrac{R_{min}}{R_{max}}|_{apparent}+(0.21\pm0.11)$.

$M_{tot}(<R_{Ein})/L_{B}=(-1.75\pm1.06)\sfrac{R_{min}}{R_{max}}|_{apparent}+(3.42\pm0.85)$
for a Chabrier IMF and a Bruzual \& Charlot SMT. The corresponding
$DMf$ is:

$DMf(<R_{Ein})=(+0.35\pm0.13)\sfrac{R_{min}}{R_{max}}|_{apparent}+(0.35\pm0.11)$.

The analysis assumes spherical symmetry but the authors argued this
to be of minimal importance. We assigned the results to group 1 reliability.

\subsection{Jackson {\it et al.} (1998)}

Jackson {\it et al. }~\cite{Jackson} studied 12 lensing systems
in the context of a  ``dark galaxy'' search. $\sfrac{M}{L}$ within $R_{Ein}$
were inferred using a Singular Isothermal Sphere model for the mass
distribution. We kept 4 galaxies out of the original 14 (2 lensing
systems are double-lenses). We rejected B2114+022 G1 as a E+A galaxy
that has in addition unreliable $\sfrac{M}{L}$ due to dust~\cite{Chae}.
B1608+656, B1600+434, B1933+507 and B0218+357 are spiral galaxies. B2045+265
may be a Sa. Galaxies B1030+074 and B1127+385 show signs of interaction.
The two lenses of B1127+385 seem to be late types galaxies~\cite{Koopmans99},
had no redshift values (they could be close and interacting) and their
$\sfrac{M}{L}$ were not given in the H-band. All these galaxies were rejected.
Our sample comprised 4 galaxies: B2114+022 G2 (ellipticity from
\cite{Chae}), MG0414+054, B0712+472 (both ellipticities
are from~\cite{Keeton}) and B1938+666 (ellipticity
from~\cite{King98}). The $\sfrac{M}{L_H}$ vs $\sfrac{R_{min}}{R_{max}}$
fit is shown in Fig.~\ref{fig:jackson} and yields: %
\begin{figure}
\centering
\includegraphics[scale=0.5]{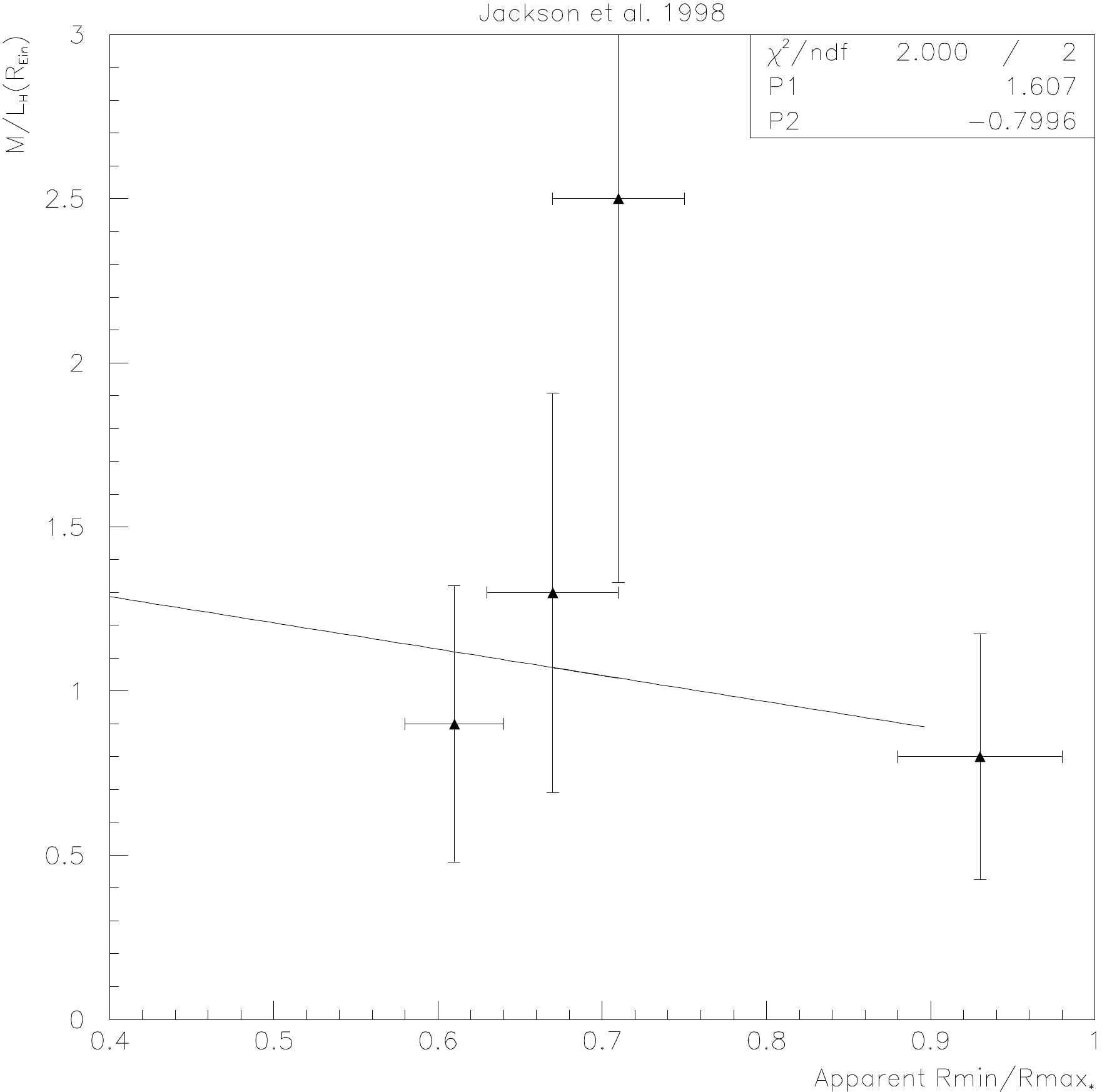}
\vspace{-0.4cm} \caption{\label{fig:jackson}$\sfrac{M}{L_H}$ vs axis ratio from Jackson {\it et al.}~\cite{Jackson}. 
}
\end{figure}

$M_{tot}(<R_{Ein})/L_{H}=(-0.80\pm1.67)\sfrac{R_{min}}{R_{max}}|_{apparent}+(1.61\pm1.31)$.

The caveats of this analysis are that the characteristics of the galaxies
are poorly known and the $\sfrac{M}{L}$, which were not the main focus of
\cite{Jackson}, were computed in a simple way without accounting
e.g. for environment interaction. The results were assigned to group
3 reliability.

{\footnotesize One may question the inclusion of B2114+022 G2 in our
sample because it belongs to a complicated 2-lenses system. G1 and G2
have different redshifts so they are not interacting with each
other. However, the modeling is more delicate. In any case, not including
G2 does not essentially change the results. The best fit gives: $M_{tot}(<R_{Ein})/L_{B}=(-0.58\pm2.36)\sfrac{R_{min}}{R_{max}}+(1.39\pm1.89)$.}{\footnotesize \par}

\subsection{Jiang and Kochanek (2007)\label{sub:Jiang-and-Kochanek}}

Jiang and Kochanek~\cite{Jiang-Kochanek} used strong lensing
data and stellar velocity dispersion measurements to model the total
and stellar mass profiles $M_{tot}(r)$ and $M_*(r)$ for 22 early-type
galaxies. It is the same galaxy sample as analyzed by Koopmans {\it et
al.}, see Section~\ref{sec:Koopmans 2006}, but the authors
argued that they employed a more physical model for the profiles (Hernquist
profile~\cite{Hernquist} for light and NFW~\cite{NFW}
profile for dark matter). The analysis is otherwise similar to that
of Section~\ref{sec:Koopmans 2006}, with spherical symmetry assumed
for the profiles and solving the spherical Jeans equations to obtain
them. We rejected 4 S0 galaxies (according to~\cite{Auger 1}),
2 galaxies with $M_{tot}>10^{12}M_{\odot}$ (using $M_{tot}$ from
\cite{Grillo09}) and 3 galaxies for which Jiang and Kochanek
did not reproduce well the measurements ( ``fit outliers'' in~\cite{Jiang-Kochanek}).
Finally, we removed PG1115+080 as it was stated that it seems peculiar
( ``low probability to belong to an homogeneous sample''~\cite{Jiang-Kochanek}). 
All the remaining galaxies are from the SLACS survey. We used axis
ratios from Bolton {\it et al.}~\cite{Bolton (SLACS)}. We used
the $M_{Ein}$ and $L_{B}$ given in~\cite{Jiang-Kochanek} to
form $\sfrac{M}{L}$. However, $M_{Ein}$ is the mass within $R_{Ein}$, with
typically $R_{Ein}\sim0.5R_{eff}$, whilst $L_{B}$ is the total luminosity.
Hence, we underestimate the $\sfrac{M}{L_B}$ and they may be subject to
systematic variation from galaxy to galaxy. We still present these
data for completeness but instead, we focus on $M_{E}/M_*$ and
$DMf$. Two classes of results were reported in~\cite{Jiang-Kochanek}:
one with adiabatic compression and another without. The authors
concluded that the model including it is strongly
favored. No uncertainties were given on $DMf$. We assumed them to
be proportional to $DMf$ with a proportionality constant so that
$\sfrac{\chi^{2}}{ndf}=1$. The fits are shown in Fig.~\ref{fig:jiang-kochanek}
and yield for sample 1: %
\begin{figure}
\centering
\includegraphics[scale=0.5]{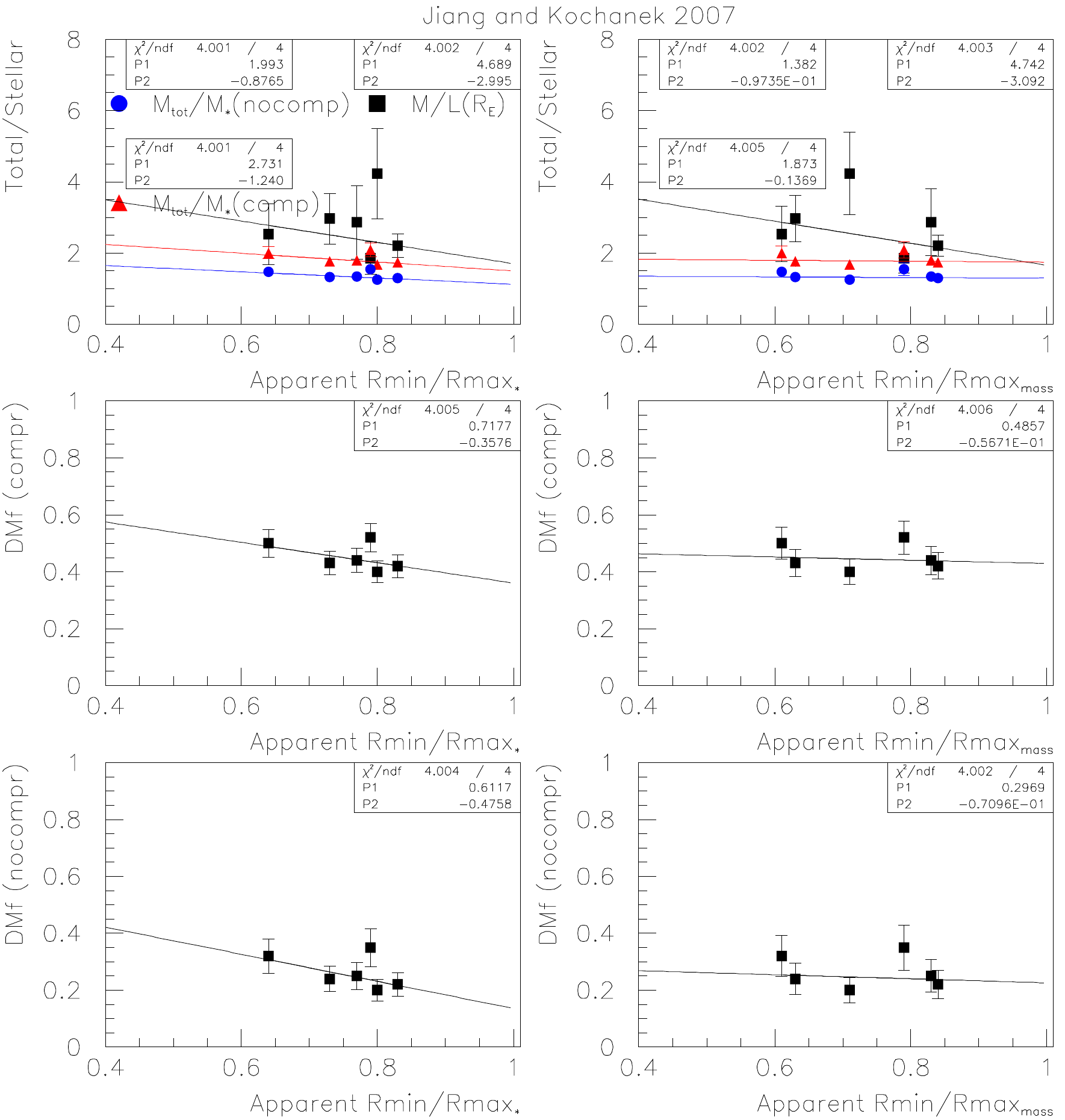}
\vspace{-0.5cm} \caption{\label{fig:jiang-kochanek}Dark matter content vs apparent axis ratio (left panels) and mass
axis ratio (right panels) for Jiang and Kochanek~\cite{Jiang-Kochanek}.
The top panels display $\sfrac{M_{tot}}{M_*}$ at $R_{Ein}$ with adiabatic
compression (red triangles) and without (blue filled circles), and
$\sfrac{M}{L_B}$ (black squares). The middle row displays the dark matter
fractions including adiabatic compression and the bottom row is for
the $DMf$s without compression.
}
\end{figure}

$\sfrac{M_{tot}}{M_*}(R_{Ein})=(-1.24\pm0.92)\sfrac{R_{min}}{R_{max}}|_{apparent}+(2.73\pm0.72)$
(with adiabatic compression),

$\sfrac{M_{tot}}{M_*}(R_{Ein})=(-0.14\pm0.68)\sfrac{R_{min}}{R_{max}}|_{mass}+(1.87\pm0.50)$
(with adiabatic compression),

$\sfrac{M_{tot}}{M_*}(R_{Ein})=(-0.88\pm0.58)\sfrac{R_{min}}{R_{max}}|_{apparent}+(1.99\pm0.45)$
(without adiabatic compression),

$\sfrac{M_{tot}}{M_*}(R_{Ein})=(-0.10\pm0.45)\sfrac{R_{min}}{R_{max}}|_{mass}+(1.38\pm0.34)$
(without adiabatic compression),

$DMf(R_{Ein})=(-0.36\pm0.30)\sfrac{R_{min}}{R_{max}}|_{apparent}+(0.72\pm0.23)$
(with adiabatic compression),

$DMf(R_{Ein})=(-0.06\pm0.23)\sfrac{R_{min}}{R_{max}}|_{mass}+(0.49\pm0.17)$
(with adiabatic compression),

$DMf(R_{Ein})=(-0.48\pm0.35)\sfrac{R_{min}}{R_{max}}|_{apparent}+(0.61\pm0.27)$
(without adiabatic compression),

$DMf(R_{Ein})=(-0.07\pm0.26)\sfrac{R_{min}}{R_{max}}|_{mass}+(0.30\pm0.20)$
(without adiabatic compression).\\
The analysis assumes spherical symmetry and no uncertainties were
provided. For our global analysis, we used the $\sfrac{M_{tot}}{M_*}$
and $DMf$ results with compression (favored in~\cite{Jiang-Kochanek})
and we assigned them to group 2 reliability. \\
{\footnotesize For information: }{\footnotesize \par}

{\footnotesize $M_{tot}(R_{Ein})/L_{B}=(-3.00\pm4.32)\sfrac{R_{min}}{R_{max}}|_{apparent}+(4.69\pm3.43)$,}{\footnotesize \par}

{\footnotesize $M_{tot}(R_{Ein})/L_{B}=(-3.09\pm2.57)\sfrac{R_{min}}{R_{max}}|_{mass}+(2.03\pm1.33)$.}{\footnotesize \par}

\subsection{Keeton, Kochanek and Falco (1997)}

\begin{figure}
\centering
\includegraphics[scale=0.5]{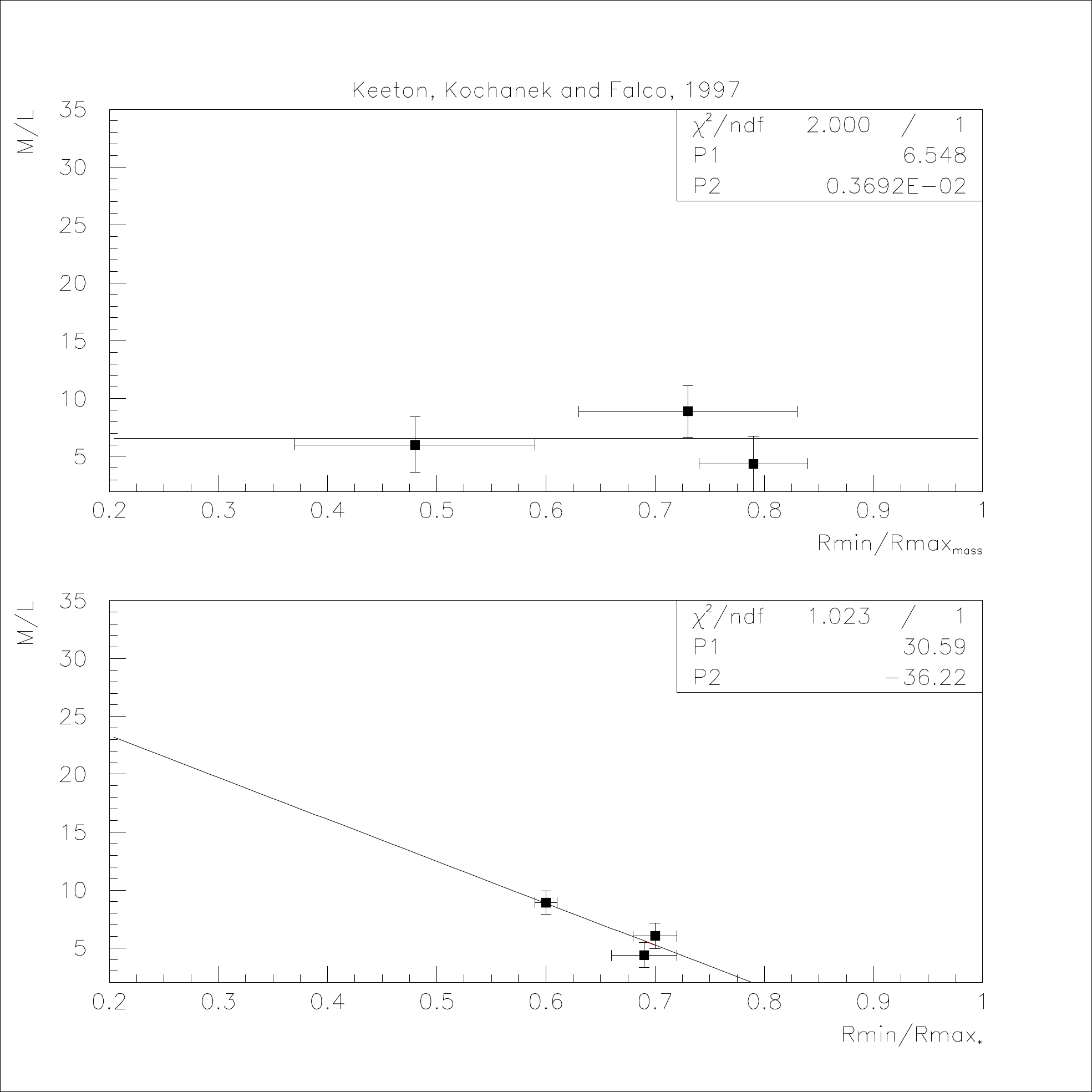}
\vspace{-0.4cm} \caption{\label{fig: keeton}$\sfrac{M}{L}$ vs mass axis ratio (top panel) and apparent axis ratio (bottom
panel) for Keeton {\it et al.}~\cite{Keeton}. 
}
\end{figure}

In~\cite{Keeton} Keeton, Kochanek and Falco analyzed 17 strong
lensing candidates. The total mass was obtained by using either a
SIE model or a Singular Isothermal
Sphere with tidal correction (SIS+Shear) model. The SIE model used to model
the 2D mass profiles accounts for ellipticity. This implies
a correlation between mass profile and ellipticity, with steeper profile
galaxies having larger ellipticity. The luminosity profiles were assumed
to follow a de Vaucouleur law. The luminosity uncertainty dominates
that of $\sfrac{M}{L}$. From the 17 galaxies considered in~\cite{Keeton},
we excluded 2 spiral galaxies (B0218+357 and B1933+503), 3 for which
$\sfrac{M}{L}$ was not given (MG0414+0534, HST12531-2914 and BRI0952-0115),
4 that are clearly influenced by their environments (MG0751+2716,
Q0957+561, B1422+231 and MG1131+0456) and 1 that was poorly understood
in~\cite{Keeton} (B1600+434). We removed galaxies B0712+472
and MG1549+304 that have a velocity dispersion $\sigma<200$~km.s$^{-1}$
($\sigma$ are from~\cite{van de Ven}). We relaxed our selection
compared to our standard $\sigma<225$~km.s$^{-1}$, otherwise only
Q0142-100 would have passed selection. We rejected PG1115+080
as it may be peculiar~\cite{Jiang-Kochanek} and
B1608+656 because it may have a companion. In all, 3 galaxies
remained: Q0142-100, HST14176+5226 and MG1654+1346. We used the $\sfrac{M}{L}$
values computed in the $\Omega_{0}=0.2,\,\lambda_{0}=0.8$ cosmological
model. The $\sfrac{M}{L}$ vs apparent and mass axis ratios are shown in Fig.
\ref{fig: keeton}. 
 The best fits yield:

$\sfrac{M}{L_B}=(-36.22\pm15.14)\sfrac{R_{min}}{R_{max}}|_{mass}+(30.59\pm9.89)$,

$\sfrac{M}{L_B}=(0.00\pm10.27)\sfrac{R_{min}}{R_{max}}|_{apparent}+(6.55\pm7.01)$.\\
The advantage of the analysis is that ellipticity is accounted for.
The specific caveats are that the analysis is dated (for a strong
lensing analysis) and due to the small number of galaxies we had to
relax our standard criteria minimizing S0 and giant elliptical contaminations.
For these reasons, results were assigned to group 3 reliability. 

\subsection{Koopmans {\it et al.} (2006) \label{sec:Koopmans 2006}}

\begin{figure}
\centering
\includegraphics[scale=0.5]{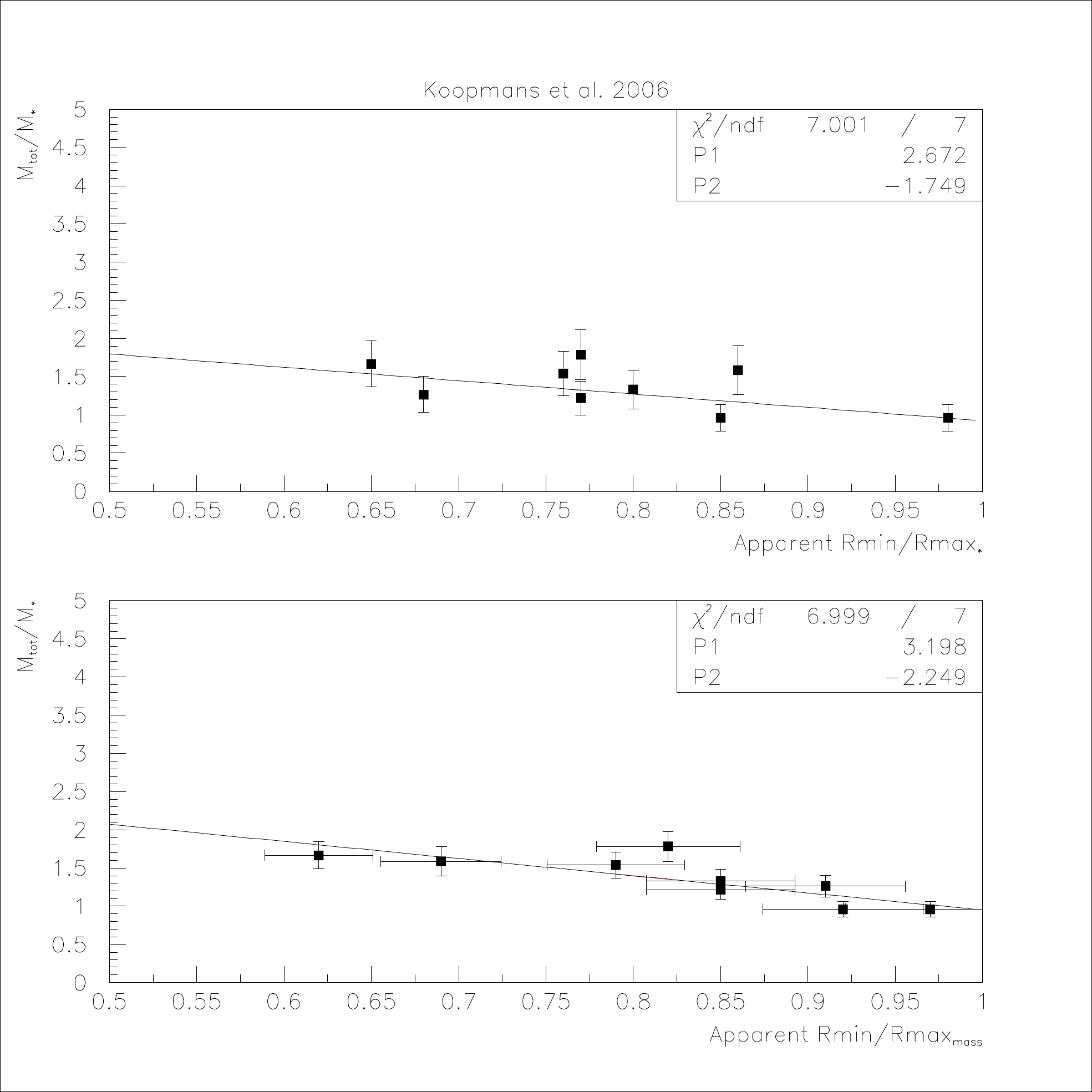}
\vspace{-0.4cm} \caption{\label{fig:koopmans}$\sfrac{M_{tot}}{M_*}$ vs apparent and mass
axis ratios  (top and bottom, respectively) from Koopmans {\it et al.}~\cite{Koopmans}. 
}
\end{figure}

Koopmans {\it et al.}~\cite{Koopmans} analyzed strong lensing
and photometry data from 15 early-type field galaxies from the SLACS
survey. The strong lensing data allowed them to determine the total
mass $M_{tot}$ within $R_{Ein}$. The photometry data was used to
determine the total density profile $\rho(r)$ by solving the spherical
Jeans equations with an assumed form $\rho(r)\propto r^{-\gamma'}$
and a Hernquist profile for the stellar density. Calculations were
repeated with a Jaffe profile for stellar density in order to assess
the model dependence on the choice of stellar profile. For mass deprojection,
a SIE model was used and the mass axis ratio was deduced with precision
better than 10\%. A systematic study of uncertainties was carried
out. There are 4 main sources of uncertainty. The unknown stellar
velocity anisotropy has small effects, as does the influence of neighboring 
galaxies. The choice of model for the stellar and mass densities produced
an uncertainty acceptable within the accuracy level of the data. The
authors argued that the effect of the halo ellipticity on their study (dark
matter fraction variation with radius) is small. We rejected 3 lenses
identified as S0 in~\cite{Auger 1}, excluded 2 galaxies with
$M_{tot}>10^{12}M_{\odot}$ and one with $\sigma<225$~km.s$^{-1}$.
In all, our final sample comprises 9 galaxies. The $\sfrac{M_{tot}}{M_*}$
vs axis ratio, is shown in Fig.~\ref{fig:koopmans}. We rescaled the uncertainty
provided by the authors so that $\sfrac{\chi^{2}}{ndf}\simeq1$ for the best
fits. they yield:

$\sfrac{M_{tot}}{M_*}=(-1.75\pm0.77)\sfrac{R_{min}}{R_{max}}|_{apparent}+(2.67\pm0.64)$, 

$\sfrac{M_{tot}}{M_*}=(-2.25\pm0.57)\sfrac{R_{min}}{R_{max}}|_{mass}+(3.20\pm0.49)$.\\
Specific caveats are some model dependence, in particular in the choice
of density profile which was argued to be unphysical in~\cite{Ferreras2}
and spherical symmetry was assumed (except for mass deprojection).
The analysis has the advantages that systematic uncertainties and
model dependence were studied and argued to be under control. Furthermore,
care was taken in choosing and verifying that the galaxies are isolated.
We assigned the results to group 1 reliability. 

\subsection{Leier (2009) \label{sub:Leier-data-(2009)}}

\begin{figure}
\centering
\includegraphics[scale=0.5]{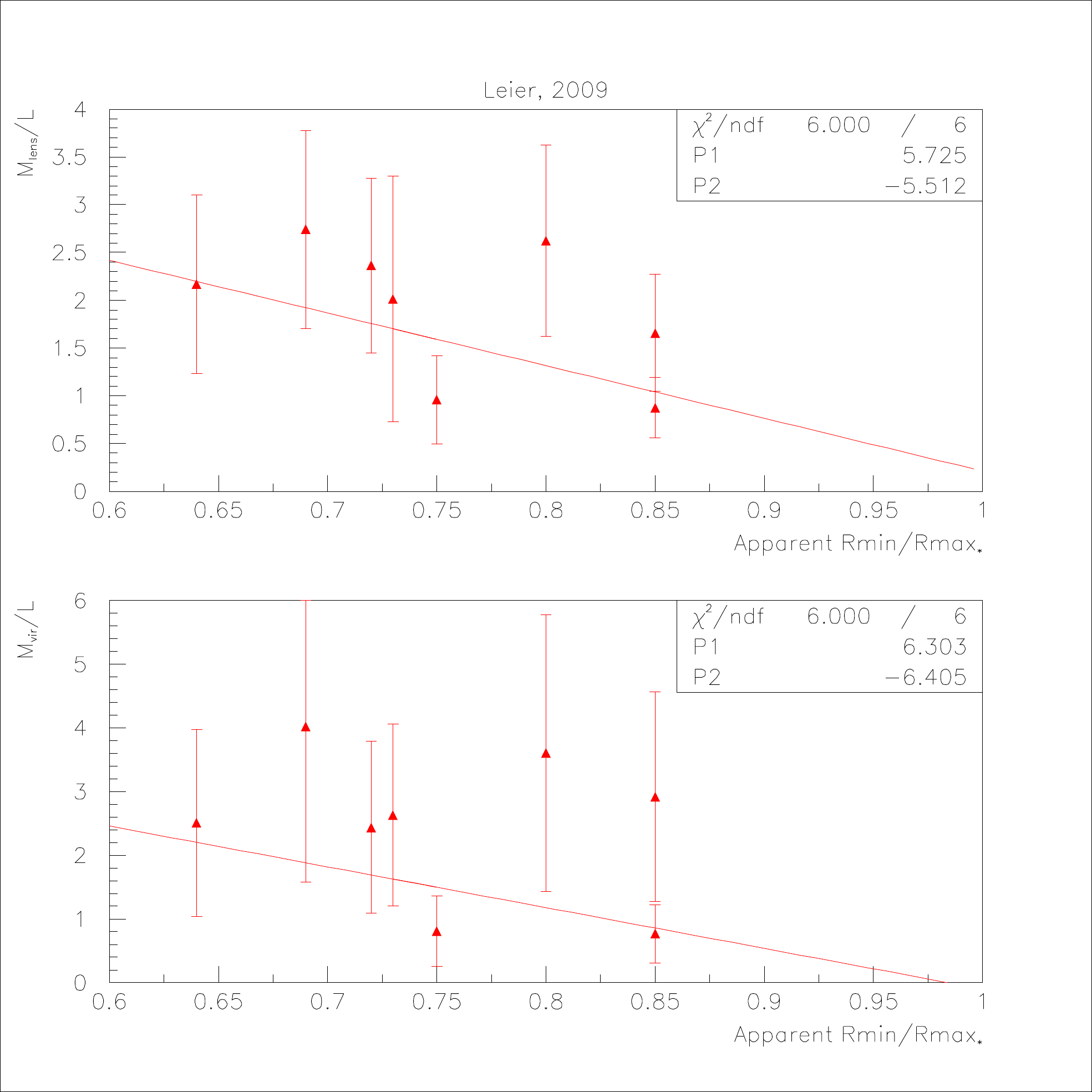}
\vspace{-0.4cm} \caption{\label{fig: leier1}$\sfrac{M_{tot}}{L}$ vs apparent axis ratio from Leier (2009)~\cite{Leier09}.
Top: $M_{tot}=M_{lens}$. Bottom: $M_{tot}=M_{vir}$.
}
\end{figure}

In~\cite{Leier09}, Leier analyzed 19 strong lenses
with available photometry from the SLACS and CASTLE surveys. The lensing
data were used to assess the projected total mass $M_{lens}$.
From it and luminosity data, $\sfrac{M_{lens}}{L_{I}}$ at $r=R _{lens}$
was extracted ($R_{lens}$ is of the same order as $R_{eff}$ for
most lenses). $M_{lens}$ was used with the virial theorem to form
a velocity dispersion $\sigma_{len}$. It agrees generally
well with the observed one, $\sigma_{obs}$, indicating
that these galaxies are virialized. We rejected two galaxies
for which data are not available (CFRS03.1077, HST14176), one known
spiral (Q2237+656)\footnote{It has also one of the two worst $\sigma_{obs}/\sigma_{len}$
ratio, $\sigma_{obs}/\sigma_{len}=1.47$, indicating it may not be
well virialized.},
two galaxies identified as S0 in~\cite{Auger 1},
one with significant interaction and classified as peculiar in~\cite{Jiang-Kochanek}
(PG1115+080, which has $\sigma_{obs}/\sigma_{len}=1.47$ and so may
not be well virialized), one with significant contribution from
its cluster (Q0957) and one with a close companion and unreliable
axis ratio (B1608+656). In addition, we applied our standard selection
to minimize contamination from giant elliptical galaxies and S0, rejecting
one high mass galaxy (J0956+510). In all, 8 galaxies constitute 
our final sample. For all of them $\sigma_{obs}$ and $\sigma_{len}$
agree within 20\%. Uncertainties on $\sfrac{M}{L}$ were not provided. We estimated
them using the uncertainties given for $M_{lens}$ and $M_{vir}$
and assuming a 20\% uncertainty on the luminosity. The resulting $\sfrac{M}{L_I}$
uncertainties were then scaled so that $\sfrac{\chi^{2}}{ndf}=1$. The $\sfrac{M}{L}(R_{lens})$
vs apparent axis ratio, shown in Fig.~\ref{fig: leier1}, have best fits: %

$\sfrac{M_{lens}}{L_{I}}(R_{lense})=(-5.51\pm3.27)\sfrac{R_{min}}{R_{max}}|_{apparent}+(5.72\pm2.62)$,

$\sfrac{M_{vir}}{L_{I}}(R_{lense})=(-6.41\pm4.97)\sfrac{R_{min}}{R_{max}}|_{apparent}+(6.30\pm3.95)$.\\
The two fits agree well, indicating that the virial
and strong lensing methods are consistent and that the galaxies are
virialized. 
A limitation of this analysis is that spherical symmetry was assumed.
 The analysis has the advantage that care was taken in choosing virialized
galaxies. We assigned the results to group 1 reliability.

\subsection{Leier {\it et al.} (2011) \label{sub:Leier 2011}}

\begin{figure}
\centering
\includegraphics[scale=0.5]{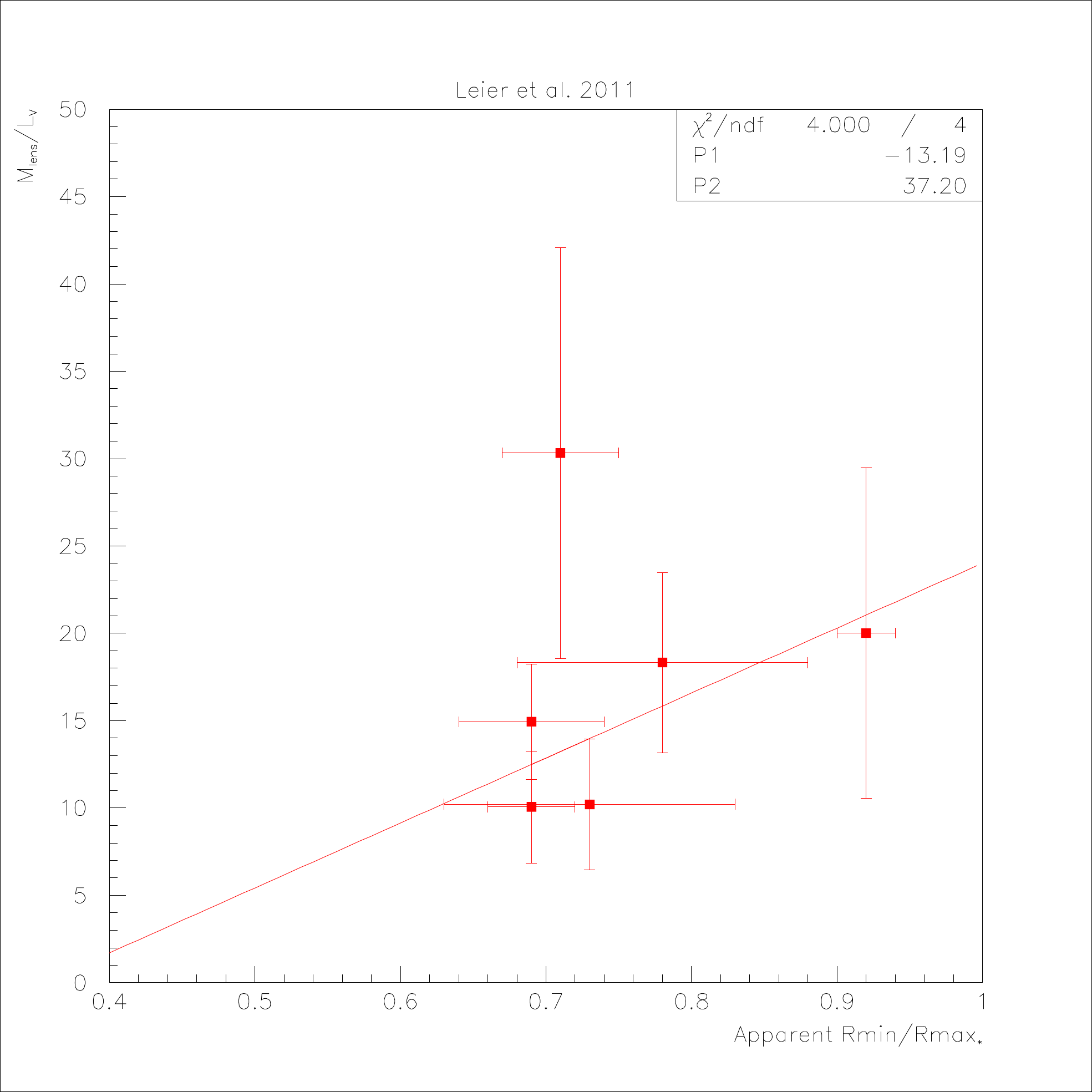}
\vspace{-0.4cm} \caption{\label{fig: leier2}$\sfrac{M_{lens}}{L_V}$ vs apparent axis ratio from Leier {\it et al.}
(2011)~\cite{Leier2011}. The line is the best fit to the data. 
}
\end{figure}

In~\cite{Leier2011}, Leier {\it et al.} computed the baryon
fraction profile for 19 strong lensing galaxies from the CASTLE survey.
Lensing and photometry were used to recover the total mass and the
mass and light profiles. A Sersic profile was used for stars and the
apparent  axis ratio was accounted for in the stellar mass calculation.
$\sfrac{M}{L_B}$ were extracted {\it{via}} a Stellar Population Synthesis model.
Various IMF were used with results agreeing within 10\%. Ignoring
the spiral bulge Q2237, the giant MG2016 and RXJ0911 that is believed
to not be a true lens, and applying our standard $M_{tot}<10^{12}M_{\odot}$
and $\sigma_{obs}\geq225$~km.s$^{-1}$ criteria, we rejected 13 of
the 19 galaxies ($\sigma_{obs}$ are from~\cite{van de Ven}).
3 others had to be rejected because of notable interactions with environment,
leaving only 3 galaxies (Q0142, Q0047 and MG1104). Slightly relaxing
our selection to $\sigma_{obs}\geq215$  km.s$^{-1}$ or $\sigma_{lens}\geq215$  km.s$^{-1}$
provides a sample of 6 galaxies. The $\sfrac{M}{L}(R_{lens})$ vs apparent
axis ratio is shown in Fig.~\ref{fig: leier2}. 
The best fit is:

$\sfrac{M_{lens}}{L_V}=(+37.2\pm36.6)\sfrac{R_{min}}{R_{max}}|_{apparent}+(-13.2\pm26.3)$.\\
The uncertainties were scaled so that the $\sfrac{\chi^{2}}{ndf}=1$. The caveats
are that in order to get a reasonably sized sample, we relaxed
our selection. Also, like in the other CASTLE lens survey analyses,
the sample is more inhomogeneous, as noted e.g. in~\cite{Leier09}.
Finally, the axis ratios came from different authors thus possibly
adding point-to-point fluctuations, although this was accounted for
by forcing $\sfrac{\chi^{2}}{ndf}=1$ for the fit. We assigned the results
to group 2 reliability.

\subsection{Ruff {\it et al.} (2011)}

\begin{figure}
\centering
\includegraphics[scale=0.57]{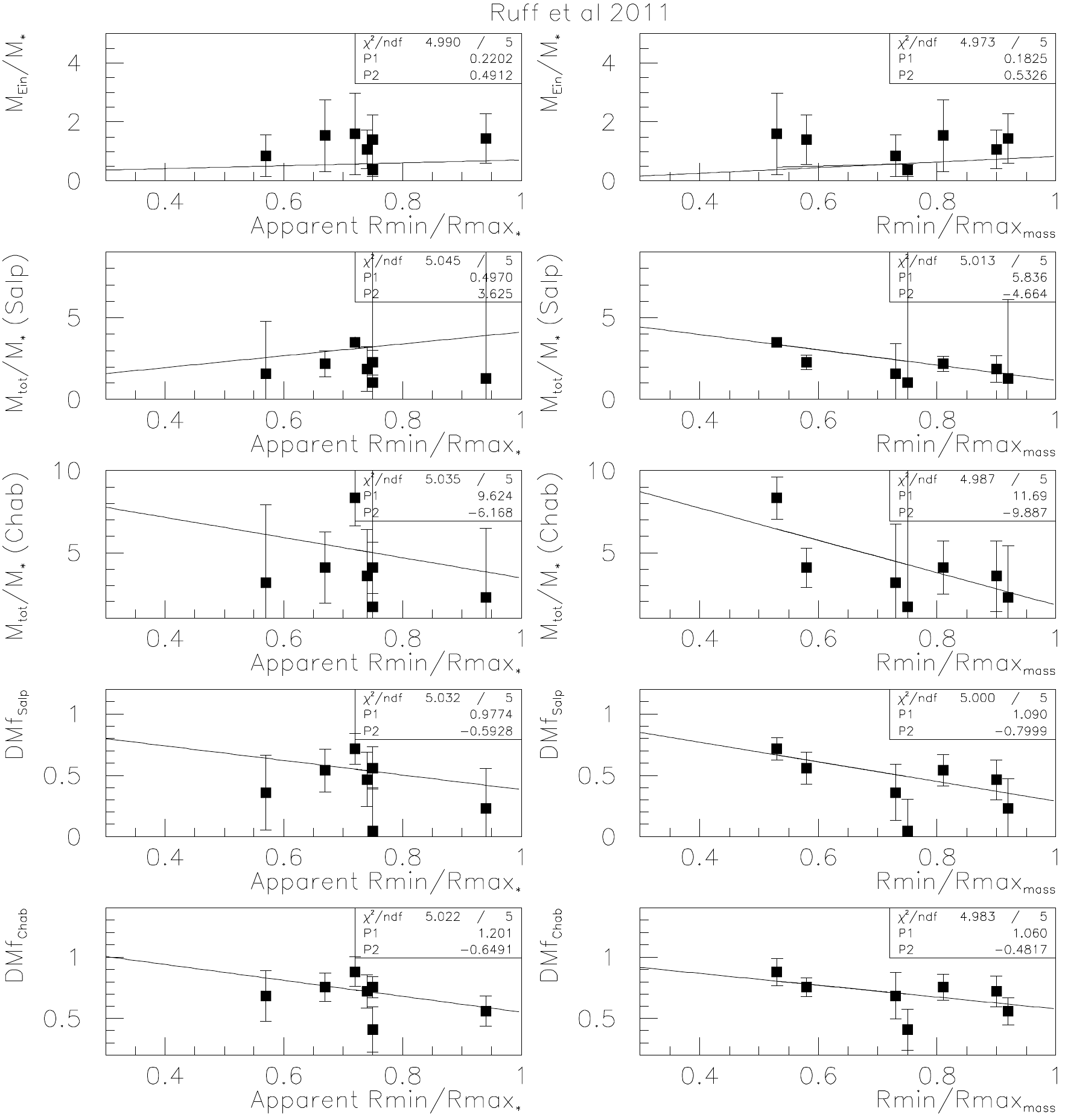}
\vspace{-0.4cm} \caption{\label{fig: ruff}Mass ratios vs apparent (left column) and mass (right column)
axis ratios from Ruff {\it et al.}~\cite{Ruff}. The top row
is for $\sfrac{M_{Ein}}{M_*}$. The 2$^{\mbox{\scriptsize{nd}}}$ and 3$^{\mbox{\scriptsize{rd}}}$
rows are for $\sfrac{M_{tot}}{M_*}$ extracted using the $DMf$ assuming
either a Chabrier or a Salpeter IMF. The last two rows show the corresponding
$DMf$. 
}
\end{figure}

Ruff {\it et al.}~\cite{Ruff} analyzed 11 early-type strong-lensing galaxies
from the SL2S survey. The lenses have an average
redshift of <$z$>=0.5. A SIE model was used to obtain the total mass
$M_{Ein}$ within $R_{Ein}$. Photometric data were used
to compute both the stellar mass $M_*$, using either a Chabrier
or a Salpeter IMF, and the size of the galaxy. An elliptical de Vaucouleur
profile was used to obtain the surface brightness. The $DMf$ was
calculated at $R_{eff}/2$ by solving the spherical Jeans equations.
No galaxy type identification is given in the article. The requirements
that $M_{tot}<5\times10^{11}M_{\odot}$ and $\sigma\geq225$~km.s$^{-1}$
were used to remove possible S0 and giants. 7 galaxies passed the
selection. Mass ratios vs apparent and mass axis ratios are
shown in Fig.~\ref{fig: ruff}. %
 The best fits are:

$\sfrac{M_{Ein}}{M_*}=(+0.49\pm2.90)\sfrac{R_{min}}{R_{max}}|_{apparent}+(0.22\pm2.17)$,

$\sfrac{M_{Ein}}{M_*}=(+0.53\pm2.47)\sfrac{R_{min}}{R_{max}}|_{mass}+(0.18\pm1.88)$,

$\sfrac{M_{tot}}{M_*}|_{Sal}=(+3.62\pm8.81)\sfrac{R_{min}}{R_{max}}|_{apparent}+(0.50\pm6.33)$,

$\sfrac{M_{tot}}{M_*}|_{Sal}=(-4.66\pm1.43)\sfrac{R_{min}}{R_{max}}|_{mass}+(5.84\pm0.85)$,

$\sfrac{M_{tot}}{M_*}|_{Chab}=(-6.17\pm14.65)\sfrac{R_{min}}{R_{max}}|_{apparent}+(9.62\pm10.70)$,

$\sfrac{M_{tot}}{M_*}|_{Chab}=(-9.89\pm4.85)\sfrac{R_{min}}{R_{max}}|_{mass}+(11.65\pm3.25)$,

$DMf|_{Sal}=(-0.59\pm1.11)\sfrac{R_{min}}{R_{max}}|_{apparent}+(0.98\pm0.80)$,

$DMf|_{Sal}=(-0.80\pm0.37)\sfrac{R_{min}}{R_{max}}|_{mass}+(1.09\pm0.26)$,

$DMf|_{Chab}=(-0.65\pm0.52)\sfrac{R_{min}}{R_{max}}|_{apparent}+(1.20\pm0.39)$,

$DMf|_{Chab}=(-0.48\pm0.29)\sfrac{R_{min}}{R_{max}}|_{mass}+(1.06\pm0.21)$,\\
The mass ratio uncertainties were scaled so that $\sfrac{\chi^{2}}{ndf}=1$.

The difference between $\sfrac{M_{Ein}}{M_*}$ and $\sfrac{M_{tot}}{M_*}$ is that
$\sfrac{M_{Ein}}{M_*}$ stands at $r=R_{Ein}$ and the models account for
the apparent ellipticity and do not need IMF input. $\sfrac{M_{tot}}{M_*}$
stands at $r=R_{eff}/2$, used models accounting for ellipticity but
was obtained solving the spherical Jeans equations with an assumed
IMF. For our sample, we have $0.73<R_{ein}/R_{eff}<2.06$, but with
$R_{ein}\simeq R_{eff}$ in most cases. 

A caveat here is that only a few tentative type identifications are
available from~\cite{Ruff} so we required $\sigma\geq225$~km.s$^{-1}$
and $M_{tot}<5\times10^{11}M_{\odot}$ to exclude possible of S0 and
giant galaxies. Little indication on the environment influence is
given in~\cite{Ruff}. Advantages are that ellipticities were
accounted for in the mass calculations and the model dependence upon
the IMF can be estimated. For these reasons, the $\sfrac{M_{Ein}}{M_*}$
results are assigned to group 2 reliability and the $\sfrac{M_{tot}}{M_*}$
and $DMf$ using a Chabrier IMF are assigned to group 3 reliability. 

\subsection{Treu and Koopmans (2004)\label{sub:Treu-and-Koopmans}}

\begin{figure}
\centering
\includegraphics[scale=0.5]{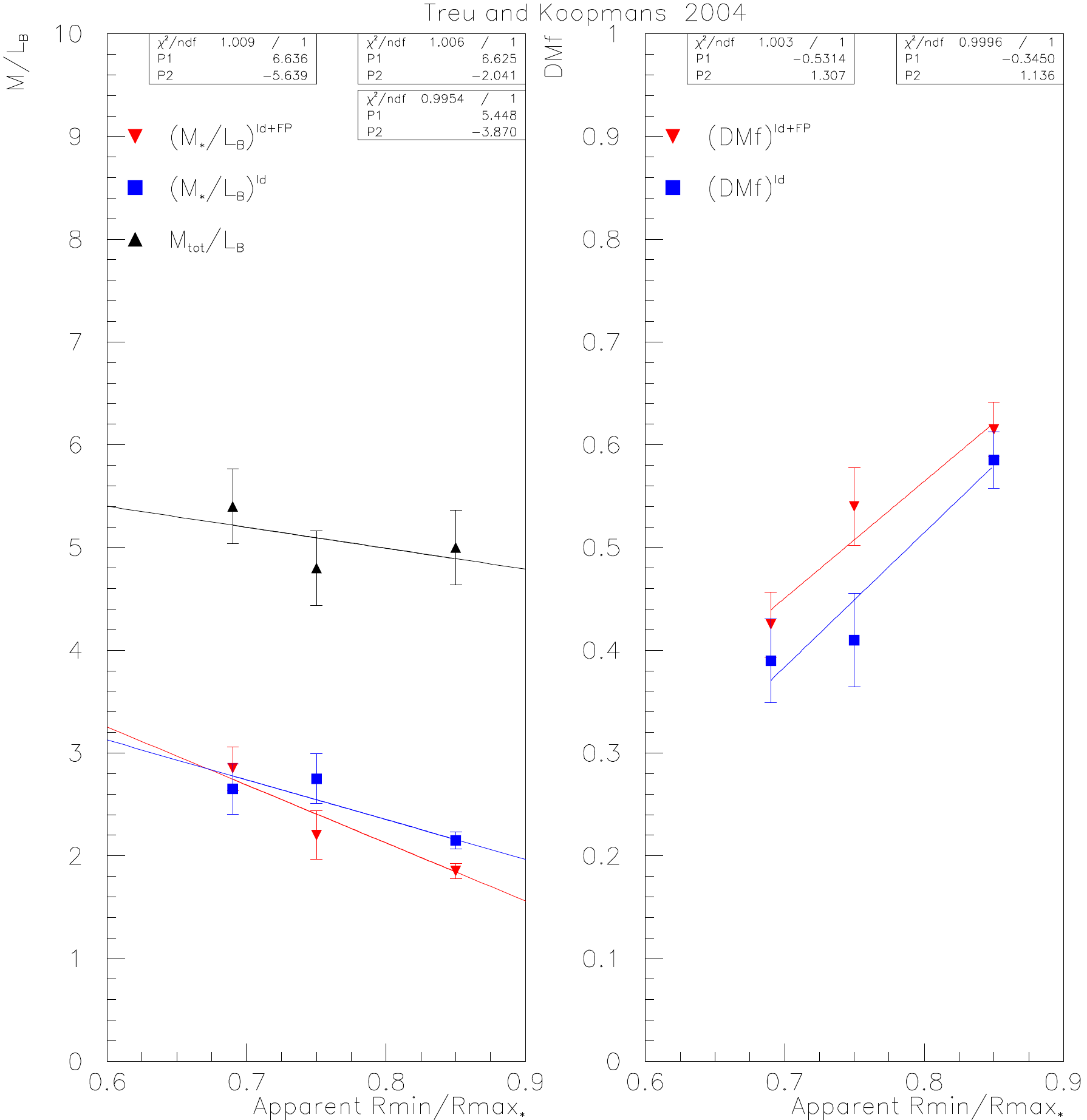}
\vspace{-0.4cm} \caption{\label{fig: treu04} $\sfrac{M_{tot}}{L}_{B}$  vs apparent axis ratio from Treu and Koopmans~\cite{Treu04}
(left column). The black up triangles are for the total mass over
B-band luminosity. The stellar mass to apparent ratios are also given
for information. Also shown is the corresponding $DMf$
(right column) vs axis ratio. The reason why $\sfrac{M_{tot}}{L}_{B}$ has a negative
slope and the $DMf$ have a positive one is because $\sfrac{M_*}{L}$ drops
faster with axis ratio than $\sfrac{M_{tot}}{L}$. 
}
\end{figure}

Treu and Koopmans~\cite{Treu04} analyzed 5 strong lenses to
obtain the distributions of luminous and dark matter within $R_{Ein}$.
The galaxies have redshifts 0.5<$z$<1. A SIE model was used to provide
a robust estimate of the total mass whilst the luminous mass profile
was deduced from stellar dynamics data. Dark and luminous profiles
were assumed to be spherical. The dark matter was assumed to follow
a NFW profile and the luminous one a Hernquist profile. The lenses
were chosen relatively isolated. The authors studied the effects of
stellar anisotropy and of neglecting ellipticity and concluded
that these effects are small. Applying standard $\sigma\geq225$
km.s$^{-1}$ and $M_{tot}<10^{12}M_{\odot}$ would have selected only
2 galaxies (0047 and C0302). We slightly relaxed the selection to
accept H1417 ($\sigma=224$~km.s$^{-1}$). We note that C0302 may belong
to a small galaxy group and so subject to environment
interaction. The 2 other available galaxies were rejected because
one is a giant (MG2016) and the other most probably a S0 (H1543).
The best fits to $\sfrac{M_{tot}}{L}_{B}$ and $DMf$ vs apparent axis ratio, shown in Fig.~\ref{fig: treu04}, are

$\sfrac{M_{tot}}{L}_{B}=(-2.04\pm3.20)\sfrac{R_{min}}{R_{max}}|_{apparent}+(6.63\pm2.45)$,

$DMf=(+1.31\pm0.30)\sfrac{R_{min}}{R_{max}}|_{apparent}+(-0.53\pm0.24)$
and 

$DMf^{+FP}=(+1.14\pm0.24)\sfrac{R_{min}}{R_{max}}|_{apparent}+(-0.35\pm0.19)$
 for which Fundamental Plan assumptions were made to further constrain
the data. 

The uncertainties from~\cite{Treu04} were symmetrized and scaled
so that $\sfrac{\chi^{2}}{ndf}=1$. The reason why $\sfrac{M_{tot}}{L}_{B}$ has a negative
slope and the $DMf$ have a positive one is because $\sfrac{M_*}{L}$ drops
faster with axis ratio than $\sfrac{M_{tot}}{L}$. 

The data were analyzed  assuming  spherical symmetry
but the authors argued that accounting for ellipticities produced
little difference. Likewise, the effect of stellar ellipticity was
studied and found unimportant. We assigned the $\sfrac{M_{tot}}{L}_{B}$ result
to group 1 reliability. 

\section{Ellipticity and projection corrections}

There are two types of ellipticity corrections. The first one is due to the ellipticity
affecting  the computation of quantities relevant to our analysis. We call this 
the ``ellipticity corrections''  proper and discuss it in the next Section. 
The second type is  correcting for the fact that 
often only the apparent ellipticity of a galaxy is known, which strongly reduces the correlation 
we study in this article. We address  this correction in Section~\ref{sec:Projection-correction}.

\subsection{Ellipticity corrections} 

The exact formula to compute $\sfrac{M}{L}$ depends on ellipticity. Generally,
assuming spherical symmetry underestimates $\sfrac{M}{L}$. Sensitivity to
ellipticity differs for different methods. It is important e.g. for
the virial method but less so for the strong lensing method when the
total mass is extracted at the Einstein radius and the luminosity
is obtained from deprojected data. Except for the strong lensing data
for which the correction is expected to be small, the data discussed
here at least approximately or partially corrected to account for
ellipticity. The caveat is that the correction may be approximate
and often uses apparent axis ratio rather than the true intrinsic axis
ratio. Hence, it may still not correct fully the $\sfrac{M}{L}$.

\subsection{Projection correction\label{sec:Projection-correction}}
This correction is independent of the technique used to extract 
$\sfrac{M}{L}$, as long as the axis ratio distribution of the galaxy sample
is also independent of the technique, e.g. a technique does not preferentially choose
galaxies with large or small apparent axis ratios. 
We assume that this is the case and apply the projection correction, unless
it was already done in the original publication as e.g. in sections
\ref{sec:Cappellari06},~\ref{sub:Bertola-et-al.} or~\ref{sec:Barnabe-et-al.}.
We use our results from Section~\ref{sub:Bacon-et-al.} (Bacon {\it et
al.}~\cite{BMS}) to estimate this correction since it is the
second largest sample of galaxies (the largest sample is that of Prugniel
\& Simien~\cite{Prugniel}. However, the authors remarked it  could be
slightly biased).

\begin{figure}[tp]
\centering
\includegraphics[scale=0.6]{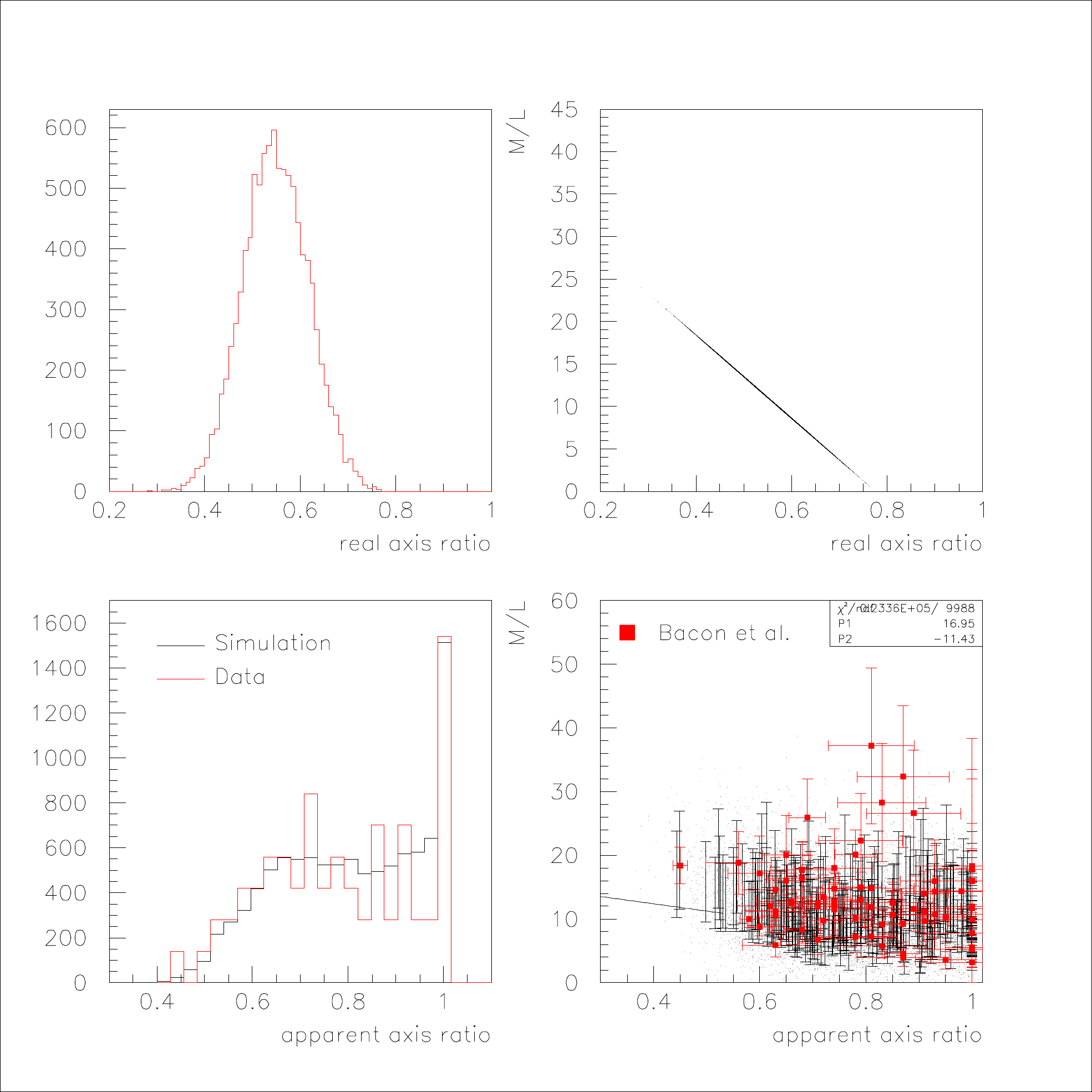}
\vspace{-0.4cm} \caption{\label{Flo:R/R projection effect1-1}Effect of projecting the
3-dimensional elliptical galaxies in our 2-dimensional observation plan. Top left plot: simulated
distribution of the real axis ratio. Top right pannel: hypothesized linear
relation between $\sfrac{M}{L}$ and real axis ratio. Bottom left panel: projected
axis ratio (black: simulation, red: data). Bottom right plot: $\sfrac{M}{L}$
vs apparent axis ratio (Black: simulation, red observed data).
}
\end{figure}

\begin{figure}[tp]
\centering
\includegraphics[scale=0.6]{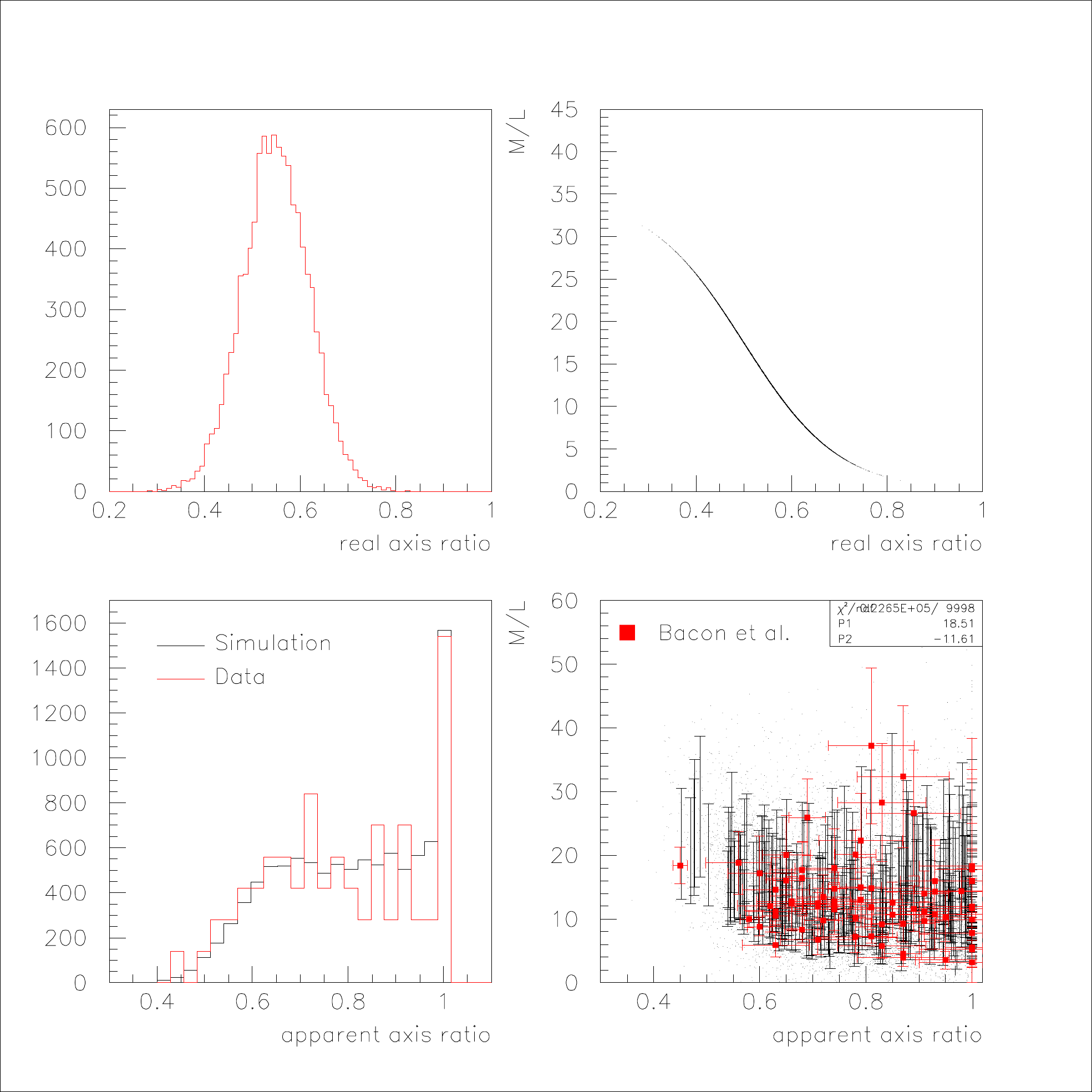}
\vspace{-0.4cm} \caption{\label{Flo:R/R projection effect2-1}Same as Fig.~\ref{Flo:R/R projection effect1-1}
but with a Fermi-Dirac relation assumed between $\sfrac{M}{L}$
ratio and real axis ratio. 
}
\end{figure}

For an oblate spheroid viewed at an angle $\theta$, its real axis ratio $\sfrac{R_{min}}{R_{max}}|_{true}$ relates to 
its projected apparent axis ratio $\sfrac{R_{min}}{R_{max}}|_{apparent}$ as~\cite{Sparke}:
\begin{equation}
\sfrac{R_{min}}{R_{max}} |_{apparent}=  \sqrt{(\sfrac{R_{min}}{R_{max}}|_{true})^{2}sin^{2}\theta+cos^{2}\theta} \label{eq:R/R proj formula}
\end{equation}
\noindent We can safely assume that the galaxies in our study are
oblate and none are prolate. We assume the $R_{min}/R_{max}|_{true}$ distribution 
 to be gaussian, see top left plot of Fig.~\ref{Flo:R/R projection effect1-1}.
The characteristics of the gaussian are determined by matching the simulated distribution
of apparent axis ratios to the actual distribution of our
data sample (bottom left plot). This results in a gaussian centered at 0.55 and with
a full width of 0.07. The top right plot shows the assumed correlation
between $\sfrac{M}{L}$ and real axis ratio. The simplest choice is
a linear relation. However, it could lead to unphysical (negative) $\sfrac{M}{L}$.
To circumvent this problem, we can choose a function that is linear
for most of the $\sfrac{R_{min}}{R_{max}}|_{true}$ values but flattens near the
$\sfrac{R_{min}}{R_{max}}|_{true}=0$ and $\sfrac{R_{min}}{R_{max}}|_{true}=1$ limits,
e.g. a Fermi-Dirac function. We do the projection correction
for both linear (Fig.~\ref{Flo:R/R projection effect1-1}) and Fermi-Dirac
functions (Fig.~\ref{Flo:R/R projection effect2-1}) cases in order
to estimate the dependence upon the choice of function. The linear 
function fitting best the observed $\sfrac{M}{L}$ vs $\sfrac{R_{min}}{R_{max}}$ (bottom
right plot) is $(\alpha x+\beta)$ with $\alpha=-49$ and
$\beta=38$. The best fit  using a Fermi-Dirac function $\big(\frac{a}{e^{\sfrac{x-b}{c}}+1}+d\big)$
is for $a=45$, $b=0.5$, $c=0.1$ and $d=0$. The bottom left panel
shows the observed (red) and simulated (black) distributions of the apparent axis ratio. 
To obtain the simulated one, the gaussian distribution is transformed using Eq.~\ref{eq:R/R proj formula}
with the viewing angle $\theta$ randomly chosen between 0 and $\pi/2$.
A random shift toward larger $\sfrac{R_{min}}{R_{max}}|_{apparent}$
values is added to reproduce the rounding effect of the resolution of the detector.
(If, due to this shift, $\sfrac{R_{min}}{R_{max}}|_{apparent}>1$, we either
set it to 1 or redraw a random value for the original $\sfrac{R_{min}}{R_{max}}|_{true}$).
The bottom right panel displays $\sfrac{M}{L}$ vs apparent axis
ratio. We use $10^{4}$ events for the fit but for clarity, only
the error bars of the first 100 events are shown. The distribution reproduces
well the observed one (shown by red symbols). This supports that the
$\sfrac{M}{L}$ dispersion is not only gaussian but mostly comes from the galaxies' random
projections. As done with the observational data, we
fit this distribution with a linear form $\sfrac{M}{L}=P_{2}(\sfrac{R_{min}}{R_{max}}|_{apparent})+P_{1}$.
In the simulation, the $\sfrac{M}{L}$ uncertainties are proportional to $\sfrac{M}{L}$,
as for the data. Using a gaussian distribution, we randomly offset
the central values of the simulated points on the bottom right figure
within their error bars. 

The correction has some model dependence due to the following assumptions:\\
~$\bullet$ No prolate galaxies in our sample;\\
~$\bullet$ gaussian distribution of the real axis ratios;\\
~$\bullet$ The assumed form for the relation between the real axis ratio and $\sfrac{M}{L}$;\\
~$\bullet$ The modeling of the detector resolution effect.\\ 
Furthermore, it is unclear how to apply the correction because the real axis ratios for
the observed data are unknown and the correction is determined
on statistical basis. Given the distribution
of the real axis ratio (top left panel), it is sufficient to only consider the range
$0.35 < \sfrac{R_{min}}{R_{max}}|_{true}<0.65$. There, the Fermi-Dirac form is
approximately linear. Averaging the results derived with both forms, the simulation of the projection effect
gives a correction of about $5\pm1$ for the slope $\sfrac{M}{L}$ vs axis ratio correlation.

\section{Correlations} \label{correlations}
Many of the quantities describing galaxies are interrelated. Therefore they may indirectly  
generate correlations with $\sfrac{R_{min}}{R_{max}}$ or $\sfrac{M}{L}$, some of them spurious:
a measurement or observation bias unrelated to $\sfrac{R_{min}}{R_{max}}$ and 
$\sfrac{M}{L}$ may propagate to one or both of them due to interrelations.
We systematically studied this possibility using mostly  the galaxy sample from Ref.~\cite{BMS}.
We summarize our findings in this section. 
The detailed analysis can be found in Appendix~\ref{sec:Detailed-analysis-of Bacon et al}.
Using the galaxy sample from~\cite{BMS} is convenient due to its large 
statistics and the relatively large number of galactic characteristics provided. They are: 
\begin{itemize}
\item The $\sfrac{M}{L}$ ratios obtained from three different virial estimates.
\item  The apparent effective radius, $Re$ (given in arcsec).
\item  The absolute effective radius, $Re$ (given in  Kpc). 
\item  The apparent axis ratio, $\sfrac{R_{min}}{R_{max}}$.
\item  The galactic surface brightness, $I_{e}$.
\item  The central velocity dispersion, $\sigma_{0}$.
\item  The distance modulus, $DM$, for each galaxy. 
\item  The galactic absolute blue magnitude, $M_{b}$. 
\item  The galactic integrated apparent blue magnitude, $B_{t}$. \end{itemize}
Furthermore, we also used galaxy samples other than that of Ref.~\cite{BMS} to check 
correlations with galactic characteristics other than those just listed.

That Ref~\cite{BMS}'s data are dated is not an issue because the selection criteria
keeps only galaxies with characteristics agreeing with the NED ones, which are current. 
Furthermore, older data are presumably more contaminated with measurement or 
observation biases, and therefore, they are better suited to study spurious correlations
and provide a useful upper limit for these biases. In other words, while
the older data may be less sensitive to an actual $\sfrac{M}{L}$ vs $\sfrac{R_{min}}{R_{max}}$
correlation because it would be diluted by jitters and worst resolutions, they are more 
sensitive to spurious correlations and if an upper limit for those is determined to not be large,
the genuine origin of the $\sfrac{M}{L}$ vs $\sfrac{R_{min}}{R_{max}}$ correlation can be validated.\\
Three types of interrelations are possible:\\
I) Relations of well-understood origin, e.g. that between $\sfrac{M}{L}$ 
and $\sigma$ originating from the virial theorem, or the reduction of apparent sizes and luminosities 
as the distances from the galaxies to Earth increase. \\
II) Relations known Phenomenologically, e.g. the Kormendy~\cite{Kormendy} and Faber-Jackson~\cite{Faber-Jackson}
relations. \\
III) Unknown relations.

\begin{figure}[tp]
\centering
\includegraphics[scale=0.7]{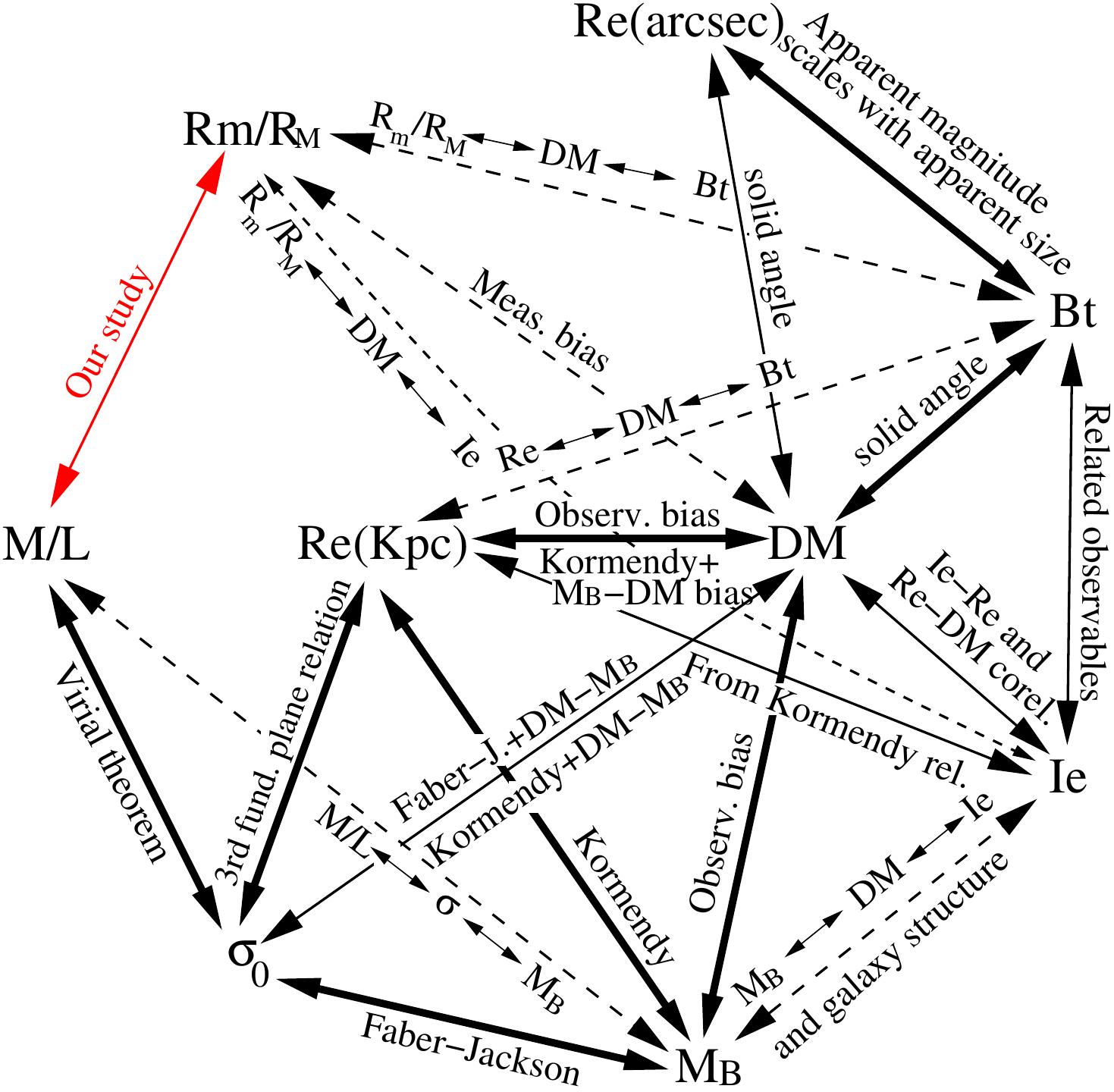}
\includegraphics[scale=0.7]{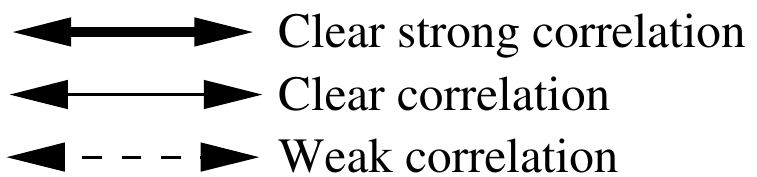}
\vspace{-0.4cm}
\caption{Observed interrelations between galaxy characteristics for the galaxy sample~\cite{BMS}. 
Thick arrows denote strong correlations ({\it viz} a clear non-zero slope for the best linear fit to 
the 2D distribution of the  quantities linked by the arrow). 
The thinner arrows denote clear correlations. The dashed arrows denote weak correlations. 
No arrow indicates that no correlation was observed, although all the combinations between pairs of characteristics  
were checked. 
The names and origins of the correlations, when available, are shown near their corresponding arrows.
The red arrow indicates the correlation investigated in this document. }
\label{fig:correlations main}
\end{figure}

We looked for correlations for all the combinations between pairs of characteristics by linearly fitting them, a non-zero
fit slope indicating a correlation. We directly used the uncertainties given in Ref.~\cite{BMS} (no rescaling
according to the {\it unbiased estimate}). We clearly confirmed all known relations (types I and II above)
except for the surface brightness $I_{e}$ vs absolute blue magnitude $M_{b}$ relation, which was
reported in~\cite{Binggeli}. The correlation is seen, yet not as strongly as for the other known relations.
We found several other relations. They are weaker and are either resulting from  known relations, or possibly due 
to measurement or observation biases. All these relations are pictured in Fig.~\ref{fig:correlations main}.

We remark that without applying the selection criteria 
described in Section~\ref{sub:General-selection-criteria}, other relations than those indicated on 
Fig.~\ref{fig:correlations main} should have been present. It is known, e.g.,
that small galaxies are prone to be flatter, indicating a correlation between $ \sfrac{R_{min}}{R_{max}}$ and 
$Re$(Kpc). Or that the $\sfrac{M}{L}$ of giant and dwarf galaxies tend to be larger, indicating a correlation between
$\sfrac{M}{L}$  and $Re$(Kpc). Furthermore, giant and dwarf galaxies tend to be
rounder indicating a correlation between $ \sfrac{R_{min}}{R_{max}}$ and $\sfrac{M}{L}$. Such correlations
would have obscured the present study and this provides another reason for implementing the selection criteria 
of Section~\ref{sub:General-selection-criteria}

Fig.~\ref{fig:correlations main} shows that there is no direct relations
between both $ \sfrac{R_{min}}{R_{max}}$ and $\sfrac{M}{L}$  and a third characteristic. This
precludes (at least for sample~\cite{BMS}) that the $\sfrac{R_{min}}{R_{max}}$
vs $\sfrac{M}{L}$ correlation is a indirect consequence of interrelations between some of the
other 7 galaxy characteristics in Fig.~\ref{fig:correlations main}. Yet, the
other relations may still somewhat contribute to the $ \sfrac{R_{min}}{R_{max}}$ vs $\sfrac{M}{L}$ 
correlation {\it via} multiple steps, e.g. $\sfrac{M}{L} \to \sigma_{0} \to DM \to \sfrac{R_{min}}{R_{max}}$. 
We discuss this possibility in section~\ref{bias cor.}. 

Beside the interrelations between the galaxy characteristics listed in~\cite{BMS},
we also checked, using the sample of Ref.~\cite{Lauer}, the effect of the known 
correlation between metallicity and $\sfrac{M}{L}$.
Since no correlation was found between $\sfrac{R_{min}}{R_{max}}$ and
$\mbox{M}_{\mbox{g}_{2}}$, the $\sfrac{M}{L}$  - $\mbox{M}_{\mbox{g}_{2}}$ correlation
should not affect $\frac{d(\sfrac{M}{L})}{d\sfrac{R_{min}}{R_{max}}}$. 
Also, a relation between $\sfrac{M}{L}$ and the luminosity density $\rho$ was signaled in~\cite{Lauer}. We
confirmed it but saw no correlation between $\sfrac{R_{min}}{R_{max}}$ $\rho_{0}$ and thus,
again, this should not affect $\frac{d(\sfrac{M}{L})}{d\sfrac{R_{min}}{R_{max}}}$.
  
 \subsection{Measurement bias study} \label{bias cor.}
Fig.~\ref{fig:correlations main} shows that all the weak correlations suggestive of
measurement biases are associated with $DM$.
This calls for studying the influence of $DM$ on $\frac{d(\sfrac{M}{L})}{d\sfrac{R_{min}}{R_{max}}}$. 
To do so we used again the galaxy sample~\cite{BMS}.  
$DM$ was binned and $\frac{d(\sfrac{M}{L})}{d\sfrac{R_{min}}{R_{max}}}$ plotted for each bins.
A linear fit of $\frac{d(\sfrac{M}{L})}{d\sfrac{R_{min}}{R_{max}}}$ vs $DM$ yields a non-zero positive 
slope of $3.42\pm2.00$ ($\chi^{2}/ndf=1.8$). This indicates that, once a $DM$ bias is corrected for, 
$\frac{d(\sfrac{M}{L})}{d\sfrac{R_{min}}{R_{max}}}$ would increase by a factor of $1.6\pm0.81$.
However, since this relies on linearly extrapolating over 31 $DM$ units using a fit just done over 4.5 units, 
we decided to not correct $\frac{d(\sfrac{M}{L})}{d\sfrac{R_{min}}{R_{max}}}$. Supporting this  conservative
choice are the facts that:\\
I) It is unclear if the bias is real, the fit slope being just 1.6$\sigma$ from zero; \\
II) the errors on the extrapolation are large; \\
III) This bias cannot be at the origin of the correlation we are studying since the correction 
would decrease it. 

\section{S0 contamination study}

It is often delicate to distinguish between S0 galaxies and elliptical galaxies with large
axis ratios. Such $\sfrac{R_{min}}{R_{max}}$-dependent contamination could 
cause the decrease of $\sfrac{M}{L}$ with $\sfrac{R_{min}}{R_{max}}$
reported in Section~\ref{sec:Global-results} because $\sfrac{M}{L}$ tends to be smaller for S0 than 
for elliptical galaxies.
If so, the correlation's origin would not be physical but a systematic bias in galaxy classification. 
The strict generic rejection criteria described in section~\ref{sub: selection criteria Local-galaxies}
suppress that contamination because they systematically exclude galaxies of unclear/transitional 
morphologies, namely E/S0, E+ or E? types (E+ are  ``late elliptical
galaxies'', a transition stage between E and S0) since those may be more
prone to misclassification. Furthermore, effects of a S0 contamination
would would emerge in the systematic studies, specifically 
varying the strictness of the  S0 rejection criteria cf. Section~\ref{Independent S0 rejection criterion}, 
the study of how the $\sfrac{M}{L}$ vs $\sfrac{M}{L}$ with $\sfrac{R_{min}}{R_{max}}$
depends on $DM$ cf. Section~\ref{Correlation with distance moduli} and
apparent magnitude cf. Section~\ref{Correlation with apparent magnitude}, 
and investigating a possible $\sfrac{M_*}{L}$ vs $\sfrac{R_{min}}{R_{max}}$ correlation (cf. 
Section~\ref{Stellar MoL ratio vs ellipticity study} and
appendix~\ref{sec:Correlation w/stellar M/L}). 

We discuss here these specifics tests. They indicate that S0 being at the origin of the correlation is unlikely.

\subsection{Independent S0 rejection criterion\label{Independent S0 rejection criterion}}

A standard method to check the efficiency the main rejection criterion 
and the consequence of a possible contamination
is to add an independent rejection criterion and varies its strictness. 
To do this, to the primary criterion (based on the NED classification)
we added a second one consisting of rejecting low velocity dispersion galaxies, as S0 tend
to have lower velocity dispersion values. These two rejection criteria are independent in
the sense that any rejection inefficiency of one criterion is unrelated
to the inefficiency of the other criterion.

Applying the second rejection criterion results in a correlation agreeing  with
the nominal analysis. This test is done on the largest sample of galaxies
(112 galaxies, of which 44 pass the second criterion) so it is statistically
significant. A possible issue is that the second criterion might
bias the correlation e.g. because of the proportionality between $\sfrac{M}{L}$ and velocity
dispersion. The average value of $ \sfrac{M}{L} $ is in fact smaller once the second criterion is
applied. To compensated for this, we increased uniformly those $\sfrac{M}{L}$ 
so that their average matches that of the nominal analysis. 
If S0 contamination was biasing the nominal analysis, 
an agreement between the nominal and second analyses 
could still occur if: \\
1) a new effect emerged from introducing the second criterion, and\\ 
2) that effect mimics the (now suppressed) S0 contamination effect. \\
Such coincidence being unlikely, this check attests the reliability of the nominal rejection criterion.
Another such misleading agreement between the nominal  and second analyses 
could also occur if S0 and E display similar
$\sfrac{M}{L}$ vs $\sfrac{R_{min}}{R_{max}}$ correlations\footnote{
Such possibility appears
to be ruled out by other checks performed in this manuscript.}. If so, our check would not
be able to assess the S0 contamination. However, this one would then have no influence on our final result.

\subsection{Consequence of a S0 contamination for galaxy census}

In this section, we assume that the  $\sfrac{M}{L}$ vs $\sfrac{R_{min}}{R_{max}}$  correlation
reported in this document stems S0
contamination, and we investigate the consequence on galaxy type census.
To reduce possible systematic bias, we conducted the analysis on large
samples of data determined either with the virial theorem~\cite{Prugniel} or from lensing~\cite{Auger 1}. Our analyses (described next)
conclude that assuming that the correlation is due to S0
contamination would lead to a proportion of E and S0 galaxies clearly conflicting 
with the observed census.

\subsubsection{Analysis using virial data (Prugniel \& Simien data set)\label{S0 Prugniel-Simien}}

We use here data from Prugniel \& Simien~\cite{Prugniel}. From those,
$\sfrac{M}{L}=(-6.58\pm1.98)\sfrac{R_{min}}{R_{max}}|_{apparent}+(8.99\pm1.68)$ (see table of
Section~\ref{Summary of the individual analyses}).
At small values of $\sfrac{R_{min}}{R_{max}}$, S0 contamination is suppressed since S0 are
highly flattened. Thus, the value of 
$\sfrac{M}{L}$ at $\sfrac{R_{min}}{R_{max}}|_{apparent}\simeq0.3$ offers a clean value of $\sfrac{M}{L_E}$. Our
work hypothesis here being that for elliptical galaxies
$\sfrac{M}{L}$ is independent of $\sfrac{R_{min}}{R_{max}}$, we have under this hypothesis 
$\sfrac{M}{L_E}\geq7.02\pm1.80$.%
\footnote{This is the value prior to the projection correction discussed in Section~\ref{sec:Projection-correction}. 
This correction should in principle be implemented but, since it is proportionate
to the {\it{physical}} correlation assumed here to not be present, the correction has no effect.} 
We have a lower bound for $\sfrac{M}{L_E}$ because, to be consistent with
the assumption that S0 contamination produces the correlation seen
over the full span of $\sfrac{R_{min}}{R_{max}}$, there must be some remaining
S0 contamination at $\sfrac{R_{min}}{R_{max}}|_{apparent}\simeq0.3$. 

To determine the mass-to-light ratio for S0, $\sfrac{M}{L_{S0}}$,  we apply a similar procedure as that for elliptical galaxies. We keep only good S0, rejecting
dwarf or peculiar S0, S0 listed in the Arp catalog, E/S0, S0?, SB0 (because a visible bar would surely identify  a S0,
even if it is face-on, which then would be rejected and unable to contaminate the elliptical galaxy sample), 
BrClg, etc. We obtain a set of 24 S0 with an average
$\sfrac{M}{L_{S0}}=-2.86\pm0.38$. Using a subset of S0 with $\sfrac{R_{min}}{R_{max}}<0.5$
to avoid possible E contamination, the average becomes $\sfrac{M}{L_{S0}}=-2.66\pm0.44$. 

\begin{figure}
\centering
\includegraphics[scale=0.55]{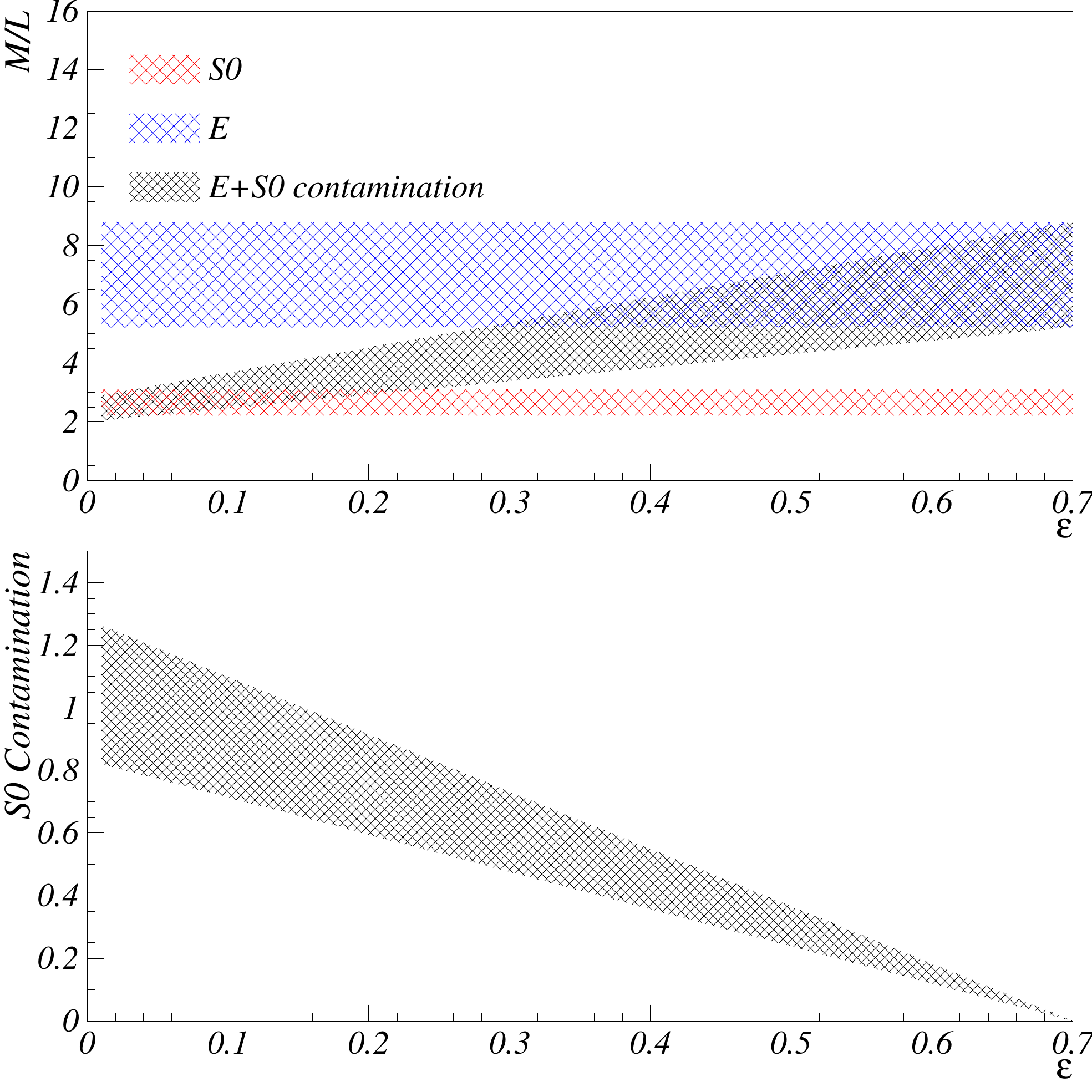}
\vspace{-0.4cm} \caption{\label{Flo:S0cont 1}Top: observed $\sfrac{M}{L}$ vs $\varepsilon \equiv 1- \sfrac{R_{min}}{R_{max}}$  correlation for the Prugniel
\& Simien data~\cite{Prugniel} (black band). Hypothesized constant $\sfrac{M}{L}$ for
elliptical galaxies (blue band), and for S0 (red band) for the same
data. Bottom: the putative S0 contamination to the elliptical galaxy set %, $\mathcal{C}(\varepsilon )$,
needed to explain the correlation seen in the top plot. %The physical range is $0<\mathcal{C}(\varepsilon) <1$.
}
\end{figure}

Supposing that $\sfrac{M}{L_E}$ and $\sfrac{M}{L_{S0}}$ are independent of $\sfrac{R_{min}}{R_{max}}$, 
and that a S0 contamination causes a  $\sfrac{M}{L}=a[\sfrac{R_{min}}{R_{max}}|_{apparent}]+b$ correlation,
imply a contamination $\mathcal{C}(\sfrac{R_{min}}{R_{max}}|_{apparent}) = 
a[\sfrac{R_{min}}{R_{max}}|_{apparent}-\sfrac{R_{min}}{R_{max}}|_{0}]/[\sfrac{M}{L_{S0}} - 
a(\sfrac{R_{min}}{R_{max}}|_{0})-b]$,
where $\sfrac{R_{min}}{R_{max}}|_{0}=0.3$ is the value of $\sfrac{R_{min}}{R_{max}}|_{apparent}$ at which
the S0 contamination is supposed to become negligible, and $\mathcal{C}(\sfrac{R_{min}}{R_{max}}|_{apparent})$
is the ratio of misidentified S0 over the total amount of assumed elliptical galaxies.
Results vs $\varepsilon \equiv 1-\sfrac{R_{min}}{R_{max}}$ 
are displayed in Fig.~\ref{Flo:S0cont 1}. Clearly, the putative contamination
would have to be implausibly large to account for the observed $\sfrac{M}{L}$ vs $\sfrac{R_{min}}{R_{max}}$
correlation. For instance, it would imply that all round
($\varepsilon \equiv 1- \sfrac{R_{min}}{R_{max}}|_{apparent}=0$) elliptical galaxies that we used 
in this analysis are in fact S0 misclassified as ellipticals. 
Even the rather conservative 2$\sigma$ lower bound of the
error band in the bottom panel of Fig.~\ref{Flo:S0cont 1} implies
a large $M(0.9<\sfrac{R_{min}}{R_{max}}|_{apparent}<1)=0.54$, signifying  that
over half of the apparently round elliptical  galaxies are misidentified 
and in fact are S0. Considering that $\sim 5\%$ of a set of disks oriented randomly would
have $\sfrac{R_{min}}{R_{max}}|_{apparent} \gtrsim 0.95$, together with the fact that in the observed 
$\sfrac{R_{min}}{R_{max}}|_{apparent}$ distribution
of ellipticals, $\sim 30\%$ of the galaxies identified as ellipticals have $\sfrac{R_{min}}{R_{max}}=0.95\pm0.05$ (see e.g. Fig.~\ref{Flo:R/R projection effect1-1}),
then the ratio of S0 to elliptical galaxies should be 7-to-1. Although estimated conservatively, this 
still disagrees with the censuses that indicate a similar number of elliptical and S0 
galaxies~\cite{Calvi:20212}.
We conclude that a S0 contamination cannot be the origin of the $\sfrac{M}{L}$ vs $\sfrac{R_{min}}{R_{max}}$  correlation.

\subsubsection{Analysis using lensing data (Auger {\it et al.} data set)\label{S0 Auger}}

\begin{figure}
\centering
\includegraphics[scale=0.55]{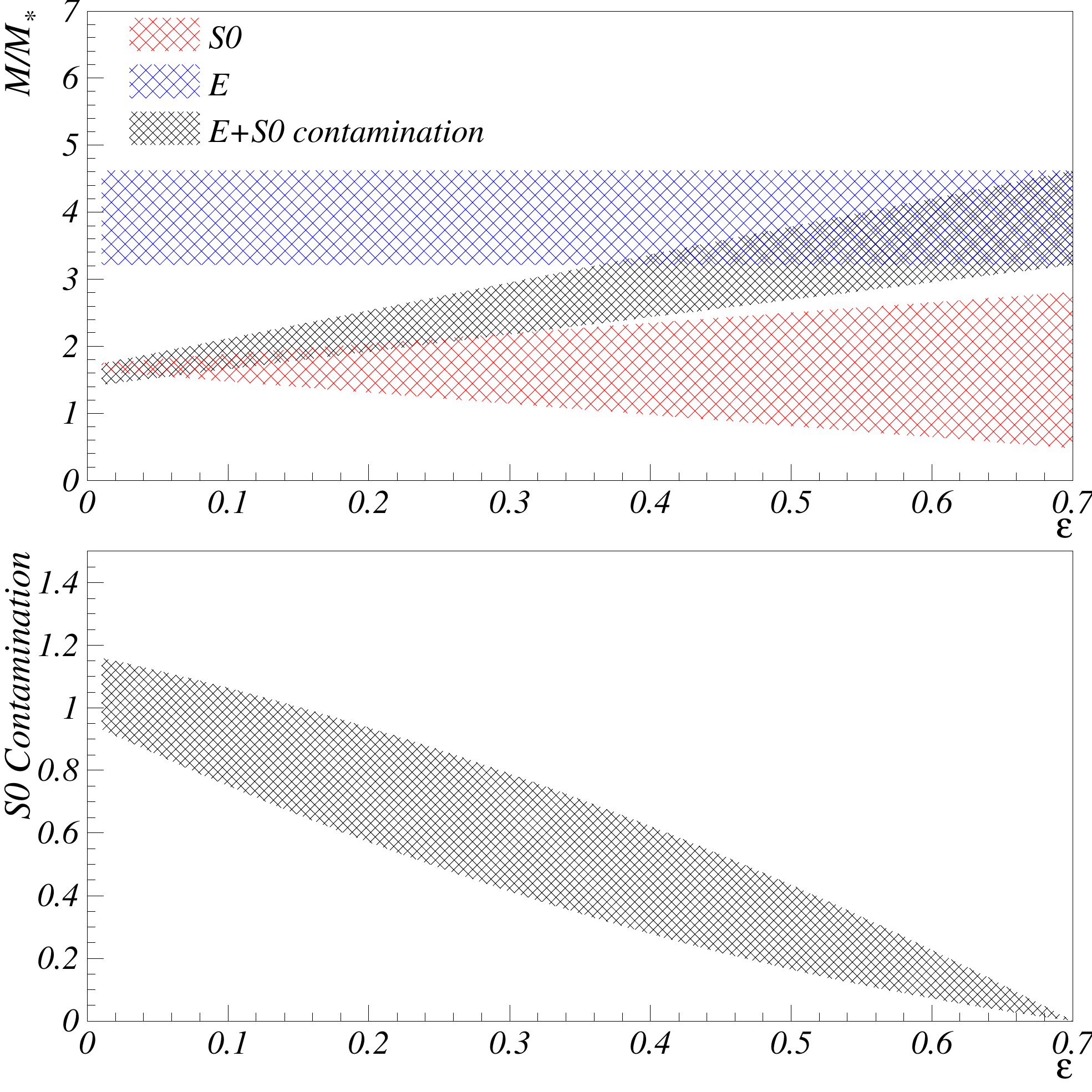}
\vspace{-0.4cm} \caption{\label{Flo:S0cont 3} Same as Fig.~\ref{Flo:S0cont 1} but for the
lensing data from Auger {\it et al}~\cite{Auger 1}.
}
\end{figure}

The Prugniel \& Simien data~\cite{Prugniel} are obtained using the virial theorem.
We repeat here the analysis with the Auger {\it et al.} data~\cite{Auger 1}, obtained
using gravitational lensing for which the mass estimate procedure
is generic to both E and S0. From these data, 
$\sfrac{M_{tot}}{M_*}=(-3.38\pm0.79)\sfrac{R_{min}}{R_{max}}|_{mass}+(4.92\pm0.66)$. We conservatively use
the largest correlation ($\sfrac{R_{min}}{R_{max}}=\sfrac{R_{min}}{R_{max}}|_{mass}$,
and $\sfrac{M}{M_*}$ extracted using a Chabrier IMF), $\sfrac{M}{M_*}(\sfrac{R_{min}}{R_{max}}=0.3)=3.92\pm0.70$. 
Selecting S0 from the set~\cite{Auger 1}, we obtain a subset of 5 S0
(we rejected 2 E/S0). From this sample, we obtain a $\sfrac{M}{L}$ vs $\sfrac{R_{min}}{R_{max}}$ correlation slope of $+0.06\pm1.60$.
The deduced putative contamination is shown in Fig.~\ref{Flo:S0cont 3}.
It is similar to the one obtained from virial data and so the same
conclusion applies.

\subsection{Case of the S0 contamination at small $\sfrac{R_{min}}{R_{max}}$ only}

In sections~\ref{S0 Prugniel-Simien} and~\ref{S0 Auger}, the S0 contamination was chosen to dependent
linearly with $\sfrac{R_{min}}{R_{max}}$ in order to reproduce the linear dependence of
$\sfrac{M}{L}$ with $\sfrac{R_{min}}{R_{max}}$. However, this linear dependence stems from our choice
of the fit function, which we chose to be linear as it is the simplest assumption, see Section~\ref{Uncertainties}. 
If  S0 contamination were in fact at the origin of the $\sfrac{M}{L}$ vs $\sfrac{R_{min}}{R_{max}}$
correlation, it could be due e.g.  to a large S0 contamination presents solely about 
$\sfrac{R_{min}}{R_{max}}|_{apparent}\sim1$. If so, higher order polynomials would fit the data better.
whilst this is not the case we nevertheless, for completude, investigate here this possibility. 

If a $\sfrac{R_{min}}{R_{max}}|_{apparent}\sim1$ (i.e, face-on S0 ) S0 contamination were significant, 
removing these
galaxies from the analysis should significantly decrease 
$d(\sfrac{M}{L})/d(\sfrac{R_{min}}{R_{max}}|_{apparent})$. This prediction was investigated
with the highest statistics data set of Prugniel \& Simien~\cite{Prugniel}. 
Excluding galaxies with $\sfrac{R_{min}}{R_{max}}|_{apparent}>0.9$
decreases the sample from 102 to 76 galaxies and yields 
$d(\sfrac{M}{L})/d(\sfrac{R_{min}}{R_{max}}|_{apparent})=-10.32\pm2.89$. 
This is compatible with the nominal value of $-6.58\pm1.98$
and is, if anything, steeper than the nominal value in contrast to what we would have expected 
from face-on S0 contamination.

It is also immediately evident from any high statistics correlation plots, e.g. Fig.~\ref{fig: bacon 1},
that excluding galaxies with high $\sfrac{R_{min}}{R_{max}}|_{apparent}$
would not modify significantly $d(\sfrac{M}{L})/d\sfrac{R_{min}}{R_{max}}|_{apparent}$.

\subsection{Stellar $\sfrac{M_*}{L}$ ratio vs ellipticity study \label{Stellar MoL ratio vs ellipticity study}}

We investigate in appendix~\ref{sec:Correlation w/stellar M/L} the
possibility of a $\sfrac{M_*}{L}$ vs ellipticity correlation. No such correlation
was found. However, just like for $\sfrac{M}{L}$, the stellar $\sfrac{M_*}{L}$
ratios tend to be smaller for S0: Ref~\cite{Cappellari 2013a},
reports that low velocity dispersion $\sigma$ early-type galaxies, i.e, galaxies
more likely to be S0, have $\sfrac{M_*}{L}$ of about 2 $\sfrac{M_{\odot}}{L_{\odot}}$. 
High $\sigma$ early-type galaxies, i.e, those more likely to be ellipticals, 
have $\sfrac{M_*}{L}$ of about 4 $\sfrac{M_{\odot}}{L_{\odot}}$.\footnote{Similarly,
in~\cite{Cappellari 2006}, it is shown that fast rotators have
a smaller $ $ $\sfrac{M_*}{L}$ (about 1.5 to 3 $\sfrac{M_{\odot}}{L_{\odot}}$)
than slow rotators (about 2.5 to 3.5 $\sfrac{M_{\odot}}{L_{\odot}}$). If fast
rotators tend to be S0, then S0 tend to have smaller $\sfrac{M_*}{L}$ than
elliptical galaxies. }
Consequently, if S0 contamination were at the origin of the $\sfrac{M}{L}$
vs ellipticity correlation, it would also produce a $\sfrac{M_*}{L}$ vs
ellipticity correlation, which is not seen. This further rules out 
the possibility that S0 contamination is at the origin of the $\sfrac{M}{L}$
correlation.

\subsection{Correlation with distance moduli \label{Correlation with distance moduli}}

If the $\sfrac{M}{L}$ vs $\sfrac{R_{min}}{R_{max}}$ correlation  were 
due to contamination from S0, then its slope $d(\sfrac{M}{L})/d(\sfrac{R_{min}}{R_{max}})$
should be steeper for distant galaxies because
they are harder to classify. However, Fig.~\ref{fig: global} shows that this is not
the case: the $\sfrac{M}{L}$ obtained using the strong lensing method pertain to distant galaxies, yet their 
$d(\sfrac{M}{L})/d(\sfrac{R_{min}}{R_{max}})$ are similar 
(in fact slightly lower in average) to that obtained with the virial
theorem, typically applied to local galaxies. 
Furthermore, a contamination from S0 would cause an increase of $d(\sfrac{M}{L})/d(\sfrac{R_{min}}{R_{max}})$
with distance modulus, yet there is no
evidence of such increase, see page~\pageref{sub:Distance-Moduli correl}. 

\subsection{Correlation with apparent magnitude \label{Correlation with apparent magnitude}}

\begin{figure}
\centering
\includegraphics[scale=0.4]{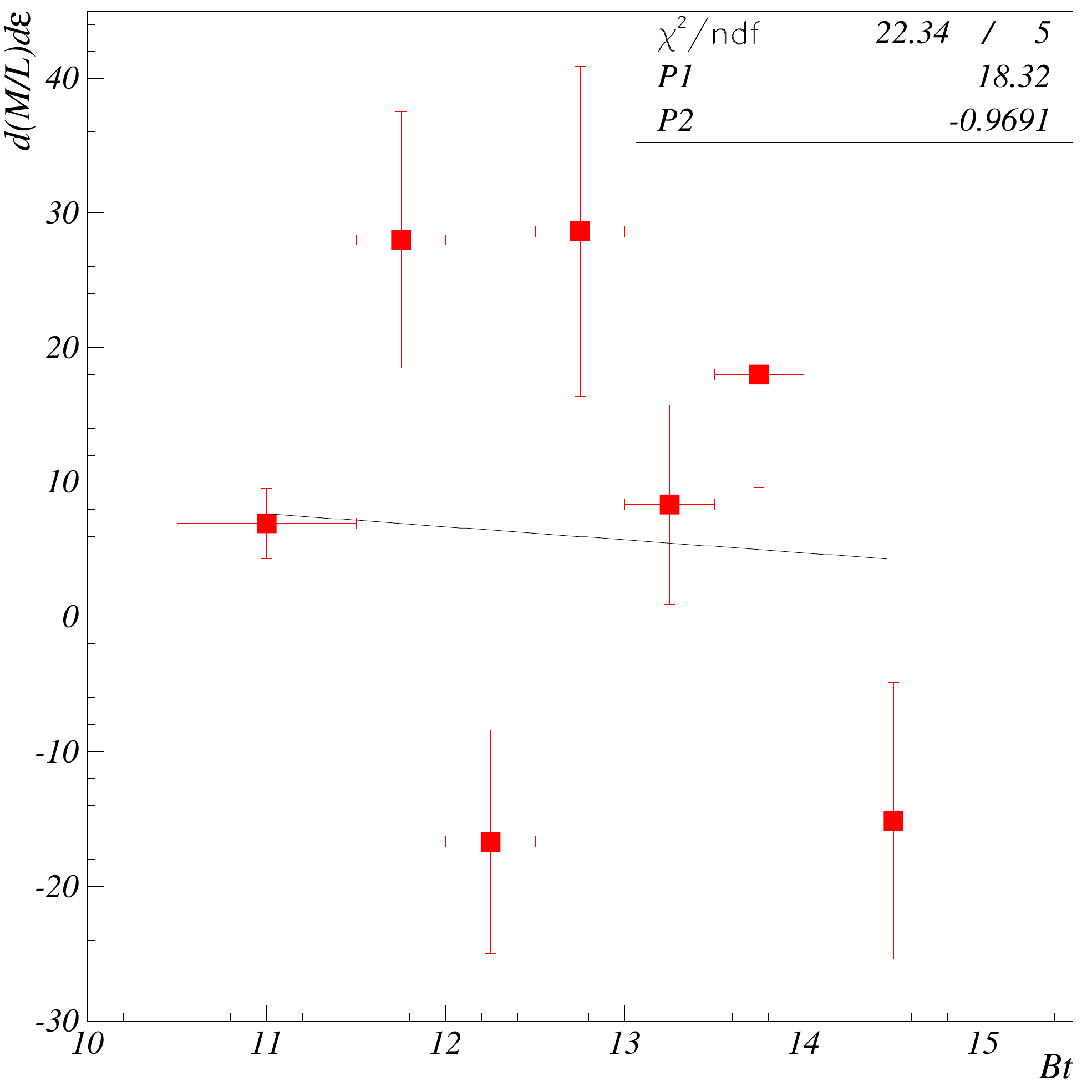}
\vspace{-0.4cm} \caption{\label{Flo:S0cont 4}  $d(\sfrac{M}{L})/d\varepsilon$ vs apparent
integrated blue magnitude. The Bacon {\it et al.} data set was used.
}
\end{figure}

Following the same argument as in the previous section, a contamination from S0 would 
cause $d(\sfrac{M}{L})/d(\sfrac{R_{min}}{R_{max}})$
to be inversely dependent on apparent magnitude because fainter galaxies
are harder to identify. The apparent magnitude being
strongly correlated with distance modulus, the conclusion of
section~\ref{Correlation with distance moduli} should directly apply. Nevertheless, we
investigate explicitly here the possible correlation between $d(\sfrac{M}{L})/d(\sfrac{R_{min}}{R_{max}})$ 
(or equivalently $d(\sfrac{M}{L})/d\varepsilon$)
and apparent magnitude. 
We grouped  the $\sfrac{M}{L}$ in seven equidistant bins of $B_{t}$, the integrated apparent blue magnitude, and then 
extracted for each bin the slope $d(\sfrac{M}{L})/d\varepsilon$. The result is shown in Fig.~\ref{Flo:S0cont 4}. There
is no increase of $d(\sfrac{M}{L})/d\varepsilon$ with $B_{t}$. The best linear fit
to the data yields $d(\sfrac{M}{L})/d\varepsilon=(-0.97\pm1.97)B_{t}+18.32\pm23.12$.
As expected from the strong correlation between $B_{t}$ and distance
modulus, there is a hint of effect opposite to what should have been
expected from a S0 contamination, i.e, a possibly negative slope $d^2(\sfrac{M}{L})/(d\varepsilon dB_{t})=-0.97\pm1.97$.

\section{Global results \label{sec:Global-results}}

\begin{figure}
\centering
\includegraphics[scale=0.45]{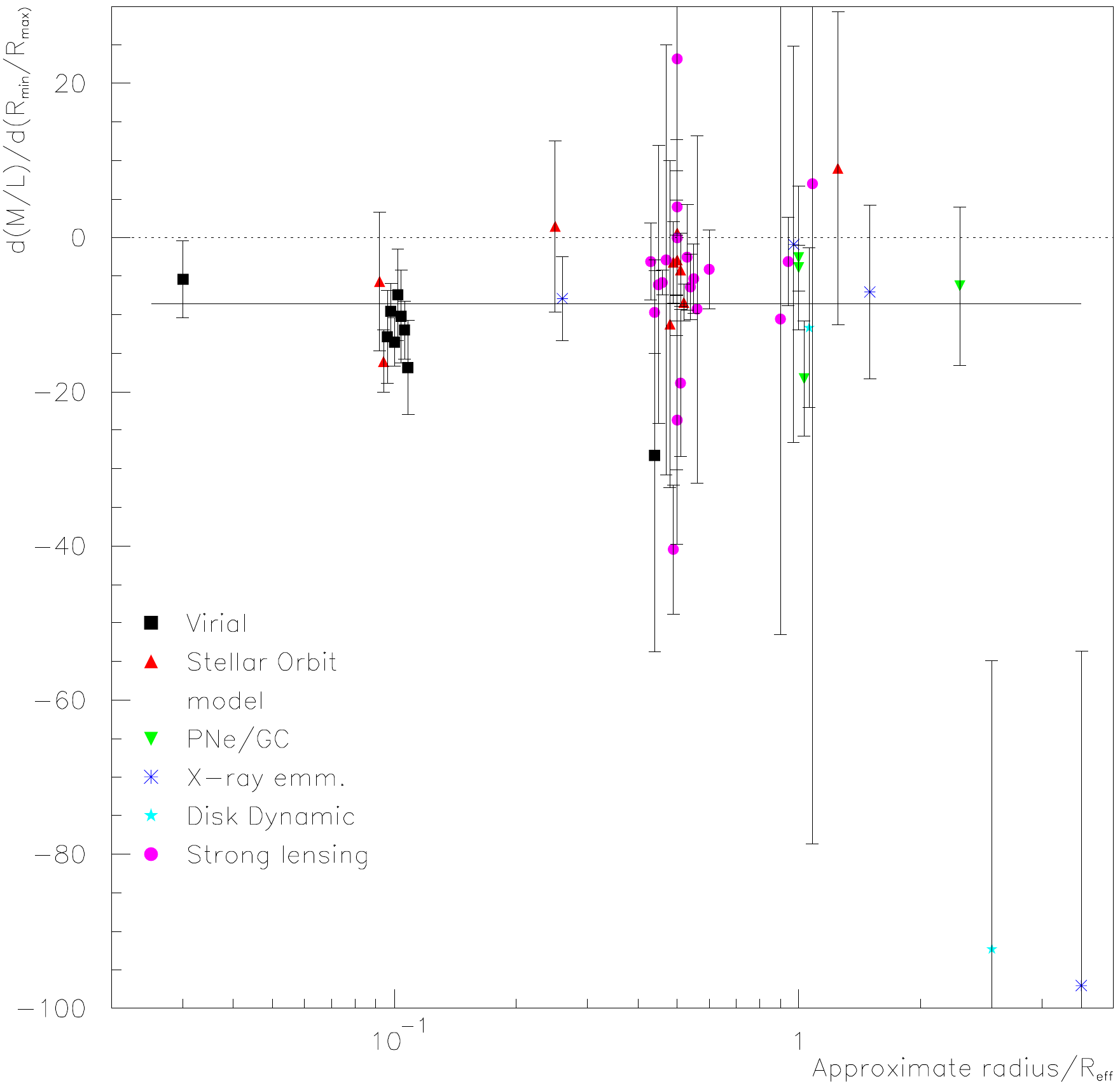}
\includegraphics[scale=0.45]{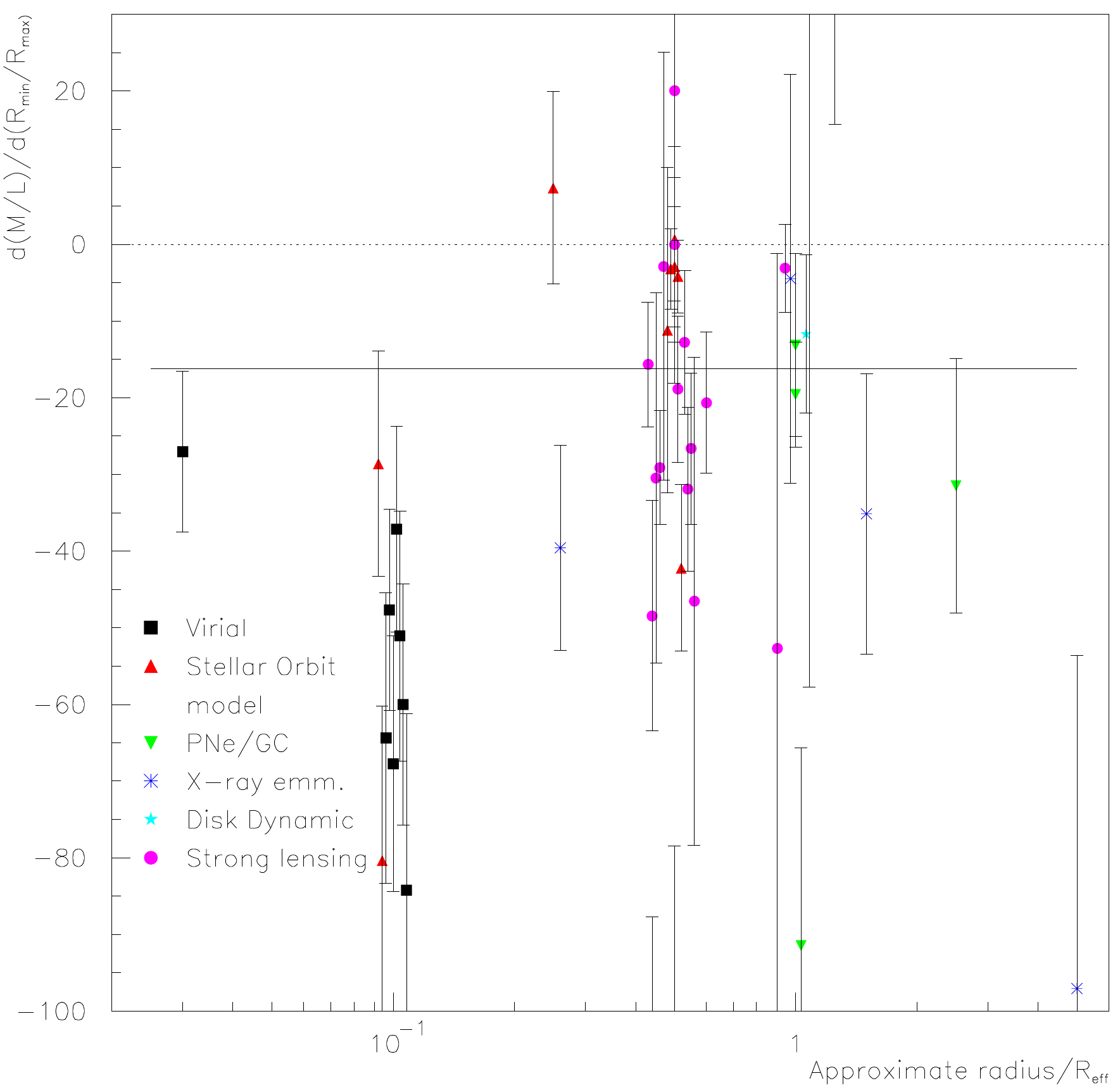}
\vspace{-0.4cm} \caption{\label{fig: global}The $d(\sfrac{M}{L})/d(\sfrac{R_{min}}{R_{max}})$ slopes vs the approximate radius
values (normalized by their $R_{eff}$) where the $\sfrac{M}{L}$ are obtained. 
Different symbols are for the different methods used to extract the $\sfrac{M}{L}$. The
plain line shows the average value of $d(\sfrac{M}{L})/d(\sfrac{R_{min}}{R_{max}})$
after accounting for the reliability of each $\sfrac{M}{L}$ extraction and
the shared statistics (the error bars shown do not account for this.
They are the values quoted in sections~\ref{sec:Data-sets-using virial theo}
to~\ref{sec:Data-sets-using strong lensing}). 
%The dashed line is set at 0 for reference. 
The left plot is before applying
projection correction (see Section~\ref{sec:Projection-correction})
unless it is directly provided in the original publication. The right
plot is after projection correction, plotted on the same scale (consequently, some
of the points are now off the graph). We added small offsets on the
radius to enhance visual clarity. 
}
\end{figure}

The results of sections~\ref{sec:Data-sets-using virial theo}
to~\ref{sec:Data-sets-using strong lensing} can be combined 
to extract $\left\langle \frac{d(\sfrac{M}{L})}{d(\sfrac{R_{min}}{R_{max}})}\right\rangle $,
the average slope of $\sfrac{M}{L}$ vs axis ratio. Several points require
caution: 
\begin{itemize}
\item whilst $\sfrac{M}{L}$ are most often estimated in the B-band, not all of them are. 
Another scaling factor occurs between $\sfrac{M}{L}$ using different Hubble
parameter values. Lastly, some results are expressed as $\sfrac{M_{tot}}{M_*}$
instead of $\sfrac{M}{L}$. We assume here that $\sfrac{M_{tot}}{M_*} \propto \sfrac{M}{L}$. 
To address these points, we impose $\sfrac{M}{L}(\sfrac{R_{min}}{R_{max}}=0.7) = 
\sfrac{M_{tot}}{M_*}(\sfrac{R_{min}}{R_{max}}=0.7)$=8.
We choose to normalize to 8 as it is a typical value for $\sfrac{M}{L}$ 
of elliptical galaxies. The choice $\sfrac{R_{min}}{R_{max}}=0.7$ corresponds to
where the distribution of elliptical galaxies peaks, which makes
extrapolations unnecessary, in contrast to choosing other
values such as e.g. $\sfrac{R_{min}}{R_{max}}=1$.
\item Some results use similar methods to obtain $\sfrac{M}{L}$ and share a number
of the same galaxies in their sample. Thus, these results can
be highly correlated, biasing the average. Furthermore, some results
are, in the context of our study, less reliable than others. This is addressed by
applying the following procedure:

\begin{itemize}
\item The uncertainty is scaled by the reliability factor from 1 (most reliable) to 4 (least reliable). 
\item For extractions using a similar method, grouped in Sections~\ref{sec:Data-sets-using virial theo}
to~\ref{sec:Data-sets-using strong lensing}, we identify the number
of shared galaxies and increase each uncertainty assuming
that the uncertainty is statistically dominated. 
This procedure accounts
also for the difference in reliability factors between analyses because
results are weighted by them when combined. Uncertainties on results
using shared galaxies are multiplied by 1.3 to 6.1 for the virial
methods, by 1.2 to 2.0 for the stellar modeling method, by 1.1 to
2.1 for the PNe and CG method, by 1.2 for the X-ray and disk methods
and 1.2 to 6.2 for the strong lensing method (except for~\cite{Keeton}
which is fully weighted out as it employs only galaxies used by more
reliable analysis). For this procedure, one must assume that analyses employing the same
method and the same galaxies are exactly correlated. This is not true
as a particular analysis uses different assumptions, IMF, profiles, 
etc... Particular analyses also vary in their details and
the quality of the data differs since the results have been published over a 30
years range. Hence, our correction is conservative, yielding overestimated
uncertainties. This is partly alleviated by fitting $d(\sfrac{M}{L})/d(\sfrac{R_{min}}{R_{max}})$
vs radius with a constant and scaling the uncertainties so that
$\sfrac{\chi^{2}}{ndf}=1$. 
\end{itemize}
We note that, whilst important, this procedure happens to numerically
have a small influence.

\item When the results of an analysis are provided for different IMF, we choose the result using a Chabrier
IMF because,  with the Salpeter one, it is the most commonly used IMF whilst being more recent than  Salpeter's. 
\item With the strong lensing method, 
$d(\sfrac{M}{L})/d(\sfrac{R_{min}}{R_{max}})$ can be extracted
using apparent or mass axis ratios. We prefered using 
the apparent axis ratio rather than the mass one for several
reasons: 

\begin{itemize}
\item The mass axis ratio is more model dependent.
\item It is not systematically calculated in the articles, although it can be obtained
from other works using the same galaxies. However, its is desirable
for consistency to use $ $mass axis ratio obtained using the same
model and assumptions as used for the $\sfrac{M}{L}$ extraction. 
%\item It is not clear what the mass axis ratio actually reflects (ellipticity of the dark h+baryonic matter?).
We remark that, apart for Grillo {\it et al.}~\cite{Grillo09},
the $\sfrac{M}{L}$ vs axis ratio correlation is always stronger for mass
axis ratio, i.e, $\left|\frac{d(\sfrac{M}{L})}{d(\sfrac{R_{min}}{R_{max}}|_{apparent})}\right|<$$\left|\frac{d(\sfrac{M}{L})}{d(\sfrac{R_{min}}{R_{max}})|_{mass}}\right|$.
It may be because $\sfrac{R_{min}}{R_{max}}|_{apparent}$ reflects better the intrinsic axis
ratio, or it could be a systematic effect of the lensing method since this one does
not fully account for ellipticity effects. 
\end{itemize}
\item We globally rescale the uncertainties so that for
$d(\sfrac{M}{L})/d(\sfrac{R_{min}}{R_{max}})$ vs radius, $\sfrac{\chi^{2}}{ndf}=1$.
(We use a one-parameter fit, viz $d(\sfrac{M}{L})/d(\sfrac{R_{min}}{R_{max}})$ 
is assumed to be independent of radius.) Forcing $\sfrac{\chi^{2}}{ndf}=1$
is justified since many different methods were used to obtain the $\sfrac{M}{L}$  with either uncorrelated 
systematic uncertainties, or, for $\sfrac{M}{L}$ using the same extraction method, we accounted
for the correlation of the results. 
\end{itemize}
Fig.~\ref{fig: global} displays the slopes $d(\sfrac{M}{L})/d(\sfrac{R_{min}}{R_{max}})$. %for the various results obtained. 
For clarity, they are shown vs the approximate average value of the radius 
at which a $\sfrac{M}{L}$ is extracted. 
%%%below is obsolete. There is indeed a r-dependence of the slope and we will take it into account.
%Since $\sfrac{M}{L}$ is approximately constant
%with radius for $r<R_{eff}$ no dependence of $d(\sfrac{M}{L})/d(\sfrac{R_{min}}{R_{max}})$
%with radius for $r<R_{eff}$ is expected as long as the galaxy has
%a constant axis ratio. At larger values, a dependence is expected since
%$\sfrac{M}{L}$ has been shown to depends on radius, since the
%galaxy  axis ratio can be different from its central value, and since
%dark matter from the environment may contaminate $\sfrac{M}{L}$. Nevertheless,
%we include the $d(\sfrac{M}{L})/d(\sfrac{R_{min}}{R_{max}})$at large $r$ in
%our averaging. We remark that due to their low precision they contribute little to
%the average. 
%
The average slope $d(\sfrac{M}{L})/d(\sfrac{R_{min}}{R_{max}})$ before the projection
correction describe in Section~\ref{sec:Projection-correction}  is:

$\left\langle \sfrac{d(\sfrac{M}{L})}{d(\sfrac{R_{min}}{R_{max}})}\right\rangle =-9.16\pm1.75$,
and

$\left\langle \sfrac{d(\sfrac{M}{L})}{d(\sfrac{R_{min}}{R_{max}})}\right\rangle =-17.42\pm4.30$
after correction\footnote{The result is not approximately $5\pm1$ times larger, as was obtained in 
Section~\ref{sec:Projection-correction}, 
since the additional uncertainty from the projection correction
is proportional to the slope. Thus, the contribution to points near 
$\left\langle \sfrac{d(\sfrac{M}{L})}{d(\sfrac{R_{min}}{R_{max}})}\right\rangle \sim0$
is smaller and their relative weight in the averaging procedure increases. }.

{\footnotesize Note: accounting for the reliability of the methods
and for correlations between the various results is methodologically important but this happens
to not change significantly the results. Without it, 
$\left\langle \sfrac{d(\sfrac{M}{L})}{d(\sfrac{R_{min}}{R_{max}})}\right\rangle =-8.52\pm0.84$
before projection correction, to be compared to $-9.16\pm1.75$ with it. }

These results assume that $\sfrac{M}{L}$ is constant with radius $r$.

Fig.~\ref{fig: global} reveals that combined results from different methods have systematic shifts,
especially between the virial and the strong lensing combined results. This
could come from the fact that different methods estimate $M/L$  at typically different radius values:
in fact, it has been reported that at smaller radii, the baryonic mass dominates over the dark 
one~\cite{Bertola93, Capaccioli, Kronawitter, Magorrian01, Nagino, Napolitano, Thomas, van der Marel 1991}. 
In other words, the $M/L$ values increase with radius. In our analysis however, all values 
$M/L(\sfrac{R_{min}}{R_{max}}=0.7)$ are normalized  to 8$\mbox{M}_{\odot}/\mbox{L}_{\odot}$, 
which effectively removes the dependence on which particular radius value is used to obtain  $M/L$.
However, this introduces an artificial radius dependence of $d(\sfrac{M}{L})/d(\sfrac{R_{min}}{R_{max}})$ since this one
scales with the factor used to normalize $M/L$ to 8$\mbox{M}_{\odot}/\mbox{L}_{\odot}$, that factor itself
depending by definition on the value of the extraction radius. 
This artificial dependence is opposite to the actual dependence of $M/L$  with radius,
i.e.  $d(\sfrac{M}{L})/d(\sfrac{R_{min}}{R_{max}})$ decreases with $r$, see Fig.~\ref{fig: global}. 
To fix this issue, we use a normalization derived from the results of Ref.~\cite{Capaccioli}, 
\begin{equation}\label{eq:r-dep-norm}
M/L(\sfrac{R_{min}}{R_{max}}_{app}=0.6)=(6+1.7r/R_{eff})\mbox{M}_{\odot}/\mbox{L}_{\odot},
\end{equation} 
to include the  $r$-dependence of $M/L$. Fig.~\ref{fig: global2} displays 
$d(M/L)/d(\sfrac{R_{min}}{R_{max}})$ obtained with Eq.~(\ref{eq:r-dep-norm})
normalization. Compared to Figs.~\ref{fig: global}, the $r$-dependence of $d(M/L)/d(\sfrac{R_{min}}{R_{max}})$ in
Fig.~\ref{fig: global2} has decreased. To completely cancel it, a factor of 3 to 4 (rather than 1.7 in 
 Eq.~\ref{eq:r-dep-norm}) would be needed.
As this larger factor remains compatible with the dependence derived from Ref.~\cite{Capaccioli}'s data,
the $r$-dependence seen in Fig.~\ref{fig: global} may originate fully  from the $r$-dependence of $M/L$.
Nevertheless, other effects could contribute to the systematic differences noticed in
Fig.~\ref{fig: global}. For example, the less stringent selection applied to the distant galaxies
used with the strong lensing method could result in a contamination that reduces the correlation. 
Furthermore, the $M/L$ extracted with the strong lensing method have no ellipticity correction since 
the method is less sensitive to $R_{min}/R_{max}$. Actually, we remark that if $d(\sfrac{M}{L})/d(\sfrac{R_{min}}{R_{max}}_{mass})$
is used instead of $d(\sfrac{M}{L})/d(\sfrac{R_{min}}{R_{max}}_{app})$, the ellipticity-corrected virial results and the strong lensing
one would agree.

\subsection{Results}
\begin{figure}
\centering
\protect\includegraphics[scale=0.45]{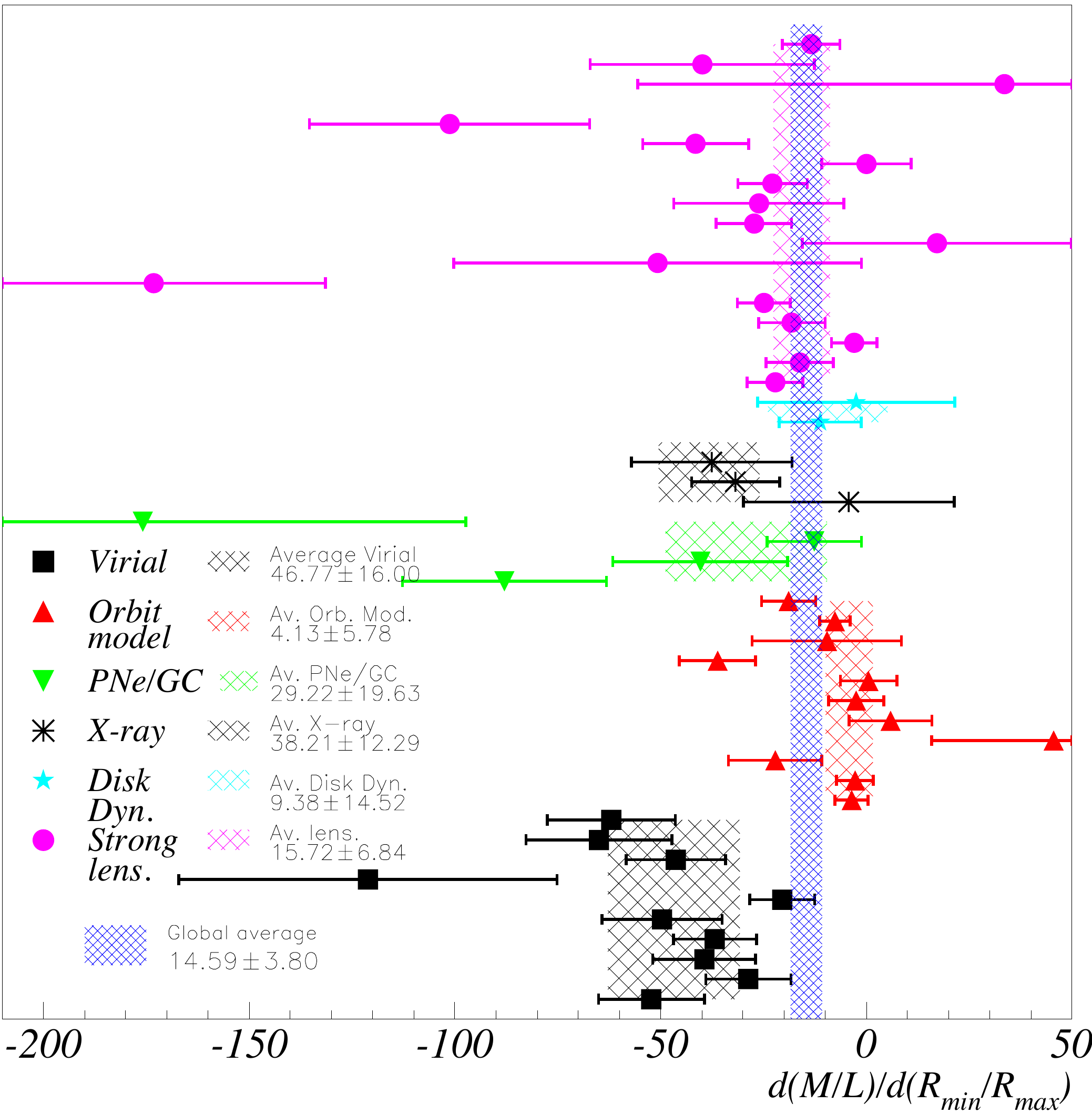}
\vspace{-0.5cm}
\caption{\label{fig: global2}\hspace{0cm}Same as Fig.~\protect\ref{fig: global}, but with the radius dependence of $M/L$ 
corrected using  Eq.~\protect(\ref{eq:r-dep-norm})}
\end{figure}

Fig.~\ref{fig: global2} shows that the averaged $\frac{d(M/L)}
{d(\sfrac{R_{min}}{R_{max}})}$ obtained using the six different methods are all negative and 
average to $\left\langle \frac{d(M/L)}{d(\sfrac{R_{min}}{R_{max}})}\right\rangle 
=(-14.53\pm3.79) \mbox{M}_{\odot}/\mbox{L}_{\odot}$. 
The uncertainty stemming from the 
$r$-dependence of the normalization is conservatively estimated from the difference between the two 
results obtained with the $r-$dependent normalization and with the constant one. 
With this additional uncertainty accounted for, the final result of the analysis is 
\begin{equation}
\boxed{\left\langle \frac{d(M/L)}{d(\sfrac{R_{min}}{R_{max}})}\right\rangle =(-14.53\pm4.77) \mbox{M}_{\odot}/\mbox{L}_{\odot}.}
\label{eq:final result}
\end{equation}
This signals a statistically meaningful correlation between 
$M/L$ and $\sfrac{R_{min}}{R_{max}}$. 
The slope~(\ref{eq:final result}) is proportional to the
normalization $M/L(r=R_{eff},\sfrac{R_{min}}{R_{max}}=0.7)=7.7\mbox{M}_{\odot}/\mbox{L}_{\odot}$, see Eq.~\ref{eq:r-dep-norm}.
The slope~(\ref{eq:final result})  is large compared to 7.7$\mbox{M}_{\odot}/\mbox{L}_{\odot}$.
We remark that if the galaxy selection 
criteria had not been applied, the correlation would vanish, diluted by large fluctuations. This 
may account for the fact that the correlation had not been identified earlier. The 
analyses of Ref~\cite{Cappellari 2013a, Cappellari 2013b} e.g. did not find evidence of this 
correlation. However, whilst they analyzed a large galaxy sample (260 galaxies), only 
$\sim10\%$ of those galaxies are bona-fide elliptical galaxies, and $\sim80\%$ are 
lenticular or spiral galaxies. Similarly, we checked using the data~\cite{BMS} that when selection is not applied, the correlation 
vanishes.

\section{Summary and conclusion}

We investigated the possibility of a correlation between the dark
matter content of elliptical galaxies and their ellipticity. Elliptical galaxies
can differ importantly  from each other, and peculiarities might bias the estimation of the dark mass, 
causing systematic and random variations. Therefore, it was important to select a large
and homogeneous sample of galaxies. Effects of the peculiarities are
then minimized by the homogeneity and suppressed statistically. 
Furthemore, since the value of $\sfrac{M}{L}$ depends on the galactic radius $r$ at which the ratio is extracted, as well as on the wavelength at 
which the galaxy luminosity $L$ is calculated (typically the B-band, but not always), it was necessary to normalize the $\sfrac{M}{L}$ at a given $r$
to a unique value. We chose $r=0.7R_{eff}$
and there, used the typical value $8\mbox{M}_{\odot}/\mbox{L}_{\odot}=\sfrac{M}{L}(0.7R_{eff}) $.
With this normalization, we found a clear correlation: 
$\left\langle \sfrac{d(\sfrac{M}{L})}{d(\sfrac{R_{min}}{R_{max}})}\right\rangle =
-14.53\pm4.77\mbox{M}_{\odot}/\mbox{L}_{\odot} $.
The dark matter information is
obtained from six different approaches (virial theorem, stellar orbit modeling, orbits of
planetary nebulae and globular clusters, embedded disk dynamics, hydrostatic equilibrium, and strong lensing) to minimize methodological bias.
Possible effects of measurement or observation biases were studied
thoroughly. Furthermore, we repeated  the same analysis on the stellar $\sfrac{M_*}{L}$,
see appendix~\ref{sec:Correlation w/stellar M/L}, and no significant
correlation with ellipticity was found, as it should be. This suggests
that our procedure is free of significant biases. Possible conclusions
are either that:     
\begin{enumerate}
\item There is a surprisingly strong influence of the dark matter halo on
a galaxy shape, possibly from the halo shape as suggested by the stronger
correlation generally seen when investigated with mass axis ratio
rather than apparent (project luminous) axis ratio. This would allow
us to experimentally address the question of the shape of the dark
halo and be critical to understand galaxy formation. 
\item The dynamical evidences from which the dark matter content of galaxy
is inferred are misinterpreted. In fact, the impulse for the present study originated
from a prediction from Ref.~\cite{Deur:2009ya} (see also Refs.~\cite{Deur:2016bwq, Deur:2020wlg}). 
In this framework, if a homogenous system is locally dense enough so that in that location, the non-linearity
of General Relativity are non negligible, this system should display a correlation between its dynamical total mass
analyzed using Newton's law of gravity (in our case, the galaxy dark mass) and its asymmetry
(in our case, the galaxy ellipticity). Beside the present correlation, this effect also explains~\cite{Deur:2019kqi} the
correlation between dynamical and baryonic matter accelerations observed in Ref.~\cite{McGaugh}.
Finally, it provides an explanation for the origin of dark energy: it emerges from the Universe inhomogeneities and anisotropies~\cite{Deur:2017aas}.
%that are neglected by the Cosmological principle.
\item There is a significant bias in the data and/or methods, in which case
they cannot be trusted to estimate accurately the dark matter content of elliptical
galaxies. However, our thorough investigation of the various inter-dependences
of the variables characterizing an elliptical galaxy reasonably suggests
that this correlation is physical rather than a methodological, observational
or measurement bias.
\end{enumerate}
Finally,  a practical use of the correlation is that once the total galactic mass is known the true ellipticity
of the galaxy can be directly deduced.

~

\textbf{Acknowledgment}  This research has made use of the NASA/IPAC
Extragalactic Database (NED) which is operated by the Jet Propulsion
Laboratory, California Institute of Technology, under contract with
the National Aeronautics and Space Administration.

%%%%bibliography*******

\appendix

\section{Other data\label{sec:Other-data}}

\begin{figure}[h]
\centering
\includegraphics[scale=0.4]{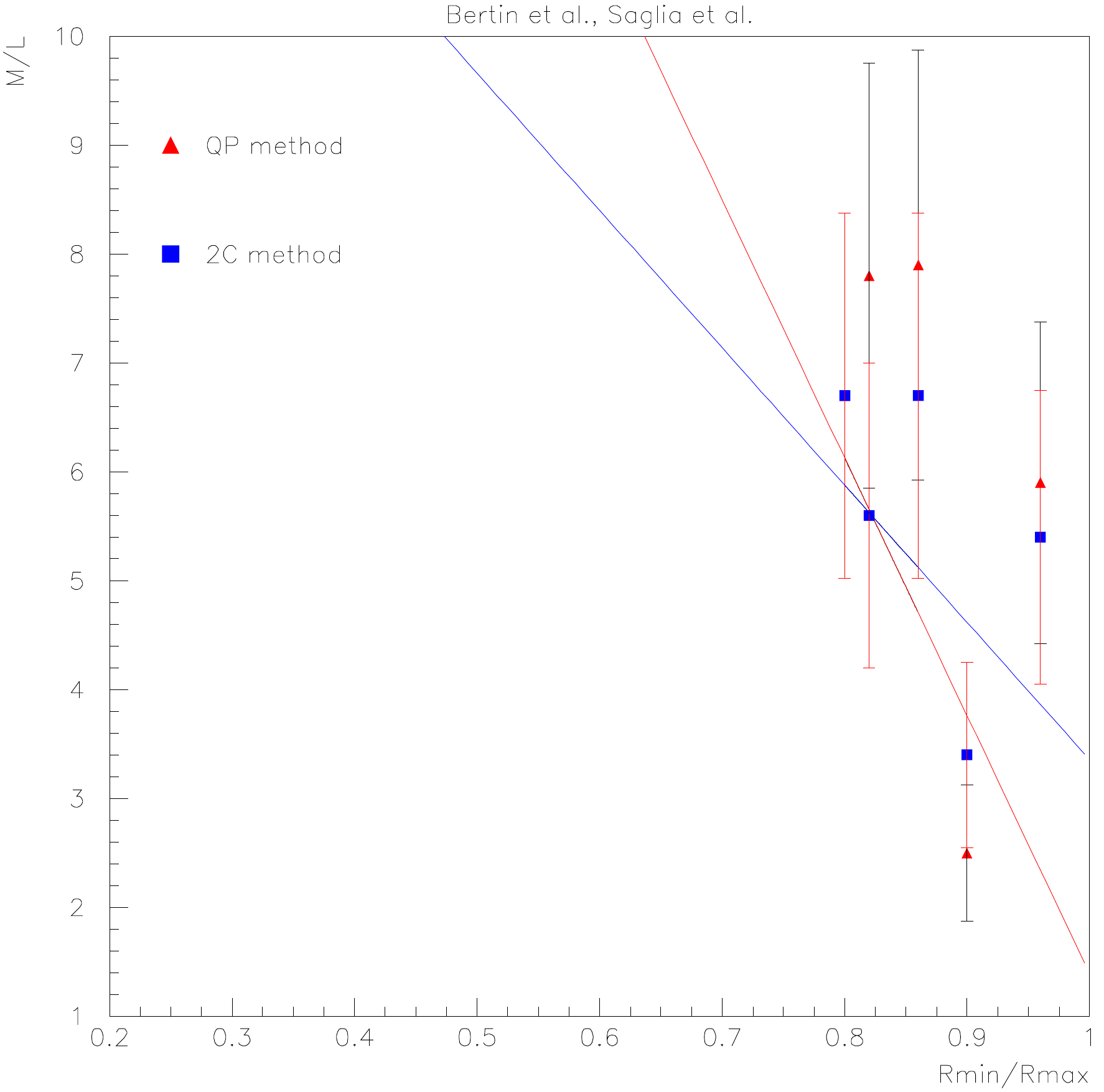}
\vspace{-0.4cm} \caption{\label{fig:Bertin}$\sfrac{M}{L}$ vs $\sfrac{R_{min}}{R_{max}}$ for the Bertin
{\it et al.} and Saglia {\it et al.} data sets. The different symbols correspond to the two
methods used to model stellar dynamics. The corresponding linear fits are
the lines of matching colors. 
}
\end{figure}

Other works studying the dark matter content for several elliptical  galaxies 
can be found in the literature. We list them here, in alphabetical order, 
for completeness and state why they were not used in the present analysis:
\begin{itemize}
\item \underline{Bertin {\it et al.} (1994)} and \underline{Saglia {\it et al.} (1993)}.
The two articles are from the same group and use the same method. We
may thus combine the two data sets without introducing a systematic
bias. The authors estimate $\sfrac{M}{L}$ using two models to fit stellar
dynamics (model 2C: 2 components method and model QP: quadratic programming
method) for 9 elliptical and lenticular galaxies. Applying our usual
selection criteria and relying on NED for the galaxy type and $\sfrac{R_{min}}{R_{max}}$,
only 2 suitable elliptical galaxies remain after rejecting E+/cD,
S0, SA0 galaxies and NGC7144 since no realistic modeling for this
galaxy was achieved. The two remaining galaxies are E0, with too little
lever arm to obtain a meaningful $\sfrac{M}{L}$ vs ellipticity relation. If
we keep in addition the 3 cD galaxies, the fits to the $\sfrac{M}{L}$ vs ellipticity
indicate a correlation, see Fig.~\ref{fig:Bertin}. The best fit result
for the 2C method gives $\sfrac{M}{L}=(-12.6\pm10.2)\sfrac{R_{min}}{R_{max}}+(16.0\pm9.0)$.
The best fit result for the QP method gives $\sfrac{M}{L}=(-23.7\pm12.1)\sfrac{R_{min}}{R_{max}}+(25.1\pm10.8)$.
The respective $\sfrac{\chi^{2}}{ndf}$ are 1.5 and 4.6. This data set has
a number of caveats: the data are dated (1994). The $\sfrac{M}{L}$ are not
estimated at the same relative radius (they are estimated at $R$ around
$R_{eff}$, with 0.8$R_{eff}<R<2.1R_{eff}$). No uncertainties are
provided. We assumed 25\% point-to-point uncertainty as this seems
to be typical for such study. This leads to 2C and QP results that
are compatible with each others, which indicate that 25\% is reasonable.
The $\sfrac{R_{min}}{R_{max}}$ span is small (0.8<$\sfrac{R_{min}}{R_{max}}$ <0.96),
making the slope extracted from the fits less accurate. Finally, the
statistics are limited (5 galaxies), the models assume spherical
symmetry -which introduces a $\sfrac{M}{L}$ vs $\sfrac{R_{min}}{R_{max}}$ bias- and they also assumme
no galaxy rotation. The main drawback is that we have cD galaxies
in the sample. It explains the strong correlation found: cD galaxies
tend to have higher $\sfrac{M}{L}$ than standard elliptical galaxies and,
for this particular sample, the cD galaxies have higher ellipticity.
For these reasons, we do not consider these results.
\end{itemize}

\begin{itemize}
\item \underline{Barnabe \it{et al.}} study 6 strong lensing early-type galaxies from the
SLACS survey in~\cite{Barnabe-superseeded}. This work is superseded 
by~\cite{Barnabe} and hence we do not use these results. However,
we summarize them briefly: out of the 6 galaxies, we reject one potential
giant that has $M_{Ein}>10^{12}M_{\odot}$ (J0216-0813) and one E/S0
(J0912+0029). The fit to the 4 remaining elliptical galaxies yields  $\sfrac{M_{tot}}{M_*}=(-1.12\pm0.85)\sfrac{R_{min}}{R_{max}}|_{true}+(1.14\pm0.62)$
and $\sfrac{M_{tot}}{M_*}=(0.11\pm0.52)\sfrac{R_{min}}{R_{max}}|_{mass}+(0.23\pm0.41)$.
\item \underline{Brighenti and Mathews}~\cite{Brighenti} determine the $\sfrac{M}{L}$ for
3 early-type galaxies. However, only one is an elliptical. The two
others are lenticular galaxies and belong to the Arp catalogue. 
\item \underline{Cappellari{\it et al.}}~\cite{Cappellari 2009} provide $\sfrac{M}{L}$
ratios for 9 early-type galaxies at large redshift $z\simeq2$. No detailed information
on the early-type structure is available. With our standard $\sigma\geq225$
km.s$^{-1}$ to minimize S0, we can retain only one galaxy. 
\item \underline{Carollo {\it et al. }}~\cite{Carollo95} study the dark matter
content of 4 elliptical galaxies but provide only the velocity dispersion profile,
without giving the total mass or $\sfrac{M}{L}$.
\item \underline{Carollo and Danziger} study 5 elliptical galaxies in~\cite{Carollo94}.
Among them, 3 are SAB, one a SA0 or SApec and one is a cD, leaving
only one suitable galaxy.
\item \underline{Coccato {\it et al.}}~\cite{Coccato} study PNe as tracers for
6 galaxies and add 10 others from the literature. However, only their rotation
curves are provided, without total mass or $\sfrac{M}{L}$. 
\item \underline{Conroy and van Dokkum}~\cite{Conroy} construct $\sfrac{M}{L}$ for 38
early type galaxies. However just a few galaxies pass our selection
criteria. Furthermore, only the stellar $\sfrac{M}{L}$ is provided.
\item \underline{Das {\it et al.}}~\cite{Das} extract mass distributions, circular
velocity curves and $DMf$ from the deprojected density
and temperature profiles of the hot gas surrounding 6 X-ray bright
elliptical galaxies. These extracted quantities are given from the
galaxy center to a few $R_{eff}$ (2.5 $R_{eff}$ for NGC4472 to 9 $R_{eff}$
for NGC 1407). The $DMf$ is extracted using the
stellar $\sfrac{M_*}{L}$ ratio from various references. However, no galaxy
passes our standard selection and we thus cannot use these data. 
\item \underline{Fassnacht and Cohen}~\cite{Fassnacht} analyze 3 strong lenses
from the CLASS survey (B0712+472, B1030+074 and B1600+434). However,
the lens velocity dispersions are below our standard $\sigma<225$
km.s$^{-1}$ criterion, two of them well below. Furthermore B1600+434
is significantly affected by its environment. Hence, the 3 galaxies
are not suitable for our study
\item \underline{Gebhardt {\it et al.}}~\cite{Gebhardt} provide $\sfrac{M}{L}$ for 12
early-type galaxies. $\sfrac{M}{L}$ is obtained for the central region ($r < \sfrac{1}{4}R_{eff}$)
where stars and central black holes dominate the gravitational potential.
Consequently, the model used to infer $\sfrac{M}{L}$ does not include a dark
matter halo and assume instead  that the total mass follows light,
i.e, $\sfrac{M}{L}$ is independent of the galaxy radius. This assumption,
ruled out by many recent studies, remains reasonable in the radius range
studied in~\cite{Gebhardt}. After removing the S0, LINERS, AGN,
galaxies for which $\sfrac{M}{L}$ correlates with the central black hole mass,
and Arp galaxies, we are left with only 2 adequate galaxies. Given
all these limitations, we did not analyze these data.
\item The data from  \underline{Grillo {\it et al.}}~\cite{Grillo08} are not used
for two reasons: 1) Grillo {\it et al.} model only stellar masses
and use the lensing masses from~\cite{Koopmans}. Since the stellar
masses from~\cite{Grillo08} and~\cite{Koopmans} agree
well, the $\sfrac{M_{tot}}{M_*}$ will be similar to those obtained in~\cite{Koopmans}.
2) The same group has done a more recent study (Grillo {\it et al.}
\cite{Grillo09}) in which they consider more galaxies and compute
both the lensing and stellar masses. This latter work supersedes the
earlier results~\cite{Grillo08}.
\item \underline{Holden{\it et al.}}~\cite{Holden} study 4 strong lenses. They
belong to a cluster and may be subject to interaction with their environment.
In addition two of the lenses are giant elliptical galaxies and a
third one has a velocity dispersion of 130 km.s$^{-1}$, below our
standard $\sigma<225$~km.s$^{-1}$ criterion. We are then left with
only one suitable galaxy.
\item The two studies from  \underline{Humphrey {\it et al.}}~\cite{Humphrey  2005},
\cite{Humphrey  2006} employ the same set of 7 early-type galaxies
with X-ray data from {\it Chandra}. After rejecting galaxies that
are peculiar, belonging to the Arp catalog, or being S0/Sy, LINER/Sy, 
3 galaxies remain, one of them a LINER. The sample
is then too small for a useful study.
\item In the work of  \underline{More  {\it et al.}}~\cite{More}, 5 strong lensing
candidates are analyzed. Out of the 5, one has a S0-like morphology,
one is a spiral lens candidate, one might not be a true lens and one
yielded inconsistent results, leaving only one potential elliptical
candidate. 
\item \underline{Napolitano {\it et al.}}~\cite{Napolitano} compile values of
$\sfrac{M}{L}$ for 21 early-type galaxies in order to study the $\sfrac{M}{L}$ dependence
with radius. At small  and large radii (typically $R_{in}\sim R_{eff}/2$ and several $R_{eff}$, 
respectively), $\sfrac{M}{L}$ is determined using PNe
or GC kinematics. $\sfrac{M}{L}$ is obtained either
by fitting modeled orbital distributions to the line of sight velocity
distribution or by fitting the velocity distribution profiles using
Jeans or similar models. After standard selection insuring undisturbed
galaxies we retain only 2 galaxies: NGC821 and 2434. This, with the
fact that the data is a compilation of the literature (implying inhomogeneities
in the $\sfrac{M}{L}$ extraction) already used in this article, makes us to
not use the results reported in~\cite{Napolitano}. 
\item \underline{Proctor {\it et al.}}~\cite{Proctor} analyze data for 5 galaxies.
However, only 2 are adequate elliptical galaxies (E0 and E1-2), with
an additional E6?. The authors provide rotation curves but no $\sfrac{M}{L}$
ratios.
\item \underline{O'Sullivan, Sanderson and Ponman}~\cite{O'sulivan} determine
$\sfrac{M}{L}$ for 3 early-type galaxies. However, only one is an elliptical galaxy,
the two others being E+ and S0 galaxies.
\item \underline{Saglia, Bertin and Stiavelli}~\cite{Saglia92} interpret with
their model the photometry and kinematic profiles of 10 galaxies to
extract $\sfrac{M}{L}$. However, after removing the NELG, cD, E+, peculiar,
Seyfert, LERG, NLRG, S0 galaxies and the ones belonging to the Arp
catalogue, we are left with only two galaxies, one of them a LINER.
This sample is too small for a meaningful study.
\item \underline{Saglia {\it et al.}}~\cite{Saglia93} investigate the presence
of dark matter for 3 galaxies. One of them is a RLG, BrClG and LINER
and another a E2/S0, Sy2 LINER galaxy appearing in the Arp catalogue.
We are then left with only one suitable galaxy.
\item \underline{di Serego Alighieri {\it et al.}}~\cite{di Serego} compute $\sfrac{M}{L_B}$
for 20 early-type galaxies from the K20 survey at large redshift $0.88<z<1.3$.
However, no details are given on the galaxies. We infered the axis ratios
visually and, to minimize S0 and giants contamination, applied our standard
$\sigma\geq225$~km.s$^{-1}$ and $M_{tot}>5\times10^{11}M_{\odot}$
criteria. Only 3 galaxies pass this selection, one of them for which we could not determine
the axis ratio. $\sfrac{M}{L_B}$ is computed using simply 
the dynamical mass $M=5\sigma^{2}R_{eff}/G$. Such estimate can be
easily done for many local galaxies with well established type and
characteristics and there is no value {\it in our context} to consider
an additional few galaxies of uncertain characteristics.
\item \underline{Trinchieri, Fabbiano and Kim}~\cite{Trinchieri} use X-ray data
to extract the temperature distribution of the hot gas in 5 early-type
galaxies. From it the obtain their $\sfrac{M}{L}$. However, after removing the cD, peculiar,
S0 galaxies and one belonging to the Arp catalogue, we are left with
only one suitable galaxy. 
\item \underline{van Dokkum and Stanford}~\cite{Van Dokkum} compute $\sfrac{M}{L}$ for
3 early-type galaxies at large redshift $z=1.27$. They belong to
the RDCS J0848+4453 cluster. However, the galaxies' characteristics of are unknown but may be giants, unsuitable
in our context. $\sfrac{M}{L_B}$ is simply computed as $M\propto\sigma^{2}R_{eff}$.
Such estimate can be easily done for many local galaxies with well
established type and characteristics. Considering an additional few
galaxies with uncertain characteristics would not add to our
study.
\end{itemize}

\newpage

\section{Correlation between stellar $\sfrac{M_*}{L}$ and axis ratio \label{sec:Correlation w/stellar M/L}}

Since no strong correlation between the stellar $\sfrac{M_*}{L}$
ratio and the axis ratio is expected, studying $\sfrac{M_*}{L}$ vs $\sfrac{R_{min}}{R_{max}}$ 
should provide a stringent test of our procedure. 
Among all the data sets considered in this article, 13 of them provide $\sfrac{M_*}{L}$.
Here, we describe these sets and the $\sfrac{M_*}{L}$ vs $\sfrac{R_{min}}{R_{max}}$ correlations derived from them.

\subsection{data sets\label{sub:data-sets for stellar M/L}}

For all the analyses below except that of Section~\ref{sub:Cappellari stellar 2013a},
the uncertainty were taken to be proportional to the $\sfrac{M_*}{L}$ values
and scaled according to the {\it unbiased estimate}. Since the
determination of $\sfrac{M_*}{L}$ depends on modeling the stellar populations
rather than being based on dynamical arguments as for $\sfrac{M}{L}$, we do not
apply ellipticity corrections to the data (this concerns only the 2
PNe/GC data sets, the other sets had no corrections in the first place).
We also do not correct for a possible radius dependence
of $\sfrac{M_*}{L}$ (see Section~\ref{sec:Global-results}) because the radius at which the
$\sfrac{M_*}{L}$ are extracted is usually unspecified (if so we assign a
value of $r\sim R_{eff}$, for plotting purpose only). 
Finally, since the object of the selection criteria is to insure the reliability of the 
 dynamical determination of $\sfrac{M}{L}$, these criteria become irrelevant to 
 for $\sfrac{M_*}{L}$. However, since the goal of the $\sfrac{M_*}{L}$ vs 
 $\sfrac{R_{min}}{R_{max}}$ analysis is to test our primary $\sfrac{M}{L}$ 
 analysis, we use the same analysis procedure, including applying the selection criteria.

\subsubsection{Capaccioli {\it et al.} }

\begin{figure}[h]
\centering
\includegraphics[scale=0.4]{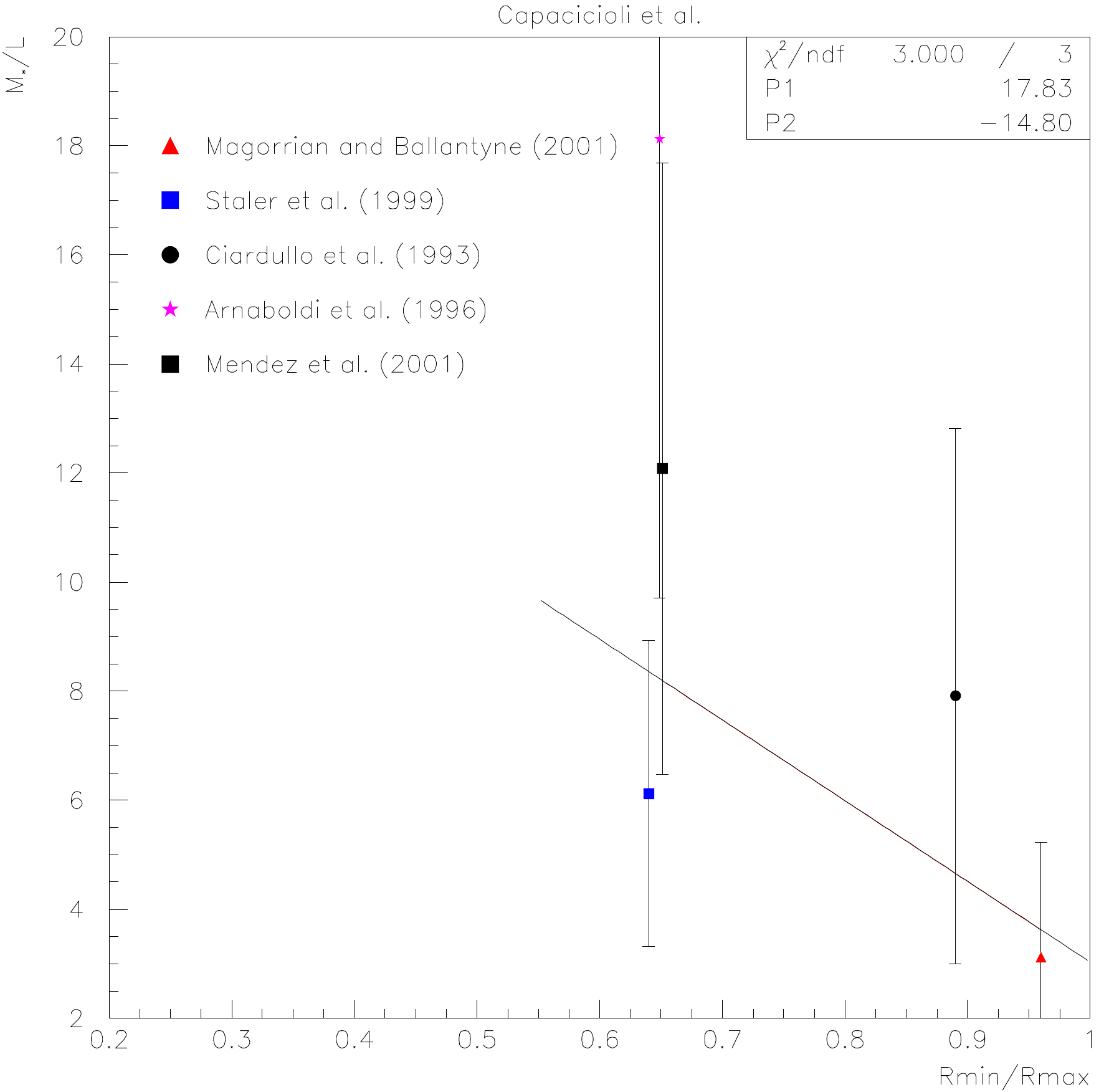}
\vspace{-0.4cm} \caption{\label{fig:capaccioli stellar M/L}Stellar $\sfrac{M_*}{L}$ vs the luminous axis ratio $\sfrac{R_{min}}{R_{max}}$
for the Capaccioli {\it et al.} data set~\cite{Capaccioli}.
}
\end{figure}

The data set is described in Section~\ref{sub:Capaccioli}. $\sfrac{M_*}{L}$
vs $\sfrac{R_{min}}{R_{max}}|_{light,true}$ is shown in Fig.~\ref{fig:capaccioli stellar M/L}.
The best fit yields:\\
$\sfrac{M_*}{L}=(-14.80\pm9.99)\sfrac{R_{min}}{R_{max}}|_{light,true}+(17.83\pm8.4)$ \\
As discussed in Section~\ref{sub:Capaccioli}, the data were assigned
to the reliability group 3. In addition, the $\sfrac{M_*}{L}$ is obtained
by assuming it to be equal to the calculated central $\sfrac{M}{L}$. Consequently,
we assign the result to group 4 reliability.

\subsubsection{Cappellari {\it et al.} SAURON project (2006)}

The data and the $\sfrac{M_*}{L}$ vs axis ratio analysis are described in
Section~\ref{sec:Cappellari06}. The fit for the stellar population yields\\ 
$\sfrac{M_*}{L}=(-0.02\pm1.72)\sfrac{R_{min}}{R_{max}}|_{true}+(2.39\pm1.16)$. \\
The analysis is assigned to group 1 reliability.

\subsubsection{Cappellari {\it et al.} ATLAS project (2013).\label{sub:Cappellari stellar 2013a}}

\begin{figure}
\centering
\includegraphics[scale=0.4]{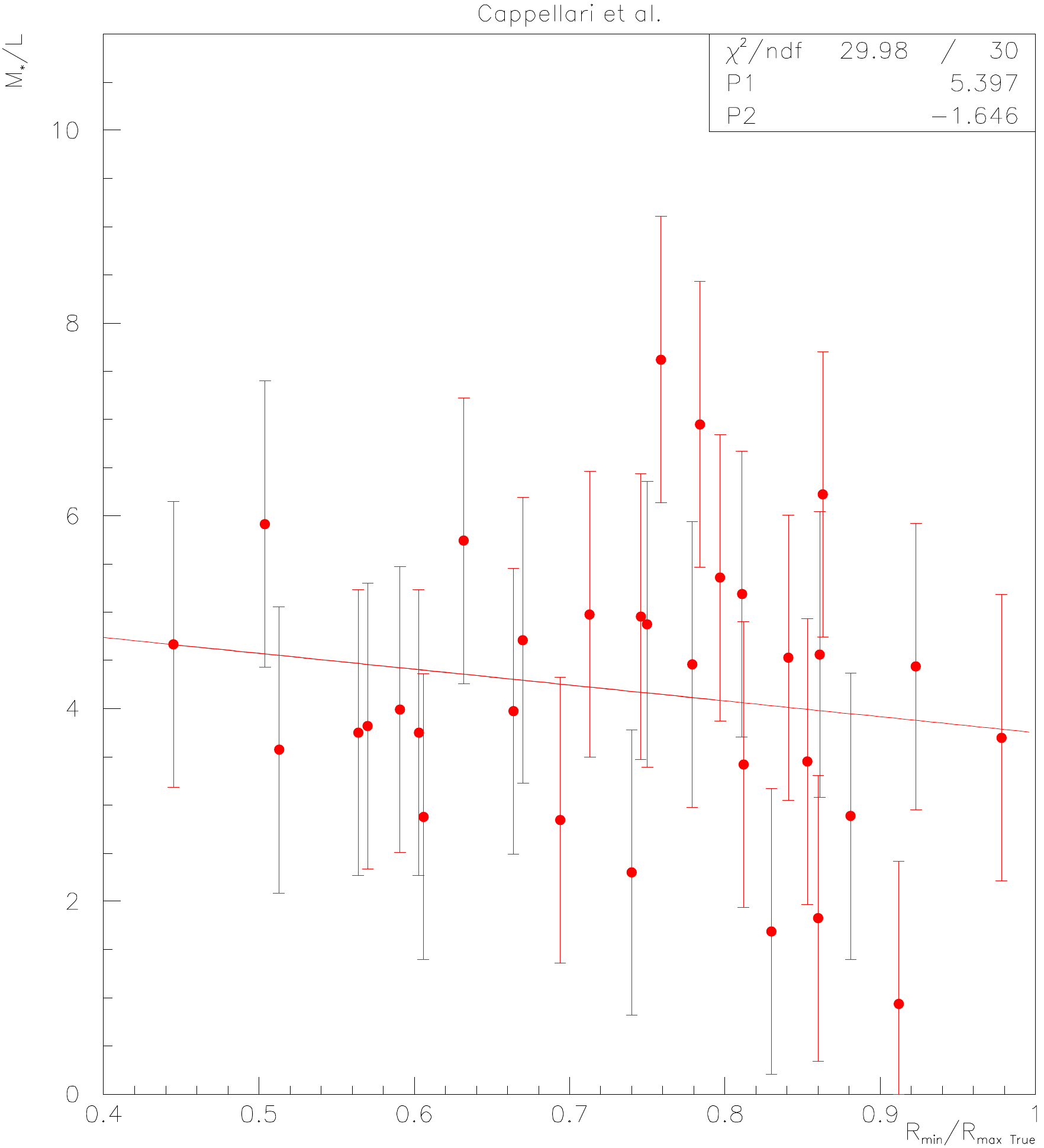}
\vspace{-0.4cm} \caption{\label{fig:cappellari 2013a stellar M/L}Stellar $\sfrac{M_*}{L}$ vs true axis ratio $\sfrac{R_{min}}{R_{max}}$
for the Cappellari {\it et al.} 2013 data set~\cite{Cappellari 2013a}. 
}
\end{figure}

The data are described in Section~\ref{sub:Cappellari-et-al. 2013}.
The $\sfrac{M_*}{L}$ data are from~\cite{Cappellari 2013a}. The best
fit for the stellar population yields\\ 
$\sfrac{M}{L}_*=(-7.17\pm1.79)\sfrac{R_{min}}{R_{max}}|_{true}+(8.30\pm1.45)$.\\
The large slope is mostly driven by one single point (NGC 4489), with
especially low $\sfrac{M_*}{L}$, which implies a low uncertainty $\Delta \sfrac{M_*}{L}$.
Removing NGC4489 would yield a much smaller slope: $\sfrac{M}{L}_*=(-3.21\pm2.21)\sfrac{R_{min}}{R_{max}}|_{true}+(5.71\pm1.69)$,
but we see no reason to remove it. To avoid the fit to be driven
by the low $\Delta \sfrac{M_*}{L}$ of NGC 4489, we  set the 
uncertainties to be constant rather than proportional to $\sfrac{M_*}{L}$.
The resulting fit gives:\\
$\sfrac{M}{L}_*=(-1.65\pm1.96)\sfrac{R_{min}}{R_{max}}|_{true}+(5.40\pm1.47)$,\\
see Fig.~\ref{fig:cappellari 2013a stellar M/L}.
The analysis is assigned to group 1 reliability.

\subsubsection{Thomas {\it et al.} (2007, 2011) and Wegner {\it et al.} (2012)}

The data are discussed in Section~\ref{sub:Thomas-et-al.}, including 
the stellar $\sfrac{M_*}{L}$ ratios. It was found that:\\
$\sfrac{M_*}{L_{Krou}}=(+2.35\pm1.75)\sfrac{R_{min}}{R_{max}}|_{apparent}+(2.01\pm1.37)$, \\
$\sfrac{M_*}{L_{Sal}}=(+2.87\pm3.82)\sfrac{R_{min}}{R_{max}}|_{apparent}+(3.69\pm2.92)$.\\
The Kroupa IMF results will be used for the global analysis. We assign
the result to group 1 reliability.

\subsubsection{Conroy and van Dokkum }

\begin{figure}[h]
\centering
\includegraphics[scale=0.4]{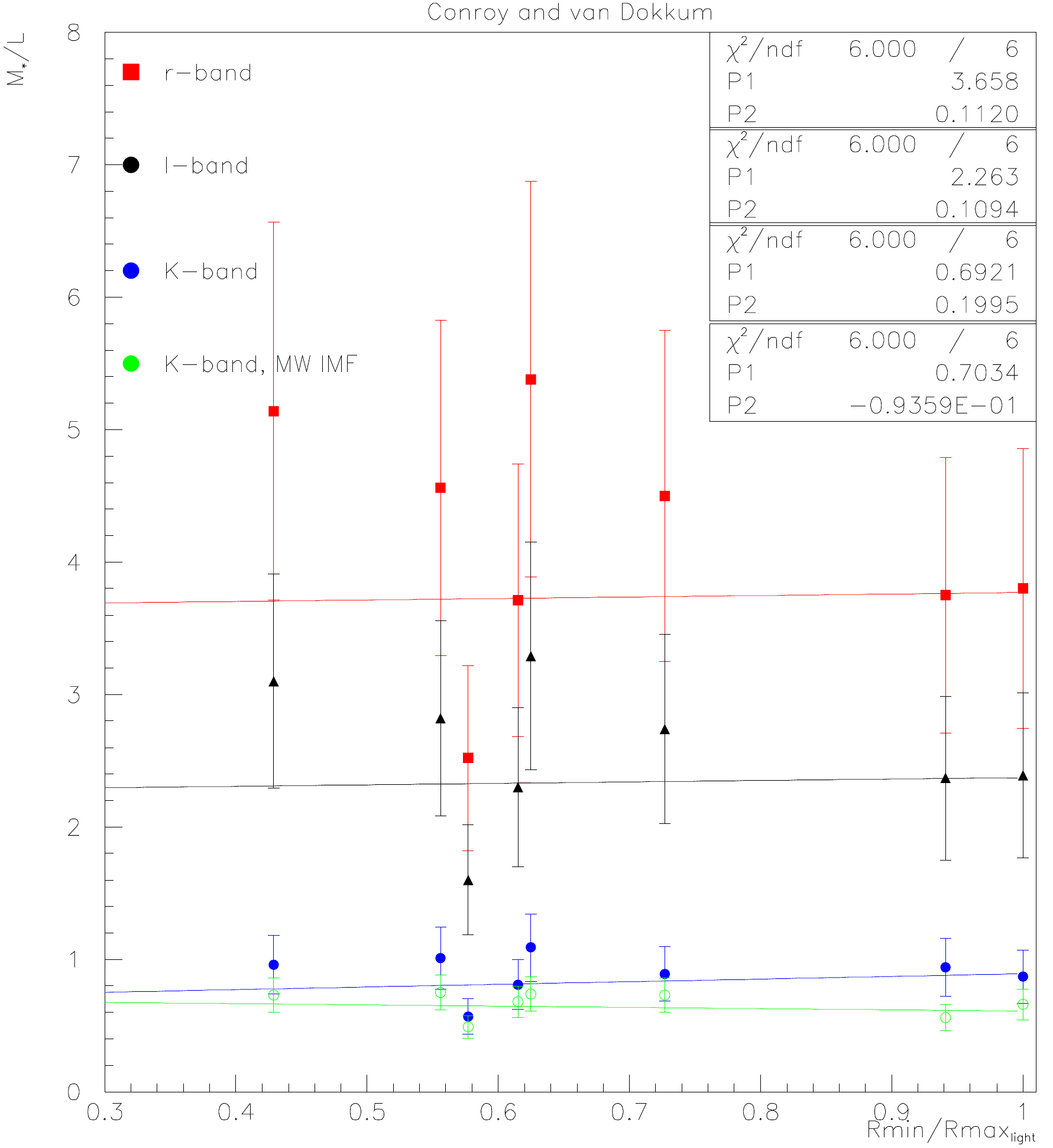}
\vspace{-0.4cm} \caption{\label{fig:conroy stellar M/L}Stellar $\sfrac{M_*}{L}$ vs axis ratio $\sfrac{R_{min}}{R_{max}}$ for
the Conroy and van Dokkum data set~\cite{Conroy}.
The different symbols correspond to different bands or IMF.
}
\end{figure}

These data, discussed in Section~\ref{sec:Other-data}, were
not used in the main analysis because only stellar $\sfrac{M_*}{L}$ are provided.
We show $\sfrac{M_*}{L}$ for different bands in Fig.~\ref{fig:conroy stellar M/L}.
The best fits for the different data are:\\
$\sfrac{M}{L}_*=(+0.11\pm2.12)\sfrac{R_{min}}{R_{max}}|_{true}+(3.66\pm1.50)$
for the r-band (with varying IMF). \\
$\sfrac{M}{L}_*=(+0.11\pm1.24)\sfrac{R_{min}}{R_{max}}|_{true}+(2.26\pm0.88)$
for the I-band (with varying IMF). \\
$\sfrac{M}{L}_*=(+0.20\pm0.40)\sfrac{R_{min}}{R_{max}}|_{true}+(0.69\pm0.28)$
for the K-band (with varying IMF). \\
$\sfrac{M}{L}_*=(-0.09\pm0.22)\sfrac{R_{min}}{R_{max}}|_{true}+(0.70\pm0.15)$
for the K-band (with a Milky Way (Kroupa 2001) IMF). \\
All the fits indicate  no ellipticity dependence. We will use
the I-band  results when combining with the other publications data.
The analysis is assigned to group 1 reliability.

\subsubsection{Deason {\it et al.} }

\begin{figure}
\centering
\includegraphics[scale=0.4]{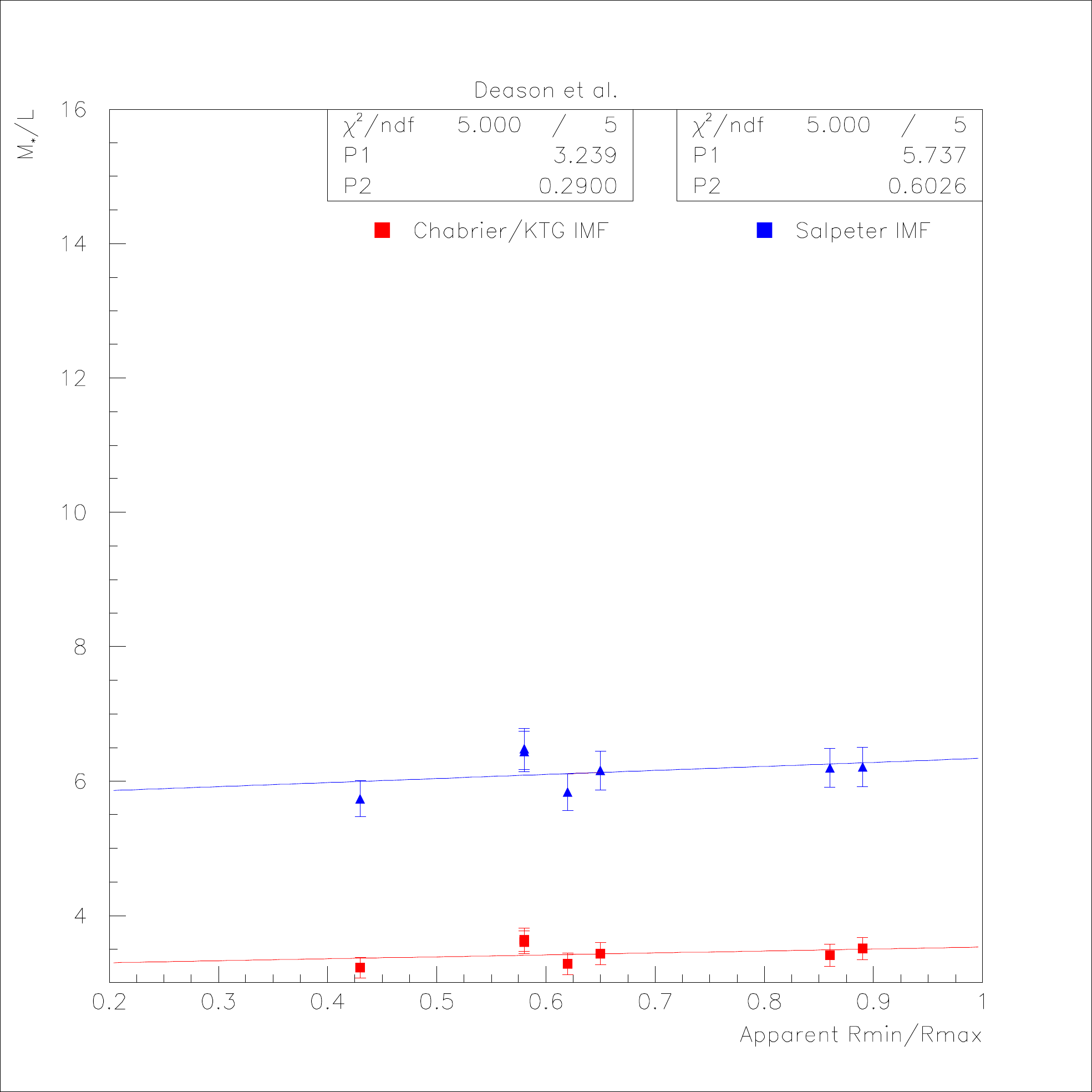}
\vspace{-0.4cm} \caption{\label{fig:Deason stellar M/L}Stellar $\sfrac{M_*}{L}$ vs apparent axis ratio $\sfrac{R_{min}}{R_{max}}$
for the Deason {\it et al.} data set~\cite{Deason}. 
}
\end{figure}

The data set is described in Section~\ref{sub:Deason}. We form $\sfrac{M_*}{L}=\sfrac{M}{L}\times(1-fdm)$.
its correlation with $\sfrac{R_{min}}{R_{max}}|_{light}$ is shown in Fig.~\ref{fig:Deason stellar M/L}. 
The best fit for the results
derived with a Chabrier IMF yields:\\
$\sfrac{M_*}{L}=(+0.29\pm0.41)\sfrac{R_{min}}{R_{max}}|_{apparent}+(3.24\pm0.28)$\\
\noindent and, for results derived with a Salpeter IMF:\\
$\sfrac{M_*}{L}=(+0.60\pm0.71)\sfrac{R_{min}}{R_{max}}|_{apparent}+(5.74\pm0.48)$.\\
When forming the global result, we will use Chabrier result and assign it to group 1 reliability.

\subsubsection{Auger {\it et al.} }

\begin{figure}
\centering
\includegraphics[scale=0.45]{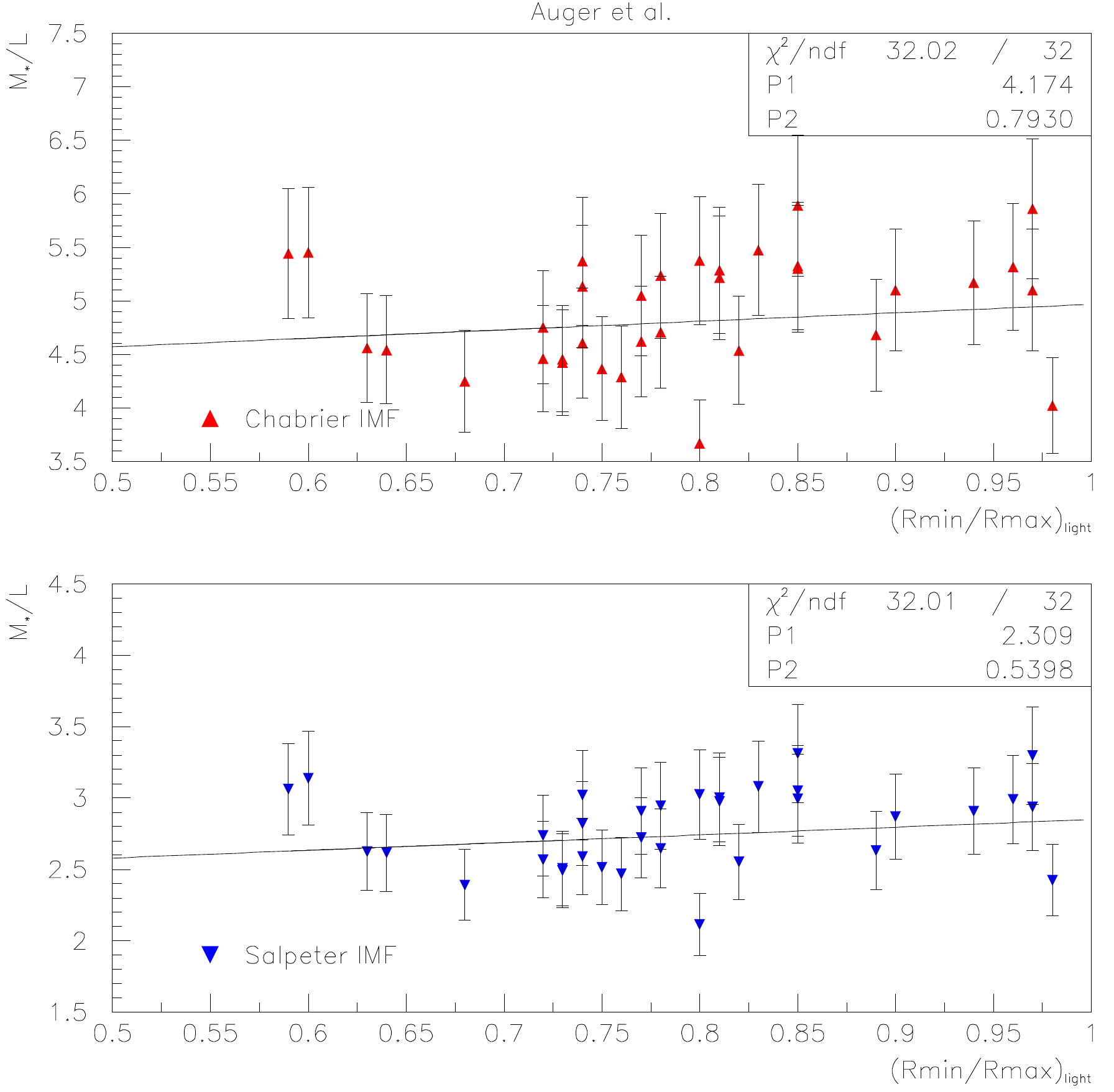}
\vspace{-0.4cm} \caption{\label{fig:Auger stellar M/L}Stellar $\sfrac{M_*}{L}$ vs apparent axis ratio $\sfrac{R_{min}}{R_{max}}$
for the Auger {\it et al.} data set~\cite{Auger 1}. The top
panel is for the mass SIE model using a Chabrier IMF, and the bottom panel
is for a Salpeter IMF. %The lines are the best linear fits to the data.
}
\end{figure}

The data are described in Section~\ref{sub:Auger-et-al.}. The stellar
mass $M_*$ is given in~\cite{Auger 1}. The luminosity $L$
is from~\cite{Bolton (SLACS)}. Forming $\sfrac{M_*}{L}$, we obtain
the correlation with $\sfrac{R_{min}}{R_{max}}|_{light}$ shown in Fig.~\ref{fig:Auger stellar M/L}.
The best fit yields, for the results derived with a SIE model using
a Chabrier IMF:\\
$\sfrac{M_*}{L}=(+0.79\pm0.93)\sfrac{R_{min}}{R_{max}}|_{apparent}+(4.17\pm0.73)$\\
\noindent and, for results derived with a Salpeter IMF:\\
$\sfrac{M_*}{L}=(+0.54\pm0.49)\sfrac{R_{min}}{R_{max}}|_{apparent}+(2.31\pm0.39)$.\\
We assign the Chabrier result, to be used in the global result, to group 1 reliability.

\subsubsection{Barnabe {\it et al.} }

\begin{figure}
\centering
\includegraphics[scale=0.4]{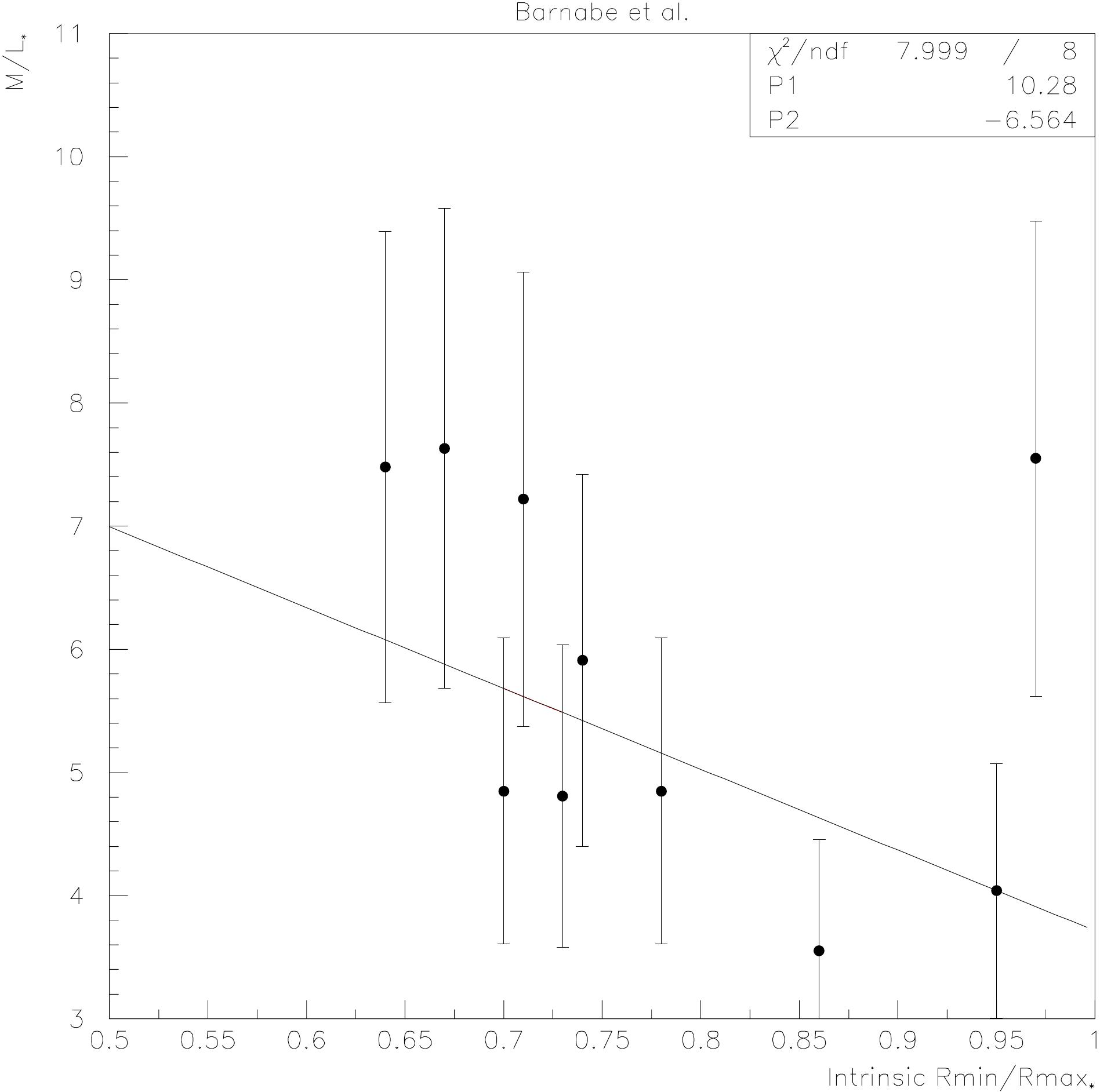}
\vspace{-0.4cm} \caption{\label{fig:Barnabe stellar M/L}Stellar $\sfrac{M_*}{L}$ vs the apparent axis ratio $\sfrac{R_{min}}{R_{max}}$
for the Barnabe {\it et al.} data set~\cite{Barnabe}.
}
\end{figure}

The data set is described in Section~\ref{sec:Barnabe-et-al.}.
$\sfrac{M_*}{L}$ vs $\sfrac{R_{min}}{R_{max}}|_{light,true}$ is shown in Fig.~\ref{fig:Barnabe stellar M/L}.
The best fit yields:\\
$\sfrac{M_*}{L}=(-6.56\pm4.08)\sfrac{R_{min}}{R_{max}}|_{light,true}+(10.28\pm3.30)$. \\
We assign the result to group 1 reliability.

\subsubsection{Cardone {\it et al.} (2009) }

\begin{figure}
\centering
\includegraphics[scale=0.4]{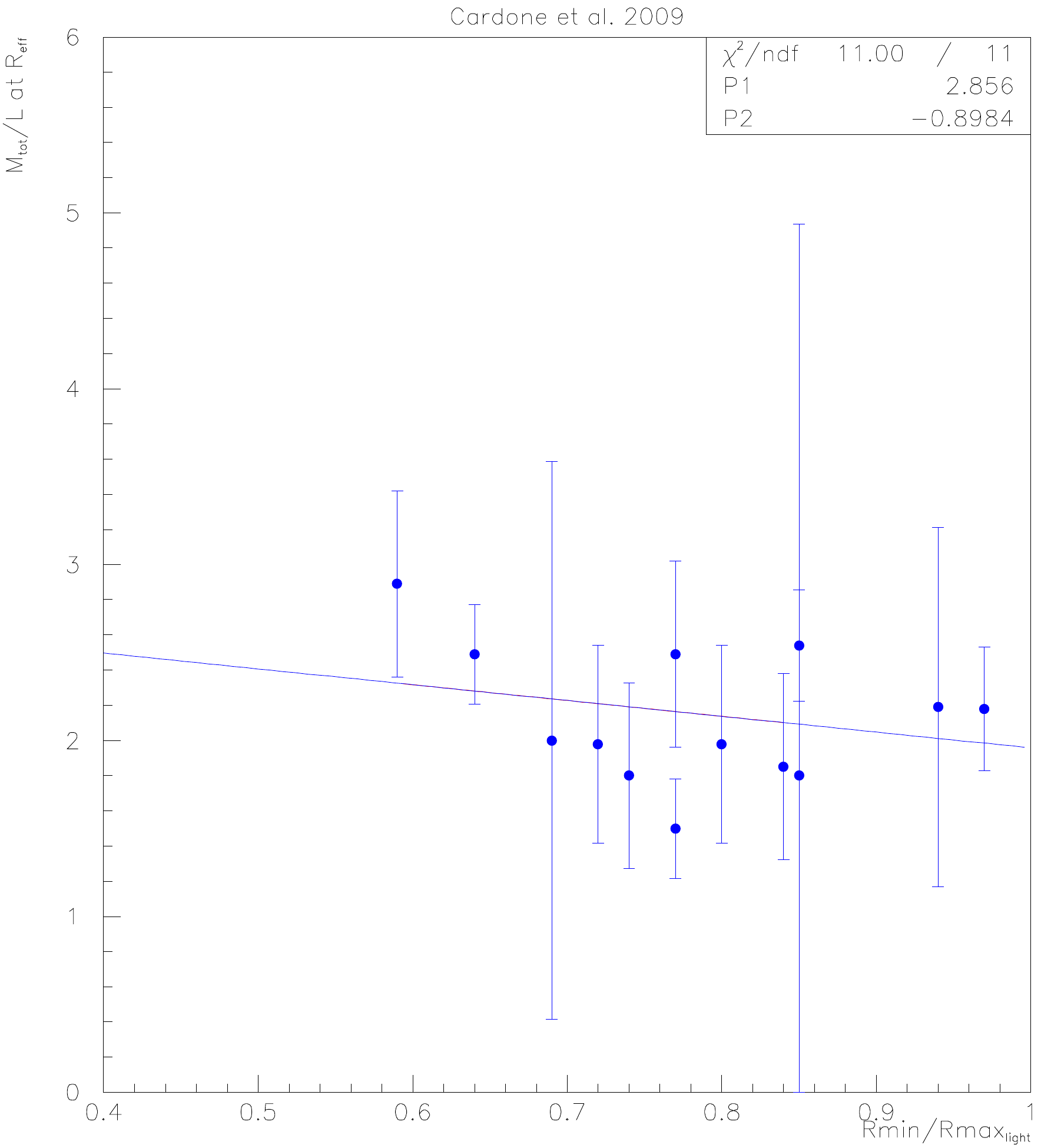}
\vspace{-0.4cm} \caption{\label{fig:Cardone09 stellar M/L}Stellar $\sfrac{M_*}{L}$ vs the apparent axis ratio $\sfrac{R_{min}}{R_{max}}$
for the Cardone {\it et al.} data set~\cite{Cardone09}.
}
\end{figure}

The Cardone {\it et al.} (2009) data are described in section
\ref{sub:Cardone-et-al.2009}. (We do not use the $\sfrac{M_*}{L}$ from
the Cardone {\it et al.} (2011) data discussed in Section~\ref{sub:Cardone-et-al.2009}
since they are from Auger {\it et al.}, which are already analyzed.)
The $\sfrac{M_*}{L}$ vs $\sfrac{R_{min}}{R_{max}}|_{light,true}$ are shown in Fig.
\ref{fig:Cardone09 stellar M/L}. The best fit yields:\\
$\sfrac{M_*}{L}=(-0.90\pm1.14)\sfrac{R_{min}}{R_{max}}+(2.85\pm0.89)$.\\
We assign the result to group 1 reliability.

\subsubsection{Grillo {\it et al.} }

\begin{figure}[h]
\centering
\includegraphics[scale=0.4]{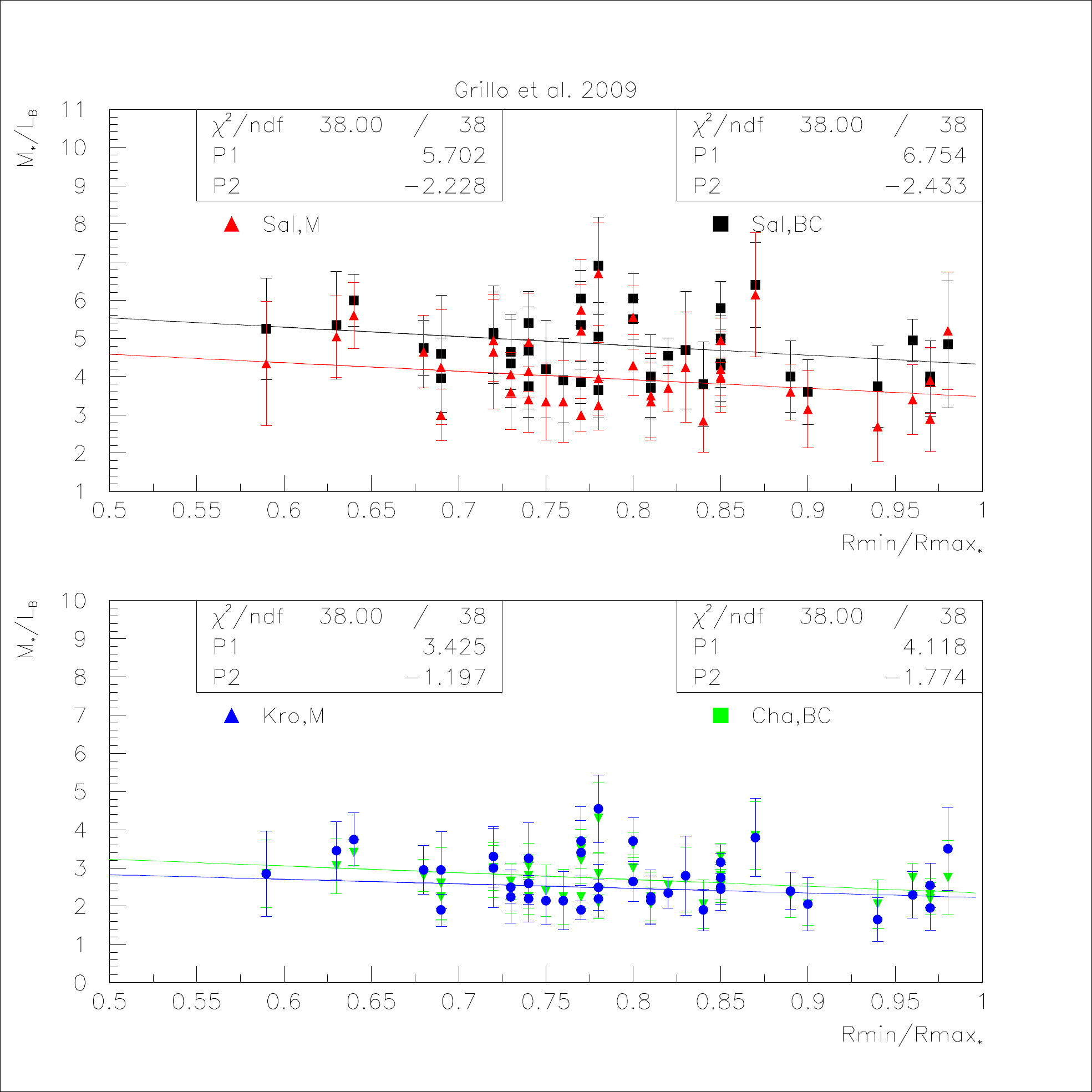}
\vspace{-0.4cm} \caption{\label{fig:Grillo stellar M/L}Stellar  $M_*/L$ vs the apparent
axis ratio $\sfrac{R_{min}}{R_{max}}$ for the Grillo {\it et al.}
data set~\cite{Grillo09}.
}
\end{figure}

The Grillo {\it et al.} data set is described in Section~\ref{sub:Grillo}.
The $\sfrac{M_*}{L}$ vs $\sfrac{R_{min}}{R_{max}}|_{light}$ are shown in Fig.~\ref{fig:Grillo stellar M/L}
for various IMF and SMT. The best fit yields:\\
$\sfrac{M_*}{L}=(-2.43\pm1.55)\sfrac{R_{min}}{R_{max}}|_{light}+(6.75\pm1.25)$
for results using a Salpeter IMF and a Bruzual-Charlot~SMT, \\
$\sfrac{M_*}{L}=(-2.22\pm1.63)\sfrac{R_{min}}{R_{max}}|_{light}+(5.70\pm1.31)$
for results using a Salpeter IMF and a Maraston SMT, \\
$\sfrac{M_*}{L}=(-1.77\pm0.86)\sfrac{R_{min}}{R_{max}}|_{light}+(4.12\pm0.69)$
for  results using a Chabrier IMF and a Bruzual-Charlot~SMT, \\
$\sfrac{M_*}{L}=(-1.20\pm1.13)\sfrac{R_{min}}{R_{max}}|_{light}+(3.42\pm0.90)$
for results using a Kroupa IMF and a Maraston SMT. \\
As in the main global analysis, we will used the results derived with
a Chabrier IMF and a Bruzual-Charlot SMT. We assign the result to
group 1 reliability.

\subsubsection{Jiang and Kochanek}

\begin{figure}[h]
\centering
\includegraphics[scale=0.45]{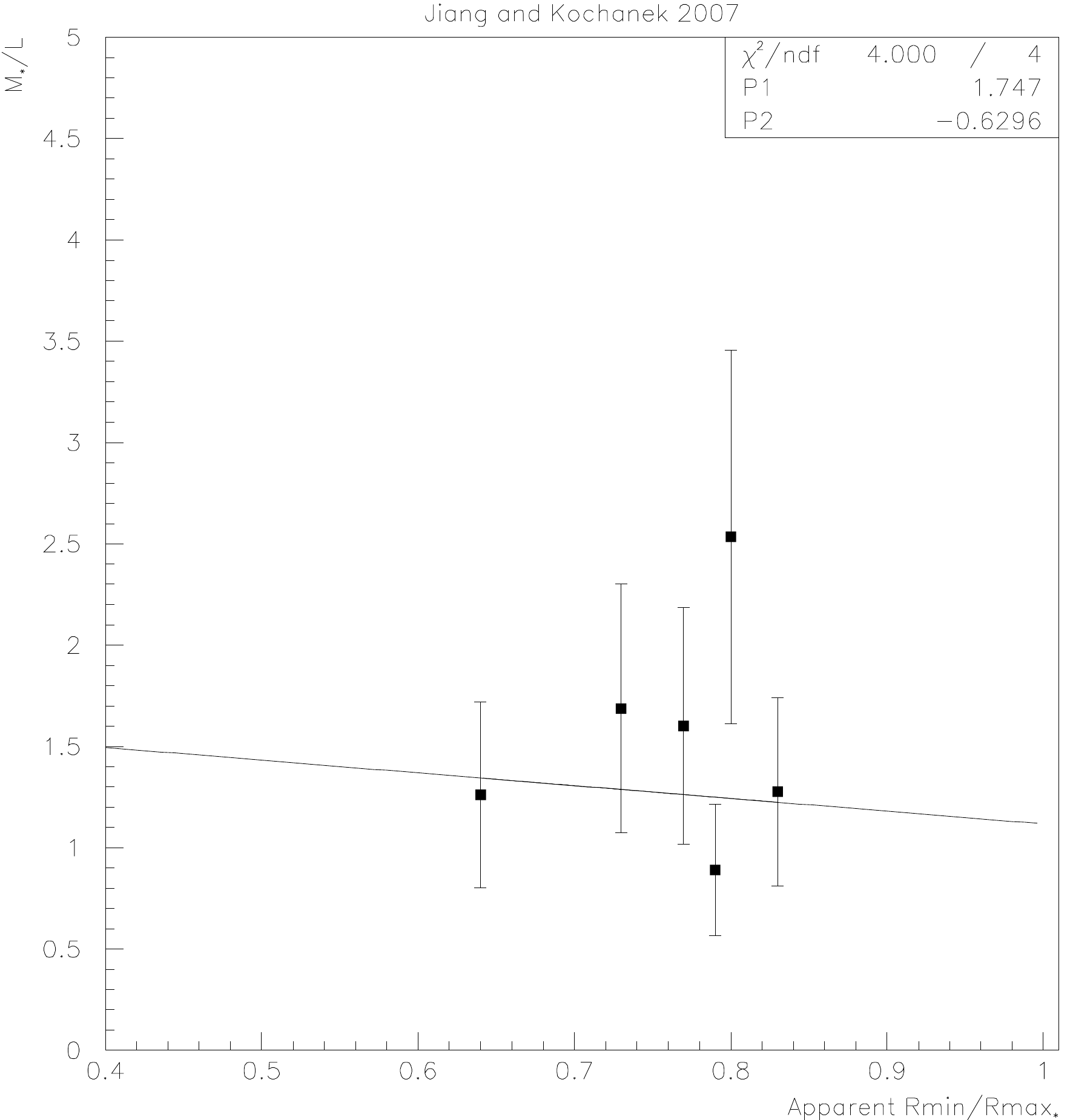}
\vspace{-0.4cm} \caption{\label{fig:Jiang and Kochanek stellar M/L}Stellar $\sfrac{M_*}{L}$ vs the apparent axis ratio $\sfrac{R_{min}}{R_{max}}$
for the Jiang and Kochanek data set~\cite{Jiang-Kochanek}.
}
\end{figure}

The Jiang and Kochanek data are described in Section~\ref{sub:Jiang-and-Kochanek}.
The $\sfrac{M_*}{L}$ vs $\sfrac{R_{min}}{R_{max}}|_{light}$ are shown in Fig.~\ref{fig:Jiang and Kochanek stellar M/L}.
The best fit yields:\\
$\sfrac{M_*}{L}=(-0.63\pm3.06)\sfrac{R_{min}}{R_{max}}+(1.74\pm2.34)$.\\ 
We assign the result to group 3 reliability.

\subsubsection{Leier {\it et al.} (2011)}

\begin{figure}[h]
\centering
\includegraphics[scale=0.4]{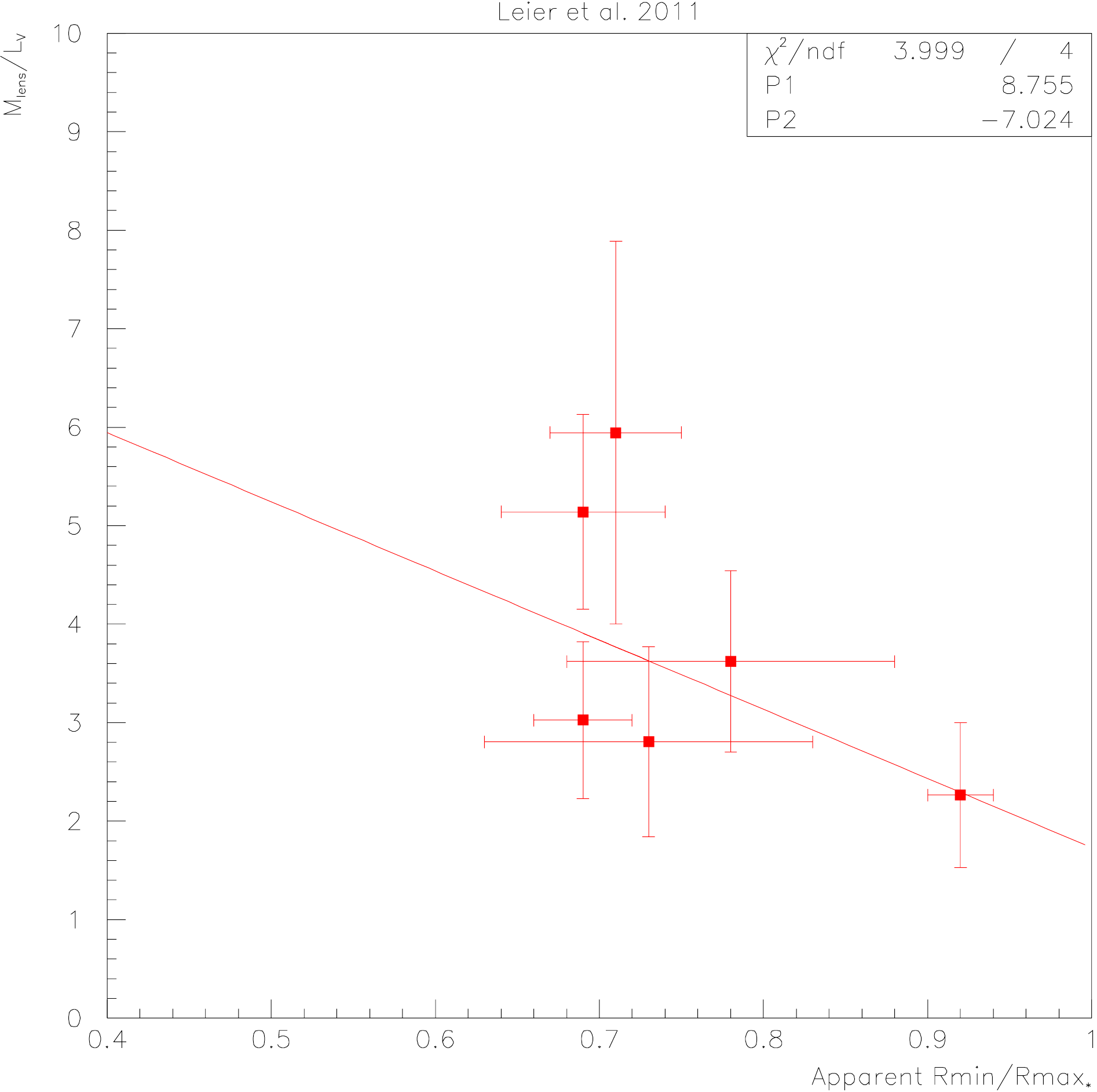}
\vspace{-0.4cm} \caption{\label{fig:Leier stellar M/L}Stellar $\sfrac{M_*}{L}$ vs the apparent axis ratio $\sfrac{R_{min}}{R_{max}}$
for the Leier {\it et al.} (2011) data set~\cite{Leier2011}.
}
\end{figure}

The Leier {\it et al.} data are described in Section~\ref{sub:Leier 2011}.
$\sfrac{M_*}{L}$ vs $\sfrac{R_{min}}{R_{max}}|_{light}$ is shown in Fig.~\ref{fig:Leier stellar M/L}.
The best fit~yields:\\
$\sfrac{M_*}{L}=(-7.02\pm4.37)\sfrac{R_{min}}{R_{max}}+(8.75\pm3.42)$. \\
We assign the result to group 2 reliability.

\subsubsection{Treu and Koopman}

The data are discussed in Section~\ref{sub:Treu-and-Koopmans}. It
was found that:\\
$\sfrac{M_*}{L}=(-3.87\pm1.44)\sfrac{R_{min}}{R_{max}}|_{apparent}+(5.45\pm1.19)$, and \\
$\sfrac{M_*}{L_{FP}}=(-5.64\pm1.27)\sfrac{R_{min}}{R_{max}}|_{apparent}+(6.63\pm1.05)$,\\
for which Fundamental Plan assumptions were made to further constrain
the data. 
The results without Fundamental Plan assumptions will be used for the
global analysis. We assign the result to group 1 reliability.

\subsection{Global results for $\sfrac{M_*}{L}$}

\begin{figure}[h]
\centering
\includegraphics[scale=0.4]{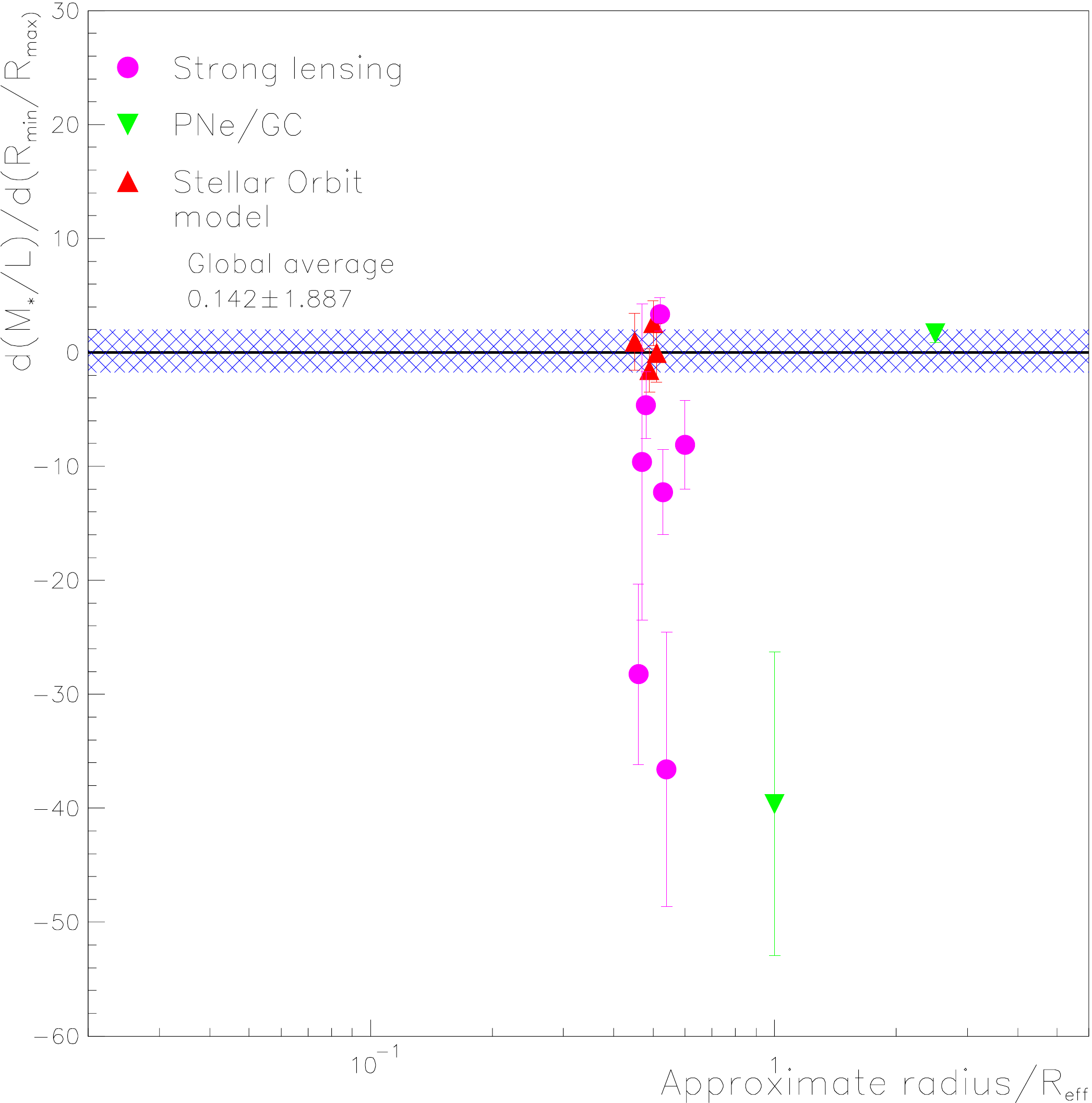}
\vspace{-0.4cm} \caption{\label{fig: global-1}
Slopes $\sfrac{d(\sfrac{M_*}{L})}{d(\sfrac{R_{min}}{R_{max}})}$ vs the approximate
radii (in $R_{eff}$ unit) at which the $\sfrac{M_*}{L}$ are obtained.
The various symbols indicate the method used to obtain $\sfrac{M_*}{L}$.
The band shows the average value of $\sfrac{d(\sfrac{M_*}{L})}{d(\sfrac{R_{min}}{R_{max}})}$
after accounting for the reliability of each $\sfrac{M_*}{L}$ extraction
and the shared statistics (the uncertainties shown do not account for
this. They are the values quoted in Section~\ref{sec:Correlation w/stellar M/L}).
The plain line indicates 0, for reference. 
}
\end{figure}

To combine the individual $\sfrac{M_*}{L}$ results, we follow the same procedure as for the main analysis (see Section~\ref{sec:Global-results}, we choose
to normalize the average $\sfrac{M_*}{L}$ to 4) and correct for projection
(see Section~\ref{sec:Projection-correction}, the same correction
factor 5$\pm1$, determined with the total $\sfrac{M}{L}$ from~\cite{BMS}).
The slopes $\sfrac{d(\sfrac{M_*}{L})}{d(\sfrac{R_{min}}{R_{max}})}$ for the
individual results  are shown in Fig.~\ref{fig: global-1}. 
For clarity, they are plotted vs the approximate average radius value
at which the $\sfrac{M_*}{L}$ are extracted. The average slope is
$$\sfrac{d(\sfrac{M_*}{L})}{d(\sfrac{R_{min}}{R_{max}})}=0.14\pm1.19, $$
\noindent a value well compatible with zero. We can compare it to
the results of the main analysis by normalizing the average mass over
light ratios to a same value. Then, the slope for $\sfrac{M_*}{L}$ is
50 times smaller than that for $\sfrac{M}{L}$. 

We note that excluding the PNe/GC data sets, in case 
not applying the ellipticity corrections is questioned  -a difference between the
$\sfrac{M_*}{L}$ and $\sfrac{M}{L}$ analyses-, we obtain $\sfrac{d(\sfrac{M_*}{L})}{d(\sfrac{R_{min}}{R_{max}})}=-3.47\pm3.19$,
which is still compatible with zero. Normalizing the average mass over
light ratios to a same value, then the slope for $\sfrac{M_*}{L}$ is
twice smaller than that for $\sfrac{M}{L}$, and they are compatible
within uncertainty. The inconclusive results, in the case of excluding
the PNe/GC data sets, underline the limitation of this test due to
its limited number of data sets and $\sfrac{M}{L}$ extraction methods.

%%%%%%%%%%%%%%%%%%%%%%%%%%%%%%%%%%%%%%%%%%
\newpage 

\section{Summary table of the NGC, IC and UGC galaxies\label{sec:Summary-table NGC}}
\vspace{-0.1in}Key: x: used. (x): used in lower reliability group. The
true $R_{t}=\sfrac{R_{min}}{R_{max}}$ is indicated when available. EXC and VCXG refer to
\cite{Fukazawa}, see Section~\ref{sec:Fukazawa}. XE has a similar
meaning as EXG and refers to~\cite{Nagino}. The symbol - means
the galaxy is available but does not pass the selection critteria. $NM$ refers to
an adequate elliptical galaxy, but unused because of mismatch between the NED numbers
and the ones listed in~\cite{BMS}.

\hspace{-1.in}{\tiny }
\begin{tabular}{| p{0.25in}| p{0.13in}| p{0.13in}| p{0.22in}| p{0.13in}| p{0.13in}| p{0.13in}| p{0.13in}| p{0.13in}| p{0.18in}| p{0.18in}| p{0.13in}| p{0.13in}| p{0.13in}| p{0.13in}| p{0.3in}| p{0.3in}| p{0.45in}| p{0.45in}| p{0.5in}|}
\hline 
{\tiny name} & {\tiny Bac} & {\tiny Ben} & {\tiny Bert} & {\tiny Capa}{\tiny \par}
 & {\tiny Dea} & {\tiny Fuk} & {\tiny Kro} & {\tiny Lau} & {\tiny Mag01} & {\tiny Mag98} & {\tiny Nag}{\tiny \par}
 & {\tiny Piz} & {\tiny Pru} & {\tiny Rom} & {\tiny SAUR} & {\tiny Thomas} & {\tiny v.d.~Mar~91} & {\tiny v.d.~Mar~07b} & {\tiny comment}\tabularnewline
\hline
\hline 
{\tiny 57} & {\tiny x} &  &  &  &  &  &  &  &  &  &  &  & {\tiny x} &  &  &  &  &  & \tabularnewline
\hline 
{\tiny 83} & {\tiny x} &  &  &  &  &  &  &  &  &  &  &  &  &  &  &  &  &  & \tabularnewline
\hline 
{\tiny 97} & {\tiny x} &  &  &  &  &  &  &  &  &  &  &  &  &  &  &  &  &  & \tabularnewline
\hline 
{\tiny 194} &  &  &  &  &  &  &  &  &  &  &  &  & {\tiny x} &  &  &  &  &  & \tabularnewline
\hline 
{\tiny 430} & {\tiny x} &  &  &  &  &  &  &  &  &  &  &  &  &  &  &  &  &  & \tabularnewline
\hline 
{\tiny 547} & {\tiny -} &  &  &  &  &  &  & - &  &  &  &  &  &  &  &  &  &  & {\tiny Arp}\tabularnewline
\hline 
{\tiny 564} &  &  &  &  &  &  &  &  &  &  &  &  & {\tiny x} &  &  &  &  &  & \tabularnewline
\hline 
\end{tabular}

{\tiny \par}
\hspace{-0.6in}{\tiny }\begin{tabular}{| p{0.25in}| p{0.13in}| p{0.13in}| p{0.22in}| p{0.13in}| p{0.13in}| p{0.13in}| p{0.13in}| p{0.13in}| p{0.13in}| p{0.13in}| p{0.13in}| p{0.13in}| p{0.13in}| p{0.13in}| p{0.3in}| p{0.3in}| p{0.2in}| p{0.13in}| p{0.65in}|}
\hline 
{\tiny name} & {\tiny Bac} & {\tiny Ben} & {\tiny Bert}{\tiny \par}
{\tiny ola} & {\tiny Capa} & {\tiny Dea} & {\tiny Fuk} & {\tiny Kro} & {\tiny Lau} & {\tiny Mag}{\tiny \par}
{\tiny 01} & {\tiny Mag}{\tiny \par}
{\tiny 98} & {\tiny Nag} & {\tiny Piz} & {\tiny Pru} & {\tiny Rom} & {\tiny SAUR-ON4} & {\tiny Thomas} & {\tiny v.d. Mar}{\tiny \par}
{\tiny 91} & {\tiny v.d. Mar}{\tiny \par}
{\tiny 07b} & {\tiny comment}\tabularnewline
\hline
{\tiny 2513} &  &  &  &  &  &  &  &  &  &  &  &  & {\tiny x} &  &  &  &  &  & \tabularnewline
\hline 
{\tiny 2663} &  &  &  &  &  &  &  &  &  &  &  &  & {\tiny x} &  &  &  &  &  & \tabularnewline
\hline 
{\tiny 2675} & {\tiny x} &  &  &  &  &  &  &  &  &  &  &  &  &  &  &  &  &  & \tabularnewline
\hline 
{\tiny 2693} & {\tiny x} & {\tiny x} &  &  &  &  &  &  &  &  &  &  & {\tiny x} &  &  &  &  &  & \tabularnewline
\hline
{\tiny 2694} &  &  &  &  &  &  &  &  &  &  &  &  & {\tiny x} &  &  &  &  &  & \tabularnewline
\hline 
{\tiny 2749} & {\tiny NM} &  &  &  &  &  &  &  &  &  &  &  & {\tiny (x)} &  &  &  & {\tiny -} &  & {\tiny LINER}\tabularnewline
\hline 
{\tiny 2778} & {\tiny x} &  &  &  &  &  &  &  &  & {\tiny x} &  &  & {\tiny x} &  &  &  &  & {\tiny x} & \tabularnewline
\hline
\end{tabular}{\tiny \par}

\hspace{-0.6in}{\tiny }\begin{tabular}{| p{0.25in}| p{0.13in}| p{0.13in}| p{0.22in}| p{0.13in}| p{0.13in}| p{0.13in}| p{0.13in}| p{0.13in}| p{0.13in}| p{0.13in}| p{0.13in}| p{0.13in}| p{0.13in}| p{0.13in}| p{0.3in}| p{0.3in}| p{0.2in}| p{0.13in}| p{0.65in}|}
\hline 
{\tiny name} & {\tiny Bac} & {\tiny Ben} & {\tiny Bert}{\tiny \par}
{\tiny ola} & {\tiny Capa} & {\tiny Dea} & {\tiny Fuk} & {\tiny Kro} & {\tiny Lau} & {\tiny Mag}{\tiny \par}
{\tiny 01} & {\tiny Mag}{\tiny \par}
{\tiny 98} & {\tiny Nag}{\tiny \par}
 & {\tiny Piz} & {\tiny Pru} & {\tiny Rom} & {\tiny SAUR-ON4} & {\tiny Thomas} & {\tiny v.d. Mar91} & {\tiny v.d. Mar}{\tiny \par}
{\tiny 07b} & {\tiny comment}\tabularnewline
\hline
\hline 
{\tiny 4213} & {\tiny x} &  &  &  &  &  &  &  &  &  &  &  &  &  &  &  &  &  & \tabularnewline
\hline 
{\tiny 4239} & {\tiny x} &  &  &  &  &  &  &  &  &  &  &  & {\tiny x} &  &  &  &  &  & \tabularnewline
\hline 
{\tiny 4278} & {\tiny -} &  & {\tiny $R _{t}=$}{\tiny \par}
{\tiny 0.71} &  &  &  & {\tiny x} &  &  &  &  &  &  &  &  &  &  &  & {\tiny Sy2, LINER}\tabularnewline
\hline 
{\tiny 4283} & {\tiny NM} &  &  &  &  &  &  &  &  &  &  &  & {\tiny x} &  &  &  &  &  & \tabularnewline
\hline 
{\tiny 4365} & {\tiny -} &  &  &  &  & {\tiny x} &  &  &  &  & {\tiny x} &  & {\tiny -} &  &  &  &  &  & {\tiny CXG, XC kinematically distinct core.}\tabularnewline
\hline 
{\tiny 4374} & {\tiny -} &  &  &  & {\tiny x} & {\tiny -} & {\tiny x} & {\tiny -} &  &  &  &  &  &  &  &  &  &  & {\tiny AGN,LERG,}{\tiny \par}
{\tiny LINER,Sy2,CXG}\tabularnewline
\hline 
{\tiny 4387} & {\tiny NM} & {\tiny x} &  &  &  &  &  & {\tiny -} &  &  &  &  & {\tiny x} &  &  &  &  &  & \tabularnewline
\hline
\end{tabular}{\tiny \par}

\hspace{-0.6in}{\tiny }\begin{tabular}{| p{0.25in}| p{0.13in}| p{0.13in}| p{0.22in}| p{0.13in}| p{0.13in}| p{0.13in}| p{0.13in}| p{0.13in}| p{0.13in}| p{0.13in}| p{0.13in}| p{0.13in}| p{0.13in}| p{0.13in}| p{0.3in}| p{0.3in}| p{0.2in}| p{0.13in}| p{0.65in}|}
\hline 
{\tiny name} & {\tiny Bac} & {\tiny Ben} & {\tiny Bert}{\tiny \par}
{\tiny ola} & {\tiny Capa} & {\tiny Dea} & {\tiny Fuk} & {\tiny Kro} & {\tiny Lau} & {\tiny Mag}{\tiny \par}
{\tiny 01} & {\tiny Mag}{\tiny \par}
{\tiny 98} & {\tiny Nag}{\tiny \par}
 & {\tiny Piz} & {\tiny Pru} & {\tiny Rom} & {\tiny SAUR-ON4} & {\tiny Thomas} & {\tiny v.d. Mar}{\tiny \par}
{\tiny 91)} & {\tiny v.d. Mar}{\tiny \par}
{\tiny 07b} & {\tiny comment}\tabularnewline
\hline
\hline 
{\tiny 4952} &  &  &  &  &  &  &  &  &  &  &  &  &  &  &  & {\tiny $R _{t}=$ 0.59} &  &  & \tabularnewline
\hline 
{\tiny 5018} &  &  &  &  &  & {\tiny x} &  &  &  &  &  &  & {\tiny x} &  &  &  &  &  & \tabularnewline
\hline 
{\tiny 5029} & {\tiny x} &  &  &  &  &  &  &  &  &  &  &  &  &  &  &  &  &  & \tabularnewline
\hline 
{\tiny 5044} &  &  &  &  &  & {\tiny -} &  &  &  &  & {\tiny -} &  & {\tiny (x)} &  &  &  &  &  & {\tiny EXG, XE, LINER}\tabularnewline
\hline 
{\tiny 5061} &  &  &  &  &  &  &  &  &  &  &  &  & {\tiny x} &  &  &  &  &  & \tabularnewline
\hline 
{\tiny 5077} &  &  & {\tiny $r_{t}=$}{\tiny \par}
{\tiny 0.50} &  &  &  &  &  &  &  &  & {\tiny $p_{0}=0.58$ $q_{0}=0.75$} &  &  &  &  &  &  & {\tiny LINER, Seyfert1.9}\tabularnewline
\hline 
{\tiny 5129} & {\tiny NM} &  &  &  &  &  &  &  &  &  &  &  & {\tiny x} &  &  &  &  &  & \tabularnewline
\hline 
\end{tabular}{\tiny \par}

\hspace{-0.6in}{\tiny }\begin{tabular}{| p{0.25in}| p{0.13in}| p{0.13in}| p{0.22in}| p{0.13in}| p{0.13in}| p{0.13in}| p{0.13in}| p{0.13in}| p{0.13in}| p{0.13in}| p{0.13in}| p{0.13in}| p{0.13in}| p{0.13in}| p{0.3in}| p{0.3in}| p{0.2in}| p{0.13in}| p{0.65in}|}
\hline 
{\tiny name} & {\tiny Bac} & {\tiny Ben} & {\tiny Bert}{\tiny \par}
{\tiny ola} & {\tiny Capa} & {\tiny Dea} & {\tiny Fuk} & {\tiny Kro} & {\tiny Lau} & {\tiny Mag}{\tiny \par}
{\tiny 01} & {\tiny Mag}{\tiny \par}
{\tiny 98} & {\tiny Nag} & {\tiny Piz} & {\tiny Pru} & {\tiny Rom} & {\tiny SAUR-ON4} & {\tiny Thomas} & {\tiny v.d. Mar91} & {\tiny v.d. Mar}{\tiny \par}
{\tiny 07b} & {\tiny comment}\tabularnewline
\hline
\hline
{\tiny 7391} & {\tiny NM} &  &  &  &  &  &  &  &  &  &  &  & {\tiny x} &  &  &  &  &  & \tabularnewline
\hline 
{\tiny 7454} & {\tiny x} &  &  &  &  &  &  & {\tiny x} &  &  &  &  & {\tiny x} &  &  &  &  &  & \tabularnewline
\hline 
{\tiny 7458} & {\tiny x} &  &  &  &  &  &  &  &  &  &  &  & {\tiny x} &  &  &  &  &  & \tabularnewline
\hline 
{\tiny 7507} & {\tiny x} &  &  &  &  &  & {\tiny x} &  & {\tiny x} &  &  &  & {\tiny x} &  &  &  & {\tiny x} & {\tiny x} & \tabularnewline
\hline 
{\tiny 7618} &  &  &  &  &  & {\tiny -} &  &  &  &  &  &  &  &  &  &  &  &  & {\tiny EXG}\tabularnewline
\hline 
{\tiny 7619} & {\tiny x} & {\tiny x} &  &  &  & {\tiny x} &  & {\tiny x} &  &  &  &  & {\tiny x} &  &  &  & {\tiny x} & {\tiny x} & {\tiny CXG}\tabularnewline
\hline 
{\tiny 7660} & {\tiny x} &  &  &  &  &  &  &  &  &  &  &  & {\tiny x} &  &  &  &  &  & \tabularnewline
\hline
\end{tabular}{\tiny \par}

\newpage

\section{Summary table for the SLACS lensing galaxies}
x=used, (x)=used but in lower reliability group, -=available but not
used (if for reasons other than standard selection, we state the reasons.
No symbol=not available. 
Masses -from~\cite{Grillo09}- indicating that the galaxy maybe a
giant are highlighted in red). Values are
given without uncertainties because only the central values  are used for
selection. The integration range for $M_{tot}$ varies from authors
to authors. 

\hspace{-0.85in}{\tiny }\begin{tabular}{| p{0.48in}| p{0.3in}| p{0.2in}| p{0.2in}| p{0.25in}| p{0.3in}| p{0.3in}| p{0.2in}| p{0.2in}| p{0.2in}|c| p{0.8in}| p{0.8in}| p{0.5in}|}
\hline 
{\tiny name} & {\tiny Ferr1} & {\tiny Aug} & {\tiny Bar} & {\tiny Card09} & {\tiny Card11} & {\tiny J\&K} & {\tiny Koop} & {\tiny Lei09} & {\tiny Gril} & {\tiny Rotator} & {\tiny $M_{tot}$ ($10^{10}M_{\odot}$)} & {\tiny $\sigma$(km/s)} & {\tiny type/}{\tiny \par}
{\tiny comment}\tabularnewline
\hline 
{\tiny J0008-0004} &  &  &  &  & {\tiny -} &  &  &  & {\tiny -} &  & {\tiny 912\cite{Cardone11}, }\textcolor{red}{\tiny $\sim105$}{\tiny~\cite{Grillo09}} & {\tiny 193\cite{Auger 1}} & {\tiny E\cite{Auger 2}}\tabularnewline
\hline 
{\tiny J0029-0055} &  & {\tiny -}{\tiny \par}
{\tiny fit out-}{\tiny \par}
{\tiny lier} &  & {\tiny x} & {\tiny x} &  &  &  & {\tiny x} &  & {\tiny 1622\cite{Cardone11},$\sim36$\cite{Grillo09}} & {\tiny 231\cite{Auger 1},229\cite{Cardone09}} & {\tiny E\cite{Auger 2}}\tabularnewline
\hline 
{\tiny J0037-0942} & {\tiny -$R_{M=1.7}$} & {\tiny x} & {\tiny x} &  & {\tiny x} & {\tiny x} & {\tiny x} & {\tiny x} & {\tiny x} & {\tiny S\cite{Barnabe}} & {\tiny 52.3~\cite{Ferreras},4677\cite{Cardone11}, 27.3\cite{Jiang-Kochanek},}{\tiny \par}
{\tiny 40\cite{Leier09}$\sim69$\cite{Grillo09}} & {\tiny 265\cite{Ferreras},282\cite{Auger 1},265\cite{Jiang-Kochanek}}{\tiny \par}
{\tiny ,230\cite{Leier09}} & {\tiny E\cite{Auger 2}}\tabularnewline
\hline 
{\tiny J0044+0113} &  & {\tiny x} &  &  & {\tiny x} &  &  &  & {\tiny x} &  & {\tiny 16596\cite{Cardone11},$\sim27$\cite{Grillo09}} & {\tiny 267\cite{Auger 1}} & {\tiny E\cite{Auger 2}}\tabularnewline
\hline 
{\tiny J0109+1500} &  &  &  &  &  &  &  &  & {\tiny x} &  & {\tiny $\sim39$\cite{Grillo09}} & {\tiny 251\cite{Auger 1}} & {\tiny E\cite{Auger 2}}\tabularnewline
\hline 
\end{tabular}{\tiny \par}

\hspace{-0.85in}{\tiny }\begin{tabular}{| p{0.48in}| p{0.3in}| p{0.2in}| p{0.2in}| p{0.25in}| p{0.3in}| p{0.3in}| p{0.2in}| p{0.2in}| p{0.2in}|c| p{0.8in}| p{0.8in}| p{0.5in}|}
\hline 
{\tiny name} & {\tiny Ferr.1} & {\tiny Aug.} & {\tiny Bar.} & {\tiny Card09} & {\tiny Card11} & {\tiny J\&K} & {\tiny Koop.} & {\tiny Leier09} & {\tiny Grillo} & {\tiny Rotator} & {\tiny $M_{tot}$ ($10^{10}M_{\odot}$)} & {\tiny $\sigma$(km/s)} & {\tiny type/}{\tiny \par}
{\tiny comment}\tabularnewline
\hline 
{\tiny J1020+1122} &  & {\tiny -} &  &  & {\tiny -} &  &  &  & {\tiny -} &  & {\tiny 1318\cite{Cardone11},}\textcolor{red}{\tiny $\sim102$}{\tiny~\cite{Grillo09}} & {\tiny 290\cite{Auger 1}} & {\tiny E\cite{Auger 2}}\tabularnewline
\hline 
{\tiny J1023+4230} &  & {\tiny x} &  & {\tiny x} & {\tiny x} &  &  &  & {\tiny x} &  & {\tiny 2951\cite{Cardone11},$\sim69$\cite{Grillo09}} & {\tiny 247\cite{Auger 1},242\cite{Cardone09}} & {\tiny E\cite{Auger 2}}\tabularnewline
\hline 
{\tiny J1029+0420} &  & {\tiny -} &  &  &  &  &  &  & {\tiny -} &  &  & {\tiny 215\cite{Auger 1}} & {\tiny S0\cite{Auger 2}}\tabularnewline
\hline 
{\tiny J1100+5329} &  &  &  &  & {\tiny -} &  &  &  & {\tiny -} &  & {\tiny 851\cite{Cardone11}}\textcolor{red}{\tiny ,$\sim141$}{\tiny~\cite{Grillo09}} & {\tiny 187\cite{Auger 2}} & {\tiny E\cite{Auger 2}}\tabularnewline
\hline 
{\tiny J1103+5322} &  &  &  & {\tiny -} &  &  &  &  &  &  &  & {\tiny 196\cite{Auger 1},196\cite{Cardone09}} & {\tiny S0/SA\cite{Auger 2}}\tabularnewline
\hline 
{\tiny J1106+5228} &  & {\tiny x} &  &  & {\tiny x} &  &  &  & {\tiny x} &  & {\tiny $\sim27$\cite{Grillo09}} & {\tiny 266\cite{Auger 1}} & {\tiny E\cite{Auger 2}}\tabularnewline
\hline 
{\tiny J1112+0826} &  & {\tiny -} &  &  & {\tiny -} &  &  &  & {\tiny -} &  & {\tiny 10000\cite{Cardone11},}\textcolor{red}{\tiny $\sim135$}{\tiny~\cite{Grillo09}} & {\tiny 328\cite{Auger 1}} & {\tiny E\cite{Auger 2}}\tabularnewline
\hline 
\end{tabular}{\tiny \par}

\hspace{-0.85in}{\tiny }\begin{tabular}{| p{0.48in}| p{0.3in}| p{0.2in}| p{0.2in}| p{0.25in}| p{0.3in}| p{0.3in}| p{0.2in}| p{0.2in}| p{0.2in}|c| p{0.8in}| p{0.8in}| p{0.5in}|}
\hline 
{\tiny name} & {\tiny Ferr1} & {\tiny Aug} & {\tiny Bar} & {\tiny Card09} & {\tiny Card11} & {\tiny J\&K} & {\tiny Koop} & {\tiny Lei09} & {\tiny Gril} & {\tiny Rotator} & {\tiny $M_{tot}$ ($10^{10}M_{\odot}$)} & {\tiny $\sigma$(km/s)} & {\tiny type/}{\tiny \par}
{\tiny comment}\tabularnewline
\hline 
{\tiny J1636+4707} & {\tiny -$R_{M=2.0}$} & {\tiny x} &  &  & {\tiny x} &  & {\tiny -} &  & {\tiny x} &  & {\tiny 52.9\cite{Ferreras},490\cite{Cardone11}, 31\cite{Leier09},$\sim54$\cite{Grillo09}} & {\tiny 221\cite{Ferreras},237\cite{Auger 1},221\cite{Leier09}} & {\tiny E\cite{Auger 2}}\tabularnewline
\hline 
{\tiny J1644+2625} &  &  &  &  & {\tiny -} &  &  &  &  &  & {\tiny 3236\cite{Cardone11}} & {\tiny 229\cite{Auger 2}} & {\tiny E\cite{Auger 2}}{\tiny \par}
{\tiny Mass axis ratio not avail.}\tabularnewline
\hline 
{\tiny J1719+2939} &  & {\tiny -} &  &  & {\tiny -} &  &  &  &  &  & {\tiny 11220\cite{Cardone11}} & {\tiny 295\cite{Auger 1}} & {\tiny S0\cite{Auger 2}}\tabularnewline
\hline 
{\tiny J2238-0754} &  & {\tiny x} & {\tiny x} &  & {\tiny x} &  &  &  & {\tiny x} & {\tiny F\cite{Barnabe}} & {\tiny 871\cite{Cardone11},$\sim39$\cite{Grillo09}} & {\tiny 200\cite{Auger 1}} & {\tiny E\cite{Auger 2}}\tabularnewline
\hline 
{\tiny J2300+0022} & {\tiny x} & {\tiny x} & {\tiny x} & {\tiny x} & {\tiny x} & {\tiny x} & {\tiny x} & {\tiny x} & {\tiny x} & {\tiny S\cite{Barnabe}} & {\tiny 40.9\cite{Ferreras},7079,30.4\cite{Jiang-Kochanek}}{\tiny \par}
{\tiny ,39\cite{Leier09},$\sim90$\cite{Grillo09}} & {\tiny 283\cite{Ferreras},284\cite{Auger 1},279\cite{Cardone09}}{\tiny \par}
{\tiny ,283\cite{Jiang-Kochanek},283\cite{Leier09}} & {\tiny E\cite{Auger 2}}\tabularnewline
\hline
\end{tabular}{\tiny \par}

\newpage

\section{Summary table for the COSMOS lensing galaxies}

x=used, -=available but not used. 

\begin{tabular}{|c| p{0.6in}|c|c|}
\hline 
name & Faure & $M_{tot}$ ($10^{10}M_{\odot}$) & $\sigma$(km/s)\tabularnewline
\hline
\hline 
COSMOS 0012+2015 & -
No $DMf$ &  & 215\cite{Faure 2008}\tabularnewline
\hline 
COSMOS 0013+2249 & -
No $DMf$ &  & \tabularnewline
\hline 
COSMOS 0018+3845 & x & 62\cite{Faure 2011} & 289\cite{Faure 2011},303\cite{Faure 2008}\tabularnewline
\hline 
COSMOS 0038+4133 & -
$DMf$<0{\footnotesize ,}{\footnotesize \par}
{\footnotesize low $\sigma$} & 14\cite{Faure 2011} & 207\cite{Faure 2011},225\cite{Faure 2008}\tabularnewline
\hline 
COSMOS 0047-5023 & -
High $\sfrac{\chi^{2}}{ndf}$ & 145\cite{Faure 2011} & 383\cite{Faure 2011},313\cite{Faure 2008}\tabularnewline
\hline 
 &  &  & \tabularnewline
\hline 
COSMOS 049+5128 & x & 73\cite{Faure 2011} & 313\cite{Faure 2011},380\cite{Faure 2008}\tabularnewline
\hline
\end{tabular}

\newpage

\section{Summary table for the CL 3C295, CL 0016+16 and CL 1601+42 cluster
galaxies}

x=used, -=available but not used (if for a reason other than the standard
selection criteria, we state the reason).

\begin{tabular}{|c| p{1.5in}|c|}
\hline 
{\footnotesize name} & {\footnotesize van der Marel} & {\footnotesize Type}\tabularnewline
\hline
\hline 
{\footnotesize 3C295-834} & {\footnotesize -}{\footnotesize \par}
{\footnotesize shows extra structures}{\footnotesize \par}
{\footnotesize S0 criteria: possibly S0 } & {\footnotesize E\cite{van der Marel 2007a}}\tabularnewline
\hline 
{\footnotesize 3C295-968} & {\footnotesize -}{\footnotesize \par}
{\footnotesize S0 criteria: possibly S0 } & {\footnotesize E\cite{van der Marel 2007a}}\tabularnewline
\hline 
{\footnotesize 0016-438} & x
{\footnotesize S0 criteria: unlikely to be S0 } & {\footnotesize E\cite{van der Marel 2007a}}\tabularnewline
\hline 
{\footnotesize 0016-461} & {\footnotesize -}{\footnotesize \par}
{\footnotesize S0 criteria: possibly  S0 } & {\footnotesize E\cite{van der Marel 2007a}}\tabularnewline
\hline 
{\footnotesize 0016-531} & {\footnotesize -}{\footnotesize \par}
{\footnotesize S0 criteria: possibly S0 } & {\footnotesize E\cite{van der Marel 2007a}}\tabularnewline
\hline 
{\footnotesize 0016-611} & {\footnotesize -}{\footnotesize \par}
{\footnotesize S0 criteria: likely to be S0 } & {\footnotesize E\cite{van der Marel 2007a}}\tabularnewline
\hline 
{\footnotesize 0016-612} & x
{\footnotesize S0 criteria: unlikely to be S0 } & {\footnotesize E\cite{van der Marel 2007a}}\tabularnewline
\hline
\end{tabular}

\newpage

\section{Summary table for SL2S lensing galaxies}

x=used, -=available but not used (if for a reason other than the standard
selection criteria, we state the reason).

\begin{tabular}{|c|c| p{1in}| p{1in}| p{1.5in}|}
\hline 
name & Ruff & $M_{tot}$ ($10^{10}M_{\odot}$) & $\sigma$(km/s) & type/comment\tabularnewline
\hline
\hline 
J02141-0405 & - (no data) &  &  & Has a companion galaxy~\cite{Gavazzi12}\tabularnewline
\hline 
J02173-0513 & - & 54\cite{Ruff} & 257\cite{Ruff} & Influenced by nearby group~\cite{Gavazzi12}\tabularnewline
\hline 
J02190-0829 & x & 19\cite{Ruff} & 305\cite{Ruff} & \tabularnewline
\hline 
J02205-0639 & x & 28\cite{Ruff} & 242\cite{Ruff} & \tabularnewline
\hline 
J02251-0454 & x & 40\cite{Ruff} & 241\cite{Ruff} & \tabularnewline
\hline 
J02261-0420 & x & 40\cite{Ruff} & 266\cite{Ruff} & \tabularnewline
\hline 
J02264-0406 & - (no data) &  &  & S0?~\cite{Gavazzi12}\tabularnewline
\hline 
J02264-0904 & -  & 65.2\cite{Ruff} & 301\cite{Ruff} & \tabularnewline
\hline 
J02325-0408 & x & 23\cite{Ruff} & 264\cite{Ruff} & \tabularnewline
\hline 
J14061-5202 & - (no data) &  &  & Influenced by nearby galaxy?~\cite{Gavazzi12}\tabularnewline
\hline 
J14061-5202 & - (no data) &  &  & \tabularnewline
\hline 
J14113-5651 & x & 15\cite{Ruff} & 228\cite{Ruff} & \tabularnewline
\hline 
J22062-0057 & - (no data) &  &  & Small companion~\cite{Gavazzi12}\tabularnewline
\hline 
J22132-0009 & - & 22\cite{Ruff} & 183\cite{Ruff} & Edge-on disk galaxy~\cite{Gavazzi12}\tabularnewline
\hline 
J22140-1807 & - & 5.7\cite{Ruff} & 167\cite{Ruff} & Probably a S0 from mass and $\sigma$~\cite{Gavazzi12}\tabularnewline
\hline 
J22160-1751 & - (no data) &  & 282\cite{Ruff} & Not a true lens.~\cite{Gavazzi12}\tabularnewline
\hline 
J221929-0017 & x & 9\cite{Ruff} & 263\cite{Ruff} & \tabularnewline
\hline
\end{tabular}

\newpage

\section{CDFS Summary table}

x=used, (x)=used but in lower reliability group, -=available but not
used, no symbol=not available. Values are given without uncertainties
because only the central values are used for selection. 

\hspace{-0.1in}{\footnotesize }\begin{tabular}{|c| p{0.5in}| p{0.5in}| p{1.5in}| p{1.5in}| p{1in}|}
\hline 
{\footnotesize Name} & {\footnotesize Rettura} & {\footnotesize van der Wel} & {\footnotesize $M_{tot}$ ($10^{10}M_{\odot}$)} & {\footnotesize $\sigma$(km/s)} & {\footnotesize type/comment}\tabularnewline
\hline
\hline 
{\footnotesize CDFS-1} & {\footnotesize -} & {\footnotesize -} & {\footnotesize 20.4\cite{Rettura},}{\footnotesize \par}
{\footnotesize 22.4\cite{van der wel}} & {\footnotesize 231\cite{Rettura},\cite{van der wel}} & {\footnotesize AGN}\tabularnewline
\hline 
{\footnotesize CDFS-2} & {\footnotesize x} & {\footnotesize x} & {\footnotesize 20.0\cite{Rettura},}{\footnotesize \par}
{\footnotesize 14.8\cite{van der wel}} & {\footnotesize 200\cite{Rettura},\cite{van der wel}} & \tabularnewline
\hline 
{\footnotesize CDFS-3} & {\footnotesize x} & {\footnotesize x} & {\footnotesize 12.6\cite{Rettura},}{\footnotesize \par}
{\footnotesize 12.6\cite{van der wel}} & {\footnotesize 300\cite{Rettura},\cite{van der wel}} & \tabularnewline
\hline 
{\footnotesize CDFS-4} & {\footnotesize -}{\footnotesize \par}
{\footnotesize (Giant)} & {\footnotesize -}{\footnotesize \par}
{\footnotesize (Giant)} & {\footnotesize 97.7\cite{Rettura},}{\footnotesize \par}
{\footnotesize 104.5\cite{van der wel}} & {\footnotesize 336\cite{Rettura},\cite{van der wel}} & {\footnotesize Large X-ray emm.}\tabularnewline
\hline 
{\footnotesize CDFS-5} & {\footnotesize -}{\footnotesize \par}
{\footnotesize (low $\sigma$)} & {\footnotesize -}{\footnotesize \par}
{\footnotesize (low $\sigma$)} & {\footnotesize 9.8\cite{Rettura},}{\footnotesize \par}
{\footnotesize 11.5\cite{van der wel}} & {\footnotesize 194\cite{Rettura},\cite{van der wel}} & \tabularnewline
\hline 
{\footnotesize CDFS-6} & {\footnotesize x} & {\footnotesize x} & {\footnotesize 8.3\cite{Rettura},}{\footnotesize \par}
{\footnotesize 5.4\cite{van der wel}} & {\footnotesize 208\cite{Rettura},\cite{van der wel}} & \tabularnewline
\hline
\end{tabular}{\footnotesize \par}

\hspace{-0.1in}{\footnotesize }\begin{tabular}{| p{0.51in}| p{0.5in}| p{0.5in}| p{1.5in}| p{1.5in}| p{1in}|}
\hline 
{\footnotesize Name} & {\footnotesize Rettura} & {\footnotesize van der Wel} & {\footnotesize $M_{tot}$ ($10^{10}M_{\odot}$)} & {\footnotesize $\sigma$(km/s)} & {\footnotesize type/comment}\tabularnewline
\hline
\hline 
{\footnotesize CDFS-25} & {\footnotesize x} & {\footnotesize x} & {\footnotesize 9.1\cite{Rettura},}{\footnotesize \par}
{\footnotesize 7.9\cite{van der wel}} & {\footnotesize 258\cite{Rettura},\cite{van der wel}} & \tabularnewline
\hline 
{\footnotesize CDFS-26} & {\footnotesize -} & {\footnotesize -} & {\footnotesize 72.4\cite{Rettura},}{\footnotesize \par}
{\footnotesize 67.60\cite{van der wel}} & {\footnotesize 249\cite{Rettura},\cite{van der wel}} & {\footnotesize Spiral}\tabularnewline
\hline 
{\footnotesize CDFS-27} & {\footnotesize -}{\footnotesize \par}
{\footnotesize (low $\sigma$)} & {\footnotesize -}{\footnotesize \par}
{\footnotesize (low $\sigma$)} & {\footnotesize 10.5\cite{Rettura},}{\footnotesize \par}
{\footnotesize 11.7\cite{van der wel}} & {\footnotesize 135\cite{Rettura},\cite{van der wel}} & \tabularnewline
\hline 
{\footnotesize CDFS-28} & {\footnotesize -} & {\footnotesize -} & {\footnotesize 74.1\cite{Rettura},}{\footnotesize \par}
{\footnotesize 147.9\cite{van der wel}} & {\footnotesize 445\cite{Rettura},\cite{van der wel}} & {\footnotesize Spiral}\tabularnewline
\hline 
{\footnotesize CDFS-29} & {\footnotesize x} & {\footnotesize x} & {\footnotesize 11.7\cite{Rettura},}{\footnotesize \par}
{\footnotesize 10.0\cite{van der wel}} & {\footnotesize 221\cite{Rettura},\cite{van der wel}} & \tabularnewline
\hline 
{\footnotesize CDFS-354} & {\footnotesize -}{\footnotesize \par}
{\footnotesize (low $\sigma$)} &  & {\footnotesize 3.6\cite{Rettura}} & {\footnotesize 99\cite{Rettura}} & \tabularnewline
\hline
\end{tabular}{\footnotesize \par}

\newpage

\section{Cluster CL1358+62 elliptical galaxies summary table}
x=used -=available but not used.

~

\hspace{-0.1in}{\footnotesize }\begin{tabular}{|c| p{0.5in}| p{1in}|}
\hline 
{\footnotesize ID} & {\footnotesize Kelson} & {\footnotesize type/comment}\tabularnewline
\hline
\hline 
212 & x & \tabularnewline
\hline 
242 & - & $M_{R}>-20.0$\tabularnewline
\hline 
256 & - & $M_{R}>-20.0$\tabularnewline
\hline 
303 & - & $M_{R}>-20.0$\tabularnewline
\hline 
360 & x & \tabularnewline
\hline 
375 & - & $M_{R}>-20.0$\tabularnewline
\hline 
409 & x & \tabularnewline
\hline 
412 & x & \tabularnewline
\hline 
531 & - & $M_{R}>-20.0$\tabularnewline
\hline 
534 & x & \tabularnewline
\hline 
536 & - & $M_{R}>-20.0$\tabularnewline
\hline
\end{tabular}{\footnotesize \par}

\pagebreak 

\section{Summary table for other lensing galaxies \label{sec:Summary-table-for other lenses}}

x=used, (x)=used but in lower reliability group, -=available but not
used (if for a reason other than standard selection criteria,
we stated the reason), no symbol=not available. Values are given
without uncertainties because the central values only are used for selection.
The integration value for $M_{tot}$ varies from authors to authors.

\hspace{-0.3in}{\tiny }\begin{tabular}{| p{0.55in}| p{0.3in}| p{0.2in}| p{0.2in}| p{0.3in}| p{0.3in}| p{0.3in}| p{0.2in}| p{0.6in}| p{0.6in}| p{0.8in}|}
\hline 
{\tiny Name} & {\tiny Kee} & {\tiny Jack} & {\tiny J\&K} & {\tiny Lei}{\tiny \par}
{\tiny 09} & {\tiny Lei}{\tiny \par}
{\tiny 11} & {\tiny Fer}{\tiny \par}
{\tiny 05} & {\tiny Treu}{\tiny \par}
{\tiny 04} & {\tiny $M_{tot}$ ($10^{10}M_{\odot}$)} & {\tiny $\sigma$(km/s)} & {\tiny type/comment}\tabularnewline
\hline
\hline 
{\tiny 0047-28108} &  &  & {\tiny -} & {\tiny x} & {\tiny x} & {\tiny x} & {\tiny x} & {\tiny 58\cite{Jiang-Kochanek}}{\tiny \par}
{\tiny 21\cite{Leier09},}{\tiny \par}
{\tiny 33\cite{Leier2011},}{\tiny \par}
{\tiny 34\cite{Ferreras2},}{\tiny \par}
{\tiny 41\cite{Treu04}} & {\tiny 250\cite{Jiang-Kochanek},}{\tiny \par}
{\tiny 229\cite{Leier09},}{\tiny \par}
{\tiny 254\cite{van de Ven}} & \tabularnewline
\hline 
{\tiny Q0142-100} & {\tiny x} &  &  &  & {\tiny x} & {\tiny x} &  & {\tiny 46\cite{Leier2011},}{\tiny \par}
{\tiny 25\cite{Ferreras2}} & {\tiny 224\cite{van de Ven}} & {\tiny E\cite{Keeton}}\tabularnewline
\hline 
{\tiny B0218+357} & {\tiny -} & {\tiny -} &  &  &  &  &  & {\tiny 2\cite{Jackson}} &  & {\tiny S\cite{Keeton},\cite{Lehar},\cite{Jackson}}\tabularnewline
\hline 
{\tiny CFRS-3.1077} &  &  &  & {\tiny -}{\tiny \par}
{\tiny (no data)} &  &  &  &  &  & \tabularnewline
\hline 
{\tiny C0302+006} &  &  & {\tiny -} &  &  &  & {\tiny x} & {\tiny 67\cite{Jiang-Kochanek},}{\tiny \par}
{\tiny 67\cite{Treu04}} & {\tiny 256\cite{Jiang-Kochanek}, }{\tiny \par}
{\tiny 251\cite{Treu04}} & {\tiny Part of small group?\cite{Treu04}}\tabularnewline
\hline 
{\tiny MG0414+0534} & {\tiny - }{\tiny \par}
{\tiny (no z values)} & {\tiny x} &  &  & {\tiny x} & {\tiny x} &  & {\tiny 103\cite{Leier2011},}{\tiny \par}
{\tiny 83\cite{Ferreras2},}{\tiny \par}
{\tiny 24\cite{Jackson}} & {\tiny 303\cite{van de Ven}} & {\tiny E\cite{Keeton}}\tabularnewline
\hline
\end{tabular}{\tiny \par}

\hspace{-0.3in}{\tiny }\begin{tabular}{| p{0.55in}| p{0.3in}| p{0.2in}| p{0.2in}| p{0.3in}| p{0.3in}| p{0.3in}| p{0.2in}| p{0.6in}| p{0.6in}| p{0.8in}|}
\hline 
{\tiny Name} & {\tiny Kee} & {\tiny Jack} & {\tiny J\&K} & {\tiny Lei}{\tiny \par}
{\tiny 09} & {\tiny Lei}{\tiny \par}
{\tiny 11} & {\tiny Fer}{\tiny \par}
{\tiny 05} & {\tiny Treu}{\tiny \par}
{\tiny 04} & {\tiny $M_{tot}$ ($10^{10}M_{\odot}$)} & {\tiny $\sigma$(km/s)} & {\tiny type/comment}\tabularnewline
\hline
\hline 
{\tiny B1127+385 G1} &  & {\tiny -} &  &  &  &  &  & {\tiny 2.7\cite{Jackson}} &  & {\tiny Late type?~\cite{Koopmans99}}\tabularnewline
\hline 
{\tiny B1127+385 G2} &  & {\tiny -} &  &  &  &  &  & {\tiny 1.1\cite{Jackson}} &  & {\tiny Late type?~\cite{Koopmans99}}\tabularnewline
\hline 
{\tiny MG1131+0456} & {\tiny -}{\tiny \par}
{\tiny (intera}{\tiny \par}
{\tiny cting)} &  &  &  &  &  &  &  &  & \tabularnewline
\hline 
{\tiny B1152} &  &  &  &  & {\tiny -}{\tiny \par}
{\tiny (intera}{\tiny \par}
{\tiny cting)} &  &  & {\tiny 30\cite{Leier2011}} &  & \tabularnewline
\hline 
{\tiny HST12531-2914} & {\tiny -}{\tiny \par}
{\tiny (no z values)} &  &  &  &  &  &  &  & {\tiny NA} & {\tiny E}\tabularnewline
\hline 
{\tiny HST14113+}{\tiny \par}
{\tiny 5211} &  &  &  &  &  & {\tiny -} &  & {\tiny 34\cite{Ferreras2}} & {\tiny 190\cite{van de Ven}} & \tabularnewline
\hline 
{\tiny HST14176+}{\tiny \par}
{\tiny 5226} & {\tiny x} &  & {\tiny -} & {\tiny x} &  & {\tiny -} & {\tiny x} & {\tiny 71\cite{Jiang-Kochanek},34.3\cite{Leier09},}{\tiny \par}
{\tiny 118\cite{Ferreras2},71\cite{Treu04}} & {\tiny 292\cite{van de Ven},212\cite{Jiang-Kochanek},}{\tiny \par}
{\tiny 245,224\cite{Treu04}} & {\tiny E\cite{Keeton}}\tabularnewline
\hline 
\end{tabular}{\tiny \par}

\hspace{-0.3in}{\tiny }\begin{tabular}{| p{0.55in}| p{0.3in}| p{0.2in}| p{0.2in}| p{0.3in}| p{0.3in}| p{0.3in}| p{0.2in}| p{0.6in}| p{0.6in}| p{0.8in}|}
\hline 
{\tiny Name} & {\tiny Kee} & {\tiny Jack} & {\tiny J\&K} & {\tiny Lei}{\tiny \par}
{\tiny 09} & {\tiny Lei}{\tiny \par}
{\tiny 11} & {\tiny Fer}{\tiny \par}
{\tiny 05} & {\tiny Treu}{\tiny \par}
{\tiny 04} & {\tiny $M_{tot}$ ($10^{10}M_{\odot}$)} & {\tiny $\sigma$(km/s)} & {\tiny type/comment}\tabularnewline
\hline
\hline 
{\tiny HE2149} &  &  &  &  & {\tiny -} & {\tiny -} &  & {\tiny 27\cite{Leier2011},14\cite{Ferreras2}} & {\tiny 203\cite{van de Ven}} & \tabularnewline
\hline 
{\tiny Q2237} & {\tiny -} &  &  & {\tiny -} & {\tiny -} &  &  & {\tiny 7\cite{Leier09},2.8\cite{Leier2011},}{\tiny \par}
{\tiny 2\cite{Ferreras2}} & {\tiny 215\cite{Leier09},}{\tiny \par}
{\tiny 168\cite{van de Ven}} & {\tiny Sb or spiral bulge\cite{Leier2011},\cite{Keeton}}\tabularnewline
\hline 
{\tiny COMBO34244} &  &  &  &  &  &  &  & {\tiny 11\cite{More}} &  & {\tiny May not be a lens\cite{More}}\tabularnewline
\hline 
{\tiny COMBO40173} &  &  &  &  &  &  &  & {\tiny 7\cite{More}} &  & \tabularnewline
\hline 
{\tiny COMBO43242} &  &  &  &  &  &  &  & {\tiny 12} &  & {\tiny Late type\cite{More}}\tabularnewline
\hline 
{\tiny COMBO46446} &  &  &  &  &  & {\tiny -} &  & {\tiny 8} &  & \tabularnewline
\hline
\end{tabular}{\tiny \par}

\newpage

\section{Summary of the individual analyses \label{Summary of the individual analyses}}

In the table below, we summarize the results of each individual analysis
described from Section~\ref{sec:Data-sets-using virial theo} to Section
\ref{sec:Data-sets-using strong lensing}. Col(1): publication
reference. Col(2): method used to compute the dark content.
Col(3): best fit result (not yet normalized to
$\sfrac{M}{L}(\sfrac{R_{min}}{R_{max}}=0.7)$=$\sfrac{M_{tot}}{M_*}(\sfrac{R_{min}}{R_{max}}=0.7)=8$).
Col(4) : number of elliptical galaxies that passed the selection
criteria of Section~\ref{sub:General-selection-criteria}. Col(5):
reliability group, see Section~\ref{sec:Global-results}, Col(6):
weight factor wf, see Section~\ref{sec:Global-results}. This a multiplicative
factor to the uncertainties: the higher wf, the less impact the data
set has on the global average. Col(7): specific notes regarding
the analysis. Here, we give the best fit result in function of ellipticity
$\varepsilon$ rather than $\sfrac{R_{min}}{R_{max}}=1-\varepsilon$.  ``-''
means that the data was not used in the global average.

\hspace{-1.1in}{\small }\begin{tabular}{| p{0.25in}| p{0.5in}| p{2.7in}| p{0.2in}| p{0.1in}| p{0.3in}| p{2.5in}|}
\hline 
{\small Ref. } & {\small Method} & {\small Best fit} & {\small stat} & {\small rg} & {\small wf} & {\small Notes}\tabularnewline
\hline
\hline 
{\small~\cite{BMS}} & {\small Virial} & {\small $\sfrac{M}{L_{B}}=(13.08\pm2.97)\varepsilon _{ap}+(16.88\pm2.32)$} & {\small 64} & {\small 2} & {\small 1.50} & {\small Added requirement that the galaxy characteristics listed in
the 1985 Ref.~\cite{BMS} are compatible with recent characteristics
given in NED~\cite{NED}.  Results using ellipticity corrections from $\sigma^{2}$ isotropic
method.}\tabularnewline
\hline 
{\small~\cite{BMS}} & {\small Virial} & {\small $\sfrac{M}{L_{B}}=(5.91\pm4.67)\varepsilon _{ap}+(10.50\pm3.48)$} & {\small 11} & {\small 2} & {\small 1.13} & {\small Same additional selection criterion  as above. Results using ellipticity
correction from $\mu^{2}$ isotropic method.}\tabularnewline
\hline 
{\small~\cite{BMS}} & {\small Virial} & {\small $\sfrac{M}{L_{B}}=(6.19\pm3.59)\varepsilon _{ap}+(9.18\pm2.64)$} & {\small 11} & {\small 1} & {\small 1.13} & {\small Same additional selection criterion  as above. Results using ellipticity
correction from $\mu^{2}$ anisotropic method.}\tabularnewline
\hline 
{\small~\cite{Bender}} & {\small Virial} & {\small $\sfrac{M}{L_{B}}=(4.03\pm1.44)\varepsilon _{ap}+(6.20\pm1.07)$} & {\small 35} & {\small 3} & {\small 2.37} & \tabularnewline
\hline 
{\small~\cite{Kelson}} & {\small Virial} & {\small $\sfrac{M}{L_{V}}=(15.53\pm7.21)\varepsilon _{ap}+(20.52\pm5.95)$} & {\small 5} & {\small 4} & {\small 1.00} & {\small Distant galaxies (cluster CL1358+62). Added $M_{B}\leq-20$
selection criterion  to minimize  contamination from boxy galaxies.}\tabularnewline
\hline 
{\small~\cite{Lauer}} & {\small Virial} & {\small $\sfrac{M}{L_{B}}=(10.94\pm10.18)\varepsilon _{ap}+(23.83\pm7.80)$} & {\small 10} & {\small 4} & {\small 6.09} & {\small $\sfrac{M}{L_B}$ extracted for galactic cores. Same added selection criterion  
as~\cite{BMS}. }\tabularnewline
\hline 
{\small~\cite{Leier09}} & {\small Virial} & {\small $\sfrac{M_{vir}}{L_{I}(R_{lense})}=(6.41\pm4.97)\varepsilon _{ap}+(6.30\pm3.95)$} & {\small 8} & {\small 2} & {\small 1.00} & {\small Distant galaxies strongly lensing. Also analyzed with strong
lensing method. }\tabularnewline
\hline 
{\small~\cite{Prugniel}} & {\small Virial} & {\small $\sfrac{M}{L_{B}}=(6.58\pm1.98)\varepsilon _{ap}+(8.99\pm1.68)$} & {\small 102} & {\small 2} & {\small 1.32} & \tabularnewline
\hline 
{\small~\cite{Rettura}} & {\small Virial} & {\small $\sfrac{M_{dyn}}{M_*}=(4.29\pm1.36)\varepsilon _{ap}+(5.04\pm1.25)$} & {\small 16} & {\small 3} & {\small 1.35} & {\small Distant galaxies. $M_*$ obtained with composite stellar
population models using a Kroupa IMF~\cite{Kroupa}}.\tabularnewline
\hline 
{\small~\cite{van der wel}} & {\small Virial} & {\small $\sfrac{M}{L_{B}}=(5.36\pm1.24)\varepsilon _{ap}+(6.42\pm1.12)$} & {\small 13} & {\small 3} & {\small 1.35} & {\small Distant galaxies.}\tabularnewline
\hline 
{\small~\cite{Cappellari 2006}} & {\small Modeling} & {\small $\sfrac{M}{L}_{Jeans}=(1.47\pm1.56)\varepsilon _{true}+(3.84\pm1.10)$}{\small \par}
{\small $\sfrac{M}{L}_{Schw}=(1.09\pm1.68)\varepsilon _{true}+(3.48\pm1.17)$} & {\small 6}{\small \par}
{\small 6} & {\small 1}{\small \par}{\small 1} & {\small 1.38}{\small \par}
{\small 1.38} & {\small Two estimates of $\sfrac{M}{L}$ provided. One based on a 2-integral
Jeans model and one on a 3-integral Schwartzchild model~\cite{Schwarzschild}.}\tabularnewline
\hline 
{\small cap. 2013a} & {\small Modeling} & {\small $\sfrac{M}{L}=(2.02\pm1.47)\varepsilon _{e}+(3.51\pm0.42)$} & {\small 31} & {\small 1} & 1.86 & \tabularnewline
\hline 
{\small cap. 2013b} & {\small Modeling} & {\small $\sfrac{M}{L}=(4.72\pm2.18)\varepsilon _{e}+(3.33\pm0.59)$} & {\small 31} & {\small 1} & 1.86 & \tabularnewline
\hline 
{\small~\cite{Kronawitter} } & {\small Modeling} & {\small $\sfrac{M}{L_{B}}|_{in}=(4.58\pm7.04)\varepsilon _{ap}+(9.58\pm6.17)$}{\small \par}
{\small $\sfrac{M}{L_{B}}|_{out}=(-6.53\pm14.48)\varepsilon _{ap}+(1.25\pm12.48)$} & {\small 10}{\small \par}
{\small 10} & {\small 1}{\small \par}{\small 1} & {\small 1.17}{\small \par}
{\small 1.17}{\small \par}
 & {\small Kept LINERS, AGN and Seyfert galaxies because isothermal or
virial equilibrium is not required.}\tabularnewline
\hline 
{\small~\cite{Magorrian98}} & {\small Modeling} & {\small $\sfrac{M}{L}=(-0.69\pm4.95)\varepsilon _{ap}+(3.26\pm4.73)$} & {\small 7} & {\small 1} & {\small 1.33} & \tabularnewline
\hline 
{\small~\cite{Thomas} and}{\small \par}
{\small~\cite{Wegner}} & {\small Modeling} & {\small $\sfrac{M}{L}=(2.57\pm6.80)\varepsilon _{true}+(8.82\pm5.30)$}{\small \par}
{\small $\sfrac{M}{L}|_{sc}=(-0.57\pm7.02)\varepsilon_{true}+(6.67\pm5.57)$} & {\small 7}{\small \par}
{\small 7} & {\small 1}{\small \par}{\small 2} & {\small 1.16}{\small \par}
{\small 1.63} & {\small Distant galaxies (Coma cluster and Abell 262 cluster). $\sfrac{M}{L}|_{sc}$
\cite{Thomas} was assumed to be independent of radius.}\tabularnewline
\hline 
{\small~\cite{van der Marel 1991}} & {\small Modeling} & {\small $\sfrac{M}{L_{R}}=(3.43\pm0.92)\varepsilon _{ap}+(5.65\pm0.84$)} & {\small 9} & {\small 1} & {\small 1.47} & \tabularnewline
\hline 
{\small~\cite{van der Marel 2007a}} & {\small Modeling} & {\small $\sfrac{M}{L_{B}}=(6.19\pm11.45)\varepsilon _{"true"}+(8.75\pm7.38)$} & {\small 19} & {\small 2} & {\small 1.00} & {\small Distant galaxies.}\tabularnewline
\hline 
{\small~\cite{van der Marel 2007b}} & {\small Modeling} & {\small $\sfrac{M}{L_{B}}=(3.50\pm2.62)\varepsilon _{ap}+(9.59\pm2.08)$} & {\small 17} & {\small 1} & {\small 2.04} & {\small Homogenized compilation of literature (local galaxies).}\tabularnewline
\hline 
{\small~\cite{Capaccioli}} & {\small PNe/GC} & {\small $\sfrac{M}{L_{B}}=(19.43\pm7.85)\varepsilon _{ap}+(22.14\pm6.82)$} & {\small 5} &  & {\small 2.14} & {\small Relaxed selection: includes all galaxies except the disrupted
ones, the ones showing possible interactions and the ones clearly
non-elliptical.}\tabularnewline
\hline 
\end{tabular}{\small \par}

\hspace{-1.1in}{\small }\begin{tabular}{| p{0.25in}| p{0.5in}| p{2.7in}| p{0.2in}| p{0.1in}| p{0.3in}| p{2.5in}|}
\hline 
{\small Ref. } & {\small Method} & {\small Best fit} & {\small stat} & {\small rg} & {\small wf} & {\small Notes}\tabularnewline
\hline
\hline 
{\small~\cite{Deason}} & {\small PNe/GC} & {\small $\sfrac{M}{L}=(10.89\pm17.71)\varepsilon _{ap}+(21.45\pm12.73)$} & {\small 7} & {\small 1} & {\small 1.24} & {\small Relaxed selection.}\tabularnewline
\hline 
{\small~\cite{Magorrian01}} & {\small PNe/GC} & {\small $\sfrac{M}{L}=(2.45\pm8.60)\varepsilon _{ap}+(9.18\pm7.57)$} & {\small 8} & {\small 1} & {\small 1.14} & \tabularnewline
\hline 
{\small~\cite{Romanowsky}} & {\small PNe/GC} & {\small $\sfrac{M}{L_{B}}=(142.47\pm63.47)\varepsilon _{true}+(111.50\pm42.86)$} & {\small 3} & {\small 1} & {\small 1.22} & {\small True axis ratios are from~\cite{Cappellari 2006} ,~\cite{Foster}
and~\cite{Statler}.}\tabularnewline
\hline
{\small~\cite{Fukazawa}} & {\small X-ray} & {\small $\sfrac{M}{L_{B}}=(0.55\pm15.44)\varepsilon _{ap}+(5.29\pm11.74)$} & {\small 7} & {\small 2} & {\small 1.17} & \tabularnewline
\hline 
{\small~\cite{Nagino}} & {\small X-ray} & {\small $\sfrac{M}{L_{B}}(0.5R_{eff})=(12.24\pm8.41)\varepsilon _{ap}+(20.94\pm6.01)$}{\small \par}
{\small $\sfrac{M}{L_{B}}(3R_{eff})=(20.85\pm33.35)\varepsilon _{ap}+(38.32\pm23.74)$}{\small \par}
{\small $\sfrac{M}{L_{B}}(6R_{eff})=(232.0\pm94.0)\varepsilon _{ap}+(182.5\pm59.9)$} & {\small 3}{\small \par}
{\small 3}{\small \par}
{\small 2} & {\small 2}{\small \par}{\small 2}{\small \par}{\small 2} & {\small 1.17}{\small \par}
{\small 1.17}{\small \par}
{\small 1.17} & \tabularnewline
\hline 
{\small~\cite{Bertola93} }{\small \par}
{\small and}{\small \par}
{\small~\cite{Bertola91}} & {\small Gas disk} & {\small $\sfrac{M}{L_{B}}=(4.21\pm3.55)(1-\sfrac{bc}{a^{2}})+(5.83\pm2.09)$ } & {\small 4} & {\small 1} & {\small 1.15} & {\small Relaxed selection (kept LINERS and/or Sy types) $a,b$ and $c$ are the radii of the elliptical galaxy triaxial
shape model.}\tabularnewline
\hline 
{\small~\cite{Pizeella}} & {\small Gas disk} & {\small $\sfrac{M}{L_{T}}=(1.39\pm13.13)(1-q_{o}p_{o})+(4.83\pm6.80)$} & {\small 4} & {\small 1} & {\small 1.15} & {\small Relaxed selection (kept LINERS and/or Sy types) $q_{0}$ and $p_{0}$ are the intrinsic axis ratios of the
triaxial galaxy.}\tabularnewline
\hline 
{\small~\cite{Auger 1}} & {\small Lensing} & {\small $\sfrac{M_{tot}}{M_*}|_{Chab}=(3.38\pm0.79)\varepsilon _{mass}+(4.92\pm0.66)$ }{\small \par}
{\small $\sfrac{M_{tot}}{M_*}|_{Sal}=(1.86\pm0.46)\varepsilon _{mass}+(2.76\pm0.38)$ }{\small \par}
{\small $\sfrac{M_{tot}}{M_*}|_{Chab}=(1.47\pm0.70)\varepsilon _{ap}+(3.30\pm0.57)$}{\small \par}
{\small $\sfrac{M_{tot}}{M_*}|_{Sal}=(0.79\pm0.40)\varepsilon _{ap}+(1.84\pm0.32)$} & {\small 34}{\small \par}
{\small 34}{\small \par}
{\small 34}{\small \par}
{\small 34} & {\small 1}{\small \par}{\small 1}{\small \par}{\small 1}{\small \par}{\small 1} & {\small -}{\small \par}
{\small -}{\small \par}
{\small 2.04}{\small \par}
{\small -} & {\small S0 galaxies are already identified in~\cite{Pizeella}. Thus standard $\sigma\geq225$~km.s$^{-1}$
criterion is not applied in this analysis (and all other analyses
using SLACS data). Mass ratios extracted using either a Chabrier~\cite{Chabrier}
or a Salpeter IMF~\cite{Salpeter}.}\tabularnewline
\hline 
{\small~\cite{Barnabe}} & {\small Lensing} & {\small $\sfrac{M}{L}|_{Chab.}=(33.1\pm16.7)\varepsilon _{true}+(37.2\pm13.9)$}{\small \par}
{\small $\sfrac{M}{L}|_{Sal}=(15.9\pm9.8)\varepsilon _{true}+(19.0\pm8.2)$ } & {\small 10}{\small \par}
{\small 10} & {\small 1}{\small \par}{\small 1} & {\small 2.05}{\small \par}
{\small -} & {\small $\sfrac{M}{L}$ formed with DMf using either a Chabrier~\cite{Chabrier}
or a Salpeter IMF~\cite{Salpeter}.}\tabularnewline
\hline 
{\small~\cite{Cardone09}} & {\small Lensing} & {\small $\sfrac{M}{L}=(5.39\pm3.40)\varepsilon _{mass}+(9.32\pm2.85)$}{\small \par}
{\small $\sfrac{M}{L}=(1.94\pm3.52)\varepsilon _{ap}+(6.37\pm2.82)$} & {\small 13}{\small \par}
{\small 13} & {\small 1}{\small \par}{\small 1} & {\small -}{\small \par}
{\small 5.12} & \tabularnewline
\hline 
{\small~\cite{Cardone11}} & {\small Lensing} & {\small $\sfrac{M}{L}(R_{Ein})=(4.57\pm2.79)\varepsilon _{mass}+(8.39\pm2.32)$}{\small \par}
{\small $\sfrac{M}{L}(R_{Ein})=(2.50\pm3.00)\varepsilon_{ap}+(6.59\pm2.40)$}{\small \par}
{\small $\sfrac{M}{L}(R_{eff})=(8.17\pm1.60)\varepsilon _{mass}+(13.00\pm1.37)$}{\small \par}
{\small $\sfrac{M}{L}(R_{eff})=(4.82\pm1.32)\varepsilon _{ap}+(10.00\pm1.09)$} & {\small 36}{\small \par}
{\small 36}{\small \par}
{\small 36}{\small \par}
{\small 36} & {\small 2}{\small \par}{\small 2}{\small \par}{\small 2}{\small \par}{\small 2} & {\small -}{\small \par}
{\small 4.92}{\small \par}
{\small -}{\small \par}
{\small 1.00} & {\small Use Secondary Infall Model and Salpeter IMF.}\tabularnewline
\hline 
{\small~\cite{Faure 2011}} & {\small Lensing} & {\small $\sfrac{M_{tot}}{M_*}=(30.21\pm6.13)\varepsilon+(27.12\pm4.88)$} & {\small 7} & {\small 1} & {\small 1.00} & \tabularnewline
\hline 
{\small~\cite{Ferreras2}} & {\small Lensing} & {\small $\sfrac{M_{tot}}{M_*}|_{V,Chab}=(1.83\pm5.78)\varepsilon _{ap}+(2.67\pm4.09)$}{\small \par}
{\small $\sfrac{M_{tot}}{M_*}|_{V,Sal}=(3.16\pm3.69)\varepsilon _{ap}+(2.61\pm3.40)$} & {\small 4}{\small \par}
{\small 4} & {\small 1}{\small \par}{\small 1} & {\small 1.29}{\small \par}
{\small -} & {\small $M_*$ determined with a Chabrier or a Salpeter IMF.}\tabularnewline
\hline 
{\small~\cite{Ferreras}} & {\small Lensing} & {\small $\sfrac{M_{tot}}{M_*}=(-0.47\pm2.63)\varepsilon _{ap}+(0.61\pm2.16)$
and}{\small \par}
{\small $\sfrac{M_{tot}}{M_*}=(-5.55\pm3.22)\varepsilon _{mass}+(-3.00\pm2.30)$} & {\small 4}{\small \par}
{\small 4} & {\small 1}{\small \par}{\small 1} & {\small 1.80}{\small \par}
{\small -} & \tabularnewline
\hline 
{\small~\cite{Grillo09}} & {\small Lensing} & {\small $\sfrac{M_{tot}}{L_{B}}|_{Mar,Sal}=(1.65\pm0.97)\varepsilon _{ap}+(3.30\pm0.78)$}{\small \par}
{\small $\sfrac{M_{tot}}{L_{B}}|_{BC,Sal}=(1.52\pm1.18)\varepsilon _{ap}+(3.27\pm0.94)$}{\small \par}
{\small $\sfrac{M_{tot}}{L_{B}}|_{Mar,Krou}=(1.12\pm1.10)\varepsilon _{ap}+(2.95\pm0.87)$}{\small \par}
{\small $\sfrac{M_{tot}}{L_{B}}|_{BC,Chab}=(1.75\pm1.06)\varepsilon _{ap}+(3.42\pm0.85)$} & {\small 40}{\small \par}
{\small 40}{\small \par}
{\small 40}{\small \par}
{\small 40} & {\small 1}{\small \par}{\small 1}{\small \par}{\small 1}{\small \par}{\small 1} & {\small -}{\small \par}
{\small -}{\small \par}
{\small -}{\small \par}
{\small 2.00} & {\small Stellar Composite Model used to obtain $M_*$ with two different
sets of metallicity template (Bruzual \& Charlot~\cite{Bruzual}
or Maraston~\cite{Maraston}) and three different IMF (Salpeter
\cite{Salpeter}, Kroupa~\cite{Kroupa} or Chabrier~\cite{Chabrier}).}\tabularnewline
\hline 
{\small~\cite{Jackson}} & {\small Lensing} & {\small $\sfrac{M_{tot}}{L_{H}}=(0.80\pm1.67)\varepsilon _{ap}+(1.61\pm1.31)$} & {\small 4} & {\small 3} & {\small 1.37} & \tabularnewline
\hline 
{\small~\cite{Jiang-Kochanek}} & {\small Lensing} & {\small $\sfrac{M_{tot}}{M_*}=(1.24\pm0.92)\varepsilon _{ap}+(2.73\pm0.72)$}{\small \par}
{\small $\sfrac{M_{tot}}{M_*}=(0.14\pm0.68)\varepsilon _{mass}+(1.87\pm0.50)$} & {\small 12}{\small \par}
{\small 12} & {\small 2}{\small \par}
{\small 2} & {\small 6.24}{\small \par}
{\small -} & {\small Use results with adiabatic compression (favored by the authors'
analysis).}\tabularnewline
\hline 
{\small~\cite{Keeton}} & {\small Lensing} & {\small $\sfrac{M}{L_{B}}=(36.22\pm15.14)\varepsilon _{mass}+(30.59\pm9.89)$}{\small \par}
{\small $\sfrac{M}{L_{B}}=(0.00\pm10.27)\varepsilon _{ap}+(6.55\pm7.01)$} & {\small 3}{\small \par}
{\small 3} & {\small 3}{\small \par}
{\small 3} & {\small -}{\small \par}
{\small $\infty$} & {\small Due to the small number of galaxies, we relaxed the S0 rejection
criterion and used $\sigma<200$~km.s$^{-1}$. Thus, this result is weighted out in the global
average.}\tabularnewline
\hline 
{\small~\cite{Koopmans}} & {\small Lensing} & {\small $\sfrac{M_{tot}}{M_*}=(1.75\pm0.77)\varepsilon _{ap}+(2.67\pm0.64)$}{\small \par}
{\small $\sfrac{M_{tot}}{M_*}=(2.25\pm0.57)\varepsilon _{mass}+(3.20\pm0.49)$} & {\small 9}{\small \par}
{\small 9} & {\small 1}{\small \par}
{\small 1} & {\small 2.52}{\small \par}
{\small -} & \tabularnewline
\hline 
{\small~\cite{Leier09}} & {\small Lensing} & {\small $\sfrac{M}{L_{I}}=(5.51\pm3.27)\varepsilon _{ap}+(5.72\pm2.62)$} & {\small 8} & {\small 1} & {\small 4.98} & \tabularnewline
\hline 
{\small~\cite{Leier2011}} & {\small Lensing} & {\small $\sfrac{M}{L_{V}}=(37.2\pm36.6)\varepsilon _{ap}+(-13.2\pm26.3)$} & {\small 3} & {\small 2} & {\small 2.83} & \tabularnewline
\hline 
{\small~\cite{Ruff}} & {\small Lensing} & {\small $\sfrac{M}{M_*}=(-0.49\pm2.90)\varepsilon _{ap}+(0.22\pm2.17)$}{\small \par}
{\small $\sfrac{M}{M_*}=(-0.53\pm2.47)\varepsilon _{mass}+(0.18\pm1.88)$}{\small \par}
{\small $\sfrac{M_{tot}}{M_*}|_{Sal}=(-3.62\pm8.81)\varepsilon _{ap}+(0.50\pm6.33)$}{\small \par}
{\small $\sfrac{M_{tot}}{M_*}|_{Sal}=(4.66\pm1.43)\varepsilon _{mass}+(5.84\pm0.85)$}{\small \par}
{\small $\sfrac{M_{tot}}{M_*}|_{Chab}=(6.17\pm14.65)\varepsilon _{ap}+(9.62\pm10.70)$}{\small \par}
{\small $\sfrac{M_{tot}}{M_*}|_{Chab}=(9.89\pm4.85)\varepsilon _{mass}+(11.65\pm3.25)$} & {\small 7}{\small \par}
{\small 7}{\small \par}
{\small 7}{\small \par}
{\small 7}{\small \par}
{\small 7}{\small \par}
{\small 7} & {\small 2}{\small \par}
{\small 2}{\small \par}
{\small 2}{\small \par}
{\small 2}{\small \par}
{\small 2}{\small \par}
{\small 2} & {\small 1.00}{\small \par}
{\small -}{\small \par}
{\small -}{\small \par}
{\small -}{\small \par}
{\small 1.00}{\small \par}
{\small -} & {\small The $M_*$ determined with a Chabrier or a Salpeter IMF
are for $\sfrac{M}{L}$ at $R_{eff}$. $\sfrac{M}{L}$ at $R_{Ein}$ does not need IMF
input.}\tabularnewline
\hline 
{\small~\cite{Treu04}} & {\small Lensing} & {\small $\sfrac{M}{L_{B}}=(2.04\pm3.20)\varepsilon _{ap}+(6.63\pm2.45)$} & {\small 3} & {\small 1} & {\small 1.23} & \tabularnewline
\hline
\end{tabular}{\small \par}

\newpage

\section{Detailed analysis of the Bacon {\it et al}. data \label{sec:Detailed-analysis-of Bacon et al}}

This section provides the details of the analysis done using the Bacon {\it et al}.
data. The results reported here are slightly different from the results given in Section~\ref{sub:Bacon-et-al.}
Since additional corrections, mostly addressing  possible
correlations, are applied. In addition, some of the NED data used
may have become  obsolete as they  are from the 2008 database. The
purpose of this section is to define the analysis method, identification
requirement and most importantly, \underline{correlation investigations}. The slight 
difference with  Section~\ref{sub:Bacon-et-al.} is thus irrelevant.

\subsection{Data quality and galaxy selection \label{sub:Selection}}

\begin{figure}[h]
\centering
\includegraphics[scale=0.4]{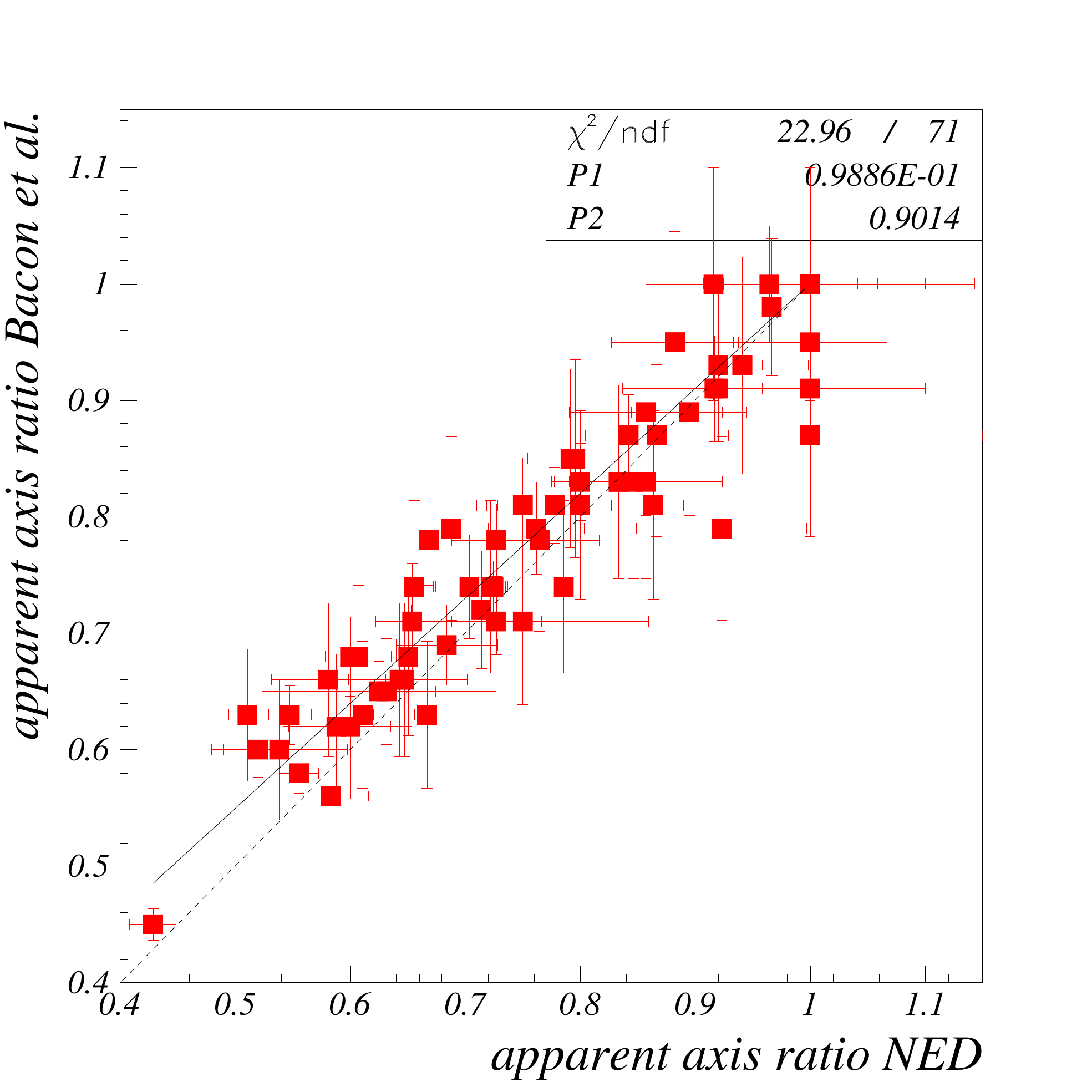} 
\vspace{-0.4cm} \caption{Apparent axis ratio $\sfrac{R_{min}}{R_{max}}$ for NED vs~\cite{BMS}.
The plain line is the best fit to the data and the dashed one indicates
$y=x$. The best fit is $y=(0.901\pm0.048)x+(0.099\pm0.034)$.
It reveals a slight bias: Ref.~\cite{BMS} tends to overestimate
$\sfrac{R_{min}}{R_{max}}$, perhaps because older data have
poorer resolution which tends to round ellipses. 
\label{fig:apratned_apratbms}
}
\end{figure}

\begin{figure}
\centering
\includegraphics[scale=0.4]{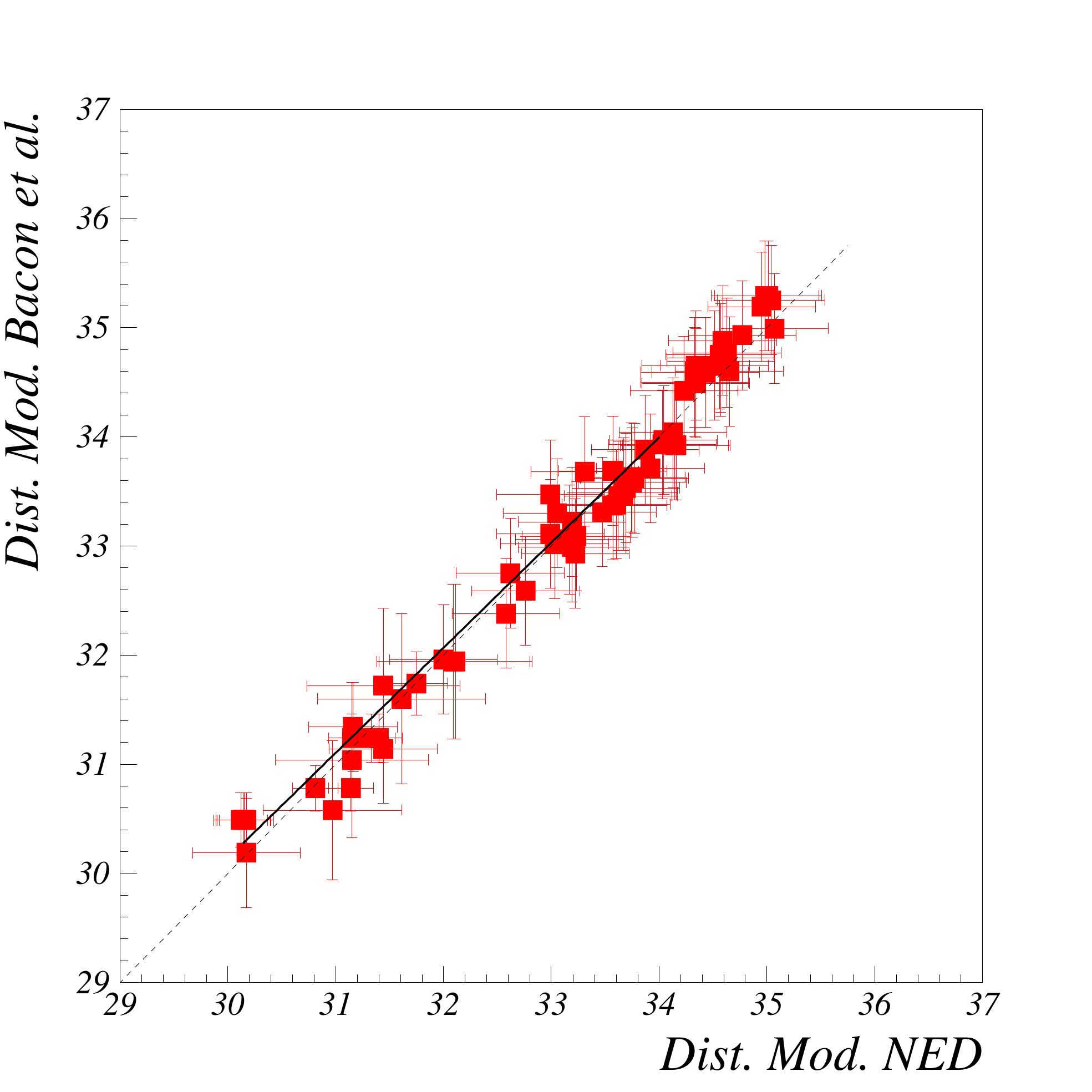} 
\vspace{-0.4cm} \caption{Distance Moduli from NED vs~\cite{BMS}. The plain line is
the best fit to the data and the dashed one indicates $y=x$. The
best fit is $y=(0.964\pm0.035)x+(1.235\pm1.126)$, with no indication
of discrepancy.
\label{fig:dmbms_dmned}
}
\end{figure}

\begin{figure}
\centering
\includegraphics[scale=0.4]{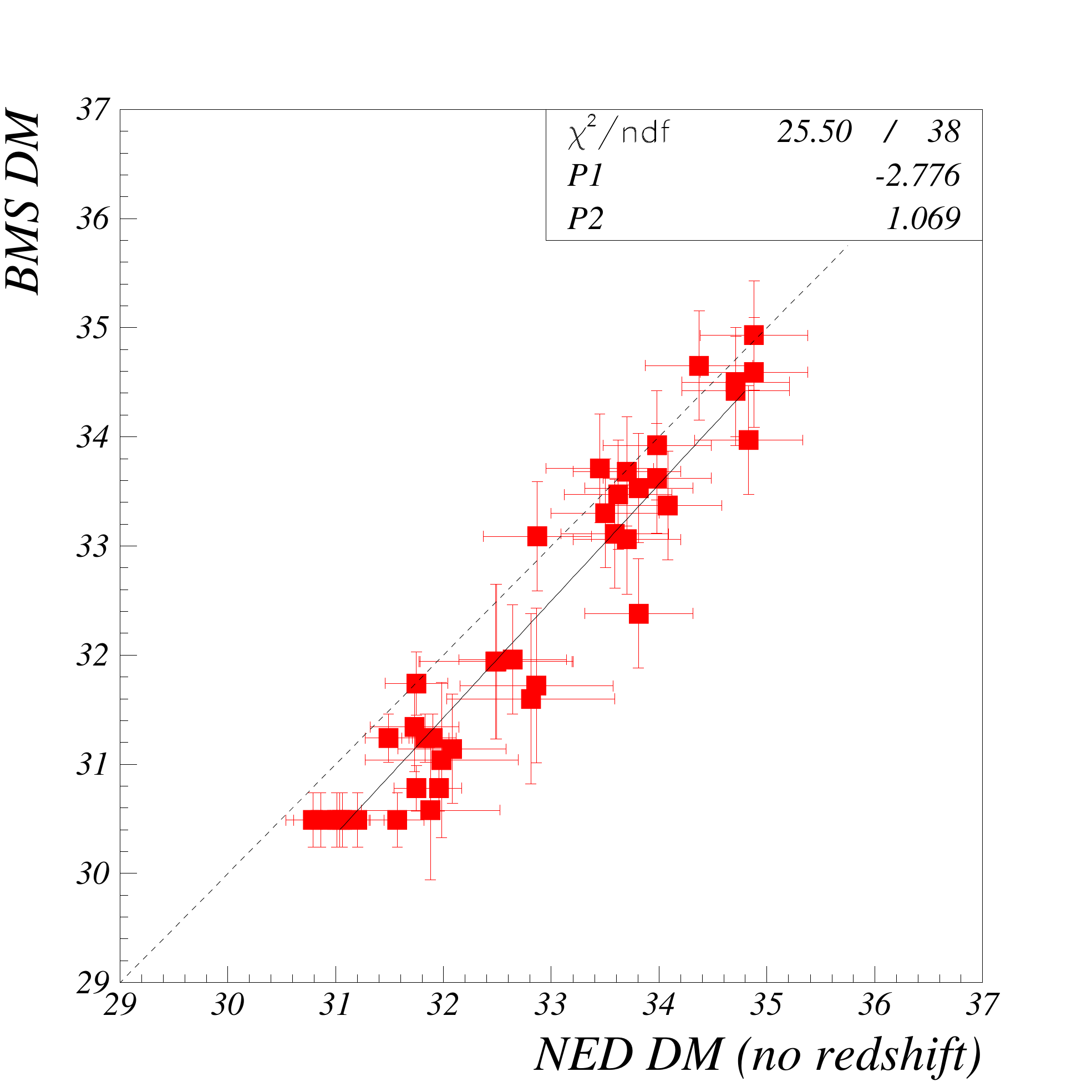} 
\vspace{-0.4cm} \caption{Distance Moduli from NED, determined without redshift information
when available, vs~\cite{BMS}. The plain line is the best
fit to the data and the dashed one indicates $y=x$. The best fit
is $y=(1.067\pm0.054)x-(2.776\pm1.744)$, indicating that the non-redshift based NED's $DM$
indicate larger distances from galaxies to Earth. 
\label{fig:dmbms_dmnednox}
}
\end{figure}

\begin{figure}
\centering
\includegraphics[scale=0.4]{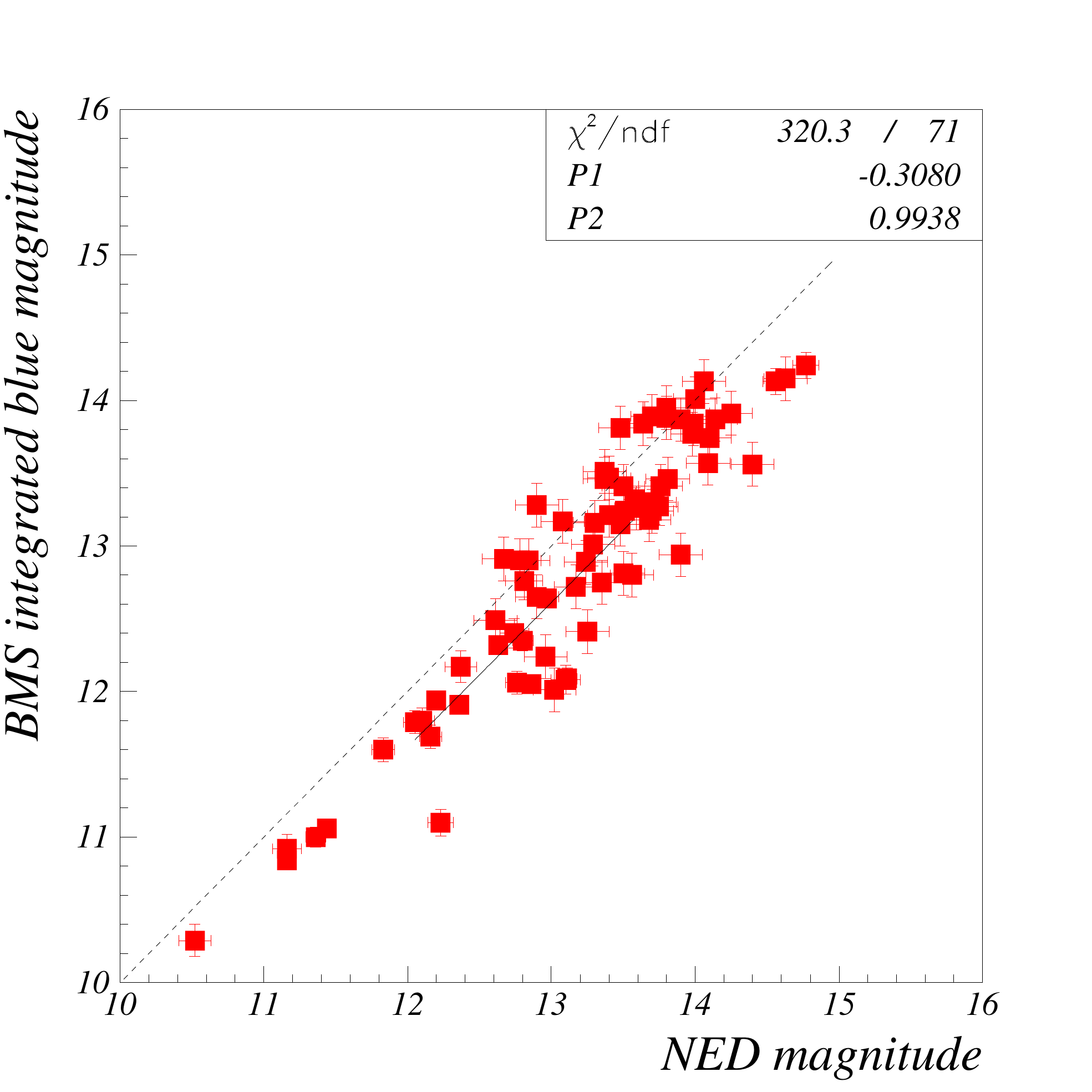} 
\vspace{-0.4cm} \caption{Apparent magnitude from NED vs integrated blue magnitude~\cite{BMS}.
The plain line is the best fit to the data and the dashed one indicates
$y=x$. The best fit is $y=(0.994\pm0.018)x-(0.308\pm0.229)$ with
possibly a slight bias: the apparent magnitudes from~\cite{BMS}
maybe slightly underestimated. Correcting for a larger apparent magnitude
would lead to a smaller surface brightness (see Fig.~\ref{Fig: Bt correlations})
and thus a slightly larger $\sfrac{M}{L}$ according the the virial formula
in~\cite{BMS}. 
\label{fig:btbms_bted}
}
\end{figure}

We apply our usual selection criteria. In addition, we reject the 8 last galaxies
of~\cite{BMS} for lack of reference. Since the data~\cite{BMS}
are old, we verified that they agree with data also available from NED (as of 2008).
We exclude%
\footnote{We exclude the data rather than correct, with the NED value, $\sfrac{M}{L}$
and other quantities: we prefer to assume that the discrepancy
reflects a difficult measurement and hence a suspicious datum, rather
than assuming that NED is correct and Ref.~\cite{BMS} wrong. %
} from our sample galaxies for which their distance to Earth (redshift based) 
disagree by more than 20\% between NED and~\cite{BMS}. We
apply the same criterion for the apparent axis ratio $\sfrac{R_{min}}{R_{max}}$.
The apparent magnitude from NED and~\cite{BMS} always agree
within 10\%, so no galaxy is excluded on the basis of an apparent
magnitude discrepancy. We note that some of the uncertainties quoted in~\cite{BMS}
may be underestimated since some of the sets are incompatible
(assuming that the NED data are more accurate). Comparing
the analysis done using the galactic characteristics from~\cite{BMS} and the analysis done using that of NED is useful
for studying  if  the correlations seen between the galactic
characteristics are due to measurement bias rather
than a physical relation or an observational bias. When no uncertainty
was available from NED, we used the uncertainty reported in~\cite{BMS}.
To check for possible biases, we fit the Ref.~\cite{BMS} vs
NED data for $\sfrac{R_{min}}{R_{max}}$, distance moduli $DM$, and apparent magnitudes
using a linear function, see Figs.~\ref{fig:apratned_apratbms}, \ref{fig:dmbms_dmned},
\ref{fig:dmbms_dmnednox} and~\ref{fig:btbms_bted}.%
\footnote{We will account here for the horizontal uncertainty
in a recursive way: we first fit the distribution without accounting
for the horizontal error bars. We then use this first fit result to
transform the horizontal errors in vertical ones and add the sum in
quadrature. This procedure assumes that all errors are gaussian.
} No significant biases are found, except for the non-redshift based
$DM$. Using the NED data, particularly its non-redshift based $DM$, 
significantly strengthen the $\sfrac{M}{L}$ vs $\sfrac{R_{min}}{R_{max}}$
correlation.

All in all, we kept 73 galaxies in the sample. The list is: NGC57,
83, 97, 430, 584, 636, 720, 741, 990, 1008, 1016, 1199, 1209, 1426,
1439, 1521, 1573, 2675, 2693, 2778, 2800, 2810, 2954, 3070, 3562,
3640, 3710, 3812, 3818, 3837, 3853, 3904, 3940, 4070, 4187, 4213,
4239, 4365, 4473, 4478, 4489, 4510, 4551, 4564, 4648, 4660, 4860,
4869, 5029, 5223, 5329, 5546, 5642, 5710, 5845, 5966, 6020, 6137,
6411, 6487, 6623, 7391, 7454, 7458, 7507, 7619, 7660, 7778, 7785,
and IC179, 948, 1152 and 1211. Among this sample, 12 galaxies have
$\sfrac{M}{L}$ available using two additional virial formulae presumably less
sensitive to unknown true flattening~\cite{BMS}. 
%This low number of galaxies makes analysis using this sub-sample barely reliable.
%We have nevertheless included the results here. The 12 galaxies
They are: NGC584, 720, 2778, 3818, 3904, 4365, 4473, 4478, 4551, 5845,
7619 and 7785.

\subsection{Data analysis\label{sec:Data-analysis}}

\begin{figure}[h]
\centering
\includegraphics[scale=0.4]{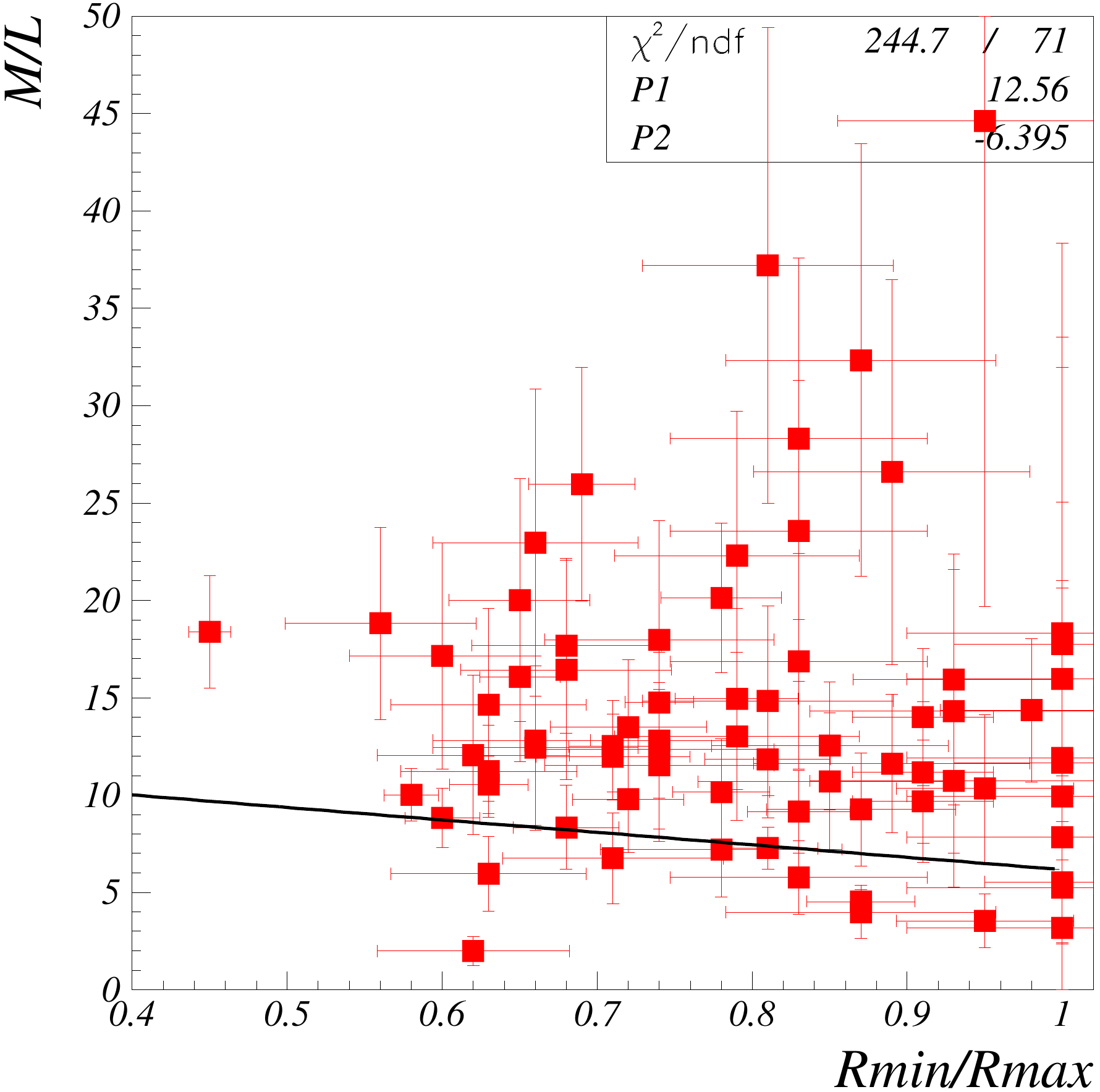}
\vspace{-0.4cm} \caption{\label{fig:ml1_apar}
Correlation between $\sfrac{M}{L}$ and the apparent  $\sfrac{R_{min}}{R_{max}}$
for 73 galaxies using the first virial formula of~\cite{BMS}.
The best linear fit (plain line) is
$y=(-6.40\pm2.32)x+(12.56\pm1.79)$, indicating a $2.8\sigma$ sigma
effect. The large $\sfrac{\chi^{2}}{ndf}$ is mostly due to ellipsoid projection
effect. A $2^{nd}$ order polynomial fit yields a similar $\sfrac{\chi^{2}}{ndf}$. 
}
\end{figure}

\begin{figure}
\centering
\includegraphics[scale=0.4]{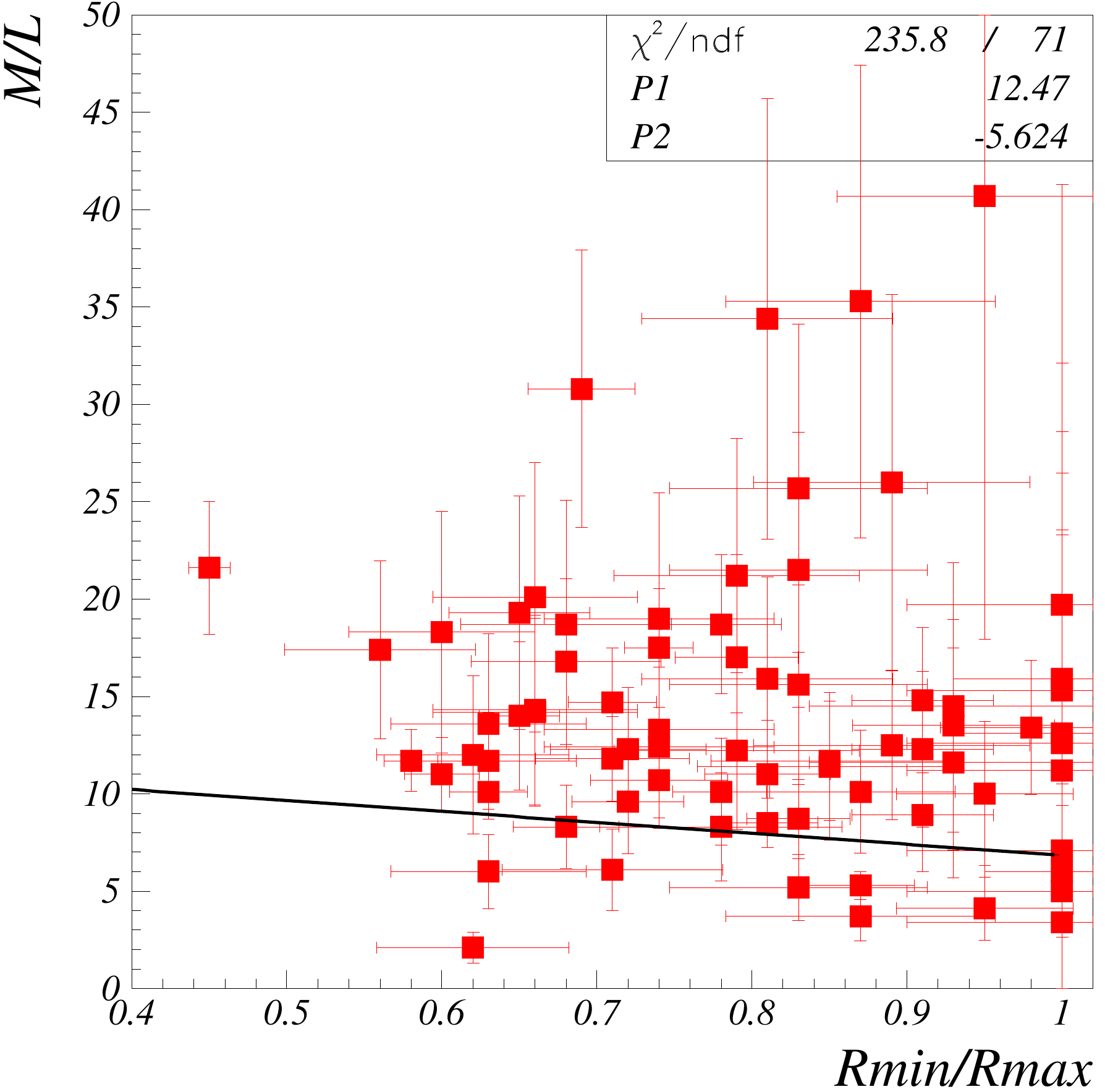}
\vspace{-0.4cm} \caption{\label{fig:ml1_apar_bmsdist}Same as Fig.~\ref{fig:ml1_apar} but without NED distance correction.
The fit is $y=(-5.62\pm2.48)x+(12.47\pm1.91)$. 
}
\end{figure}

\begin{figure}
\centering
\includegraphics[scale=0.4]{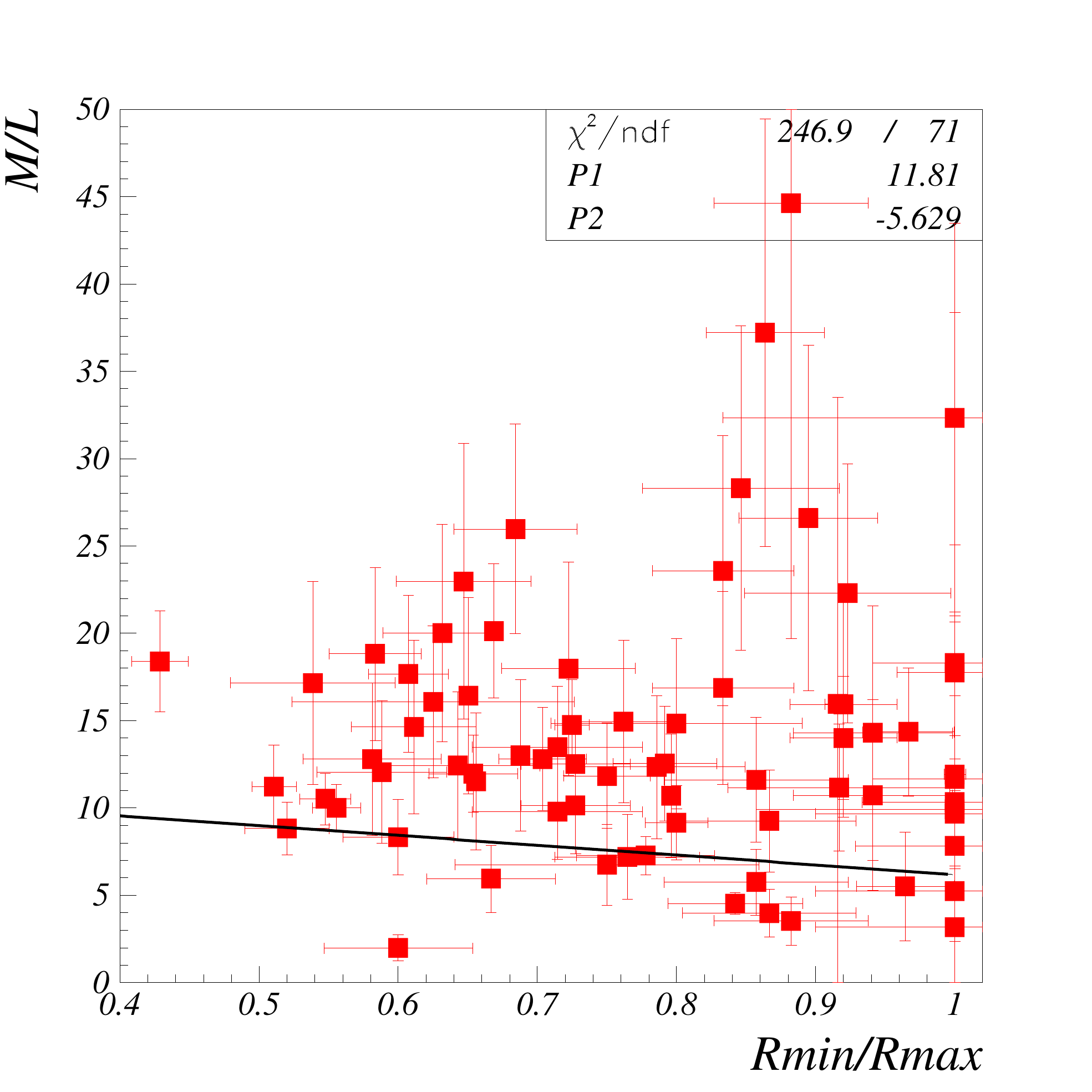}
\vspace{-0.4cm} \caption{\label{fig:ml1_nedapar}Same as Fig.~\ref{fig:ml1_apar} but with NED $\sfrac{R_{min}}{R_{max}}$.
The  fit is $y=(-5.63\pm2.15)x+(11.81\pm1.60)$. 
}
\end{figure}

Selection is applied to the data sample. In addition, the ratio redshift
distance from NED over distance from~\cite{BMS} is applied to
$\sfrac{M}{L}$ to correct these ratios on a galaxy per galaxy basis. We use
$\sfrac{R_{min}}{R_{max}}$ from~\cite{BMS} for consistency with the
$\sfrac{M}{L}$ calculations, as it is an important ingredient of the calculation.
The results for the three $\sfrac{M}{L}$ are shown in Figs.~\ref{fig:ml1_apar},
\ref{fig:ml2_apar} and~\ref{fig:ml3_apar}. 
Correcting with the NED distance is not significant: not correcting
yields similar results (Fig.~\ref{fig:ml1_apar_bmsdist}). Similarly,
using NED's $\sfrac{R_{min}}{R_{max}}$ does not change significantly the results
(Fig.~\ref{fig:ml1_nedapar}). All the fits give a slope of $6\pm2$.
The best linear fit for the data in Fig.~\ref{fig:ml1_apar} is $y=(-6.40\pm2.32)x+(12.56\pm1.79)$.
A non-zero slope indicates a correlation, so the fit yields a significantly
negative slope with a $2.8\sigma$ sigma confidence%
\footnote{The point at $\sfrac{R_{min}}{R_{max}}=0.45$ stands outside the distribution.
Although there is no reason to remove it since it passes the selection criteria, 
%discussed in Section~\ref{sub:Selection}, 
we checked that this datum
is not solely responsible for the observed correlation. After removing
the datum, the best fit slope becomes $4.01\pm2.35$, a 1.7 sigma
effect. Anticipating upcoming corrections, the final result (Fig.
\ref{Flo:ML vs R/R after Hubble correction}) becomes $10.69\pm2.28$
after removing the datum, that is a 4.7 sigma effect.%
}, but with a $\sfrac{\chi^{2}}{ndf}$ of 3.4. A $2^{nd}$ order polynomial yields
a similar $\sfrac{\chi^{2}}{ndf}$.  As we will show, the large $\sfrac{\chi^{2}}{ndf}$
of the  {\it linear} fit comes mostly from the effect of the projection of the 3D elliptical galaxy shapes into flat
ellipses on the observation plane. This transforms a, e.g., {\it linear}
$\sfrac{M}{L}$ dependence with {\it apparent} axis ratio $\sfrac{R_{min}}{R_{max}}$
into a {\it non-linear} dependence with the {\it real} axis ratio.

\begin{figure}
\centering
\includegraphics[scale=0.26]{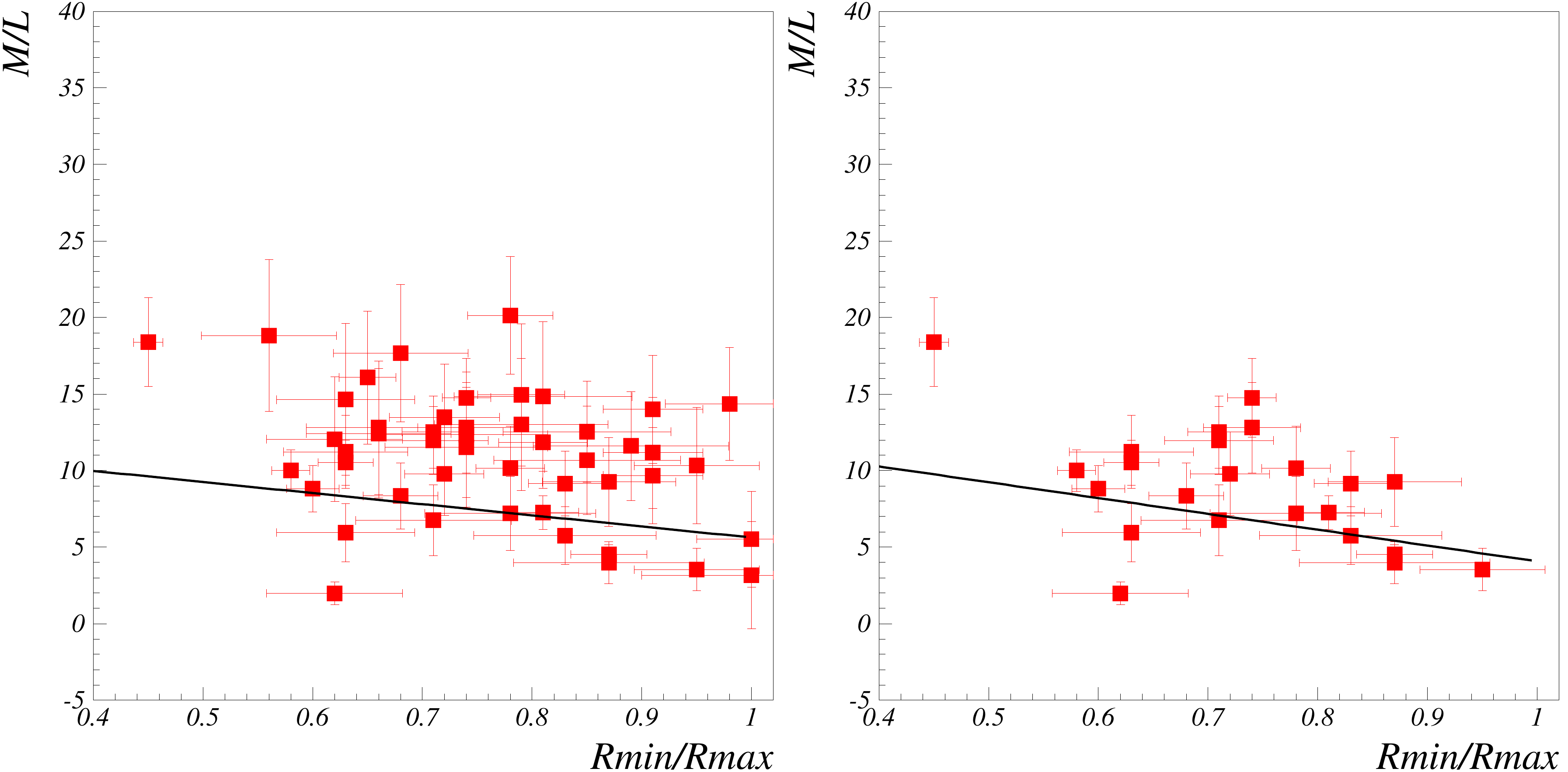}
\vspace{-0.4cm} \caption{\label{fig:pearsoncuts}Same as Fig.~\ref{fig:ml1_apar} but using only data 
with uncertainties $\Delta \sfrac{M}{L}<5$ (left panel) or $\Delta \sfrac{M}{L}<3$ (right panel).
These selections enhance the correlation. Since Ref.~\cite{BMS}
gives {\it relative} uncertainties of similar magnitude for most
data, the condition on the {\it absolute} uncertainty $\Delta \sfrac{M}{L}<5$
or 3 removes mostly points at large $\sfrac{M}{L}$.
}
\end{figure}

\begin{figure}
\centering
\includegraphics[scale=0.4]{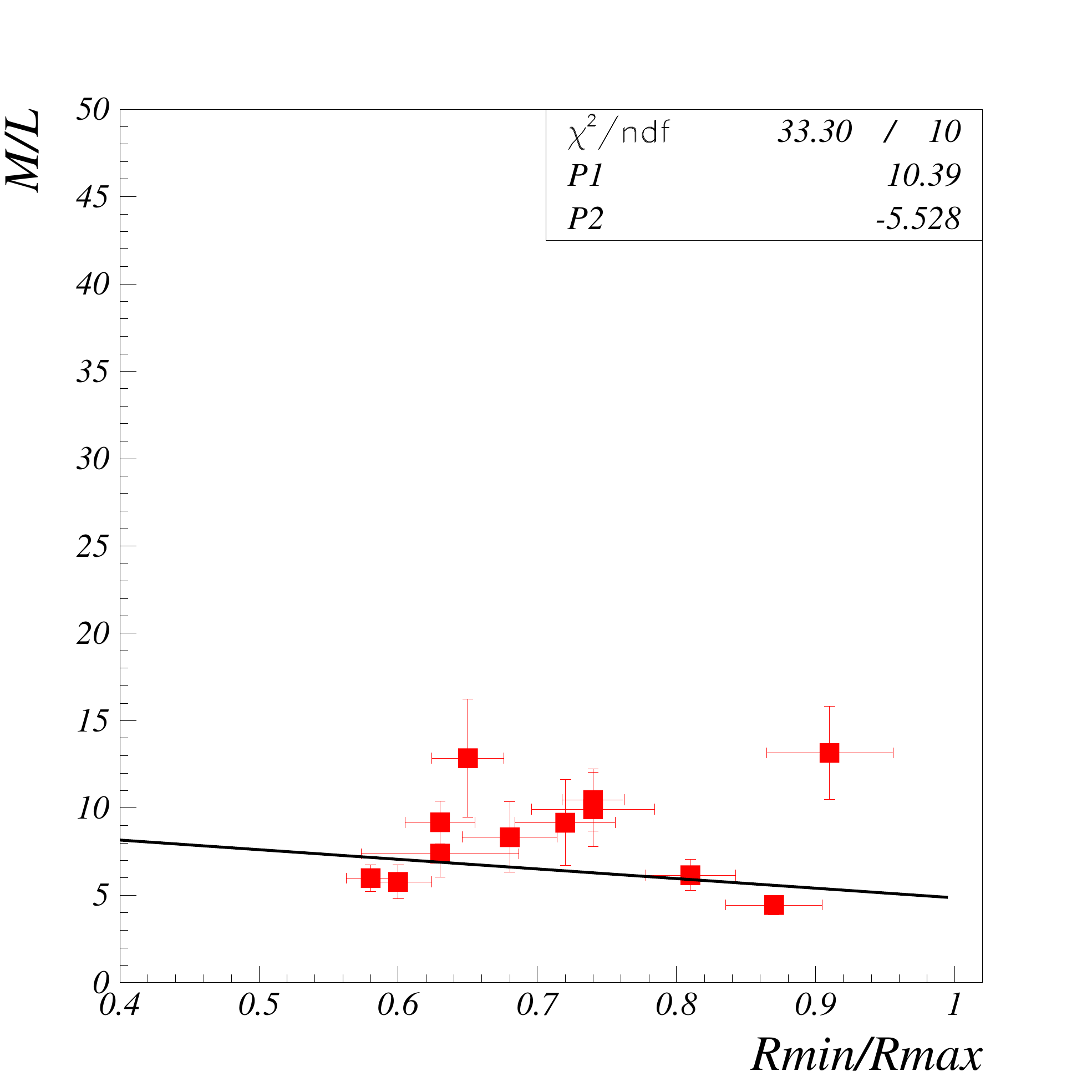}
\vspace{-0.4cm} \caption{\label{fig:ml2_apar}Correlation between $\sfrac{M}{L}$ and the apparent axis ratio $\sfrac{R_{min}}{R_{max}}$
for 12 galaxies using the $2^{nd}$ virial formula of~\cite{BMS}.
The best linear fit is $y=(-5.53\pm2.70)x+(2.00\pm3.71)$. 
}
\end{figure}

\begin{figure}
\centering
\includegraphics[scale=0.4]{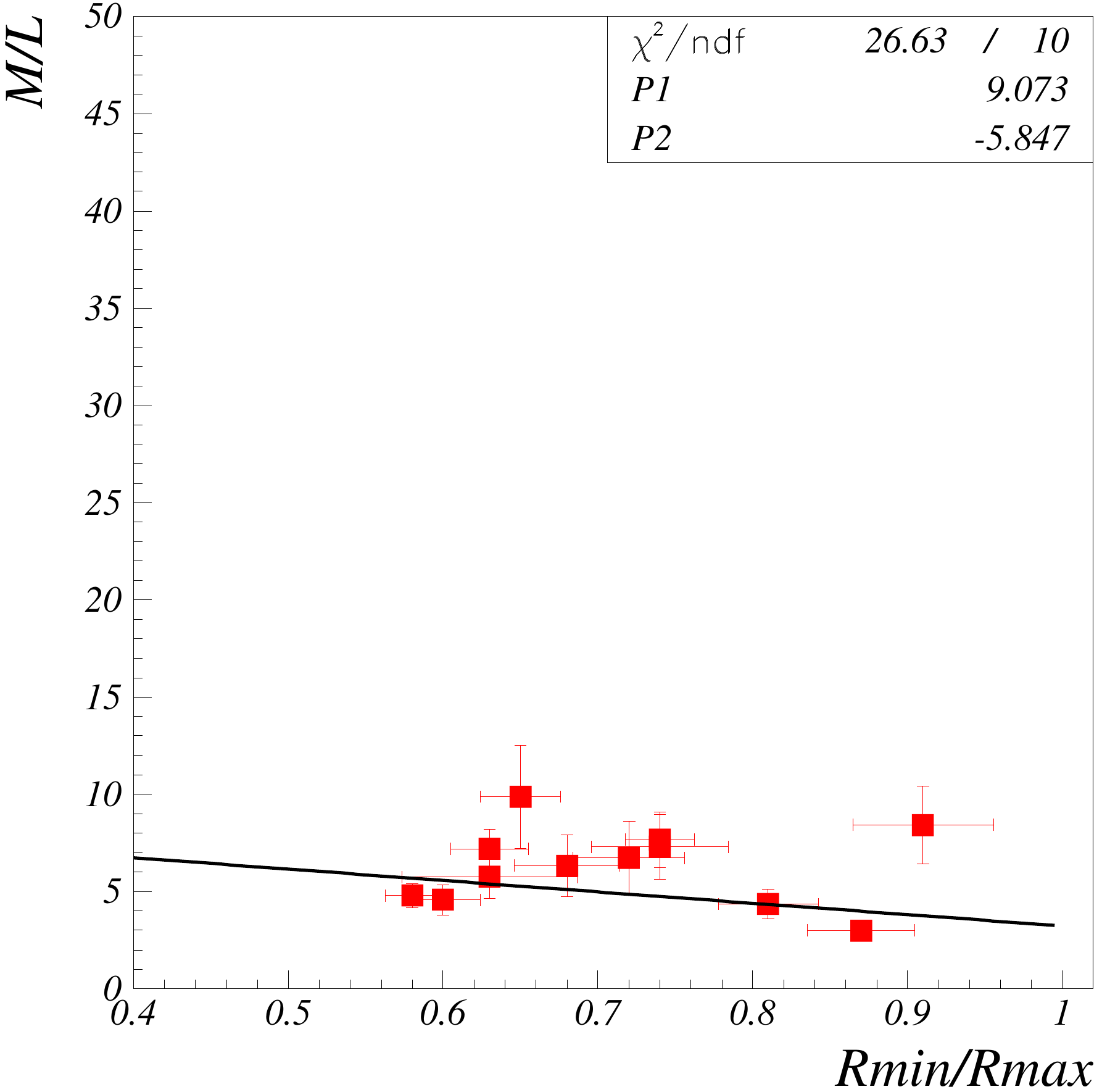}
\vspace{-0.4cm} \caption{\label{fig:ml3_apar}Correlation between $\sfrac{M}{L}$ and the apparent axis ratio $\sfrac{R_{min}}{R_{max}}$
for 12 galaxies using the $3^{rd}$ $\sfrac{M}{L}$ results from~\cite{BMS}.
The best linear fit is $y=(-5.847\pm2.32)x+(9.07\pm1.70)$. 
}
\end{figure}

Alternatively, to assess the correlation, one can compute the Pearson
correlation coefficient $r$ given by the covariance of $\sfrac{R_{min}}{R_{max}}$
and $\sfrac{M}{L}$ divided by their standard deviations: 
$r=\sfrac{\mbox{cov}(\sfrac{R_{min}}{R_{max}},\sfrac{M}{L})}{\sigma_{Rmin/Rmax}\sigma_{M/L}}$.
We have $|r|\leq1$ and larger values of $|r|$ indicates clearer
(small dispertion) and/or stronger (steeper slope) correlations. However,
since such statistical analysis does not account for statistical
weights, contrarily to a fit, it is ill-suited for our sample that
displays a large range in uncertainties. This problem can be partly
circumvented by keeping data of a given absolute precision, see Fig.
\ref{fig:pearsoncuts}. 
Keeping data for which the uncertainty on $\sfrac{M}{L}$ is smaller than 5
yields $r=-0.367$ (the sample is reduced to 48 galaxies. For this
sample, the best linear fit is $y=(-7.26\pm2.39)x+(12.88\pm1.88)$,
to be compared to the results in Fig.~\ref{fig:ml1_apar}). Applying
a tighter selection $\Delta \sfrac{M}{L}<3$ reduces the sample to 23 galaxies
and yields $r=-0.511$ (the best linear fit is $y=(-10.34\pm2.71)x+(14.41\pm2.06)$).
Such values of $r$ reveal a medium to large correlation between $\sfrac{M}{L}$
and the axis ratio and confirm the conclusion from the fit method.
Interestingly, selecting the highest precision data strengthen the
$\sfrac{M}{L}$ vs $\sfrac{R_{min}}{R_{max}}$ correlation, both for the determination
using of the Pearson criterion and for the determination from the
linear fit.

Results using the two other determinations of $\sfrac{M}{L}$ that employ an
additional observable (maximum stellar rotation velocity) confirm
in both cases (Figs.~\ref{fig:ml2_apar} and~\ref{fig:ml3_apar}) the
significant negative slope, with similar $\sigma$. 

\begin{figure}
\centering
\includegraphics[scale=0.4]{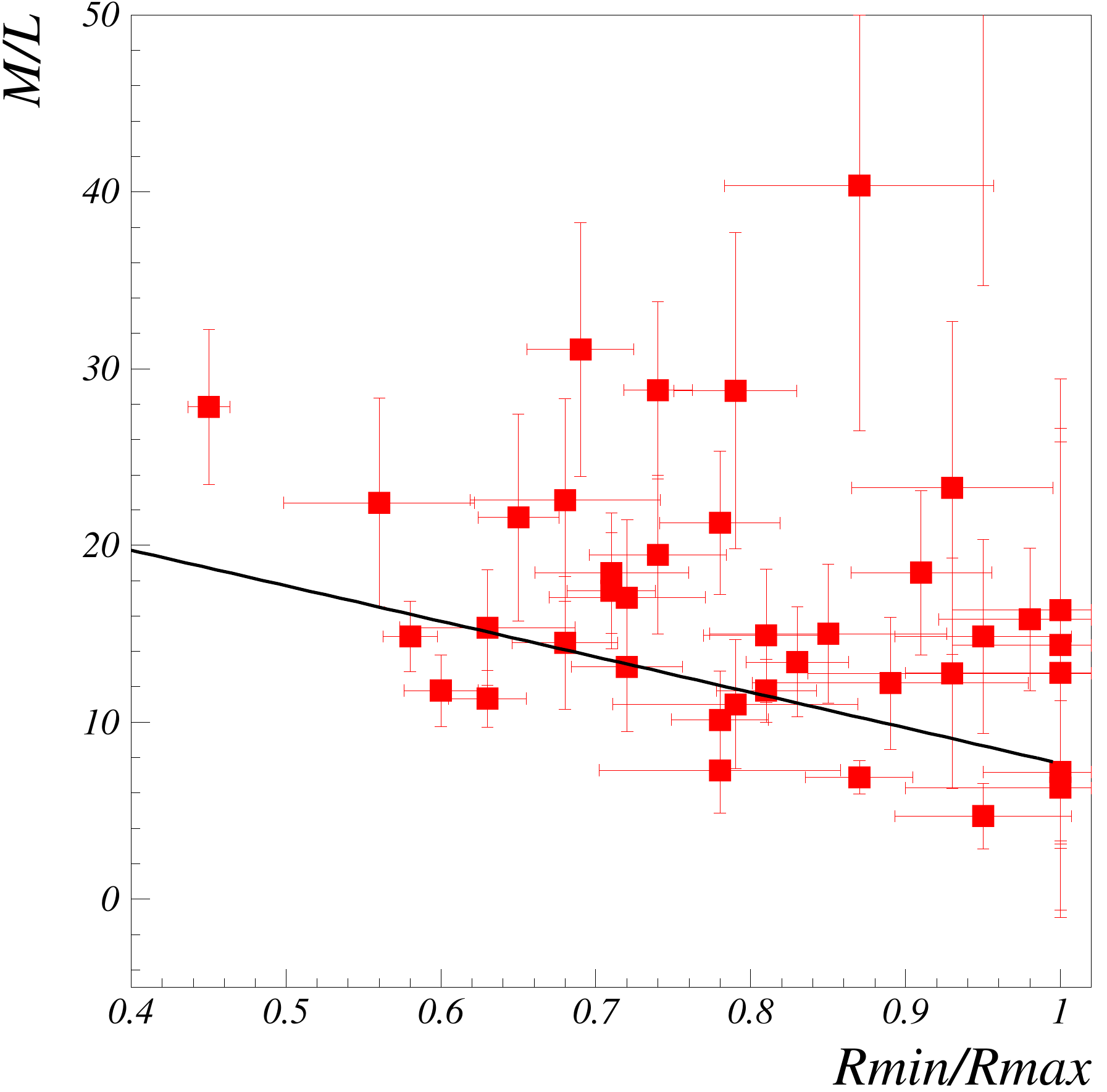}
\vspace{-0.4cm} \caption{\label{fig:ml1_apar_noz}Correlation between $\sfrac{M}{L}$ and the apparent axis ratio $\sfrac{R_{min}}{R_{max}}$
for 40 galaxies for which NED distances not based on redshift are
available. The linear correlation slope is $-20.10\pm4.14$.
}
\end{figure}

Although we assume in this appendix that the correct distances are
given by the redshifts, we show in Fig.~\ref{fig:ml1_apar_noz} the
result using NED distances not based on redshifts. The vertical scale
changes following $\sfrac{M}{L}\rightarrow \sfrac{M}{L}\times$(distance without redshift)/(distance
from redshift). In that case, since the no-redshift distances
are systematically larger than the distances estimated from redshifts
(see Fig.~\ref{fig:dmbms_dmnednox}), the correlation is even stronger,
with a slope of $-20.10\pm4.14$ i.e, a 5$\sigma$ effect. (We note
that the 40 galaxies that have no-redshift distances available   already had a larger correlation, even before the distance
correction is applied to $\sfrac{M}{L}$.) The Pearson coefficient, after removing
data with uncertainty of $\sfrac{M}{L}$ greater than 5 (this reduces the sample
to  25 galaxies), is $r=-0.440$.%

\subsection{Correlations\label{sec:Correlations}}

Studying correlations is critical since they can bias 
the studied correlation. Possible measurement or observation biases not
directly related to $\sfrac{M}{L}$ and $\sfrac{R_{min}}{R_{max}}$ can propagate
to them {\it{via}} correlations. Furthermore, a (not understood) correlation
can be removed either by sample selection  or be mathematically corrected
for. However, this must be done carefully: if this correlation is
actually a consequence of a $\sfrac{M}{L}$ dependence with $\sfrac{R_{min}}{R_{max}}$
propagating {\it{via}} other correlations, correcting or selecting it could
wrongly suppresses the $\sfrac{M}{L}\longleftrightarrow \sfrac{R_{min}}{R_{max}}$ correlation.
Thus, it is important to check the correlations between variables
characterizing an elliptical galaxy and make sure they do not originate
from measurement or observation biases. It should then be checked
what is the effect of these correlation on $\sfrac{M}{L}$ vs $\sfrac{R_{min}}{R_{max}}$. 

Except in one occasion, we do not investigate the possibility that
characteristics other than the ones given in~\cite{BMS} are correlated
to both $\sfrac{M}{L}$ and $\sfrac{R_{min}}{R_{max}}$. 
%and thus could introduce a misleading $\sfrac{M}{L}$ vs $\sfrac{R_{min}}{R_{max}}$ correlation. 
This could certainly
occur and it is a limitation of our analysis to keep in mind. (The
exception mentionned above is the metallicity content of elliptical
galaxies, see Section~\ref{sub:Absolute-magnitude-Mb corel}.)

\begin{figure}
\centering
\includegraphics[scale=0.5]{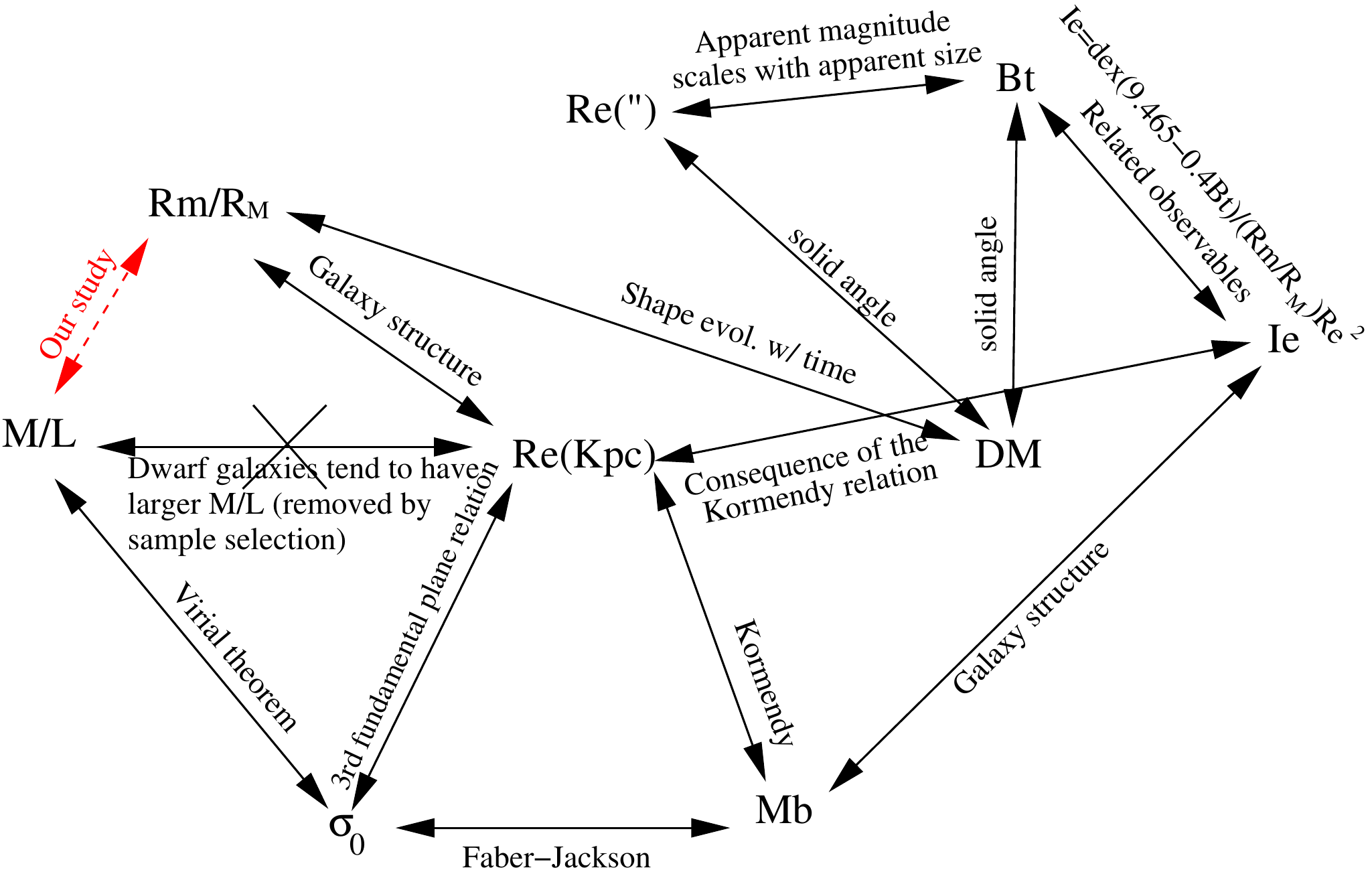}
\vspace{-0.4cm} \caption{\label{fig:correlations}Expected correlations. The red dashed arrow is the possible correlation
investigated in this study.
}
\end{figure}

\subsubsection{Expected correlations}
\begin{itemize}
\item Effective radius $Re$ (Kpc) vs velocity dispersion $\sigma_{0}$;
effective radius $Re$ (Kpc) vs absolute blue magnitude $M_{b}$;
central velocity dispersion $\sigma_{0}$ vs absolute blue magnitude
$M_{b}$: those are the empirical fundamental plane relations~\cite{Binney},
respectively the third fundamental plane relation; the Kormendy relation~\cite{Kormendy}
and the Faber-Jackson relation~\cite{Faber-Jackson}. 
\item Distance Moduli $DM$ vs effective radius $Re(")$ and integrated
blue magnitude $B_{t}$: they are trivial reductions of the apparent intensities
of quantities with their distance to the observer%
\footnote{If the $DM$ span is large, i.e, we observe galaxies over a large
time span over which they have time to evolve, we would also expect
a correlation between $DM$ and the galaxy characteristics, e.g. the
absolute blue magnitude $M_{b}$, effective radius $Re(Kpc)$, axis
ratio $\sfrac{R_{min}}{R_{Max}}$ or velocity dispersion $\sigma_{0}$. Decrease
of galaxy magnitudes ($M_{b}$) with time is established. Structure
evolution with time ($Re$) has been observed by comparing high redshift
galaxies to local ones~\cite{Re evol. w/ time}: older massive
elliptical galaxies are more compact than younger ones. This would
imply a $DM$-dependence of the axis ratio as well since the decrease of the galaxy
size  with time increases the rotation speed. Finally, it
is known that relaxation with time reduces the central velocity
dispersion. However, in our sample, $DM$ varies between $\sim30$
and $\sim35.5$., which corresponds to distances of 33 Megalight-years
(Mly) and 411 Mly. The time span considered is thus 378 Myears. Since
the galaxies were formed long before 411 Myears ago,
and since we removed from our sample galaxies displaying signs of
mergers or disturbances, galaxies had time to relax to their equilibrium.
Consequently, we do not expect the sample galaxies to evolve significantly
during the relatively short 378 Myears time span.%
}.
\item Apparent blue magnitude $B_{t}$ vs apparent radius $Re(")$:
the larger the apparent radius is, the higher the apparent magnitude
tends to be.
\item Surface Brightness $I_{e}$ vs absolute magnitude $M_{b}$. {\it Central}
surface brightness vs absolute blue magnitude correlation has
been observed~\cite{Binggeli}. 
\item Integrated blue magnitude $B_{t}$ vs surface brightness $I_{e}$:
these two quantities describing the luminosity of a galaxy are related
by the equation given in Fig.~\ref{fig:correlations}, where $Re$
is in arcsec. 
\item Axis ratio $\sfrac{R_{min}}{R_{Max}}$ vs effective radius $Re(Kpc)$: this
reflects the galaxy structure, with small galaxies tending to be more elongated.
\item Mass to light ratio $\sfrac{M}{L}$ vs effective radius $Re(Kpc)$: 
at both ends of the mass spectrum, dwarf and giant (e.g. BrClg)
galaxies tend to have higher $\sfrac{M}{L}$. However, those are excluded from
our sample. Consequently, we do not expect a strong correlation here. 
\item Mass to light  ratio $\sfrac{M}{L}$ vs velocity dispersion $\sigma_{0}$.
The virial theorem links potential and kinetic energy. 
\end{itemize}

\subsubsection{Correlation study }

In this section we used arrow symbols with the following meanings:
\begin{itemize}
\item $\Longleftrightarrow$: strong correlation;
\item $\longleftrightarrow$: clear correlation;
\item $\dashleftarrow\dashrightarrow$: clear but weaker correlation; 
\item $\leftarrow?\rightarrow$: unclear correlation. Weak if it exists.
\end{itemize}
The correlation strengths are summarized in Figs.~\ref{correlation table. } (summary
table),~\ref{fig:correlations 2} (clear correlations) and~\ref{fig:correlations 3} (unclear
weak correlations).

\paragraph{Correlations with apparent $\sfrac{R_{min}}{R_{max}}$ from
\cite{BMS} \label{sub:Correlation R/R}}
\begin{figure}[H]
\centering
\includegraphics[scale=0.24]{apratned_apratbms}\includegraphics[scale=0.24]{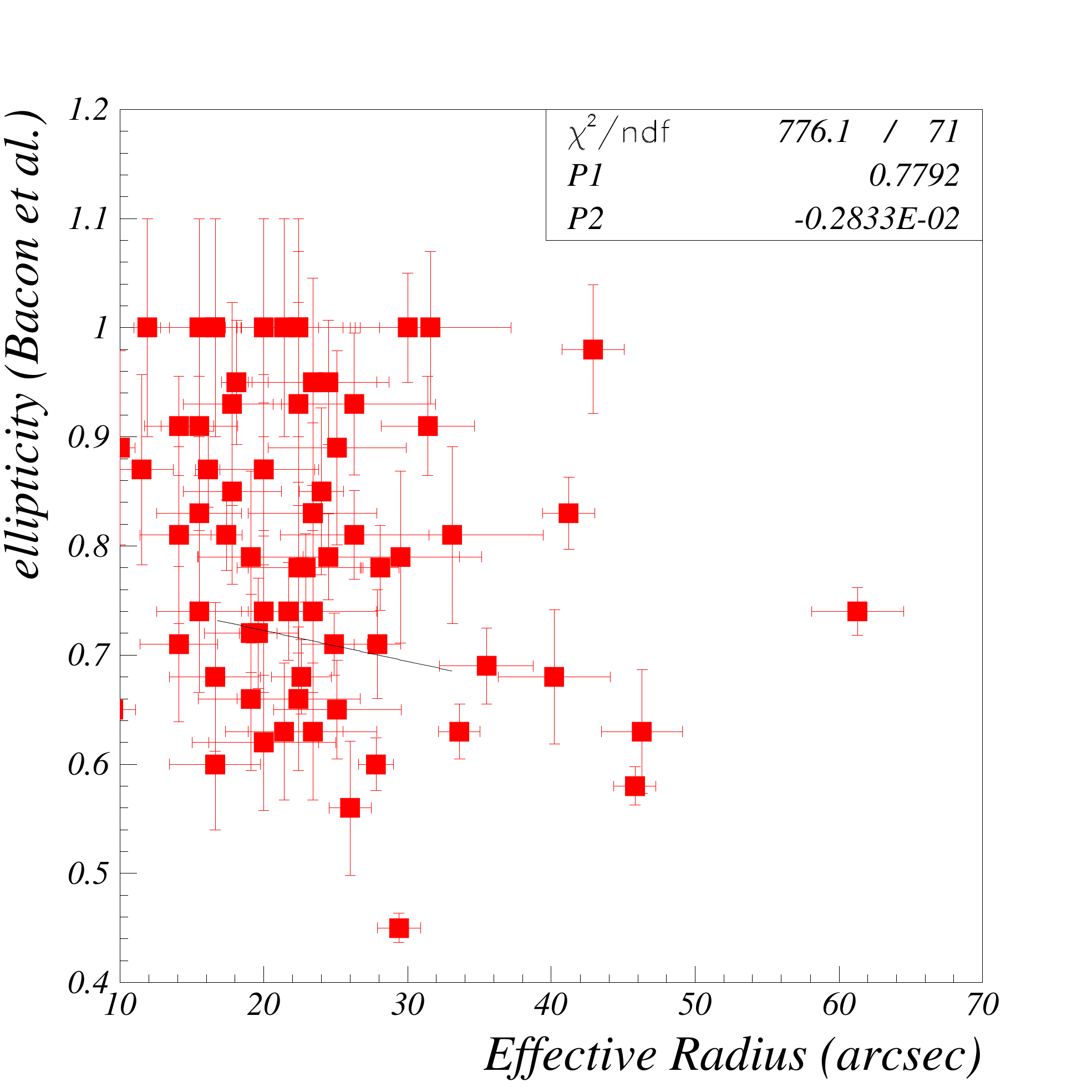}\includegraphics[scale=0.24]{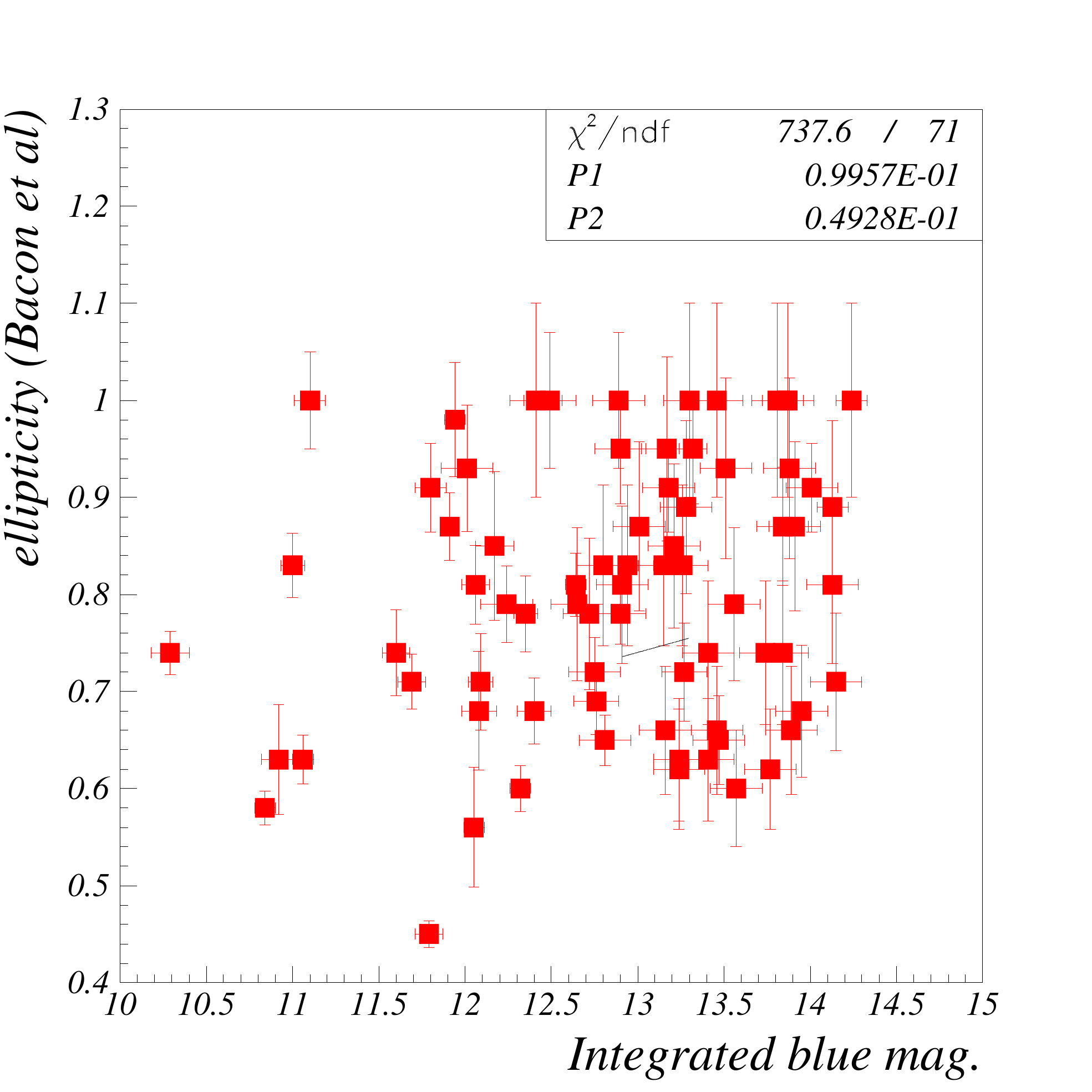}\includegraphics[scale=0.24]{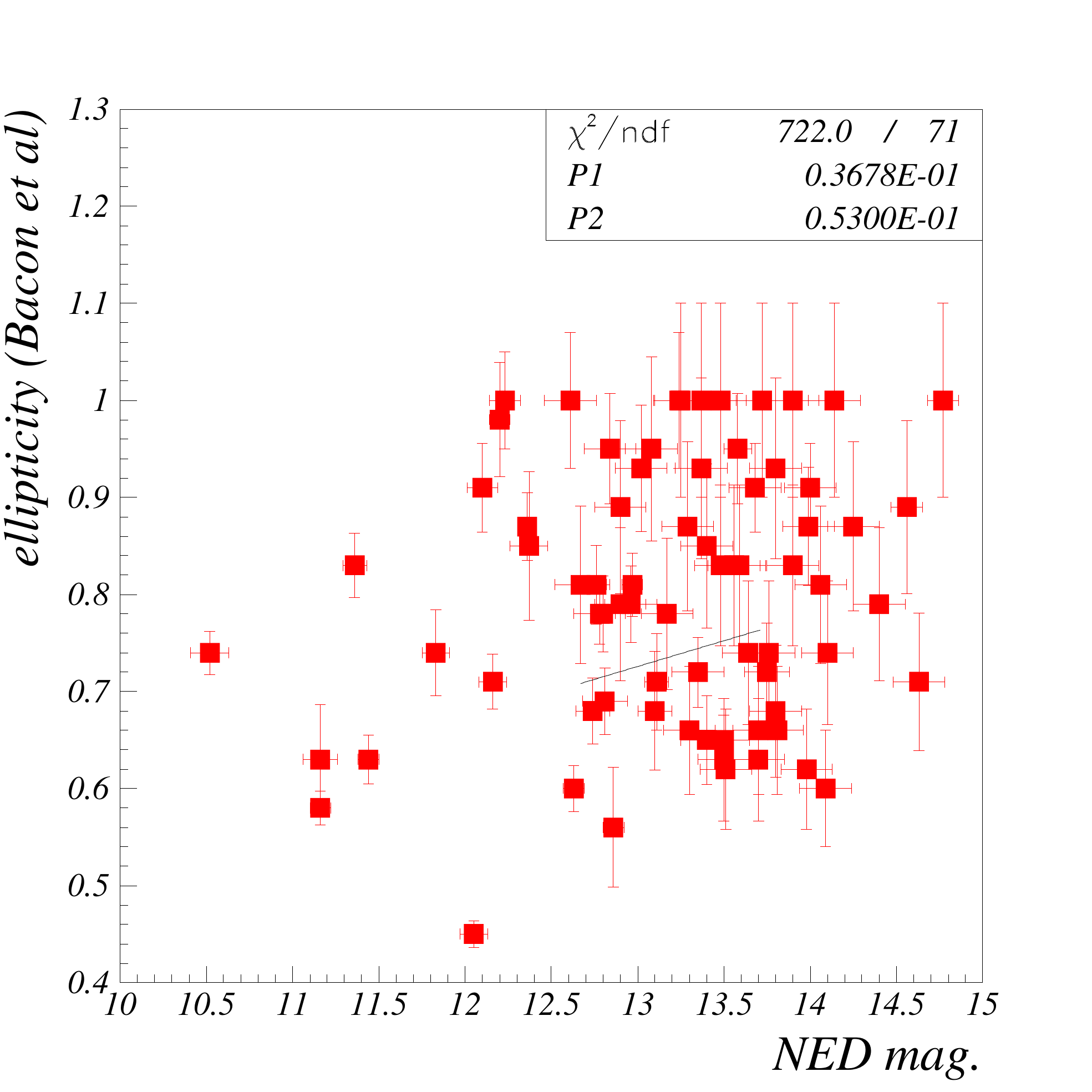}\protect \\
\includegraphics[scale=0.24]{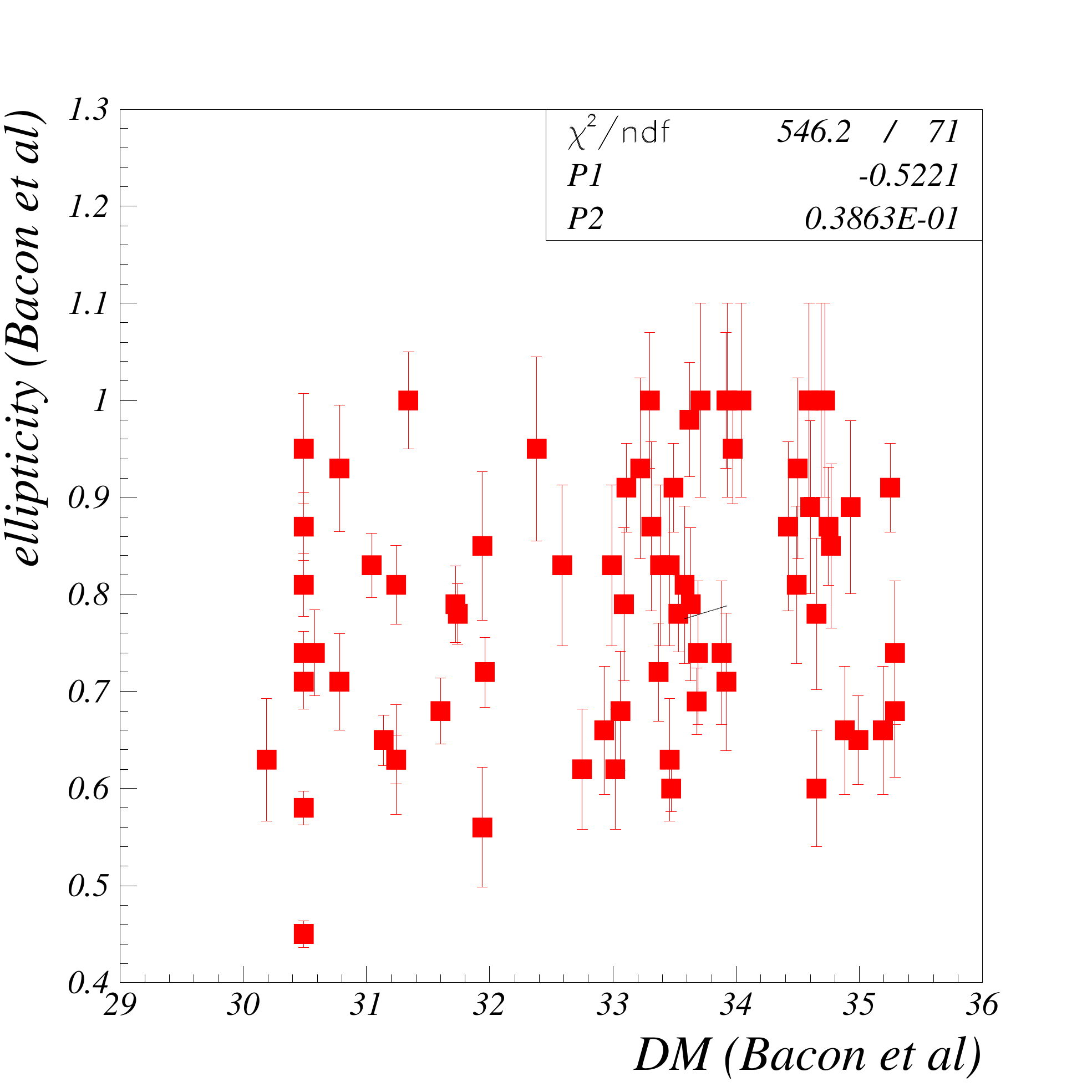}\includegraphics[scale=0.24]{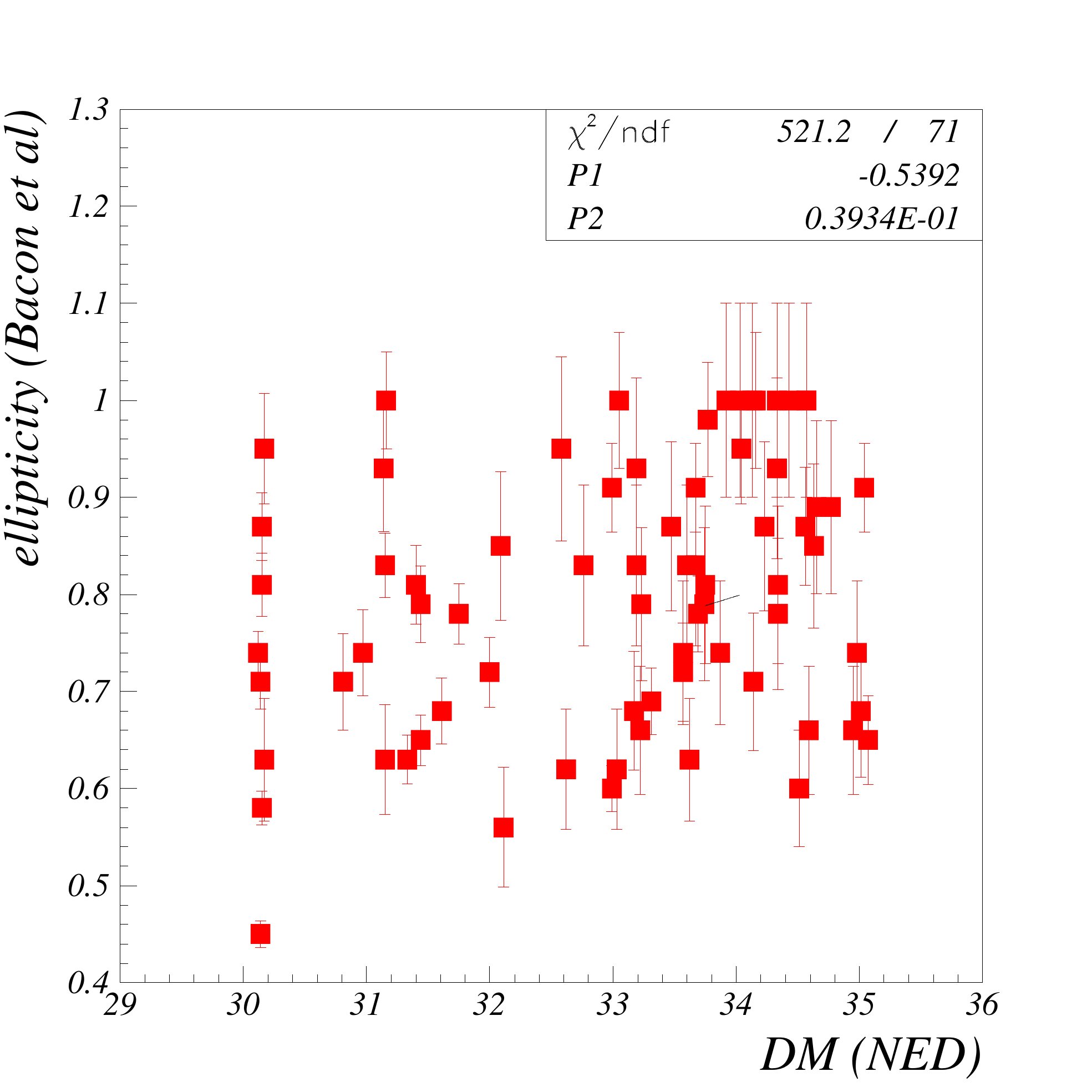}\includegraphics[scale=0.24]{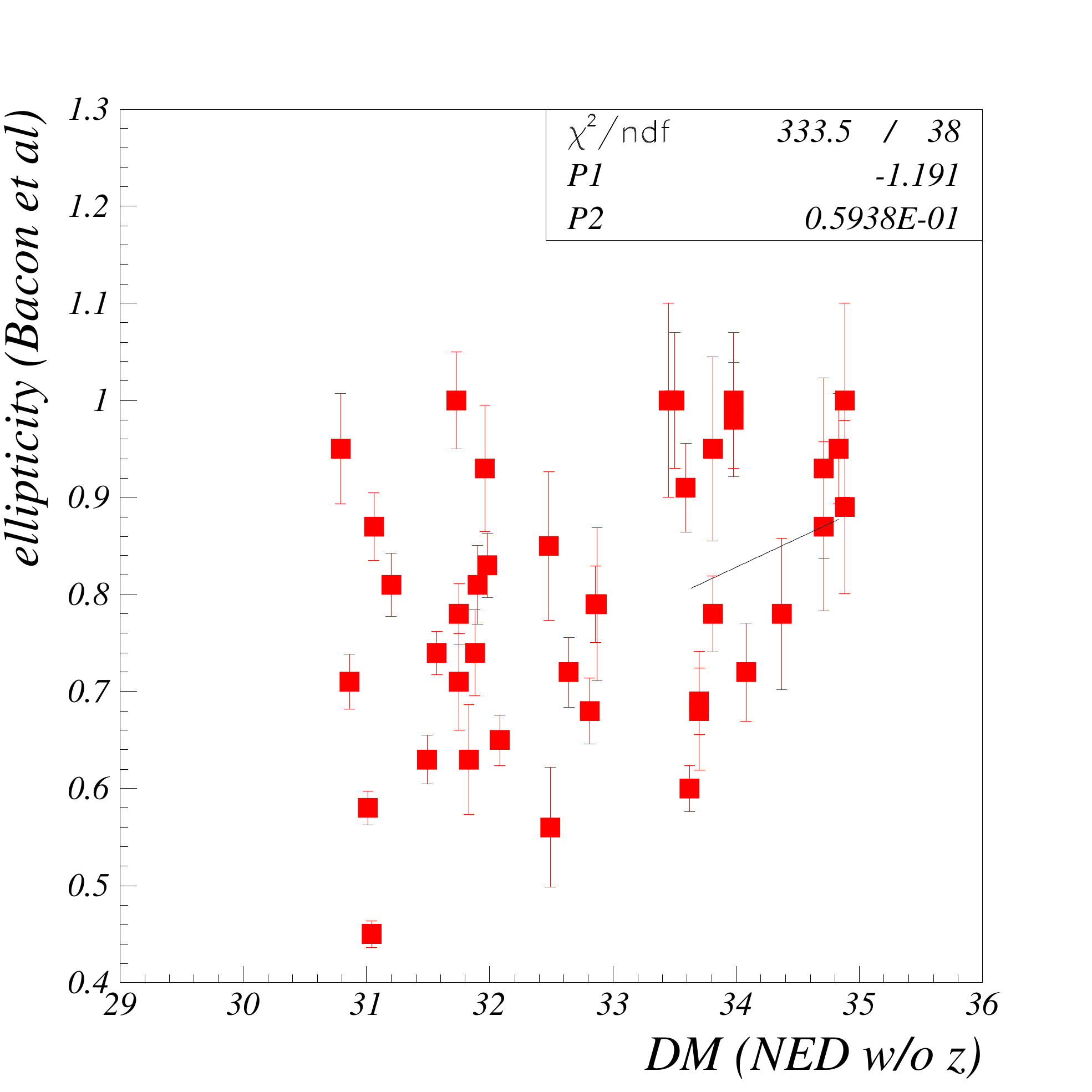}\includegraphics[scale=0.24]{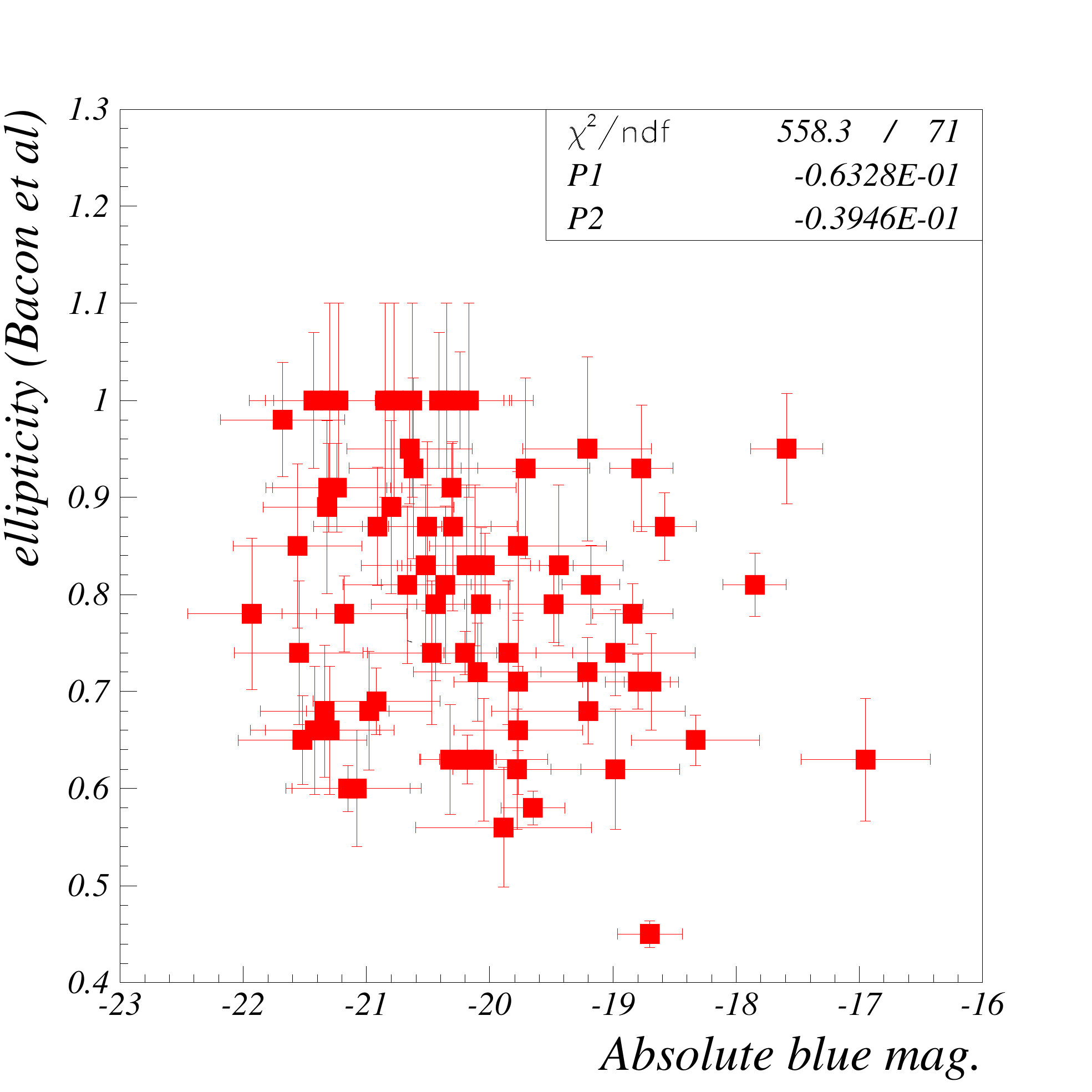}\protect \\
\includegraphics[scale=0.24]{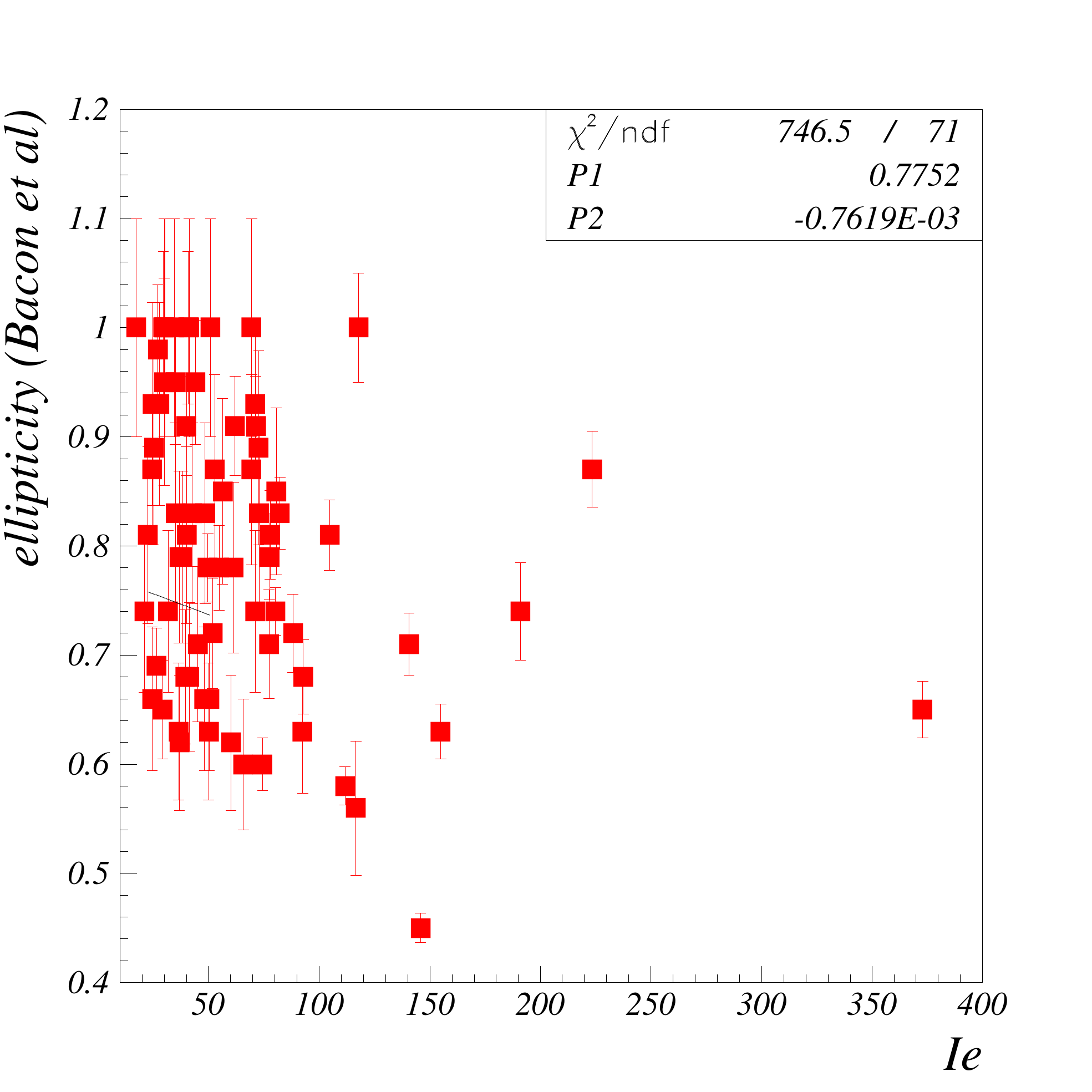}\includegraphics[scale=0.24]{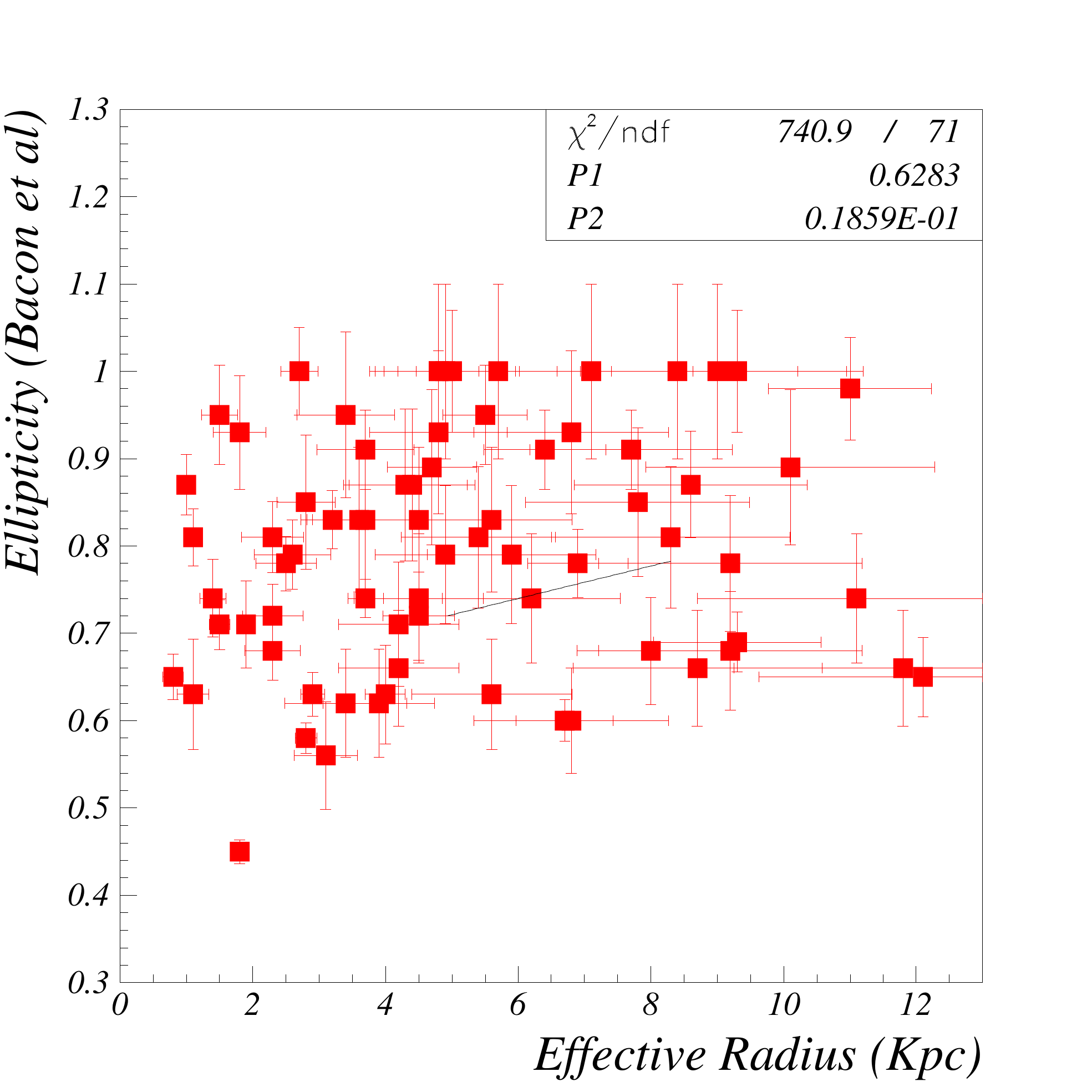}\includegraphics[scale=0.24]{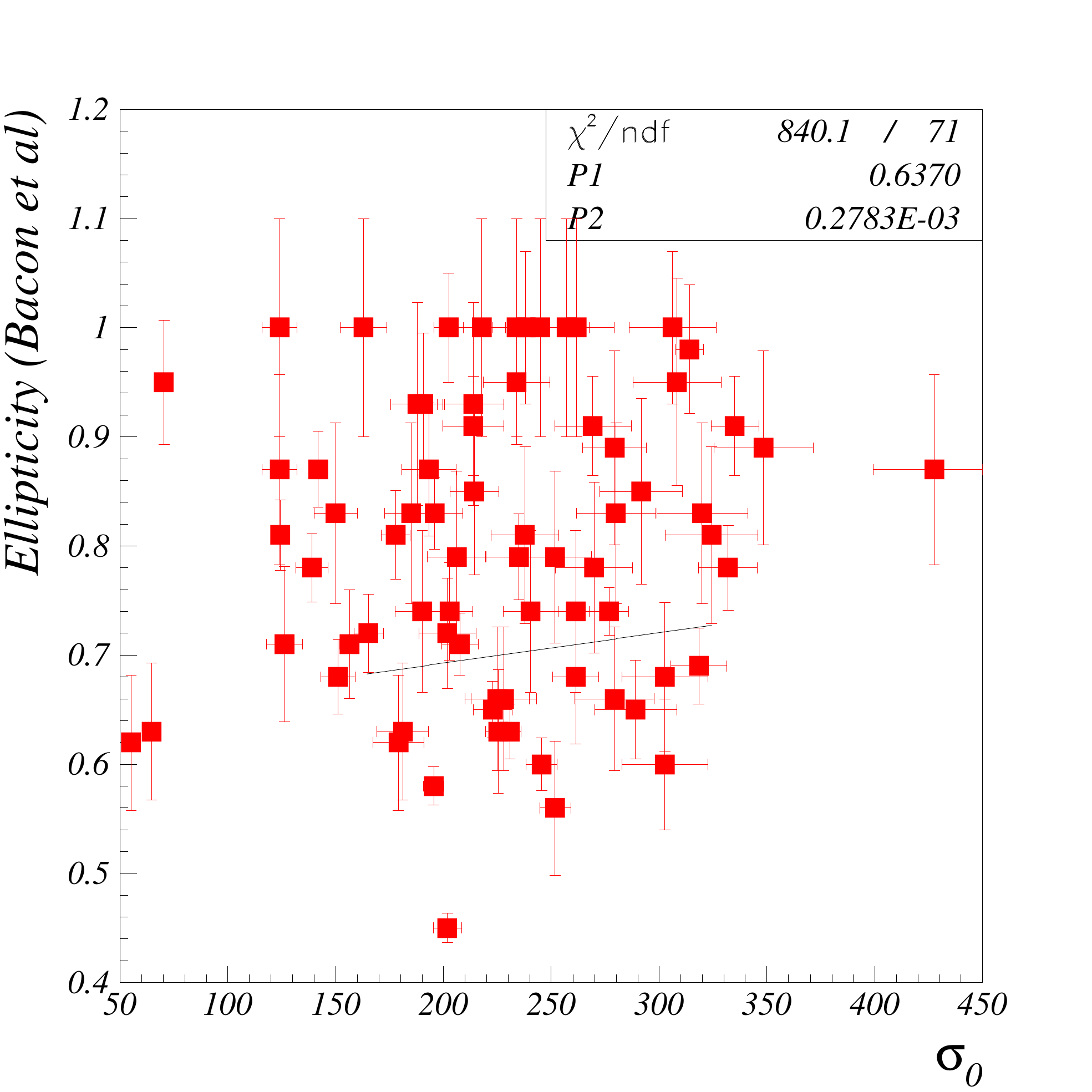}
\vspace{-0.4cm} \caption{\label{Fig: ellipticity correlations}Correlations between 
apparent $\sfrac{R_{min}}{R_{max}}$ from~\cite{BMS} and (from top
left to bottom right): $\sfrac{R_{min}}{R_{max}}$ from NED, apparent effective
radius $Re(")$, integrated blue magnitude $B_{t}$, magnitude from
NED,  $DM$ from~\cite{BMS}, $DM$
from NED using redshift information, $DM$ from NED without
redshift information, absolute blue magnitude, surface brightness
$I_{e}$, effective radius (parsec) and velocity distribution $\sigma_{0}$.
}
\end{figure}
The origins of the correlations are, from the top left plot to the
bottom right one: 
\begin{enumerate}
\item Apparent axis ratio from~\cite{BMS} vs NED apparent axis
ratio: trivial correlation between different measurements of the same
quantity. This is already discussed in Fig.~\ref{fig:apratned_apratbms}.
\item Apparent axis ratio from~\cite{BMS} vs apparent effective
radius $Re(")$: this possible weak correlation may be
a measurement/observation bias. The more recent data from NED display
a similar, even slightly stronger correlation. It could
be a consequence of the $\sfrac{R_{min}}{R_{Max}}\longleftrightarrow$$DM\Longleftrightarrow$$Re(\lyxmathsym{\textquotedblleft})$
correlations. Using the fit results%
\footnote{This type of calculation is merely indicative, given the large $\sfrac{\chi^{2}}{ndf}$
of the fits. %
} $(3.9\pm0.4)E^{-2}DM=\sfrac{R_{min}}{R_{Max}}+c$ and $(-2.1\pm0.2)DM=Re(")+c'$
yields $\sfrac{R_{min}}{R_{Max}}=(-1.9\pm1.6)E^{-2}Re(")+c"$, in agreement with
the observed $\sfrac{R_{min}}{R_{Max}}=(-0.28\pm0.04)E^{-2}Re(")+C"$. Since $ $$\sfrac{R_{min}}{R_{Max}}\longleftrightarrow DM$
is a measurement bias, $\sfrac{R_{min}}{R_{Max}}\dashleftarrow\dashrightarrow$
$Re(")$ is an indirect measurement bias.
\item Apparent axis ratio from~\cite{BMS} vs integrated blue magnitude
$B_{t}$: this weak correlation may be  a measurement/observation
bias. The more recent data from NED, in particular the NED apparent
axis ratio vs magnitude from NED in Fig.~\ref{Fig: NED ellipticity correlations},
display similar correlations. This correlation could be a consequence
of the $\sfrac{R_{min}}{R_{Max}}\dashleftarrow\dashrightarrow DM$$\Longleftrightarrow$
$B_{t}$ correlations. Using the fit results $(3.9\pm0.4)E^{-2}DM=\sfrac{R_{min}}{R_{Max}}+c$
and $(1.2\pm0.0)B_{t}=DM+c'$ yields $\sfrac{R_{min}}{R_{Max}}=(4.7\pm0.5)E^{-2}B_{t}+c"$,
in  agreement with the observed $\sfrac{R_{min}}{R_{Max}}=(4.9\pm0.5)E^{-2}B_{t}+C"$.
Thus, $\sfrac{R_{min}}{R_{Max}}\dashleftarrow\dashrightarrow B_{t}$ is an indirect
measurement bias.
\item Apparent axis ratio from~\cite{BMS} vs magnitude from NED:
same as above.
\item Apparent axis ratio from~\cite{BMS} vs 
$DM$ from~\cite{BMS}: this weak correlation could be a measurement
bias: the further the galaxy, the harder it is to observe and so the
rounder it tends to appear due to resolution. The more recent data
from NED, in particular the NED apparent axis ratio vs magnitude
from NED in Fig.~\ref{Fig: NED ellipticity correlations}, displays
a somewhat smaller correlations, supporting that it is a measurement
bias. 
\item Apparent axis ratio from~\cite{BMS} vs $DM$
from NED using redshift information: see above.
\item Apparent axis ratio from~\cite{BMS} vs $DM$
from NED without redshift information: see above.
\item Apparent axis ratio from~\cite{BMS} vs absolute blue magnitude:
the possible weak correlation could come from a measurement bias (the
result from NED in Fig.~\ref{Fig: NED ellipticity correlations} has
somewhat smaller correlations). However, it seems to be a consequence
of the $\sfrac{R_{min}}{R_{Max}}\dashleftarrow\dashrightarrow DM\Longleftrightarrow M_{b}$
and $\sfrac{R_{min}}{R_{Max}}\longleftrightarrow \sfrac{M}{L}\dashleftarrow\dashrightarrow M_{b}$
correlations: using the fit results $(3.9\pm0.4)E^{-2}DM=\sfrac{R_{min}}{R_{Max}}+c$
and $(-0.53\pm0.03)DM=M_{b}+c'$ yields $\sfrac{R_{min}}{R_{Max}}=(-7.4\pm1.2)E^{-2}M_{b}+c"$.
Using the fit results $(-6\pm2)\sfrac{R_{min}}{R_{Max}}=\sfrac{M}{L}+c$ and $(-1.7\pm0.3)M_{b}=\sfrac{M}{L}+c'$
yields $\sfrac{R_{min}}{R_{Max}}=(0.28\pm0.14)M_{b}+c"$. Adding both results yields
$\sfrac{R_{min}}{R_{Max}}=(0.20\pm0.16)M_{b}+c"$, in reasonable agreement with
the observed $\sfrac{R_{min}}{R_{Max}}=(-3.9\pm0.6)E^{-2}M_{b}+c"$, suggesting
that the correlation is an indirect measurement bias. Alternatively,
we note that a correlation between axis ratio and mass is known: heavier
(hence more luminous) galaxies tend to be rounder. Although we have
rejected the heaviest elliptical galaxies, we may be seing here the
same correlation. 
\item Apparent axis ratio from~\cite{BMS} vs surface brightness
$I_{e}$: this clear correlation cannot be a (direct) measurement
bias: it does not have the expected pattern (the dimer, the more difficult
the measurement so the larger $\sfrac{R_{min}}{R_{Max}}$ would be). In addition,
the newer NED data indicate a stronger correlation, see Fig.~\ref{Fig: NED ellipticity correlations}.
This correlation could be a consequence of the $I_{e}\longleftrightarrow DM$$\dashleftarrow\dashrightarrow \sfrac{R_{min}}{R_{Max}}$
correlations. Using the fit results $(-1.7\pm0.1)E^{-2}I_{e}=DM+c$
and $(3.9\pm0.4)E^{-2}DM=\sfrac{R_{min}}{R_{Max}}+c'$ yields $\sfrac{R_{min}}{R_{Max}}=(-6.6\pm1.1)E^{-4}I_{e}+c"$,
in  agreement with the observed $\sfrac{R_{min}}{R_{Max}}=(-7.6\pm0.7)E^{-4}I_{e}+C"$.
This supports that the correlation is an indirect measurement
bias.
\item Apparent axis ratio from~\cite{BMS} vs effective radius
(parsec): this possible correlation would agree with the observation
that smaller galaxies tend to be more elongated. However, this correlation
appears to be a consequence of the $Re(Kpc)\longleftrightarrow DM$$\dashleftarrow\dashrightarrow \sfrac{R_{min}}{R_{Max}}$
correlations. Using the fit results $(1.18\pm0.06)DM=Re+c$ and $(3.9\pm0.4)E^{-2}DM=\sfrac{R_{min}}{R_{Max}}+c'$
yields $\sfrac{R_{min}}{R_{Max}}=(3.3\pm0.5)E^{-2}Re+c"$, in  agreement
with the observed $\sfrac{R_{min}}{R_{Max}}=(1.9\pm0.2)E^{-2}I_{e}+C"$. This is
an indirect measurement bias.
\item Apparent axis ratio from~\cite{BMS} vs velocity distribution
$\sigma_{0}$. No significant correlation is seen. The newer data
from NED show no correlation. We expected that $\sfrac{R_{min}}{R_{Max}}\longleftrightarrow \sfrac{M}{L}\Longleftrightarrow\sigma_{0}$,
$\sfrac{R_{min}}{R_{Max}}\dashleftarrow\dashrightarrow DM\longleftrightarrow\sigma_{0}$,
$\sfrac{R_{min}}{R_{Max}}\dashleftarrow\dashrightarrow Re(Kpc)\Longleftrightarrow\sigma_{0}$
and $\sfrac{R_{min}}{R_{Max}}\dashleftarrow\dashrightarrow M_{b}\Longleftrightarrow\sigma_{0}$,
would contribute to secondary correlations: using the fit results
$(-6\pm2)\sfrac{R_{min}}{R_{Max}}=\sfrac{M}{L}+c$ and $(5.4\pm0.4)E^{-2}\sigma_{0}=\sfrac{M}{L}+c'$
yields $\sfrac{R_{min}}{R_{Max}}=(-9.0\pm3.7)E^{-3}\sigma_{0}+c"$. Using the fit
results $(3.9\pm0.4)E^{-2}DM=\sfrac{R_{min}}{R_{Max}}+c$ and $(1.24\pm0.08)E^{-2}\sigma_{0}=DM+c'$
yields $\sfrac{R_{min}}{R_{Max}}=(4.8\pm0.8)E^{-4}\sigma_{0}+c"$. Using the fit
results $(1.9\pm0.2)E^{-2}Re=\sfrac{R_{min}}{R_{Max}}+c$ and $(1.67\pm0.08)E^{-2}\sigma_{0}=Re+c'$
yields $\sfrac{R_{min}}{R_{Max}}=(3.17\pm0.5)E^{-4}\sigma_{0}+c"$. Using the fit
results $(-4.0\pm0.6)E^{-2}M_{b}=\sfrac{R_{min}}{R_{Max}}+c$ and $(-1.29\pm0.08)E^{-2}\sigma_{0}=M_{b}+c'$
yields $\sfrac{R_{min}}{R_{Max}}=(5.2\pm1.1)E^{-4}\sigma_{0}+c"$. Naively adding
linearly the four results yields $\sfrac{R_{min}}{R_{Max}}=(-7.7\pm4.0)E^{-3}\sigma_{0}+c"$,
a 2 sigma disagreement with the observed $\sfrac{R_{min}}{R_{Max}}=(0.3\pm0.1)E^{-3}\sigma_{o}+c"$.
However, these results cannot be simply added linearly because of
the strong correlation between $Re(Kpc)$ \& $M_{b}$. Also, $\sfrac{M}{L}$
\& $M_{b}$, $DM$ \& $M_{b}$ and $Re(Kpc)$ \& $DM$ show significant
correlations. We assume that overall the four results mostly cancel
each other as suggested by the opposite signs of the correlations.
\end{enumerate}

\paragraph{Correlation with apparent  $\sfrac{R_{min}}{R_{max}}$ from
NED}
The origins of the correlations are discussed in the previous section
(Section~\ref{sub:Correlation R/R}).
\begin{figure}[H]
\centering
\includegraphics[scale=0.24]{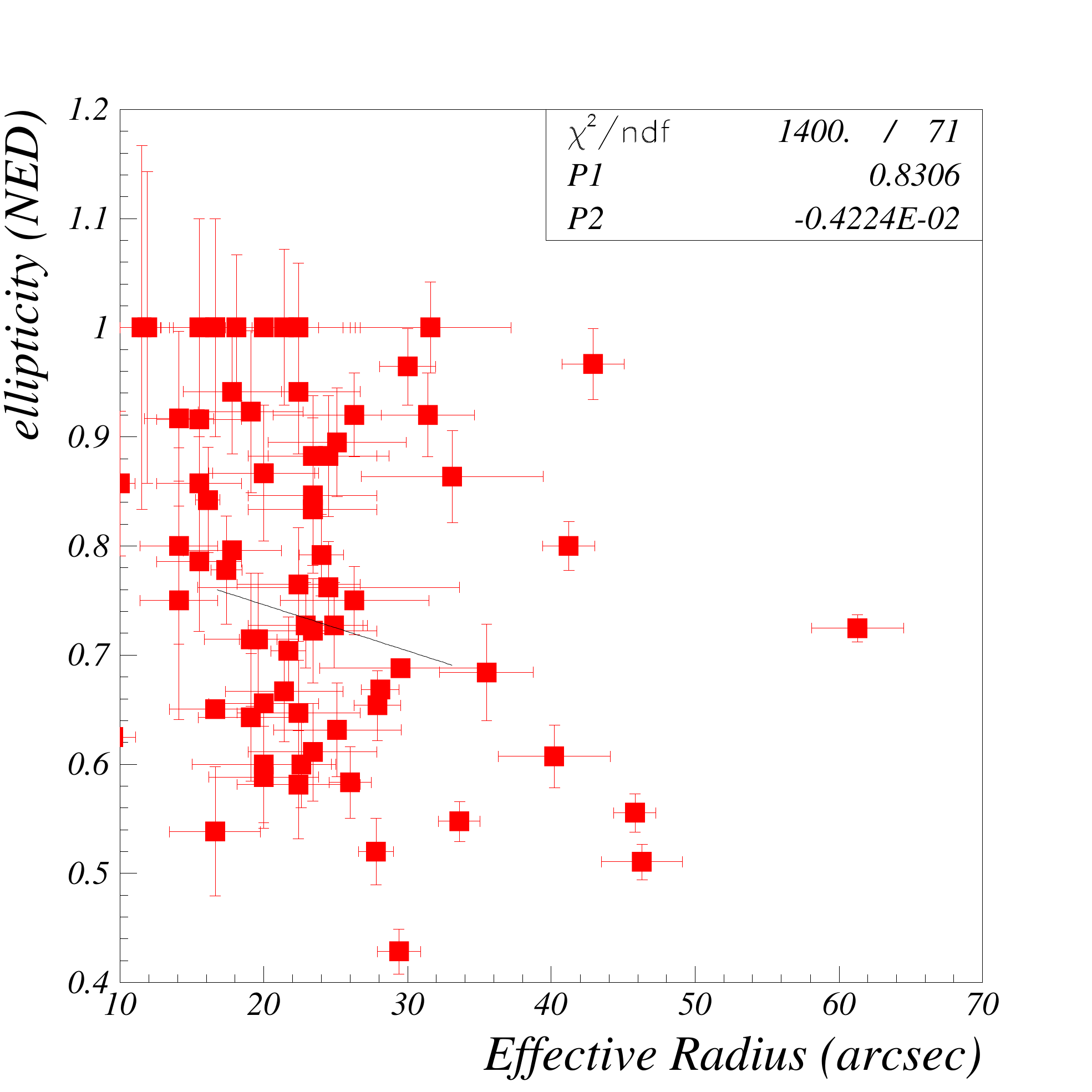}\includegraphics[scale=0.24]{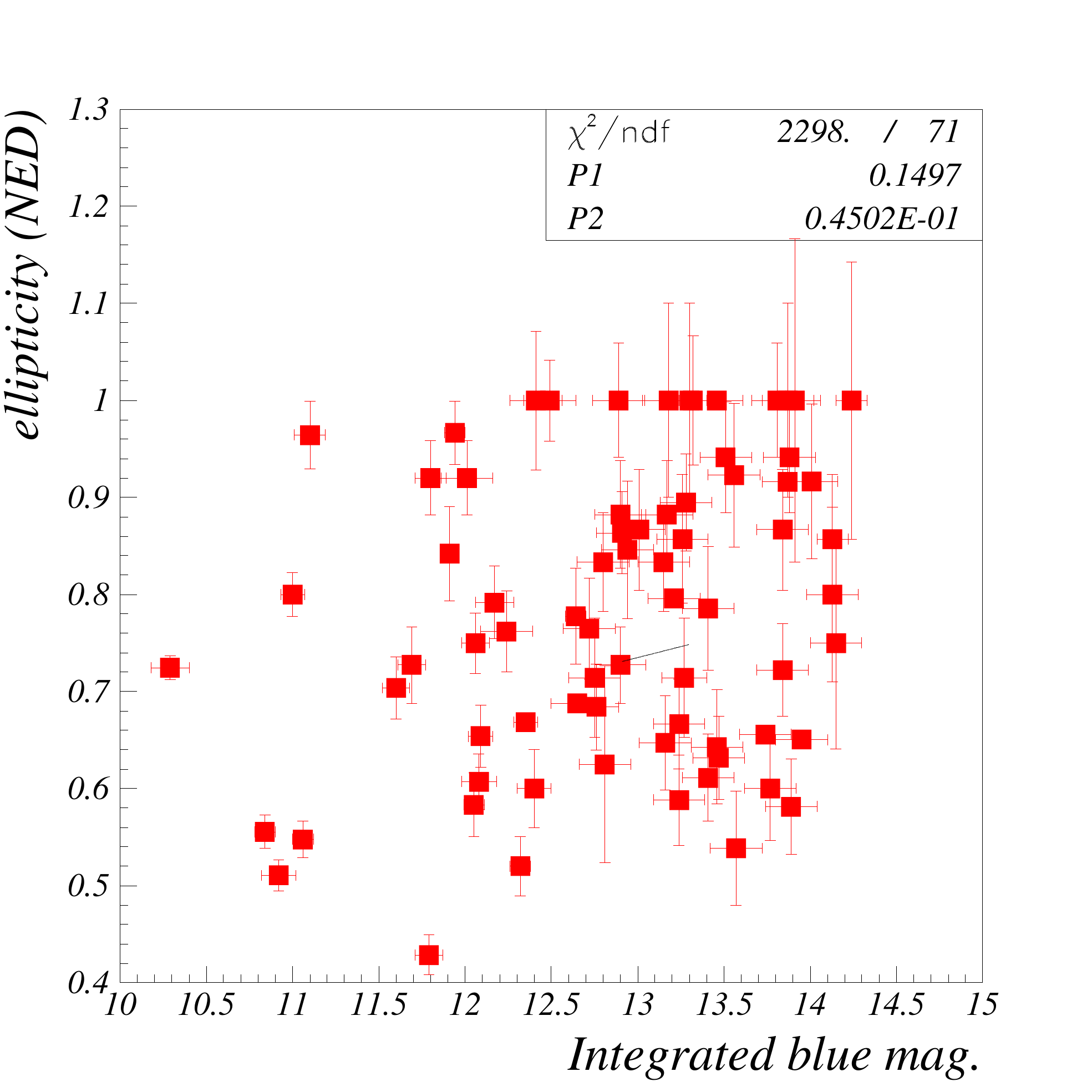}\includegraphics[scale=0.24]{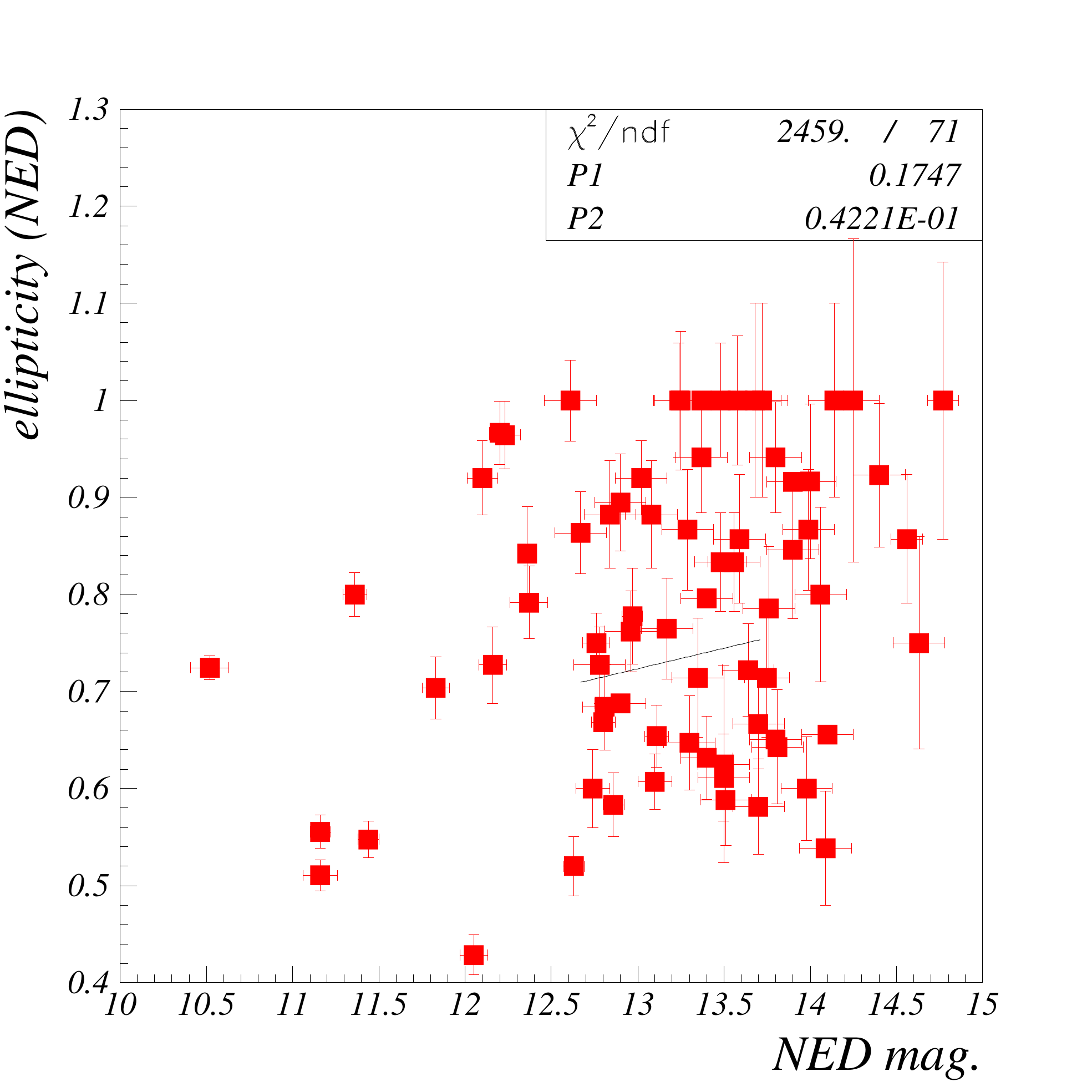}\includegraphics[scale=0.24]{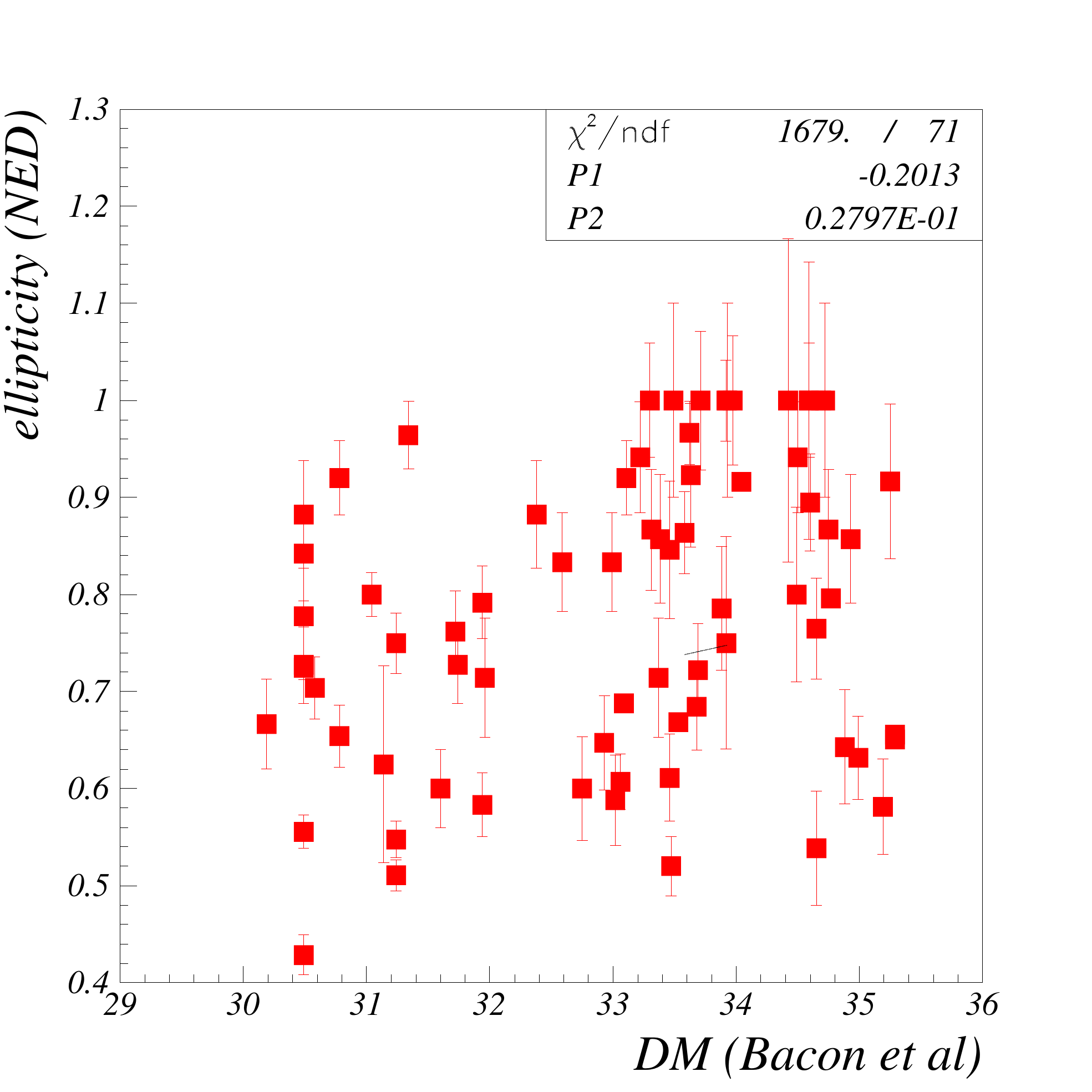}\protect \\
\includegraphics[scale=0.24]{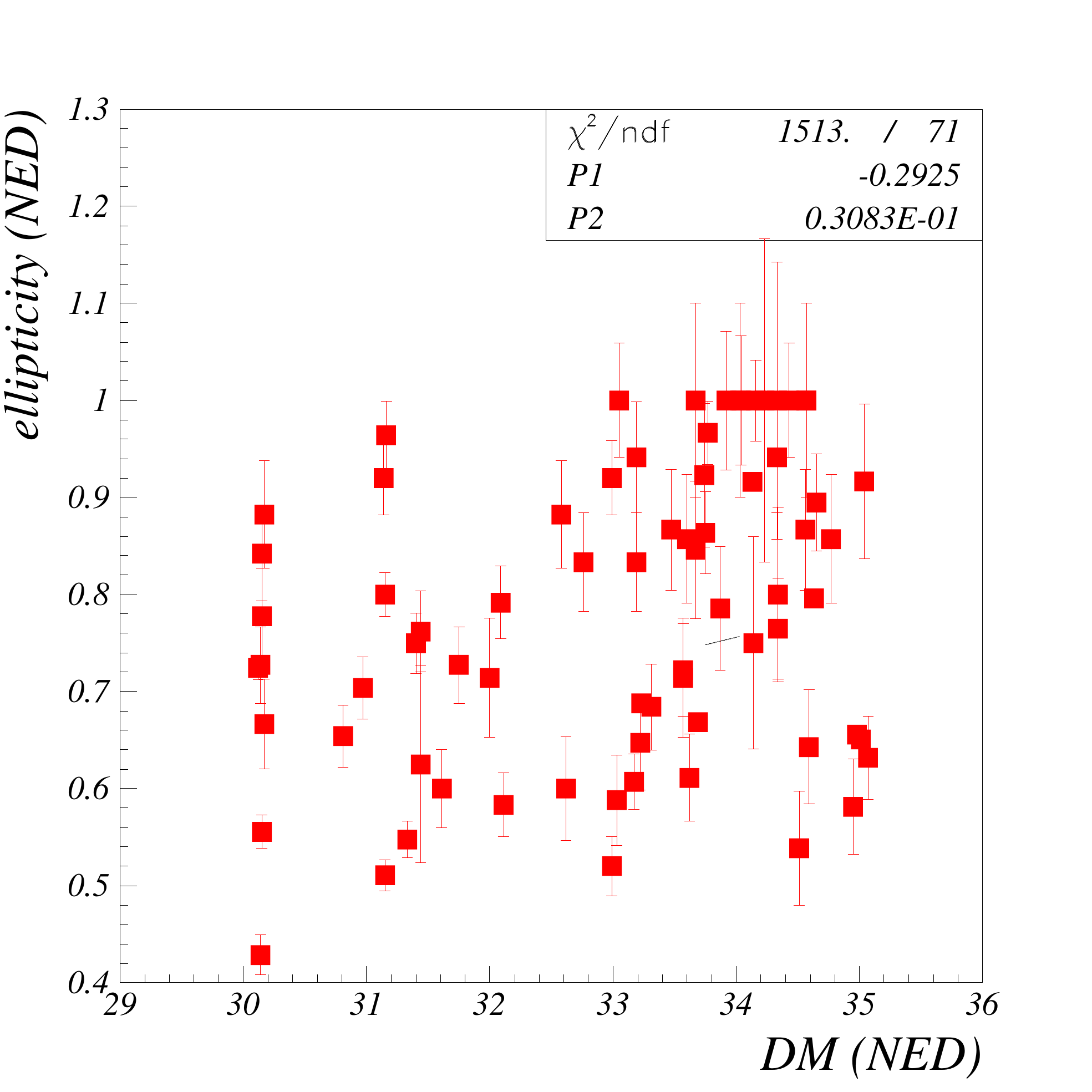}\includegraphics[scale=0.24]{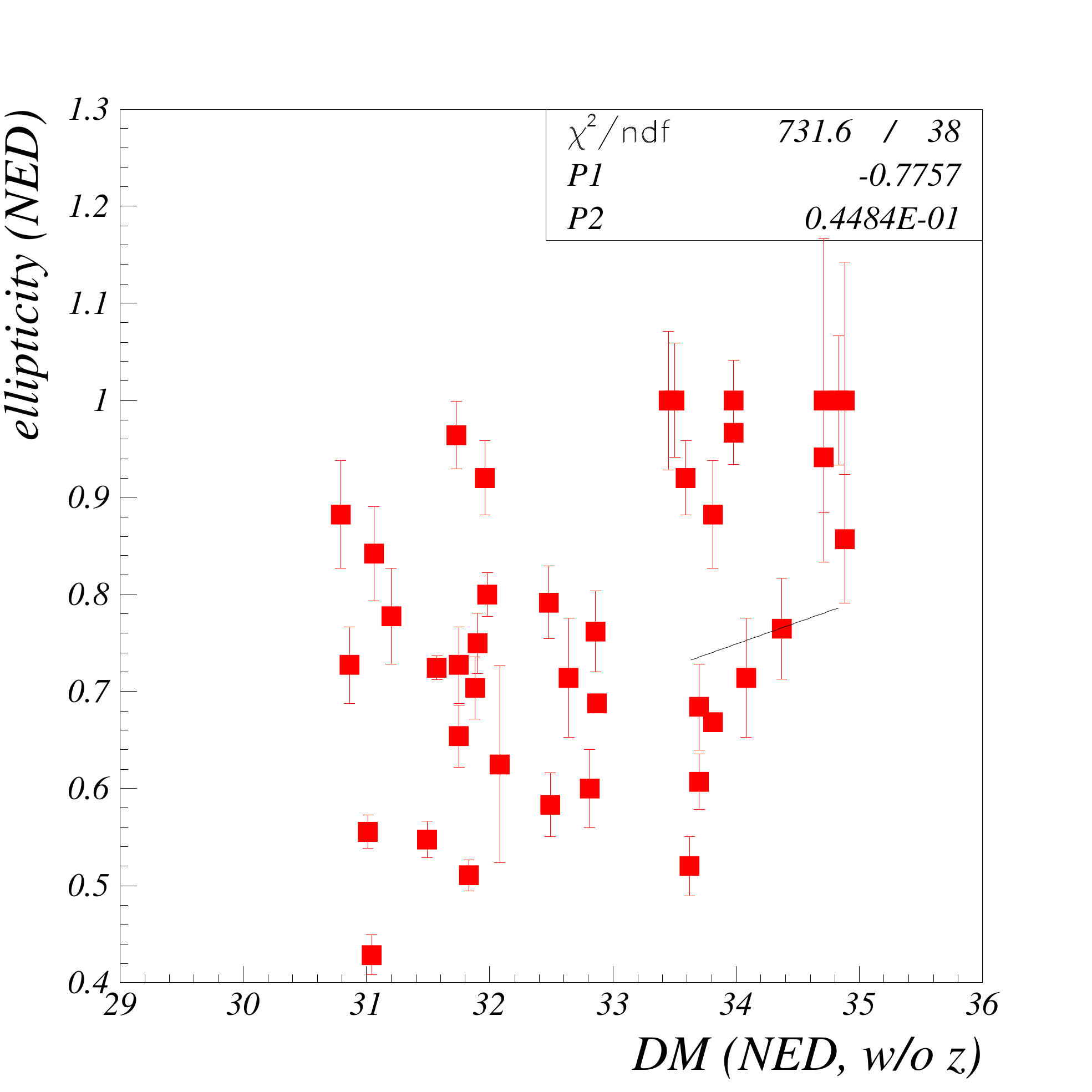}\includegraphics[scale=0.24]{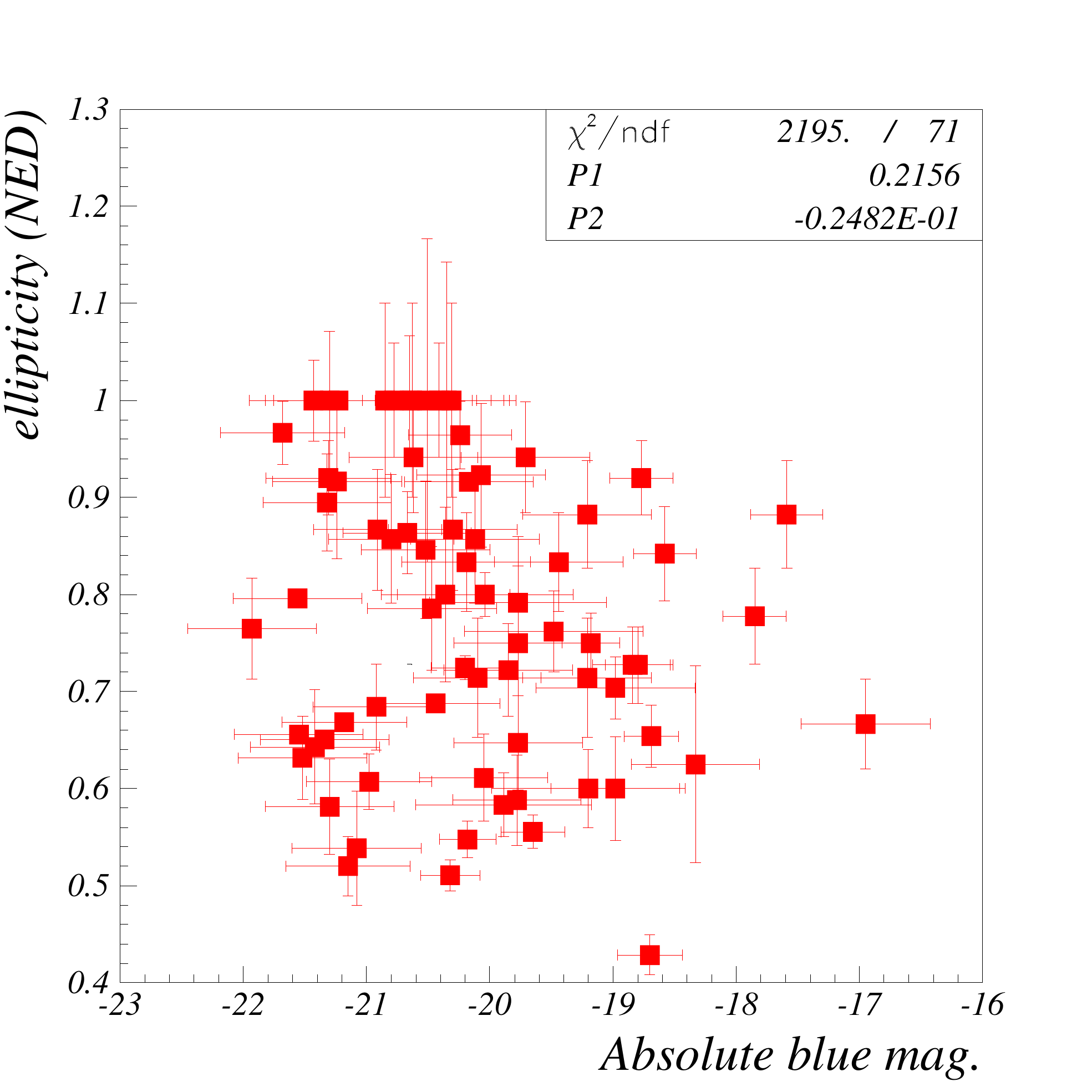}\includegraphics[scale=0.24]{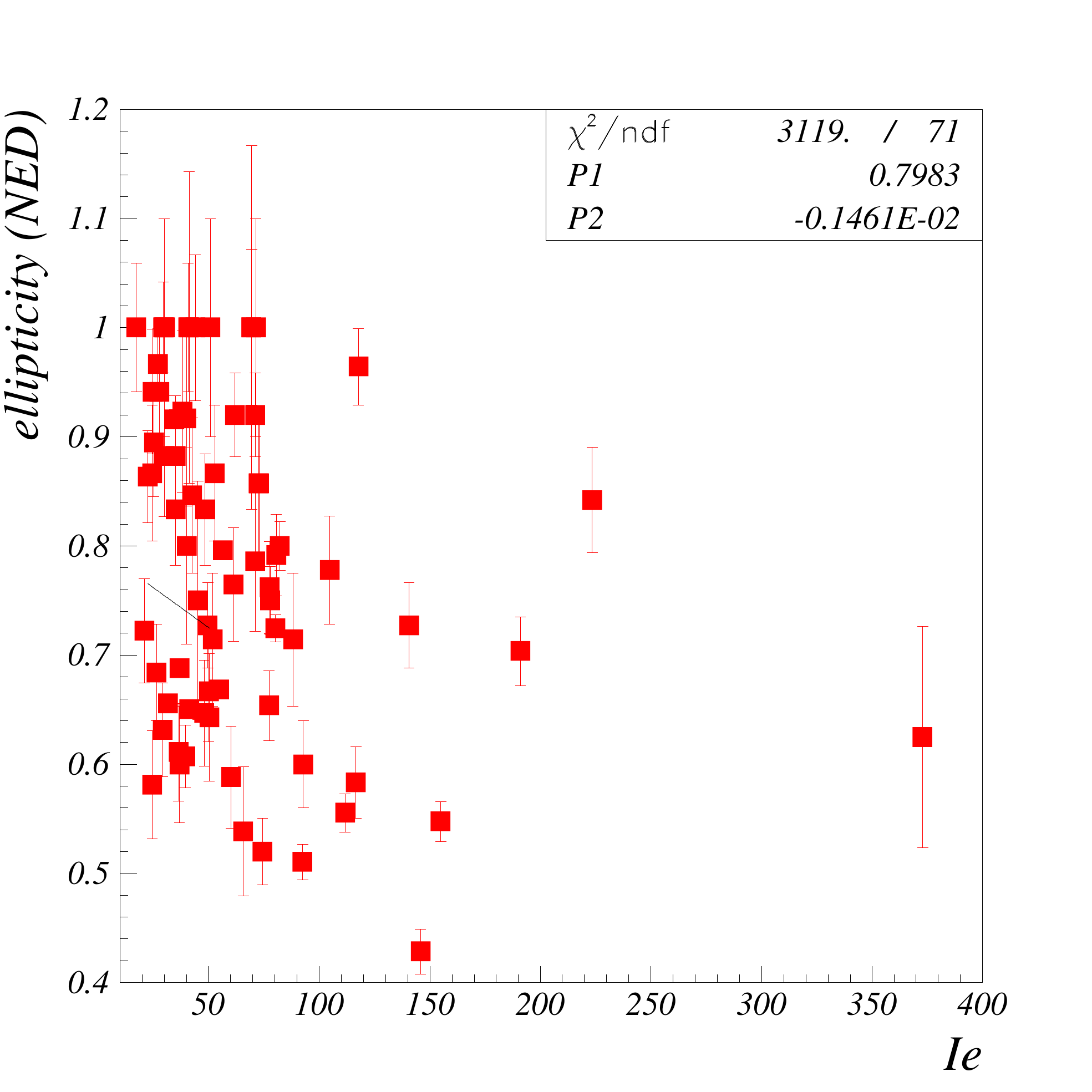}\protect \\
\includegraphics[scale=0.24]{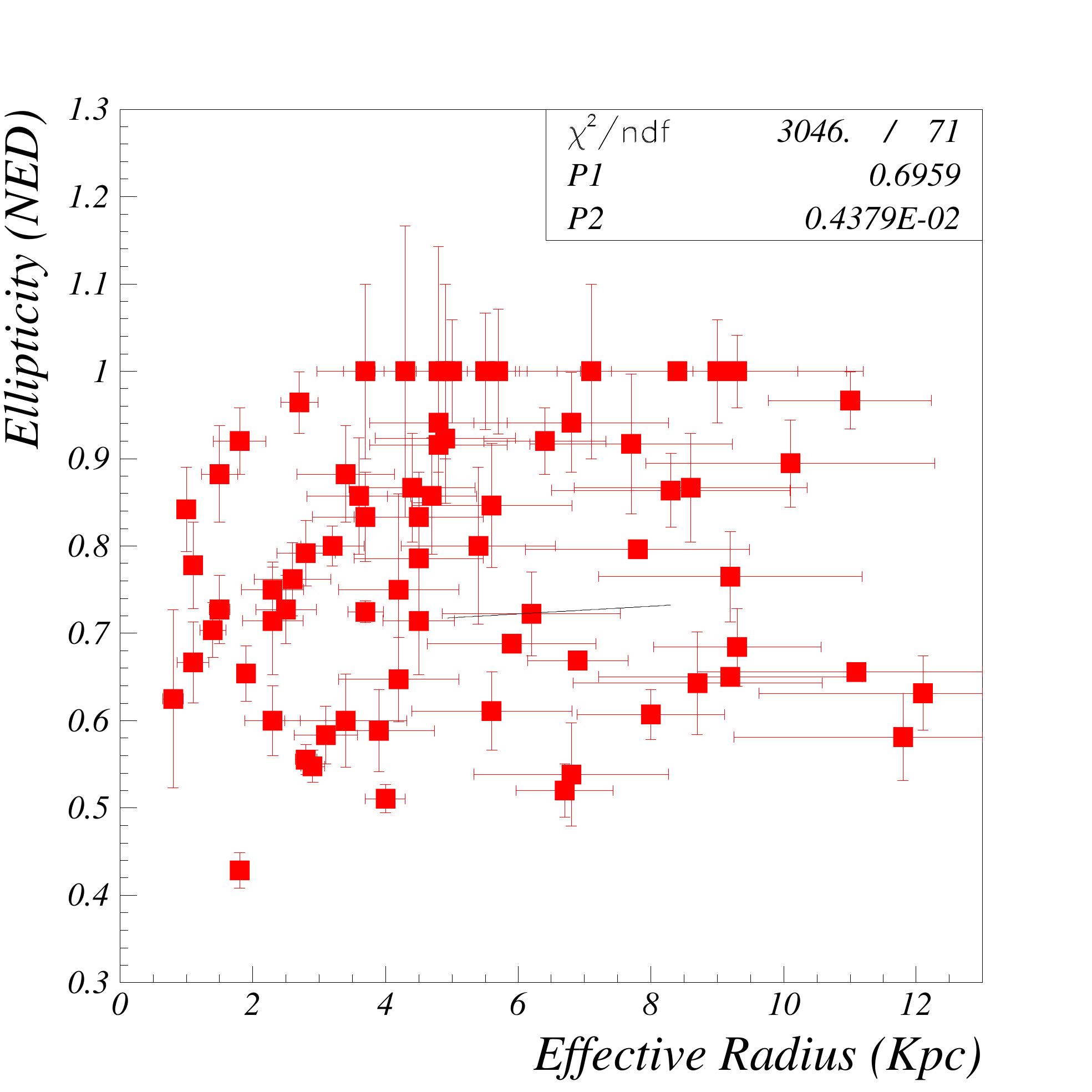}\includegraphics[scale=0.24]{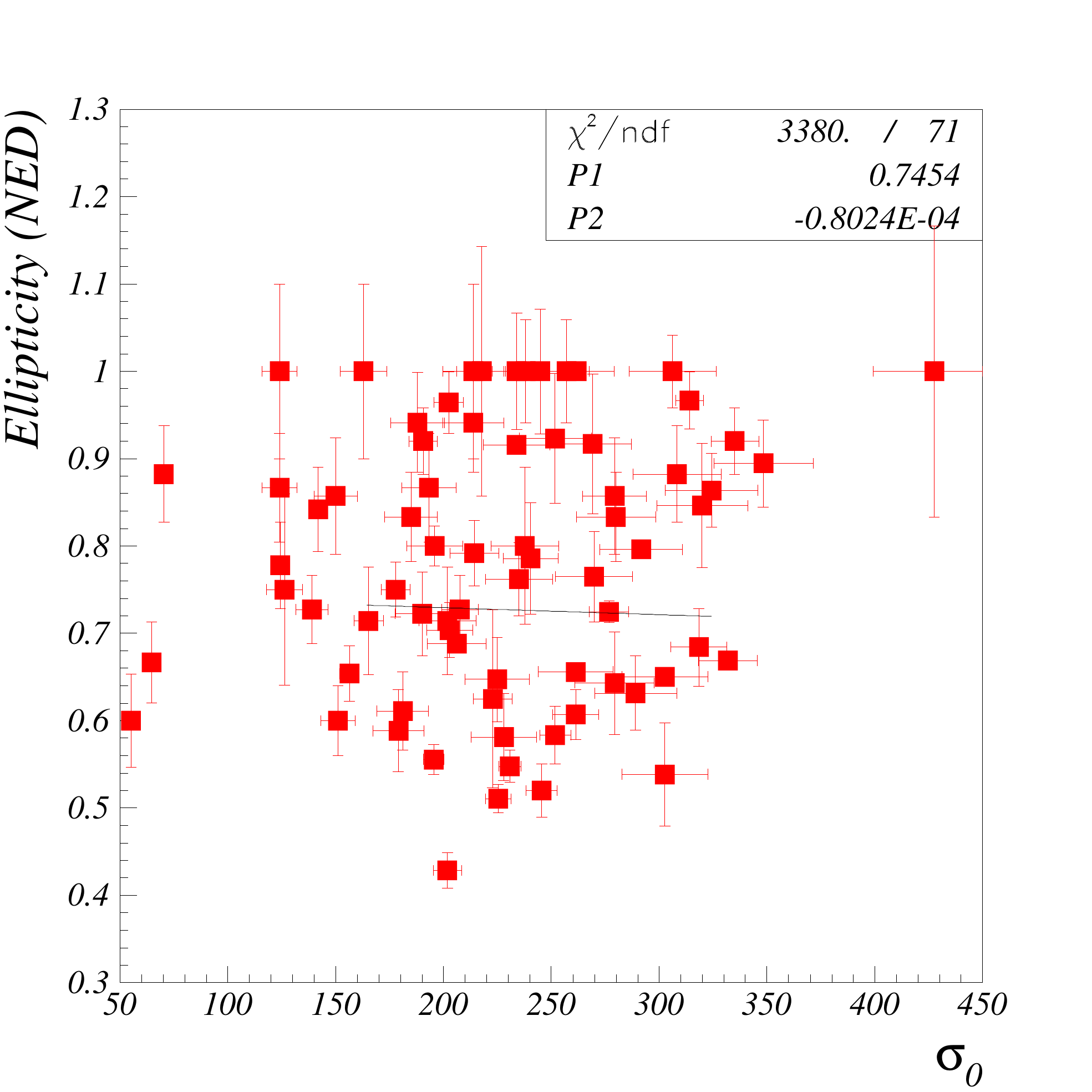}
\vspace{-0.4cm} \caption{\label{Fig: NED ellipticity correlations}Correlations between the
apparent $\sfrac{R_{min}}{R_{Max}}$ from NED and (from top left to
bottom right): apparent effective radius $Re(")$, integrated blue
magnitude $B_{t}$, magnitude from NED, $DM$ from
\cite{BMS}, $DM$ from NED using redshift information,
$DM$ from NED without redshift information, absolute blue
magnitude, surface brightness $I_{e}$, effective radius (parsec)
and velocity distribution $\sigma_{0}$.
}
\end{figure}

\paragraph{Apparent effective radius $Re( ``)$ correlations}
\begin{figure}[H]
\centering
\includegraphics[scale=0.24]{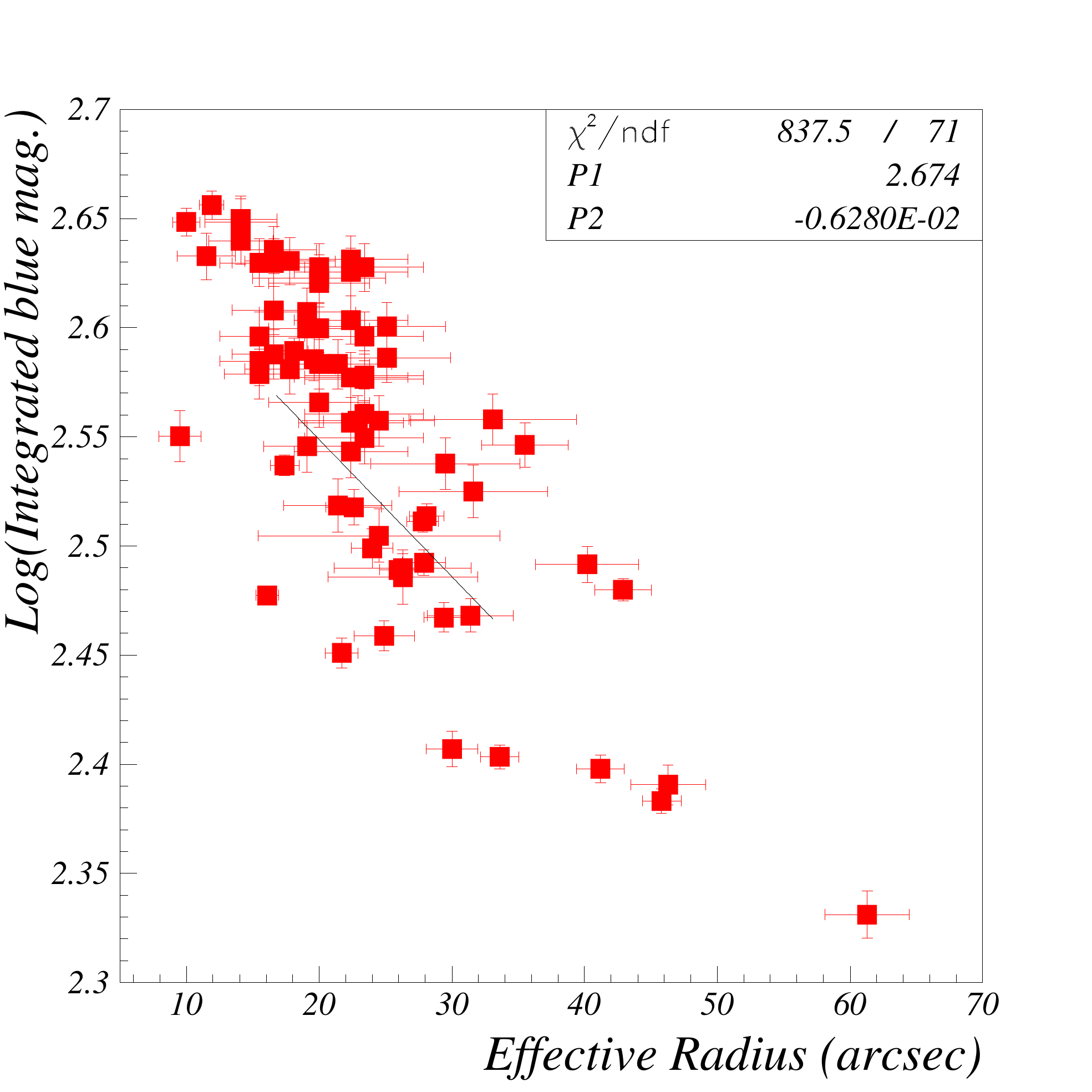}\includegraphics[scale=0.24]{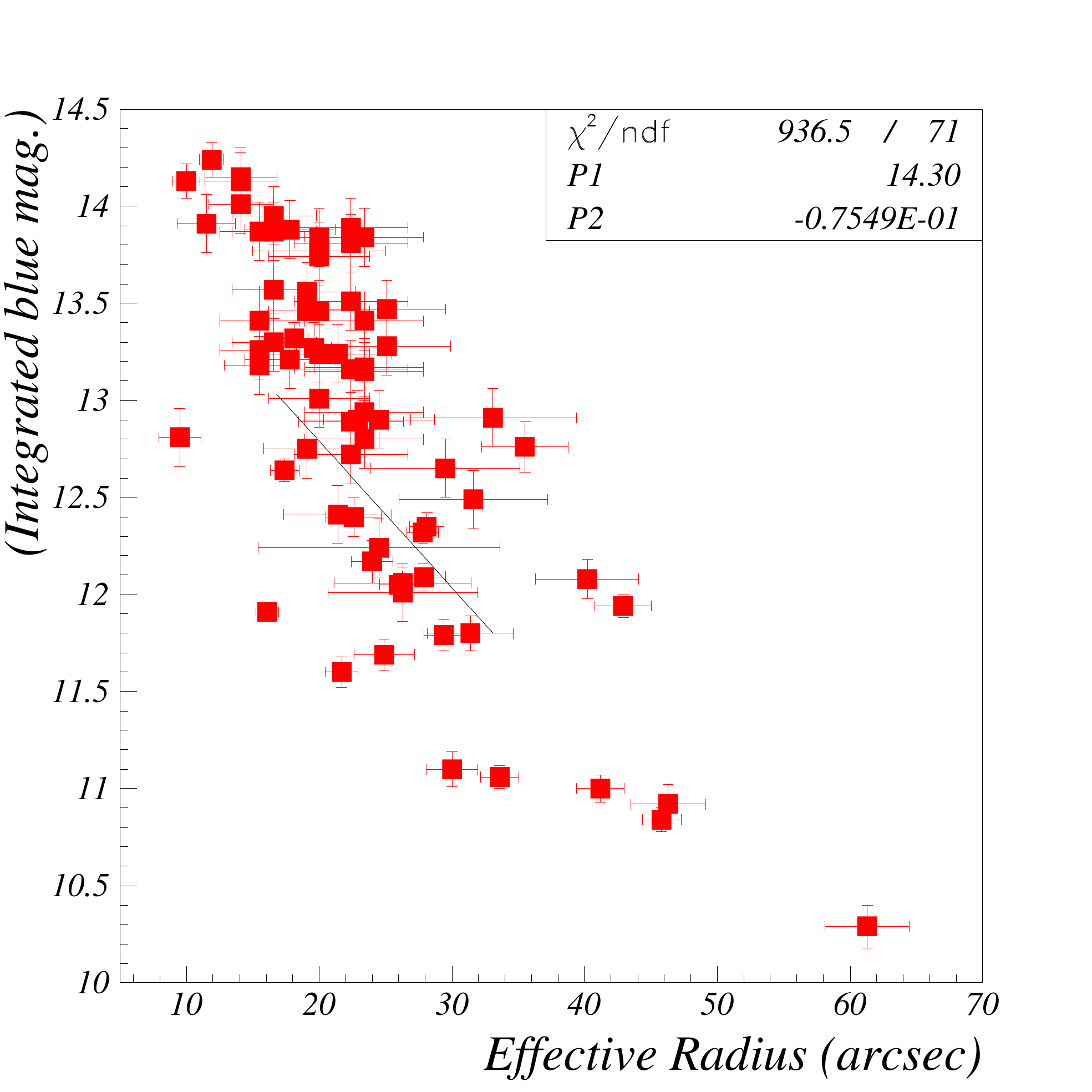}\includegraphics[scale=0.24]{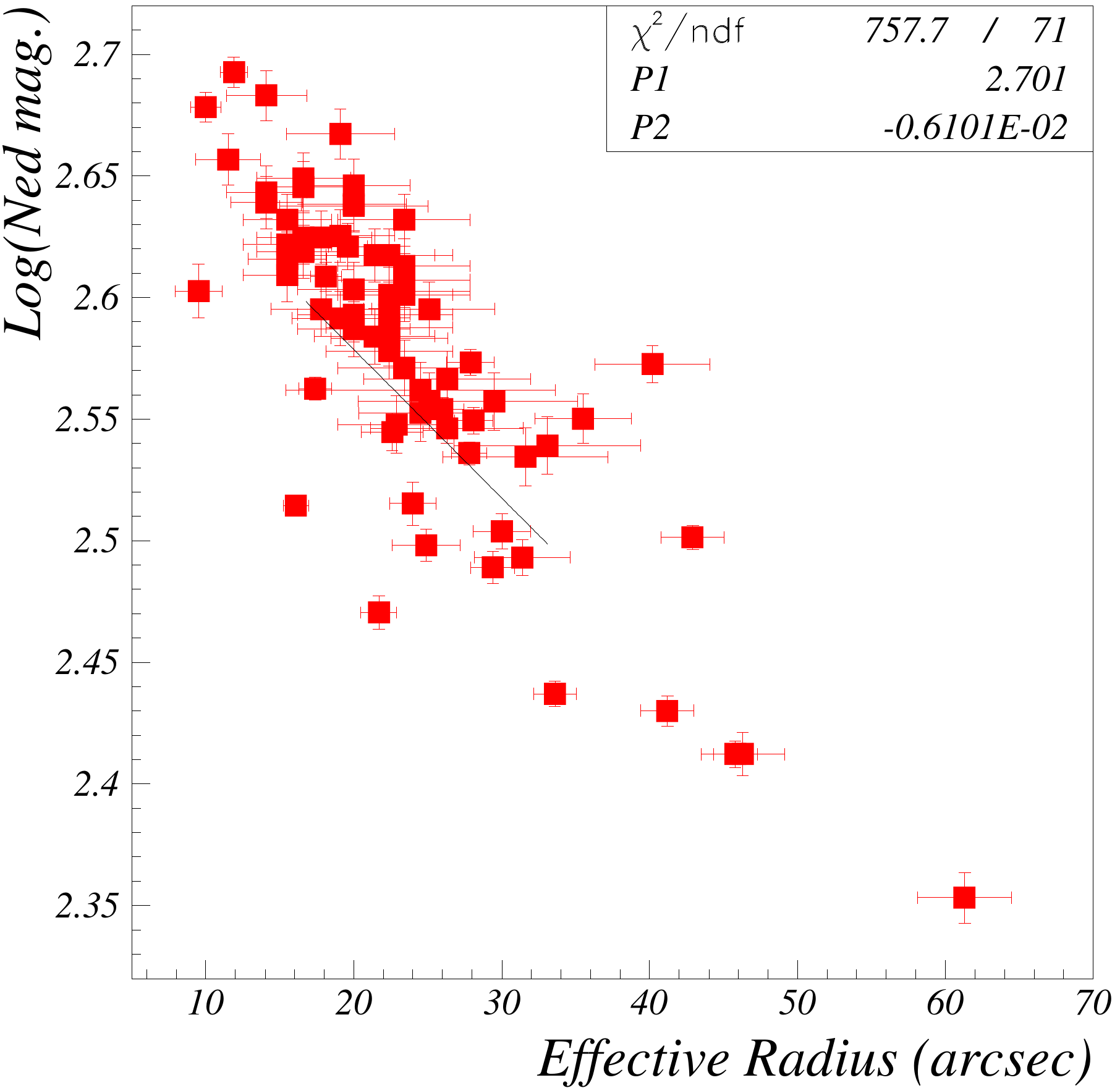}\includegraphics[scale=0.24]{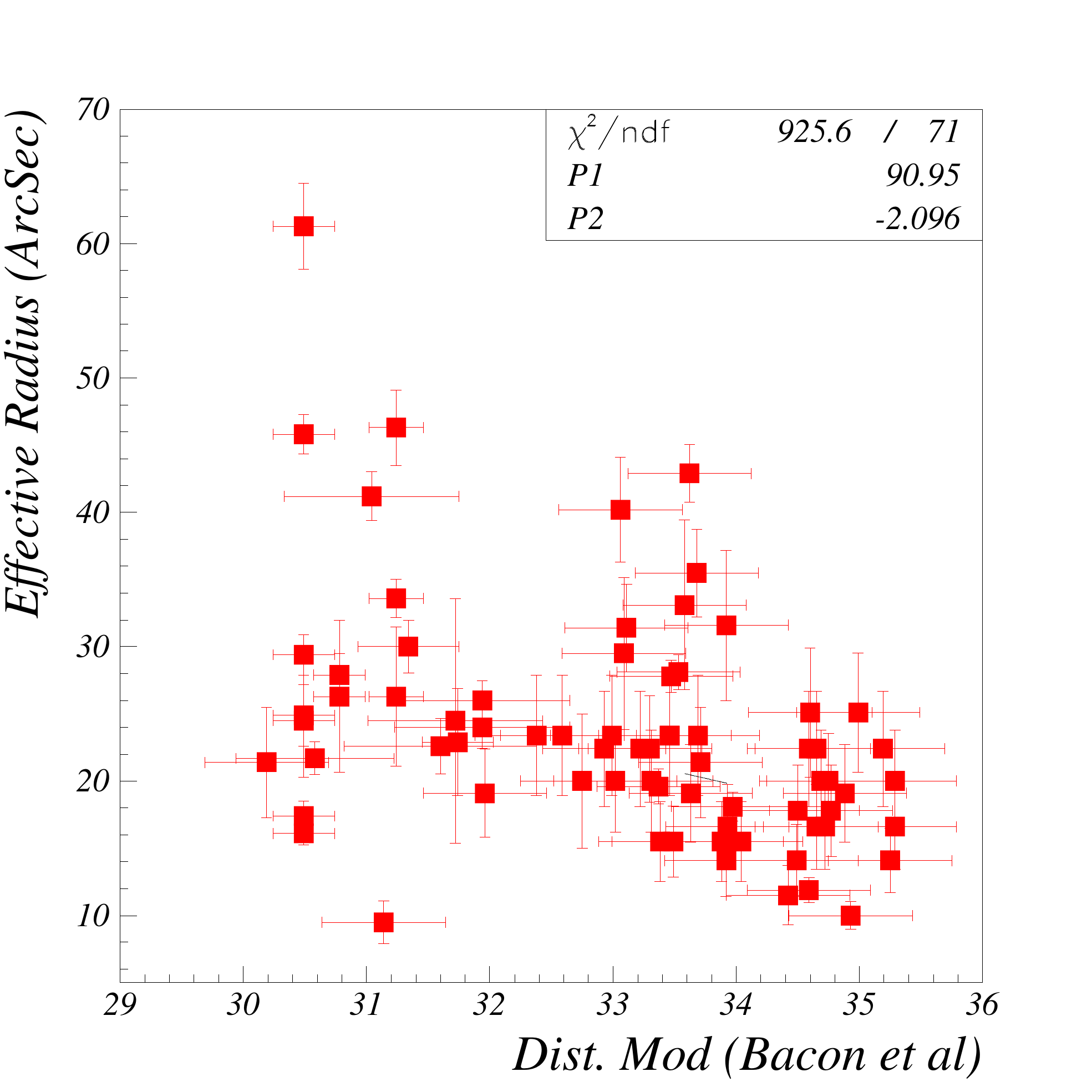}\protect \\
\includegraphics[scale=0.24]{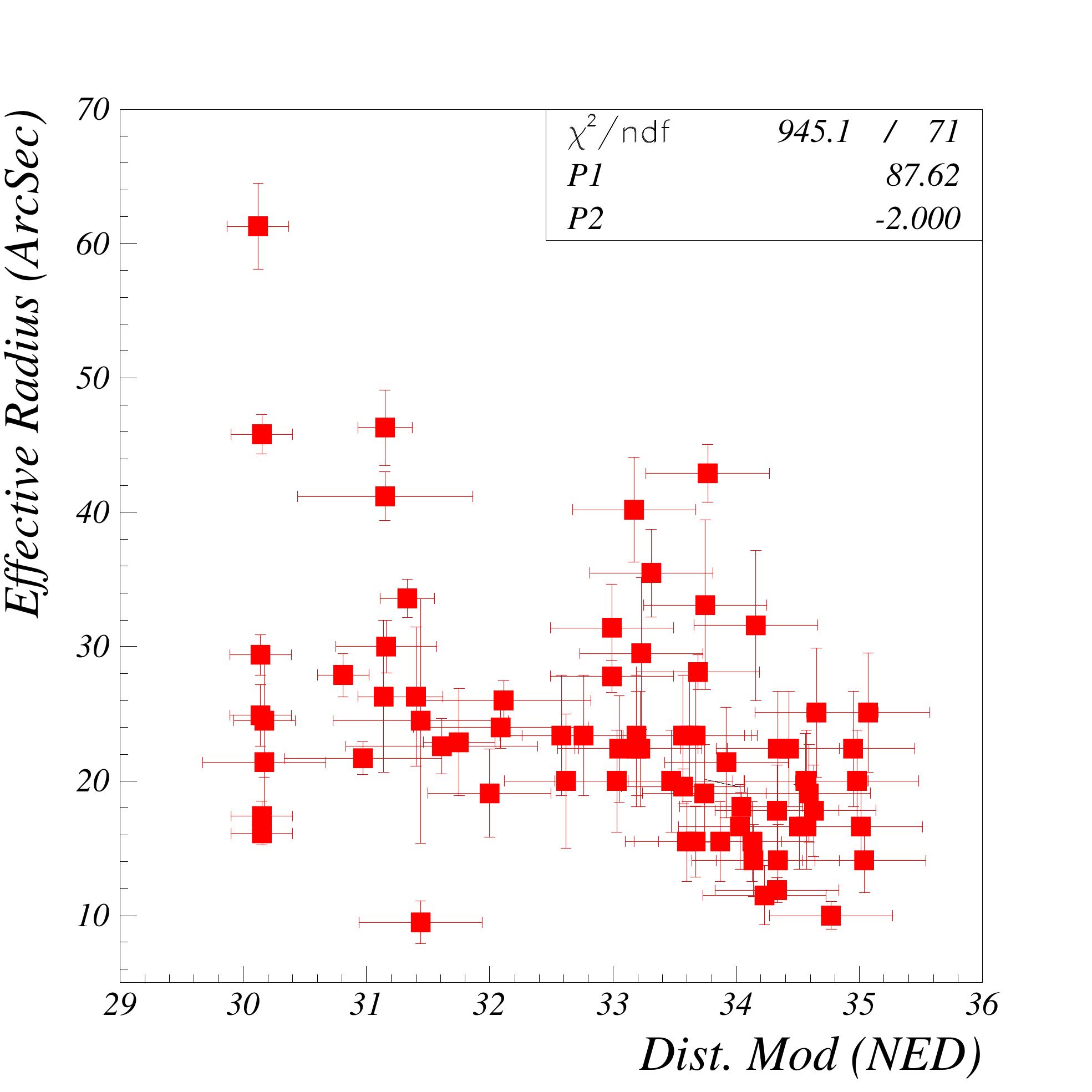}\includegraphics[scale=0.24]{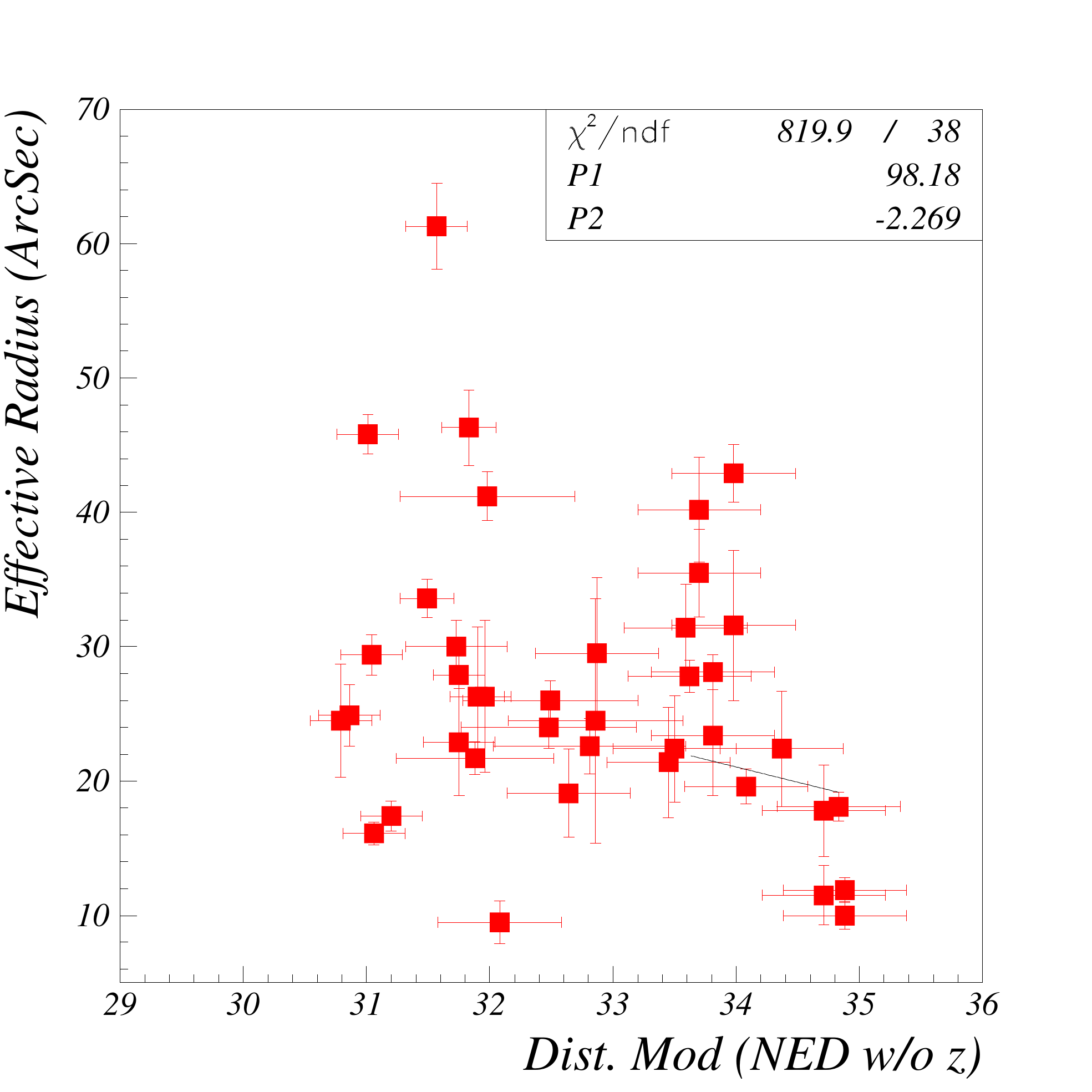}\includegraphics[scale=0.24]{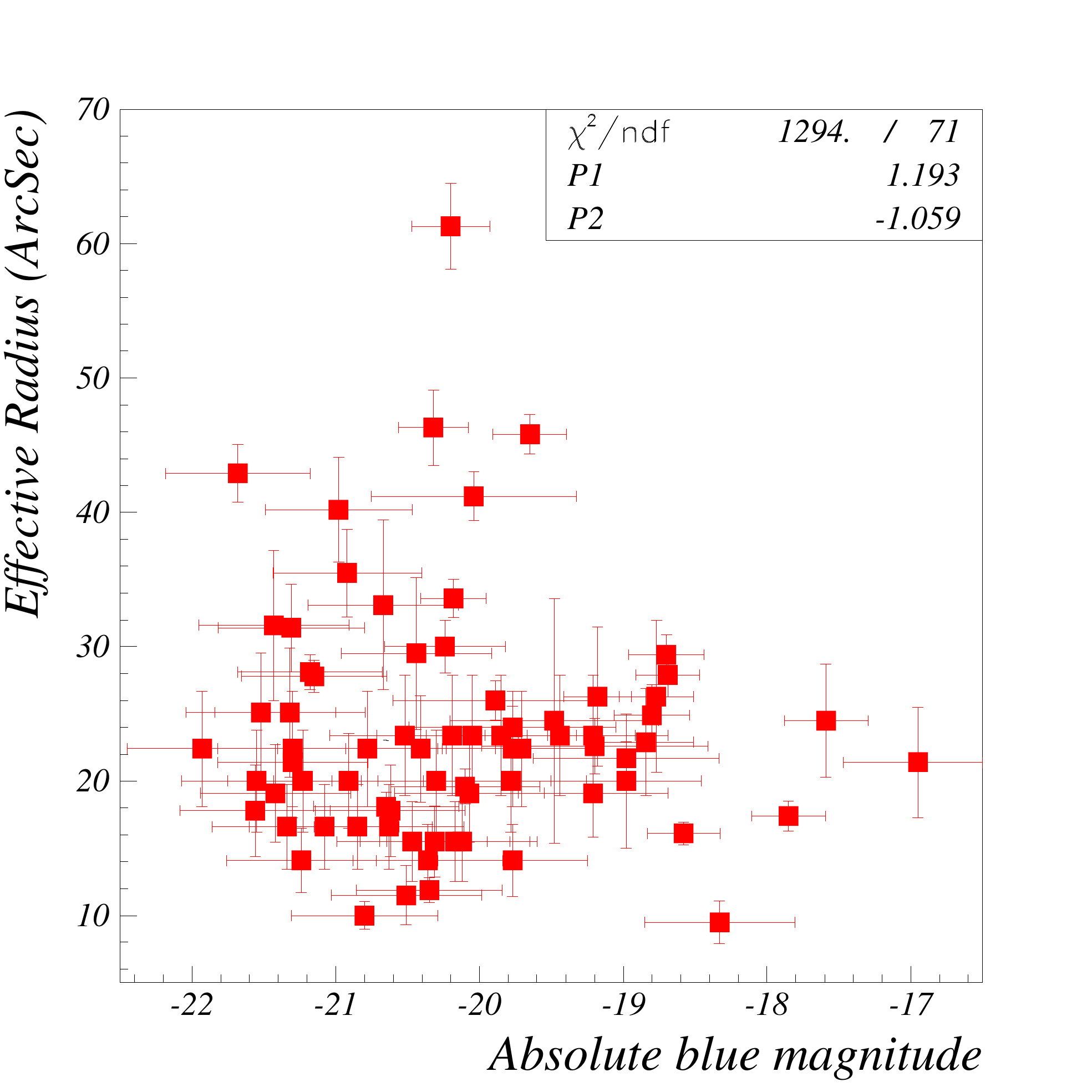}\includegraphics[scale=0.24]{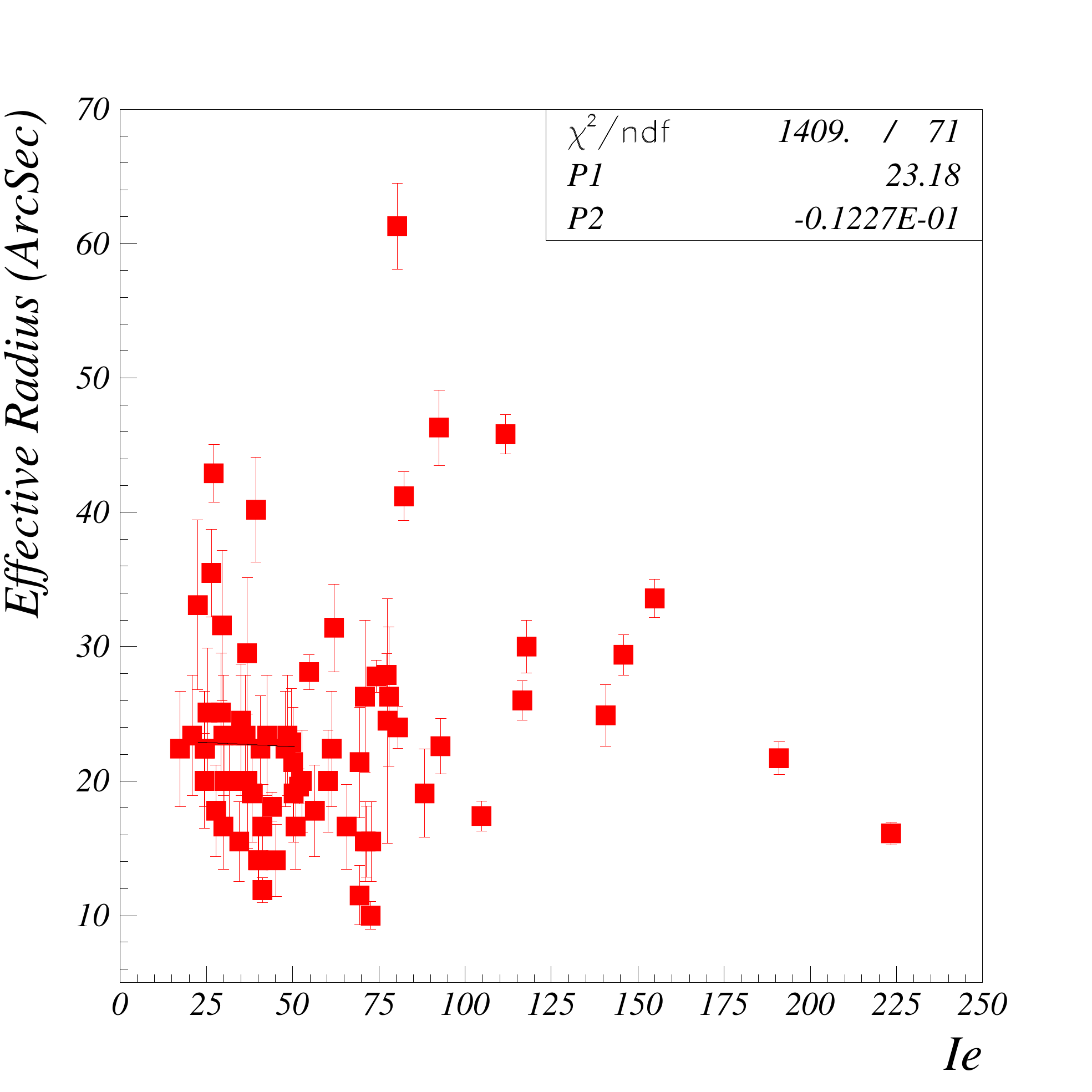}\protect \\
\includegraphics[scale=0.24]{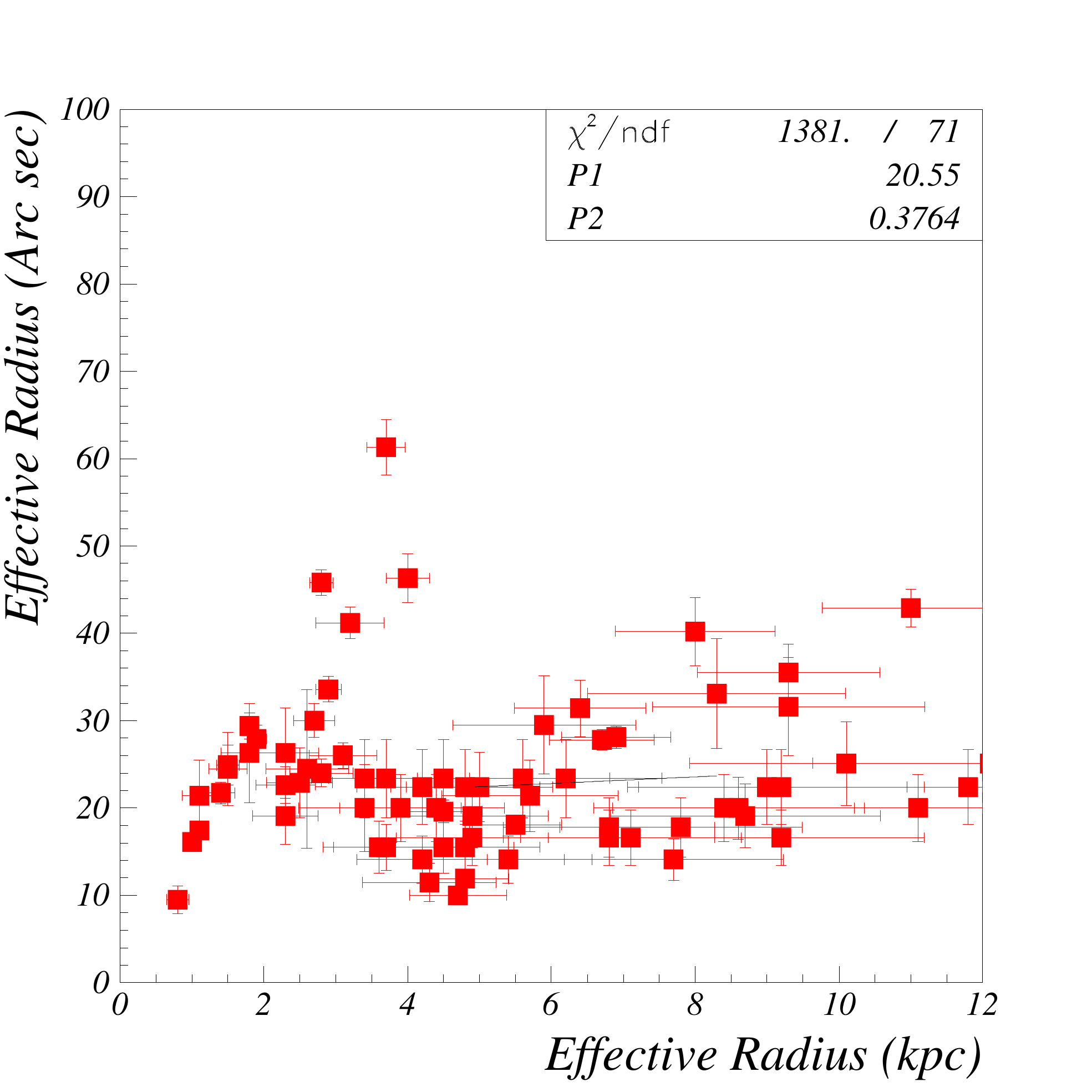}\includegraphics[scale=0.24]{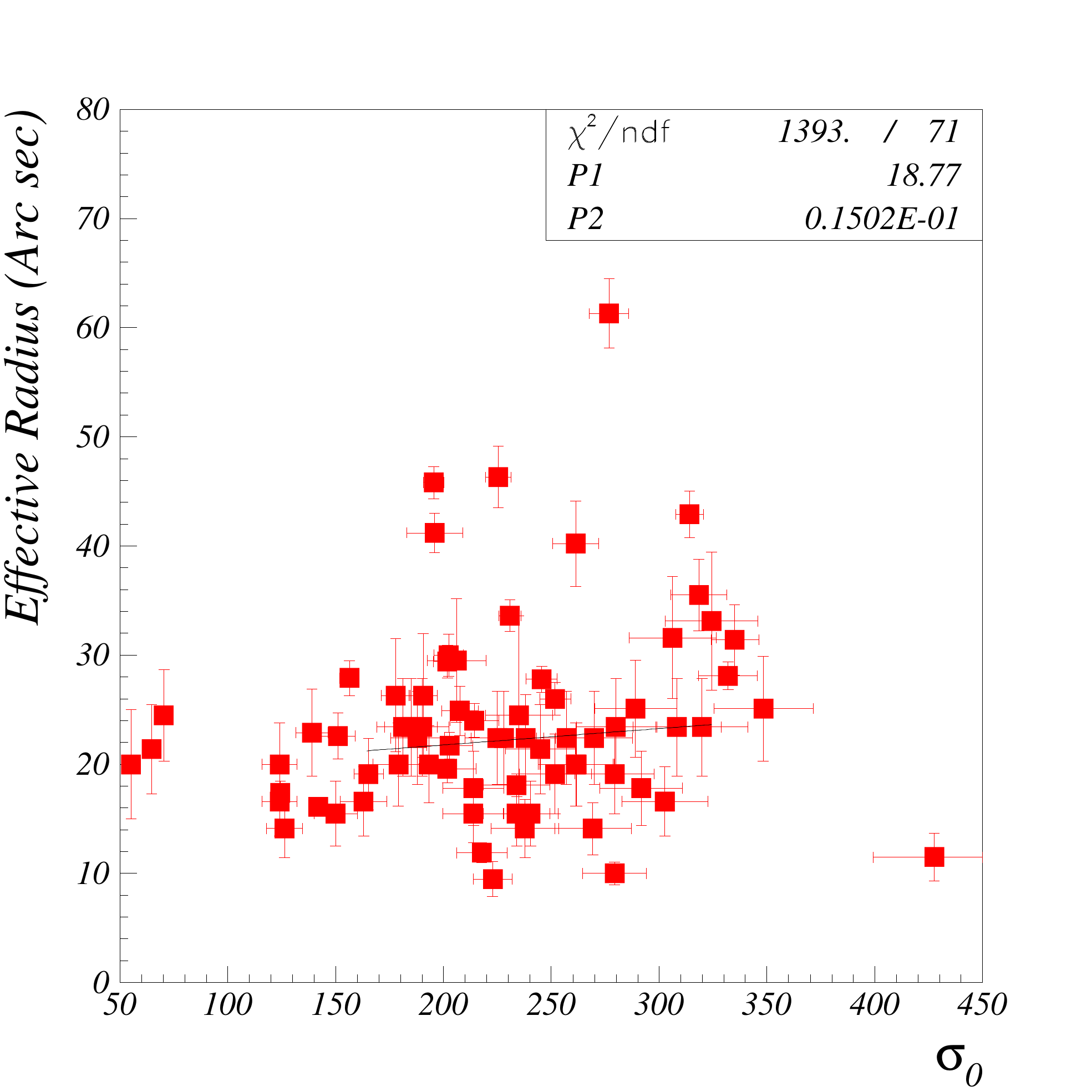}\includegraphics[scale=0.24]{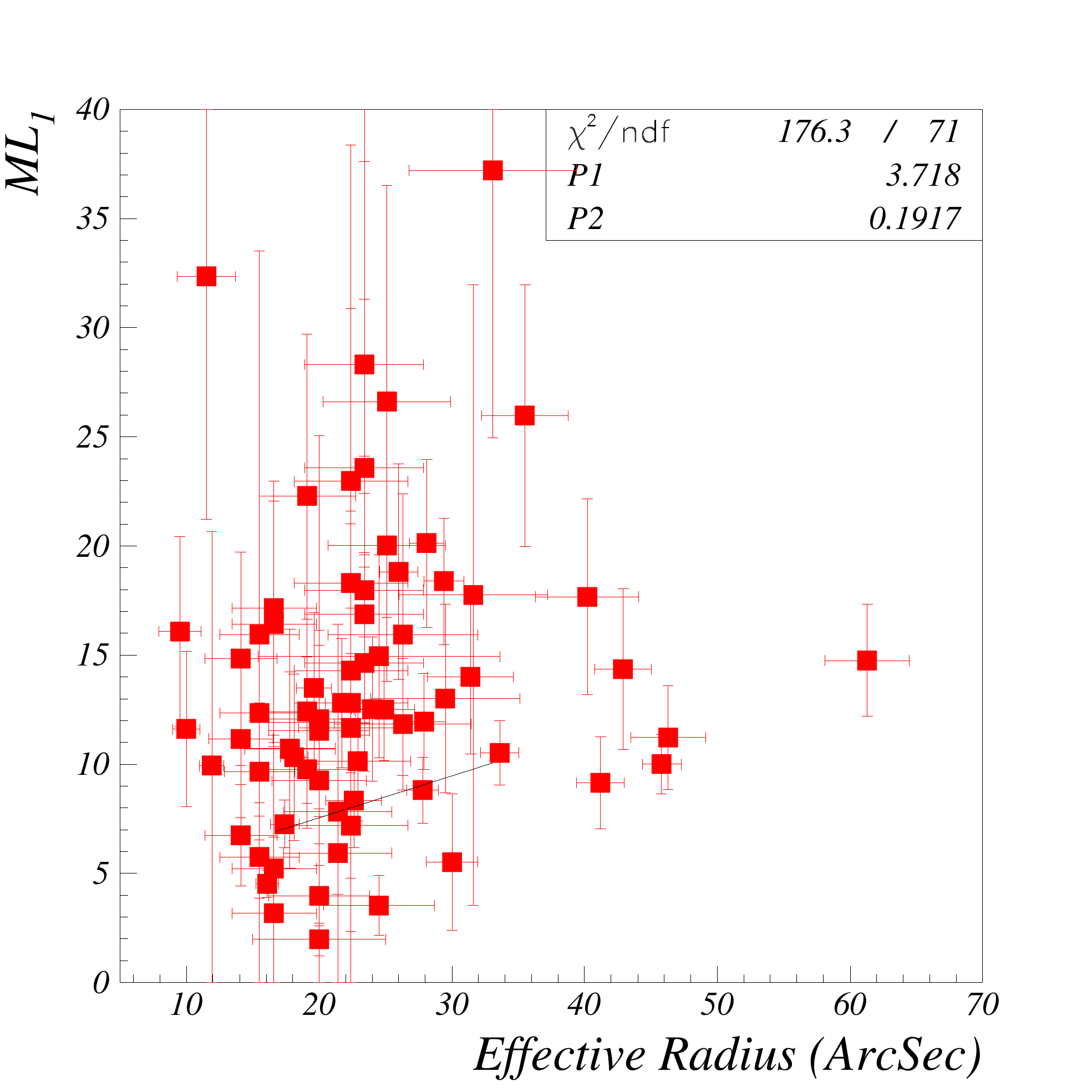}\includegraphics[scale=0.24]{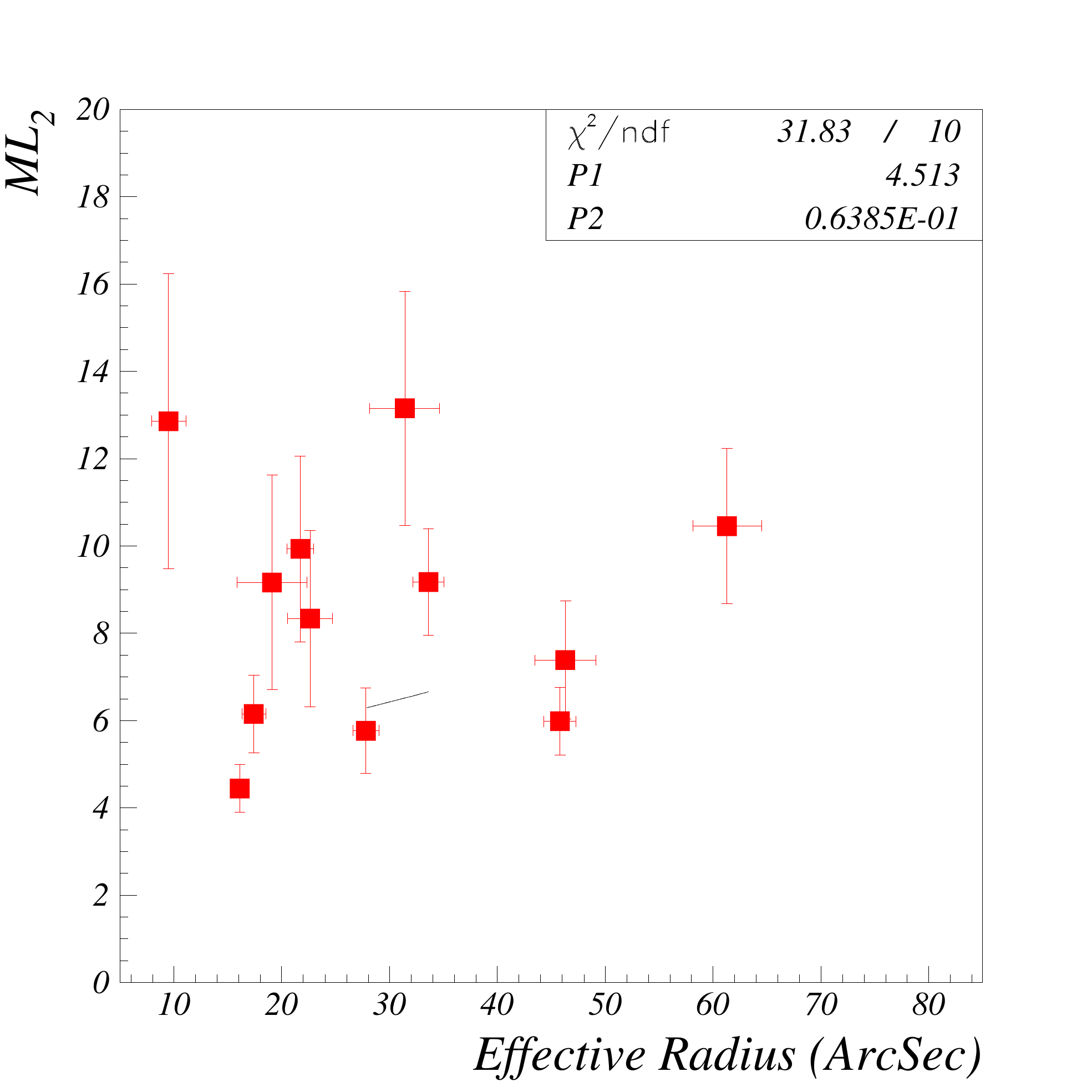}\protect \\
\includegraphics[scale=0.24]{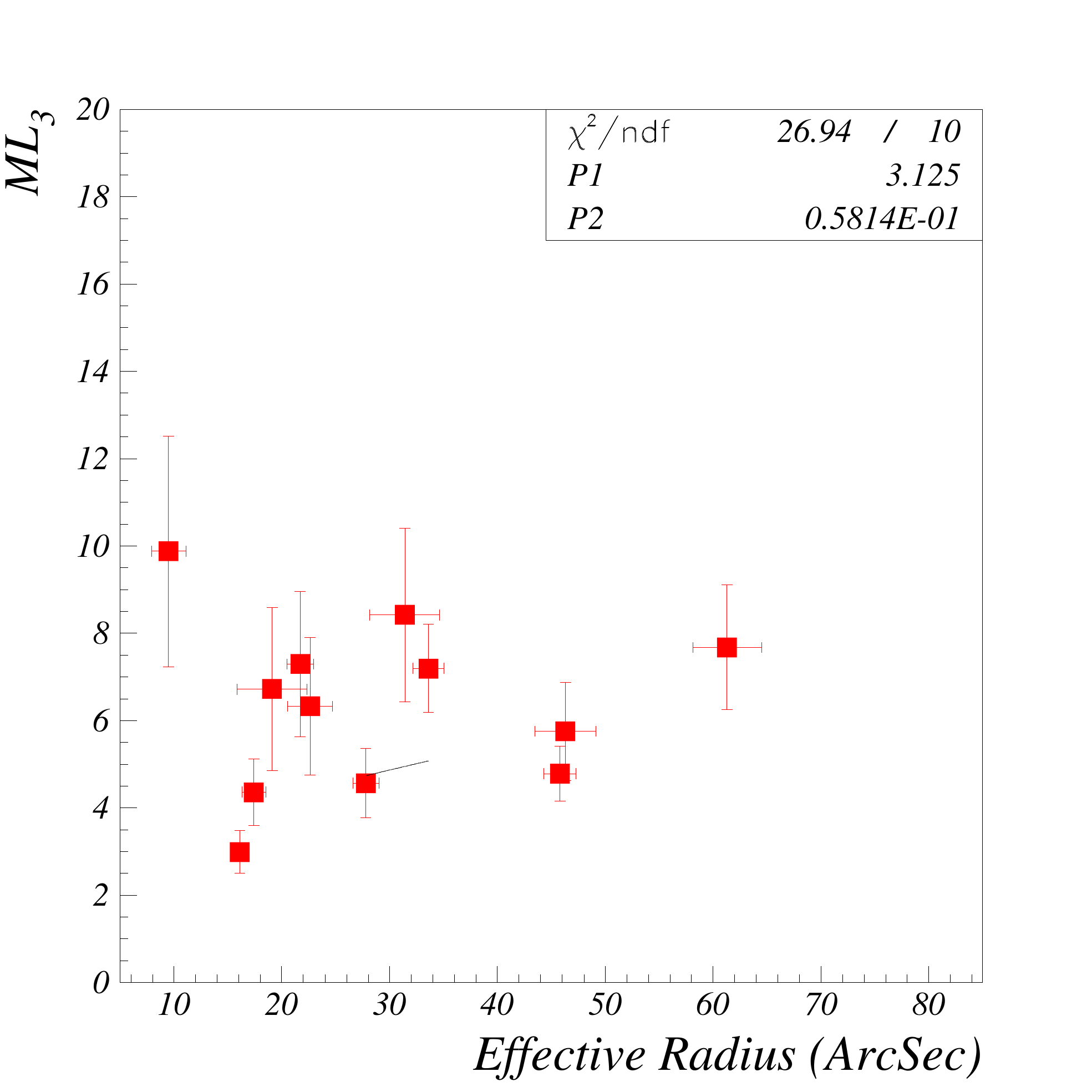}
\vspace{-0.4cm} \caption{\label{Fig: Re(") correlations}Correlations between the apparent
effective radius $Re(\lyxmathsym{\textquotedblleft})$ and (from top
left to bottom right): integrated blue magnitude $B_{t}$ (with log
and linear scales), magnitude from NED, $DM$ from
\cite{BMS}, $DM$ from NED using redshift information,
$DM$ from NED without redshift information, absolute blue
magnitude, surface brightness $I_{e}$, effective radius (parsec),
velocity distribution $\sigma_{0}$ and $\sfrac{M}{L}$.
}
\end{figure}
The origins of the correlations are, from the top left plot to the
bottom right one: 
\begin{enumerate}
\item Effective radius $Re(")$ vs integrated blue magnitude $B_{t}$:
this strong correlation is expected from physics since the larger the apparent
radius, the larger the apparent magnitude tends to be.
\item Same as above but with a linear vertical scale.
\item $Re(")$ vs magnitude from NED: same as above.
\item $Re(")$ vs $DM$ from~\cite{BMS}:
this is the trivial decrease of apparent size with 
distance to the observer.
\item $Re(")$ vs $DM$ from NED using redshift
information: same as above.
\item $Re(")$ vs $DM$ from NED without
redshift information: same as above.
\item $Re(")$ vs absolute blue magnitude: no clear
correlation is seen. A linear (or any functional) fit fails to account
for the data distribution. We were expecting a clear secondary correlation
of opposite sign from the $M_{b}\Longleftrightarrow DM\longleftrightarrow Re(")$
correlations. Using the fit results $(-0.53\pm0.03)DM=M_{b}+c$ and
$(-2.1\pm0.2)DM=Re(")+c'$ yields $Re(")=(4.0\pm0.6)M_{b}+c"$. The
other important contribution is from the $M_{b}\Longleftrightarrow Re\mbox{ (Kpc)}\longleftrightarrow Re(")$
correlations. Using the fit results $(-3.4\pm0.1)M_{b}=Re\mbox{ (Kpc)}+c$
and $(0.4\pm0.1)Re\mbox{(Kpc)}=Re(")+c'$ yields $Re(")=(-1.4\pm0.4)M_{b}+c"$.
Again, these two results cannot be added linearly because of correlations
between $DM$ \& $Re(Kpc)$, but the opposite signs of the two correlations
suggest a cancellation.
\item $Re(")$ vs surface brightness $I_{e}$: a weak
possible correlation is seen. The correlations $Re(")\Longleftrightarrow Bt\longleftrightarrow I_{e}$
and $Re(")\longleftrightarrow DM\longleftrightarrow I_{e}$ can both
contribute to create it. Using the linear fit results $(-7.6\pm0.2)E^{-2}Re(")=Bt+c$
and $(-7.5\pm0.2)E^{-3}I_{e}=B_{t}+c'$ yields $Re(")=(9.8\pm0.1)E^{-1}I_{e}+c"$
. Using the linear fit results $(-2.1\pm0.2)DM=Re(")+c$ and $(-1.7\pm0.1)E^{-2}I_{e}=DM+c'$
yields $Re(")=(3.6\pm0.1)E^{-2}I_{e}+c"$. As $B_{t}$ \& $DM$ are
strongly correlated, we cannot simply add these results, but their
same sign suggests a strong correlation, in disagreement with the
sign of the observed possible correlation. However, this disagreement
may stem from the unreliable fit. The correlation itself, if it exists,
is weak.
\item $Re(")$ vs effective radius (parsec): because
a distance in Kpc is given by $10^{0.2DM-2}$, then $Re(")=Re\mbox{(Kpc)}/10^{0.2DM-2}/4.85E^{-6}$
($4.85E^{-6}$ converts rad in arcsec). 
\item $Re(")$ vs velocity distribution $\sigma_{0}$:
no, or unclear, weak correlation. 
\item $Re(")$ vs $\sfrac{M}{L}$: this weak correlation
may arise from the correlations $Re(")\dashleftarrow\dashrightarrow \sfrac{R_{min}}{R_{Max}}\longleftrightarrow \sfrac{M}{L}$,
$Re(")\dashleftarrow\dashrightarrow\sigma_{0}\Longleftrightarrow \sfrac{M}{L}$,
$Re(")\longleftrightarrow DM\dashleftarrow\dashrightarrow \sfrac{M}{L}$ and
$Re(")\Longleftrightarrow B_{t}\longleftrightarrow \sfrac{M}{L}$. Using the linear
fit results $(-4.2\pm0.3)E^{-3}Re(")=\sfrac{R_{min}}{R_{Max}}+c$ and $(-6\pm2)\sfrac{R_{min}}{R_{Max}}=\sfrac{M}{L}+c'$
yields $\sfrac{M}{L}=(2.5\pm1.0)E^{-2}Re(")+c"$. Using the linear fit results
$(1.5\pm0.4)E^{-2}\sigma_{0}=Re(")+c$ and $(5.4\pm0.4)E^{-2}\sigma_{0}=\sfrac{M}{L}+c'$
yields $\sfrac{M}{L}=(3.6\pm1.2)Re(")+c"$. Using the linear fit results $(-2.0\pm0.2)DM=Re(")+c$
and $(0.5\pm0.2)DM=\sfrac{M}{L}+c'$ yields $\sfrac{M}{L}=(-2.5\pm1.3)E^{-1}Re(")+c"$.
Using the linear fit results $(-7.6\pm0.2)E^{-2}Re(")=B_{t}+c$ and
$(-1.3\pm0.3)B_{t}=\sfrac{M}{L}+c'$ yields $\sfrac{M}{L}=(9.9\pm2.5)E^{-2}Re(")+c"$.
$\sfrac{R_{min}}{R_{max}}$ \& $DM$, $\sfrac{R_{min}}{R_{max}}$ \& $B_{t}$, $\sigma_{0}$
\& $DM$ and $DM$ \& $B_{t}$ are correlated. However, adding the
contributions, we obtain $\sfrac{M}{L}=(3.47\pm1.4)Re(")+C"$, just 2.3$\sigma$
away from the observed $\sfrac{M}{L}=(0.19\pm0.03)Re(")+C"$. This suggests
that this weak correlation may originate from the four correlations
$Re(")\dashleftarrow\dashrightarrow \sfrac{R_{min}}{R_{Max}}\longleftrightarrow \sfrac{M}{L}$,
$Re(")\dashleftarrow\dashrightarrow\sigma_{0}\Longleftrightarrow \sfrac{M}{L}$,
$Re(")\longleftrightarrow DM\dashleftarrow\dashrightarrow \sfrac{M}{L}$ and
$Re(")\Longleftrightarrow B_{t}\longleftrightarrow \sfrac{M}{L}$. 
\end{enumerate}

\paragraph{Integrated blue magnitude correlations\label{sub:Integrated-blue-magnitude correl}}
\begin{figure}[H]
\centering
\includegraphics[scale=0.24]{btbms_btned.pdf}\includegraphics[scale=0.24]{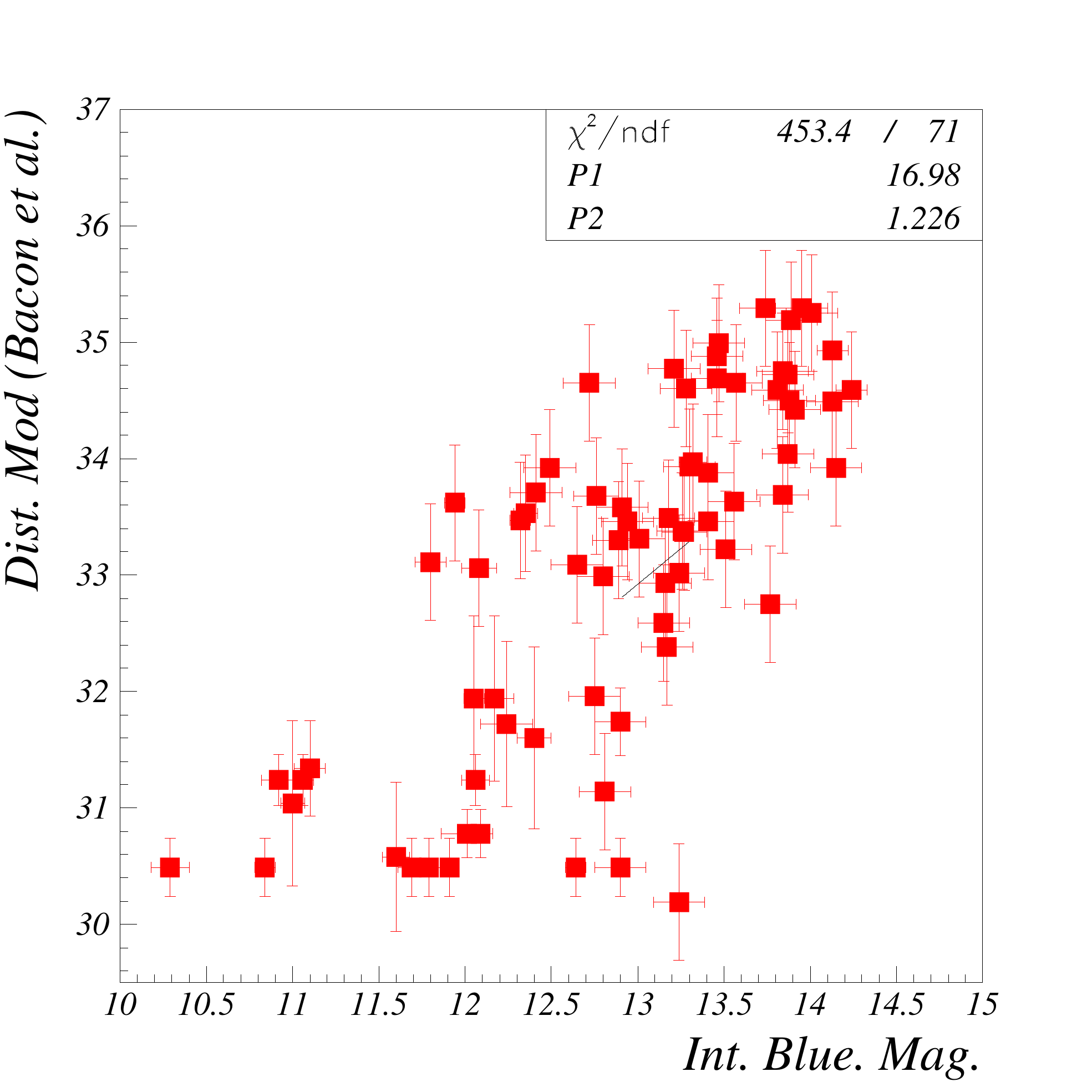}\includegraphics[scale=0.24]{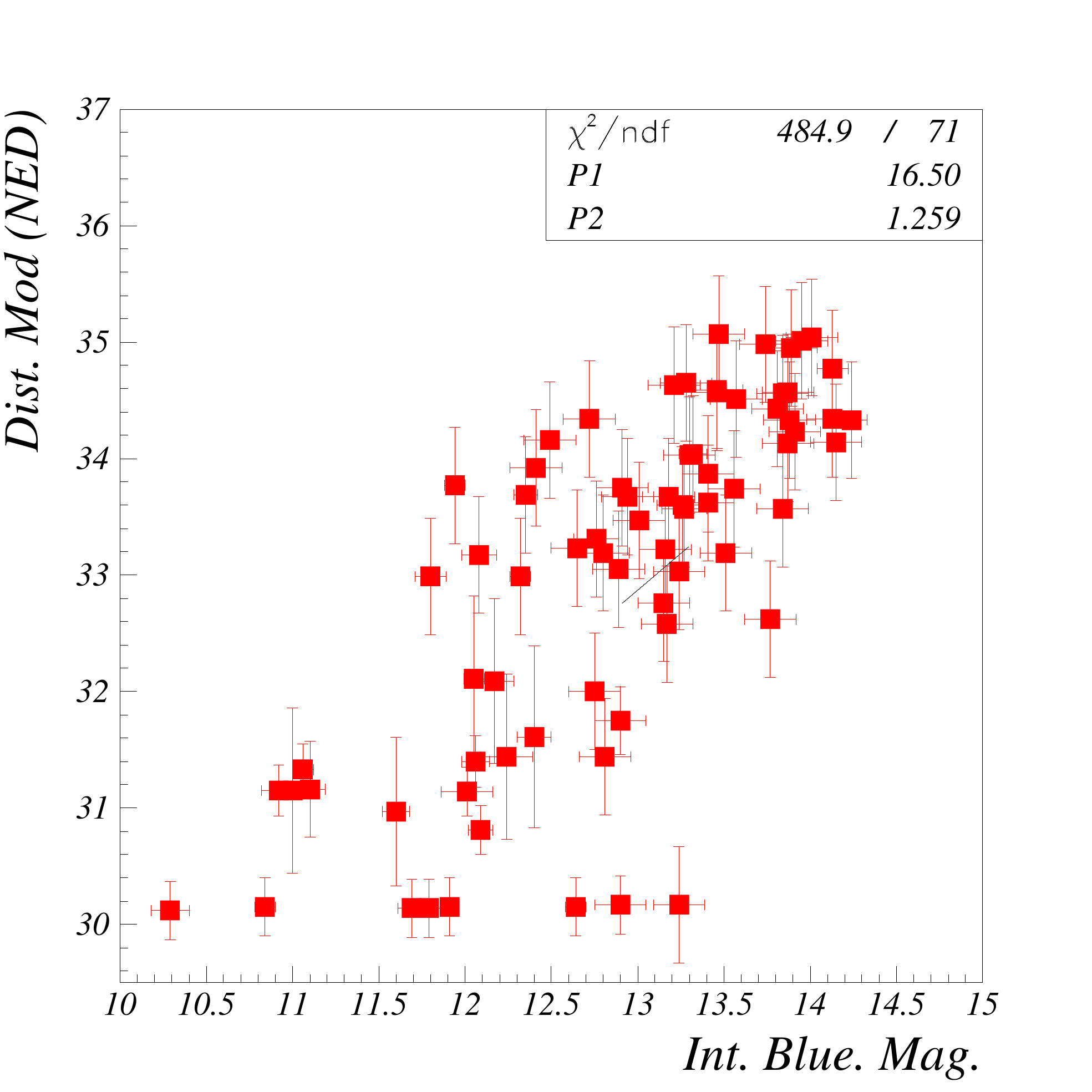}\includegraphics[scale=0.24]{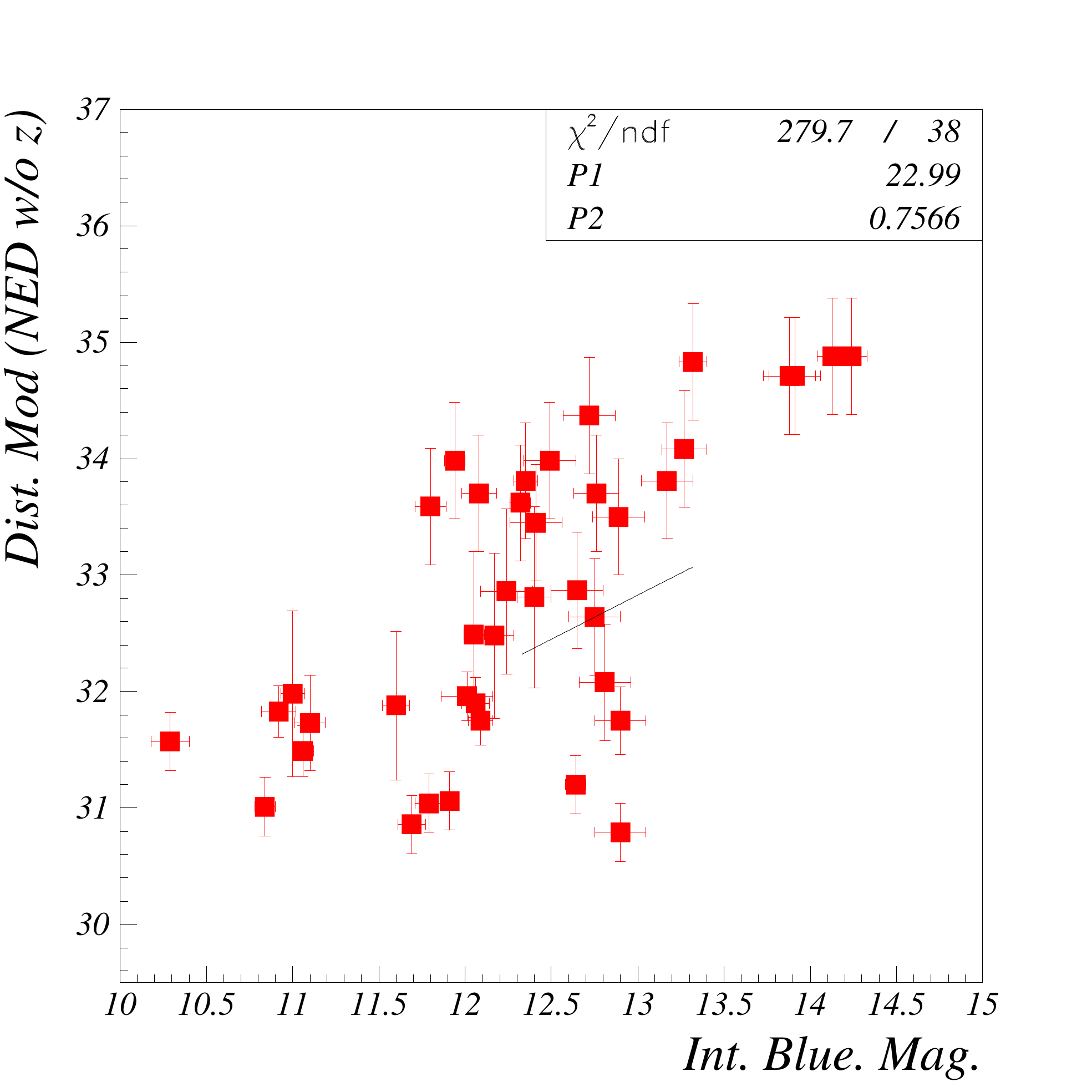}\protect \\
\includegraphics[scale=0.24]{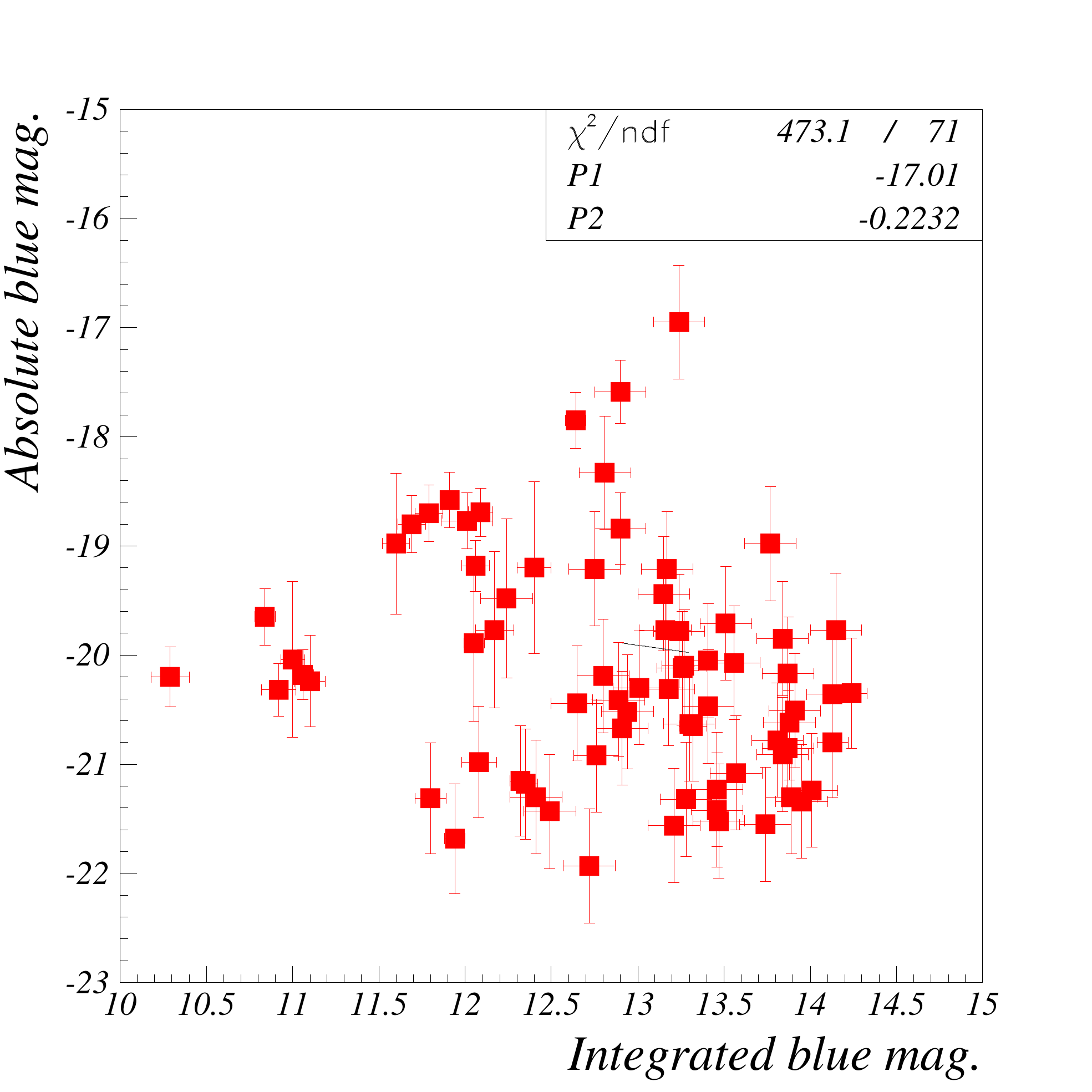}\includegraphics[scale=0.24]{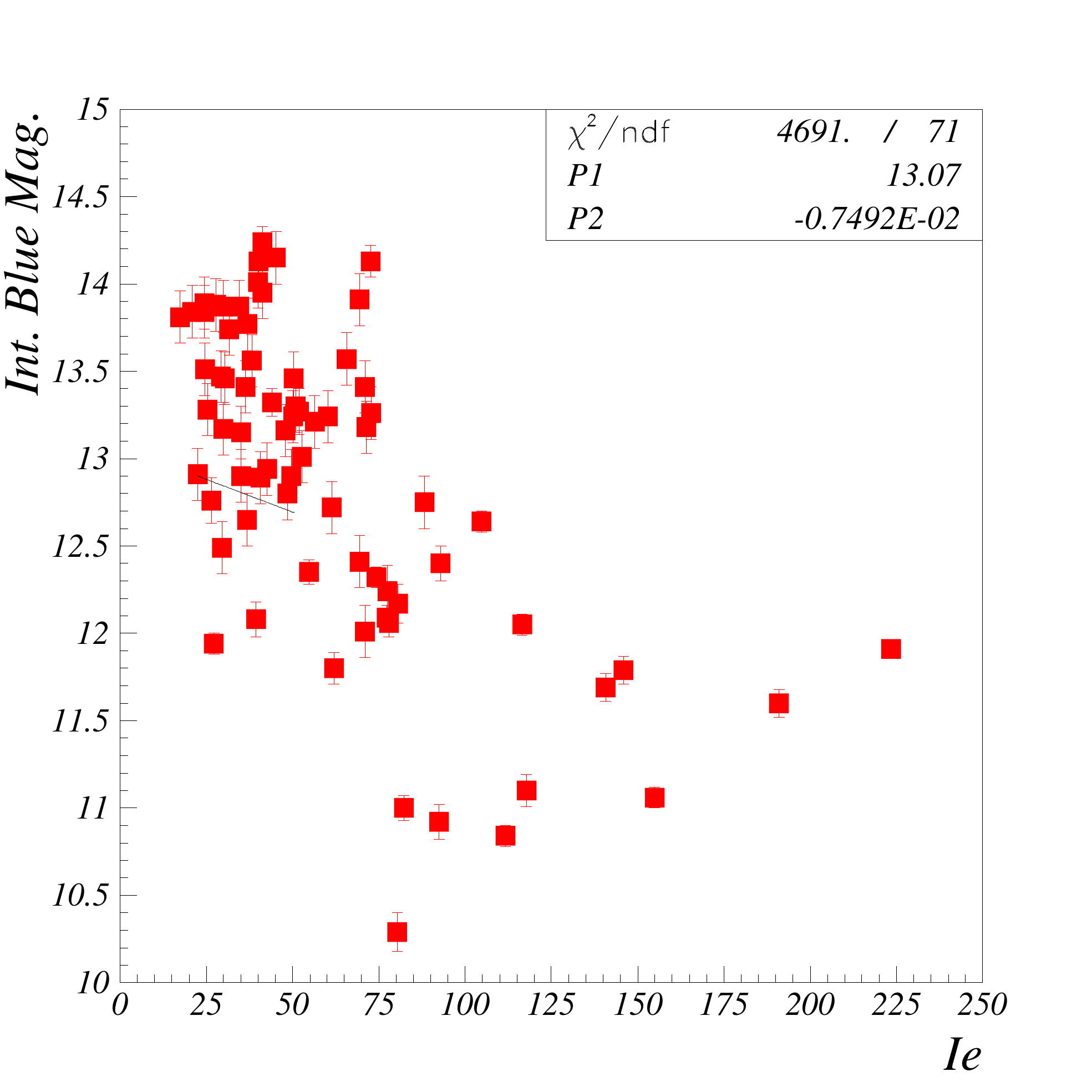}\includegraphics[scale=0.24]{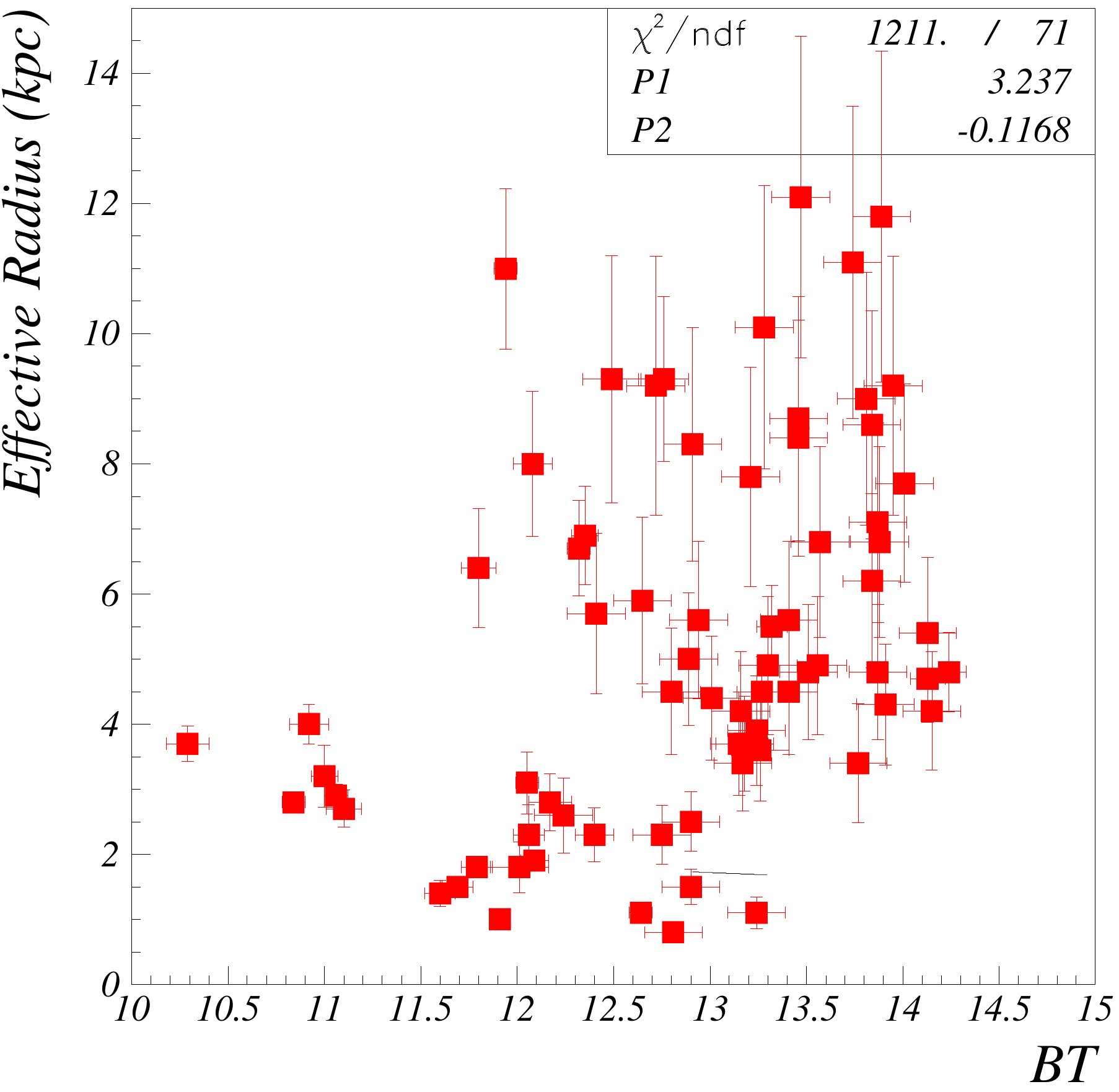}\includegraphics[scale=0.24]{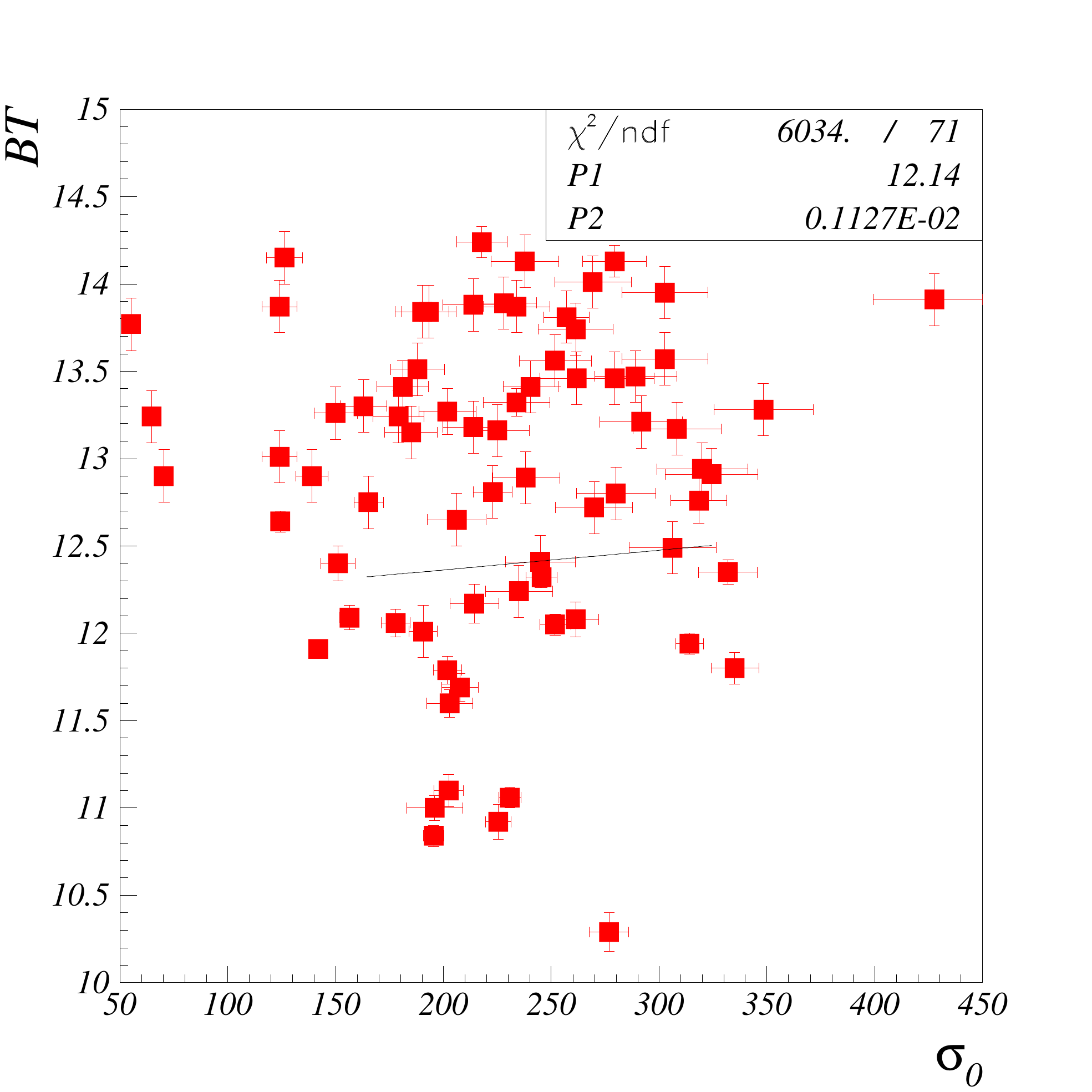}\protect \\
\includegraphics[scale=0.24]{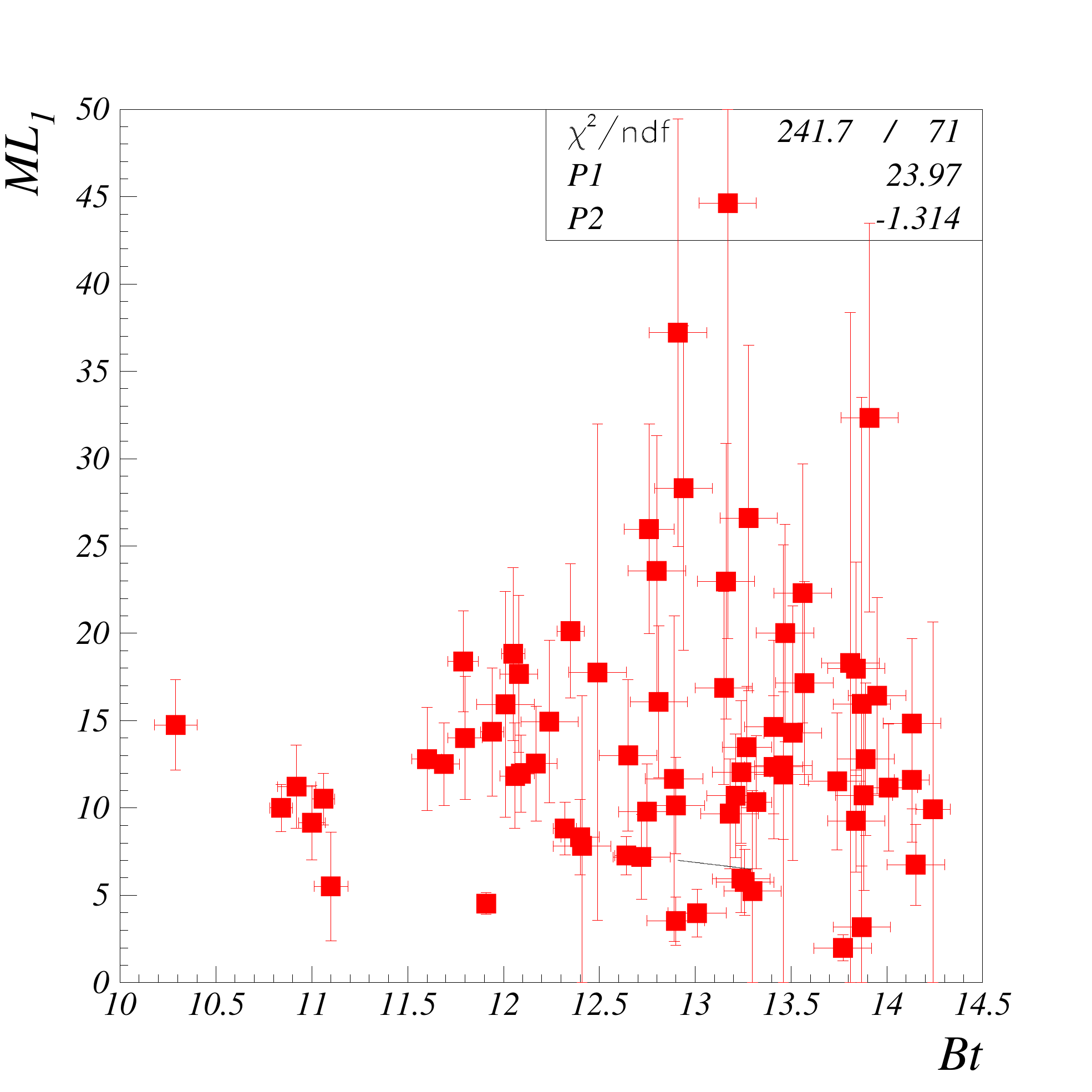}\includegraphics[scale=0.24]{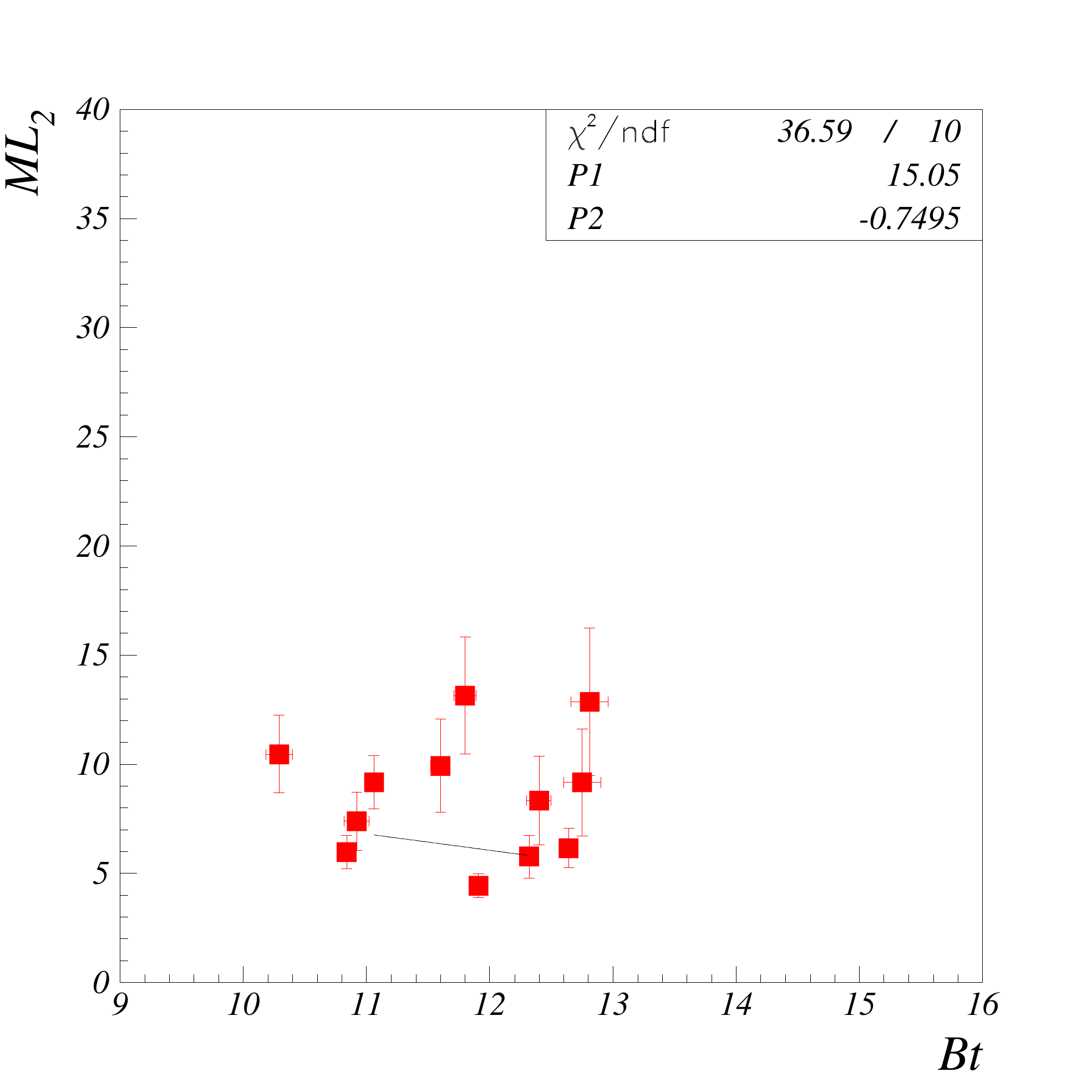}\includegraphics[scale=0.24]{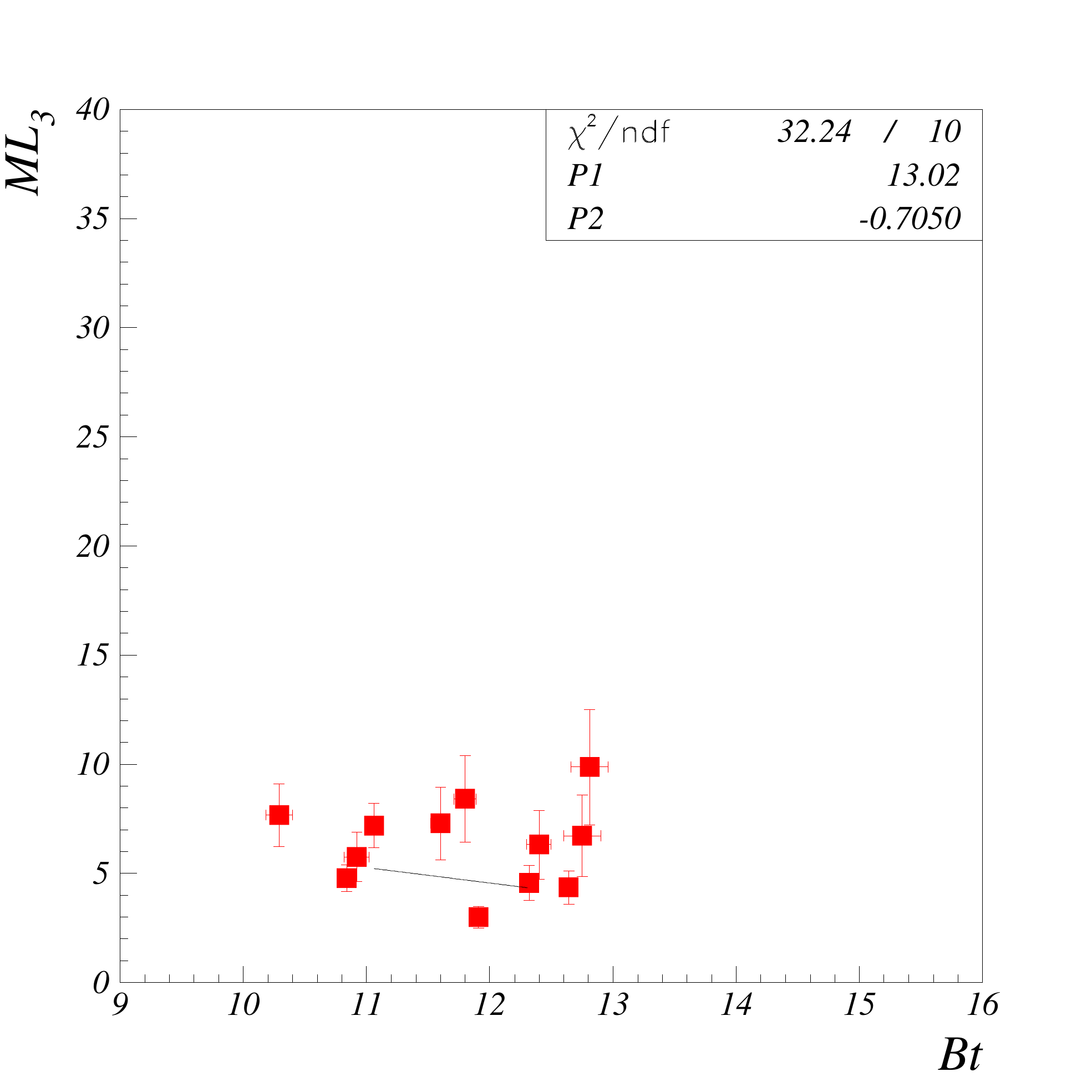}
\vspace{-0.4cm} \caption{\label{Fig: Bt correlations}Correlations between the integrated blue
magnitude $B_{t}$ and (from top left to bottom right): magnitude
from NED,  $DM$ from~\cite{BMS}, $DM$
from NED using redshift information, $DM$ from NED without
redshift information, absolute blue magnitude, surface brightness
$I_{e}$, effective radius (parsec), velocity distribution $\sigma_{0}$
and $\sfrac{M}{L}$.
}
\end{figure}
The origins of the correlations are, from the top left plot to the
bottom right one: 
\begin{enumerate}
\item Integrated blue magnitude $B_{t}$ vs magnitude from NED: this
was already discussed (see Fig.~\ref{fig:btbms_bted}).
\item $B_{t}$ vs  $DM$ from
\cite{BMS}: this clear strong correlation is due to the trivial
decrease of apparent brightness with distance to
the observer.
\item $B_{t}$ vs $DM$ from NED
using redshift information: same as above.
\item $B_{t}$ vs $DM$ from NED
without redshift information: same as above.
\item $B_{t}$ vs absolute blue magnitude:
no or little correlation is observed. However we expect 3 strong secondary
correlations to contribute: $B_{t}\Longleftrightarrow DM\Longleftrightarrow M_{b}$,
$B_{t}\longleftrightarrow I_{e}\dashleftarrow\dashrightarrow M_{b}$
and $B_{t}\dashleftarrow\dashrightarrow Re(Kpc)\Longleftrightarrow M_{b}$.
Using the linear fit results $(1.23\pm0.04)B_{t}=DM+c$ and $(-0.53\pm0.03)DM=M_{b}+c'$
yields $M_{b}=(-0.65\pm0.06)B_{t}+c"$. Using the linear fit results
$(-7.5\pm0.2)E^{-3}Ie=B_{t}+c$ and $(8.1\pm0.9)E^{-3}Ie=M_{b}+c'$
yields a contribution $M_{b}=(-1.08\pm0.15)B_{t}+c"$. Using the linear
fit results $(-0.12\pm0.05)B_{t}=Re(Kpc)+c$ and $(-1.56\pm0.09)M_{b}=Re(Kpc)+c'$
yields a contribution $M_{b}=(0.08\pm0.04)B_{t}+c"$. These contributions
cannot be linearly added since $DM$ \& $I_{e}$, $DM$ \&$M_{b}$,
$DM$ \& $Re(Kpc)$ and $I_{e}$ \& $Re(Kpc)$ display significant
correlations.
\item $B_{t}$ vs surface brightness $I_{e}$:
these observables are related by $Ie=dex(9.465-0.4Bt)/(\sfrac{R_{min}}{R_{max}})Re(")^{2}$.
\item $B_{t}$ vs effective radius (parsec):
we expect a positive correlation from $Re(Kpc)\Longleftrightarrow DM\Longleftrightarrow Bt$.
This is observed. However, we cannot verify it numerically due to
the failure of the $Re(Kpc)$ vs $B_{t}$ fit.
\item $B_{t}$ vs velocity distribution $\sigma_{0}$:
no correlation is seen. We expected some contribution from $\sigma_{0}\longleftrightarrow DM\Longleftrightarrow B_{t}$,
$\sigma_{0}\Longleftrightarrow Re(Kpc)\dashleftarrow\dashrightarrow B_{t}$
and $\sigma_{0}$. Using the linear fit results $(1.23\pm0.04)B_{t}=DM+c$
and $(1.24\pm0.08)E^{-2}\sigma_{0}=DM+c'$ yields a contribution $B_{t}=(1.01\pm0.10)E^{-2}\sigma_{0}+c"$.
The second correlation cannot be computed due to the failure of the
$Re(Kpc)$ vs $B_{t}$ fit. However, we can estimate that it yields
a contribution $B_{t}\sim6E^{-3}\sigma_{0}+c"$. Overall, the two
small contributions agree with the observed absence of correlation
(the smaller $B_{t}=(1.1\pm0.2)E^{-3}\sigma_{0}+C"$ is unreliable
due to the large $\sfrac{\chi^{2}}{ndf}$ of the fit).
\item $B_{t}$ vs $\sfrac{M}{L}$. No correlation
is expected (all secondary correlations involve a weak/unclear correlation).
This is what is observed.
\end{enumerate}

\paragraph{Magnitude from NED correlations}
\begin{figure}[H]
\centering
\includegraphics[scale=0.24]{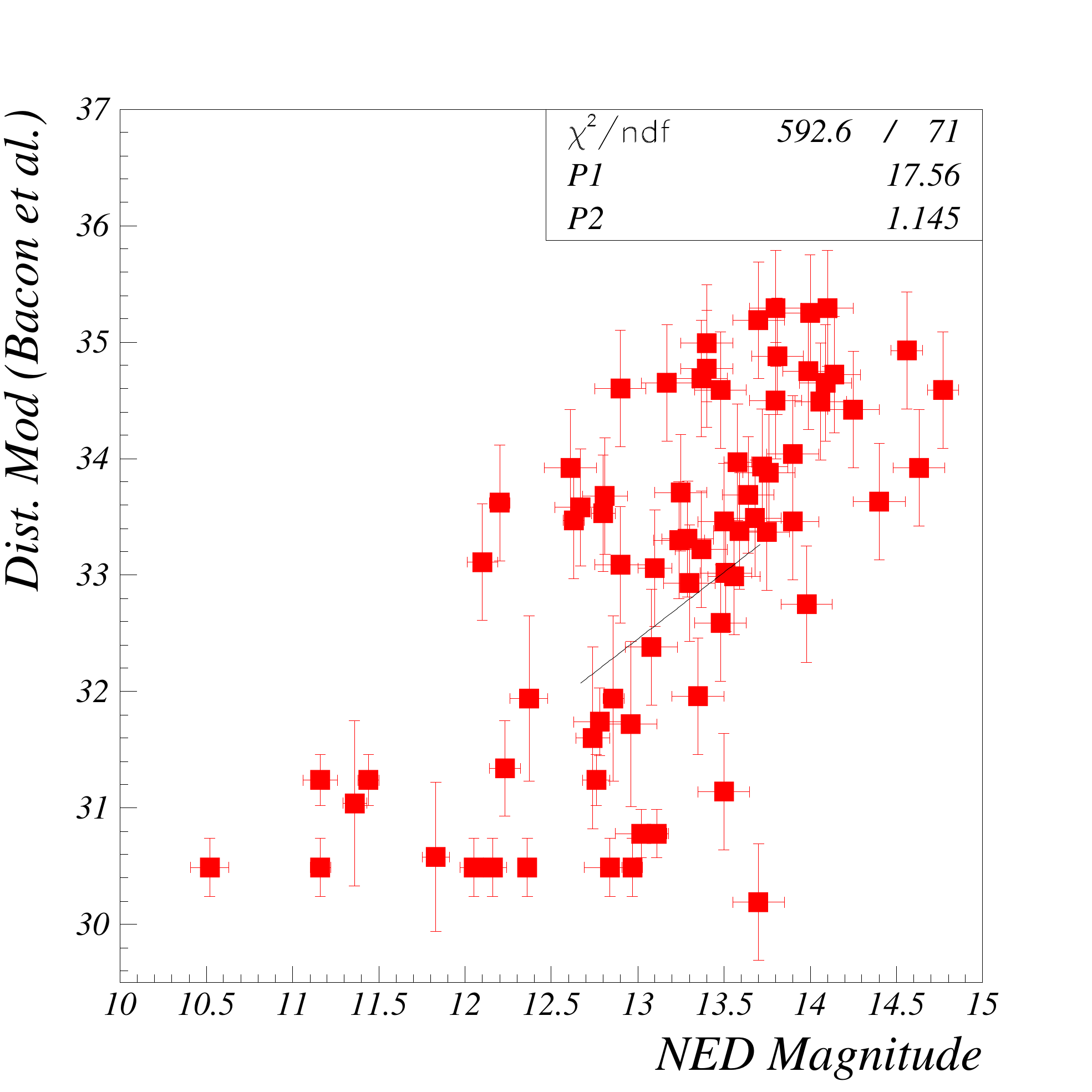}\includegraphics[scale=0.24]{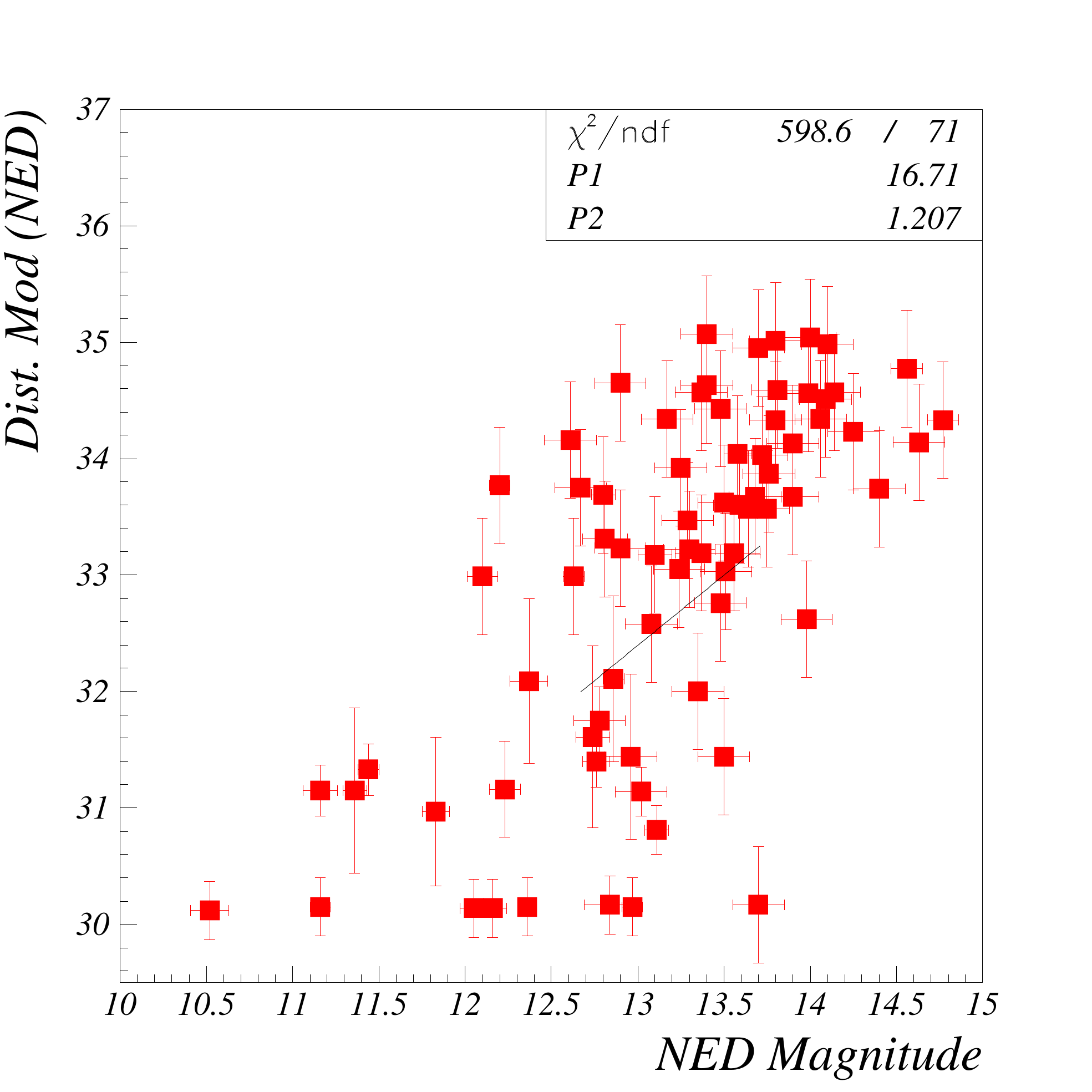}\includegraphics[scale=0.24]{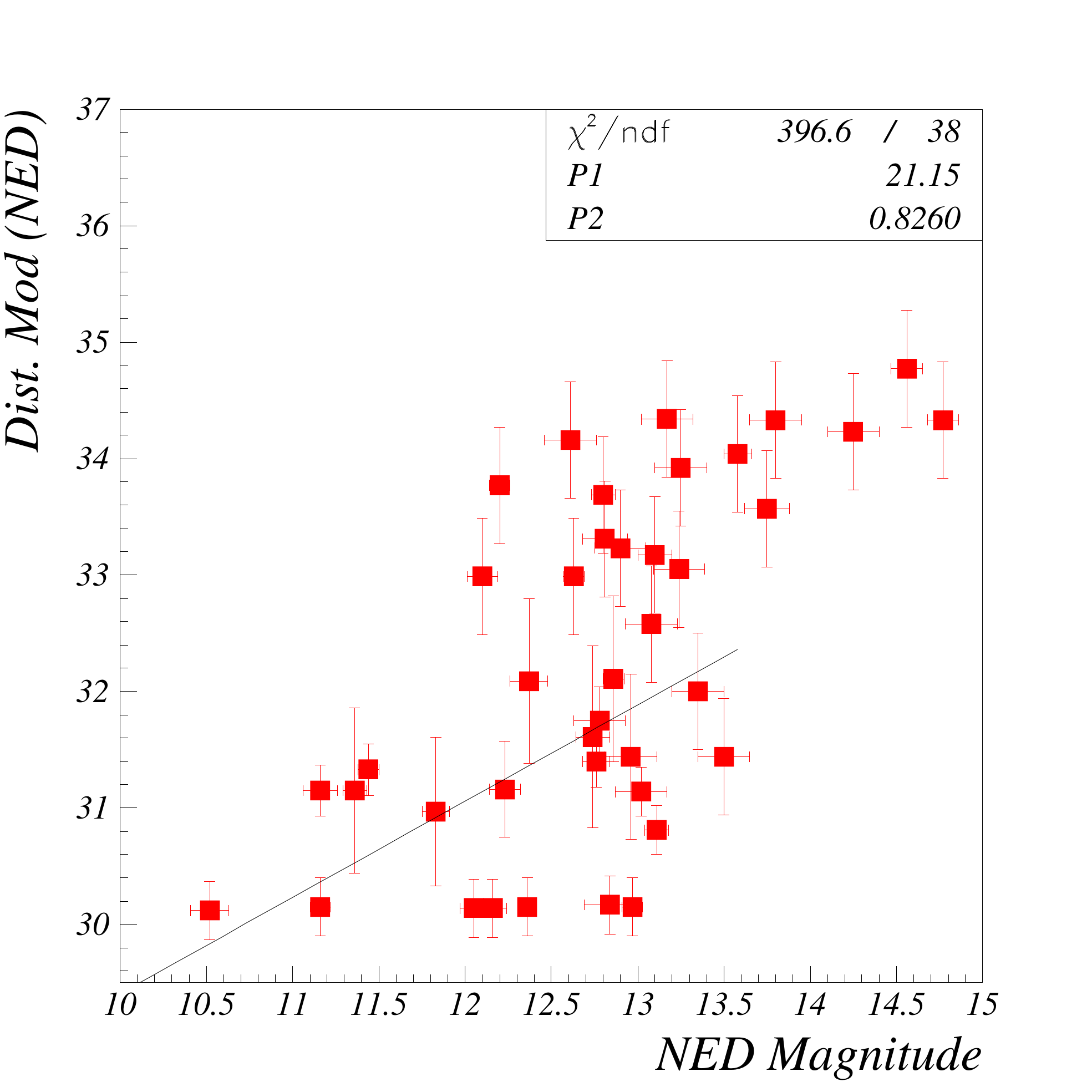}\includegraphics[scale=0.24]{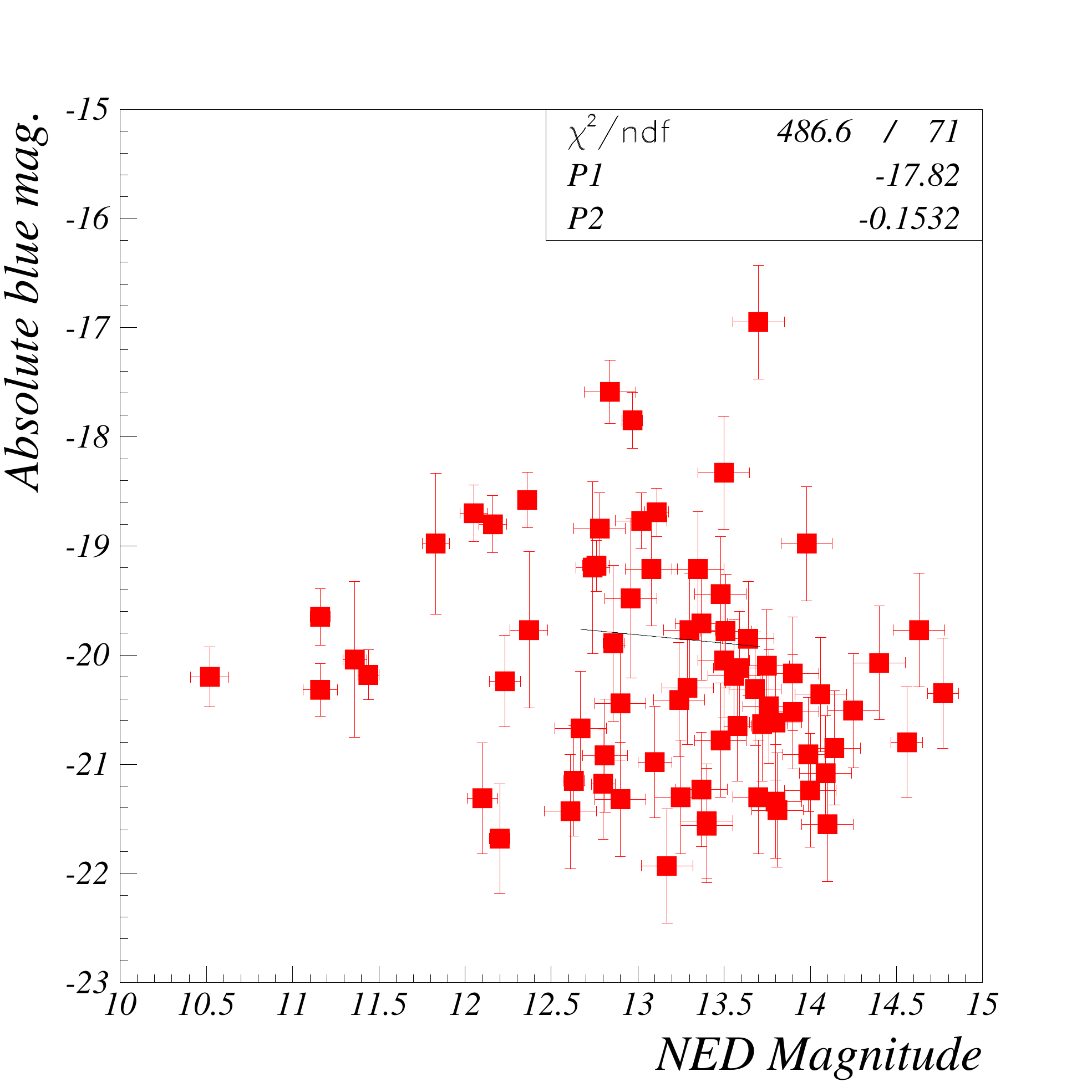}\protect \\
\includegraphics[scale=0.24]{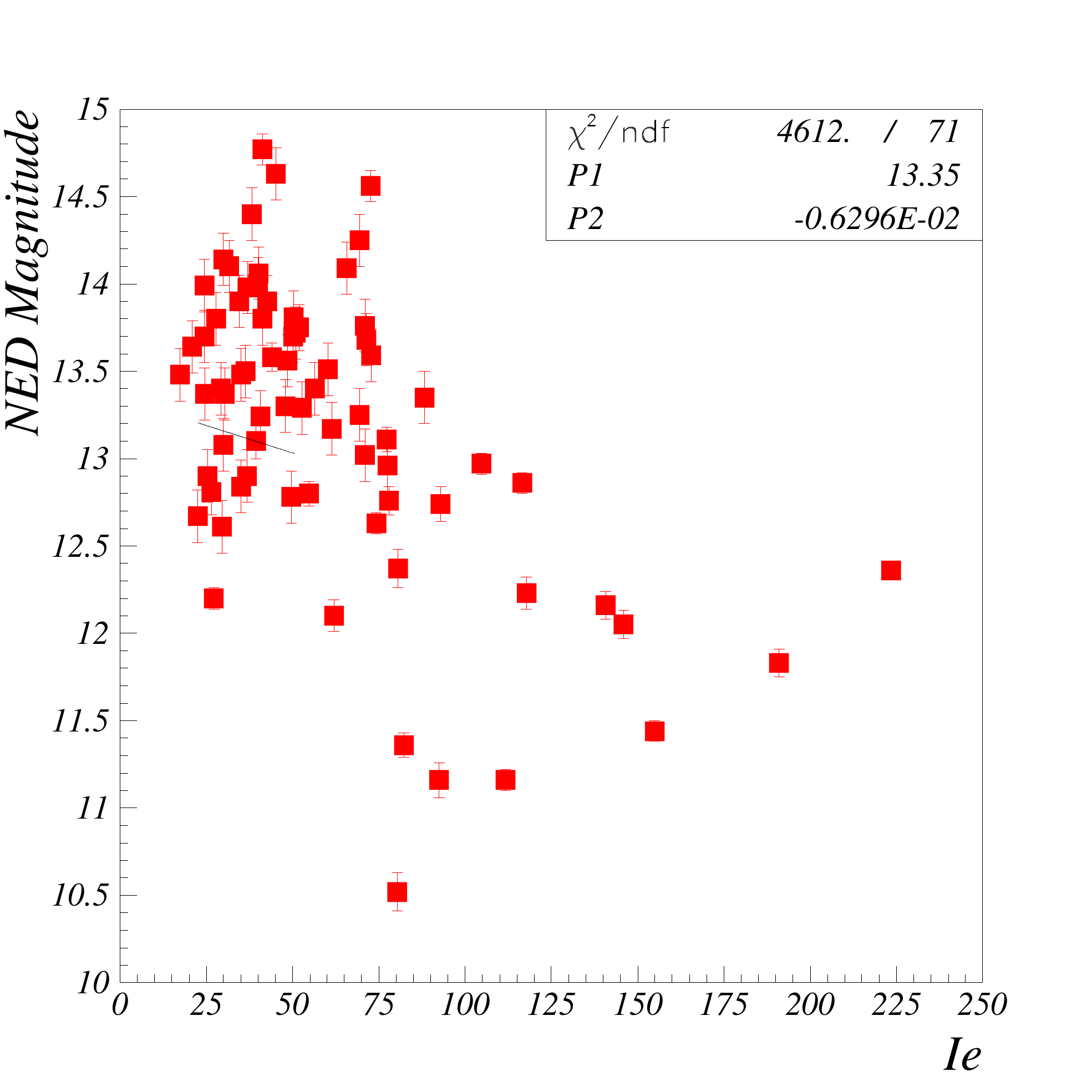}\includegraphics[scale=0.24]{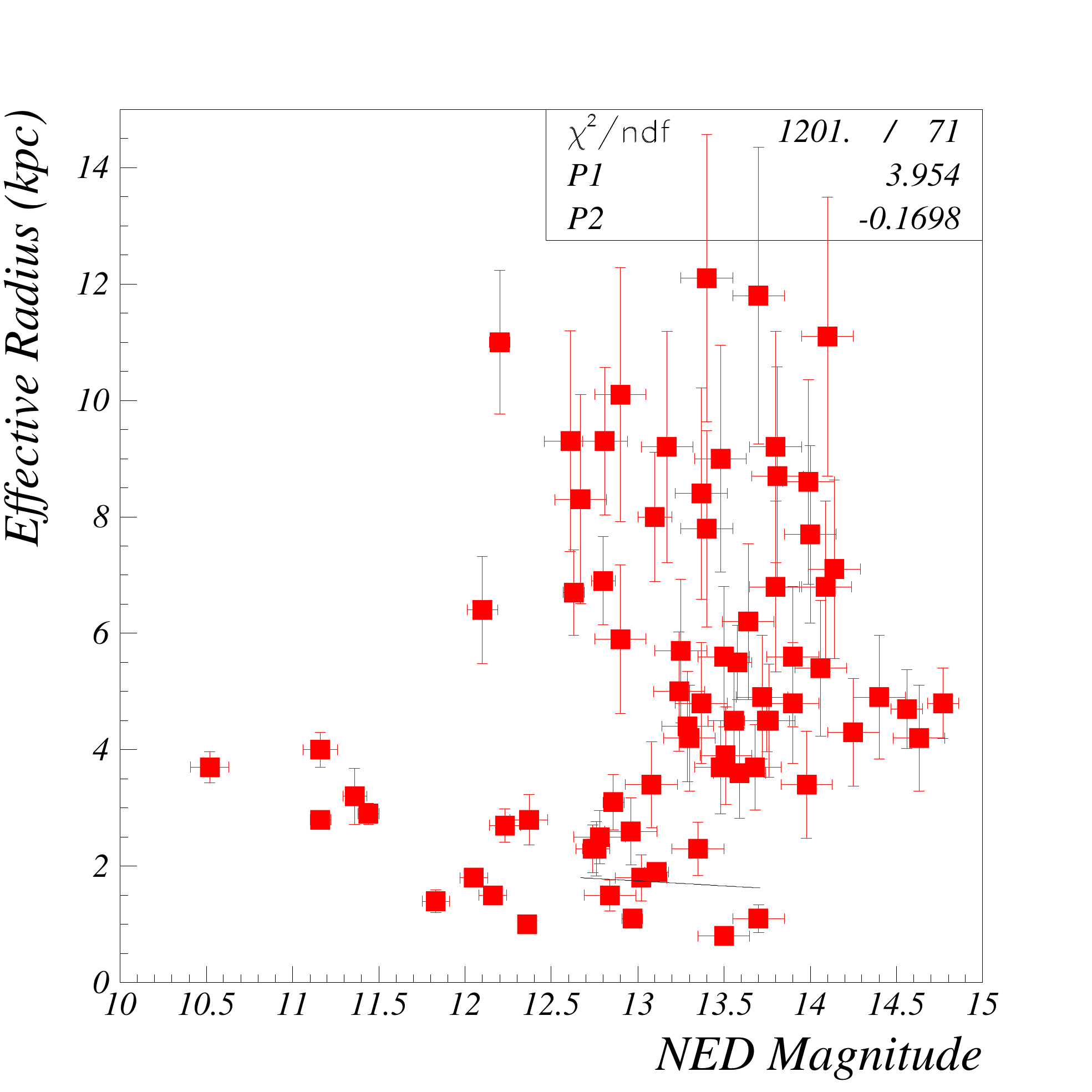}\includegraphics[scale=0.24]{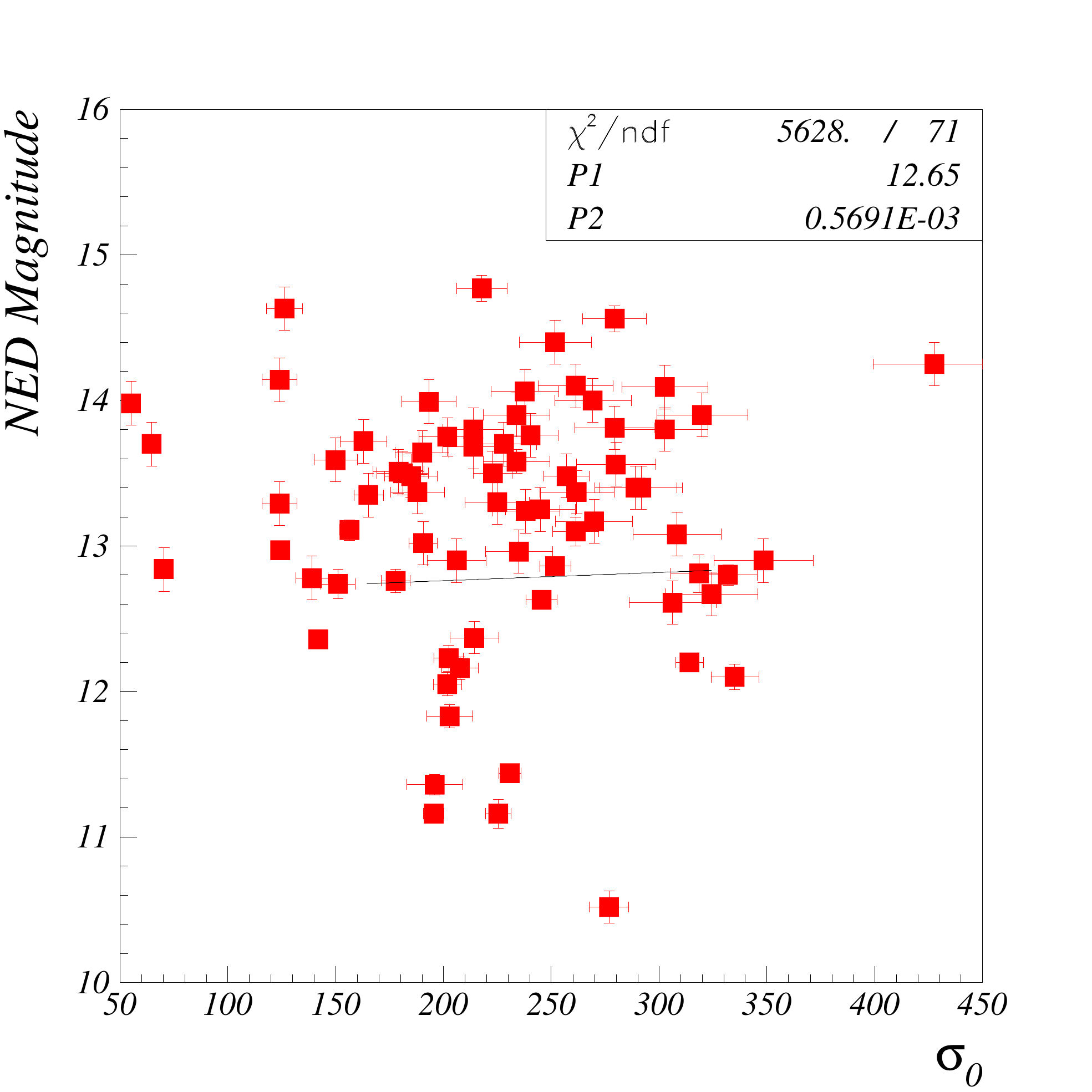}\includegraphics[scale=0.24]{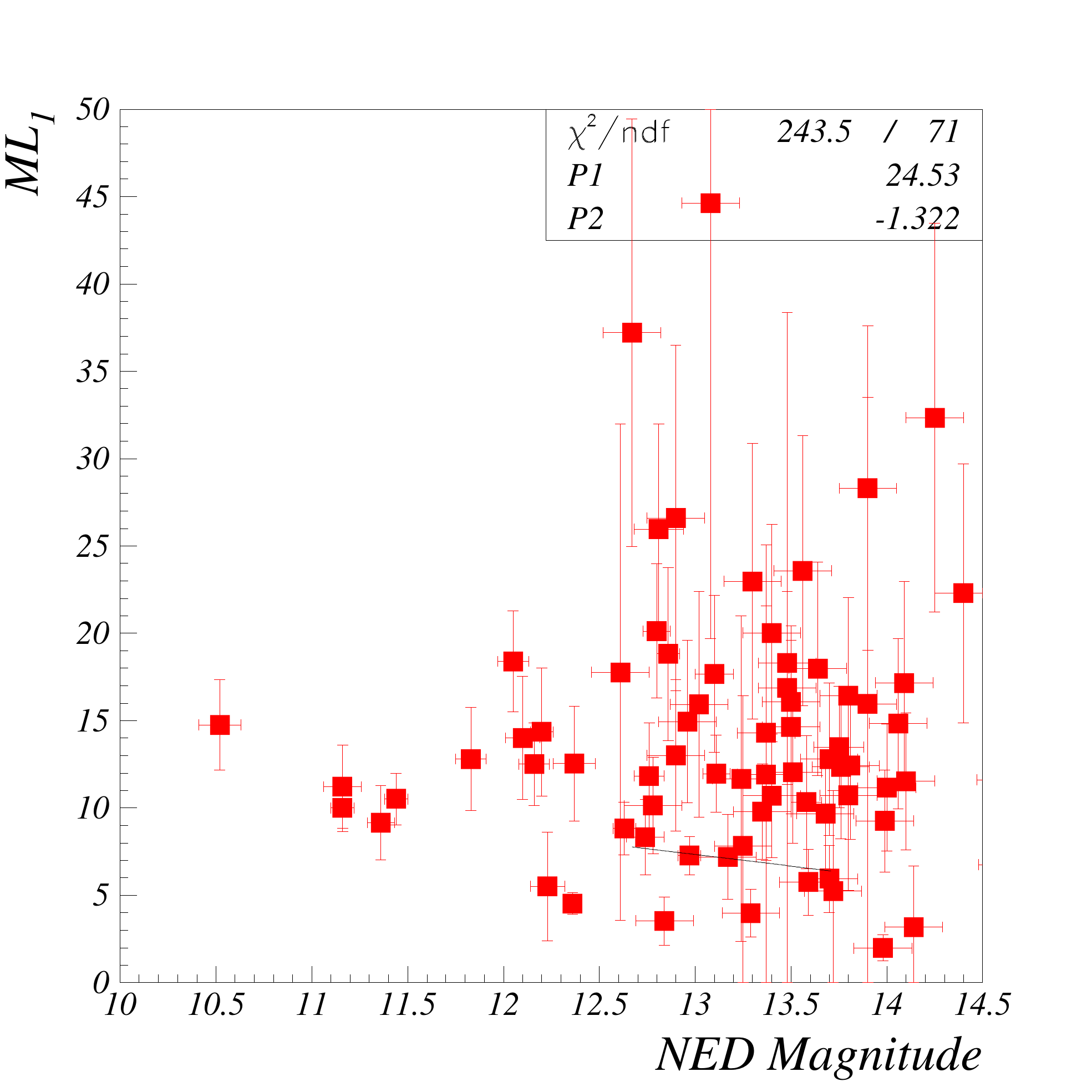}\protect \\
\includegraphics[scale=0.24]{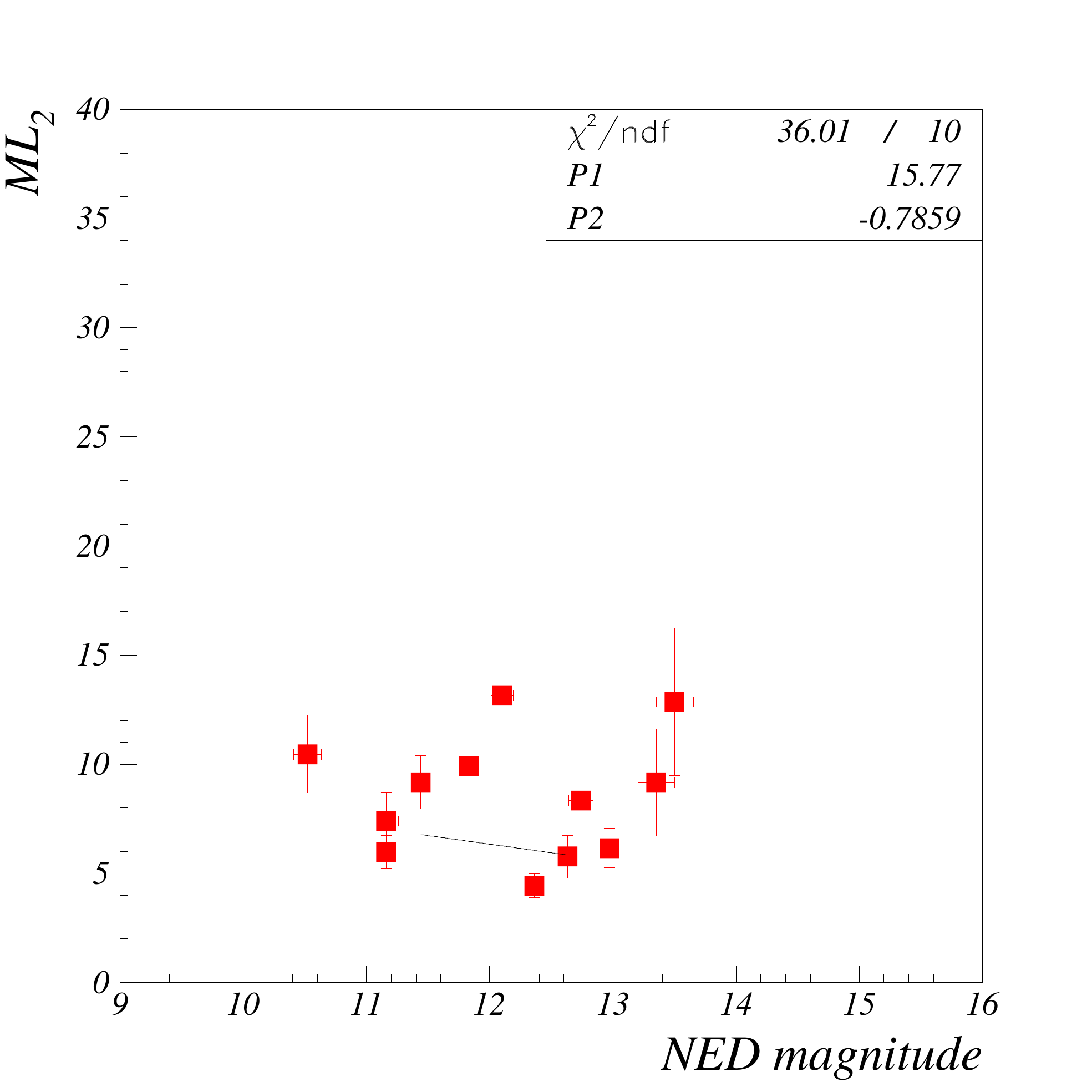}\includegraphics[scale=0.24]{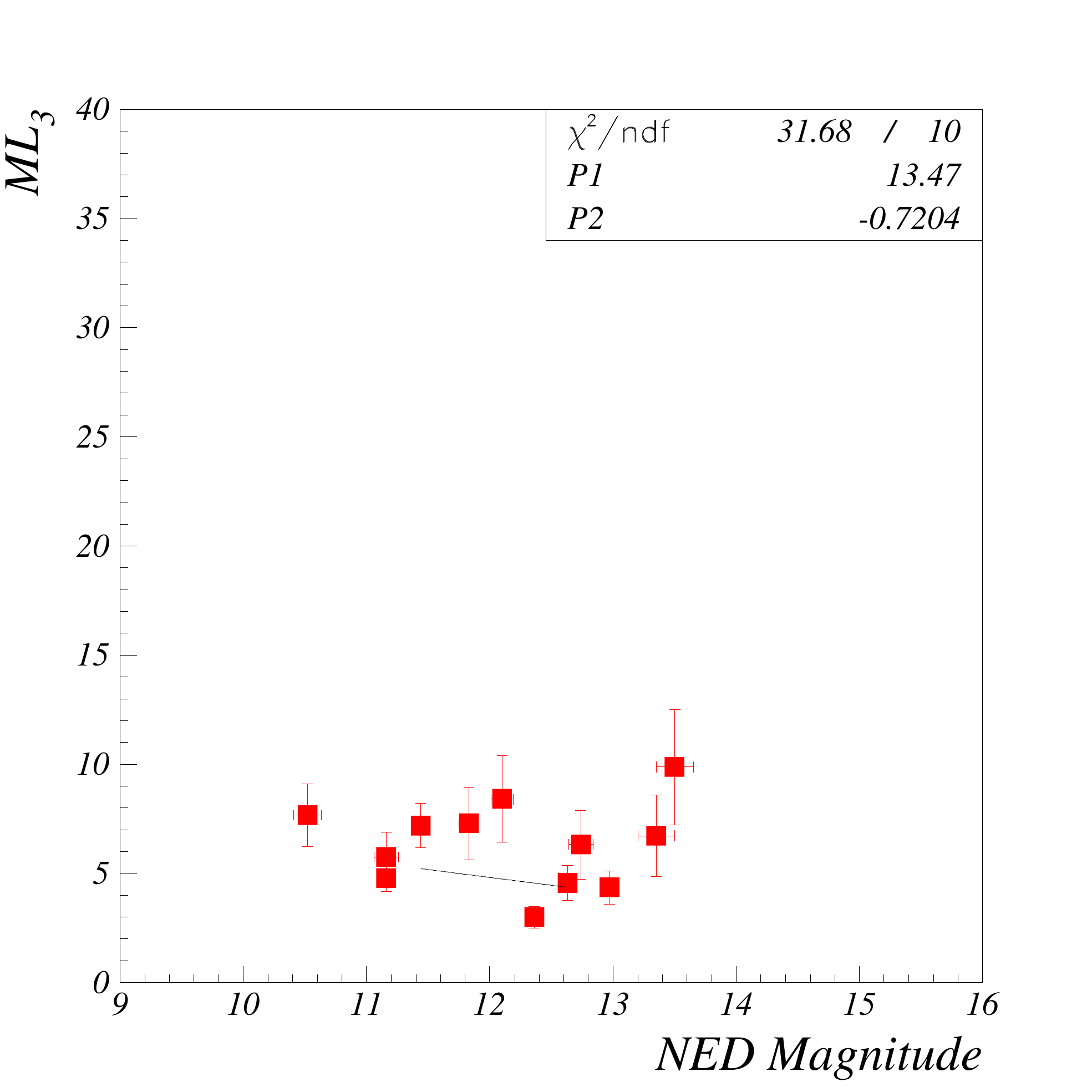}
\vspace{-0.4cm} \caption{\label{Fig: NED mag. correlations}Correlations between the NED apparent
magnitude and (from top left to bottom right):  $DM$
from~\cite{BMS}, $DM$ from NED using redshift information,
$DM$ from NED without redshift information, absolute blue
magnitude, surface brightness $I_{e}$, effective radius (parsec),
velocity distribution $\sigma_{0}$ and $\sfrac{M}{L}$.
}
\end{figure}
The origins of the correlations are similar to the ones discussed
in Section~\ref{sub:Integrated-blue-magnitude correl}.

\paragraph{Distance Moduli (\cite{BMS}) correlations\label{sub:Distance-Moduli correl}}

\begin{figure}[H]
\centering
\includegraphics[scale=0.24]{dmbms_dmned.pdf}\includegraphics[scale=0.24]{dmbms_dmnednox.pdf}\includegraphics[scale=0.24]{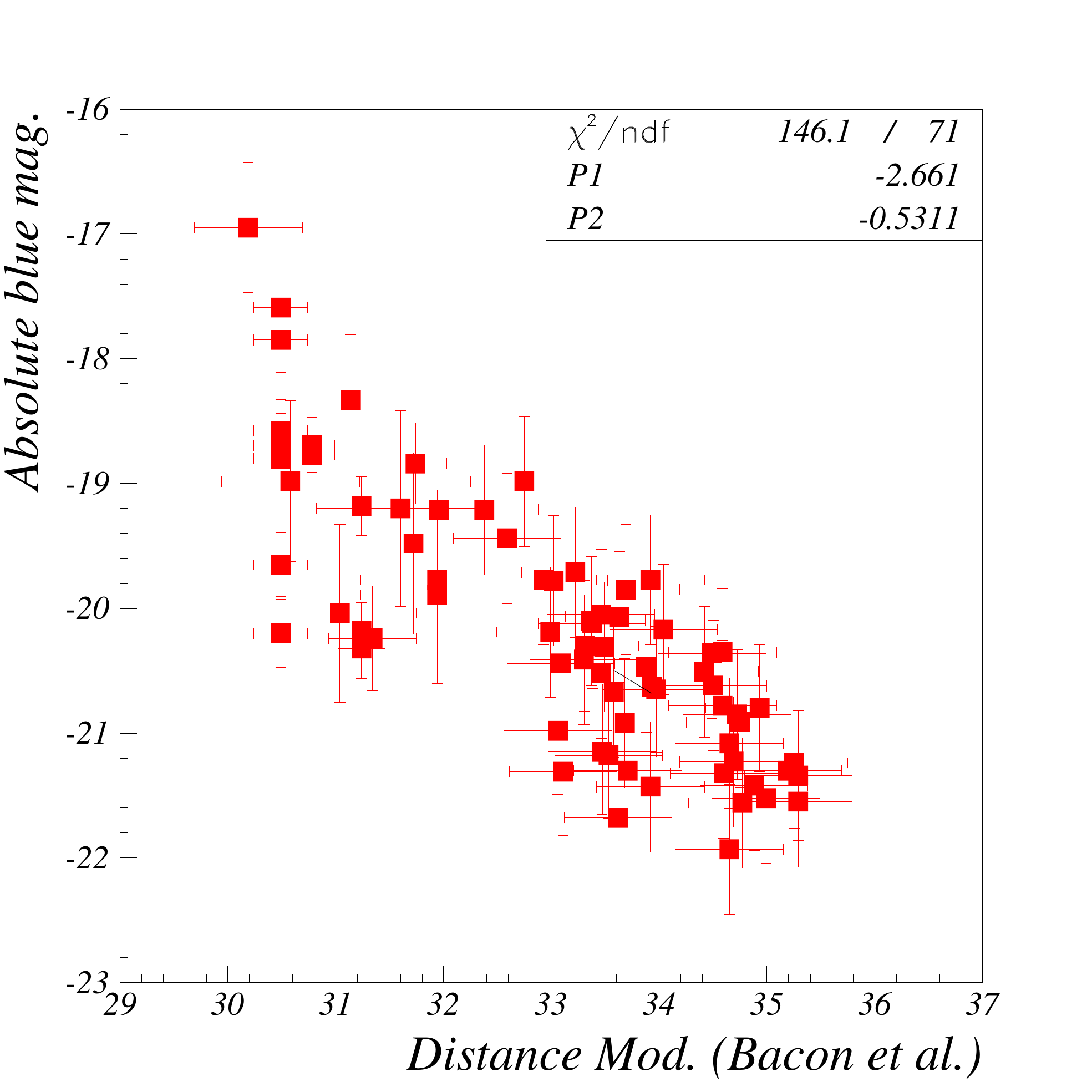}\includegraphics[scale=0.24]{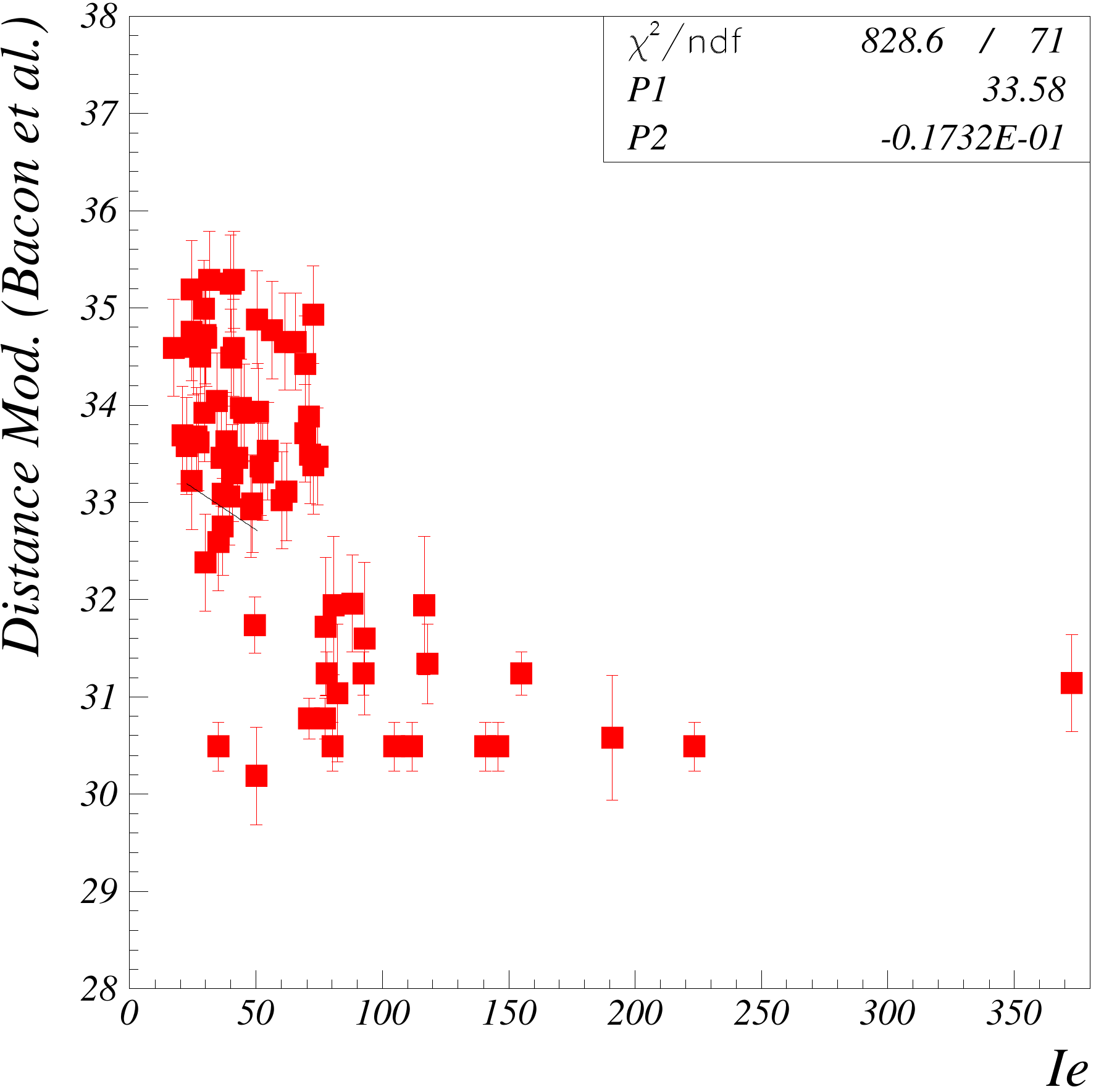}\protect \\
\includegraphics[scale=0.24]{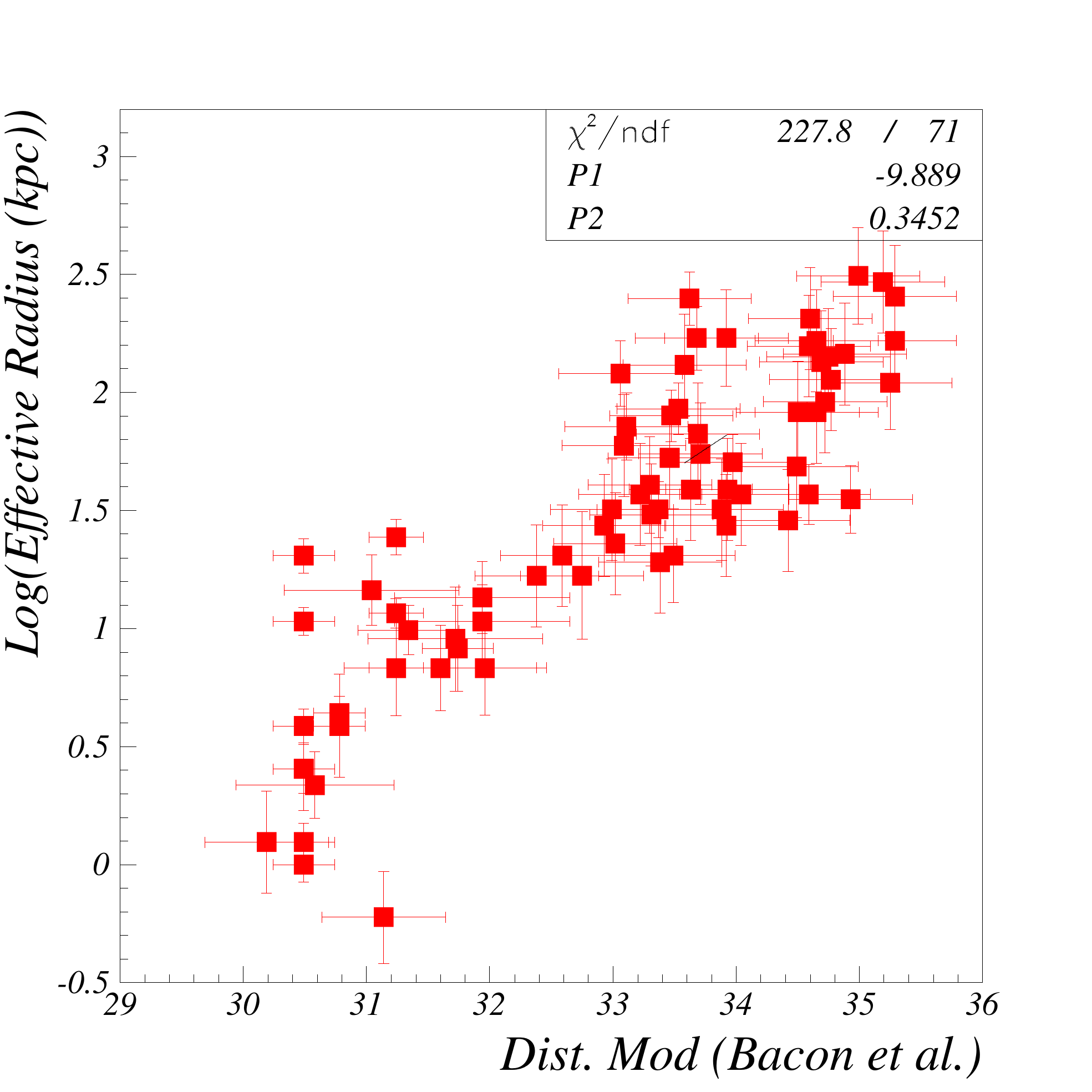}\includegraphics[scale=0.24]{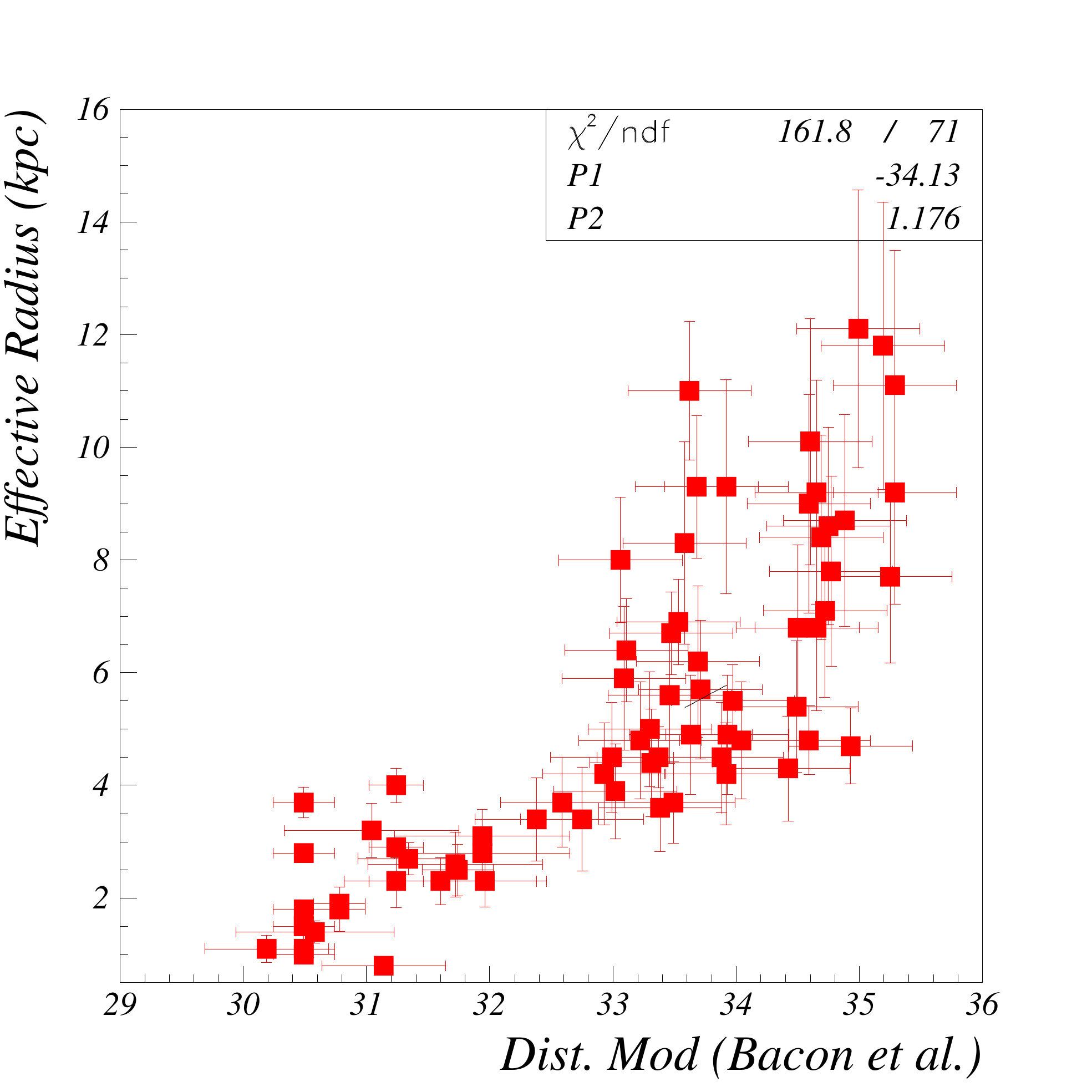}\includegraphics[scale=0.24]{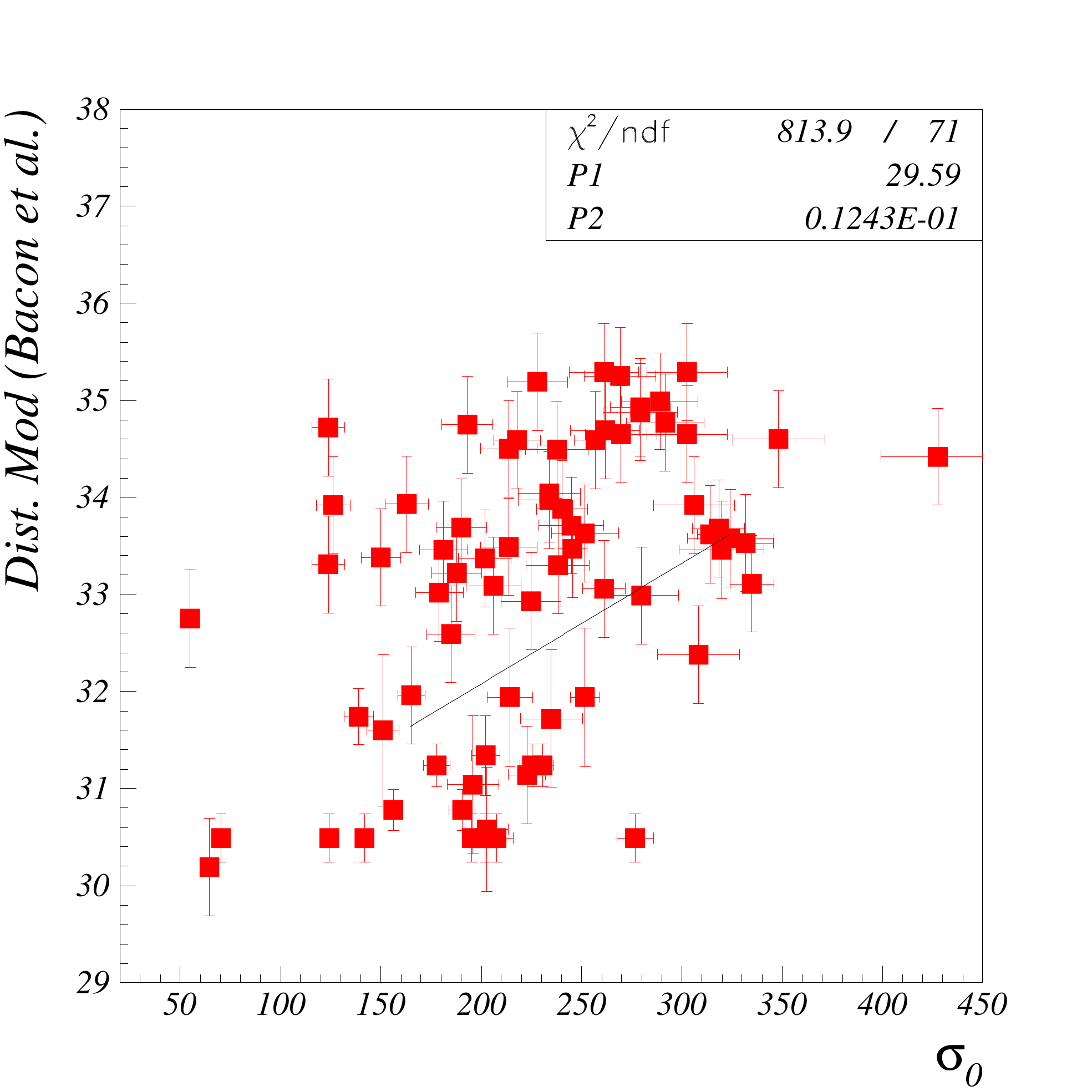}\includegraphics[scale=0.24]{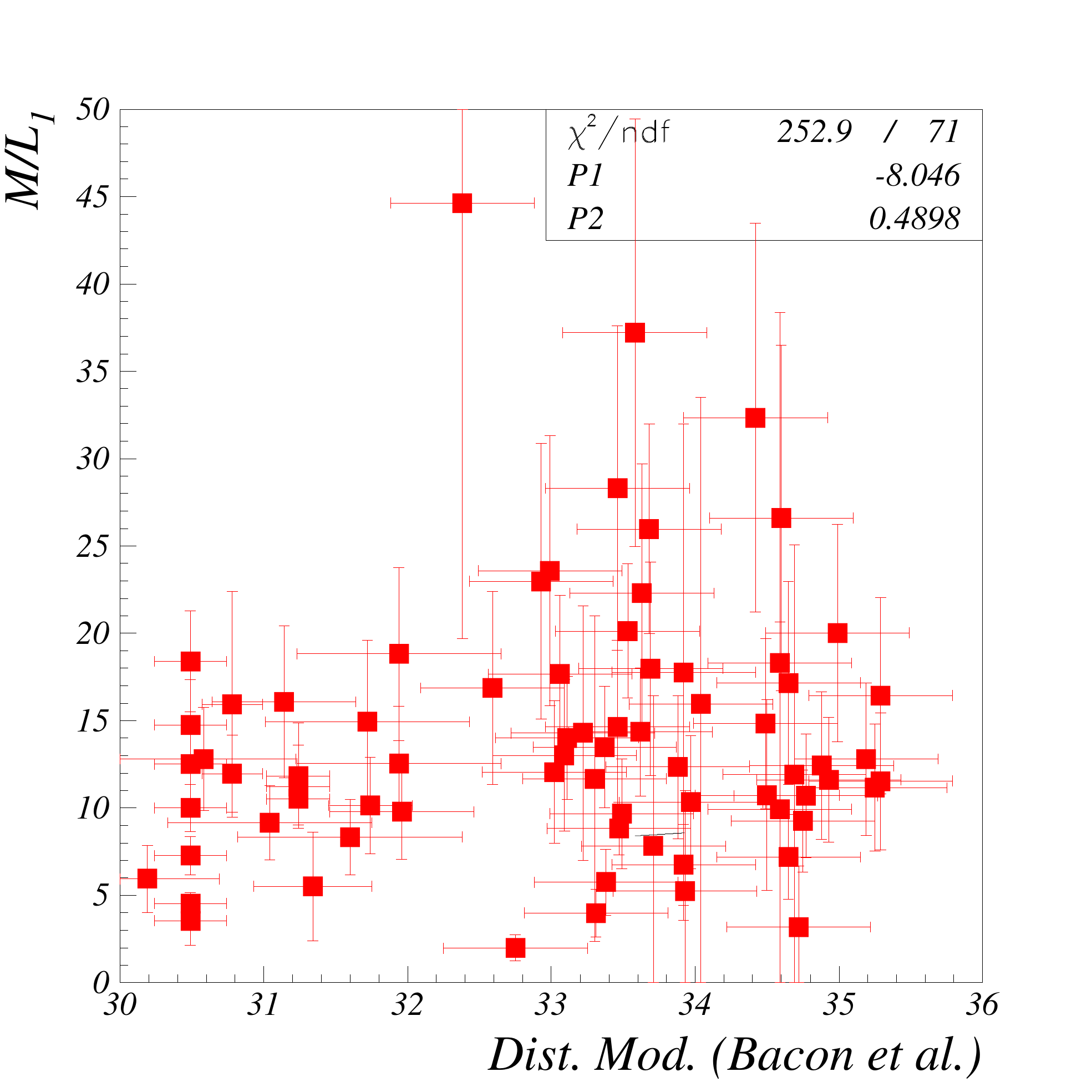}\protect \\
\includegraphics[scale=0.24]{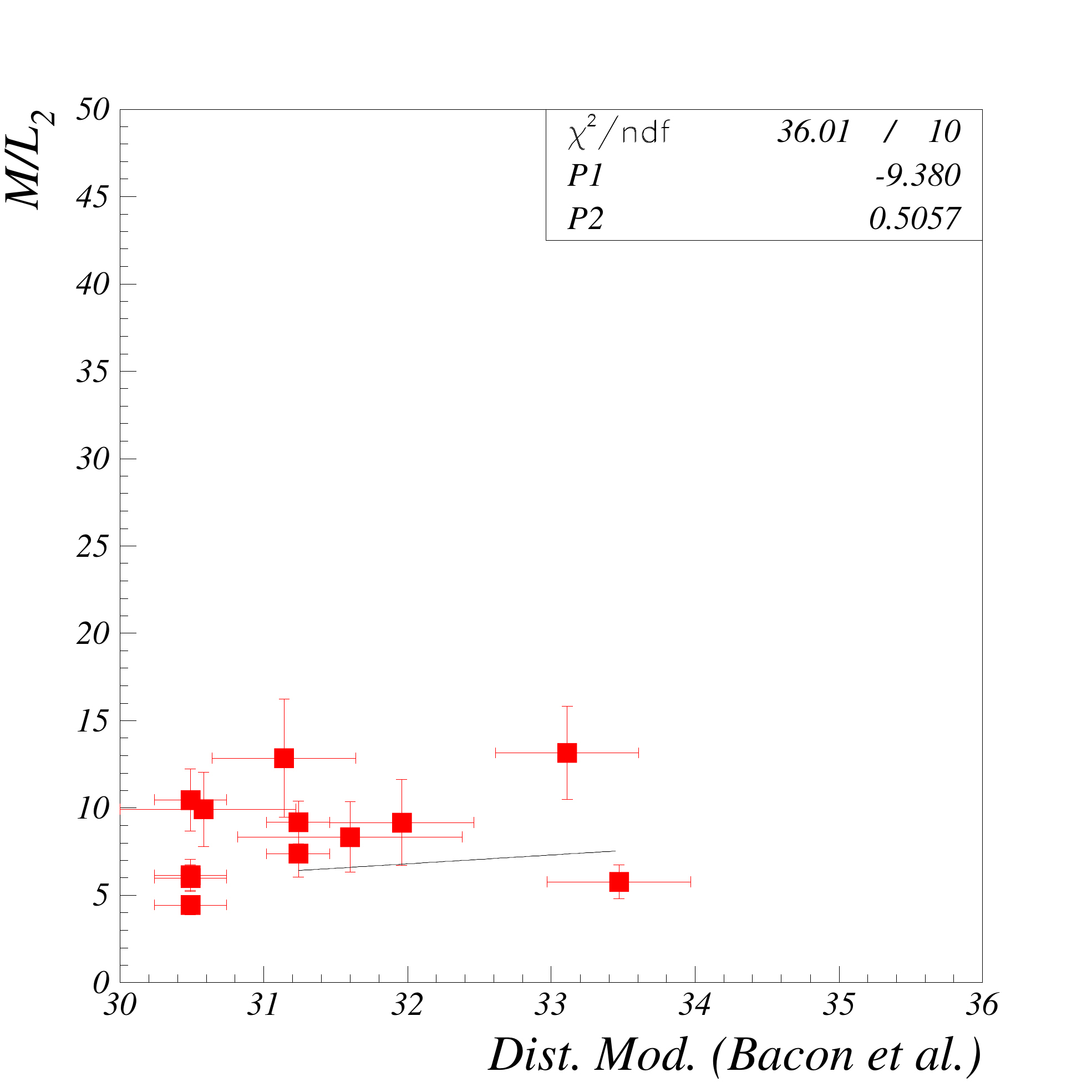}\includegraphics[scale=0.24]{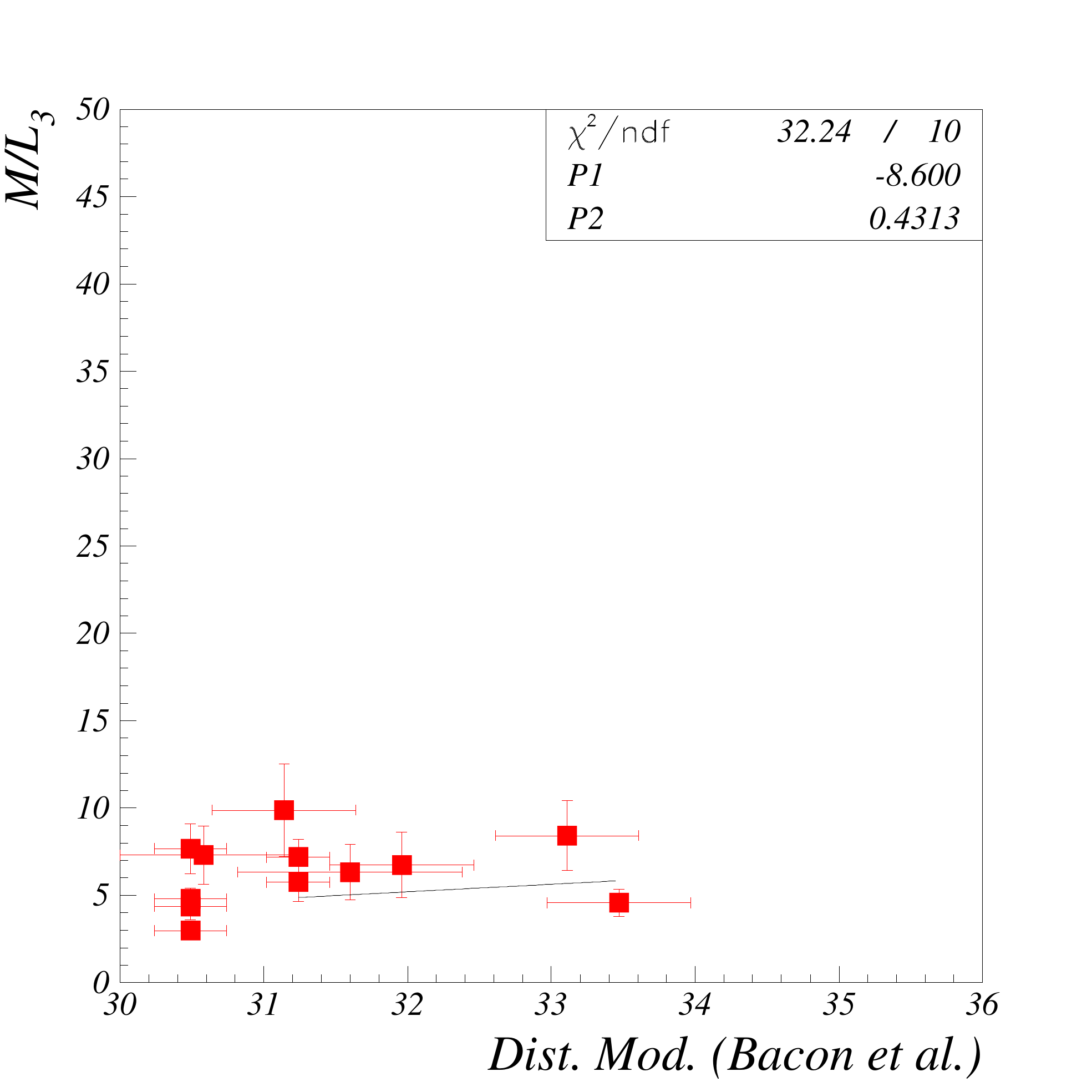}
\vspace{-0.4cm} \caption{\label{Fig: DM BMS correlations}Correlations between the distance
modulus $DM$ from~\cite{BMS} and (from top left to bottom right):
$DM$ from NED using redshift information, $DM$
from NED without redshift information, absolute blue magnitude, surface
brightness $Ie$, $\log$(effective radius (parsec)), effective radius
(parsec), velocity distribution $\sigma_{0}$ and $\sfrac{M}{L}$.
}
\end{figure}
The origins of the correlations are, from the top left plot to the
bottom right one: 
\begin{enumerate}
\item Distance modulus $DM$ from~\cite{BMS} vs $DM$
from NED using redshift information: this expected strict correlation
was already discussed, see Fig.~\ref{fig:dmbms_dmned}.
\item $DM$ from~\cite{BMS} vs $DM$
from NED without redshift information: this expected strict correlation
was already discussed, see Fig.~\ref{fig:dmbms_dmnednox}.
\item $DM$ from~\cite{BMS} vs absolute blue
magnitude: we observe a strong correlation. Since we do not expect
significant time evolution, this must be an observational bias: at
large $DM$, more luminous galaxies are easier to observe. At smaller $DM$, both
luminous and dimmer galaxies can be seen, but the presence probability
of luminous elliptical galaxies is smaller since the corresponding
volume is smaller and luminous galaxies are rarer than
dimmer ones. In all, luminous galaxies tend to be seen at large
distances while dimmer ones are seen at shorter distances. That
this is a measurement bias and not a physical characteristic is confirmed
by the absence of  $DM$ vs $\sfrac{M}{L}$ correlation,
see point 8. below.
\item $DM$ from~\cite{BMS} vs surface brightness
$I_{e}$: surface brightness is independent of distances. The correlation
seen is expected from the fact that $Re(Kpc)$ and $DM$ are almost
linearly related (see the two next plots) and that $I_{e}$ and $Re(Kpc)$
are also correlated (consequence of the Kormendy relation~\cite{Kormendy}). Indeed,
the pattern of $I_{e}$ vs $DM$ strongly resembles the one of
of $I_{e}$ vs $Re(Kpc)$. This is an indirect consequence of
the $M_b\Longleftrightarrow DM$ observation bias.
\item $DM$ from~\cite{BMS} vs effective radius
(parsec): a significant secondary (indirect) correlation is expected
from the Kormendy correlation $Re(Kpc)\Longleftrightarrow M_b$ and
$M_b\Longleftrightarrow DM$ correlation. Using the linear fit results
$(-1.58\pm0.09)M_b=Re+c$ and $(-0.53\pm0.03)DM=M_b+c'$ yields $Re=(0.84\pm0.1)DM+c"$.
Another indirect correlation is expected from the $3^{rd}$ fundamental
plan correlation $Re(Kpc)\Longleftrightarrow\sigma_{0}$ and the $\sigma_{0}\Longleftrightarrow$
$DM$ correlation. Using the fit results $(1.67\pm0.08)E^{-2}\sigma_{0}=Re+c$
and $(1.24\pm0.08)E^{-2}\sigma_{0}=DM+c'$ yields $Re=(1.35\pm0.15)DM+c"$.
These secondary correlations are similar to the observed $Re=(1.18\pm0.06)DM+c"$
(linear fit). In addition, we expect the same type of observational
bias as for the $DM$ vs $M_{b}$ correlation. All in all, this
is a consequence of the $M_b\Longleftrightarrow DM$ observation bias.
\item Same as above but with a linear vertical scale. 
\item $DM$ from~\cite{BMS} vs velocity distribution
$\sigma_{0}$: we observe a significant correction of $\sigma_{0}$
with $DM$. We expect strong secondary correlations due to the{\it 
}Kormendy relation, the $DM$ vs $M_{b}$ observational bias,
the Faber-Jackson and the $3^{rd}$ fundamental plane relations. This
is again an indirect consequence of the $M_b\Longleftrightarrow DM$
and $Re(Kpc)\Longleftrightarrow DM$ observation biases.
\item $DM$ from~\cite{BMS} vs $\sfrac{M}{L}$:
we observe little correlation. This confirms that the $DM$ vs
absolute blue magnitude correlation is an unphysical bias: if the
blue magnitude truly depended on $DM$, then, the $\sfrac{M}{L}$ would
display a similar dependence since we do not expect the mass 
of an undisturbed galaxy to evolve with time.
\end{enumerate}

\paragraph{Distance Moduli (NED with redshift information) correlations}

\begin{figure}[H]
\centering
\includegraphics[scale=0.24]{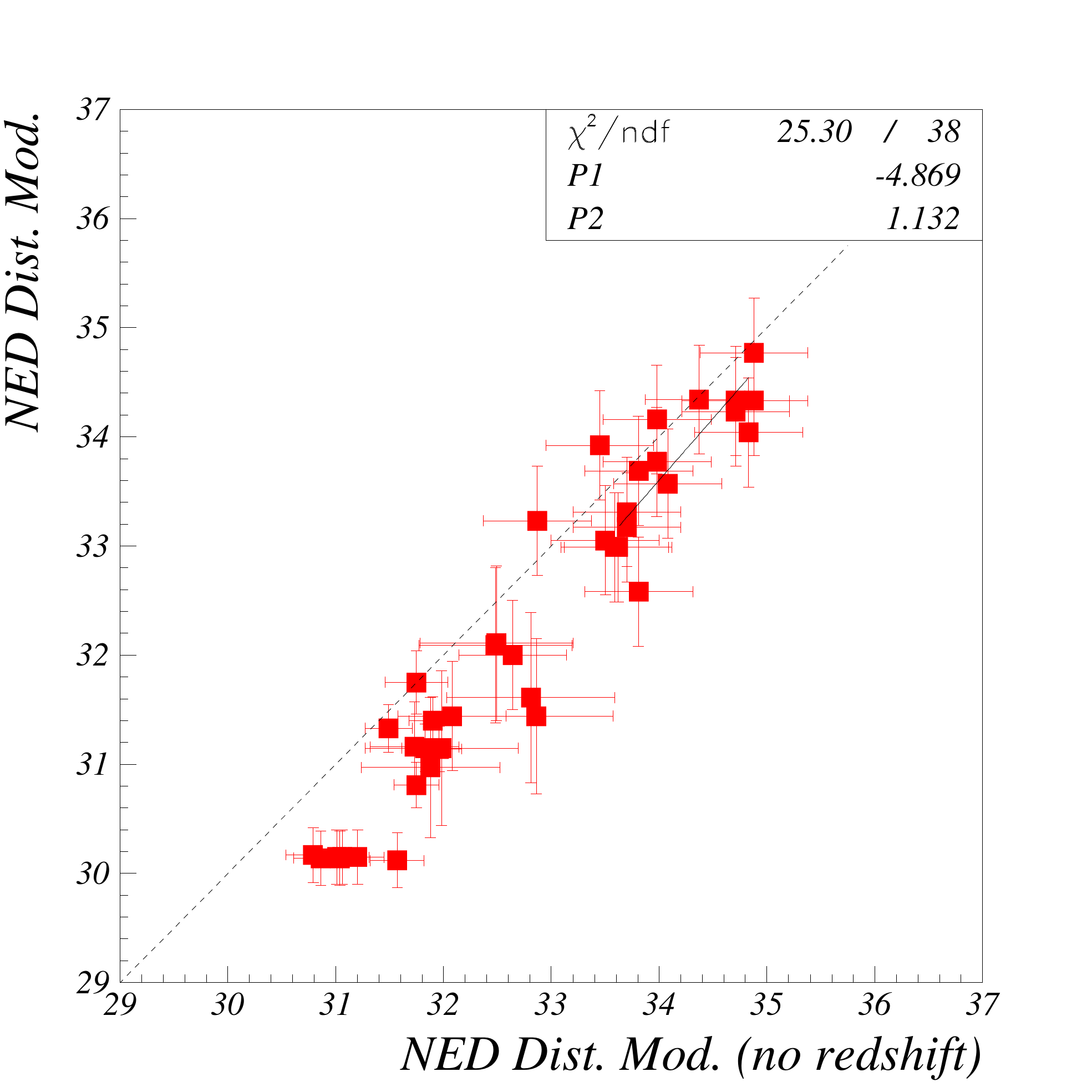}\includegraphics[scale=0.24]{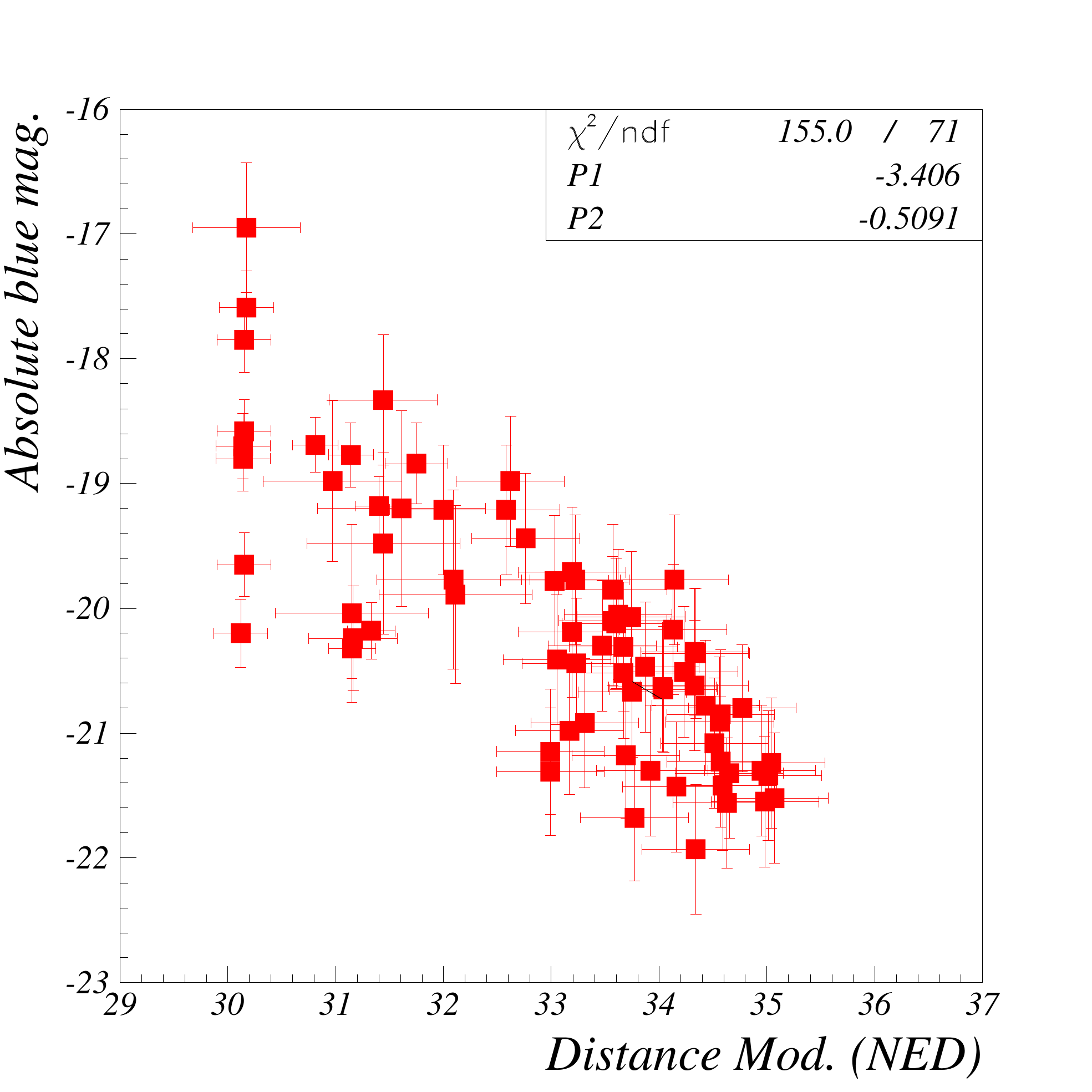}\includegraphics[scale=0.24]{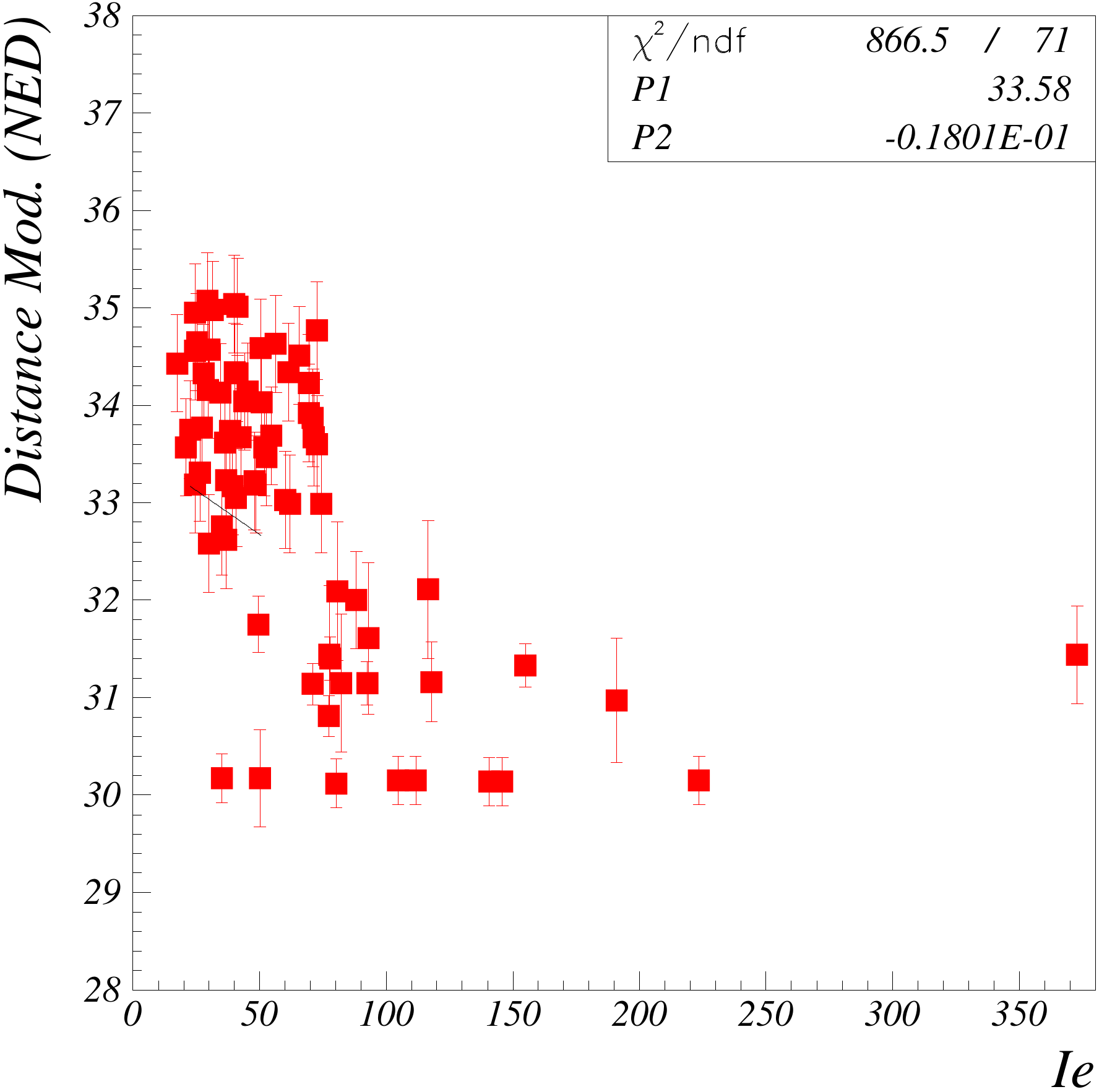}\includegraphics[scale=0.24]{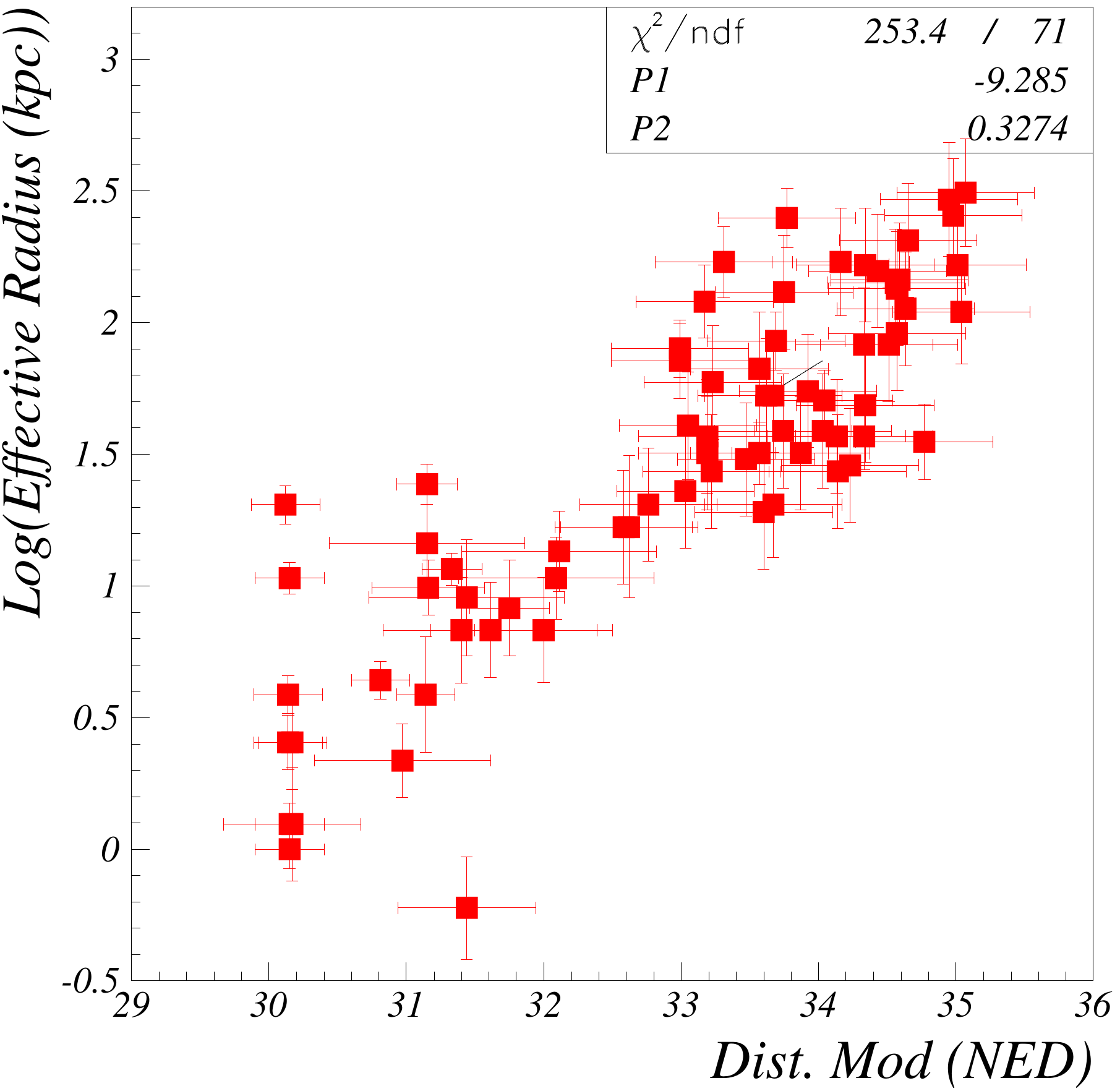}\protect \\
\includegraphics[scale=0.24]{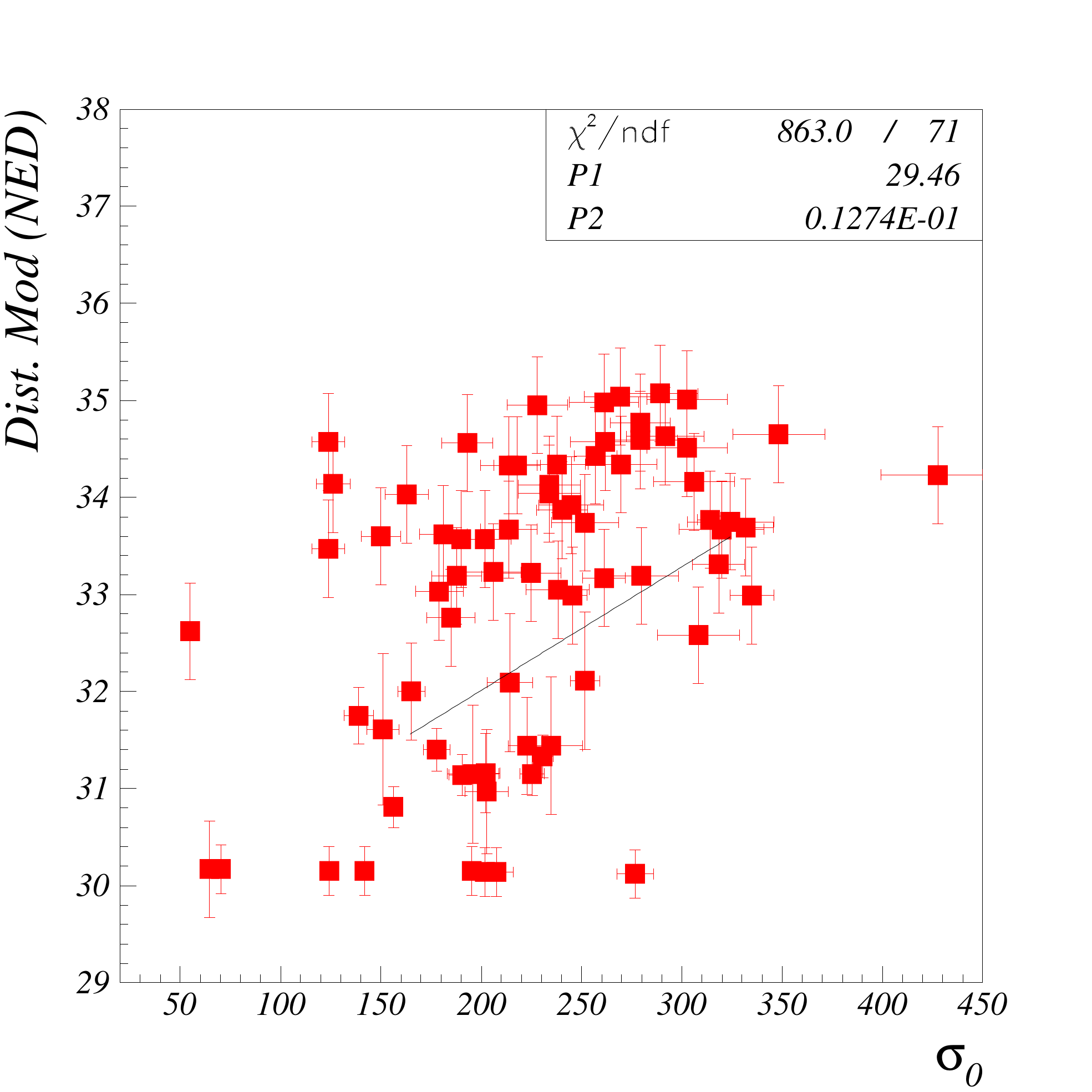}\includegraphics[scale=0.24]{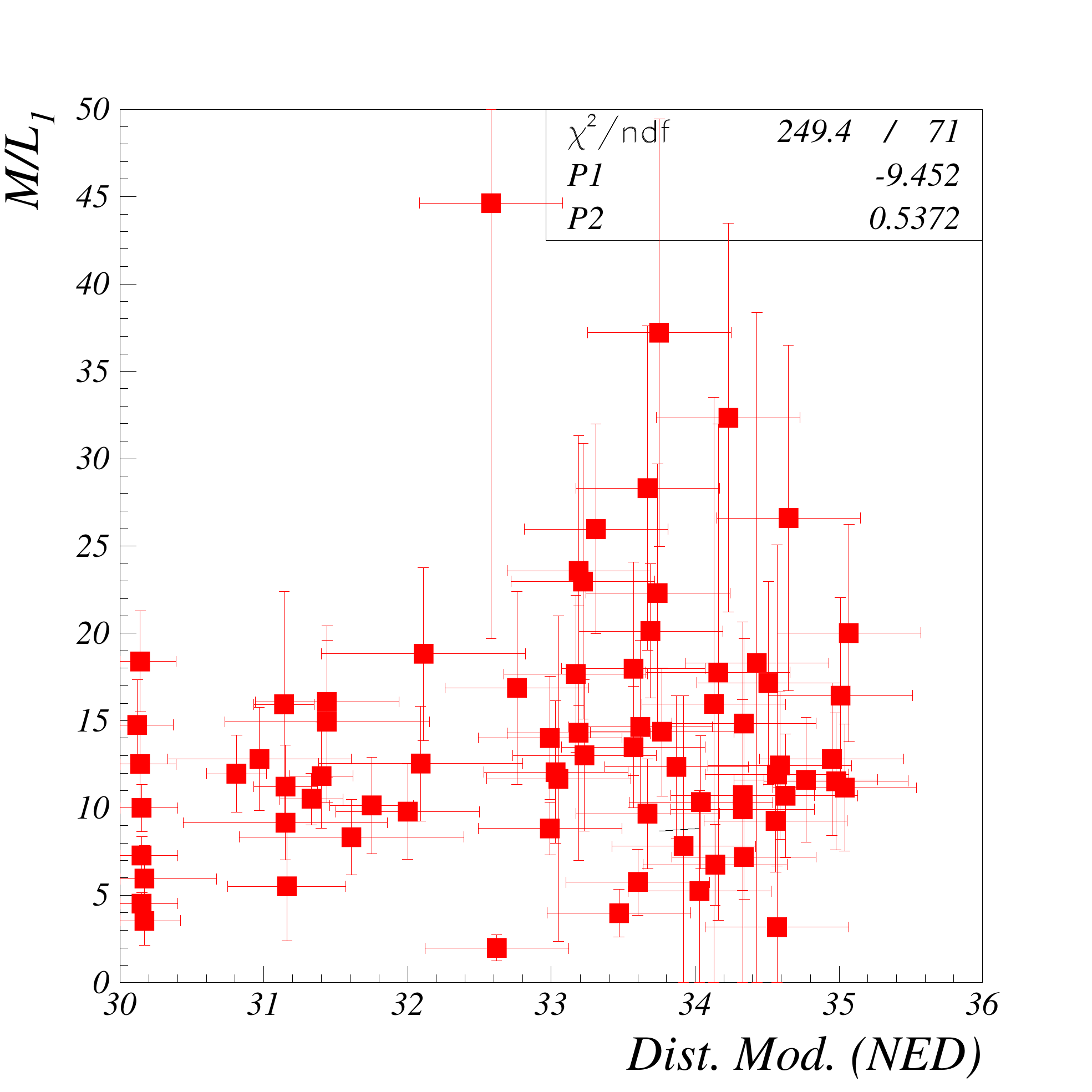}\includegraphics[scale=0.24]{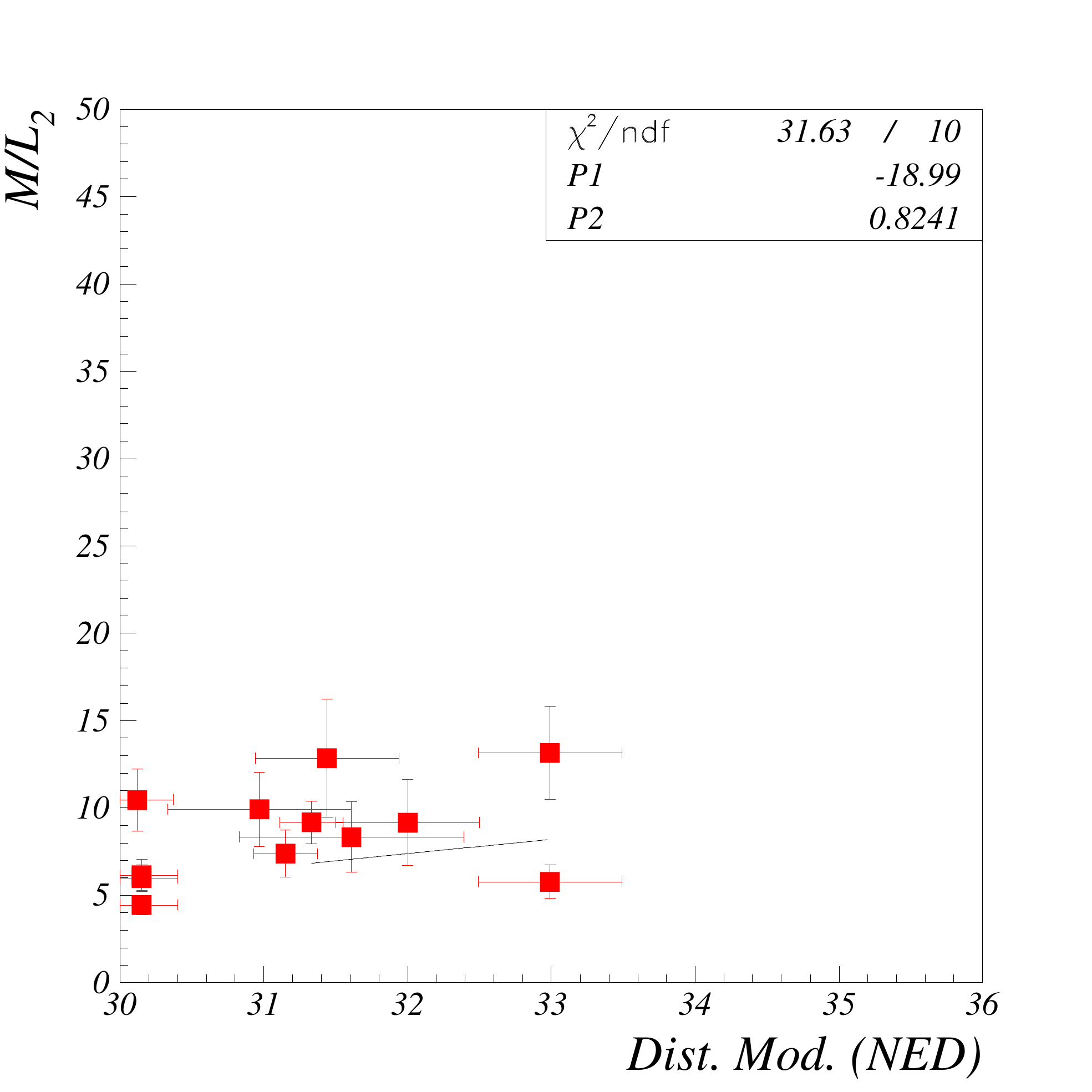}\includegraphics[scale=0.24]{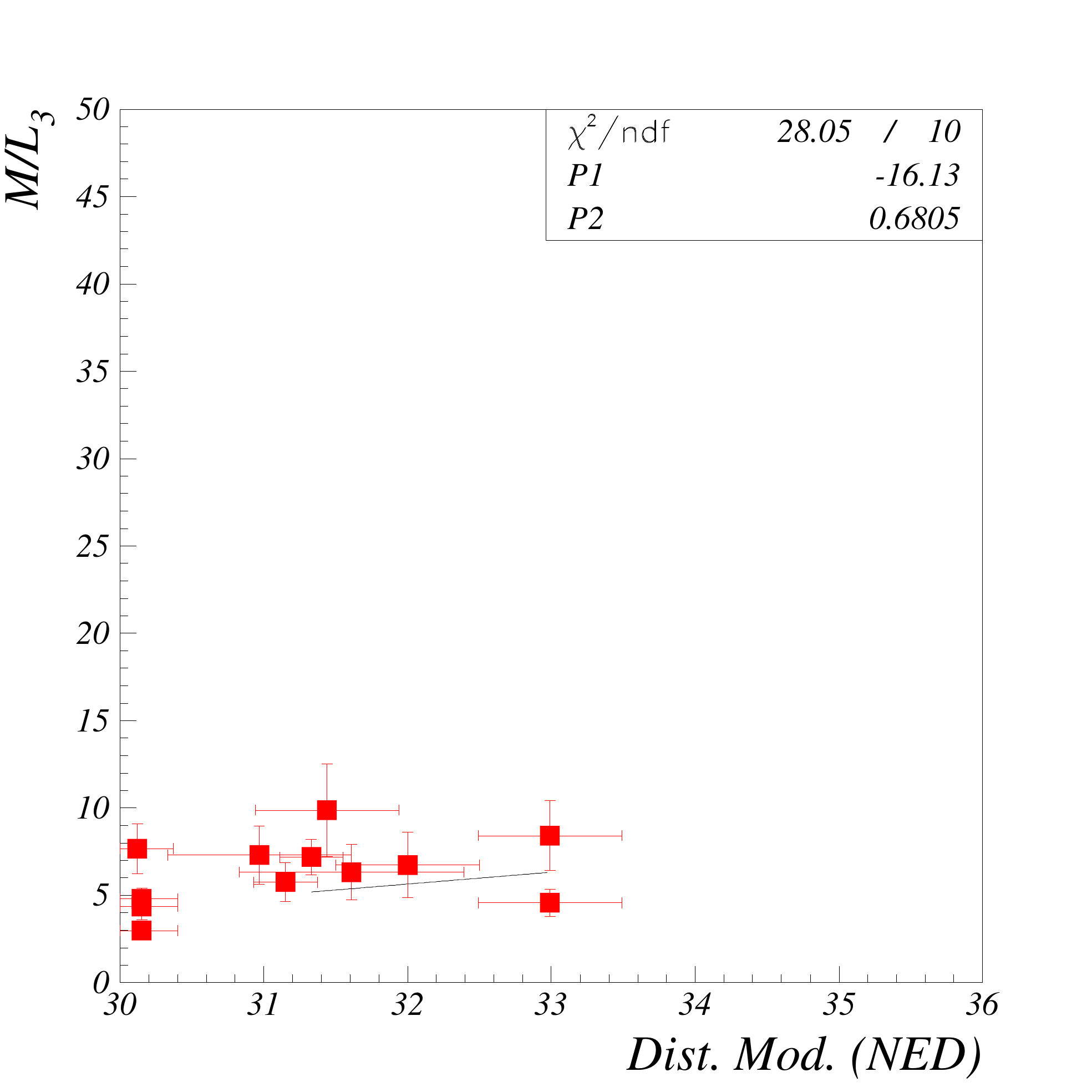}\protect \\
\vspace{-0.4cm} \caption{\label{Fig: DM NED correlations}Correlations between the distance
modulus from NED using redshift information and (from top left to
bottom right): distance modulus from NED without redshift information,
absolute blue magnitude, surface brightness $I_{e}$, effective radius
(parsec), velocity distribution $\sigma_{0}$ and $\sfrac{M}{L}$.
}
\end{figure}

The origins of the correlations are the same as the ones discussed
in Section~\ref{sub:Distance-Moduli correl}.

\paragraph{Distance Moduli (NED without redshift information) correlations}

\begin{figure}[H]
\centering
\includegraphics[scale=0.24]{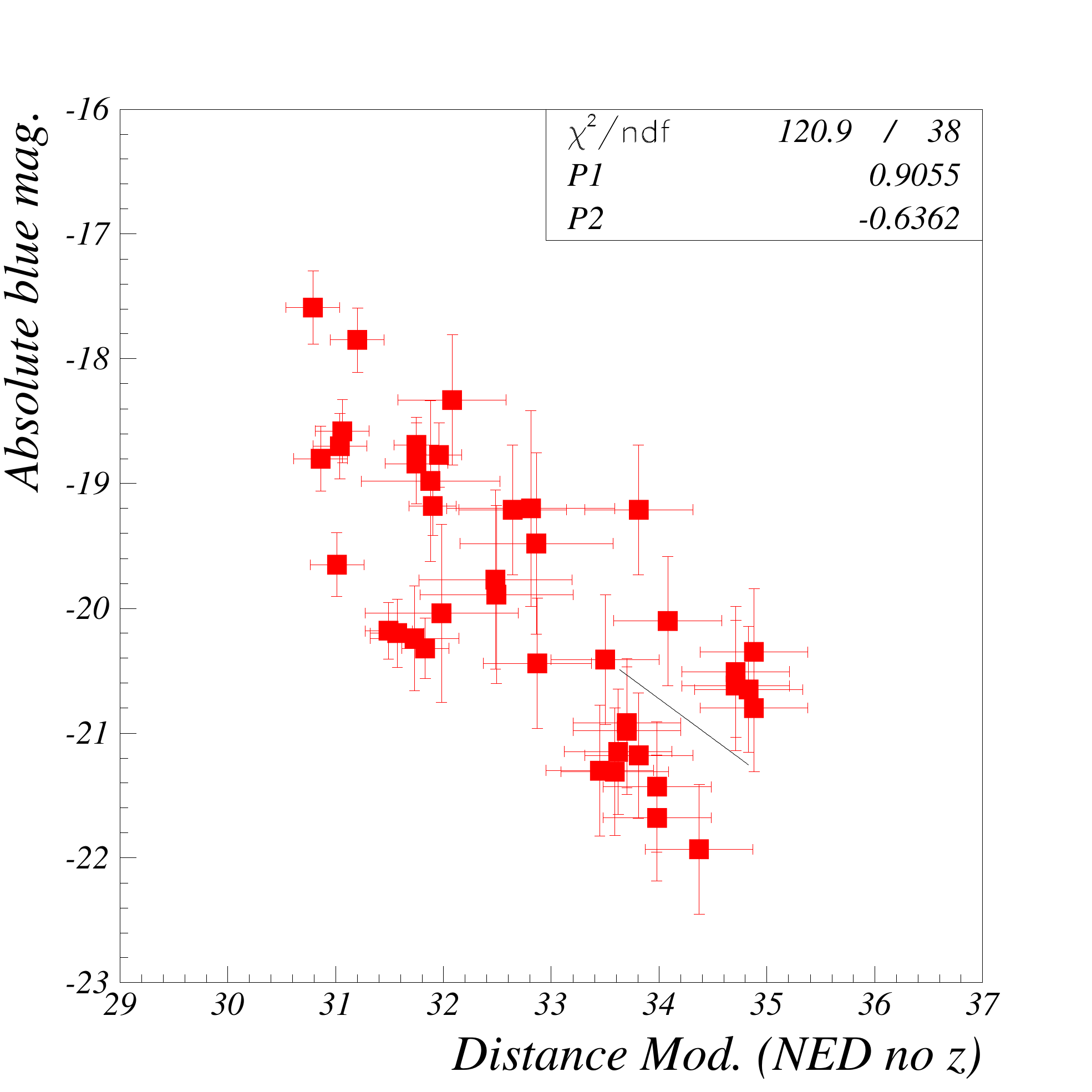}\includegraphics[scale=0.24]{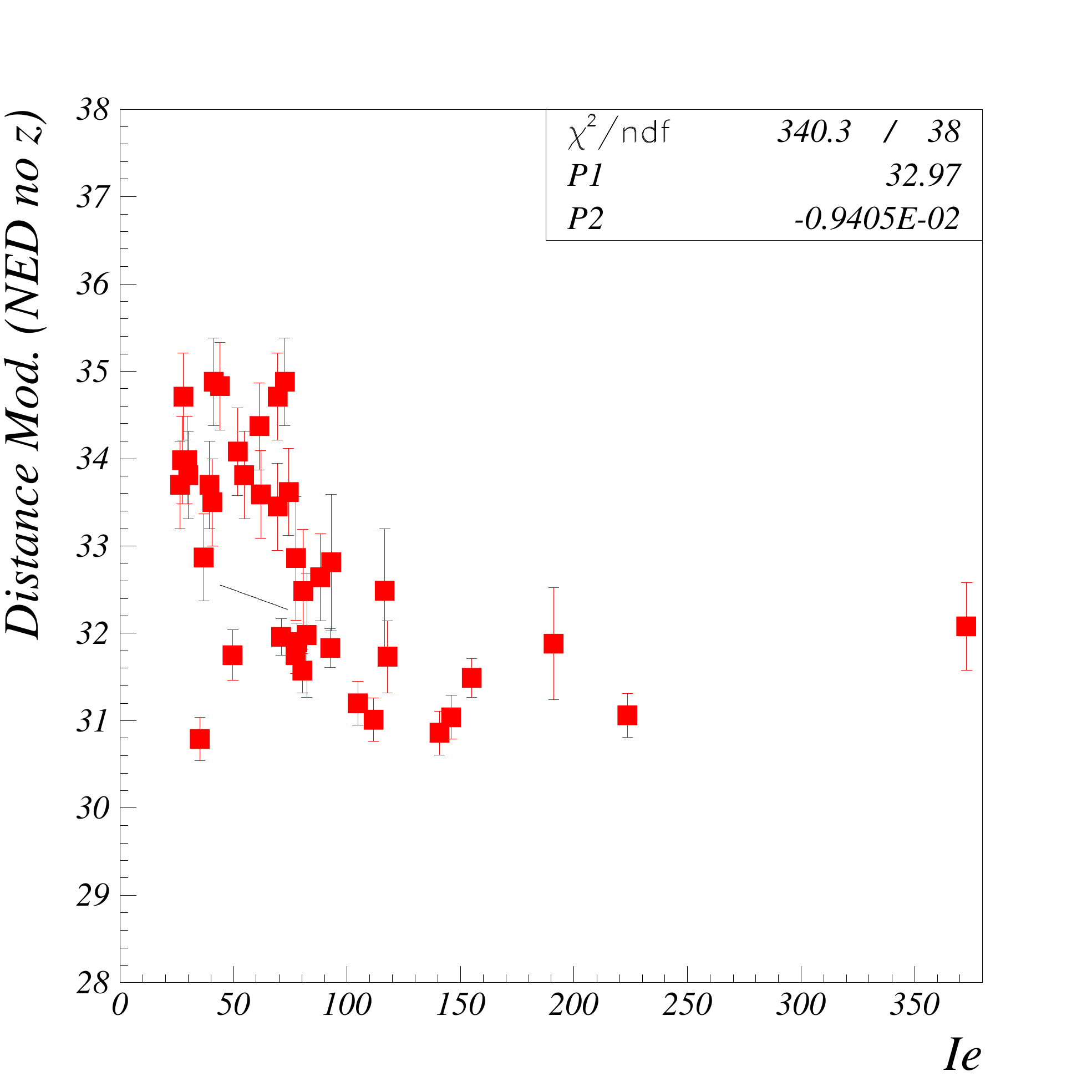}\includegraphics[scale=0.24]{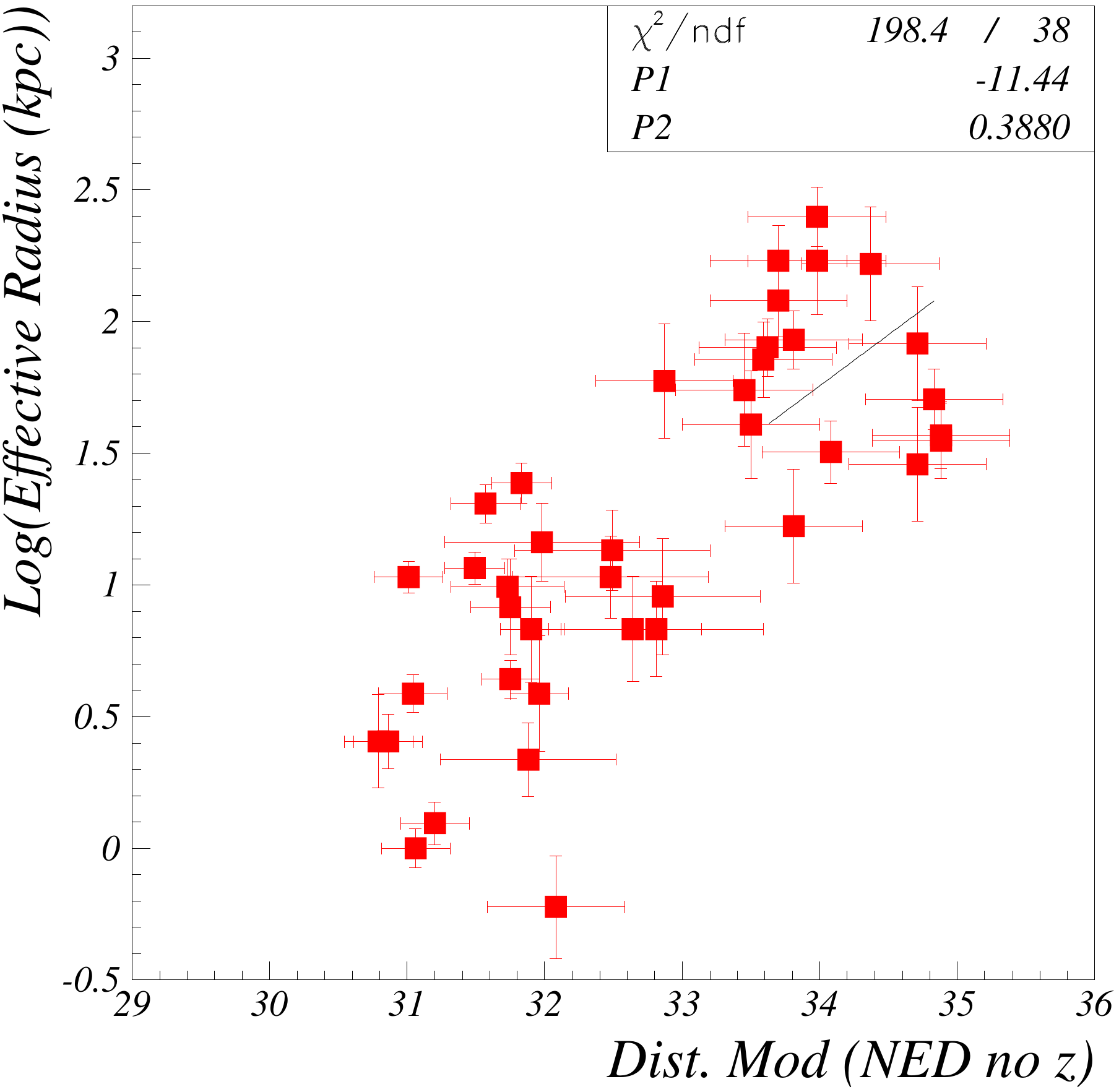}\includegraphics[scale=0.24]{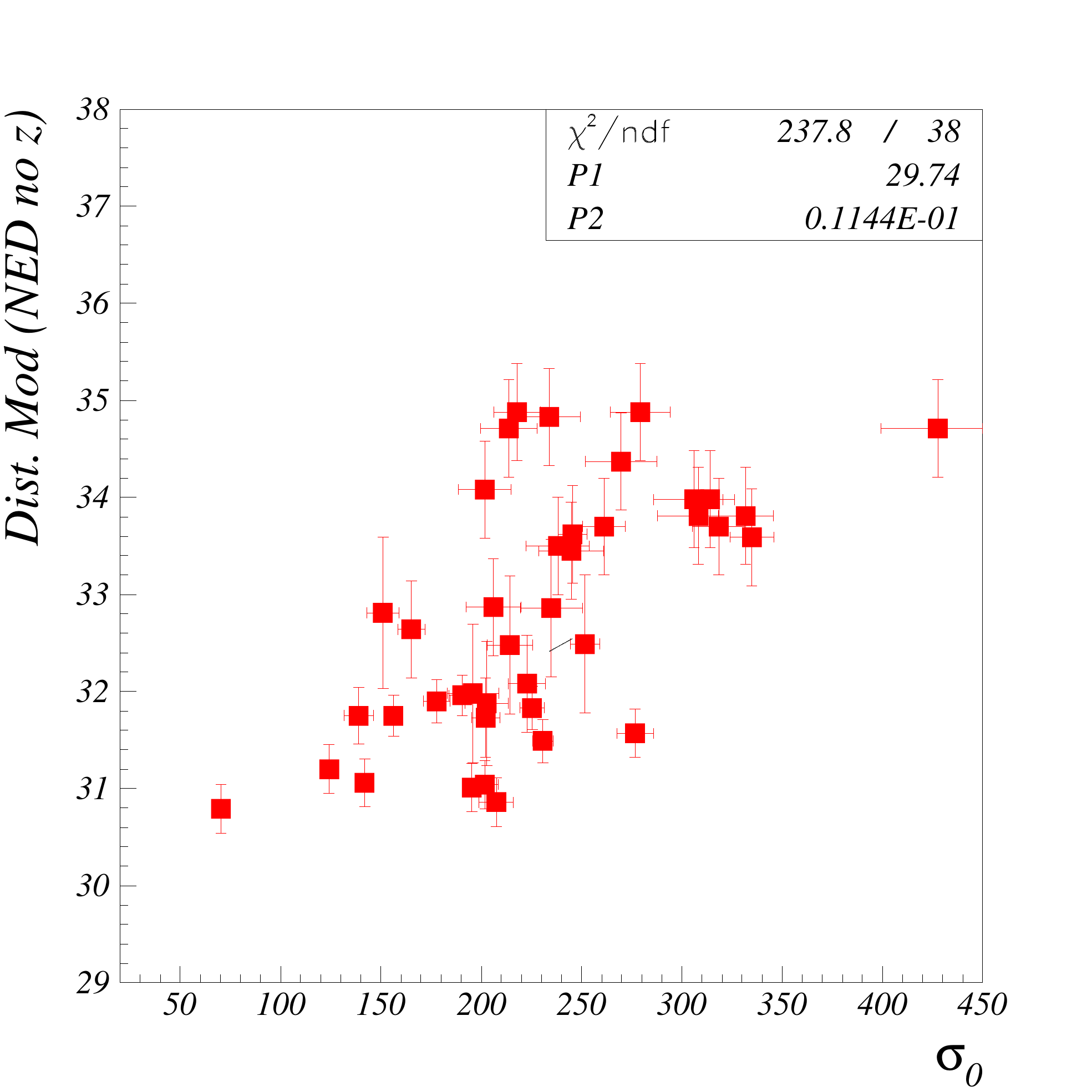}\protect \\
\includegraphics[scale=0.24]{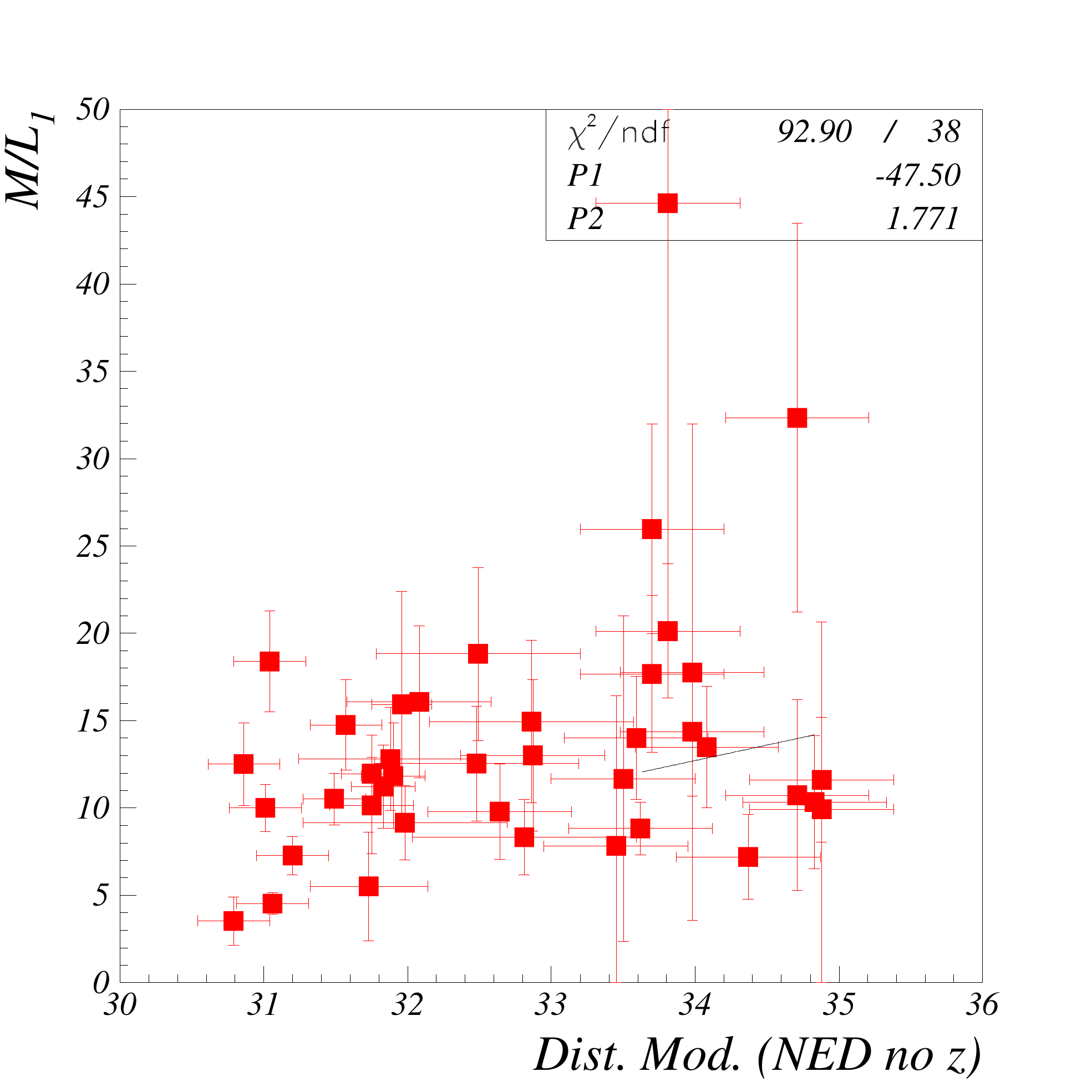}\includegraphics[scale=0.24]{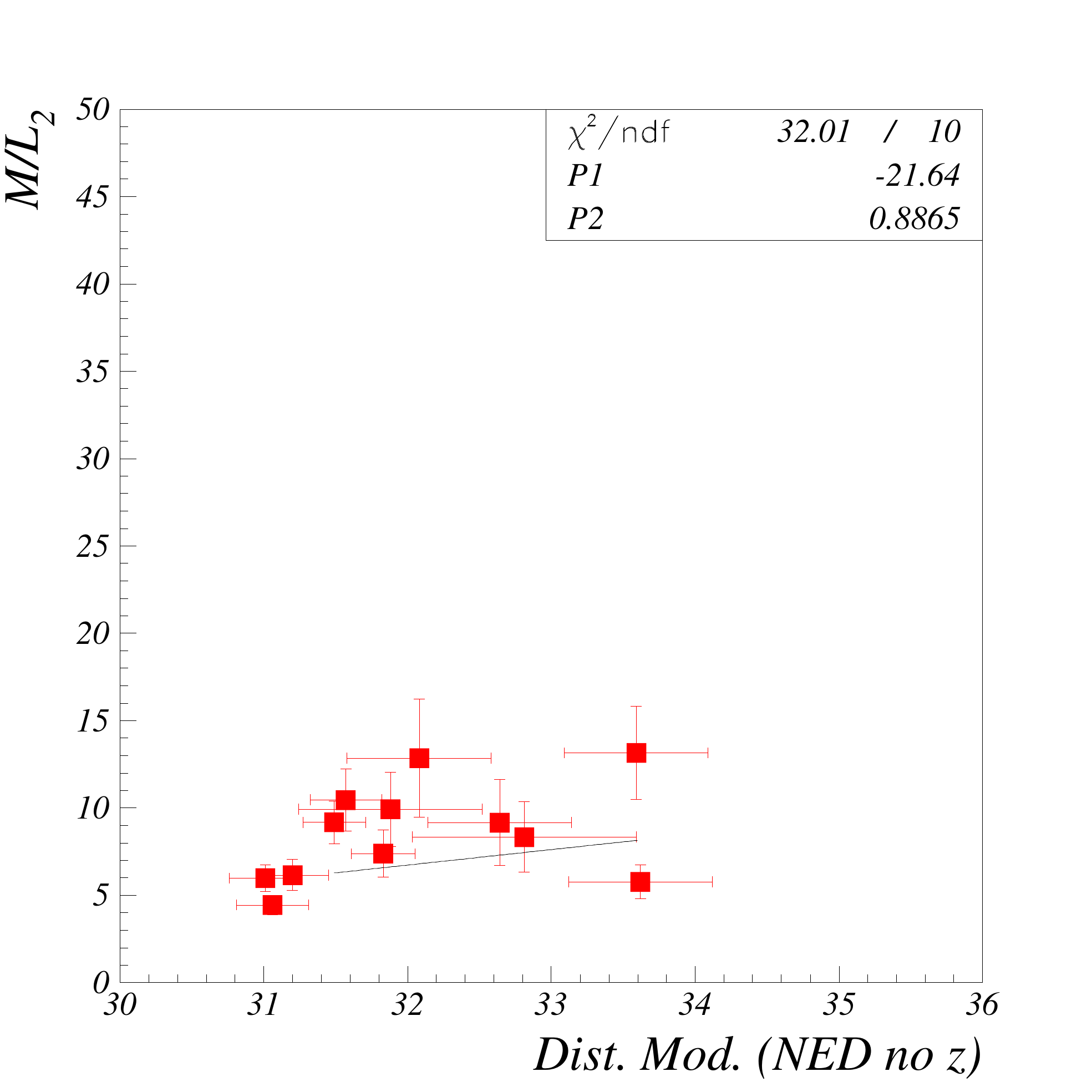}\includegraphics[scale=0.24]{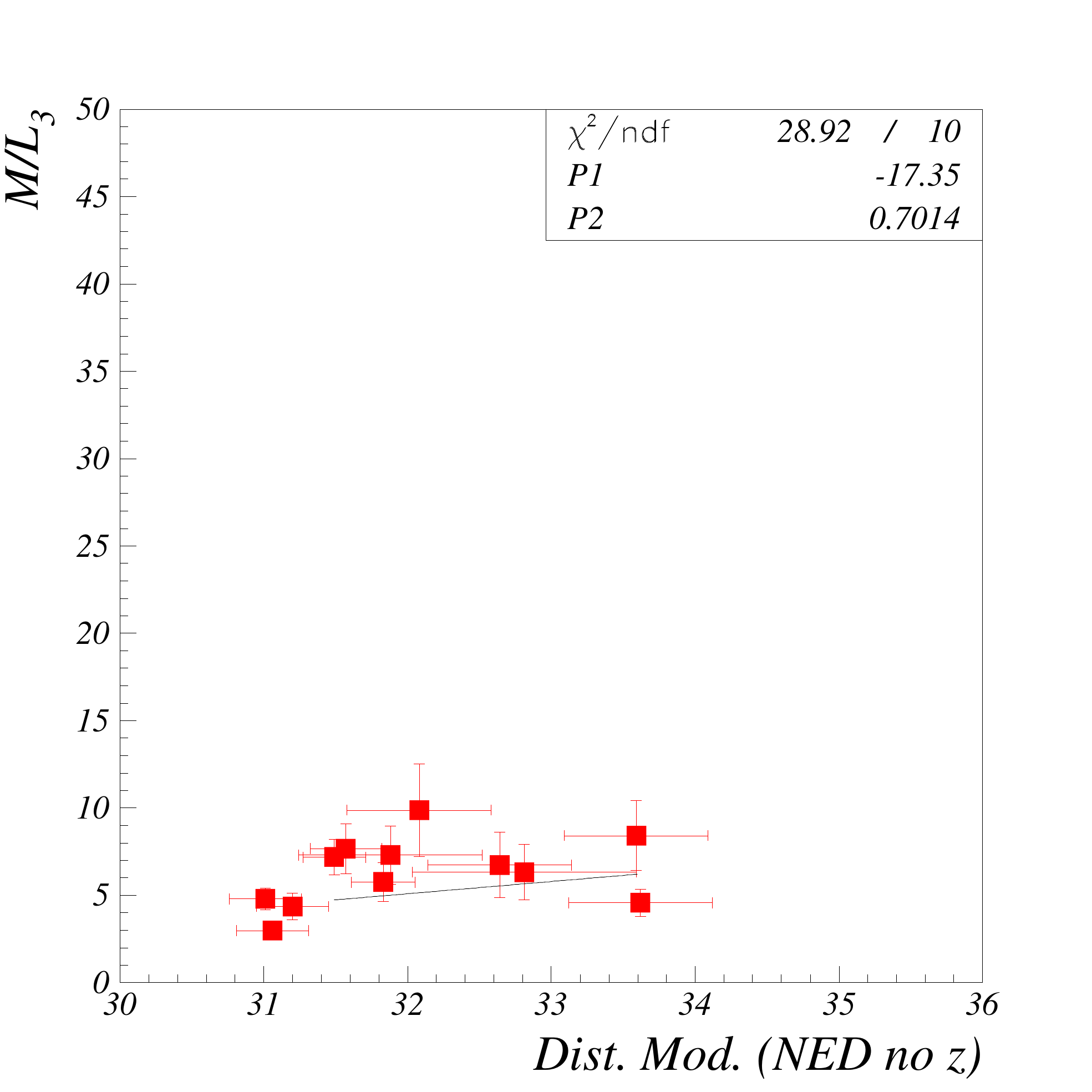}
\vspace{-0.4cm} \caption{\label{Fig: DM NED no z correlations}Correlations between the distance
modulus from NED without redshift information and (from top left to
bottom right): absolute blue magnitude, surface brightness $I_{e}$,
effective radius (parsec), velocity distribution $\sigma_{0}$ and
$\sfrac{M}{L}$.
}
\end{figure}

The origins of the correlations are the same as the ones discussed
in Section~\ref{sub:Distance-Moduli correl}.

\paragraph{Absolute magnitude $M_{b}$ correlations\label{sub:Absolute-magnitude-Mb corel}}

\begin{figure}[H]
\centering
\includegraphics[scale=0.24]{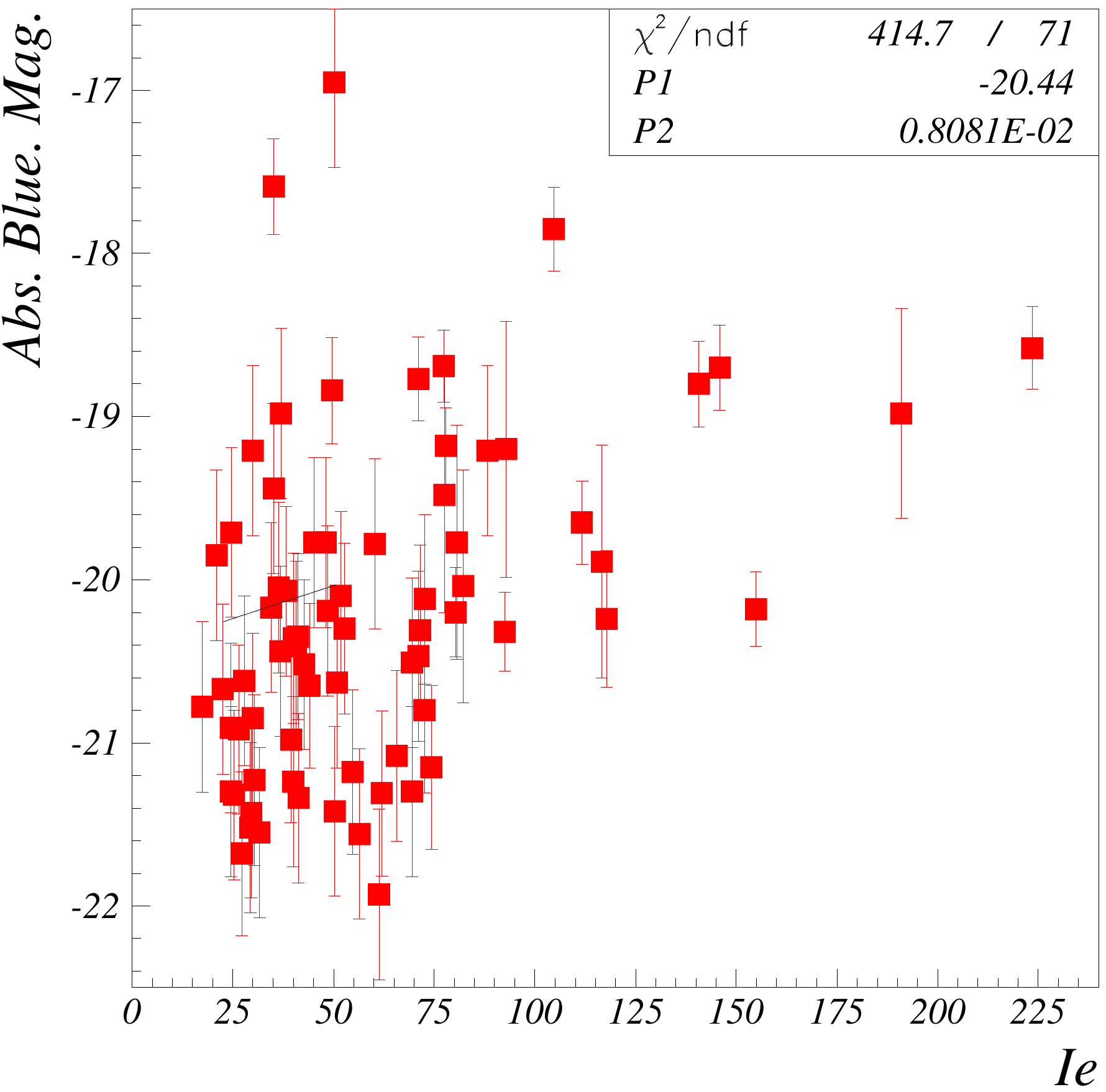}\includegraphics[scale=0.24]{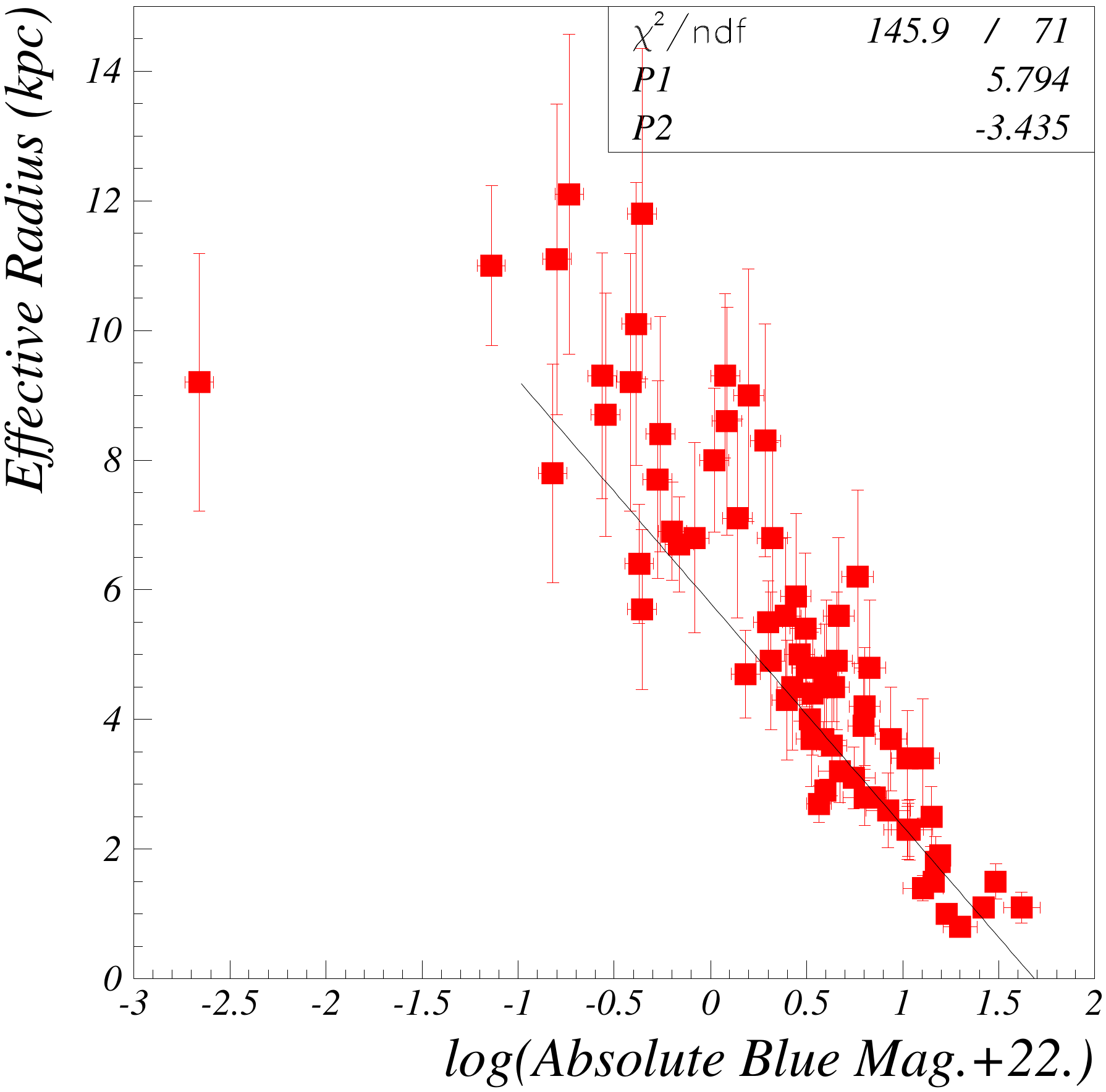}\includegraphics[scale=0.24]{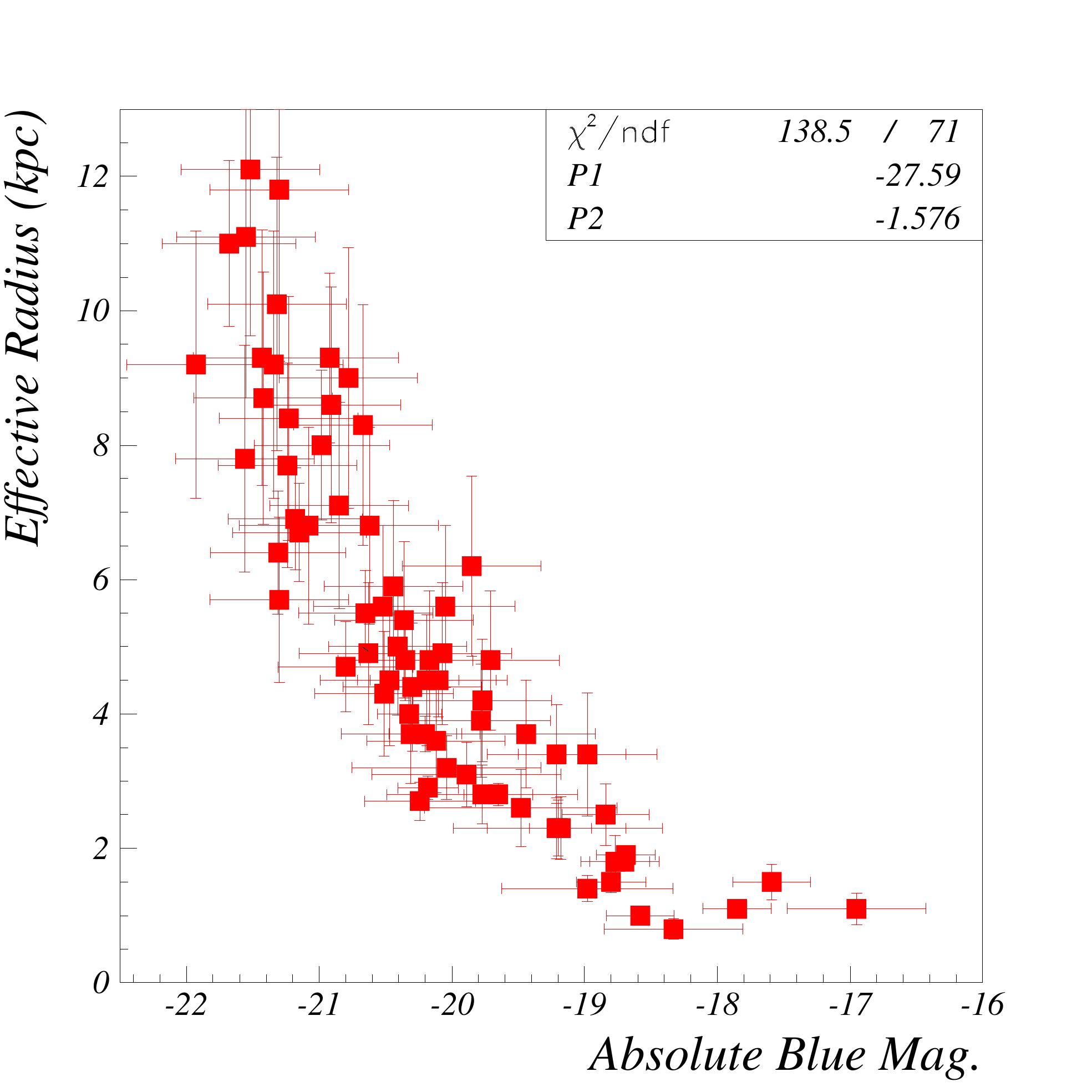}\includegraphics[scale=0.24]{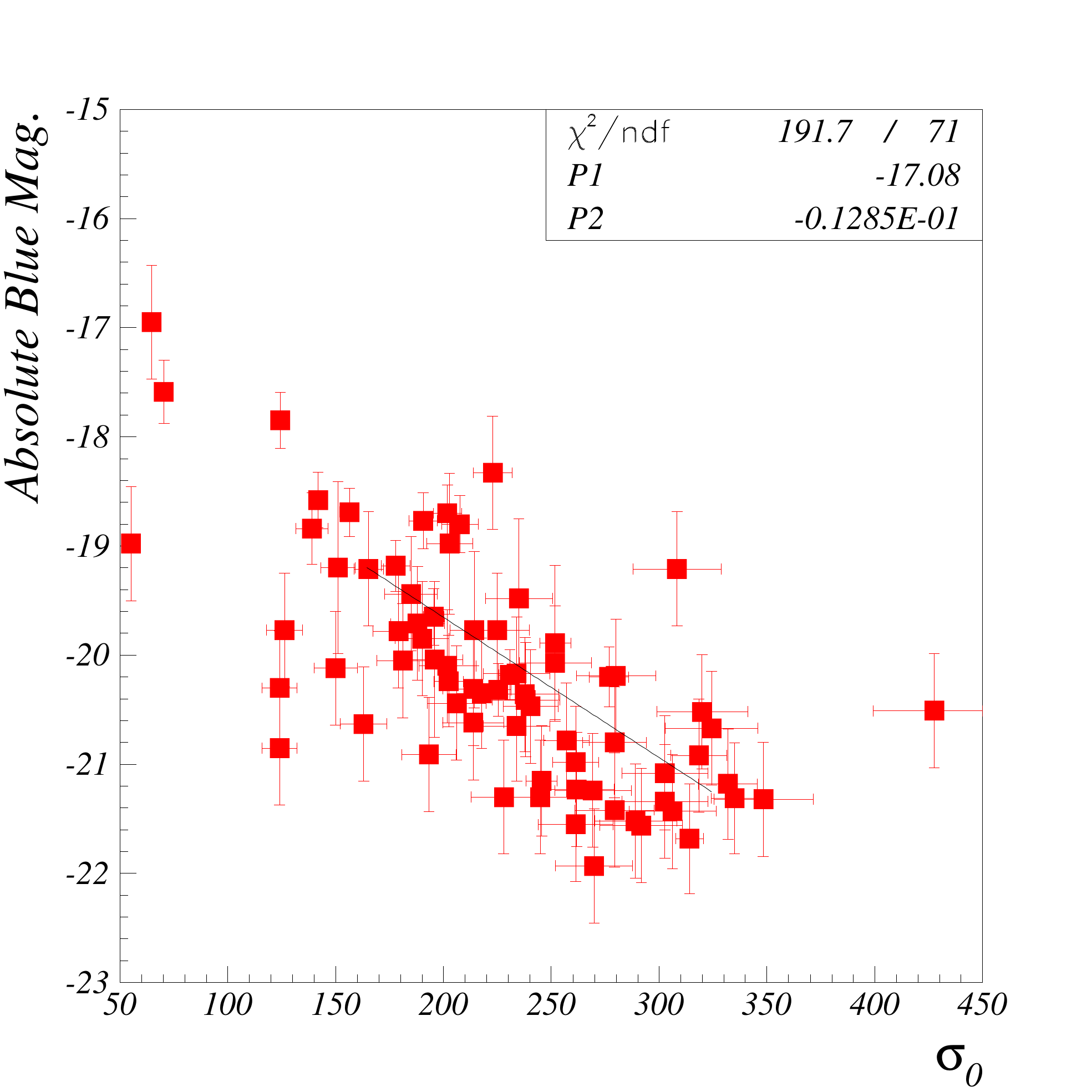}\protect \\
\includegraphics[scale=0.24]{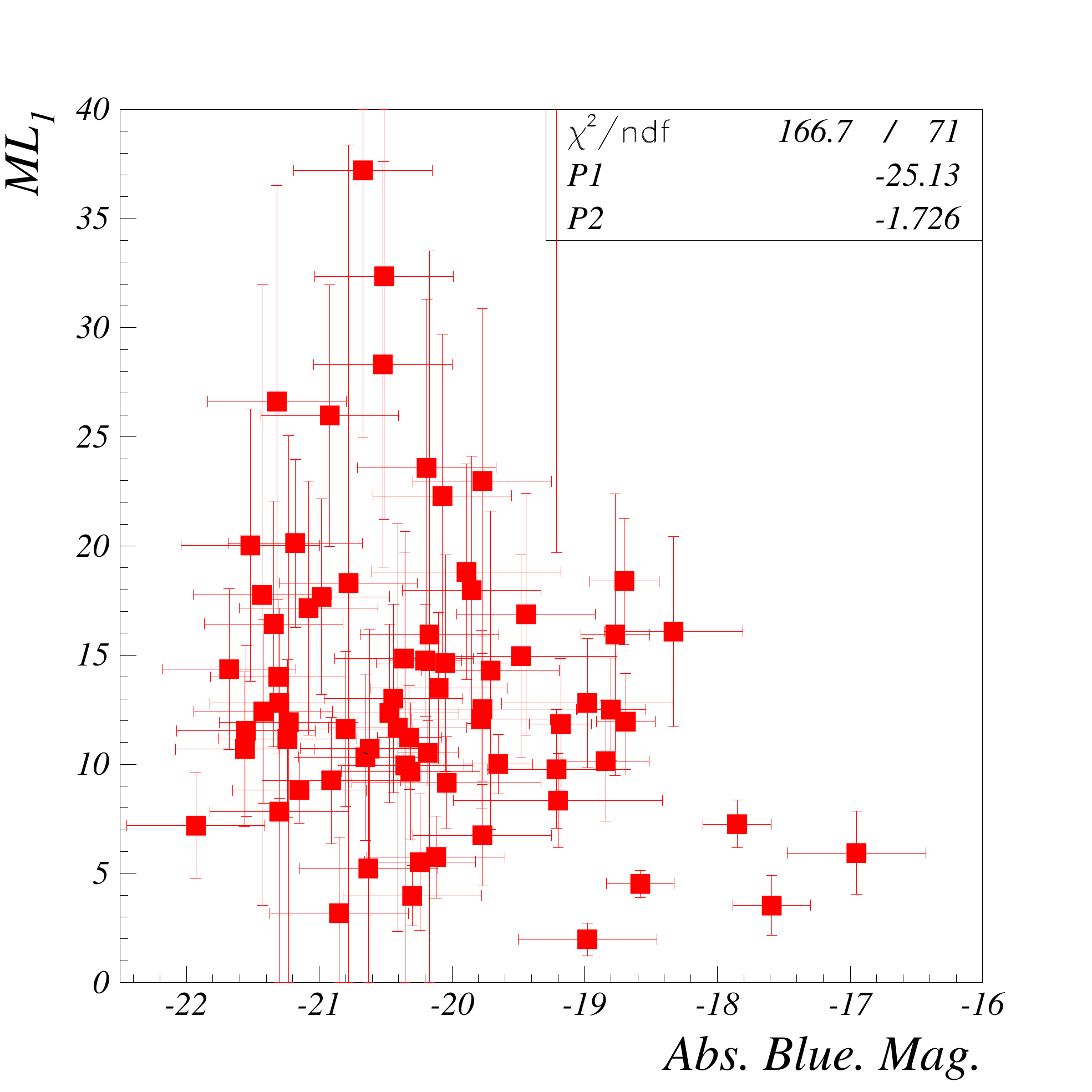}\includegraphics[scale=0.24]{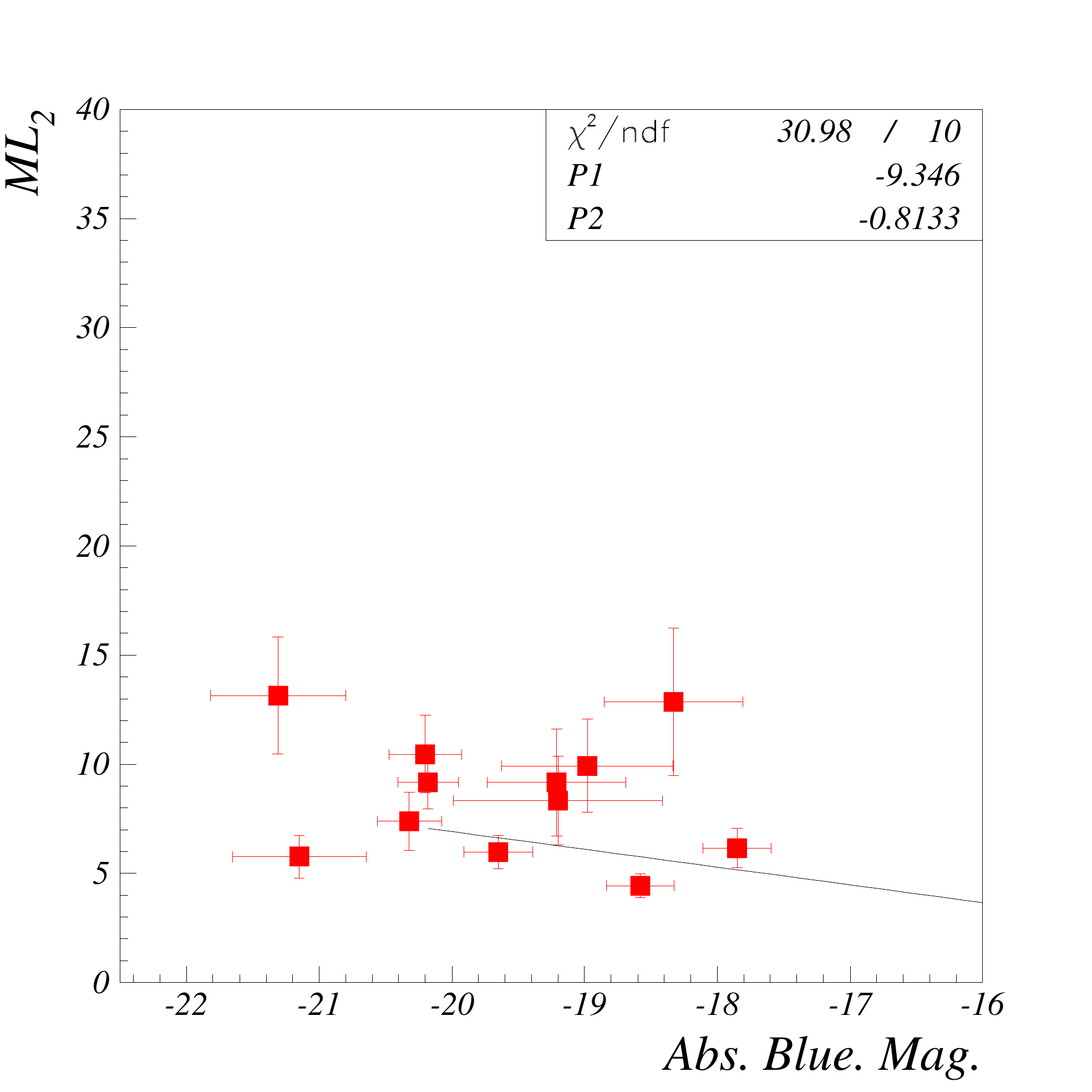}\includegraphics[scale=0.24]{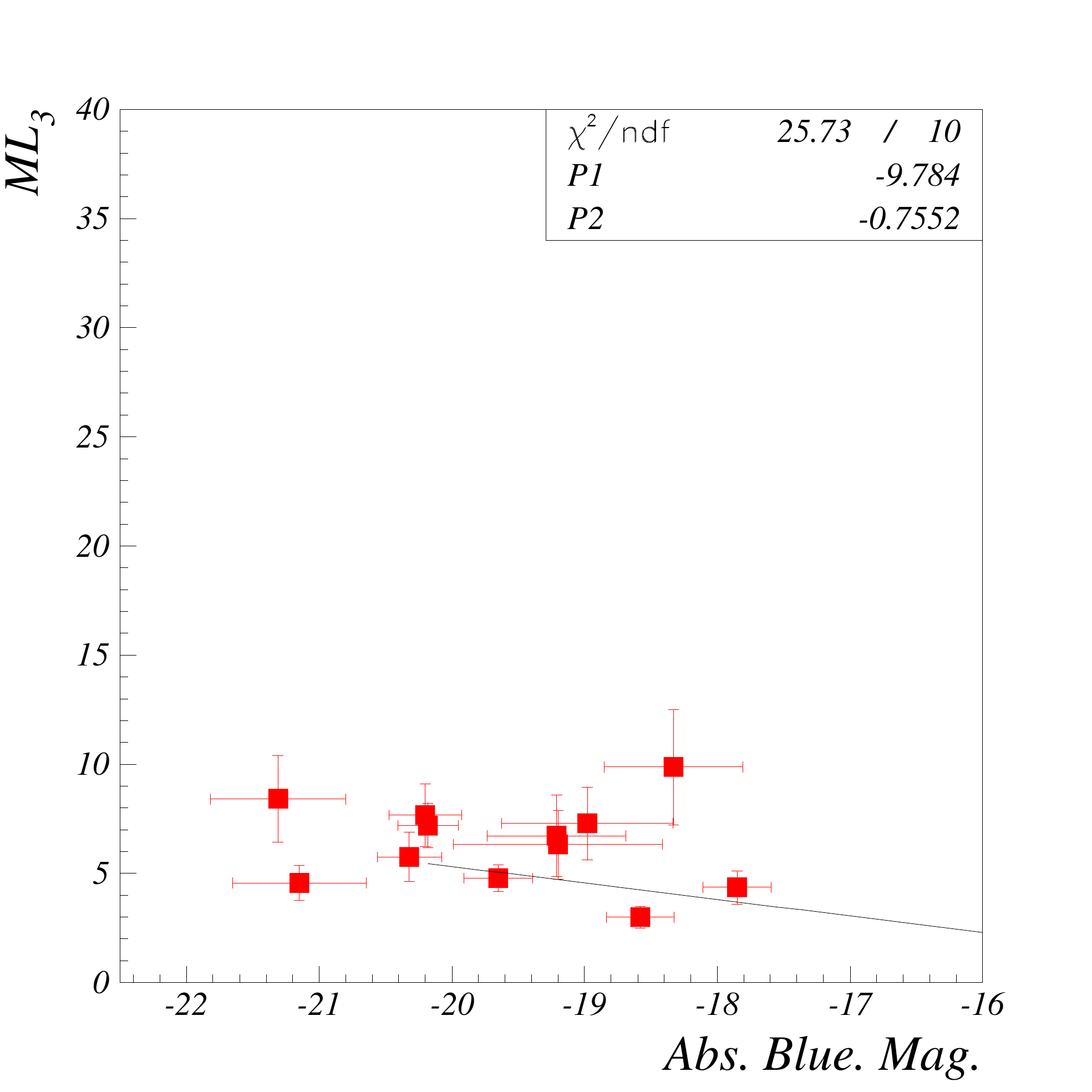}
\vspace{-0.4cm} \caption{\label{Fig: MBT correlations}Correlations between the absolute blue
magnitude and (from top left to bottom right): surface brightness
$I_{e}$, effective radius (parsec), velocity distribution $\sigma_{0}$
and $\sfrac{M}{L}$.
}
\end{figure}

The origins of the correlations are, from top left to bottom right: 
\begin{enumerate}
\item Absolute blue magnitude $M_{b}$ vs surface brightness $I_{e}$:
this pattern agrees with the observed central surface brightness vs
$M_b$ relation~\cite{Binggeli}. However, it
could also arise from two secondary correlations: $M_{b}\Longleftrightarrow DM\longleftrightarrow I_{e}$
and possibly $M_{b}\Longleftrightarrow Re(kpc)\dashleftarrow\dashrightarrow I_{e}$.
Using the linear fit results $(-0.53\pm0.03)DM=M_{b}+c$ and $(-1.73\pm0.09)E^{-2}I_{e}=DM+c'$
yields $M_{b}=(0.92\pm0.10)E^{-2}I_{e}+c"$. Using the linear fit
results $(-8.1\pm0.5)E^{-3}I_{e}=Re+c$ and $(-1.58\pm0.09)M_{b}=Re+c'$
yields a negligible contribution $M_{b}=(-5.1\pm0.6)E^{-3}I_{e}+c"$.
This agrees with the observed $M_{b}=(0.81\pm0.09)I_{e}+C"$.
This plot, however, suggests to reject the two lowest $M_{b}$ points
since they are close to the dwarf elliptical locus~\cite{Binggeli}.
\item $M_{b}$ vs effective radius (parsec):
this is the well known Kormendy relation~\cite{Kormendy}.
\item $M_b$  $M_{b}$ vs velocity distribution $\sigma_{0}$:
this is the well known Faber-Jackson relation~\cite{Faber-Jackson}. 
\item $M_{b}$ vs $\sfrac{M}{L}$: a clear weak
correlation is seen. It seems to be due to the virial theorem and
the Kormendy relation: $\sfrac{M}{L}\Longleftrightarrow\sigma_{0}\Longleftrightarrow M_{b}$.
Using the linear fit results $(5.4\pm0.4)E^{-2}\sigma_{0}=\sfrac{M}{L}+c$ and
$(-1.3\pm0.1)E^{-2}\sigma_{0}=M_{b}+c'$ yields $\sfrac{M}{L}=(-4.2\pm0.6)M_{b}+c"$.
This is in qualitative agreement with the observed correlation. The
value from the fit: $\sfrac{M}{L}=(-1.7\pm0.3)M_{b}+C"$ is underestimated since
a linear form is not adapted for the fit. Another contribution could
result from the fact that luminous/large elliptical galaxies tend
to have stars that are more metal-rich than less luminous elliptical
galaxies~\cite{Sparke}. This tends to increase the $\sfrac{M}{L}$
with increasing absolute luminosity $M_{b}$. In principle, this possibility
has to be ruled out in order to cleanly interpret a dependence of
$\sfrac{M}{L}$ with $\sfrac{R_{min}}{R_{max}}$. In practice, there is only an uncertain
weak correlation between $\sfrac{R_{min}}{R_{max}}$ and $M_{b}$ so it would
have little influence on the $\sfrac{M}{L}$ vs $\sfrac{R_{min}}{R_{max}}$ dependence.
Furthermore, as explained in the Section~\ref{sub:Selection}, the
most luminous galaxies are excluded from our data set. All in all,
we can ignore the consequence that large elliptical galaxies are more
metal rich.
\end{enumerate}

\paragraph{Surface Brightness $I_{e}$ correlations.}

\begin{figure}[H]
\centering
\includegraphics[scale=0.24]{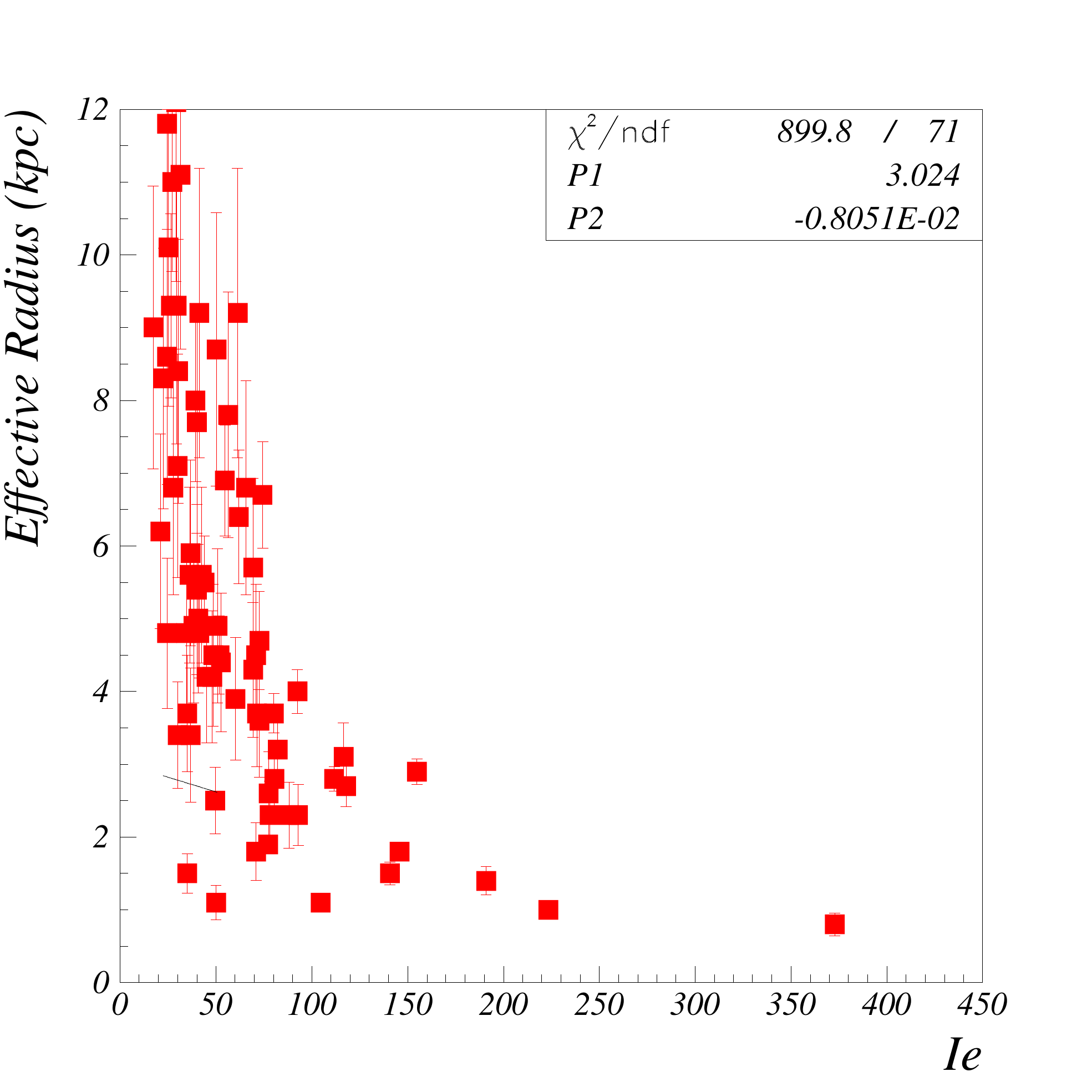}\includegraphics[scale=0.24]{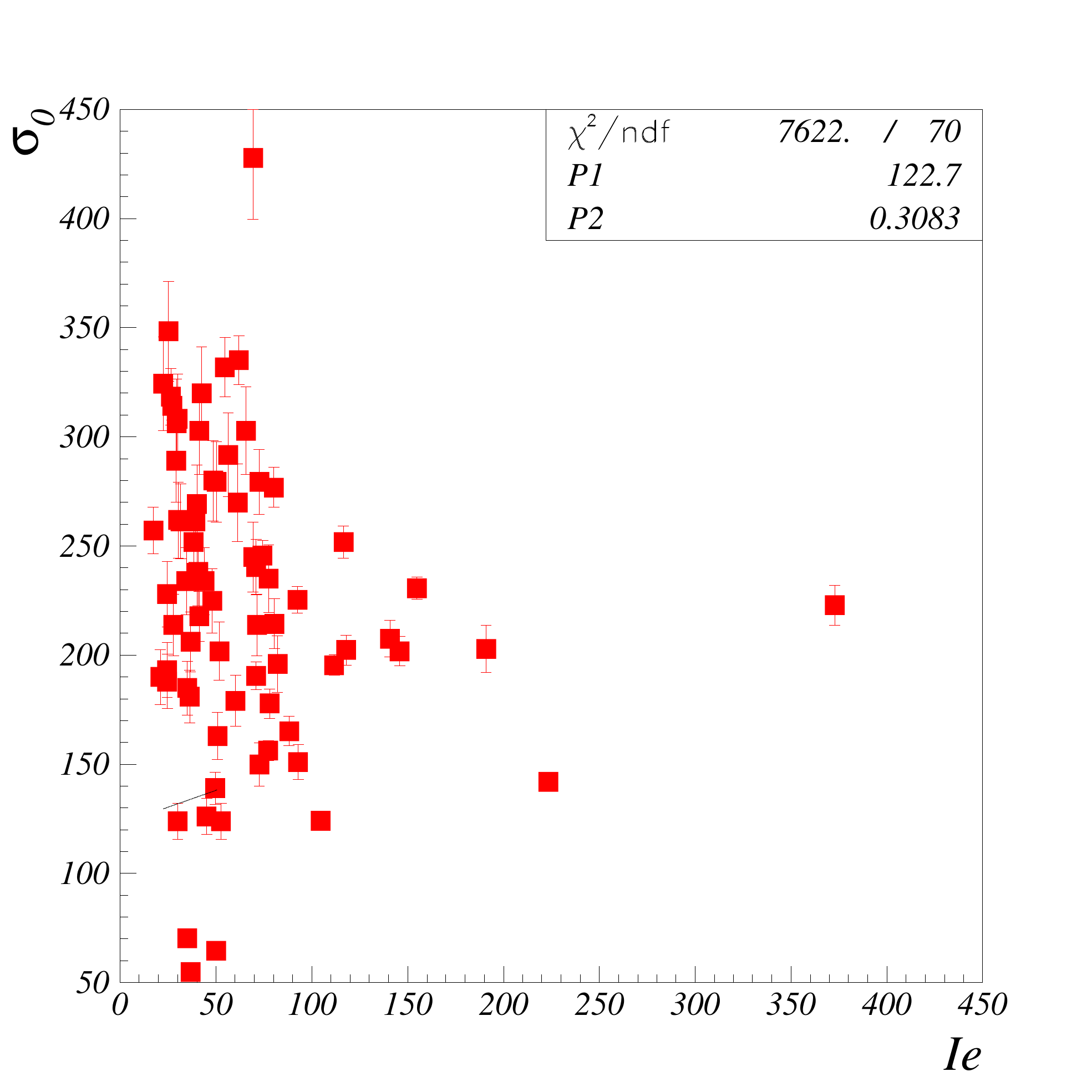}\includegraphics[scale=0.24]{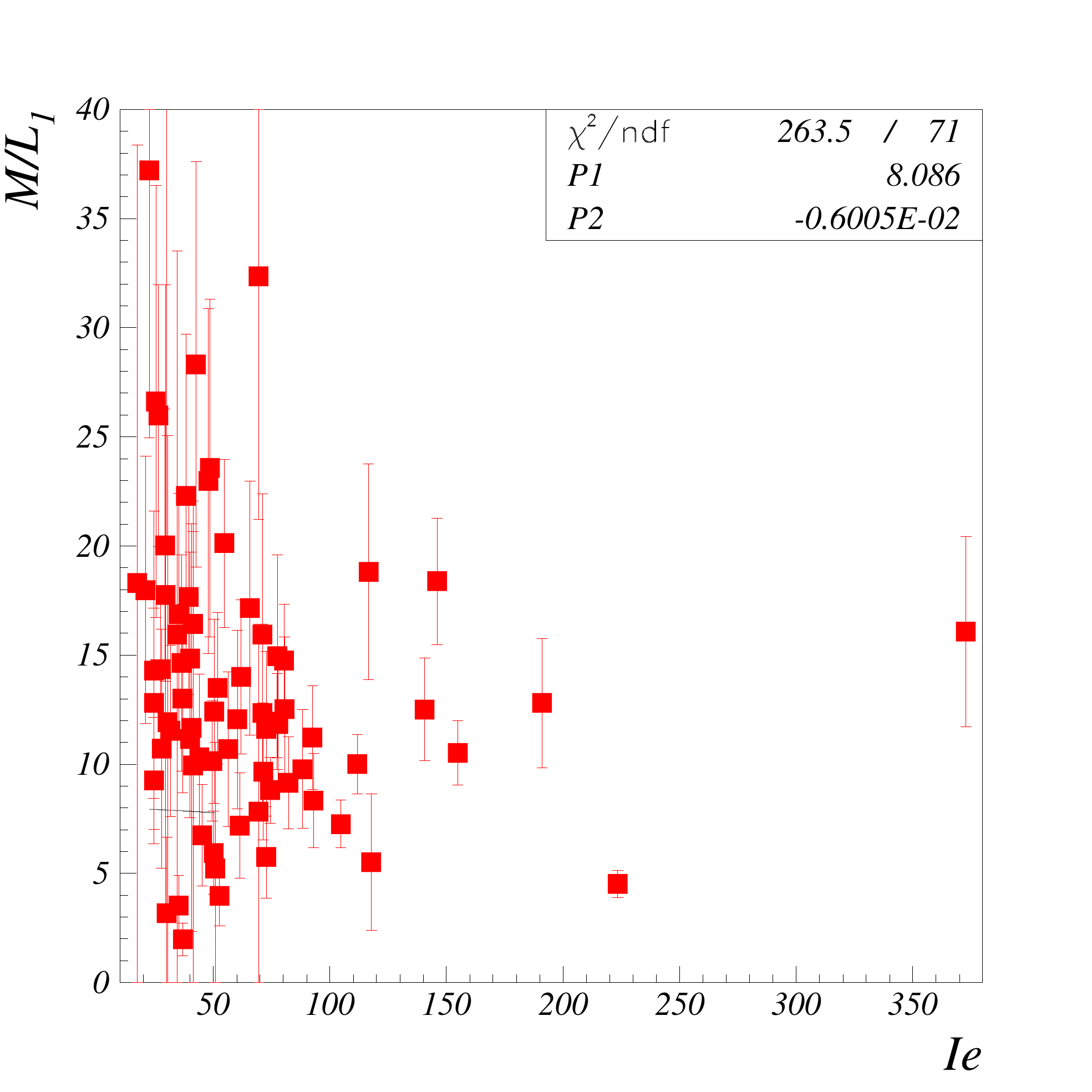}\includegraphics[scale=0.24]{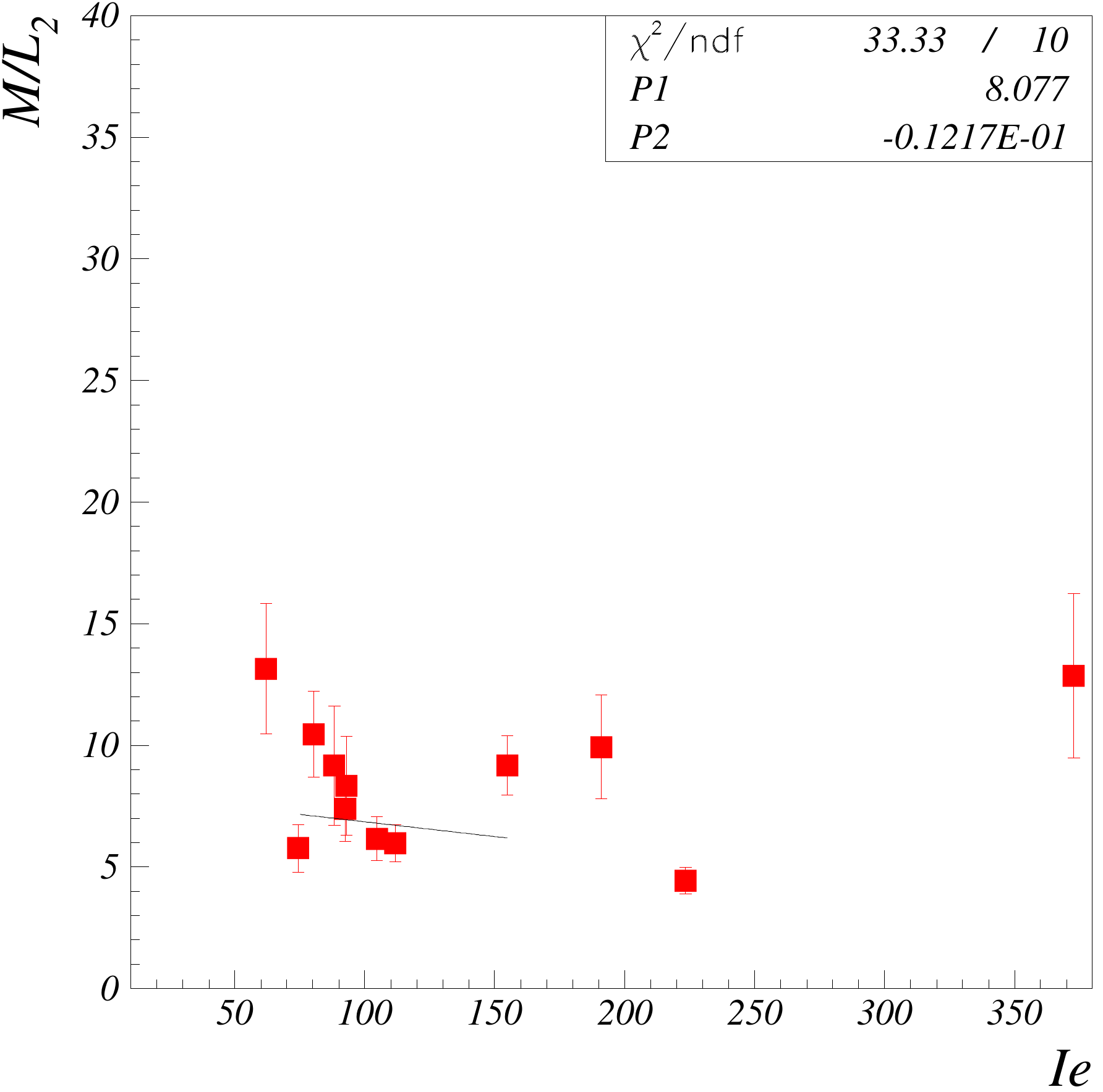}\protect \\
\includegraphics[scale=0.24]{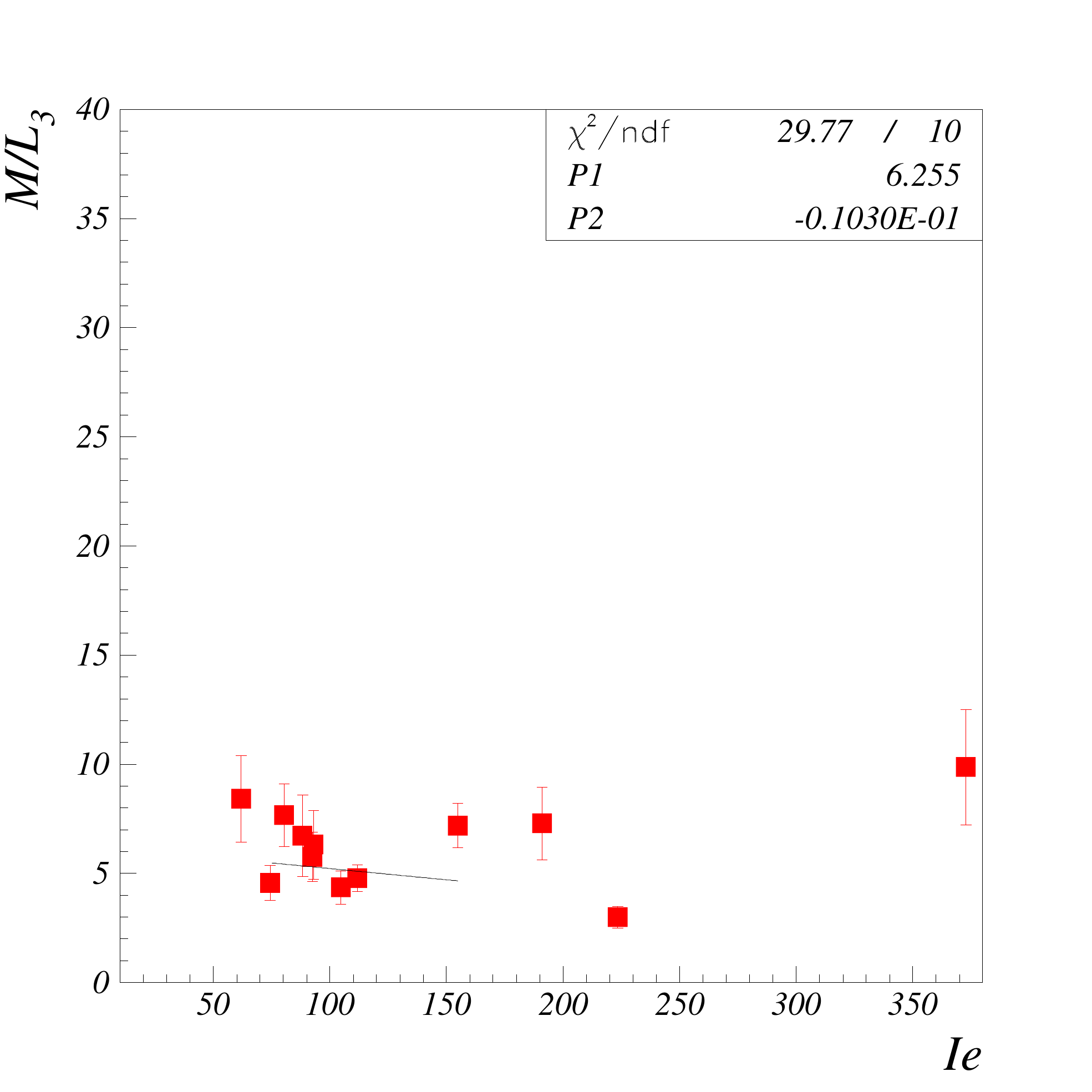}
\vspace{-0.4cm} \caption{\label{Fig: ie correlations}Correlations between the surface brightness
$I_{e}$ and (from top left to bottom right): effective radius (parsec),
velocity distribution $\sigma_{0}$ and $\sfrac{M}{L}$.
}
\end{figure}

The origins of the correlations are, from top left to bottom right: 
\begin{enumerate}
\item Surface brightness $I_{e}$  vs effective radius (parsec): this
 is expected from the Kormendy relation: it
implies that the larger the galaxy, the lower its surface brightness
(see e.g.~\cite{Binney} page 24), as seen on the plot.
\item $I_{e}$ vs velocity distribution $\sigma_{0}$:
this  is expected from the $Re(Kpc)\Longleftrightarrow\sigma_{0}\dashleftarrow\dashrightarrow I_{e}$
distributions. 
\item $I_{e}$  vs $\sfrac{M}{L}$: this
is similar to the  just discussed $I_{e}$ vs $\sigma_{0}$ distribution
due to the strong $\sigma_{0}\Longleftrightarrow \sfrac{M}{L}$ correlation.
\end{enumerate}

\paragraph{Absolute effective radius $Re$ correlations}

\begin{figure}[H]
\centering
\includegraphics[scale=0.24]{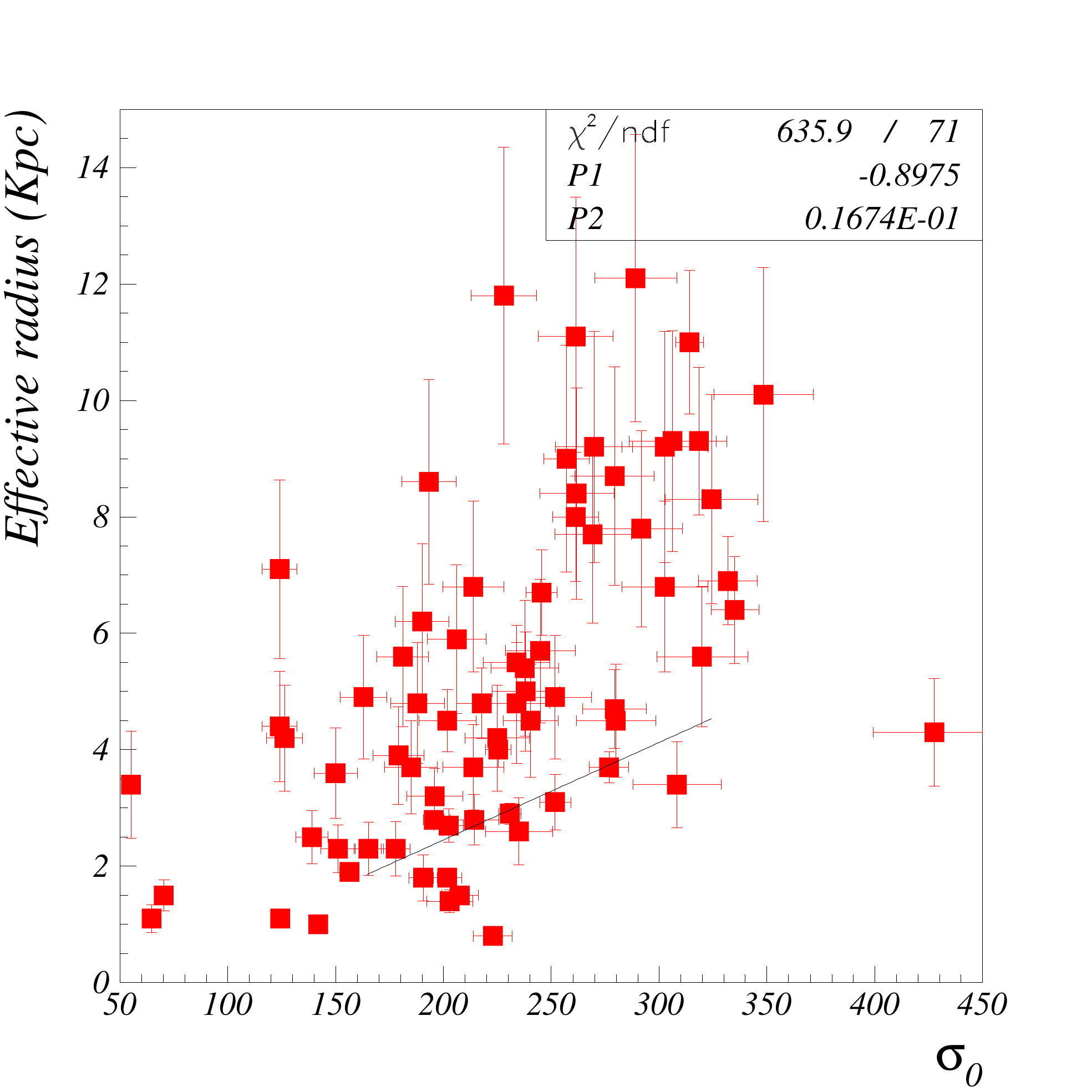}\includegraphics[scale=0.24]{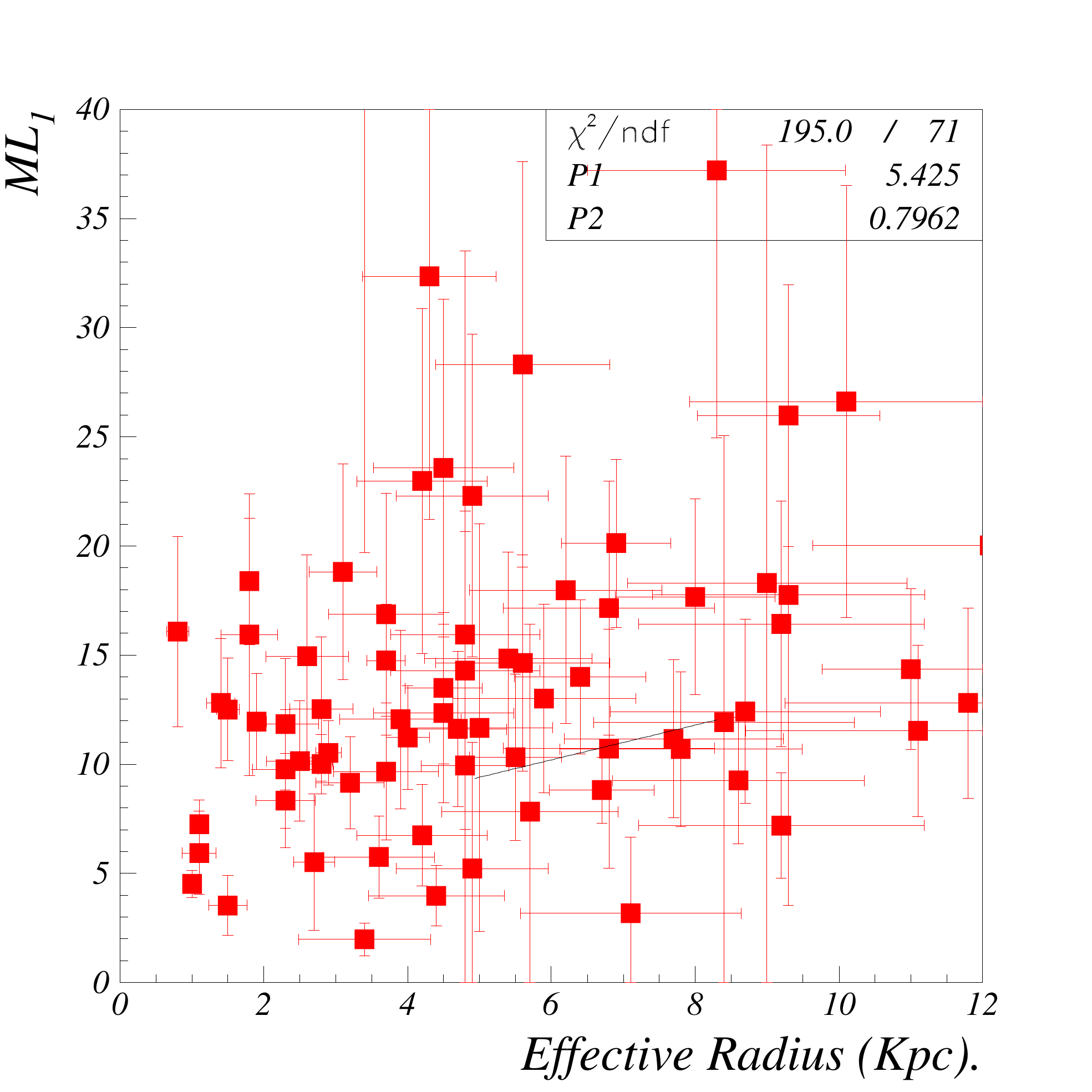}\includegraphics[scale=0.24]{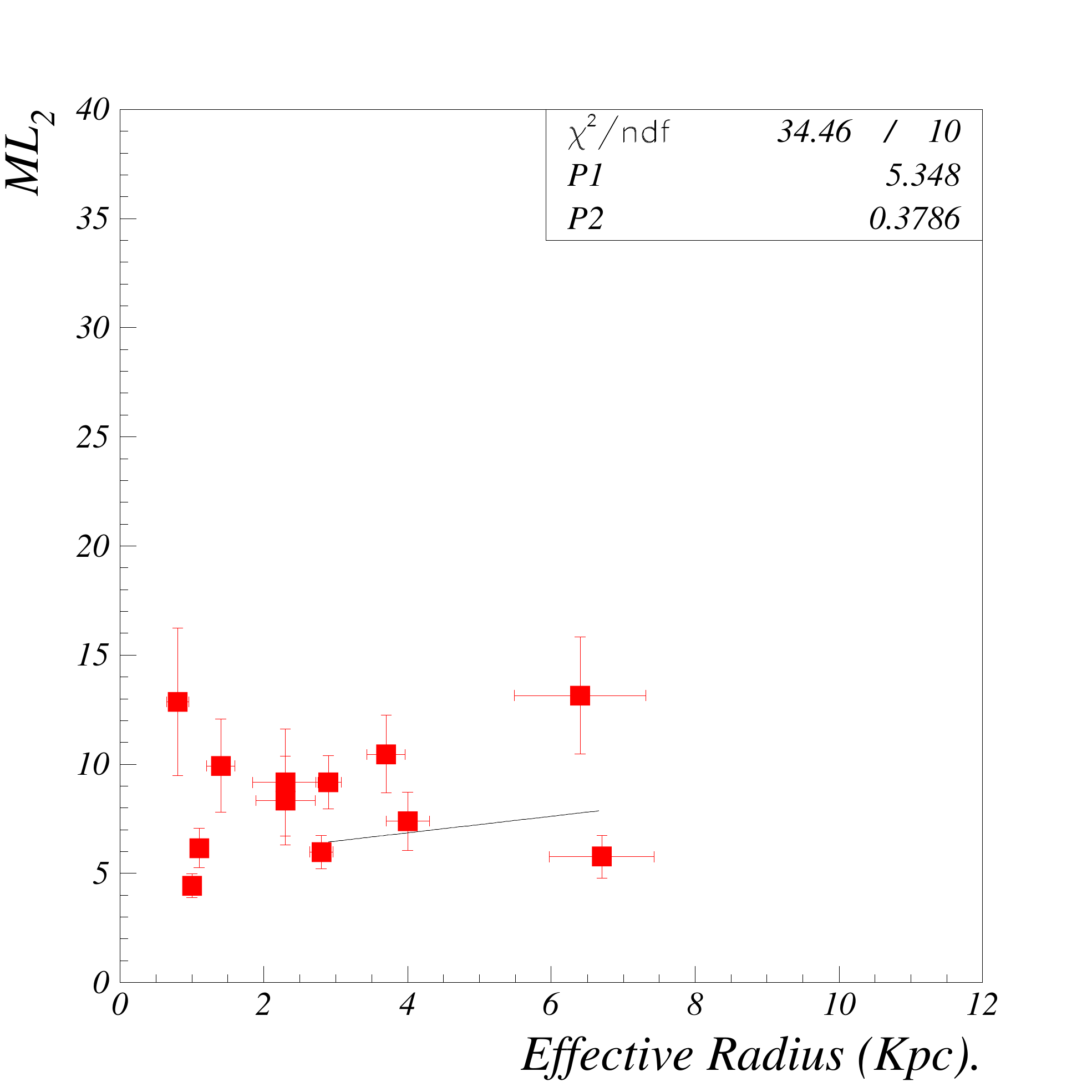}\includegraphics[scale=0.24]{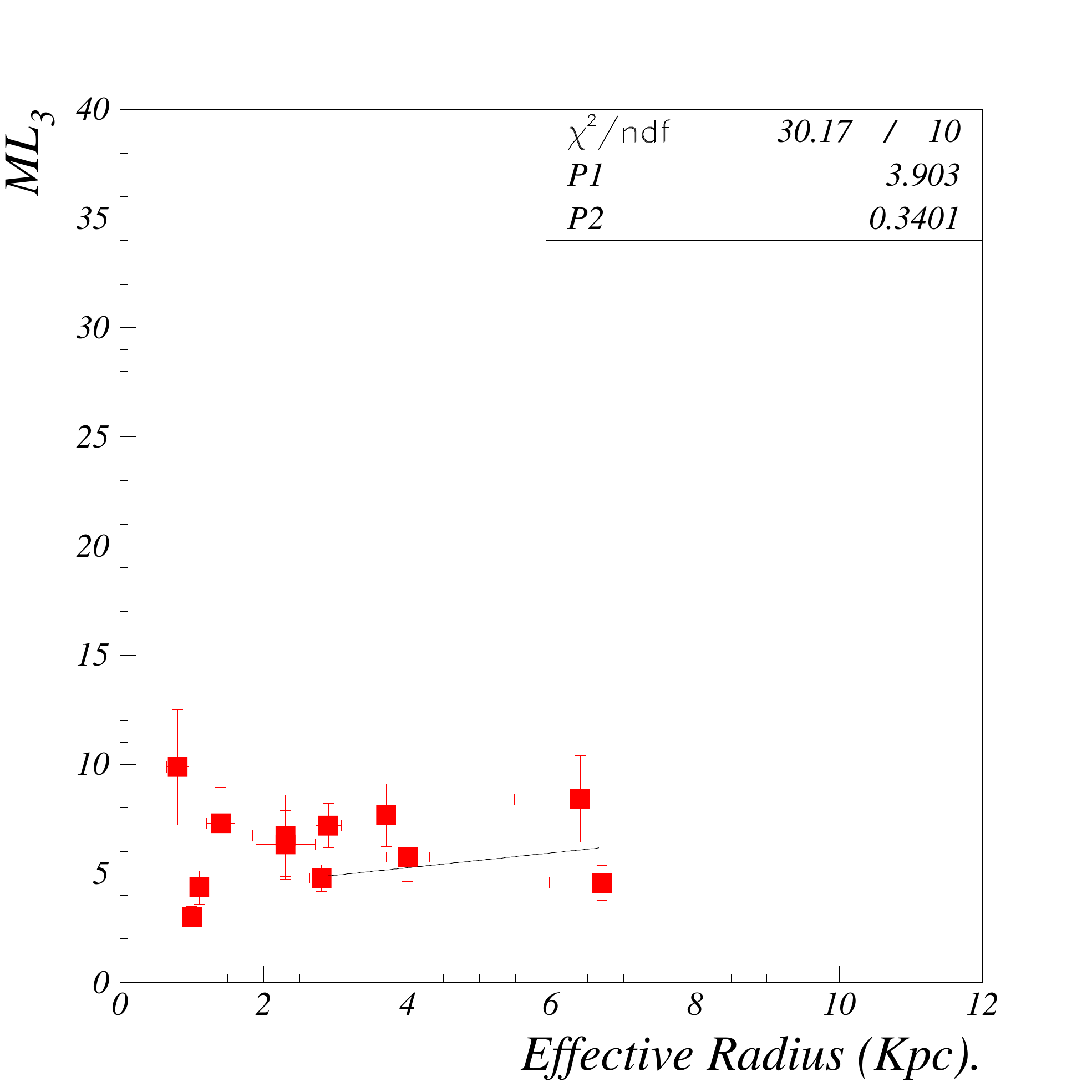}\protect \\
\vspace{-0.4cm} \caption{\label{Fig: Re (Kpc) correlations}Correlations between the absolute
effective radius and (from left to right): velocity distribution $\sigma_{0}$
and $\sfrac{M}{L}$.
}
\end{figure}

The origins of the correlations are, from left to right: 
\begin{enumerate}
\item Effective radius (parsec)  vs velocity distribution $\sigma_{0}$:
this correlation is the well known $3^{rd}$ plane relation.
\item Effective radius (parsec)  vs $\sfrac{M}{L}$: there is no or little
correlation. This is opposite to what would have been expected if
our sample had included dwarf elliptical galaxies, since they tend
to have larger $\sfrac{M}{L}$. This validates our choice of selection
critteria.
\end{enumerate}

\paragraph{Central velocity distribution $\sigma_{0}$ correlations}

\begin{figure}[H]
\centering
\includegraphics[scale=0.24]{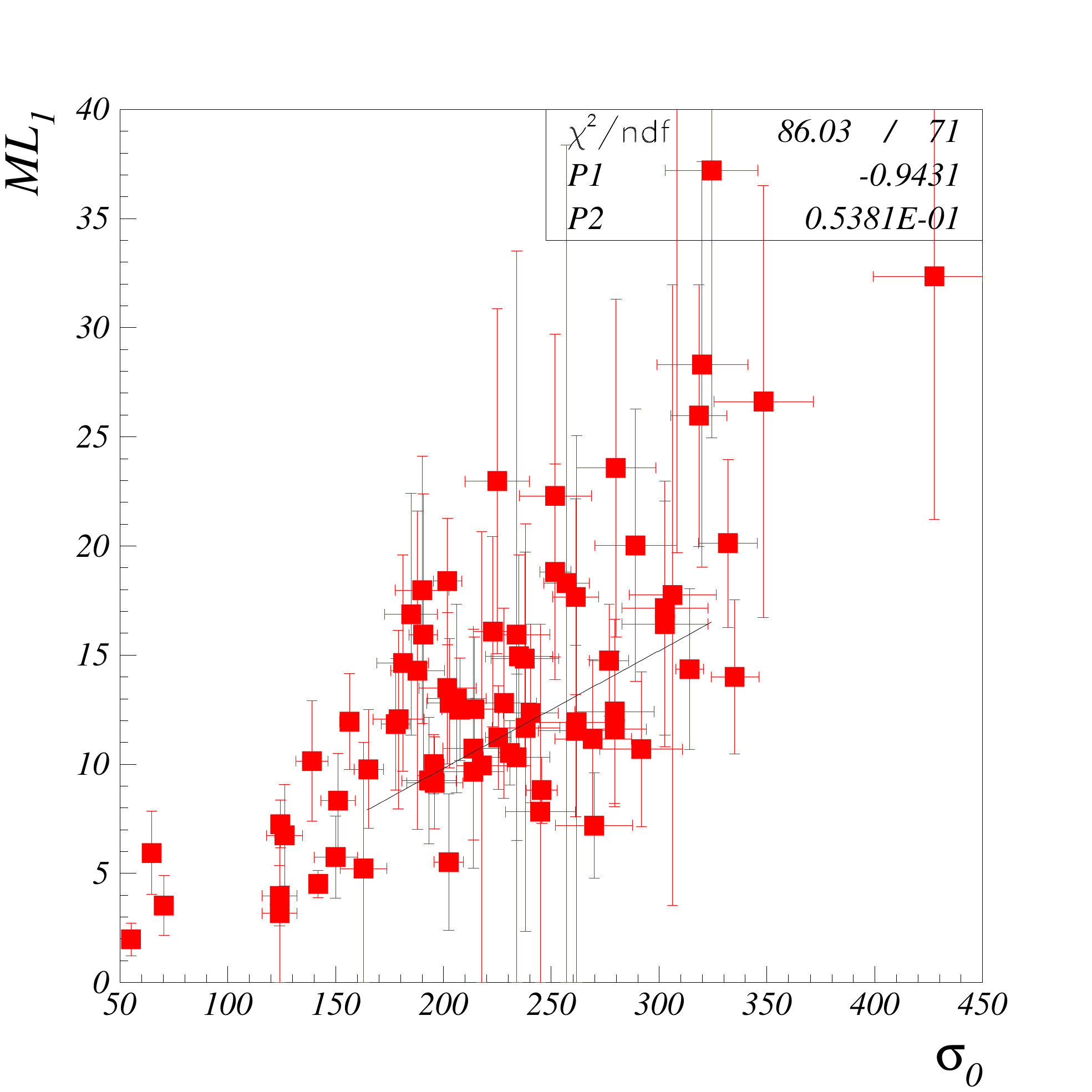}\includegraphics[scale=0.24]{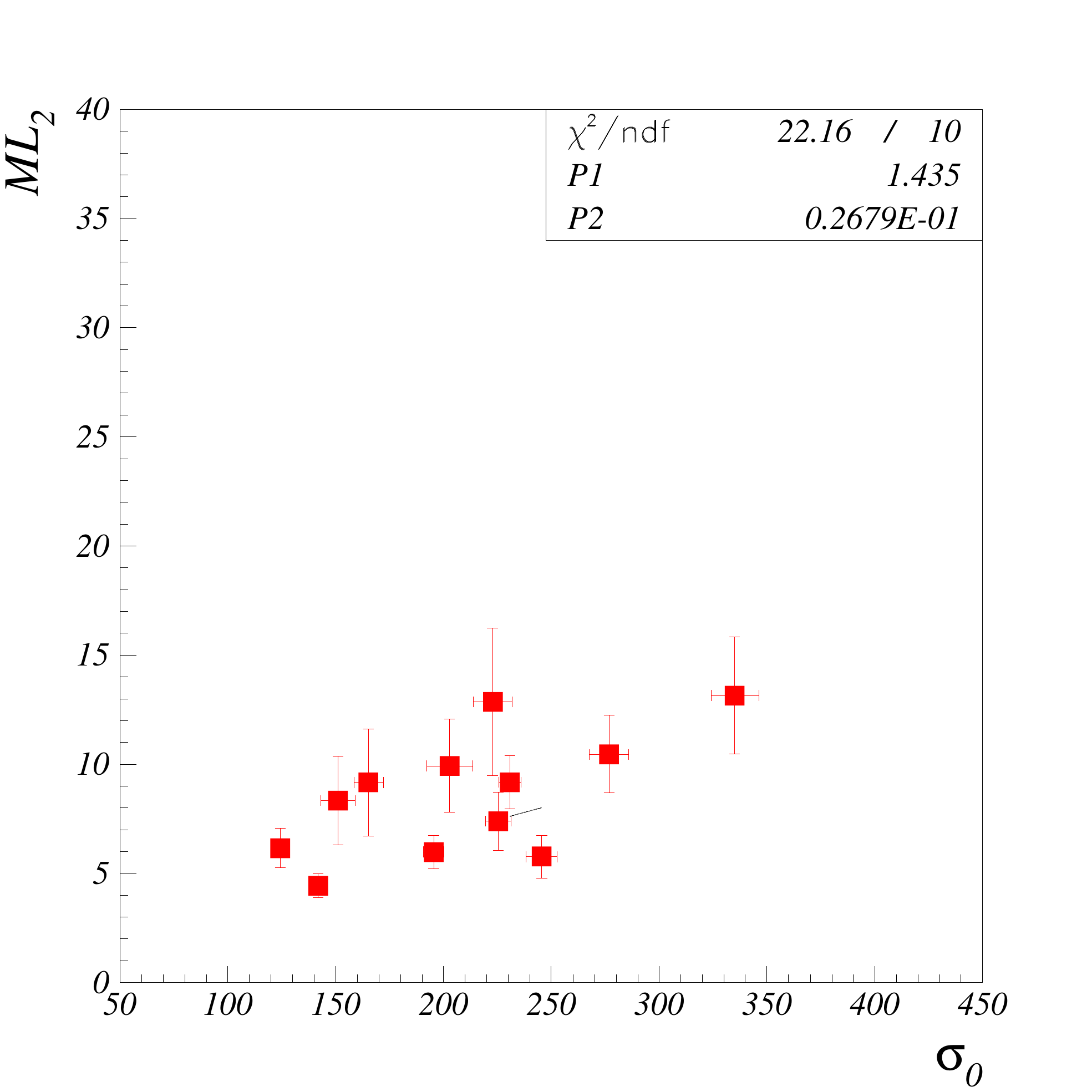}\includegraphics[scale=0.24]{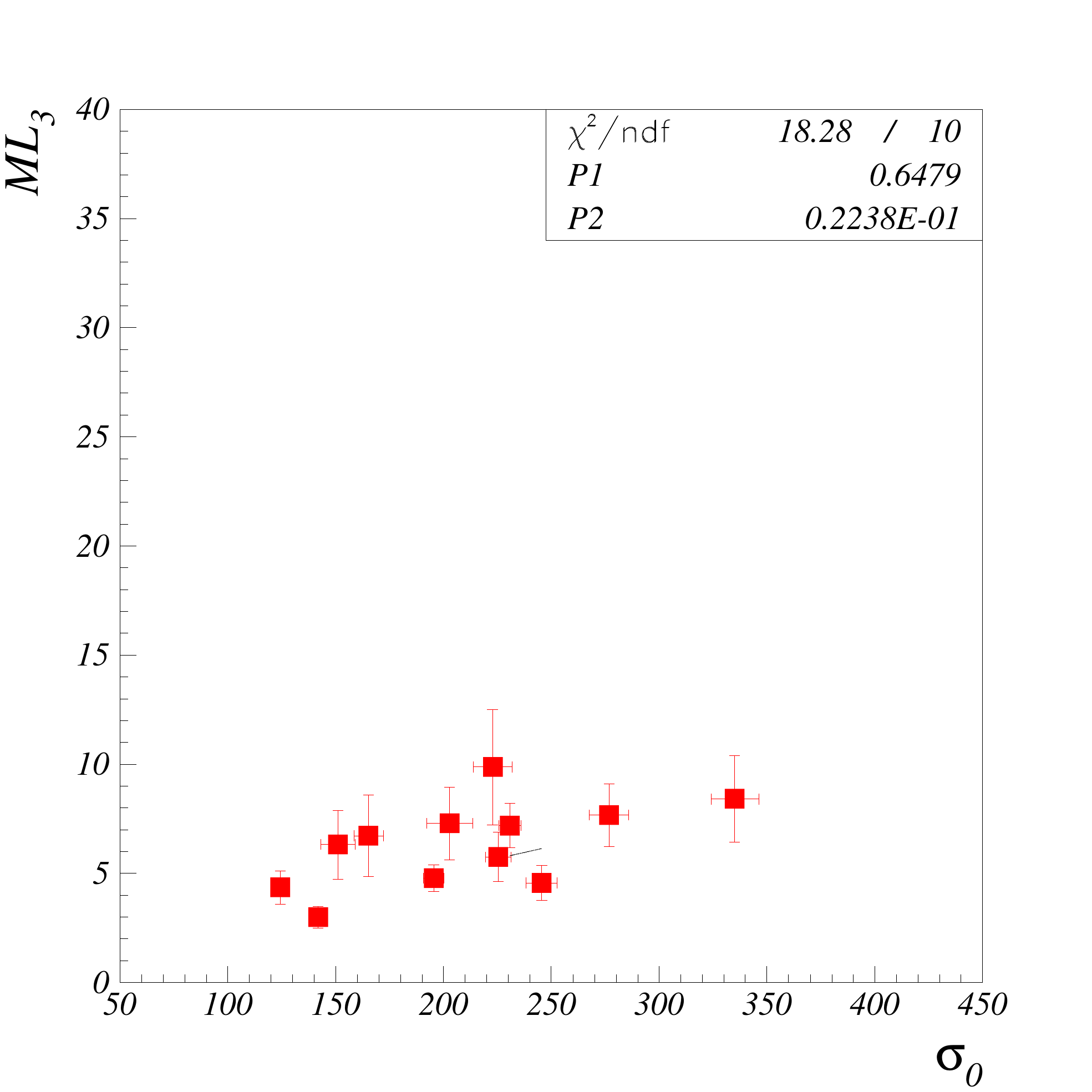}
\vspace{-0.4cm} \caption{\label{Fig: s0 correlations}Correlations between the velocity distribution
$\sigma_{0}$ and $\sfrac{M}{L}$.
}
\end{figure}

The origin of the velocity distribution $\sigma_{0}$ vs $\sfrac{M}{L}$
ratios correlation is the virial theorem.

\paragraph{Mass to light ratio $\sfrac{M}{L}$ correlations}

\begin{figure}[H]
\centering
\includegraphics[scale=0.24]{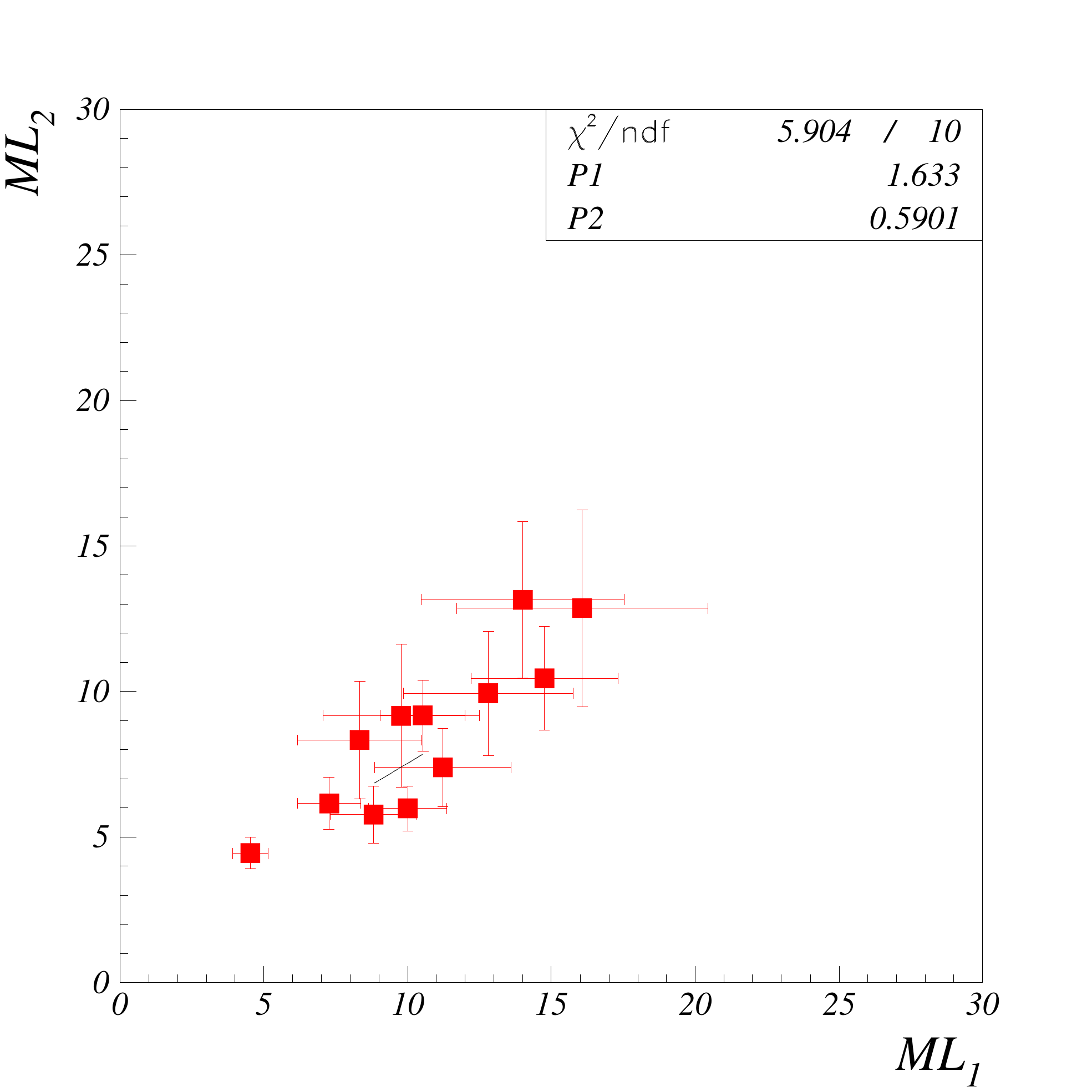}\includegraphics[scale=0.24]{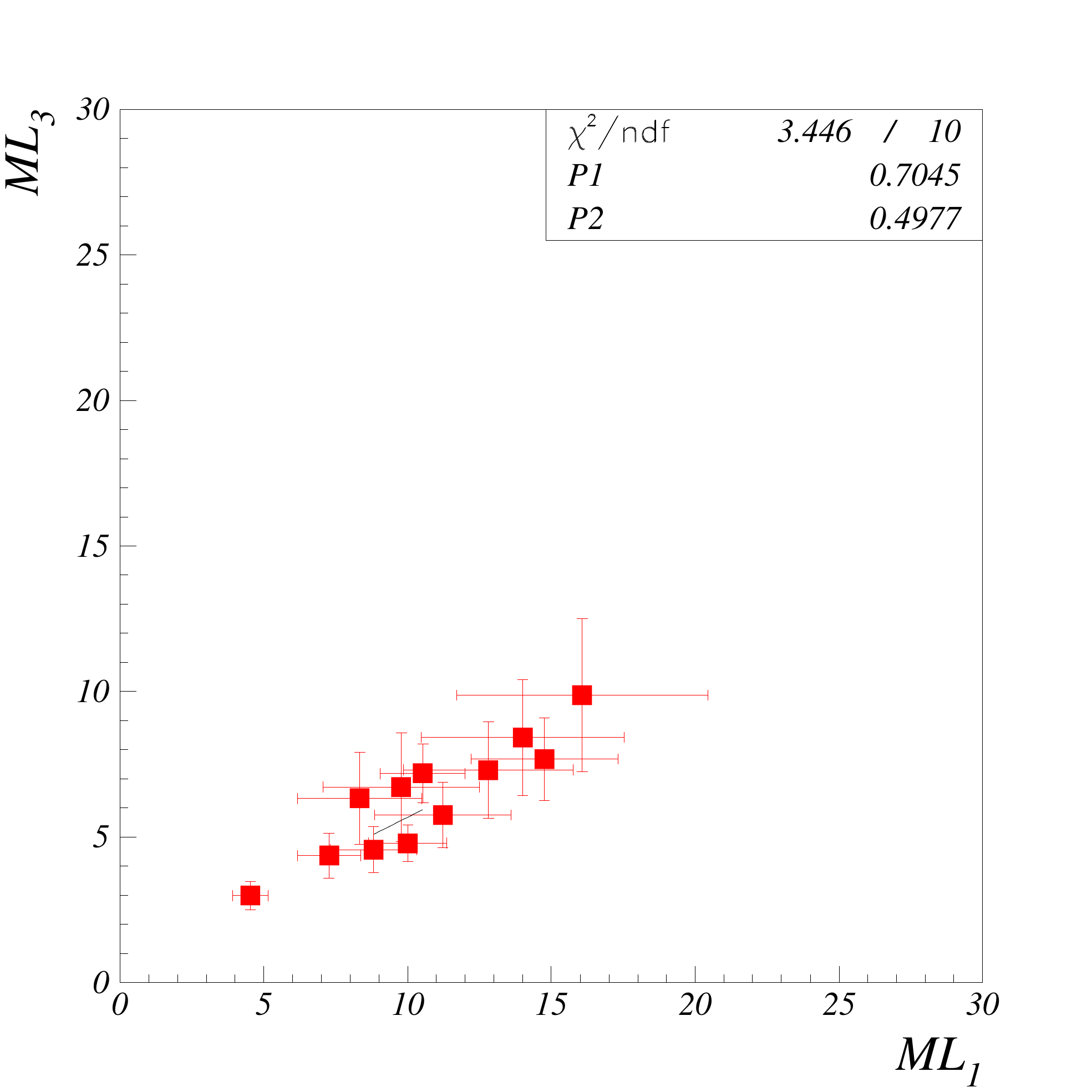}\includegraphics[scale=0.24]{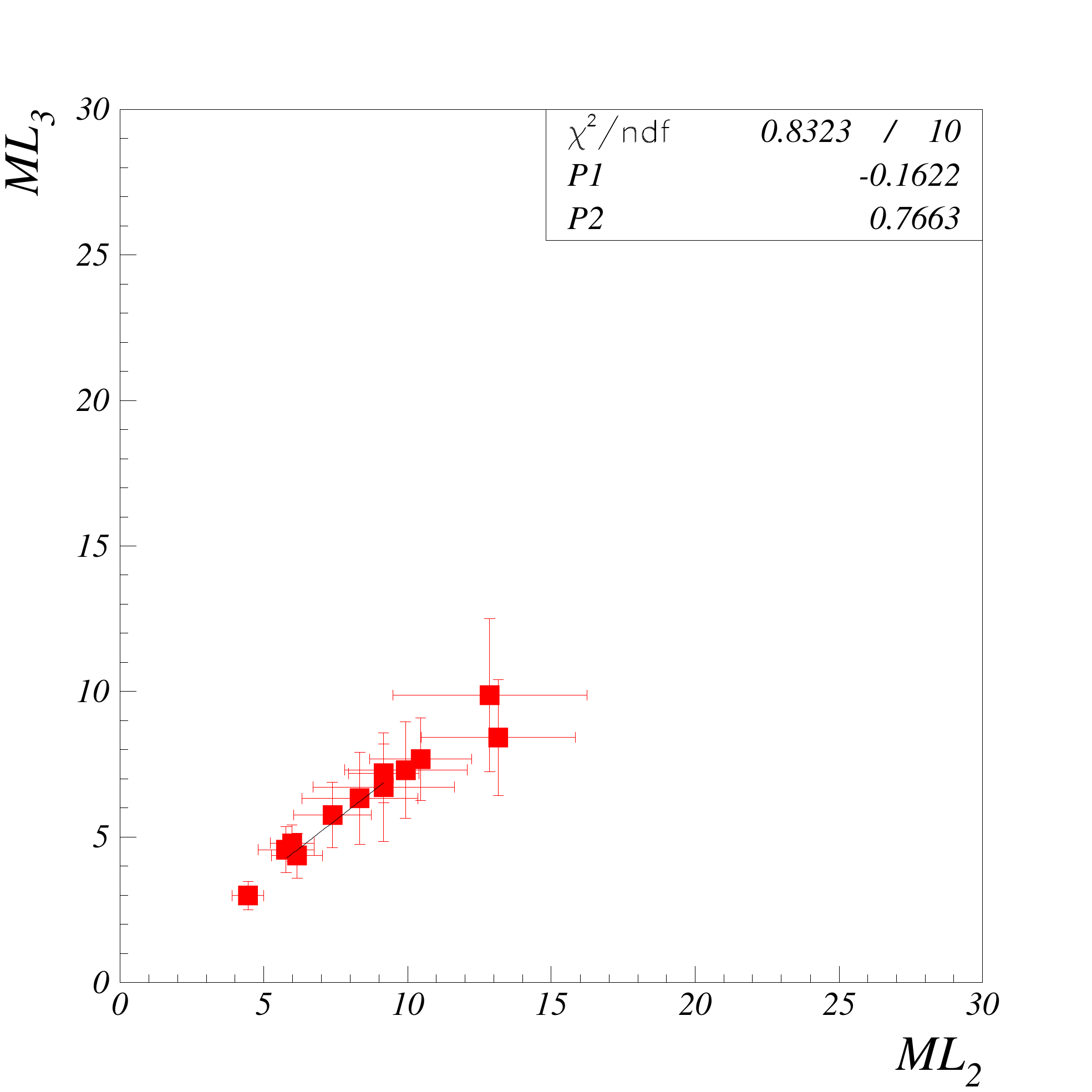}
\vspace{-0.4cm} \caption{\label{Fig: M/L correlations}Correlations between the three methods
used to obtained the $\sfrac{M}{L}$.
}
\end{figure}

The too low $\sfrac{\chi^{2}}{ndf}$, especially for the $\sfrac{M}{L}_{2}$ vs
$\sfrac{M}{L}_{3}$ plot reveal that the three methods are not independent.

\begin{figure}

\centering
\vspace{-2.5cm}
\includegraphics[angle=90,scale=0.75]{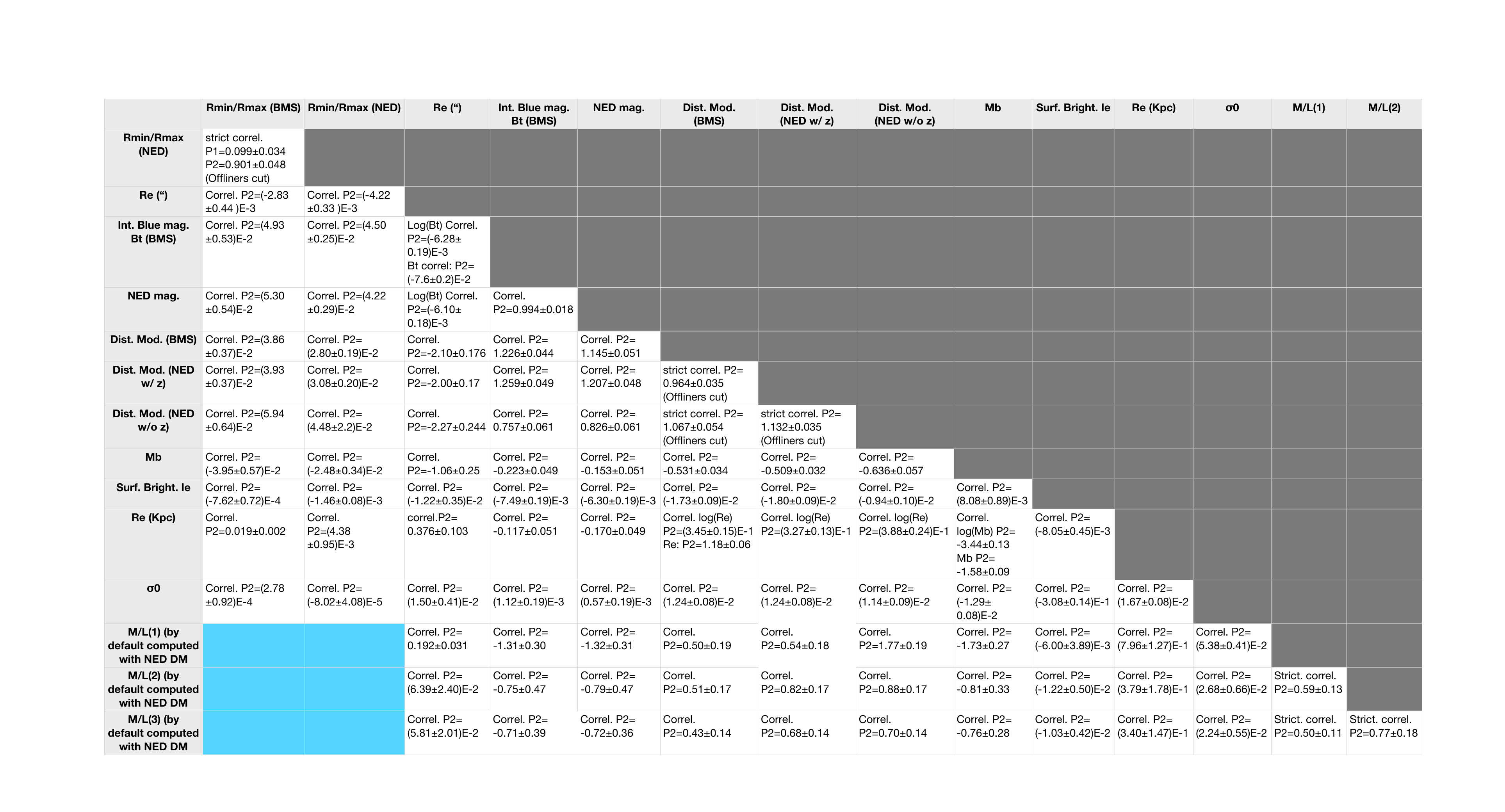}
\vspace{-2.5cm} \caption{\label{correlation table. } Correlation summary.
}
\end{figure}

\subsubsection{Discussion}

\begin{figure}[H]
\centering
\includegraphics[scale=0.65]{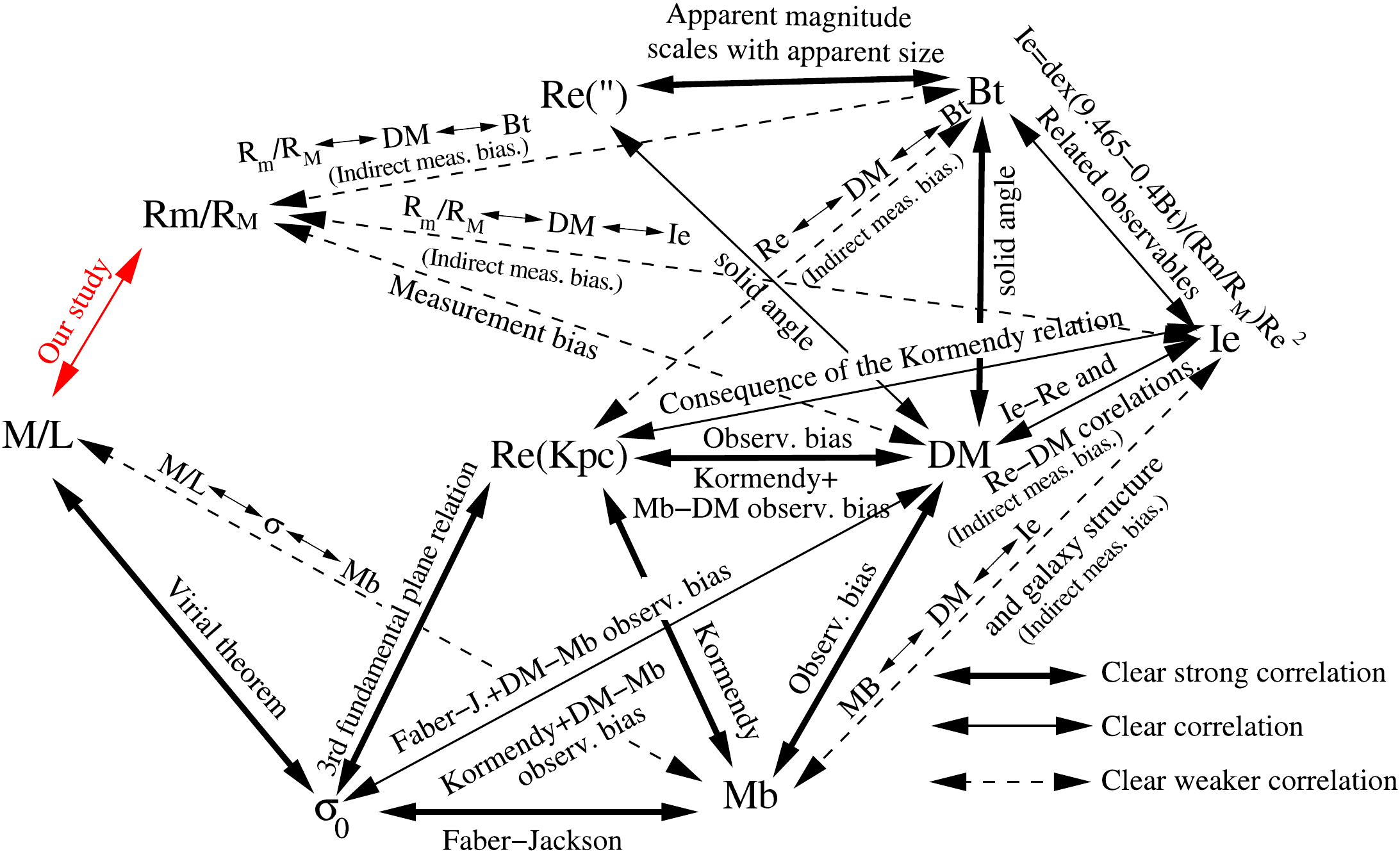}
\vspace{-0.4cm} \caption{\label{fig:correlations 2}Observed correlations. 
}
\end{figure}

\begin{figure}[H]
\centering
\includegraphics[scale=0.55]{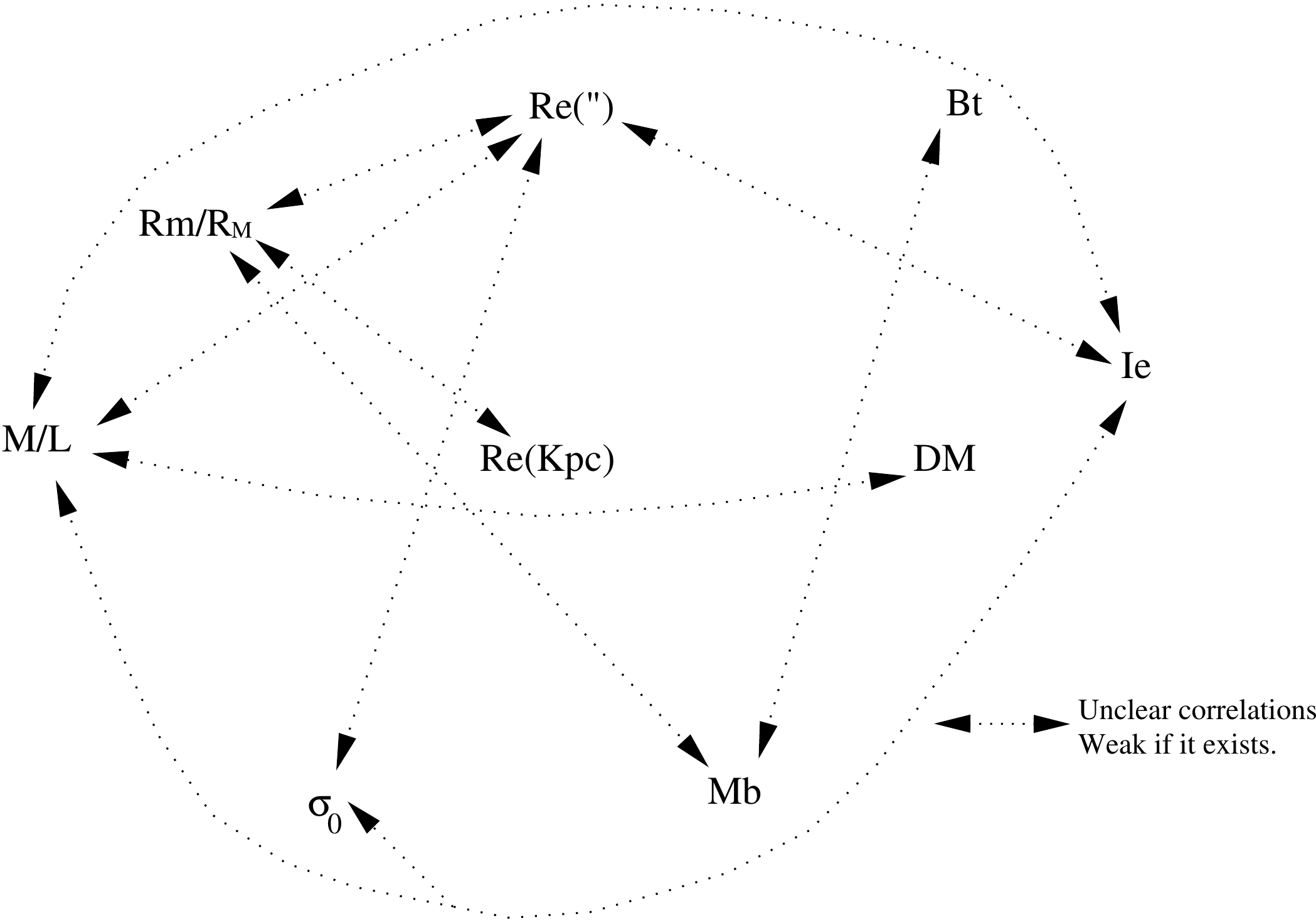}
\vspace{-0.4cm} \caption{\label{fig:correlations 3}Possible weak correlations. 
}
\end{figure}

A correlation summary is given in the table on page \pageref{correlation table. }.
The observed clear correlations are shown in Fig.~\ref{fig:correlations 2}.
Possible weak correlations are shown in Fig.~\ref{fig:correlations 3}.
In most cases the $\sfrac{\chi^{2}}{ndf}$ are larger than 1. This is mostly
because a linear fit is ill-suited to fit the data, and because 
two correlated variables can depend= on other variables,
which adds an additional non-gaussian dispersion. For example we
already mentioned that the $\sfrac{M}{L}$ vs $\sfrac{R_{min}}{R_{max}}$ correlation
has an additional variable, the (unknown) projection angle of the
ellipsoid to the observed ellipse, which creates an additional dispersion. 

All in all, the clear correlations seen in this section can all be
classified as:
\begin{itemize}
\item Physical correlations;
\item Observational biases;
\item Unexplained but known correlations related to galaxy structure (e.g.
Kormendy, Faber-Jackson relations...).
\end{itemize}
This is  satisfactory since this clarifies how to account,
if necessary, for the important correlations.

\subsection{Corrections }

\subsubsection{Corrections for measurement biases}

We will ignore the weak uncertain correlations shown in Fig.~\ref{fig:correlations 3}.
Measurement biases seem to be {\it directly} at the origin of the
$M_{b}\Longleftrightarrow DM$ and $\sfrac{R_{min}}{R_{max}}\dashleftarrow\dashrightarrow DM$
correlations. Measurement biases seem to be {\it indirectly} at the
origin of the $\sfrac{R_{min}}{R_{max}}\dashleftarrow\dashrightarrow B_{t}$,
$\sfrac{R_{min}}{R_{max}}\longleftrightarrow Re(Kpc)$ and $\sfrac{R_{min}}{R_{max}}\dashleftarrow\dashrightarrow I_{e}$
correlations. Finally, measurement biases seem to be {\it indirectly}
contributing {\it partly} to the $Re(Kpc)\Longleftrightarrow DM$,
$Re(Kpc)\dashleftarrow\dashrightarrow B_{t}$, $M_{b}\dashleftarrow\dashrightarrow I_{e}$,
$I_{e}\longleftrightarrow DM$, and $DM\longleftrightarrow\sigma_{0}$
correlations. Unsurprisingly, $DM$ is involved
in all the measurement biases. 

\begin{figure}[H]
\centering
\includegraphics[scale=0.24]{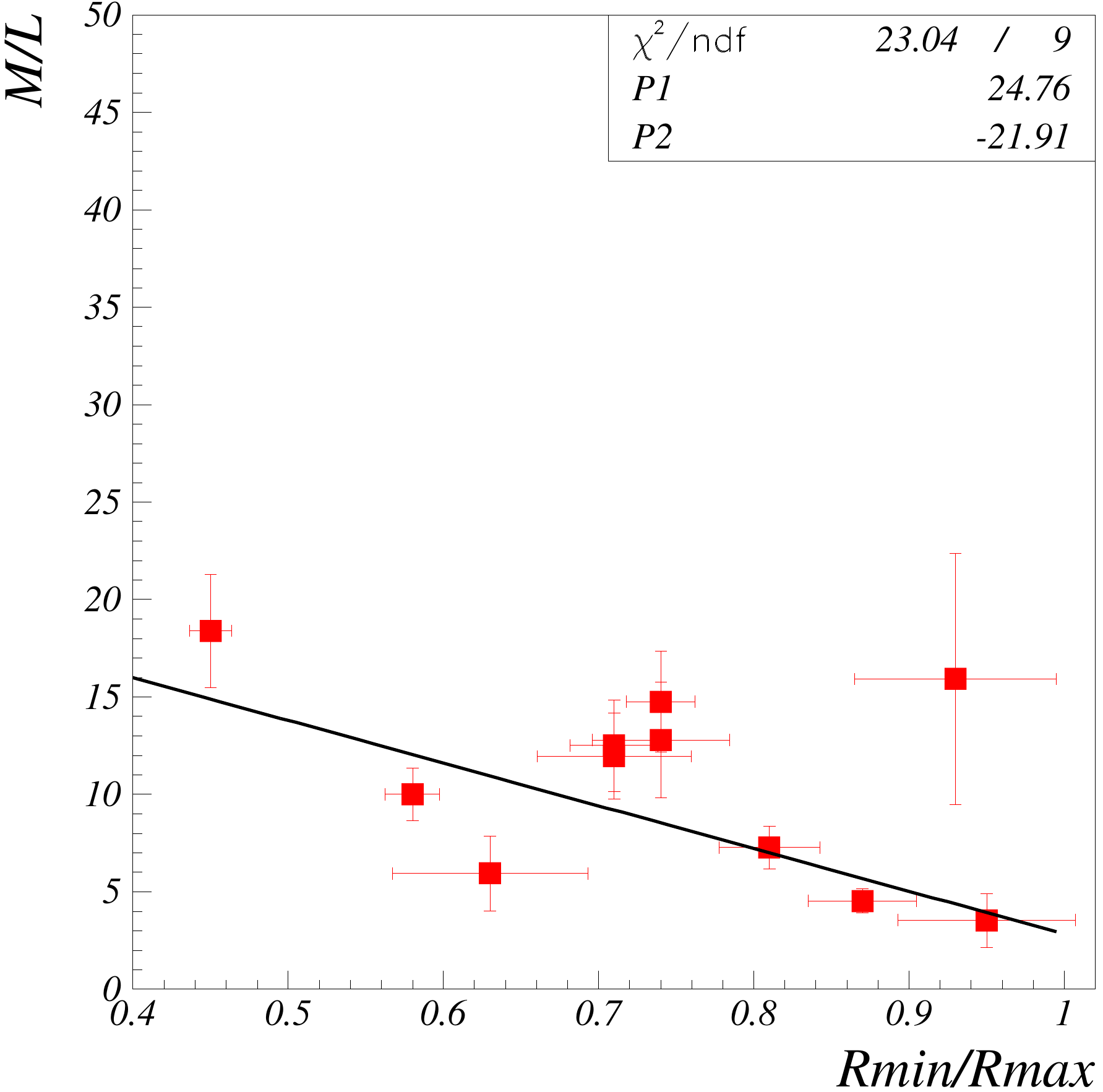}\includegraphics[scale=0.24]{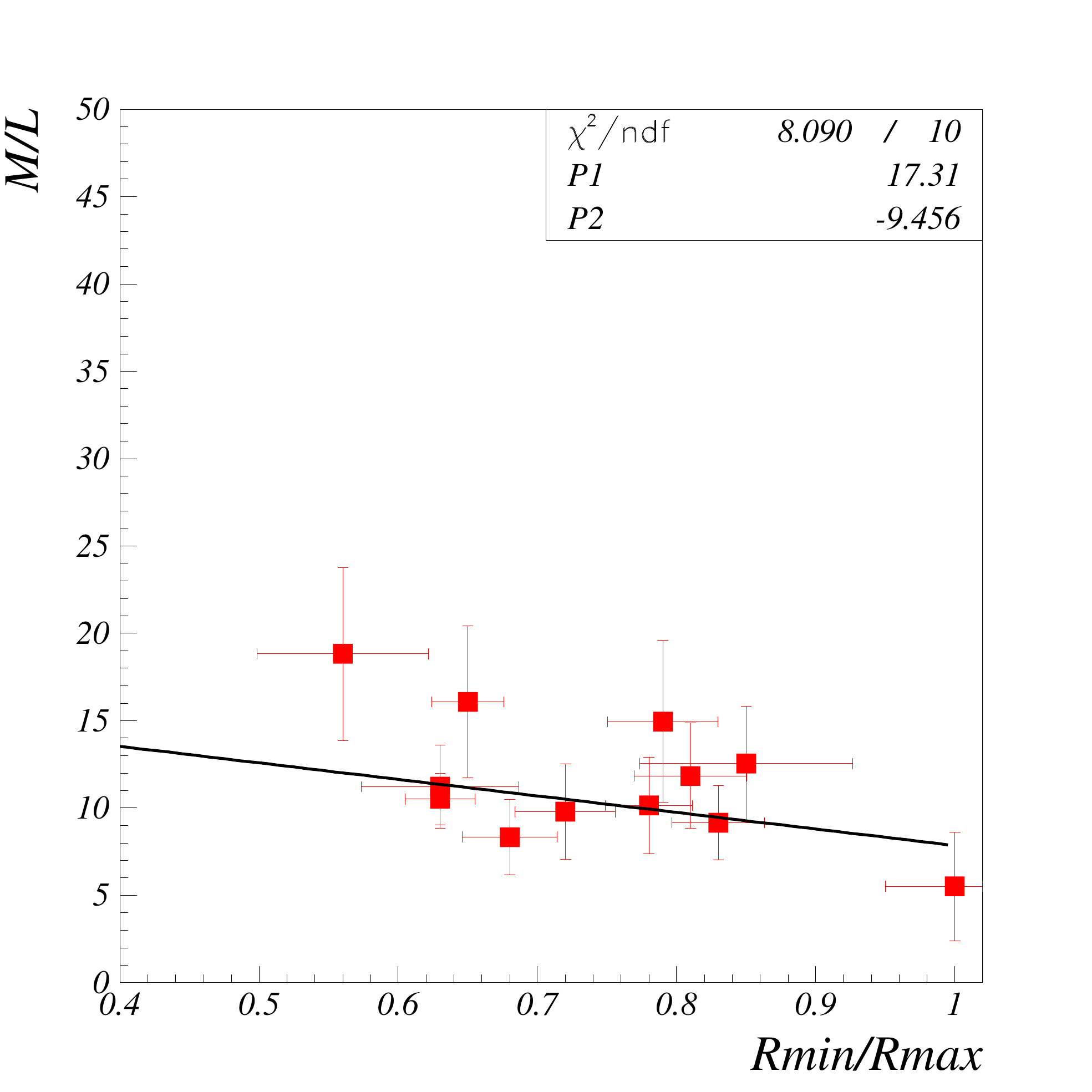}\includegraphics[scale=0.24]{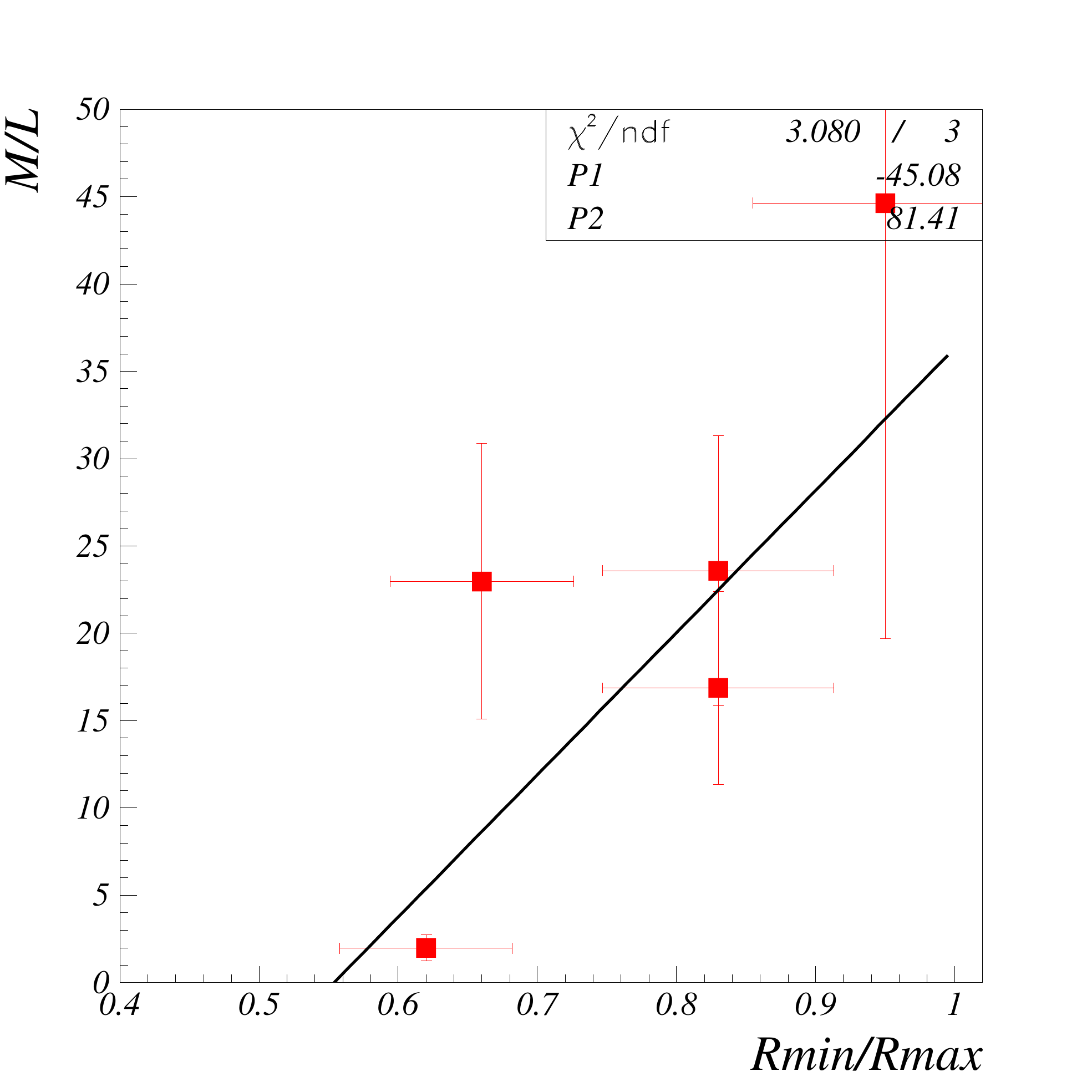}\includegraphics[scale=0.24]{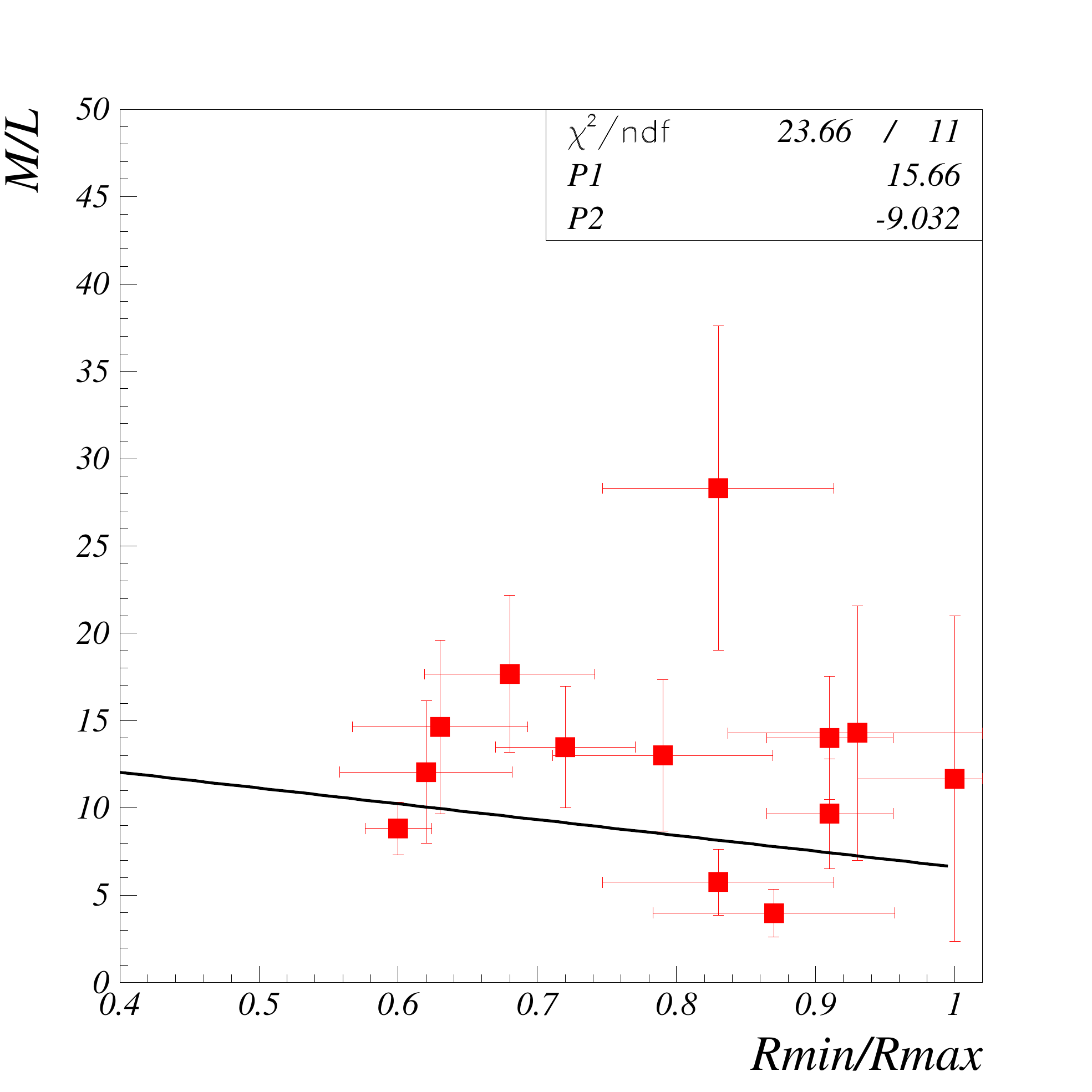}\protect \\
\includegraphics[scale=0.24]{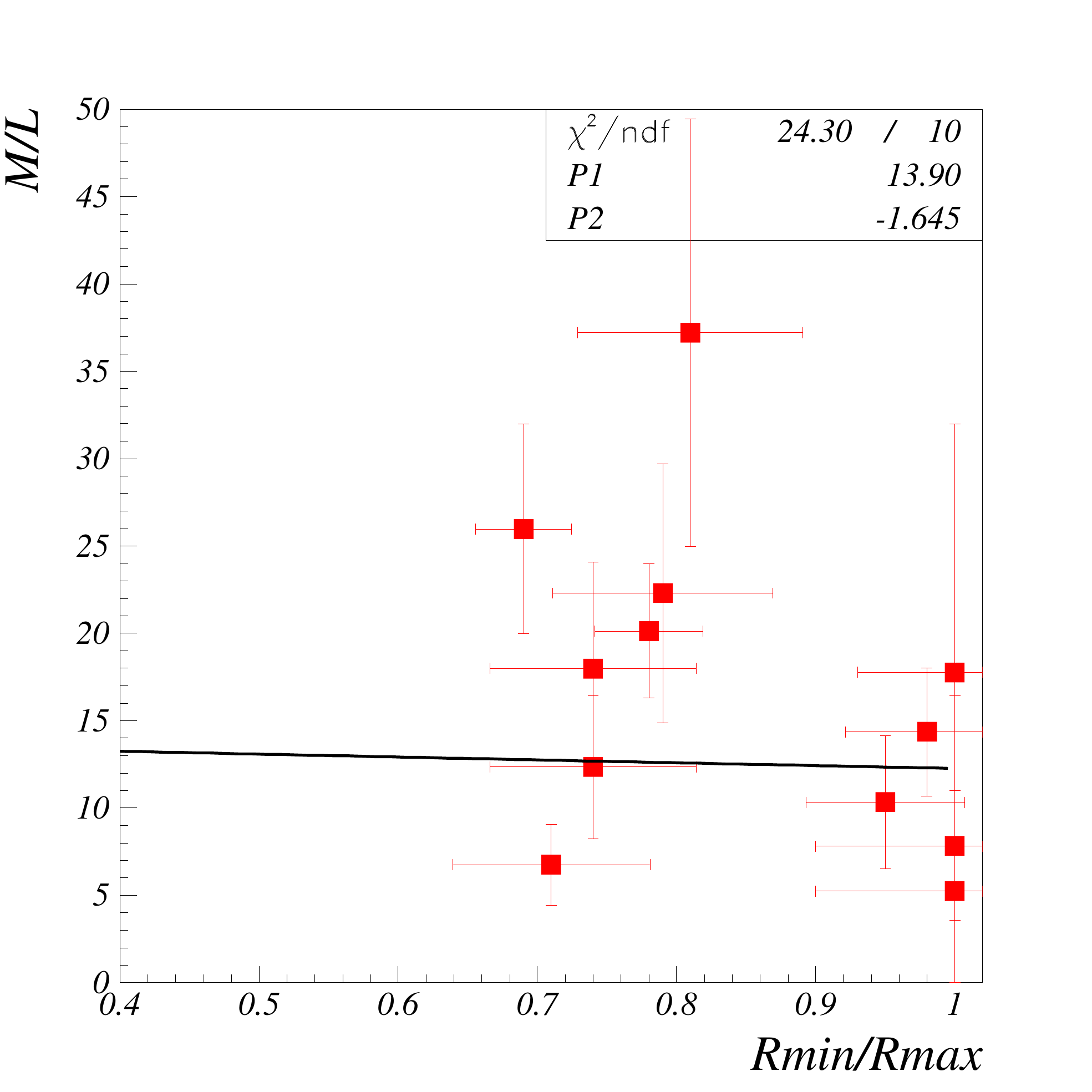}\includegraphics[scale=0.24]{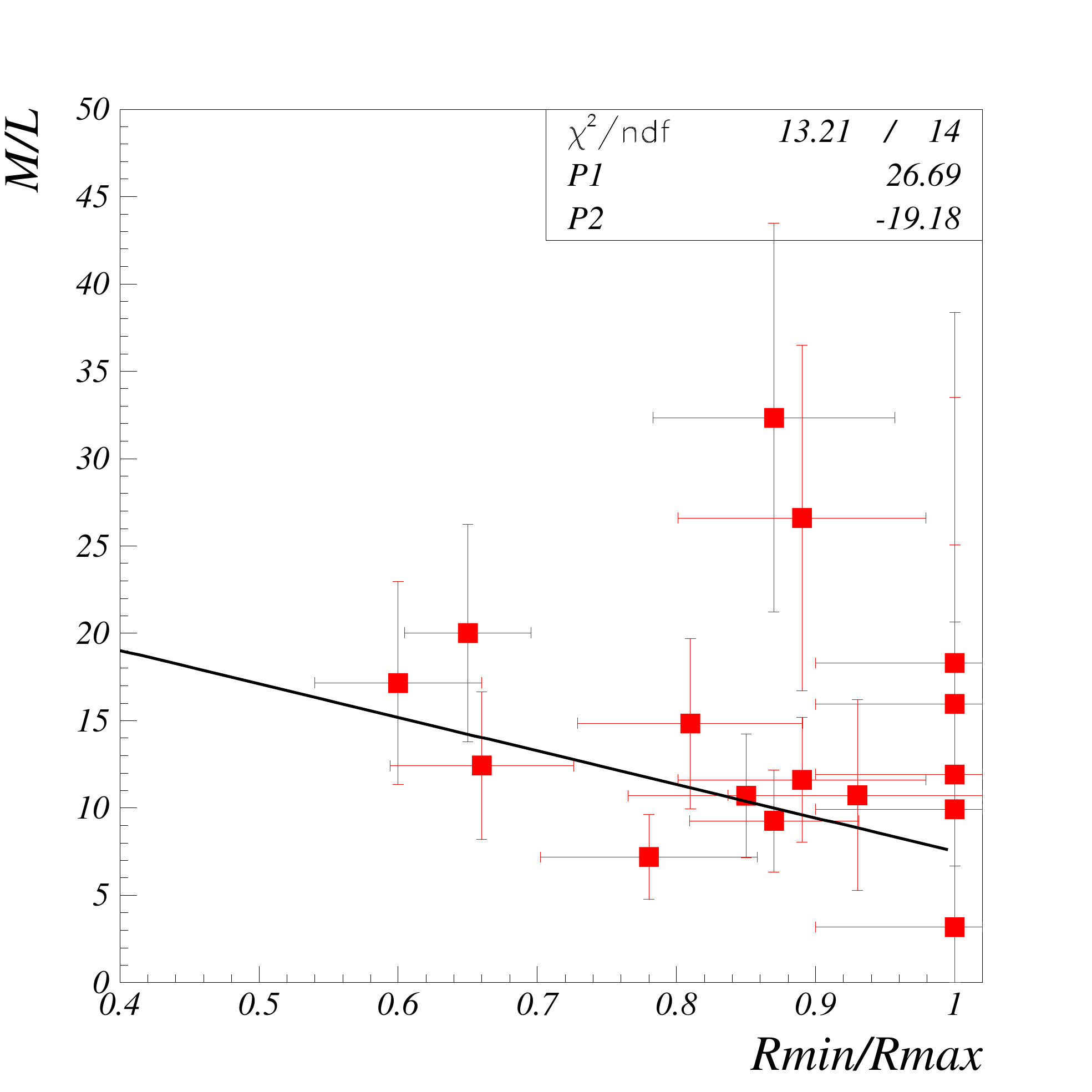}\includegraphics[scale=0.24]{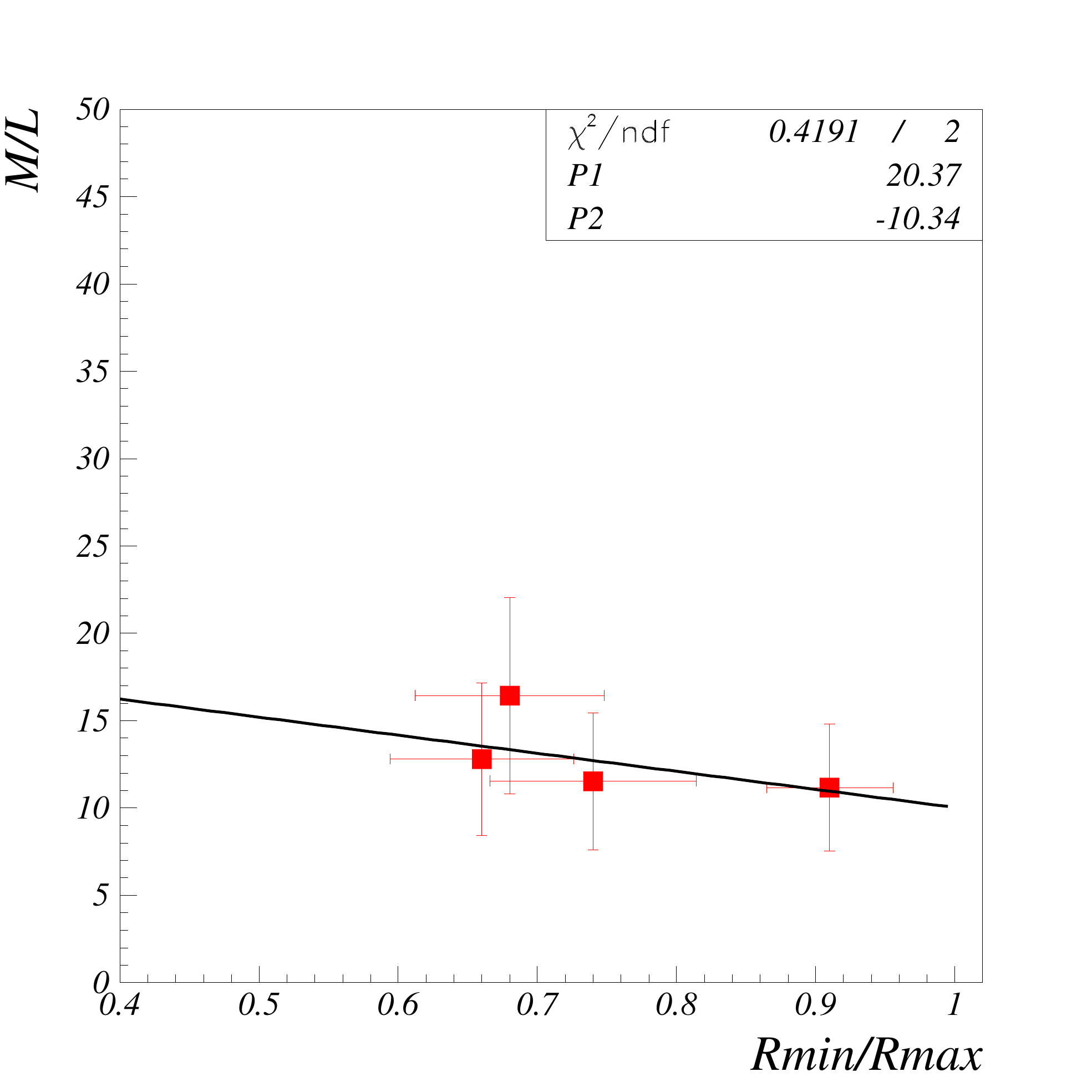}\protect \\
\vspace{-0.4cm} \caption{\label{Flo: M/L vs R/R for various DM bins}Linear fits for $\sfrac{M}{L}$
vs $\sfrac{R_{min}}{R_{max}}$ for various distance modulus $DM$ bins.
Those are, from top left to bottom right: $30\leq DM<31$, $31\leq DM<32$,
$32\leq DM<33$, $33\leq DM<33.5$, $33.5\leq DM<34$, $34\leq DM<35$
and $35\leq DM<35.5$. 
}
\end{figure}

\begin{figure}[H]
\centering
\includegraphics[scale=0.4]{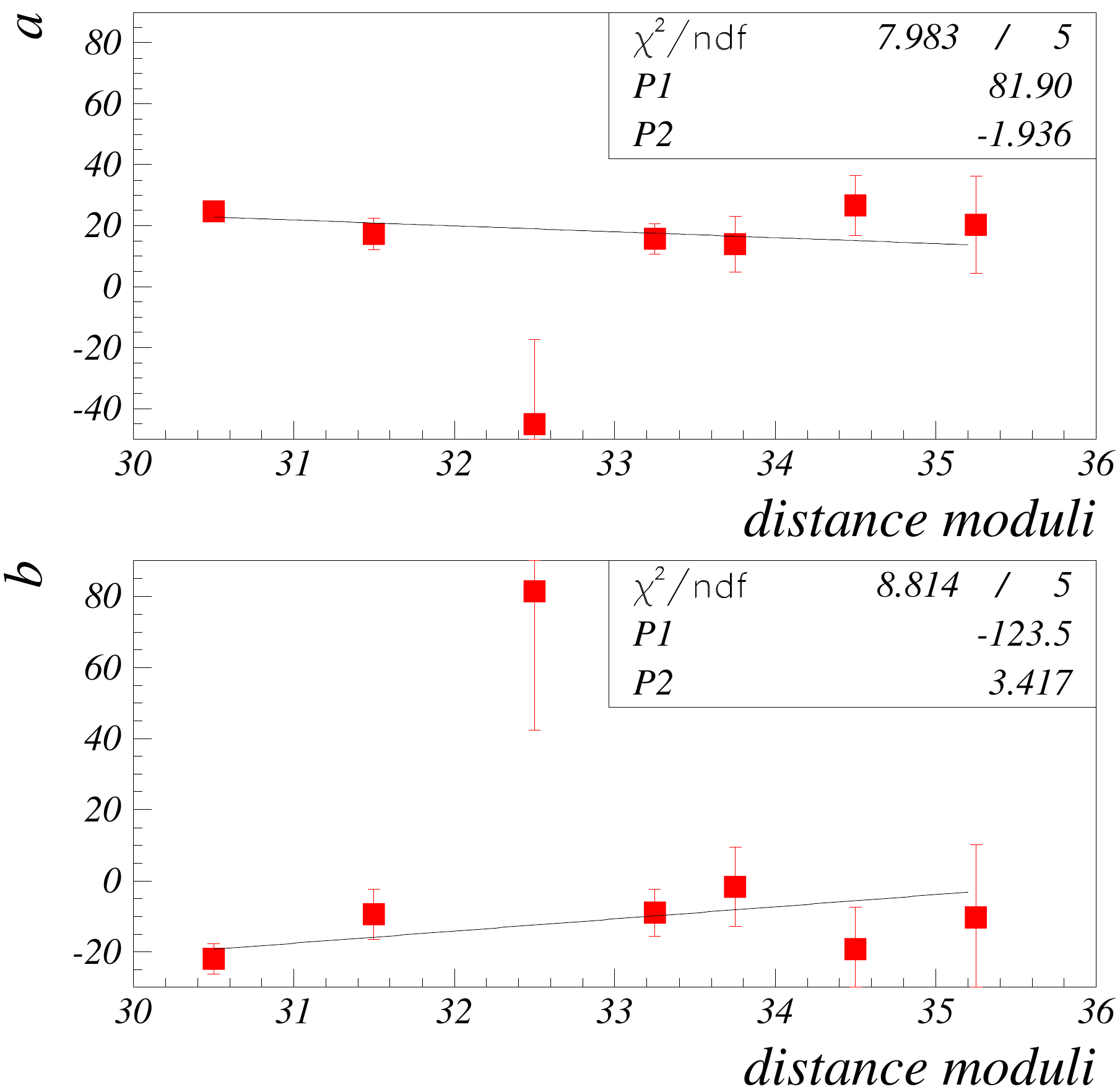}
\includegraphics[scale=0.4]{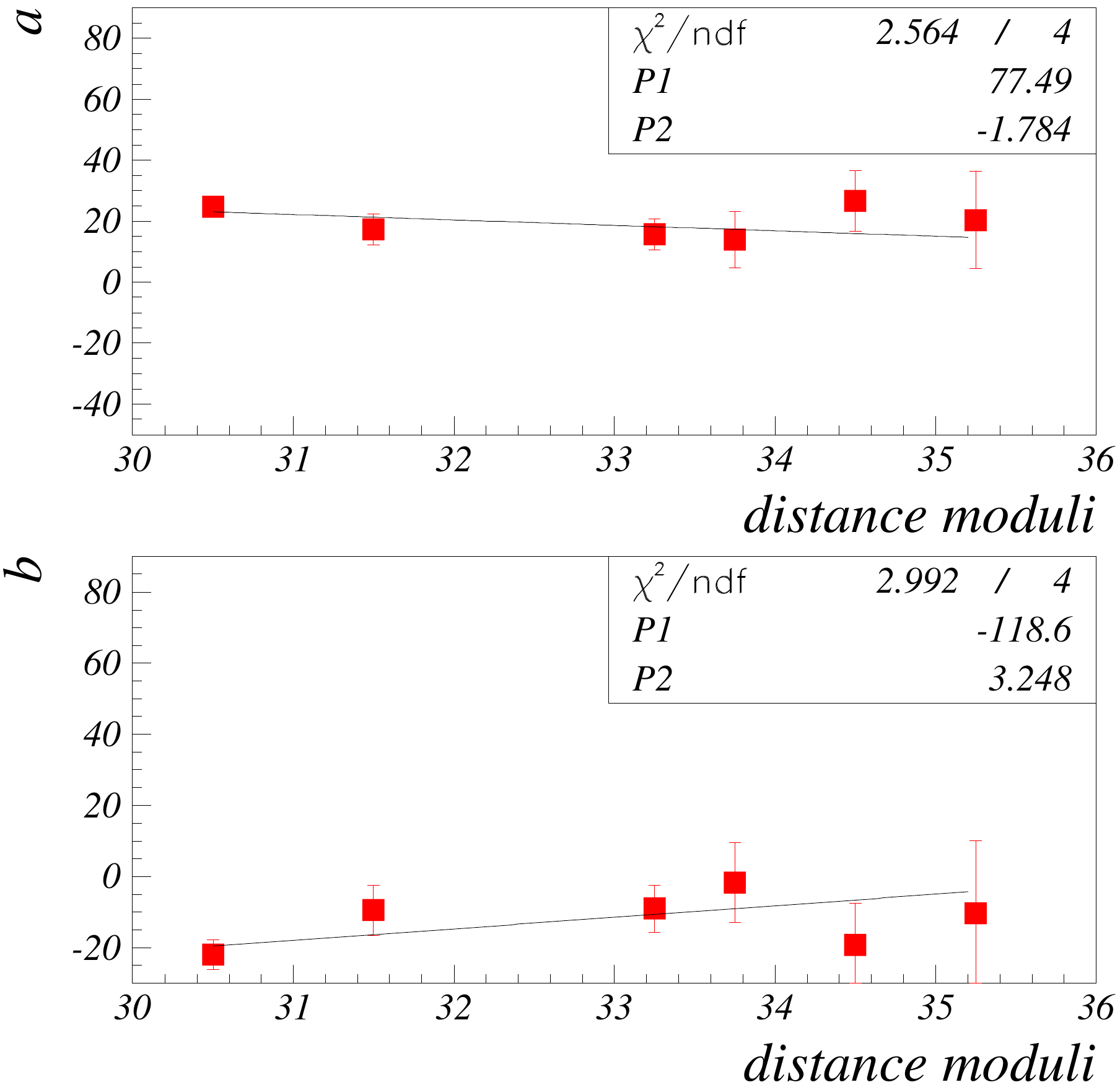}
\vspace{-0.4cm} \caption{\label{Flo: M/Lslope in function of DM}Left panel: fit coefficients
$a$ (top) and $b$ (bottom) from Fig.~\ref{Flo: M/L vs R/R for various DM bins}
in function of $DM$. The $P_{1}$ and $P_{2}$ are the results of
fits $a=P_{1}+P_{2}\times DM$ (top) and $b=P_{1}+P_{2}\times DM$ (bottom). Right
panel: same but with bin $32\leq DM<33$ excluded.
}
\end{figure}

In order to study the effect of  $DM$ on the $\sfrac{M}{L}$ dependence
with $\sfrac{R_{min}}{R_{max}}$, we bin $DM$ and plot $\sfrac{M}{L}$ in function
of $\sfrac{R_{min}}{R_{max}}$ for each of the $DM$ bins. The results for
the linear fits $\sfrac{M}{L}=P_{1}+P_{2}(\sfrac{R_{min}}{R_{max}})$ made for each
$DM$ bin are shown in Fig.~\ref{Flo: M/L vs R/R for various DM bins}.
We choose $DM$ bins of size 1, or smaller if the statistics is large.
We use $\sfrac{R_{min}}{R_{max}}$ and $DM$ values from~\cite{BMS}.
The values of $a=P_{1}$ and $b=P_{2}$ in function of the $DM$ bins
can be seen on the left panel of Fig.~\ref{Flo: M/Lslope in function of DM}.
Within the approximation of a linear dependence of $\sfrac{M}{L}$ in function
of $\sfrac{R_{min}}{R_{max}}$, then $b=\partial(\sfrac{M}{L})/\partial(\sfrac{R_{min}}{R_{max}})$.
Thus, $b$ is our quantity of interest. Except for one outlying point,
there is no strong dependence of $a$ and $b$ with $DM$: we find
$b=(3.42\pm2.00)DM-123.5\pm63.59$ for a reduced $\sfrac{\chi^{2}}{ndf}=1.8$.
This would suggest that the correction for the $DM$ bias would increase
the significance of our quantity of interest $b=\partial(\sfrac{M}{L})/\partial(\sfrac{R_{min}}{R_{max}})$.
Figs.~\ref{Flo: M/L vs R/R for various DM bins} and~\ref{Flo: M/Lslope in function of DM}
also suggest that there is something wrong with bin $32\leq DM<33$
(possibly because of the lowest point (galaxy NGC 4510) that has very
small error%
\footnote{This small error is partly an artifact: Ref.~\cite{BMS} gives
relative errors, hence small $\sfrac{M}{L}$ values have small absolute errors,
which might not reflect fully the uncertainty. Apart from re-performing
a full error analysis, there is no easy way to correct this effect.
We also refrain from removing this particular galaxy (NGC 4510) after
close-up study: most of galaxies would have a particularity after
close examination that one could use to justify excluding it from
our sample. It would be thus possible to bias the result of our study
in any arbitrary way. Obeying our general criteria and basic statistics
rules protects us from such bias. %
}). The right panel of Fig.~\ref{Flo: M/Lslope in function of DM}
shows the fit result with bin $32\leq DM<33$ excluded. We find similar
fit values: $b=(3.25\pm2.00)DM-118.7\pm62.94$ for a reduced $\sfrac{\chi^{2}}{ndf}=0.8$.
The value of the reduced $\sfrac{\chi^{2}}{ndf}$, closer to the expected 1,
supports excluding bin $32\leq DM<33$. 

\begin{figure}
\centering
\includegraphics[scale=0.4]{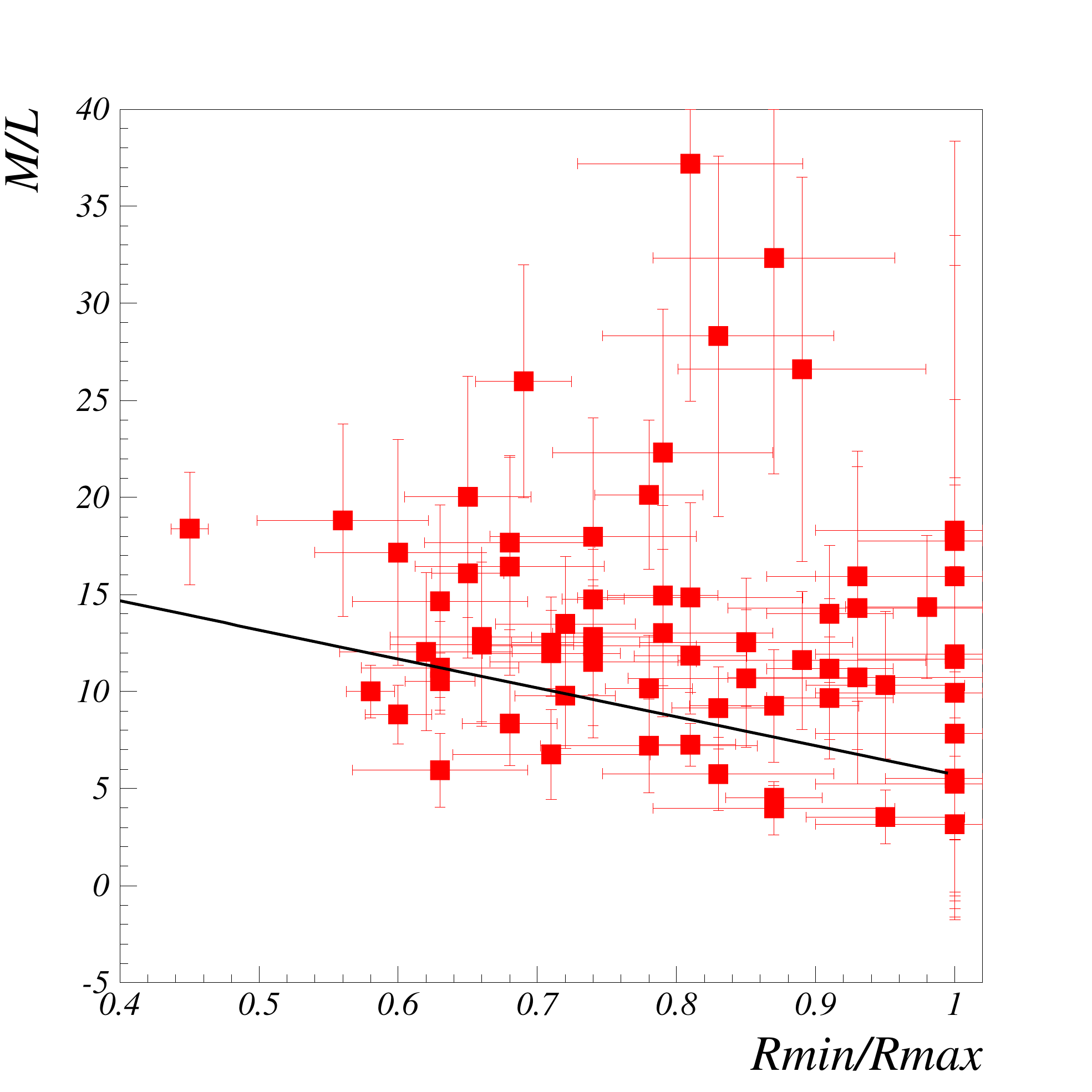}
\vspace{-0.4cm} \caption{\label{Flo:ML vs R/R after DM study}Correlation between $\sfrac{M}{L}$
and the apparent axis ratio $\sfrac{R_{min}}{R_{max}}$ for 68 galaxies (after
removing bin $32\leq DM<33$). The computed Pearson coefficient reveals
a large correlation (Pearson coefficient > 0.5).
}
\end{figure}

The correction for the $DM$ bias would increase $b$ by a factor
$(-118.65\pm62.94)/(-13.00)=9.13\pm4.84$ (here, -13.00 is the value
of $b$ at $<DM>$=32.5). However, this assumes the reliability of
a linear extrapolation over 31 units of $DM$, based on a fit performed
over a $DM$ range of 4.5 units. Such a procedure would yield average
values of $\sfrac{M}{L}\simeq80$ ($P_{1}$ coefficient of the top plots of
Fig.~\ref{Flo: M/Lslope in function of DM}), a value 6 times larger
than the average $\langle \sfrac{M}{L}\rangle=$13.3 for our data sample. Assuming similar
effect, the $DM$ bias correction would be reduced to $\sim9.13/6=1.6$.
This is still a large correction and given the uncertainties attached
to it and the fact that ignoring it would reduce the signature of
the correlation we are investigating, we choose to not apply the correction.
However, we do retain the fact that bin $32\leq DM<33$ should be
excluded from our analysis. The resulting linear fit of $\sfrac{M}{L}$ vs
$\sfrac{R_{min}}{R_{max}}$ is $\sfrac{M}{L}=(-14.92\pm2.81)\sfrac{R_{min}}{R_{max}}+20.63\pm2.19$
(Fig.~\ref{Flo:ML vs R/R after DM study}), to be compared to the
result shown in Fig.~\ref{fig:ml1_apar} ($\sfrac{M}{L}=(-6.40\pm2.32)\sfrac{R_{min}}{R_{max}}+12.56\pm1.79$). 
The Pearson coefficient
calculated when keeping data with uncertainties $\Delta \sfrac{M}{L}<5$ is -0.44 (the linear
fit becomes $\sfrac{M}{L}=(-15.69\pm2.92)\sfrac{R_{min}}{R_{max}}+20.86\pm2.27$). When
keeping data with $\Delta \sfrac{M}{L}<3$ the Pearson coefficient becomes -0.64 
(the linear fit becomes $\sfrac{M}{L}=(-19.56\pm3.46)\sfrac{R_{min}}{R_{max}}+22.73\pm2.62$).
Those indicate a large correlation. 

\subsubsection{Surface brightness vs absolute blue magnitude}

\begin{figure}
\centering
\includegraphics[scale=0.4]{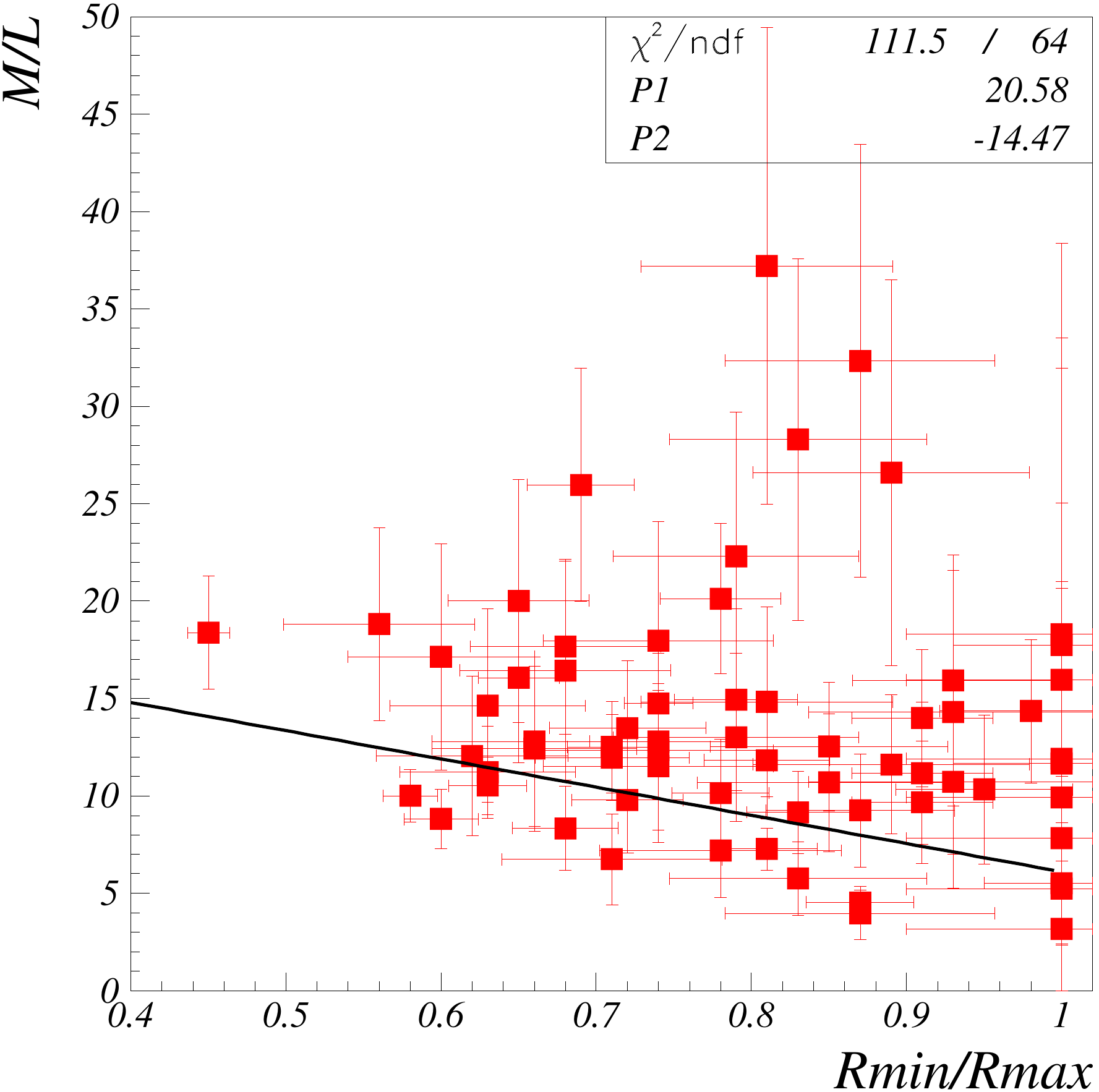}
\vspace{-0.4cm} \caption{\label{Flo:ML vs R/R after removing 2 possible dwarfs}Correlation
between $\sfrac{M}{L}$ and the apparent axis ratio $\sfrac{R_{min}}{R_{max}}$
for 66 galaxies (after removing bin $32\leq DM<33$ and two possible
compact elliptical galaxies). 
}
\end{figure}

We exclude from the sample the two galaxies that are close to the dwarf
elliptical locus, see Section~\ref{sub:Absolute-magnitude-Mb corel}
(we apply a selection at $M_{b}>-17.8$). This has no significant
consequence on the $\sfrac{M}{L}$ vs $\sfrac{R_{min}}{R_{max}}$ relation: compare
Fig.~\ref{Flo:ML vs R/R after removing 2 possible dwarfs} to 
Fig.~\ref{Flo:ML vs R/R after DM study}. The linear
fit result is $\sfrac{M}{L}=(-14.48\pm3.00)(\sfrac{R_{min}}{R_{max}})+20.58\pm2.325$.
Because this correction has little effect, we do not apply it in the
rest of the analysis. 

\subsubsection{Hubble parameter correction}

We must scale the slope of our assumed relation between $\sfrac{M}{L}$ and
$\sfrac{R_{min}}{R_{max}}$ to account for the fact that~\cite{BMS} uses
a large value of the Hubble parameter $H_{0}$. This should be corrected
for since redshift distances scale inversely to $H_{0}$ and $\sfrac{M}{L}$
scales inversely with distances. Using $H_{0}$=70
km s$^{-1}$ Mpc$^{-1}$ instead of the 95 km s$^{-1}$ Mpc$^{-1}$
used in~\cite{BMS} yields a $\sfrac{M}{L}$ correction of $70/95=0.74$.
The linear fit becomes $\sfrac{M}{L}=(-11.00\pm2.07)\sfrac{R_{min}}{R_{max}}+15.20\pm1.61$.

\begin{figure}
\centering
\includegraphics[scale=0.4]{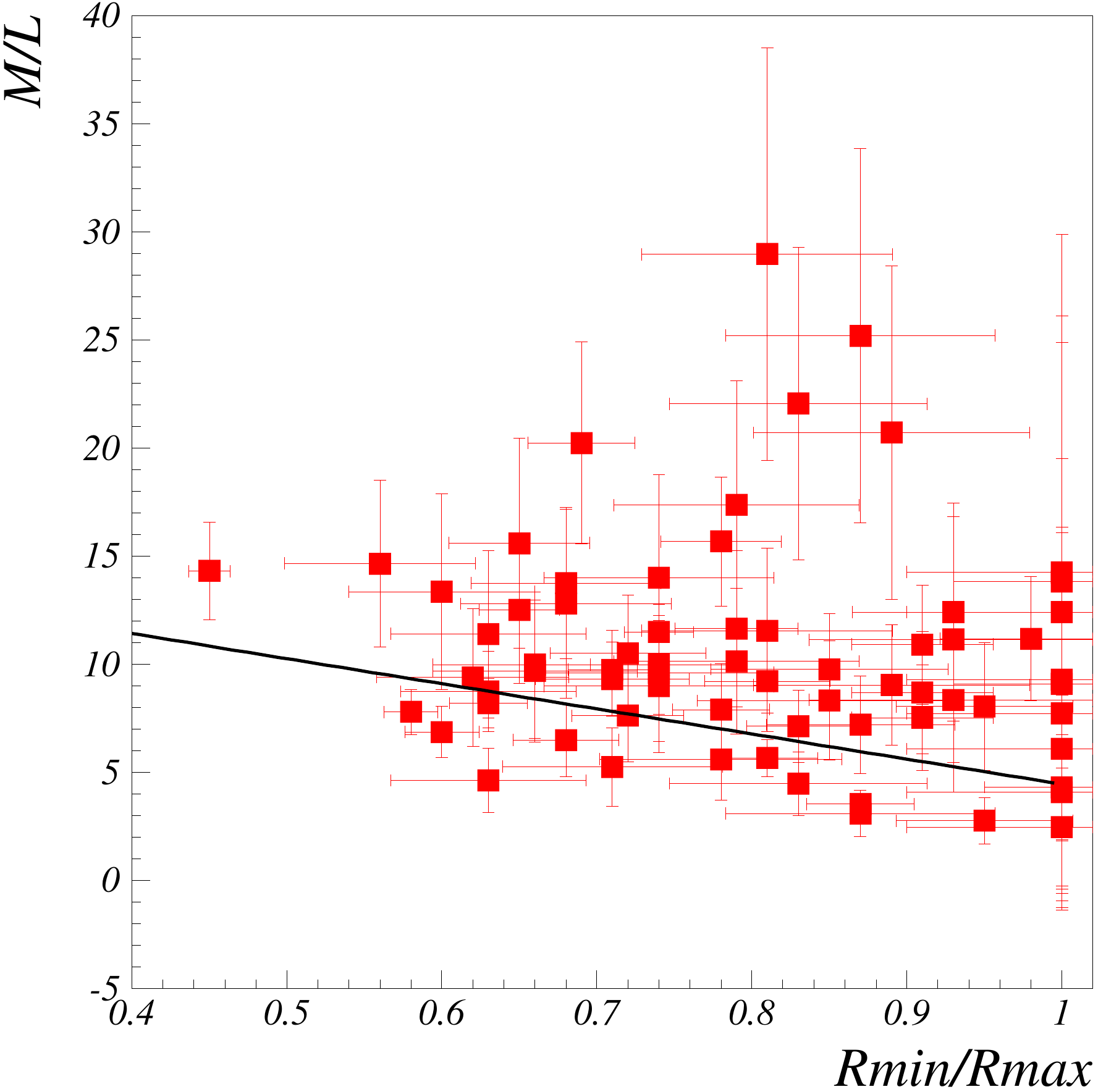}
\vspace{-0.4cm} \caption{\label{Flo:ML vs R/R after Hubble correction} Same as Fig.~\ref{Flo:ML vs R/R after DM study}
but after Hubble coefficient correction.
}
\end{figure}

\subsubsection{Projection correction\label{sub:Projection-correction}}

This correction is discussed in detail for the Bacon {\it et al.}
data~\cite{BMS} in Section~\ref{sec:Projection-correction}.

\subsection{Final result}

\begin{figure}
\centering
\includegraphics[scale=0.4]{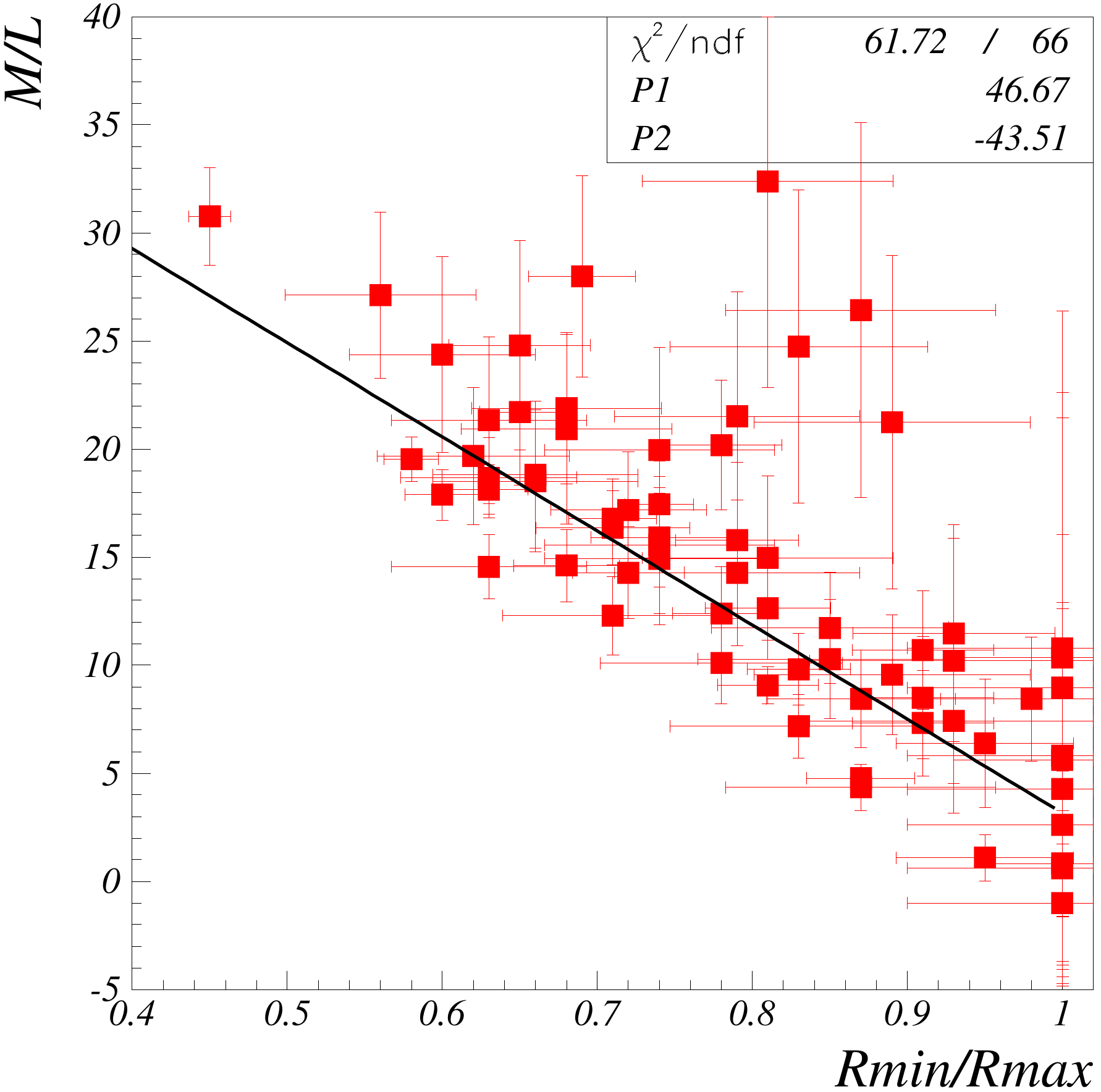}
\vspace{-0.4cm} \caption{\label{fig:ML vs real axis ratio}The mass to light ratio $\sfrac{M}{L}$ in function of the galaxy axis ratios
from our 68 galaxies sample, after correction for ellipticity projection
and the Hubble parameter. The Pearson correlation coefficient is -0.64.
}
\end{figure}

\begin{figure}
\centering
\includegraphics[scale=0.4]{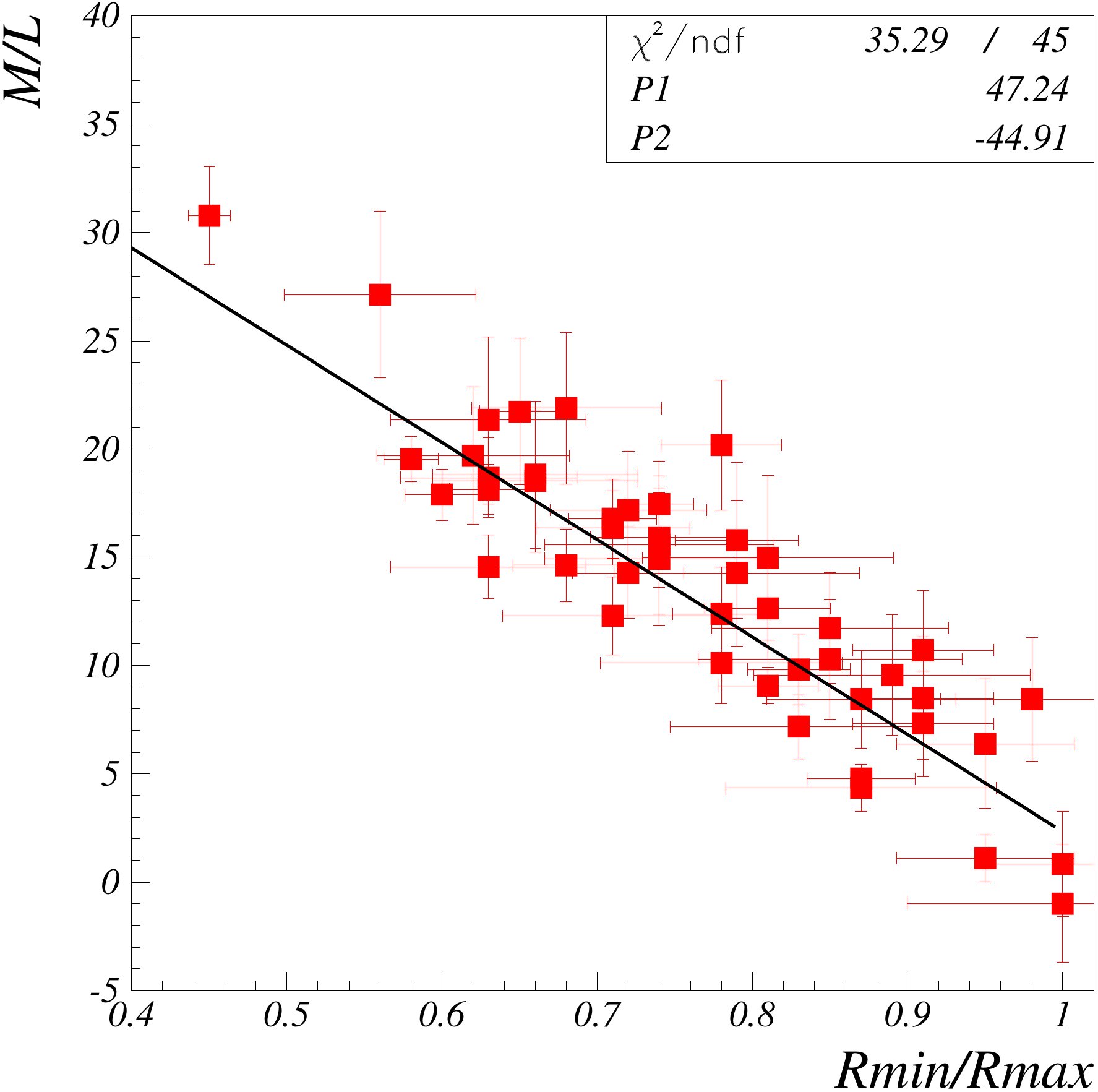} 
\label{fig:ml vs real axis ratio for pearson cc}
\includegraphics[scale=0.4]{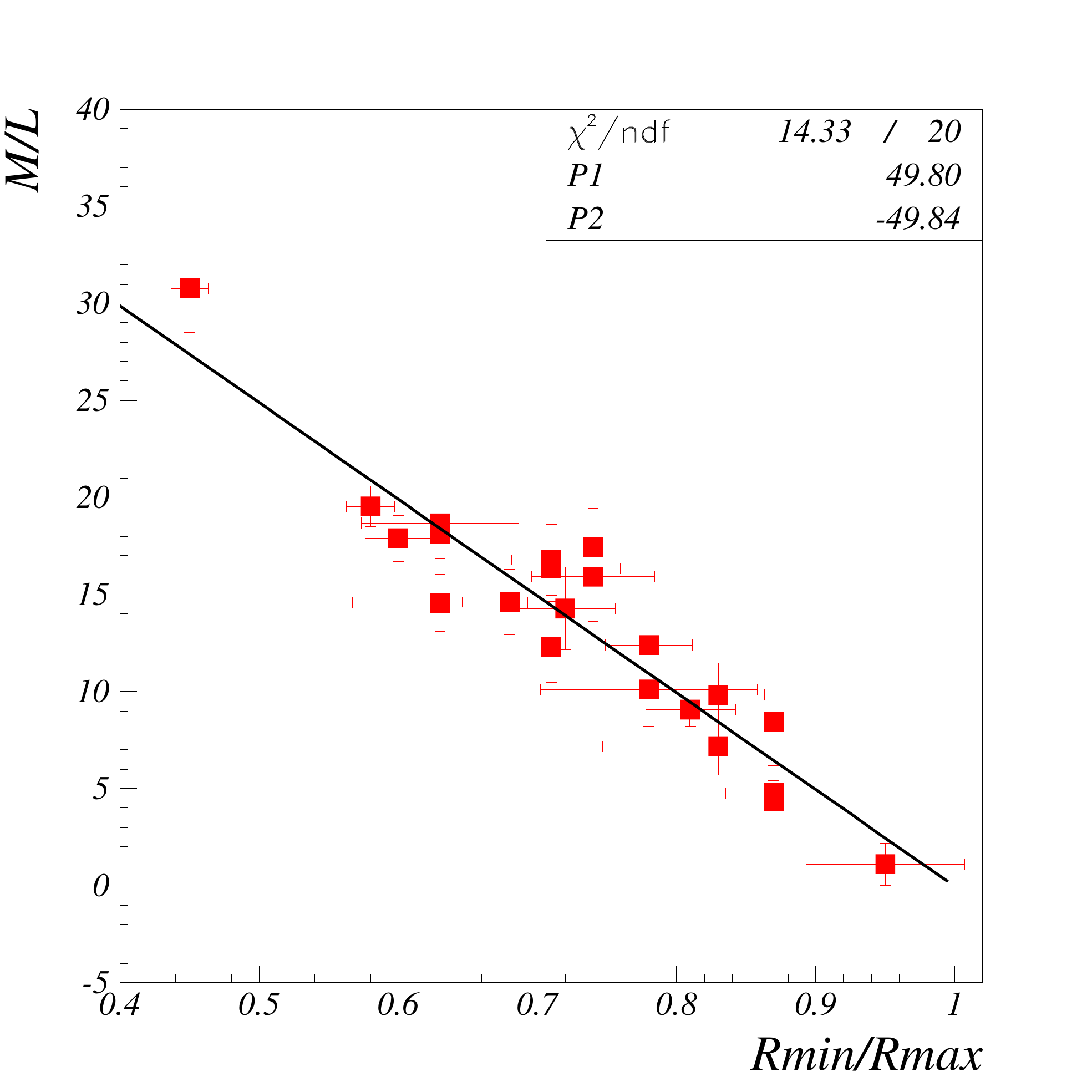}
\vspace{-0.4cm} \caption{Left plot: Mass to light ratio $\sfrac{M}{L}$ in function of the galaxy axis
ratios after removing the galaxies with uncertainties $\Delta \sfrac{M}{L}>5$. The Pearson
correlation coefficient is -0.87. Right plot: same as left but after
removing the galaxies with uncertainties $\Delta \sfrac{M}{L}>3$. The Pearson correlation
coefficient is -0.91. (Here, the values for $\Delta \sfrac{M}{L}$ refer
to before applying the correction for the ellipticity projection.)
}
\end{figure}

The final result including the projection correction, the Hubble parameter correction and
the exclusion of bin $32\leq DM<33$ but without applying the distance
modulus bias correction (that would enhance further the $\sfrac{M}{L}$ dependence
with $\sfrac{R_{min}}{R_{max}}$) is shown in Fig.~\ref{fig:ML vs real axis ratio}.
Also excluded are the possible corrections discussed in Figs.~\ref{fig:dmbms_dmnednox},
\ref{fig:btbms_bted} and~\ref{fig:ml1_apar_noz} (this is a conservative choice
since this would enhance further the $\sfrac{M}{L}$ dependence with $\sfrac{R_{min}}{R_{max}}$).
Fig.~\ref{fig:ML vs real axis ratio} can be compared to Figs.
\ref{fig:ml1_apar}, \ref{Flo:ML vs R/R after DM study} and~\ref{Flo:ML vs R/R after Hubble correction}
that show the different stages of the analysis. The best linear fit
yields $\sfrac{M}{L}=-(43.5\pm3.2)(\sfrac{R_{min}}{R_{max}})+46.7\pm2.4$, with a $\sfrac{\chi^{2}}{ndf}=0.9$.
This is a strong positive signature with a 13$\sigma$ signal. The
Pearson correlation coefficient is 0.87, or 0.91 after removing the
galaxies with (uncorrected) uncertainty above $\Delta \sfrac{M}{L}=5$ or
$\Delta \sfrac{M}{L}=3$ respectively, see Fig.~\ref{fig:ml vs real axis ratio for pearson cc}.
In all cases, it indicates a strong correlation. Again, it is interesting
to notice that if we select the highest precision data, then the $\sfrac{M}{L}$
vs $\sfrac{R_{min}}{R_{max}}$ correlation is enhanced, both for the determination
using of the Pearson criterion and for the determination from the
linear fit parameter $p_{2}$.

\begin{thebibliography}{106}
\bibitem{w/o selection }Without such selection criteria, the best
linear fit of the~\cite{BMS} data, as discussed in Fig.~\ref{fig:ml1_apar}, to the full
data set has no significant slope: $1.9\pm1.4$. This is understood as  coming from
the additional jitter caused by the wider characteristics
of the galaxies and by systematic biases. For example, a systematic
bias stems from galaxies now known to be disc galaxies. The $\sfrac{M}{L}$ of those 
mis-identified galaxies is wrong because the virial method does not  apply. 
Their $\sfrac{M}{L}$ is generally very
small, e.g. 3.6 for the SA0 galaxy NGC6909, but their apparent $\sfrac{R_{min}}{R_{max}}$
is likely to be small e.g. 0.5 for NGC6909. Thus, these disc galaxies
introduce an unphysical  bias to the correlation and should be excluded.
\end{thebibliography}
\end{document}